\documentclass[10pt,twoside]{book}
\usepackage[twoside,hcentering,dvips,nofoot,ignorehead]{geometry}
\usepackage{amsmath}
\usepackage{amssymb}
\usepackage{amscd}
\usepackage{fleqn}
\usepackage{graphics}
\usepackage{pifont}
\usepackage{latexsym}
\usepackage[small,rm]{caption}
\usepackage{graphicx}
\usepackage{boxedminipage}
\usepackage{pstricks}
\usepackage{pst-node}
\usepackage{multicol}
\usepackage{subfigure}
\usepackage{amsfonts}
\usepackage{amsthm}
\usepackage{fancyhdr}
\usepackage{dsfont}
\usepackage{bm,epic,eepic}
\usepackage[mathscr]{euscript}
\usepackage{booktabs} 
\usepackage{textcomp}
\usepackage{epsfig}
\usepackage{epstopdf}

\usepackage{natbib} 
\newcommand{\ara}[1]{#1}
\newcommand{\clearemptydoublepage}{\clearpage{\pagestyle{empty}\cleardoublepage}}
\newcommand{\openone}{\leavevmode\hbox{\normalsize1\kern-3.8pt\large1}}
\newcommand{\BEQ}{\begin{eqnarray}}
\newcommand{\EEQ}{\end{eqnarray}}
\newcommand{\ENQ}{\end{eqnarray}}
\newcommand{\BE}{\begin{eqnarray}}
\newcommand{\EE}{\end{eqnarray}}
\newcommand{\bq}{\begin{quote}}
\newcommand{\eq}{\end{quote}}
\newcommand{\nn}{\nonumber}
\newcommand{\Tr}{\mbox{Tr~}}
\newcommand{\inhoud}[2]{\hbox to #1{\hss #2 \hss}}
\newcommand{\forget}[1]{}
\newcommand{\ket}[1]{| \, #1 \rangle}

\newcommand{\bra}[1]{\langle \, #1  |}
\newcommand{\beq}{\begin{equation}}
\newcommand{\eeq}{\end{equation}}
\newcommand{\be}{\begin{equation}}
\newcommand{\ee}{\end{equation}}
\newcommand{\enq}{\end{equation}}

\newcommand{\1}{\mathds{1}}

\newcommand{\av}[1]{\langle #1 \rangle}
\newcommand{\Eq}[1]{(\ref{#1})}

\newcommand{\half}{\frac{1}{2}}
\newcommand{\goesto}{\longrightarrow}
\newcommand{\equivto}{\Longleftrightarrow}
\renewcommand{\vec}[1]{\mathbf{#1}}

\renewcommand{\H}{\mathcal{H}}

\def\be{\begin{equation}}
\def\ee{\end{equation}}
\def\eea{\end{eqnarray}}
\def\bea{\begin{eqnarray}}

\newcommand{\integer}[1]{\lceil #1 \rfloor}
\newcommand{\C}{\mathbb{C}}

\usepackage[dutch,american,activeacute]{babel} 

\renewcommand{\cite}{\citep}

\usepackage{thesisArxiv}    

\usepackage{caption}

\bibpunct{[}{]}{;}{a}{,}{,} 

\newcommand{\captionfonts}{\small}
\makeatletter  
\long\def\@makecaption#1#2{%
  \vskip\abovecaptionskip
  \sbox\@tempboxa{{\captionfonts #1: #2}}%
  \ifdim \wd\@tempboxa >\hsize
    {\captionfonts #1: #2\par}
  \else
    \hbox to\hsize{\hfil\box\@tempboxa\hfil}%
  \fi
  \vskip\belowcaptionskip}
\makeatother   

\newcounter{pagebackup}
\makeatletter
\newcommand*\std@endpart{}
\let\std@endpart\@endpart
\renewcommand*\@endpart{%
\setcounter{pagebackup}{\value{page}}%
\thispagestyle{empty}\std@endpart%
\setcounter{page}{\value{page}}}
\makeatother

\begin{document}

\frontmatter
\pagenumbering{roman}
\selectlanguage{american}

\pagestyle{empty}~\vskip2cm
\begin{center}
\includegraphics[scale=0.88]{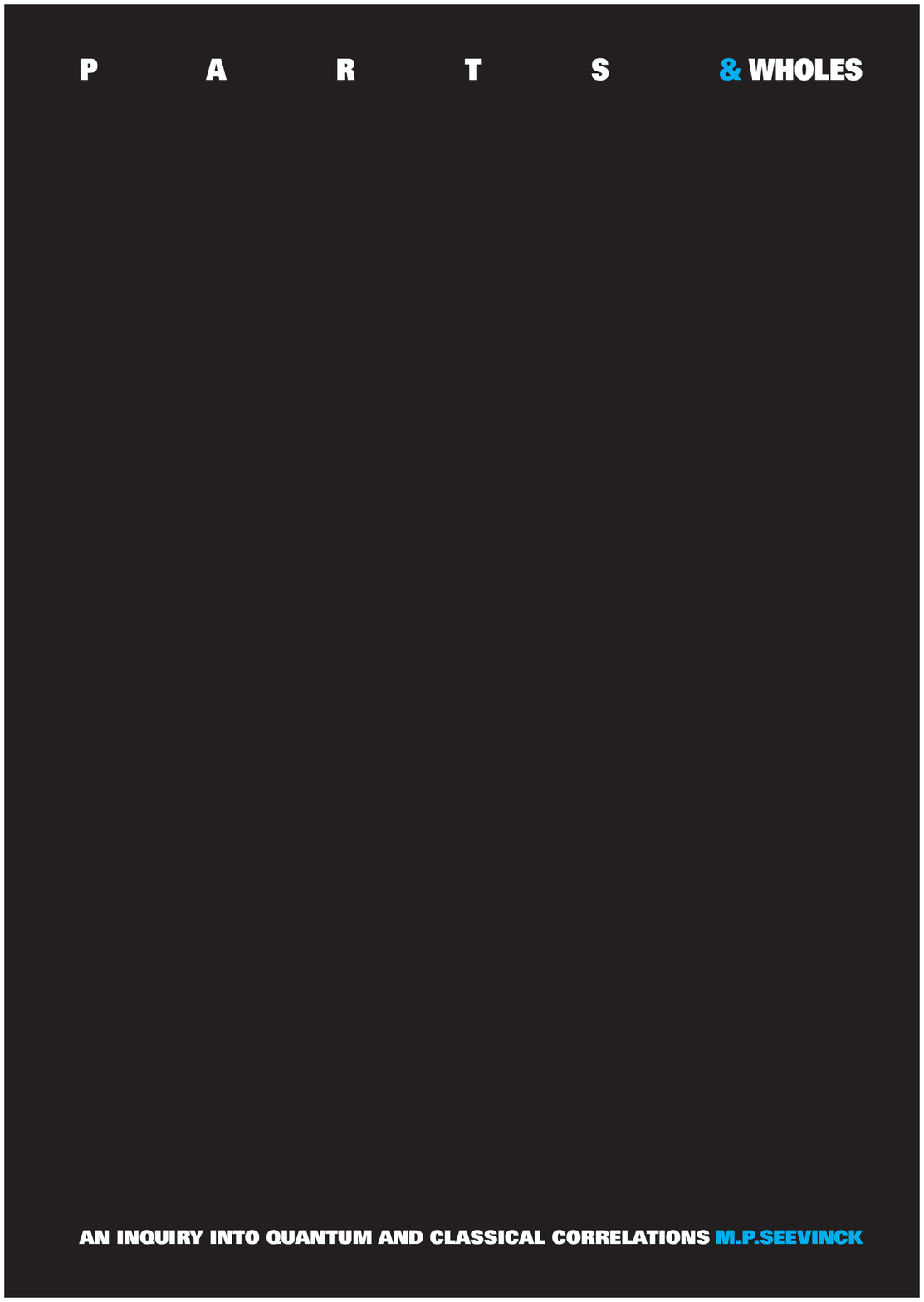}
\end{center}
%
  \cleardoublepage
\pagestyle{empty}
\noindent~\vskip2cm\noindent
\begin{center}{\huge{{\bf{{\sc{Parts and Wholes}}}}}}
\end{center}
\vskip0.7cm  
\noindent
\begin{center}
{\Large{\bf \sc{An Inquiry into\\ Quantum and Classical Correlations }}}
\end{center}

\forget{
\vskip4cm  
\noindent{{\small Alternative title:}}\\\\
\noindent{\Large \bf On correlations and relations between parts and whole in quantum and classical physics}
}

   \vskip11cm
     
   \noindent{\large{{{\bf{\Large \begin{center} M.P. Seevinck \end{center}}}

%
%
%
%
\newpage
\pagestyle{empty}
\noindent
~\vskip2cm\noindent
{\bf \normalsize Note on the different arXiv versions:}\\\\\  
{\normalsize This version arXiv/0811.1027v3  (24 April 2009)  has exactly the same content as the version arXiv/0811.1027v2 (7 November 2008). However, the page layout has been changed so that it is the same as the distributed hard copy version of the Dissertation which is on B5 format.}
   ~
\noindent
\vskip7cm
\noindent 
\emph{Colofon}\\\\
{\normalsize
Financial support by the Institute for History and Philosophy of \mbox{Science}, Utrecht University. \\\\
Copyright \copyright\ 2008  by Michael Patrick Seevinck. All rights reserved.\\\\
Cover design by Ivo van Sluis.\\
Printed in the Netherlands by PrintPartners Ipskamp, Enschede. \\
Printed on FSC certified paper.\\\\
ISBN 978-90-3934916-8
}

\pagestyle{empty}
\thesistitle  {Parts and Wholes\titlebreak}
\thesisundertitle {An Inquiry into Quantum and Classical Correlations\titlebreak}
\translatedthesistitle  {Delen en Gehelen\titlebreak}
\translatedthesisundertitle{Een Onderzoek naar Quantum en Klassieke Correlaties\titlebreak}
\dday         {maandag 27 oktober 2008}
\daypart      {des middags}
\hhour        {4.15}
\rector       {prof.dr. J.C. Stoof}
\myfullname   {Michael Patrick Seevinck}
\dateofbirth  {27 februari 1977}
\placeofbirth {Pretoria, Zuid-Afrika}

\promotor     {\normalsize{Prof.dr.~D.G.B.J. Dieks}}
\copromotor   {Dr.~J.B.M. Uffink}

\coverdesign{Ivo van Sluis}
\myshortname{M.P. Seevinck}
\ISBN{978-90-3934916-8}
\support{IGG}
\printer{PrintPartners Ipskamp, Enschede}

\makeopeningpage

\pagestyle{empty}

        \cleardoublepage
        %
        %
        %
        $~$
\noindent

\vskip4cm

\noindent
\begin{quote}
Now it is precisely in cleaning up 
intuitive ideas for mathematics
that one is likely to throw out the 
baby with the bathwater.\\

$~$\hskip1cm \emph{J.S. Bell\,;  `La nouvelle cuisine', 1990.}
\end{quote}
     \cleardoublepage
     \pagestyle{plain}
\chapter*{Preface}

Not to be found in this dissertation is a love story -- the story of the genesis of this dissertation.
Just like any love story it cannot but be a tragic one. Full of happiness and despair, joy and sorrow. 
I believe a few words about this story are in order here. 

The story began with love at first sight, but it took many years for this to become a true love and develop into a somewhat steady relationship. 
This rather slow start is due to the circumstance that when I started with the current project I was rehabilitating from a previous similar love affair, and this made me hesitant and rather uncertain of how to proceed. 
But luckily, things changed.  The great intellectual freedom I granted myself,  and which was also made possible by the \emph{carte blanche} handed to me 
at the beginning 
of the project, 
provided ideal circumstances for falling in love,  
and thus for rediscovering the truth-lover 
inside of me.
%
The present dissertation stems from that love of truth -- but this occurred not without difficulty, be it mentioned.

The intellectual freedom may further explain the fact that this dissertation is not concerned with a single research question,  but with a handful of different, though related subjects, and also that 
my love produced many ideas, only some of which turned out to be promising, whereas many 
were in fact utter nonsense. 
The latter is not to be regretted for I am convinced that a love of truth can only really be creative 
when it does not fear misfortune, 
 nor mistake or confusion.
%

I invite the reader to try and taste some of the fruits of this love.  I believe -- and sincerely  hope -- that some may taste delicious, but I realize that   others 
may very well be rather bland and tasteless. 


With these final words an episode has come to an end. Fortunately, the story continues --- forever learning how to truly love.

\begin{flushright} 
M.P. Seevinck\\
Nijmegen, September 2008
\end{flushright} 
%
%
%
%
\pagestyle{empty}
  \cleardoublepage
\pagestyle{plain}
%
%
%
%
%
\chapter*{Acknowledgements}
This dissertation has benefitted from advice of a number of people, and many helped to complete it. I would like to take the opportunity here to express my thanks and gratitude to them. 

First of all, I am grateful to my co-authors and especially to Jos Uffink, together with whom five papers have appeared.  As regards the contents of this dissertation the greatest debt by far is to Jos.  Not only is Jos is responsible for getting me interested in the foundations and philosophy of physics in the first place, he also taught me most of what I know.  
 As a dissertation adviser Jos was always available to sound out my arguments, sharpen my understanding of the pertinent issues, and stimulate me to express my ideas clearly. He never hesitated to share his insights with me. Jos and I have worked together very closely for the past years and this gave me the opportunity not only to benefit from his clear thinking but also to learn from his great style of writing. 

I would also like to thank my \emph{promotor} Dennis Dieks for being so patient with me 
and supporting me in many ways. I'm especially grateful for the editorial position with the journal Foundations of Physics that Dennis arranged for me and which allowed me to earn a living\forget{necessary money},  while maintaining enough freedom to write this dissertation. I am further grateful to my colleagues in Utrecht, and especially to Fred (F.A.) Muller for sharing numerous hotel rooms with me to cut down on costs as well as for greatly enhancing my interest in analytic  philosophy and philosophy of science and for kindly helping me enter this academic field;  to Jeroen van Dongen for convincing me to write this dissertation at all; and to Remko Muis for being a great office mate during some long four years. 

The willingness of Prof.dr.~F.~Verstraete (Vienna), Prof.dr.~H.R.~Brown (Oxford), Prof.dr. ~I. ~Pitowsky (Jerusalem), Prof.Dr.~R.D.~Gill (Leiden) and Prof.dr.\ N.P.~Landsman (Nijmegen) to be my examiners and to accept this dissertation for doctorate  is appreciated with honour and gratitude.

Thanks are also due to various people with whom I had fruitful scientific correspondence and discussion to sharpen certain points of my analysis. I especially want to thank Sven Aerts, Christian de Ronde, George Svetlichny, Chris Timpson,  Harvey Brown, Jeremy Butterfield, Marek \.Zukowski,  Geza T\'oth, Otfried G\"uhne, Hans Westman, Victor Gijsberts, Jon Barrett  and N. David Mermin. 
Most of the work described in this dissertation has already been published\forget{judged by peers} and I want to mention the anonymous reviewers of my papers for their valuable input, and the editors that accepted my submissions for their efforts. And I extend my thanks to the organizers of the many conferences I attended, as well as to the audiences that were present at the talks I presented for putting up with my ideas, which at that stage were most likely only half-baked.

I want to thank Klaas Landsman for a providing a \emph{pied a terre} for me in the old science building at the Radboud University Nijmegen. After that building had been demolished Harm Boukema kindly gave me the opportunity to use his room at the Philosophical Institute in Nijmegen for almost two years -- \emph{sans papiers}. Not only am I very grateful to him for providing me an office so close to home to work in, but also for his inspiration and moral support.
I also very much enjoyed the company of Luuk Geurts during the time I spent there on the 16$^{\textrm{th}}$ floor. 
 In addition, I had the opportunity to enjoy a very pleasant research stay at the Perimeter Institute in Waterloo, Canada. Their hospitality to host me as a short term visitor is very much appreciated, and I thank Owen Maroney for inviting me to come to PI in the first place. 


The staff of the former physics library have helped tremendously in obtaining the literature I requested. Nieneke Elsenaar deserves to be mentioned separately for her amazing ability to find requested documents that cannot be found, and for an occasional cup-a-soup.  Mentioning food, I must thank Tricolore for necessary pizzas, \emph{Diavola con gorgonzola}, which I have come to regard as the best of the Netherlands.  Even more important for my physical well-being (except for the occasional injury) has been Obelix. They provided great and necessary stress relief on the rugby pitch and amazing comradeship.
\\\\
I am grateful to my parents, family and friends. Jochem and Maarten deserve to be especially mentioned because of the gifts of great friendship; needed in general, but also in completing this dissertation.  Special thanks to Wim. \\\\
\noindent
It gives me great pleasure to dedicate this work to my beloved Tineke.  For being there in the first place and believing in me, for loving support and welcome and necessary distraction. \forget{ I could not have done this without your love.}  I will never be able to thank you adequately for bearing this burden with me.

  \cleardoublepage
          \makeatletter \renewcommand{\@dotsep}{8}  \makeatother \tableofcontents
          
          \addtolength{\footskip}{0pt}
          
\mainmatter 
\pagestyle{thesisheadingsfull}

\selectlanguage{american}
\thispagestyle{empty}
 \part{Introduction}
\thispagestyle{empty}
\chapter{Introduction and overview}
\label{introduction_chapter}
Philosophy of physics encompasses many different sorts of enquiry. At one extreme, one encounters metaphysical investigations that make use of some facts or ideas  delivered to us by modern physics, but that are not of a technical nature. At the other extreme, one finds almost pure mathematical investigations that might have their original motivation in some philosophical question on some aspect of modern physics, but which in fact have as their sole purpose  to clarify the structure of some physical theory.
Both sorts of enquiry are essential for grasping the foundations of physics  \citep[][p. 1]{halverson}, though they are not sufficient. For this to be the case, the results of both sorts of enquiry should meet somewhere and somehow. 

These enquiries have been part and parcel of \forget{the study of}the foundations of quantum \mbox{theory} right from the beginning, for example in the writings of two founding fathers of the \mbox{theory}: J. von Neumann and A. Einstein. Von Neumann gave quantum mechanics a mathematically rigorous structure whereas Einstein reflected upon the same theory in terms of  philosophical questions about the nature of physical reality and on \emph{a priori} requirements for doing any physics at all.
 Fortunately, the work of these two founding fathers met somewhere and somehow in the work of J.S. Bell when he produced his 1966 and 1964 masterpieces\footnote{Bell cites in both these works von Neumann's  monograph \cite{vonneumann} as well as Einstein's autobiographical notes and reply to critics from the Schilp volume \cite{schilp}.}. Two works that  paved the way for great progress  in the philosophy of quantum mechanics.  
  However, given the extreme sorts of enquiries that fall under the heading of philosophy of physics, as mentioned above, it is not surprising that some people in the field think Bell's work was not mathematical enough, whereas others would want a more philosophical and interpretational discussion.  But despite the fact that indeed more formal rigor was needed and more philosophy had to be done to fully appreciate Bell's insights, the spirit and style of Bell's work have been a leading example to me in writing this dissertation.
 
\forget{Maybe because of this, but}Therefore, I expect that similar complaints as those raised against Bell's work will also befall this dissertation. Some probably want more mathematics, others more philosophy.  However, with respect to the first, rest assured I will present sound results, although I do not survey all mathematical aspects completely, and with respect  to the second, I  give these results foundational and philosophical relevance, although\forget{we agree that} probably some of the philosophical fruits still need to be reaped, something I would like to pursue in the near future.\forget{I believe that even the most mathematical chapter (chapter 8) in this dissertation  contains a deep foundational insight, and that  even the most philosophical chapter contains a mathematical result  (chapter 11).} But above all, in cleaning up intuitive ideas for mathematics  I have striven for the right sort of balance of throwing only the water out while keeping the baby inside.

\section[Historical and thematic background to this dissertation]{Historical and thematic background to this\\ \mbox{dissertation}}

This dissertation derives from a series of eleven articles I wrote over the last few years,
jointly authored with J. Uffink, G. Svetlichny, G. T\'oth, and O. G\"uhne.
Most articles have already appeared in print and they are listed at the end of this dissertation.
What connects these articles and therefore the primary topic of this dissertation is,  firstly, the study of the correlations between outcomes of measurements on the subsystems of a composed system as predicted by a particular physical theory;  secondly, the study of what this physical theory predicts for the \mbox{relationships} these subsystems can have to the composed system they are a part of;  and thirdly, the comparison of different physical theories with respect to these two aspects. The physical theories I will investigate and compare are generalized probability theories in a quasi-classical physics framework and non-relativistic quantum theory.

The motivation for these enquiries is that a comparison of the relationships between parts and wholes as described by each theory, and of the correlations predicted by each theory between separated subsystems yields a fruitful method to investigate what these physical theories say about the world.\forget{
The motivation for performing these enquiries is that a fruitful method for comparing physical theories and to 
investigate what each of them say about the physical world, is through comparing the relationships between parts and whole as described by each theory, as well as through the characteristics of the correlations predicted by each theory among separated (in some specified sense, e.g., spacelike) observations on the parts of the whole.
} One then finds, independent of any physical model, relationships and constraints\forget{between parts and whole and  for the correlations involved,} that capture the essential physical assumptions and structural aspects of the theory in question.  As such one gains a larger and deeper understanding of the different physical theories under investigation and of what they say about the world.

Indeed, many enquiries in physics that have provided us such understanding are of this sort
\footnote{For example, Einstein's study of Mach's ideas about the origin of inertia and its alleged relationship to the far-away stars that eventually culminated in his relativity theories; or the study of the behavior of a few-body system as predicted by deterministic non-linear dynamics that gave rise to chaos theory.\forget{the discovery and explanation of nuclear fusion and fission.
	Kepler's study of the behavior of planets within our solar system that was later to be explained by Newton;}\forget{Kinetic gas theory and its phenomenological predictions such as phase transitions ingenious studies of remarkable properties of solids, fluids and gases such as fission, fusion, superfluidity and superconductivity, etc.}}, but many of the unresolved longstanding problems in physics are too\footnote{For example,\forget{the problem of how to reconcile time-symmetric statistical physics and time-asymmetric thermodynamics; the cosmological problem of the mismatch between total gravitational force in our galaxy and its matter density;  } the problem of how to account for the classical macro-world given the quantum micro-world.}.
Here I will use a famous example of such a problem from the history of the foundations of quantum mechanics that allows me to introduce further the background to this dissertation.  This problem was formulated in 1935 by \citet*{epr} who considered a \emph{Gedankenexperiment} (i.e., thought experiment) that bears the by now famous name of `the EPR argument'\footnote{Often referred to as `the EPR paradox', but this is a misnomer since no paradox whatsoever is proposed, but merely a sound \emph{Gedankenexperiment}. Let us incidentally note that Einstein himself seems to have preferred a simpler \emph{Gedankenexperiment}, but this discussion is not relevant for this dissertation. The full details of the EPR argument are not needed, the interested reader is directed to, for example, \citet{bub}.}. They attempted to show that quantum mechanics is incomplete.
The argument uses a \emph{reductio ad absurdum} (cf. \citet[p. 141]{brown}) whereby the completeness  of quantum mechanics can only be upheld if a form of non-locality or action-at-a-distance exists between spatially separated and non-interacting quantum systems. This is unacceptable, hence the claim must be false.
\forget{
The motivation for the argument is Einstein's realist philosophy of physics that contained two independence principles: a separability principle and a locality principle. These metaphysical principle are not part of quantum theory. The argument is set up to demonstrate an inconsistency between these two principles, a 'sufficient criterion for an element of physical reality', a necessary criterion for when a physical theory is complete and certain structural and interpretative principles of quantum theory such as the projection postulate. 
}

This argument was promptly countered by Bohr in a reply that is well-known for its difficult and unclear reasoning, and which could even be read as a refusal to accept the problem.  Nevertheless his argument effectively persuaded the majority of physicists -- they went back to business -- and this closed the classic era of debate and discussion between Einstein and Bohr.  Bohr was declared the winner, resulting in nearly thirty years of silence where Copenhagen orthodoxy reigned\footnote{A noteworthy exception is the important work by \citet{bohm52} that \citet[p. 990]{bell82} later referred to as:``But in 1952 I saw the impossible done. It was in the papers by David Bohm."
}. Another factor responsible for this was von Neumann's 1932 proof of the `no-go theorem' for introducing a more complete specification of the state of a  system than that provided by quantum mechanics \cite{vonneumann}.  It was thought by the majority at the time that this proved the impossibility of so-called hidden variables in quantum mechanics once and for all\footnote{However, an interesting exception is Grete Hermann who published in 
 1935 an argument that criticized a crucial assumption upon which von Neumann based his proof. The interested reader is directed to the English translation of her work which can be found at: {\footnotesize{\tt http://www.phys.uu.nl/igg/seevinck/gretehermann.pdf}}. This criticism seems to have gone largely unnoticed at the time. Thirty years later \citet{bell66} criticized this same assumption of von Neumann, although using a different argument.}.

A new phase in the history of the foundations of quantum mechanics started in the mid-1960s when \citet{bell66} examined the von Neumann proof carefully to see what it had \mbox{exactly} established. He famously exposed its defect and also examined the defects in other proofs that purported to have the same impact.  In this  review paper he also showed a contradiction for non-contextualist hidden-variable theories describing single systems associated with state spaces of dimension greater than two, thereby anticipating\footnote{For this reason \citet{brown} prefers to refer to this as the 'Bell-Kochen-Specker paradox' instead of the 'Kochen-Specker paradox'; the latter being the term generally used in the literature.} the more well-known Kochen-Specker theorem \cite{kochenspecker}. 
Bell also showed in detail how Bohm's hidden-variable model of the early 1950s actually worked and how it circumvented the `no hidden-variable theorems':  by being non-local, i.e., by incorporating a mechanism whereby the arrangement of one piece of apparatus may affect the outcomes of distant measurements. He next went on to examine whether ``any hidden-variable account of quantum mechanics must have this extraordinary character'' \cite[p. 452]{bell66}). \citet{bell64} answered his own question positively by proving his by now famous inequality that was used to prove that any deterministic local hidden-variable theory must be in conflict with quantum mechanics. (The 1966 paper was submitted before the 1964 paper.) \citet[p.141]{brown} puts this state of affairs strikingly as follows: ``The \emph{absurdum} [i.e., a form of non-locality. MPS] can not be avoided, even when the completeness thesis is relaxed and the possibility of `hidden variables' of the deterministic variety is entertained''.  

After the 1964 inequality variants of Bell's inequality were obtained that generalize his result that a local hidden-variable account of quantum mechanics is impossible, most notably the Clauser-Horne-Shimony-Holt (CHSH) inequality \cite{chsh} and Clauser-Holt inequality \cite{ch}. Then in 1981 \citet{aspect} performed an experiment using photons emitted by an atomic cascade that many took as providing conclusive evidence for Bell's theorem because it  showed a violation of the CHSH inequality, although it was soon realized that loopholes remained.  

In the mid-eighties the plot thickened when \citet{jarrett} showed that two conditions together imply the factorisability  condition (that Bell had called Local Causality) and which was used in deriving the CHSH inequality. \citet{shimony} used two related variants of the conditions that are now well-known under the name of Outcome and Parameter Independence. This carving up of the factorisability condition led to a new activity under the name of experimental metaphysics were it was argued that  Outcome Independence should take the blame in violations of Bell-type inequalities and that this was not `action-at-a-distance' but merely `passion at the distance' (or because of some other newly devised metaphysical circumstance), thereby allowing for peaceful coexistence  between relativity theory and quantum mechanics.

A new line of research in the study of this `quantum non-locality' was introduced in the late 1980's. Responsible for this was not sophisticated philosophical analysis but further technical results in the study of what \citet[p. 823]{schrodinger} had called  \emph{Verschr\"ankung} back in 1935 in his reply to the EPR paper and which we now know as quantum entanglement.  It had long been realized that these `spooky correlations'\footnote{In a letter to M. Born dated March 3rd 1947 \citet[p. 158]{einsteinborn} first coined the term \emph{spukhafte Fernwirkungen} for such correlations.} are responsible for violations of Bell-type inequalities and they were philosophically interpreted to be of a holistic character where the whole is more than the sum of the  parts \cite{teller86,teller87,teller89,healey91}. But it turned out that much of the structure and nature of entanglement was still to be discovered. Indeed, only as late as 1989 Werner gave the general definition of this concept as we use it now \cite{werner}. He also obtained the surprising result that local hidden-variable models exist for all measurements on some entangled bi-partite states.  In the same year a new type of Bell-theorem appeared: the so-called Greenberger-Horne-Zeilinger argument against local hidden-variable theories \cite{ghz,GHZ}. It uses a three-partite entangled state and used only perfect correlations not needing a Bell-type inequality. Inspired by this result \citet{mermin} derived the first multi-partite Bell-type inequality. Quantum mechanics violates this inequality by an exponentially large amount for increasing number of parties.  These results initiated a whole new field of study:  that of entanglement and its relation to Bell-type inequalities, both for bi-partite and multi-partite scenarios.

A second line of research started at about the same time with the work of \citet{bb84} and \citet{deutsch} who showed that 
quantum systems can be used as remarkable computational machines, and a few years later \citet{ekert} showed that violations of Bell-type inequality by entangled states can be used for quantum cryptography. This marked a paradigm change where entanglement was no longer seen as mysterious (e.g., some `spooky correlation') but as a resource that can be used for computational and information theoretic tasks. Using entanglement one can perform many such tasks more efficiently than when using only classical resources, and some such tasks are even impossible when using only classical resources. Examples include quantum computation, superdense coding, teleportation and quantum cryptography (cf. \citet{nielsen}). 

Since entanglement was central to both these new fields of research, we could welcome the marriage between quantum foundations and quantum information theory in the 1990s. This marriage has produced a lot of fruitful offspring in the last 15 years or so. It would be too much to discuss all of this, so I will only highlight the new research themes relevant for this dissertation.

(I) Bell-type inequalities have come to serve a dual purpose. Originally, they were designed in order to answer a foundational question: to test the predictions of quantum mechanics against those of a local hidden-variable  theory. However, these inequalities have been shown to also provide a test to distinguish entangled from separable (unentangled) quantum states. This problem of entanglement detection is crucial in all experimental applications of quantum information processing.  However, the gap pointed out by Werner between quantum states that are entangled and those that violate Bell-type inequalities shows that violations of Bell-type inequalities, while a good indicator for the presence of  entanglement in some composite system, by no means captures all aspects of entanglement.  \citet{POPESCU} was the first to show that this gap could be narrowed by showing that local operations and classical communication can be used to `distill' entanglement that once again suffices to violate a Bell-type inequality. However, even today the gap has not been closed completely. 
 Therefore, entanglement has been studied via other means such as non-linear separability inequalities, entanglement witnesses, 
 and many different kinds of measures of entanglement (see, e.g., the recent review paper by \citet{entanglement}).
 
(II) There has been a renewed interest in the ways in which quantum mechanics is different from classical physics.  This originated from 
the realization that in order to increase understanding of quantum mechanics, it is fruitful to distinguish it, not just from classical physics, but from non-classical theories as well. So one started to study quantum mechanics `from the outside' by demarcating those phenomena that are essentially quantum, from those that are more generically non-classical.\forget{ Investigate theories that are neither classical nor quantum; explore the space of possible theories. \`Is quantum mechanics an island in theory space?'' (Aaronson, 2004). If indeed so, where is it? What are its coordinates? A different question: what is essentially and uniquely quantum? See A. Grinbaum, BJPS Vol. 58: 387-408 (2007).} I will highlight three such investigations relevant for this dissertation: 
\begin{enumerate}
\item{} The study of non-local no-signaling correlations. This research started with Popescu and Rohrlich's question ``Rather than ask why quantum correlations violate the CHSH inequality, we might ask why they do not violate it more.''\cite[p. 382]{prbox}. Here one investigates correlations that are stronger than quantum mechanics yet that are still no-signaling and thus do not allow for any spacelike communication. Surprisingly, such  correlations can violate the CHSH expression up to its absolute maximum. But their full characteristics are still being investigated.
\forget{no signaling correlations questions of sharing/monogamy of the correlations (relevant for quantum cryptography)  
New insights in quantum foundations.}
\item{} The study of the classical content of quantum mechanics. For a long time it was thought that the question what the classical content of quantum mechanics is was answered by the distinction between separable and entangled states: separable states are `classical', entangled states are `non-classical', and the same was thought of the correlations in such states.  However, not only is it unclear whether all entangled states must be regarded as non-classical (as we have seen\forget{not all of them can be used to violate a Bell-type inequality  and} the correlations of some entangled states can have a local hidden variable -- and therefore arguably a classical  -- account)   \citet{groisman} even argued for `quantumness' of separable states. For example, they show how to obtain quantum cryptography using only separable states.  
 \item{} 
 Providing interpretations of quantum mechanics. In the last two decades we have been witnessing 
 a renewed interest in both improving existing interpretations of quantum mechanics as well as providing new ones. 
 The results of the study of entanglement and quantum information theory play a great role  in this revival  and 
two different kinds of interpretational study can be distinguished. 
  
   The first kind deals with  (i) investigating traditional interpretations such as modal interpretations, Everett's many worlds interpretation and Bohmian mechanics,  and (ii) providing new ones of a similar character such as so-called Quantum Bayesianism \cite{caves} and  the Ithaca interpretation \cite{merminithaca1,merminithaca2,merminithaca3}.
 
 The second kind has a different character  and is best characterized as reconstruction of quantum mechanics  \cite{grinbaum}.\forget{ Theorems and major results of physical theory are formally derived from simpler mathematical assumptions. These assumptions or axioms, in turn, appear as a representation in the formal language of a set of physical principles.}  Reconstruction consists  of three stages: first, give
a set of physical principles, then formulate their mathematical representation, and finally rigorously derive the formalism of the theory.
  As a result of advances in quantum information theory most of these reconstructions have used  information-theoretic foundational principles such as the  Clifton-Bub-Halverson reconstruction \cite{cbh}.

\end{enumerate}
In this dissertation I will contribute to research in the areas mentioned above under (I) and (II). However, I will not provide a conclusive analysis in any of these areas;\forget{ these research areas are still under great investigation.} this dissertation provides many new results, but it leaves us also with a lot of open questions.

\forget{
 To end this historical introduction it is tempting to speculate on future developments. 
 But I resist this temptation and merely mention that he last word has not been spoken. Indeed this dissertation provides us a lot of new results, but it leaves us also with a lot of open questions.
}

\section{Overview of this dissertation}
 
To give the reader a better idea of what can be found in this dissertation, I will give a short outline. In the next chapter, {\bf chapter \ref{definitionchapter}}, I will present  the necessary definitions, concepts and mathematical structures that will be used in later chapters. 
Most importantly, the definitions of four different kinds of correlation (local, partially-local, no-signaling and quantum mechanical) are presented as well as tools that will be used to discern them. 

 Throughout this chapter it is more precisely indicated what technical results are to be found in this dissertation. 
Here a less technical overview is given  that focuses on the issues involved, as well as on the foundational impact of the results that will be obtained. 
However, because, on the one hand I have not concerned myself with one central question, but rather with many different topics within the same field, and on the other hand many new results have been obtained instead of a few major conclusions, this introduction must necessarily be rather brief and cannot go too much into depth.

In {\bf part II}, I limit my study to systems consisting of only two subsystems where I consider correlations between outcomes of measurement of only two possible dichotomous observables on each subsystems. This is the simplest relevant situation; but the structure of the correlations that one can find for such a scenario is far from being completely understood. {\bf Chapter \ref{chapter_CHSHclassical}} investigates 
the well-known CHSH inequality for such bi-partite correlations.  I first review the fact that the doctrine of Local Realism with some additional technical  assumptions allows only local correlations and therefore obeys this inequality.  
It is then shown that one can allow for dependence of the hidden variables on the settings (chosen by the different parties) as well as explicit non-local setting and outcome dependence in the determination of the local outcomes of experiment, and still derive the CHSH inequality.  
Violations of the CHSH inequality thus rule out a broader class of hidden-variable models than is generally thought. 
 Some other foundational consequences of this result are also explored.\forget{
 
As a result of this the conditions of Outcome Independence and Parameter Independence \cite{shimony}, that taken together imply the well-know condition of Factorisability used in the derivation of the CHSH inequality, can both be violated in deriving this inequality. In the light of experimental violations of the CHSH inequality I argue that this shows that the question of whether it is Outcome Independence or Parameter Independence that should be abandoned is misplaced.  We have no reason to expect either one of them to hold solely on the basis of the CHSH inequality because they are shown not to be necessary conditions. Instead, we should look at the weakest set of assumptions that give this inequality and ask which one of these is to be abandoned. I argue that this undermines the activity called experimental metaphysics. 
}

Further, the relationship between two sets of conditions, those of \citet{jarrett} and of \citet{shimony} is commented upon. Each set implies a certain form of factorisability of joint probabilities for outcomes that is used in derivations of the CHSH inequality. It is argued that those of Jarrett are more general and more natural. I furthermore comment on the non-uniqueness of the Shimony conditions that give factorisability by proving that the conjunction of a third set of conditions, those of \citet{maudlin}, suffice too. This has been claimed before, but since no proof has been offered in the literature I provide one myself.  In order to be evaluated in quantum mechanics it is shown that the Maudlin conditions need supplementary non-trivial assumptions that are not needed by the Shimony conditions.  It is argued that  this undercuts the argument that one can equally well chose either set (Maudlin's or Shimony's). 

The non-local derivation of the CHSH inequality is compared to Leggett's inequality \cite{leggett} and Leggett-type models, which have recently drawn much attention.  The analysis and discussion of Leggett's model shows surprising relationships between different conditions at different hidden-variable levels. It turns out that which conditions are obeyed and which are not depends on the level of consideration and thus on which hidden-variable level is taken to be fundamental. This study is extended to also include the so-called surface level, where one does not consider any hidden variables.
\forget{This study is extended to also include surface probabilities. Further analogies between different inferences that can be made on the level of subsurface and surface probabilities are presented.  An interesting corollary is that any deterministic hidden-variable theory that obeys no-signaling and gives non-local correlations must show randomness on the surface, i.e., the surface probabilities cannot be deterministic. Bohmian mechanics is a striking example of this. }
 I also investigate bounds on the no-signaling correlations in terms of Bell-type inequalities that use both product (joint) and marginal expectation values. After showing that an alleged no-signaling  Bell-type inequality as proposed by \citet{roysingh} is in fact trivial (it holds for any possible correlation), a new set of non-trivial no-signaling inequalities is derived.

In {\bf chapter \ref{chapter_CHSHquantumorthogonal} and \ref{chapter_CHSHquantumtradeoff}} I consider many aspects of the CHSH inequality in quantum mechanics for the case of two qubits (two level systems such as spin-$\frac{1}{2}$ particles). 
 This inequality not only allows for discerning quantum mechanics from local hidden-variable models, it also allows for discerning separable from entangled states.  In chapter 4, significantly stronger bounds on the CHSH expression are obtained for separable states in the case of locally orthogonal observables, which, in the case of qubits, correspond to anti-commuting operators.  Some novel stronger inequalities -- not of the CHSH form -- are also obtained.  These new separability inequalities, which are all easily experimentally accessible,  provide stronger criteria for entanglement detection and they are shown to have experimental advantages over other such criteria.\forget{For the case of pure states a set of only six inequalities is indicated that already give a necessary and sufficient criterion for entanglement.}
 
Chapter \ref{chapter_CHSHquantumtradeoff}, the condition of anti-commutation (i.e., orthogonality) of the local observables is relaxed. Analytic expressions are obtained for the tight bound on the CHSH inequality for the full spectrum of non-commuting observables, i.e., ranging from commuting to anti-commuting observables, for  both entangled and separable qubit states. These bounds are shown to have experimental relevance, not shared by ordinary entanglement criteria, namely that one can allow for some uncertainty about the observables one is implementing in the experimental procedure. 
 
 The results of these two  chapters also have a foundational relevance    because these separability inequalities turn out to be not to applicable to the testing of local hidden-variable theories.  This provides a more general instance of Werner's (1989) discovery that  some entangled
two-qubit states allow a local realistic model for all correlations in a standard Bell experiment. 
 In chapter \ref{Npartsep_entanglement} this discrepancy between  correlations allowed for by local hidden-variable theories and those achievable by separable qubit states is shown to increase exponentially with the number of particles.  
It seems that the question what the classical correlations of quantum mechanics are, has still not been resolved.
\forget{
I exhibit a `gap' between the
correlations that can be obtained by separable two-qubit quantum states and
those obtainable by local hidden-variable models.  In fact, as will be shown shown in chapter \ref{Npartsep_entanglement}, the gap between the correlations allowed for by local hidden-variable theories and those achievable by separable qubit states increases exponentially with the number of particles. Therefore, local hidden-variable theories are able to give correlations for which quantum mechanics, in order to reproduce them in qubit states, needs recourse to entangled states; and even more and more so when the number of particles increases. 

===========
}

In {\bf part III} I extend the investigation to the multi-partite case which turns out to be non-trivial. Indeed, when making the transition from two to more than two parties, one finds that almost always an unexpected  richer structure arises. Again I restrict myself to the simplest case of two dichotomous observables per party, but this already gives a lot of new results.

{\bf Chapter \ref{Npartsep_entanglement}} investigates multi-partite quantum correlations with respect to their entanglement and separability properties.  
 A classification of partially separable states for multi-partite systems is proposed, extending the classification
introduced by \citet{duer2,duer22}.  This classification consists of a hierarchy of levels corresponding to different forms of partial separability, and within each level various classes are distinguished by specifying under which partitions of the system the state is separable or not. Partial separability and multi-partite entanglement are shown to be non-trivially related by presenting some counterintuitive examples. This  asks for a further refinement of the notions involved, and therefore the notions of a $k$-separable entangled state and a $m$-partite entangled state are distinguished and the interrelations of these kinds of entanglement are determined.

By generalizing the two-qubit separability inequalities of chapter \ref{chapter_CHSHquantumorthogonal} to the multi-qubit setting I obtain necessary conditions for distinguishing all types of partial separability in the full hierarchic separability classification.\forget{ Violations of the separability inequalities provide, for all $N$-qubit states, strong criteria for  the entire hierarchy of $k$-separable entanglement, ranging from the levels $k$=1 (full or genuine
$N$-particle entanglement) to $k=N$ (full separability, no entanglement), as well as for the specific classes within each level.
} These separability inequalities are all readily experimentally accessible and violations give strong criteria for different forms of non-separability and entanglement. \forget{Their strength is shown in two ways. Firstly, they imply and strengthen  other such criteria and, secondly, the number of settings needed for their implementation is not too high and they have great noise robustness.}

{\bf Chapter \ref{chapter_monogamy}} investigates correlations from a different point of view, 
namely whether they can be shared to other parties. If this is not the case the correlations are said to be monogamous.  This is a new field of study that is closely related to the study of monogamy and shareability of entanglement, although I show some crucial differences between the two. Known results are reviewed, in particular that quantum and no-signaling non-local correlations cannot be shared freely, whereas local ones can. It is next shown that unrestricted correlations as well as partially-local correlations can also be shared freely. To quantify the issue, I study the monogamy trade-offs on bounds on Bell-type inequalities that hold for different, but overlapping subsets of the parties involved. I limit myself to three parties, but this already yields many new results.
\forget{An independent simpler proof of the monogamy relation of \citet{tonerverstraete} is provided, and also  a different strengthening of this constraint than was already given by them. 

 which I will here briefly mention. 

 I present an independent simpler proof of the monogamy relation of \citet{tonerverstraete}, and provide a different strengthening of this constraint than was already given by them.  For the case of two parties, the relationship between sharing non-local quantum correlations and sharing mixed entangled states is shown to depend on the number of observables per party used. For no-signaling correlations I argue that the monogamy constraint of \citet{toner2} can be interpreted as  a non-trivial bound on the set of three-partite no-signaling correlations. \forget{, although using only product (joint) expectation values only. }
 Lastly, I derive monogamy constraints of three-qubit bi-separable quantum correlations, which is a first example of investigating monogamy of quantum correlations using a three-partite Bell-type inequality.
 }

{\bf Chapter \ref{chapter_svetlichny}} returns to the task of discerning the different kinds of multi-partite correlations using Bell-type inequalities.
  In this chapter a new family of Bell-type inequalities is constructed in terms of product (joint) expectation values that discern partially-local from quantum mechanical correlations. \forget{Also, the issue of discerning multi-partite no-signaling correlations is discussed.}  This chapter  generalizes the three-partite Svetlichny inequalities \cite{svetlichny} to the multi-partite case, thereby providing criteria to discern partially-local from stronger correlations. These inequalities are violated by quantum mechanical states and it can thus be concluded that they contain fully non-local correlations. 
However, the inequalities cannot discern no-signaling correlations from more general correlations. \forget{The question what set of inequalities,  in terms of expectation values only,  would discern the multi-partite no-signaling correlations, is left as an interesting open problem.}


{\bf Part IV} deals with more philosophical matters. I consider the ontological status of quantum correlations.
{\bf Chapter \ref{chapter_quantumworldcorrelations}} uses a Bell-type inequality argument to show that despite the fact that quantum correlations suffice to reconstruct the quantum state,
they cannot be regarded as objective local properties of the composite system in question, i.e., they cannot be given a local realistic account. 
Together with some other arguments, this is used to argue against the ontological robustness of entanglement.
  
{\bf Chapter \ref{chapter_holism}} is devoted to the idea of holism in classical and quantum physics.  I review  the well-known supervenience approach to holism developed by \citet{teller86,teller89} and \citet{healey91}, and provide an alternative approach, which uses an epistemological criterion to decide whether a theory is holistic. This approach is compared to the supervenience approach and shown to involve a shift in emphasis from ontology to epistemology. Further, it is argued that this approach better reflects the way properties and relations are in fact determined in physical theories. In doing so it is argued that holism is not a thesis about the state space a theory uses. When applying the epistemological criterion for holism  to classical physics and Bohmian mechanics it is rigorously shown that they are non-holistic, whereas quantum mechanics is shown to be holistic even in absence of any entanglement.

{\bf Part V} ends this dissertation with {\bf chapter \ref{sumout}} that contains a summary of the results obtained and  a discussion of a number of open problems and avenues for future research  inspired by the work in this dissertation.
\\\\\\
{\bf To the reader}: \\\\
(i) At the beginning of each chapter I list the article(s) on which that particular chapter is (partly) based. All these articles are listed at the  end of this dissertation on page \pageref{listpubli}.\forget{ I (jointly) wrote }\\\\
\forget{
(i) I expect that my results will interest both the practicing theoretical physicist in quantum information theory as well as the philosopher of physics that tries to understand quantum theory and its relationship to other physical theories better. Part IV is the most philosophical, part II and III contain 
results at the frontier of theoretical physics.  But each chapter (except perhaps chapter 5) discusses foundational results as well.\\\\}
(ii) Chapter 11 gives a summary that can be read independently from the rest of the dissertation and also gives suggestions for future research.  The prospective reader might want to consult this chapter since it gives a more detailed, though non-technical introduction of the results obtained in this dissertation that supplements the -- perhaps somewhat short -- introduction presented above.
%
%
%
%
\newpage
\thispagestyle{empty}
\cleardoublepage
\thispagestyle{empty}
\chapter[On correlations: Definitions and general framework]{On correlations:\\\vskip0.25cm Definitions and general framework}
\label{definitionchapter}\thispagestyle{empty}
\noindent This chapter is in part based on \citet{seevchen}.

\section{Introduction}

In this chapter we give the necessary background for discussing the technical results of this dissertation. We will present the definitions, notation and techniques  that will be used in later chapters, as well as several clarifying examples. We also discuss relevant results already obtained by others.  Along the way we will take the opportunity to indicate more precisely than was done in the previous chapter what technical results are contained in later chapters. The foundational relevance of the results will be discussed later. For conciseness and clarity of exposition we will for now refrain from any interpretational discussion.
 
We start in section \ref{allcorr} by defining the different kinds of correlation that will be studied, as well as several useful mathematical characteristics of these correlations. We follow the approach by \citet{barrett05} and \citet{masanes06} in discussing the no-signaling, local and quantum correlations, and we will supplement their presentation with a new type of correlations, the partially-local correlations\footnote{\citet{tsirelsonhodronic} also distinguished most of these types of correlations (but not the partially-local ones). He called them different kinds of `behaviors'. However, we will not follow his exposition.}.
 Discerning these different kinds of correlations  is in general a hard task. In section \ref{comparecorr} we will argue that Bell-type inequalities form a useful tool for this task. We present a general scheme for describing such inequalities in terms of bounds on the expectation values of so-called Bell-type polynomials.  After discussing this scheme we present the well-known bi-partite Clauser-Horne-Shimony-Holt (CHSH) inequality \cite{chsh} as an example. This inequality discriminates some of the bi-partite correlations, but, as we will show, not all of them. We will then further comment on the task of obtaining multi-partite Bell-type inequalities in order to discriminate the different types of multipartite correlations.
 
We next pay special attention to the issue of discriminating quantum correlations, because here the distinction between  entanglement and separability of quantum states becomes relevant. In the bi-partite case we discuss the feature of separability and entanglement, and the (non-)locality of these states. 

Lastly, in section \ref{pitfall} we discuss a possible pitfall connected to the use of Bell-type polynomials for obtaining Bell-type inequalities.  We trace the problem back to the fact that Bell-type inequalities always  use different combinations of  incompatible observables.  This re-teaches an old lesson from J.S. Bell, namely, that one should be extremely careful when considering incompatible observables, and not be lured into neglecting this issue because quantum mechanics deals so easily with incompatible observables via the non-commutativity structure that is part and parcel of its formalism.

\section{Correlations}
\label{allcorr}
\subsection{General correlations}
Consider $N$ parties, labeled by $1,2,\ldots,N$, each holding a physical system that are to be measured using a finite set of different observables. Denote by $A_j$ the observable (random variable)  that party $j$ chooses (also called the setting $A_j$)  and by $a_j$ the corresponding measurement outcomes. We assume there to be only a finite number of discrete outcomes. 
The outcomes can be correlated in an arbitrary way. 
A general way of describing this situation, independent of the underlying physical model, is by a set of
joint probability distributions for the outcomes, conditioned on the settings chosen by the $N$ parties\footnote{Here, and throughout, we conditionalize on the settings for simplicity. This brings with it a commitment to probability distributions over settings, but all our probabilistic conditions can be reformulated without that commitment, see \citet[p. 117]{butterfield1}. Such a reformulation treats settings as parameters and not as random variables. Only in a single instance, when discussing Maudlin's conditions in chapter \ref{chapter_CHSHclassical}, it is necessary to introduce a probability distributions over settings.}, where the correlations  are captured in terms of these joint probability distributions\footnote{We describe correlations in terms of the conditional joint probability distributions. An alternative way to study correlations is to  consider a measure of correlation between random variables called the correlation coefficient.  The correlation coefficient between two random variables $x$ and $y$ is given by  the covariance of $x$ and $y$, cov$(x,y):=\av{(x-\av{x})(y-\av{y})}$, divided by the square root of the product of the variances, $\sqrt{\textrm{var}(x)\textrm{var}(y)}$, with var$(x):=\av{x^2}-\av{x}^2$ and analogously for var$(y)$. If the random variables are statistically independent their joint probability distribution factorises, i.e.,  $P(x,y)=P(x)P(y)$, and then cov$(x,y)=0$, so the variables can be said to be uncorrelated. 

However, the correlation coefficient does not deal well with deterministic scenarios, since there the variances and the covariance are always zero resulting in an ill-defined correlation coefficient. However, in a deterministic case the probabilities are either $0$ or $1$, and such deterministic scenarios are thus included in the joint probability formalism used here.

In quantum mechanics only non-product states (when expressed on a local basis $\{ \ket{i}\otimes\ket{j}\otimes\ldots\}$) have a non-zero correlation coefficient. These can however be pure. Indeed, a set of random variables (observables) exists such that a pure entangled state always  gives rise to a non-zero correlation coefficient for these random variables. Classically this is never the case. Pure classical states correspond to points in a phase or configuration space and they give rise to deterministic scenarios where the correlation coefficient for any set of random variables is ill-defined.}. They are denoted by 
\begin{align}\label{generalcorr}
P(a_1,\ldots,a_N|A_1,\ldots,A_N).
\end{align}
These probability distributions are assumed to be positive 
\begin{align}\label{posgen}
P(a_1,\ldots,a_N|A_1,\ldots,A_N)\geq 0,~~ \forall a_1,\ldots,a_N,~~ \forall A_1,\ldots,A_N,
\end{align}
and obey the normalization conditions
\begin{align}\label{norma}
\sum_{a_1,\ldots,a_N}P(a_1,\ldots,a_N|A_1,\ldots,A_N)=1,~~ \forall A_1,\ldots,A_N.
\end{align}
We need not demand that the probabilities should not be greater than 1 because this follows from them being positive and from the normalization conditions.

The set of all these probability distributions has a nice structure. First, it is a convex set: convex combinations of correlations are still legitimate correlations. Second, there are only a finite number of extremal correlations. This means that every correlation can be  decomposed into a (not necessarily unique) convex combination of such extremal correlations. 

A total of $D=\prod_{i=1}^N m_{A_i} m_{a_i}$ different probabilities exist (here $m_{A_i}$ and $m_{a_i}$ are the number of different observables  and outcomes for party $i$ respectively)\forget{ where it is assumed that each observable $A_i$ (of which there are a total of $m_{A_i}$) has the same number of $m_{a_i}$ possible outcomes}. When these conditional probability distributions (\ref{generalcorr}) are considered as points in a $D$-dimensional real space\forget{\footnote{This space has dimension $D=\prod_i m_{A_i} m_{a_i}$ with $m_{A_i}, m_{a_i}$ the number of different observables  and outcomes for party $i$ respectively (here it is assumed that each observable $A_i$ (of which there are a total of $m_{A_i}$) has the same number of $m_{a_i}$ possible outcomes).}},  this set of points forms a convex polytope with a finite number of extreme points which are the vertices of the polytope. This polytope is  the convex hull of the extreme points. It  belongs to the subspace defined by (\ref{norma}) and it is bounded by a set of facets, linear inequalities that describe the halfplanes that bound it. Every convex polytope has a dual description, firstly in terms of its vertices, and secondly in terms of its facets, i.e., hyperplanes that bound the polytope uniquely. In general each facet can be described by linear combinations of joint probabilities which reach a maximum value at the facet, i.e., 
\begin{align}\label{facet}
\sum_{a_1,\ldots,a_N, A_1,\ldots,A_N} c_{a_1,\ldots,a_N,A_1,\ldots,A_N}P(a_1,\ldots,a_N|A_1,\ldots,A_N)\leq I,
\end{align}
with real coefficients $c_{a_1,\ldots,a_N,A_1,\ldots,A_N}$ and a real bound $I$ that is reached by some extreme points. 
 For each facet some extreme points of the polytope lie on this facet and thus saturate the inequality (\ref{facet}), while the other extreme points cannot violate it.   In general, when the extreme points are considered as vectors,  a hyperplane is a facet of a $d$-dimensional polytope iff $d$ affinely independent extreme points satisfy the equality that characterizes the hyperplane\footnote{\label{affine}In case the null vector belongs to the polytope, the condition of the existence of $d$ affinely independent vectors is equivalent to the existence of $(d-1)$ linearly independent vectors; otherwise it requires the existence of $d$ linearly independent vectors \cite{masanes02}.}. Consequently, for the case of general correlations    (\ref{generalcorr}) the set of extreme points that lie on a facet  must contain a total of $D$ affinely independent vectors.   For this case the facets are determined by equality in (\ref{posgen}).  The probability distributions  (\ref{generalcorr}) correspond to any normalized vector of positive numbers in this polytope.  For an excellent overview of the structure of these polytopes, see \cite{masanes02}, \cite{barrett05} and \cite{ziegler}.

The extreme points are the probability distributions that saturate a maximum of the positivity conditions (\ref{posgen}) while satisfying the normalization condition (\ref{norma}).  They are characterized by \citet{jones} to be the probability distributions such that for each set of settings $\{A_1,\ldots,A_N\}$ there is a unique set of outcomes $ \{ a_1[A_1,\ldots A_N],\ldots, a_N[A_1,\ldots A_N]\}$ for which $P(a_1,\ldots,a_N|A_1,\ldots,A_N)=1$,  with  $a_1[A_1,\ldots,A_N]$  the deterministic determination of outcome $a_1$ given the settings $A_1, \ldots, A_N$, etc. 
 There is thus a one-to-one correspondence between the extreme points and the sets of functions 
$\{a_1[A_1,\ldots A_N],\ldots, a_N[A_1,\ldots A_N]\}$ from the settings to the outcomes. Any such set defines an extreme point.  The extreme points thus correspond to deterministic scenarios: each outcome is completely fixed by the totality of all settings and consequently there is no randomness left: $P(a_1,\ldots, a_N|A_1,\ldots,A_N)= \delta_{a_1,a_1[A_1,\ldots,A_N]}\cdots  \delta_{a_N,a_N[A_1,\ldots,A_N]}$. Finding all the facets of a polytope knowing its vertices is called the hull problem and this is in general a computationally hard task \cite{pitowsky}.  The facet descriptions (\ref{facet}) will be called Bell-type inequalities, and these will be further introduced later.

The marginal probabilities are obtained in the usual way from the joint probabilities by summing over the outcomes of the other parties. It is important to realize that for general correlations these marginals may depend on the settings chosen by the other parties. For example, in the case of two parties that 
each choose two possible settings $A_1,A_1'$ and $A_2,A_2'$ respectively, the marginals for party 1 are given by 
\begin{subequations}
\label{M1}
\begin{align}
P(a_1|A_1)^{A_2}&:= \sum_{a_2}P(a_1,a_2|A_1,A_2),\label{m1}\\
P(a_1|A_1)^{A_2'}&:=\sum_{a_2}P(a_1,a_2|A_1,A_2'),\label{m2}
\end{align}
\end{subequations}
and analogously for setting $A_1'$ and for the marginals of party 2. The marginal $P(a_1|A_1)^{A_2}$ may thus in general be different from 
$P(a_1|A_1)^{A_2'}$.
 
We will now put further restrictions besides normalization on the probability distributions  (\ref{generalcorr})  that are motivated by physical considerations. We will here not be concerned with arguing for the plausibility of these physical considerations, nor what 
violations of these physically motivated restrictions amounts to, but merely give the definitions that will be used in future chapters. There we will comment on the foundational content of the restrictions and their possible violations.

\subsection{No-signaling correlations}\label{nosigsec}
Let us first consider the case of two parties that each choose two possible settings. A no-signaling correlation\footnote{We want to distinguish no-signaling from the impossibility of superluminal signaling, for the latter requires a notion of spacetime structure whereas the first does not.} for two parties is a correlation such that party $1$ cannot signal to party $2$ by the choice of what observable is measured by party $1$ and vice versa. This means that the marginal probabilities $P(a_1|A_1)^{A_2}$ (see \eqref{m1}) and $P(a_2|A_2)^{A_1}$ are independent of $A_2$ and $A_1$ respectively: 
\begin{subequations}
\label{22nosig}
\begin{align}
P(a_1|A_1)^{A_2}=P(a_1|A_1)^{A_2'}&:=P(a_1|A_1)
\forget{&\Longleftrightarrow \sum_{a_2}P(a_1,a_2|A_1,A_2)=\sum_{a_2}P(a_1,a_2|A_1,A_2'):=P(a_1|A_1)},~~ \forall a_1,A_1,A_2,A_2',\\
P(a_2|A_2)^{A_1}=P(a_2|A_2)^{A_1'}&:=P(a_2|A_2)
\forget{\\
&\Longleftrightarrow\sum_{a_1}P(a_1,a_2|A_1,A_2)=\sum_{a_1}P(a_1,a_2|A_1',A_2):=P(a_2|A_2)},~~ \forall a_2,A_1,A_1',A_2.
\end{align}
\end{subequations}
In a no-signaling context the marginals can thus be defined as $P(a_1|A_1)$, etc., i.e., without any dependence on far-away settings.

 Let us generalize this to the multi-partite setting: a no-signaling correlation  is a correlation 
  $P(a_1,\ldots,a_N|A_1,\ldots,A_N)$ such that one subset of parties, say parties $1,2,\ldots,k$, cannot signal to the other parties $k-1,\ldots,N$ by changing their measurement device settings $A_1,\ldots,A_k$. Mathematically this is expressed 
 as follows. The marginal probability distribution for each subset of parties only depends on the corresponding observables measured by the parties in the subset, i.e.,
for all outcomes $a_{k+1},\ldots, a_N$: $P(a_1,\ldots, a_k|A_1,\ldots,A_N)=P(a_1,\ldots, a_k|A_1,\ldots,A_k)$. These conditions can all be derived from the following condition \cite{barrett05}. For each $k\in \{1,\ldots,N\}$ the marginal distribution that is obtained when tracing out $a_k$ is independent of what observable ($A_k$ or $A_k'$) is measured by party $k$:
  \begin{align}\label{nosignalingdistr}
  \sum_{a_k} P(a_1,\ldots,a_k,\ldots, a_N|&A_1,\ldots,A_k,\ldots,A_N)=\nonumber\\& \sum_{a_k} P(a_1,\ldots,a_k,\ldots, a_N|A_1,\ldots,A_k',\ldots,A_N),
  \end{align}
   for all outcomes $a_1,\ldots,a_{k-1},a_{k+1},\ldots,a_N$ and all settings $A_1,\ldots,A_k,A_k',\ldots A_N$.   This set of conditions ensures that all marginal probabilities are independent of the settings corresponding to the outcomes that are no longer considered\footnote{To see that this is sufficient, let us consider the three-partite case as treated by \citet{barrett05}. Various types of communication exist that give different forms of signaling. These should all be excluded.  Party 1 should not be able to signal to either party 2 or 3 (and cyclic permutations). Also if party 2 and 3 are combined to form a composite system then party 1 should not be able to signal to this system. This is expressed by 
   \begin{align}
   \label{3nos}
   \sum_{a_1}P(a_1,a_2,a_3|A_1,A_2,A_3)=\sum_{a_1}P(a_1,a_2,a_3|A_1',A_2,A_3), ~~\forall a_2,a_3, A_1,A_1',A_2,A_3.
   \end{align}
   From this it also follows that party 1 cannot signal to either party 2 or 3 (this is easily seen by summing over outcomes $a_2$ and $a_3$ respectively). Conversely,  if systems 2 and  3 are combined they should not be able to signal to party 1. However this need not be separately specified since it already follows from condition \eqref{3nos} and its cyclic permuted versions, as we will now show. From the fact that party 2 cannot signal to  the composite system of party 1 and 3, and party 3 cannot signal  to the composite system of party 2 and 3 it follows that 
   \begin{align}
   \label{3nos1}
    \sum_{a_2,a_3}P(a_1,a_2,a_3|A_1,A_2,A_3)&=\sum_{a_2,a_3}P(a_1,a_2,a_3|A_1,A_2',A_3), ~~\forall a_1, A_1,A_2,A_2',A_3\nn\\
&=\sum_{a_2,a_3}P(a_1,a_2,a_3|A_1,A_2',A_3'), ~~\forall a_1, A_1,A_2,A_2',A_3,A_3'.
   \end{align}
   This is the condition that the composite system of party 2 and 3 cannot signal to party 1.  Hence, condition \eqref{3nos} and its cyclic permutations are the only conditions that need to be required to ensure that no-signaling obtains.}.
   In particular, \eqref{nosignalingdistr} the defines the marginal
   \begin{align}\label{marginalp}
   P(a_1,\ldots,a_{k-1},a_{k+1},\ldots, a_N|A_1,\ldots,A_{k-1},A_{k+1},\ldots,A_N),
   \end{align} for the $N-1$ parties not including party $k$. No-signaling ensures that it is not needed to specify whether $A_k$ or $A_k'$ is being measured by party $k$.
   
   These linear equations \eqref{nosignalingdistr} characterize an affine set \cite{masanes06}. The intersection of this set with the polytope of distributions 
(\ref{generalcorr}) gives another convex polytope: the no-signaling polytope. 
The vertices of this polytope can be split into two types: vertices that correspond to deterministic scenarios,  where all probabilities are either $0$ or $1$, and those that correspond to non-deterministic scenarios.  All no-signaling deterministic correlations are in fact local \cite{masanes06}, i.e., they can be written in terms of the local correlations defined below on page \pageref{sectionlocal}. But all non-deterministic vertices correspond to non-local scenarios.

The complete set of vertices of the no-signaling polytope is in general unknown.  However, in the bi-partite and three-partite case some results have been obtained: For the bi-partite case of two settings and any number of outcomes they are determined by \citet{barrett05} and for any number of settings and two possible outcomes by \citet{jonesmasanes}. For three parties, two outcomes and two settings the vertices are given in \cite{barrett05}.

\forget{[example 3 parties from barrett05 ? different kinds of no-signaling, but only one condition suffices. as a footnote]}

The facets of the no-signaling polytope follow from the defining conditions for no-signaling correlations. These are thus the trivial facets that follow from the positivity conditions as well as the non-trivial ones that follow from the no-signaling requirements  \eqref{nosignalingdistr}. In section \ref{chsintrotech} the latter will be explicitly dealt with in the two-partite case.  The importance of the non-trivial facets of the no-signaling polytope is that if a point, representing some experimental data, lies within the polytope, then a model that uses no-signaling correlations exists that reproduces the same data. On the contrary, if the point lies outside the polytope and thus violates some Bell-type inequality describing a facet of the no-signaling polytope, then the data cannot be reproduced by a no-signaling model only, i.e., including some signaling is necessary.

\forget{
   Although the vertices of some of the no-signaling polytopes are known, finding the facets is in general  a computationally hard task.
   Indeed, the facets of the no-signaling polytope are still to be described, except for the trivial ones that derive from the normalization condition (\ref{norma}).   The importance of finding the non-trivial facets of this correlation polytope is that if a point, representing some experimental data, lies within the polytope, then a model that uses no-signaling correlations exists that reproduces the same data. On the contrary, if the point lies outside the polytope and thus violates some Bell-type inequality describing a facet of the no-signaling polytope, then the data cannot be reproduced by a no-signaling model only, i.e., including some signaling is necessary. 
    }
   \subsection{Local correlations} \label{sectionlocal}
Local correlations are those that can be obtained if the parties are non-communica-ting and share classical information, i.e.,  they only have local operations and local hidden variables (also called shared randomness) as a resource. We take this to mean that these correlations can be written as
\begin{align}\label{localdistr}
P(a_1,\ldots, a_N|A_1,\ldots,A_N)=\int_\Lambda d\lambda p(\lambda) P(a_1|A_1,\lambda) \ldots P(a_N|A_N,\lambda),
\end{align}
where $\lambda\in\Lambda$ is the value of the shared local hidden variable, $\Lambda$ the space of all hidden variables and $p(\lambda)$ is the probability that a particular value of $\lambda$ occurs\footnote{Opinions differ on how to motivate (\ref{localdistr}). In chapter \ref{chapter_CHSHclassical} we will come back to this issue. The technical results of this dissertation do not depend on such a motivation and whether it is physically plausible and/or sufficient.}. Note that $p(\lambda)$ is independent of the outcomes $a_j$ and settings $A_j$. This is a `freedom' assumption, i.e., the settings are assumed to be free variables (we will discuss this assumption in the next chapter). Furthermore, $P(a_1|A_1,\lambda)$ is the probability that outcome $a_1$ is obtained by party $1$ given that the observable measured was $A_1$ and the shared hidden variable was $\lambda$, and similarly for the other terms $P(a_k|A_k,\lambda)$. 
Since these probabilities are conditioned on the hidden variable $\lambda$ we will call them subsurface probabilities, in contradistinction to the probabilities $P(a_j|A_j)$, etc., that only conditionalize on the settings, which we call surface probabilities\footnote{This terminology is partly due to \citet{fraassen82}.}.

Condition (\ref{localdistr})  is supposed to capture the idea of locality in a hidden-variable framework and it is called Factorisability, and models that give only  local correlations are called local hidden-variable (LHV) models.  These notions will be further discussed in the next chapter. Correlations that cannot be written as (\ref{localdistr}) are called non-local.  Local correlations are no-signaling thus the marginal probabilities derived from local correlations are defined in the same way as was done for no-signaling correlations, cf. \eqref{marginalp}.   

Let us review what is known about the set of local correlations. First, it is also a polytope with vertices (extremal points) corresponding to local deterministic distributions \cite{werwolf}, i.e., $P(a_1,\ldots, a_N|A_1,\ldots,A_N)= \delta_{a_1,a_1[A_1]}\cdots  \delta_{a_N,a_N[A_N]}$ where  the function $a_1[A_1]$ gives the deterministic determination of outcome $a_1$ given the setting $A_1$, etc. Thus for  each set of settings $\{A_1,\ldots,A_N\}$ there is a unique set of outcomes $ \{ a_1[A_1],\ldots, a_N[A_N]\}$ for which $P(a_1,\ldots,a_N|A_1,\ldots,A_N)=1$.  
  All these vertices 
are also vertices of the no-signaling polytope \cite{barrett05}. The  local polytope is known  to be constrained by two kinds of facets \cite{werwolf}. The first are trivial facets and derive from the positivity conditions (\ref{posgen}). Note that these are also trivial  facets of the no-signaling polytope. The second kind of facets are non-trivial and can be violated by non-local correlations. These are not facets of the no-signaling polytope. All facets are mathematically described by Bell-type inequalities \eqref{facet}, that will be further introduced below.   Determining whether a point lies within the local polytope, i.e., whether it does not violate a local Bell-type inequality, is in general very hard as \citet{pitowsky} has shown this to be related to some known hard problems in computational complexity (i.e., it is an NP-complete problem). Furthermore, determining whether a given inequality is a facet of the local polytope is of similar difficulty (i.e., this problem is co-NP complete \cite{pitowsky91}).

\subsection{Partially-local correlations}

Partially local correlations are those  that can be obtained from an $N$-partite system in which subsets of the $N$ parties form extended  systems, whose internal states can be correlated in any way (e.g., signaling),
    which however behave local with respect to each other. Suppose provisionally that parties $1,\ldots,k$ form such a subset and the remaining parties $k+1,\ldots,N$ form another subset. The partially-local correlations can then be written as 
\begin{align}\label{partiallocaldistr}
P(a_1,\ldots,&\, a_N|A_1,\ldots,A_N)=\\&\int_\Lambda d\lambda p(\lambda) P(a_1,\ldots,a_k|A_1,\ldots,A_k,\lambda) P(a_{N-k},\ldots,a_N|A_{N-k},\ldots,A_N,\lambda),\nn
\end{align}
We also refer to this condition as partial factorisability\footnote{Partial factorisability is sometimes also called partial separability.
   Indeed, in the few papers that have appeared
on this subject \cite{svetlichny, seevsvet, collins, uffink} this is the case. However, for consistency in the terminology 
we prefer the term partial factorisability. In this dissertation separability is a concept defined only in terms of the structure of quantum states on a Hilbert space and not in terms of the structure of classical probability distributions.}. The subsurface probabilities on the right hand side need not factorise any further. In case they would all fully factorise we retrieve the set of local correlations described above.

Formulas  similar to (\ref{partiallocaldistr}) with different partitions of the $N$-parties into two subsets, i.e., for different choices of the composing  parties and different values of $k$, describe other possibilities to give partially-local correlations. Convex combinations of these possibilities are also admissible.   We need not consider decomposition into more than two subsystems since any two subsystems in such  a decomposition can be considered jointly as parts of one subsystem still uncorrelated with respect to the others.   
    
    We define a model to have partially-local correlations when the correlations are of the form (\ref{partiallocaldistr}) or when they can be written as convex combinations of similar expressions on the right hand side of (\ref{partiallocaldistr}) for the different possible partitions of the $N$ parties into two subsets. Such a model is called a partially-local hidden-variable (PHLV) model\footnote{\citet{collins} have called this a `local-nonlocal model'.}. Models whose correlations cannot be written in this partially-local form are fully non-local, i.e., they are said to contain full non-locality.

The set of partially-local correlations has a finite number of extreme points and is thus also a polytope \cite{jones}, called the partially-local polytope. It is also convex since it can be easily seen that if two distributions satisfy (\ref{partiallocaldistr}) then their convex mixture will too.   For each extreme point of this convex polytope  there is a partition into two subsets, say $\{1,\ldots,k\}$ and $\{k+1,\ldots,N\}$, such that  for 
 each set of settings $\{A_1,\ldots,A_N\}$ there is a unique set of outcomes $ \{ a_1\ldots,a_N\}$ for which $P(a_1,\ldots,a_k|A_1,\ldots,A_k,\lambda)=1$ and $P(a_{N-k},\ldots,a_N|A_{N-k},\ldots,A_N,\lambda)=1$.
There is thus a one-to-one correspondence between the extreme points corresponding to a partition of the parties into two subsets and the set of functions from the two corresponding subsets of settings to the two corresponding subsets of outcomes.  Just as was the case for general and local correlations, we again see the deterministic scenario arising for the extreme points.

Let us consider this in more detail and take the example where $N=3$, first studied by \citet{svetlichny}.  Only three different partitions into two subsets are possible.
The three-partite partially-local correlations are thus of the form
\begin{align}
P(a_1,a_2,a_3|A_1,A_2,A_3)=\int_\Lambda d\lambda 
[&p_1\,\rho_1(\lambda) P_1(a_1|A_1,\lambda)P_1(a_2,a_3|A_2,A_3,\lambda)\nn\\
&+p_2\,\rho_2(\lambda) P_2(a_2|A_2,\lambda)P_2(a_1,a_3|A_1,A_3,\lambda)\nn\\
&+p_3\,\rho_3(\lambda) P_3(a_3|A_3,\lambda)P_3(a_1,a_2|A_1,A_2,\lambda)
].
\label{localnon-local3intr}
\end{align}
where  $P_1(a_2,a_3|A_2,A_3,\lambda)$ can be any probability distribution; it need not factorise into $P_1(a_2|A_2,\lambda)P_1(a_3|A_3,\lambda)$. Analogously for the other two joint probability terms. 
The $\rho_i(\lambda)$ are the hidden-variable distributions. Models whose correlations cannot be written in this form are fully non-local, i.e., they are said to contain full  three-partite non-locality.

Because the correlations between subsets of particles are allowed to be signaling, the marginal probabilities may depend on the settings corresponding to the outcomes that are no longer considered.  This must be explicitly accounted for. For example, the marginal $P(a_1,a_2,|A_1,A_2)^{A_3}$ derived from \eqref{localnon-local3intr} may depend on the setting chosen by party $3$, and the marginal $P(a_1|A_1)^{A_2,A_3}$ for party $1$ may depend on the setting chosen by both party 2 and 3, etc.  Because we must allow for convex combinations of different partially-local configurations, as  in \eqref{localnon-local3intr}, the marginals can depend on the settings chosen by all other parties, despite the fact that at the hidden variable level there can not be signaling between all three parties.

\subsection{Quantum correlations}\label{qmcorrsection}
Lastly, we consider another class of correlations: those that are obtained by general measurements on quantum states (i.e.,  those that can be generated if the parties share quantum states). These can be written as 
\begin{align}\label{quantumcorre}
P(a_1,\ldots, a_N|A_1,\ldots,A_N)=\textrm{Tr}[M_{a_1}^{A_1} \otimes\cdots\otimes M_{a_N}^{A_N} \rho].
\end{align}
Here $\rho$ is a quantum state (i.e., a unit trace semi-definite positive operator) on a Hilbert space $\H=\H_1\otimes\cdots\otimes\H_N$, where $\H_j$ is the quantum state space of the system held by party $j$. The sets $\{M_{a_1}^{A_1},\ldots,M_{a_N}^{A_N} \}$ define what is called a positive operator valued measure\footnote{Note that POVM measurements include as a special case the ordinary von Neumann measurements that use so called projection valued measures (PVM) where all positive operators are orthogonal projection operators.} (POVM), i.e., a set of positive operators $\{ M_{a_j}^{A_j}\}$ satisfying 
$\sum_{a_j}M_{a_j}^{A_j}=\1,\forall A_j$.  
Of course, all operators $ M_{a_j}^{A_j}$ must commute for different $j$
in order for the joint probability distribution to be well defined, but this is ensured since for different $j$  the operators are defined for different subsystems (with each their own Hilbert space) and are therefore commuting. Note that \eqref{quantumcorre} is linear in both $M_{a_j}^{A_j}$ and $\rho$, which is  a crucial feature of quantum mechanics.

Quantum correlations are no-signaling and therefore the marginal probabilities derived from such correlations are defined in the same way as was done for no-signaling correlations (cf. \eqref{marginalp}).  For example, the marginal probability for party $1$ is given by $P(a_1|A_1)=\textrm{Tr}[M_{a_1}^{A_1}\rho^1]$, where $\rho^1$ is the reduced state for party 1.

The set of quantum correlations has been investigated by, e.g., \citet{pitowsky}, \citet{tsirelsonhodronic},  and \citet{wernerwolf2} and is shown to be convex\forget{\footnote{This set is known to be convex for the case of all possible measurements on all possible quantum states \cite{pitowsky,wernerwolf2}. However, convexity is not proven if one restricts oneself to the measurements on a given state and to projector valued measurements (PVM) on a Hilbert space with given dimension.}}. It is not a polytope because the number of extremal points is not finite and consequently it has an infinite number of bounding halfplanes. Therefore we will refer to this set as the quantum body, in contradistinction to the sets of the other types of correlations which are referred to as polytopes.

We note that in order to describe the full measurement process it is necessary to specify the set of so-called Kraus operators $\{K_{a_i}^{A_i}\}$ that correspond to the POVM elements $\{M_{a_i}^{A_i}\}$, where $M_{a_i}^{A_i}=K_{a_i}^{A_i}(K_{a_i}^{A_i})^\dagger$. In general many different sets of Kraus operators correspond to the same POVM element. The reason for including the Kraus operators is that the description of a POVM as a set of positive operators  
$\{M_{a_i}^{A_i}\}$ is incomplete because it does not specify uniquely what the state of the system is after the measurement.  By including the Kraus operators one is able to retrieve the Projection Postulate: if a POVM measurement is performed on system $i$ then the state  $\rho_i$
directly after the measurement will be given by $\tilde{\rho}_i=K_{a_i}^{A_i}\,\rho_i\,(K_{a_i}^{A_i})^\dagger\,/\,{\rm Tr}[ K_{a_i}^{A_i}\,\rho_i\, (K_{a_i}^{A_i})^\dagger]$.

\section{On comparing and discriminating the different kinds of correlations}\label{comparecorr}

Let us present the relationships between the correlations of the previous section, some of which are already known, some of which are proven in this dissertation. The polytope of general correlations strictly contains the no-signaling polytope, which in turn contains the quantum body, which in turn contains the partially-local polytope, which in turn contains the local polytope.  See Figure \ref{figsets}. These results are obtained by comparing the facets of the relevant polytopes and halfplanes that bound the quantum bodies.  These facets  (i.e., bounding hyperplanes in the case of quantum correlations) are of course implicitly determined by the defining restrictions on the different types of correlations, but to find explicit experimentally accessible expressions for them is a hard job. A fruitful way of doing so is using so-called Bell-type inequalities. This will be discussed next.

\begin{figure}[h]
\includegraphics[scale=0.6]{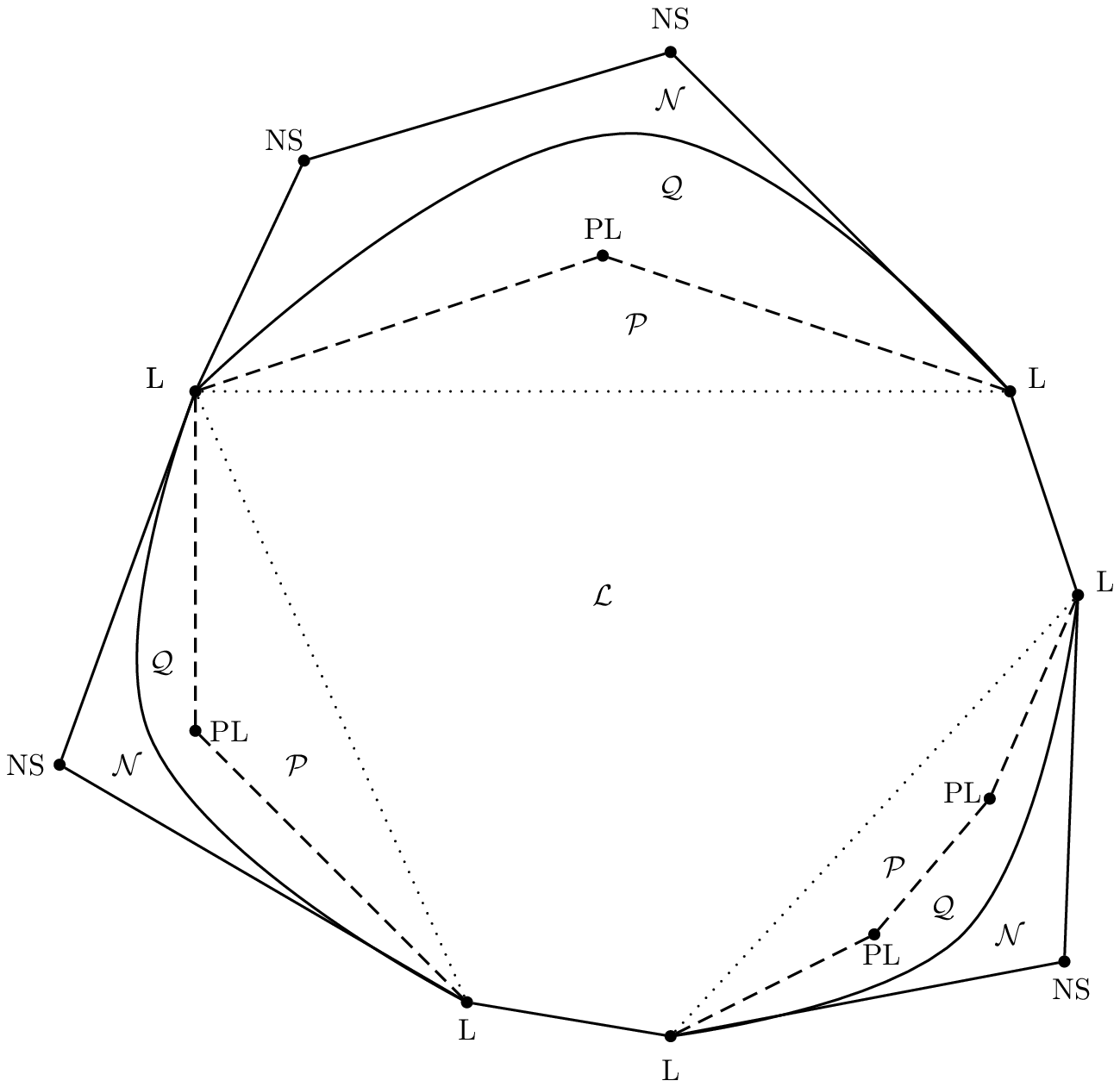}
\caption{Schematic representation of the space of correlations, after \citet{barrett05}. The vertices are labeled L, PL and NL for the local, partially-local and no-signaling polytope. The region inside each of these polytopes is denoted by $\mathcal{L}$, $\mathcal{P}$, and  $\mathcal{N}$ respectively. The accessible quantum region is denoted by $\mathcal{Q}$.
} \label{figsets}
\end{figure}

\subsection[Bell-type inequalities: bounds that discriminate between\\ different types of correlations]{Bell-type inequalities: finding experimentally\\ accessible bounds that discriminate between\\ different types of correlations}
\label{techbellineq}
In this dissertation we will investigate all of the above types of correlations by deriving experimentally accessible conditions that 
distinguish them from one another. In particular we will study Bell-type inequalities for the case where each party chooses between two alternative observables and where each observable is dichotomic, i.e., the observable has two possible outcomes which we denote by $\pm1$. 

Bell-type inequalities denote a specific bound on a linear sum of joint probabilities as in \eqref{facet}. The bound is characteristic of the type of correlation under study. However, frequently they are formulated not in terms of probabilities but in terms of product expectation values\footnote{These are also known as `joint expectation values' or `correlation functions', see e.g., \citet{zukow2002}, but we will not use this terminology.}, i.e., expectation values of products of observables, which we will denote by $ \av{A_1A_2 \cdots A_N}$. These are defined in the usual way as the weighted sum of the products of the outcomes:
 \begin{align}\label{expectation}
\av{A_1A_2 \cdots A_N}:=\sum_{a_1,\ldots,a_N} a_1a_2\cdots a_N P(a_1,\ldots, a_N|A_1,\ldots,A_N).
\end{align}
 Since we are restricting ourselves to dichotomic observables with outcomes $\pm1$ all expectation values are bounded by: $-1\leq\av{A_1A_2 \cdots A_N}\leq 1$, for all $A_1,A_2, \ldots, A_N$.

The probabilities $P(a_1,\ldots, a_N|A_1,\ldots,A_N)$  in (\ref{expectation}) are determined using the different kinds of correlations we have previously defined. If they are of the local form (\ref{localdistr})  we denote the product expectation values they give rise to by $\av{A_1A_2 \cdots A_N}_{\textrm{lhv}}$, and analogously for other types of correlation. This is captured in  table \ref{tableexpect}. 
\begin{table}[!h]
\begin{centerline}{
\begin{tabular}{ |c|c|c|}
\hline &&\\ [-2ex]
type&notation& $P(a_1,\ldots, a_N|A_1,\ldots,A_N)$ in (\ref{expectation}) given by\\[0.5ex]
\hline\hline &&\\ [-2ex]
no-signaling &$\av{A_1A_2 \cdots A_N}_{\textrm{ns}}$&(\ref{nosignalingdistr})\\[0.5ex]
local &$\av{A_1A_2 \cdots A_N}_{\textrm{lhv}}$&(\ref{localdistr})\\[0.5ex]
partially-local& $\av{A_1A_2 \cdots A_N}_{\textrm{plhv}}$&(\ref{partiallocaldistr})\\[0.5ex]
quantum &$\av{A_1A_2 \cdots A_N}_{\textrm{qm}}$&(\ref{quantumcorre})\\[0.5ex]
\hline
\end{tabular}}
\end{centerline}
\caption{The different kinds of product expectation values that arise from the different kinds of correlations.} 
\label{tableexpect}
\end{table}

\forget{
\begin{table}[!h]
\begin{centerline}{
\begin{tabular}{ |c|c|c|}
\toprule
type&notation& $P(a_1,\ldots, a_N|A_1,\ldots,A_N)$ in (\ref{expectation}) given by\\
\midrule
no-signaling &$\av{A_1A_2 \cdots A_N}_{\textrm{ns}}$&(\ref{nosignalingdistr})\\
local &$\av{A_1A_2 \cdots A_N}_{\textrm{lhv}}$&(\ref{localdistr})\\
partially-local& $\av{A_1A_2 \cdots A_N}_{\textrm{plhv}}$&(\ref{partiallocaldistr})\\
quantum &$\av{A_1A_2 \cdots A_N}_{\textrm{qm}}$&(\ref{quantumcorre})\\
\bottomrule
\end{tabular}}
\end{centerline}
\caption{The different kinds of expectation values that arise from the different kinds of correlations.} 
\end{table}
}

We will investigate the different possible correlations using Bell-type inequalities  in terms of product expectation values as given in table \ref{tableexpect}. We will not investigate them directly in terms of the joint probabilities.  
The main reason for this is that using the product expectation values simplifies the investigation considerably.  For example, consider the case of two parties that each measure two dichotomous observables each. We denoted them as $A_1,A_1'$ and $A_2,A_2'$ respectively, with outcomes $a_1,a_1'$ and $a_2,a_2'$.  Instead of dealing with the $16$-dimensional space of vectors  with components 
$P(a_1,a_2|A_1,A_2), P(a_1',a_2|A_1,A_2), \ldots, P(a_1',a_2'|A_1',A_2')$ we only have to deal with  the $4$-dimensional vectors that have as components the quantities $\av{A_1,A_2}$,$\av{A_1,A_2'}$,$\av{A_1',A_2}$,$\av{A_1',A_2'}$. To transform a vector from the $16$-dimensional space  to its corresponding $4$-dimensional space, one needs to perform a projection as given in (\ref{expectation}). It is known that the projection of a convex polytope is always a convex polytope \cite{masanes02}. Therefore, the convex polytopes we have considered previously for general, no-signaling, partially-local and local correlations in the higher dimensional joint probability space correspond to convex polytopes in the lower dimensional space of product expectation values. The set of vectors with components $\av{A_1,A_2}, \av{A_1,A_2'},\av{A_1',A_2},\av{A_1',A_2'}$  that are attainable by general, no-signaling, partially-local and local correlations are thus also characterized by a finite set of extreme points and corresponding facets. 

Dealing with the expectation values $\av{A_1A_2 \cdots A_N}$ is much simpler than dealing with the joint probabilities 
$P(a_1,\ldots, a_N|A_1,\ldots,A_N)$, although in general, the projection (\ref{expectation}) is not structure preserving. For example some non-local correlations could be projected into locally achievable expectation values of products of observables.
But for the case of two parties that  each choose two dichotomous observables, as in the set-up of the CHSH inequality, this does not happen. Indeed,  in the next subsection we will see that the CHSH inequalities describe all non-trivial facets of the local polytope.
  The $4$-dimensional vectors with components $\av{A_1,A_2}, \av{A_1,A_2'},\av{A_1',A_2}, \av{A_1',A_2'}$  and the $16$-dimensional vectors  with components $P(a_1,a_2|A_1,A_2), P(a_1',a_2|A_1,A_2), \ldots, P(a_1',a_2'|A_1',A_2')$ thus contain the same information concerning the existence of a LHV model accounting for them.

For simplicity, in this dissertation we study the correlations 
in the lower dimensional space of product expectation values, despite the fact that some information about the correlations might be lost\footnote{There is a sole exception, however.  For the case of two parties that each choose two dichotomous observables the no-signaling polytope in the four-dimensional space of product expectation values has only trivial facets.  
We will we therefore consider a larger dimensional space  
 in order to obtain Bell-type inequalities that are non-trivial for the no-signaling correlations. We will comment further on this in section \ref{chsintrotech}.}. We thus consider Bell-type inequalities that denote halfplanes in this space. For this purpose it is useful to define the so-called Bell polynomials. These are linear combinations of products of $N$ observables, one for each party, and have the generic form
\begin{align}\label{bellpoly}
B_N(\vec{c})= \sum_{j_1,\ldots,j_N} c(j_1,\ldots,j_N) A_1^{j_1}\cdots A_N^{j_N},
\end{align}
where the coefficients\footnote{To avoid confusion we note that $j_1$,$j_2$, etc., are not some numbers that indicate an exponent but labels that distinguish various measurement settings for parties $1$,$2$, etc. (i.e., the observables  $A_i^{j_i}$ for party $i$ are different for each $j_i$).}  $c(j_1,\ldots,j_N)$ are taken to be real numbers and together make up a vector $\vec{c}$ in a real dimensional space of dimension $\prod_i m_{A_i}$.   For example, for the case of two parties and two observables per party (i.e., $j_1,j_2\in\{1,2\}$ one obtains the polynomial $c(1,1)A_1^1A_2^1 +c(1,2)A_1^1A_2^2+c(2,1)A_1^2A_2^1+c(2,2)A_1^2A_2^2$, where the coefficients $c(1,1),\ldots, c(2,2)$ are still to be specified.
The quantum counterpart of the Bell polynomials, where the observables are POVM operators, will be called Bell operators.

Bell-type inequalities are  now obtained by finding non-trivial numerical  bounds $I^{N,\vec{c}}>0$ on the expectation value of $ B_N(\vec{c})$, denoted as $\av{B_N(\vec{c})}$, for each of the different types of correlations defined above. Because of linearity of the mean $\av{B_N(\vec{c})}$ can be expressed in terms of the different expectation values $\av{A_1A_2 \cdots A_N}$ of table \ref{tableexpect} for the different types of correlation.
  For example, a Bell-type inequality for local correlations reads
\begin{align}\label{lhvbell}
|\av{B_N(\vec{c})}_{\textrm{lhv}}|=| \sum_{j_1,\ldots,c_N} c(j_1,\ldots,c_N) \av{A_1^{j_1}\cdots A_N^{j_N}}_{\textrm{lhv}}| \leq I_{\textrm{lhv}}^{N,\vec{c}},
~~\forall \,A_1^{j_1},\ldots, A_N^{j_N},\end{align}
and analogous for the other types of correlations  so as to give table  \ref{tableBellineq}. These Bell-type inequalities will be called no-signaling, partially-local, local, and quantum Bell-type inequalities. 
\begin{table}[!h]
\begin{centerline}{
\begin{tabular}{ |c|c|c|}
\hline
&\\ [-2ex]
type of correlation&notation of Bell-type inequality\\ [0.5ex]
\hline\hline&\\ [-2ex]
no-signaling &$|\av{B_N(\vec{c})}_{\textrm{ns}}|\leq I_{\textrm{ns}}^{N,\vec{c}}$ \\ [0.5ex]
local &$|\av{B_N(\vec{c})}_{\textrm{lhv}}|\leq I_{\textrm{lhv}}^{N,\vec{c}}$\\ [0.5ex]
partially-local&$|\av{B_N(\vec{c})}_{\textrm{plhv}}|\leq I_{\textrm{plhv}}^{N,\vec{c}}$\\ [0.5ex]
quantum &$|\av{B_N(\vec{c})}_{\textrm{qm}}|\leq I_{\textrm{qm}}^{N,\vec{c}}$\\ [0.5ex]
\hline
\end{tabular}}
\end{centerline}
\caption{The notation of Bell-type inequalities for the different kinds of correlations.} \label{tableBellineq}
\end{table}

In order to obtain Bell-type inequalities one thus has to specify the vector $\vec{c}$ of coefficients  
$c(j_1,\ldots,c_N)$ as well as one or more of the bounds $I_{\textrm{ns}}^{N,\vec{c}}$, $I_{\textrm{lhv}}^{N,\vec{c}}$, $I_{\textrm{plhv}}^{N,\vec{c}}$, $I_{\textrm{qm}}^{N,\vec{c}}$.
This latter task is obtained by maximizing the expectation value of the Bell polynomial while obeying the restrictions that define a specific type of correlation. For example, to obtain $I_{\textrm{lhv}}^{N,\vec{c}}$ one must maximize $|\av{B_N(\vec{c})}_{\textrm{lhv}}|$ with the restriction that (\ref{localdistr}) must be obeyed for the joint probabilities that are used to obtain the expectation values $\av{A_1^{j_1}\cdots A_N^{j_N}}_{\textrm{lhv}}$.

Let us denote the absolute maximum of the expression (\ref{bellpoly}) by  $|B_N(\vec{c})|_{\textrm{max}}$  (this is also called the `algebraic maximum' or the `algebraic bound', but we will not follow this terminology).   
General unrestricted correlations  always exist that attain this absolute maximum since one can always choose each $\av{A_1^{j_1}\cdots A_N^{j_N}}$ to be either $+1$ or $-1$, depending of the sign of the coefficient $c(j_1,\ldots,c_N)$ so that it contributes positively to $\av{B_N(\vec{c})}$ so that $\av{B_N(\vec{c})}=|B_N(\vec{c})|_{\textrm{max}}$.

It remains to indicate what is meant by a non-trivial bound.  A non-trivial bound is any value $I_{\textrm{ns}}^{N,\vec{c}}$, $I_{\textrm{lhv}}^{N,\vec{c}}$, $I_{\textrm{plhv}}^{N,\vec{c}}$, $I_{\textrm{qm}}^{N,\vec{c}}$ that is strictly smaller than the absolute maximum $|B_N(\vec{c})|_{\textrm{max}}$. A bound is called a tight bound when it can be reached by the correlations under study.
Even more desirable would be obtaining a so-called tight Bell-type inequality. The tight inequalities  correspond to facets (\ref{facet}) of the relevant correlation polytopes in the larger joint probability space when the expectation values in the Bell inequality are expressed in terms of the joint probability distributions via the inverse of the projections (\ref{expectation}).  Violating a tight Bell-type inequality means precisely that the point lies above the facet, i.e., outside of the polytope\footnote{A possible confusion may arise here.  Non-trivial Bell-type inequalities are possible that can be saturated by some extremal correlations (of the type under study), but which are nevertheless not indicating facets of the relevant correlation polytope. The possible confusion arises because these inequalities can be said to be `tight' in the sense of not having a tighter upper bound.  However, we will in general not call such inequalities tight Bell-type inequalities because they do not indicate a facet. For a facet it is necessary that at least $d$ affinely independent extreme points lie on the facet, and not less (cf. footnote \ref{affine}).}. A complete set of tight Bell-type inequalities for a specific type of correlation thus gives precisely all facets of the corresponding correlation polytope. This of course does not hold for the quantum case  whose set of correlations (i.e., the quantum body)  is not a polytope. However since this set is still convex it can be described by an infinite set of bounding hyperplanes, each of which is described by a corresponding Bell-type inequality that has a tight bound. 

In this dissertation many new non-trivial bounds are obtained for novel Bell-type inequalities (of which some are tight) for different types of correlations.

\subsubsection{Bi-partite example: the CHSH inequality}\label{chsintrotech}

The best-known Bell-type inequality is the CHSH inequality for local correlations \cite{chsh} that assumes a situation of two parties and two dichotomous observables per party (with possible outcomes $\pm1$). 
We will first review this well-known result after which we consider this inequality when evaluated using quantum and no-signaling correlations.

Consider the CHSH polynomial where $c(1,1)=c(1,2)=c(2,1)=-c(2,2)=1$ in \eqref{bellpoly}:\forget{(for $N=2$, and two dichotomous observables per party):}
\begin{align} \label{CHSHpolynomial}
B_{\textrm{chsh}}=A_1A_2 +A_1A_2' +A_1'A_2 -A_1'A_2',
\end{align}  where $A_1$, $ A_1'$ denote the two different observables for party $1$,  and $A_2$, $ A_2'$ those for party $2$. 
 The product expectation values are easily obtained, e.g., $\av{A_1A_2}= P(+1,+1|A_1,A_2) + P(-1,-1|A_1,A_2)-P(+1,-1|A_1,A_2) -P(-1,+1|A_1,A_2)$, etc.  
 
 \subsubsection*{Local correlations}
 \noindent\citet{chsh} showed that all local correlations obey the tight bound\forget{\footnote{An equivalent and  frequently used way of writing the CHSH inequality is  $|\av{A_1A_2}_{\textrm{lhv}}
 +\av{A_1A_2'}_{\textrm{lhv}}|
 +|\av{A_1'A_2}_{\textrm{lhv}}
 -\av{A_1'A_2'}_{\textrm{lhv}}|\leq2$.\forget{For a proof see \cite{redhead}.}}} 
\begin{align}\label{chshineqintro}
|\av{B_{\textrm{chsh}}}_{\textrm{lhv}}|&=|\av{A_1A_2 +A_1A_2' +A_1'A_2 -A_1'A_2'}_{\textrm{lhv}}|\nn\\
&=|\av{A_1A_2}_{\textrm{lhv}}
 +\av{A_1A_2'}_{\textrm{lhv}}
 +\av{A_1'A_2}_{\textrm{lhv}}
 -\av{A_1'A_2'}_{\textrm{lhv}}|
\leq2.
\end{align}
The local polytope is the subset in the four dimensional real space $\mathbb{R}^4$ of all vectors  $(\av{A_1A_2}, \\\av{A_1A_2'}, \av{A_1'A_2},  \av{A_1'A_2'})$\forget{(i.e., whose elements are all product expectation values)} that can be attained by local correlations. It is the convex hull in $\mathbb{R}^4$ of the 8 extreme points (vertices) that are of the form \begin{align}\label{extrloc}
(1,1,1,1),(-1,-1,-1,-1),(1,1,-1,-1),(-1,-1,1,1),\nn\\(1,-1,1,-1),(-1,1,-1,1),(1,-1,-1,1),(-1,1,1,-1).
\end{align}
 This polytope is the four-dimensional octahedron and has 8 trivial facets as well as 8 non-trivial ones. The trivial ones are the inequalities of the form 
 \begin{align}\label{triv}
 -1\leq\av{A_1A_2}_{\textrm{lhv}}\leq1 ,~~~~
  -1\leq\av{A_1A_2'}_{\textrm{lhv}}\leq 1,\nn\\
   -1\leq\av{A_1'A_2}_{\textrm{lhv}}\leq 1,~~~~
    -1\leq\av{A_1'A_2'}_{\textrm{lhv}}\leq 1.
  \end{align}The non-trivial facets are all equivalent to the CHSH inequality (\ref{chshineqintro}), up to trivial symmetries, giving a total of 8 equivalent inequalities, as first proven by \citet{fine}, cf. \citet{collinsgisin}. These eight are \cite{barrett05}: 
  \begin{align}\label{8chsh}
(-1)^\gamma \av{A_1A_2}_{\textrm{lhv}} +&  (-1)^{\beta +\gamma}\av{A_1A_2'}_{\textrm{lhv}} +\nn\\ &
(-1)^{\alpha +\gamma} \av{A_1'A_2}_{\textrm{lhv}} +(-1)^{\alpha+\beta+\gamma+1}\av{A_1'A_2' }_{\textrm{lhv}}\leq 2,
\end{align} with $\alpha,\beta,\gamma \in \{0,1\}$. These are the necessary and sufficient conditions for a LHV model to exist.
  Note that for the bi-partite case there is no distinction between partially-local and local correlations, and hence the partially-local polytope and the local polytope coincide.  

 \subsubsection*{Quantum correlations}\noindent
 In terms of the CHSH polynomial a non-trivial tight quantum bound is given by the Tsirelson inequality \cite{cirelson} 
\begin{align}\label{tsirelsonintro}
|\av{B_{\textrm{chsh}}}_{\textrm{qm}}| \leq 2\sqrt{2},
\end{align}
which can be reached by entangled states.
This shows that the local polytope is strictly contained in the quantum body, which can be regarded a concise statement of Bell's theorem \cite{bell64}. In Part II we will further investigate quantum correlations using the CHSH polynomial and obtain some interesting new results.

 \subsubsection*{No-signaling correlations}\noindent
No-signaling correlations are able to violate the Tsirelson inequality (\ref{tsirelsonintro}). A well known example of this is the joint distribution known as the Popescu-Rohrlich distribution \cite{prbox}, also known as the PR box, defined by:
\begin{alignat}{2}\label{prdistr}
&P(a_1,a_2|A_1,A_2) = \frac{1}{2}\delta_{a_1,a_2}~,~~~~~~
&&P(a_1,a_2'|A_1,A_2') = \frac{1}{2}\delta_{a_1,a_2'}~,\nn\\
&P(a_1',a_2|A_1',A_2) = \frac{1}{2}\delta_{a_1',a_2}~,
&&P(a_1',a_2'|A_1',A_2') = \frac{1}{2}- \frac{1}{2}\delta_{a_1',a_2'}~,
\end{alignat}
This correlation gives $\av{B_{\textrm{chsh}}}_{\textrm{ns}} = 4$, which is the absolute maximum $|B_{\textrm{chsh}}|_{\textrm{max}}$. In fact, it is an extreme point of the no-signaling polytope for the case of two dichotomous observables per party. Furthermore, all the no-signaling extreme points of this polytope have a such a form. They can all be written as \cite{barrett05}
\begin{align}\label{extremens}
P(a_1,a_2|A_1,A_2)  = \left \{ \begin{array}{l} 1/2 ,~~\textrm{if}~~a_1\oplus a_2=A_1A_2 , \\  0,~~\textrm{otherwise}, \end{array} \right.
\end{align}
where $\oplus$ denotes addition modulo $2$. Here the outcomes $a_1,a_2$ and the settings $A_1,A_2$ are labeled by $0$ and $1$ respectively\forget{, i.e., $a_1,a_2,A_1,A_2\in\{0,1\}$}, where $0$ corresponds to outcome $+1$ and the unprimed observable respectively; and $1$ corresponds to outcome $-1$ and the primed observable respectively. It is not hard to see that (\ref{prdistr}) is indeed a member of the class (\ref{extremens}).

There is a one-to-one correspondence between the non-local extreme points and the facets of the local polytope that are given by the CHSH inequalities  \eqref{8chsh}. To show this we note that the CHSH inequalities in the larger 16-dimensional space of correlations are equal to:
\begin{align}\label{chshlarger}
1\leq P(a_1=a_2)+P(a_1=a_2')+P(a_1'=a_2)+P(a_1'\neq a_2')\leq3
\end{align}
where  $P(a_1=a_2):= P(+1,+1|A_1,A_2)+P(-1,-1|A,B)$,  $P(a_1'\neq a_2'):= P(+1,-1|A_1',A_2')+P(-1,+1|A_1',A_2')$, etc.
  This gives two inequalities and the other 6 are obtained by permuting  the  primed and unprimed quantities for system 1 and 2 respectively. A total of 8 local extreme points saturate each of these inequalities. They are deterministic, i.e., $P(+1,+1|A_1,A_2)=P(+1|A_1)P(+1|A_2)$, etc., where $P(+1|A_1)$ and $P(+1|A_2)$ are either $0$ or $1$. 
Because these 8 extreme points are also linearly independent the inequalities \eqref{chshlarger} (and the equivalent ones) give the facets of the 8-dimensional local polytope in the larger space of correlations.

The 8 local extreme points that lie on each of the local facets are also extreme points of the no-signaling polytope. Only one extreme no-signaling correlation \eqref{extremens} is on top of each local facet, and it violates the CHSH inequality associated to this local facet maximally\forget{
its associated CHSH inequality is maximally violated by this no-signaling correlation} \cite{barrett05}. This is the one-to-one correspondence referred to above. This is depicted in Figure \ref{figchshnl}. 
\begin{figure}[!h]
\includegraphics[scale=0.6]{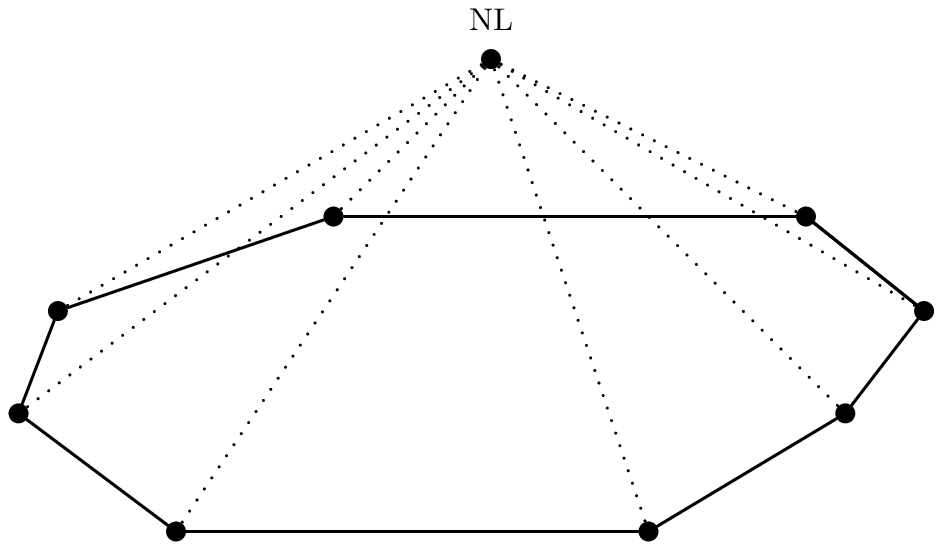}
\caption{The local facet is the hyperplane through the closed line connecting the 8 local extreme points.  Above this facet exactly one no-signaling extreme point is situated (after \citet{acin06}).} \label{figchshnl}
\end{figure}

The non-trivial facets of the no-signaling polytope are given by the defining equalities on the left hand side of \eqref{22nosig} and read in the dichotomic case 
\begin{align}\label{2nosig2}
\sum_{a_2=+1,-1}P(a_1,a_2|A_1,A_2)&=\sum_{a_2=+1,-1}P(a_1,a_2|A_1,A_2'),
\end{align}
for $a_1=+1,-1$, and analogous equalities are obtained by permutations of settings and outcomes so as to give a total of eight equalities.
The tight Bell-type inequalities   corresponding to \eqref{2nosig2} are easily obtained: 
\begin{subequations}
\label{ineqnos}\begin{align}
\sum_{a_2=+1,-1}P(a_1,a_2|A_1,A_2)\leq \sum_{a_2=+1,-1}P(a_1,a_2|A_1,A_2'),\\
\sum_{a_2=+1,-1}P(a_1,a_2|A_1,A_2)\geq\sum_{a_2=+1,-1}P(a_1,a_2|A_1,A_2').
\end{align}
\end{subequations}
In terms of expectation values we obtain non-trivial inequalities for the marginals\footnote{\label{wrongsignaling}In case 
 no-signaling obtains we can define $\av{A}_{\textrm{ns}}:=\av{A}_{\textrm{ns}}^{B}=\av{A}_{\textrm{ns}}^{B'}$  because the marginal for party 1 does not depent on the setting chosen by party 2 (cf. \eqref{22nosig}).  Inserting this in \eqref{ineqnos} gives the trivial inequalities
$\av{A}_{\textrm{ns}}\leq\av{A}_{\textrm{ns}}$ and $\av{A}_{\textrm{ns}}\geq\av{A}_{\textrm{ns}}$. However, this misses the point. Because the non-trivial tight no-signaling Bell-type inequalities are supposed to discern the no-signaling correlations from more general correlations one must allow for the most general framework in which signaling is in principle possible., i.e, where the marginals can depend on the settings corresponding to the outcomes that are no longer considered.  This cannot be excluded from the start.}:\forget{\footnote{\label{trivnontrivmarginal}
An alternative formulation that uses a different notion of marginal expectation value is the following.
Let us define
\begin{align}
\label{altern}
P(a|A)^{B}:=\sum_{b}P(a,b|A,B),~~~\textrm{and}~~~P(a|A)^{B'}:=\sum_{b}P(a,b|A,B').
\end{align}
In terms of expectation values one obtains the no-signaling inequalities:
\begin{align}
\label{nontrivial}
\av{A}^{B}\leq\av{A}^{B'}, ~~~\textrm{and} ~~~ \av{A}^{B}\geq\av{A}^{B'},
\end{align}
 where we have defined the marginal expectation value  $\av{A}^{B}:=\sum_{a} P(a|A)^{B}$  with $P(a|A)^{B}$ as defined in \eqref{altern}. Analogous inequalities follow after permutations of the settings. Note thate these inequalities are non-trivial.

However, in this formulation the marginal probabilities $P(a|A)^{B}$ and marginal expectation values $\av{A}^{B}$ now have an explicit dependence on the far-away setting. This is an unwelcome feature when considering no-signaling correlations. Whereas it is to be expectated that to evaluate product expectation values (e.g., $\av{AB}$) one needs to use global information, i.e., from both measurement stations, it is not to be expected that in evaluating marginal expectation values one needs such 
global information, local statistics should suffice.\forget{Of course,  $\av{A}^{B}$ and $\av{A}$ are fundamentally different objects. But it is the second that is generally taken to be the marginal expectation value.} Furthermore, no-signaling ensures that marginals do not depend on far-away settings chosen, and it is desirable that this is reflected in the definition too. This favors adopting the quantity $\av{A}$ over $\av{A}^{B}$ to be the marginal expectation value; something we will do here. This follows standard practise in the literature.

Only in a single instance, in section \ref{discerningno-signalingsection}, we adopt a signaling context and therefore can no longer define marginals independent of the far-away setting. There we must thus resort to the quantities  $\av{A}^{B}$, etc.
}}
\forget{\begin{align}
\label{Mnos}
\av{A_1}^{A_2}\leq\av{A_1}^{A_2'} , ~~~\textrm{and} ~~~ \av{A_1}^{A_2}\geq\av{A_1}^{A_2'},
\end{align}
where we have used $\av{A_1}^{A_2}:=\sum_{a_1} a_1P(a_1|A_1)^{A_2}$  and $P(a_1|A_1)^{A_2}$ as defined in \eqref{M1}. 
}\begin{align}
\label{Mnos}
\av{A_1}_{\textrm{ns}}^{A_2}\leq\av{A_1}_{\textrm{ns}}^{A_2'} , ~~~\textrm{and} ~~~ \av{A_1}_{\textrm{ns}}^{A_2}\geq\av{A_1}_{\textrm{ns}}^{A_2'},
\end{align}
where we have used $\av{A_1}_{\textrm{ns}}^{A_2}:=\sum_{a_1} a_1P(a_1|A_1)^{A_2}$  and $P(a_1|A_1)^{A_2}$ as defined in \eqref{M1} and obeying the  no-signaling constraint \eqref{22nosig}.

If we consider product expectation values instead of the marginal ones we only obtain trivial inequalities. In the space $\mathbb{R}^4$ of vectors with components $(\av{A_1A_2},\\ \av{A_1A_2'}, \av{A_1'A_2}, \av{A_1'A_2'})$ the 8 no-signaling extreme points \eqref{extremens} give the following vertices   \begin{align}\label{extrns}
(-1,1,1,1),(1,-1,-1,-1),(1,-1,1,1),(-1,1,-1,-1),\nn\\(1,1,-1,1),(-1,-1,1,-1),(1,1,1,-1),(-1,-1,-1,1).
\end{align} In this space the no-signaling polytope is the convex hull of the 16 local extreme points (\ref{extrloc}) and of those given by (\ref{extrns}). Its facet inequalities are just the 8 trivial inequalities in (\ref{triv}) and therefore it is in fact just the four-dimensional unit cube \cite{pitowsky08}. 
We thus obtain only trivial facet inequalities.

In the next chapter, section \ref{discerningno-signalingsection},  we derive non-trivial no-signaling inequalities in terms of the product and marginal expectation values. Although these cannot be tight inequalities, i.e., they cannot be facets of the no-signaling polytope, we show them to do useful work nevertheless.
%
In order to obtain these inequalities we will have to consider a larger dimensional space than the four-dimensional of vectors $(\av{A_1A_2}, \av{A_1A_2'},\\\av{A_1'A_2}, \av{A_1'A_2'})$.\forget{We show that the eight dimensional space that also includes the marginal expectation values $\av{A_1}$, $\av{A_1'}$, $\av{A_2}$, $\av{A_2'}$ already suffices.}  This is the only instance in this dissertation where we will have to go outside the smaller space of product expectation values. 
\forget{or using a terms of expectation values these become: $\av{A_1}_{A_2}=\av{A_1}_{A_2'}$, which, from an experimental point of view is problematic.  In an experiment one merely infers the expectation values  }

\subsubsection*{Comparing the different correlations}
\noindent
The no-signaling correlation (\ref{prdistr}) was discovered already in 1985 independently by \citet{khalfin} and \citet{rastall} who also showed it to give the algebraic maximum for the CHSH expression. However, \citet{prbox} presented this correlation in order to ask an interesting question, not asked previously: Why do quantum correlations not violate the CHSH expression by a larger amount? Such a larger violation would be compatible with no-signaling, so why is quantum mechanics not more non-local? This paper by Popescu and Rohrlich marked  the start of a new research area, that of investigating no-signaling distributions and their relationship to quantum mechanics.  

For the bi-partite case and two dichotomous observables per party the above results show how the different sets of correlations are related: Since some quantum correlations turn out to be non-local in the sense of not being of the local form (\ref{localdistr}), the set of quantum correlations is a proper superset of the set of local correlations. But it is a proper subset of the set of no-signaling correlations which are able to violate the Tsirelson inequality up to the absolute maximum.  
In summary, the CHSH polynomial gives inequalities that give a non-trivial tight bound for local and quantum correlations but not so for no-signaling correlations. Indeed,  the latter can reach the absolute maximum  $|B_{\textrm{chsh}}|_{\textrm{max}}$.

A useful way of visualizing the bounds on the CHSH inequality for the different types of correlations ---one that we will frequently use in this dissertation--- is the following. Consider another Bell-type polynomial 
$B_{\textrm{chsh}}'$ that is obtained from $B_{\textrm{chsh}}$ by permuting the primed and unprimed observables so as to give  $B_{\textrm{chsh}}'=A_1'A_2' +A_1'A_2 +A_1A_2' -A_1A_2$. 
For this Bell-type polynomial the same bounds on the Bell-type inequalities for the different types of correlations are obtained. We can now depict the accessible regions for the correlations in the $(\av{B_{\textrm{chsh}}},\av{B_{\textrm{chsh}}'})$-plane, as in \ref{chshplaatjeintro}. This figure shows the inclusion relations mentioned above. We will use many similar figures later on. They provide a useful tool to compare the different correlations via the bounds on Bell-type inequalities they admit.
\begin{figure}[h]
\includegraphics[scale=1]{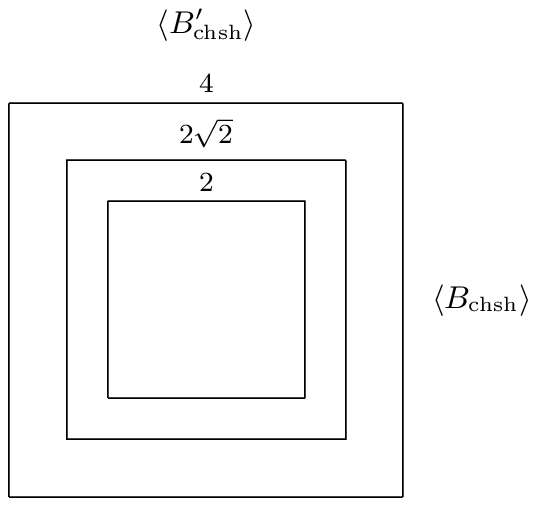}
\caption{Comparing the regions  in the $(\av{B_{\textrm{chsh}}},\av{B_{\textrm{chsh}}'})$-plane. General unrestricted and no-signaling correlations are confined to the largest square, quantum correlations to the middle square, and local correlations to the smallest square.}
\label{chshplaatjeintro}
\vspace{\baselineskip}\end{figure} \forget{
 \setlength{\unitlength}{0.20 mm}
 \begin{figure}[h]\begin{center}
\begin{picture}(200,260)(-100,-100)
\linethickness{0.133mm} 
\put(-100,-100){\line(0,1){200}}
\put(-100,-100){\line(1,0){200}}
 \put(100,100){\line(0,-1){200}}
\put(100,100){\line(-1,0){200}}
 \put(-71,-71){\line(0,1){142}}
 \put(-71,-71){\line(1,0){142}}
 \put(71,71){\line(0,-1){142}}
\put(71,71){\line(-1,0){142}}
\put(-50,-50){\line(0,1){100}}
 \put(-50,-50){\line(1,0){100}}
 \put(50,50){\line(0,-1){100}}
\put(50,50){\line(-1,0){100}}
  \put(140,0){\makebox(0,0){$\av{B_{\textrm{chsh}}}$}}
 \put(0,140){\makebox(0,0){$\av{B_{\textrm{chsh}}'}$}}
 \put(0,110){\makebox(0,0){\footnotesize{$4$}}}
 \put(0,85){\makebox(0,0){\footnotesize{$2\sqrt{2}$}}}
   \put(0,60){\makebox(0,0){\footnotesize{$2$}}}
 \end{picture}\end{center}
\caption{Comparing the regions  in the $(\av{B_{\textrm{chsh}}},\av{B_{\textrm{chsh}}'})$-plane. General unrestricted and no-signaling correlations are confined to the largest square, quantum correlations to the middle square, and local correlations to the smallest square.}
\label{chshplaatjeintro}
\vspace{\baselineskip}
\end{figure}}

\forget{Alternative forms of the CHSH ineq.: In terms of joint probabilities 
a) CH form:
b) Gill: $P(a_1\geq a_2|A_1,A_2)\leq P(a_1\geq a_2'|A_1,A_2')+
P(a_1'\geq a_2|A_1',A_2)+
P(a_1'\geq a_2'|A_1',A_2')
 $
 3)  $1\leq uitdrukking \leq 3$
 }

\subsubsection{Multipartite Bell-type inequalities}

We briefly review some known multi-partite Bell-type inequalities relevant for this dissertation for the four types of correlations we have distinguished. This gives an opportunity to further  introduce some of the results that are obtained in this dissertation.

(1) \emph{Local correlations}: For two dichotomic observables per party the full set of necessary and sufficient Bell-type inequalities for local correlations is known. These are the Werner-Wolf-\.Zukowski-Brukner(WWZB) inequalities \cite{wernerwolf2, zukowskibrukner}. They give all facets of the local polytope. A special form of the WWZB inequalities are the so called Mermin-type inequalities \cite{mermin,roysingh91,ardehali92,belinskiiklyshko93} which were the first multi-partite Bell-type inequalities for all $N$ that gave bounds on local correlations and which were shown to be exponentially violated by quantum correlations.  For more than two outcomes and settings, many partial results exist, but no full set of local inequalities is found.\forget{Obtaining Bell-type inequalities  for two parties where one allows for more than two observables and outcomes per party.} For a recent overview, see \citet{gisinbell}.

(2) \emph{Partially-local correlations}:
For $N$=$3$ \citet{svetlichny} obtained partially-local inequalities for two dichotomous observables per party. In this dissertation we give the generalization of this result to $N$-parties, thereby obtaining the so-called Svetlichny inequalities for all $N$.

(3) \emph{Quantum correlations}: 
The quantum body in the space of multi-partite correlations is not well investigated.  For the two-qubit case some results have been obtained for two dichotomous observables and two parties: the well-known Tsirelson inequality and some non-linear inequalities \cite{navascues,uffink,pitowsky08} that strengthen this inequality.  For more observables with a finite number of outcomes for the case of two parties \citet{navascues} gave a hierarchy of conditions where each condition is formulated as a semi-definite program. 

For more parties but two dichotomous observables per party one can often use the Bell-polynomials that feature in Bell-type inequalities for local correlations to obtain non-trivial inequalities for the set of quantum correlations as well.  If the quantum bound on the expectation value of the Bell-polynomial is less than the absolute maximum of polynomial, one has a non-trivial inequality for bounding the quantum correlations. Only a subset of the Bell-polynomials used to obtain the WWZB inequalities give such inequalities, but not all of them. We will show that the Bell-polynomials of the generalized Svetlichny inequalities for partial locality also give non-trivial quantum bounds. Some non-linear strengthening of these quantum bounds are known \cite{uffink,roy,nagata2002}. In this dissertation this will be strengthened even further using state-dependent upper bounds.

Apart from using linear or non-linear Bell-type inequalities in terms of Bell-polynomials where all parties are involved, we will also investigate another fruitful way of studying the different kinds of correlations via the question whether the correlations can be shared.  Here one focuses on subsets of the particles and asks whether their correlations can be extended to parties not in the original subsets. This can be done either directly in terms of joint probability distributions or in terms of relations between Bell-type inequalities that hold for different, but overlapping subsets of the parties involved. When a correlation cannot be shared it  is said to have monogamy constraints.\forget{In chapter \ref{chapter_monogamy} this will be further introduced.}  For three-partite quantum and no-signaling correlations such monogamy is shown to exist using a Bell-type inequality. These monogamy results give non-trivial bounds on multi-partite quantum and no-signaling correlations thereby discriminating them from each other and from more general correlations.
Because we will use this technique only in a single chapter, chapter \ref{chapter_monogamy}, we will not introduce the technical details of this issue here, but will do so in the introduction to that chapter.

(4) \emph{No-signaling correlations}: The facets of the convex no-signaling polytope follow from the defining 
conditions \eqref{nosignalingdistr} on the space of correlations\\  $P(a_1,\ldots, a_N|A_1,\ldots,A_N)$. 
In terms of marginal expectation values  they are of the form: for each $k\in\{1,\ldots,N\}$
 \begin{align}
    \av{A_1,\ldots,A_{k-1},A_{k+1},\ldots, A_N}^{A_k}=\av{A_1,\ldots,A_{k-1},A_{k+1},\ldots, A_N}^{A_k'},
     \end{align}
     for all settings $A_1,\ldots,A_{k-1},A_k,A_k',A_{k+1},\ldots,A_N$. This ensures that all marginal expectation values are independent of the settings that are no longer considered. The tight Bell-type inequalities corresponding to this equality are easily obtained: replace all occurrences of $=$ by $\leq$ and $\geq$.
     
          This procedure gives only restrictions on the marginal expectation values. However, it is sometimes useful to have non-trivial no-signaling Bell-type inequalities in terms of the product expectation values  $\av{A_1\cdots A_N}$, despite the fact that they cannot be tight inequalities, i.e., they cannot be facets of the no-signaling polytope.  For $N=2$ and two dichotomous observables per party we will present such non-trivial Bell-type inequalities in the next chapter, section \ref{discerningno-signalingsection}.  Unfortunately, for $N>2$ both the WWZB and the generalized Svetlichny inequalities (which give non-trivial bounds on the local and quantum correlations in terms of product expectation values) do not give non-trivial bounds for no-signaling correlations, since these correlations can attain the absolute maximum of the corresponding Bell-type polynomials.  However, for $N=3$ we will argue in chapter \ref{chapter_monogamy} that an already existing monogamy constraint gives a non-trivial bound for the no-signaling correlations in terms of product expectation values only. 

For completeness we note that non-trivial Bell-type inequalities exist  (i) for specific forms of non-local no-signaling resources \cite{brunner}, but these do not hold for general no-signaling correlations, and (ii) for some specific forms of signaling resources that employ a finite amount of auxiliary communication \cite{tonerbacon}.  In this dissertation we will strive to be as general as possible and therefore do not study such specific no-signaling or signaling resources.

\forget{
\subsubsection{Three-partite case}
ga alle correlaties na.

Of plaatje van convex polytopes voor drie deeltjes geval?

Svetlichny showed.. We will generalise the three partite case.

local: mermin
PLHV: svetlichny
qm case: both mermin ineq+svetlichny ineq.

no-signaling ineq. not known before. Svetlihcny en memrin ineq do not give a non-trivial bound for nosignaling ineq.
}
\subsection{Further aspects of quantum correlations}

The structure of the set of correlations in a general quantum system can be studied in at least two different ways. The first way looks at the state space of quantum mechanics and investigates bounds on the accessible quantum body in the space of quantum correlations (e.g., quantum Bell-type inequalities) as well as the separability and entanglement properties of the states that live in this space.
The second way investigates the non-locality and no-signaling characteristics of the quantum states. In this case one investigates if the quantum mechanical correlations these states give rise to can be described by local, partially-local or no-signaling models.  Bell-type inequalities are the main tool here.
     In this thesis we will use both methods to investigate $N$-partite quantum correlations. 
     
       These two investigations are not independent, as the following example shows.  Consider the CHSH polynomial $B_{\textrm{chsh}}$ and the corresponding local, quantum and no-signaling inequalities $|\av{B_{\textrm{chsh}}}_\textrm{lhv}|\leq 2$,  $|\av{B_{\textrm{chsh}}}_\textrm{qm}| \leq 2\sqrt{2}$ and  $|\av{B_{\textrm{chsh}}}_\textrm{ns}| \leq 4$ respectively, whose bounds are all tight. Since the quantum bound is strictly greater than the local bound but strictly smaller than the no-signaling bound one concludes that states that reach the quantum bound are both no-signaling and non-local.
       Furthermore, it is well known that two types of quantum states exist: entangled states and non-entangled states, i.e., separable states (see the next subsection \ref{sepentangintro} for a formal definition).  The correlations these two types of states give rise to also have different characteristics. For example, entangled states can give rise to non-local correlations in the sense that they violate $|\av{B_{\textrm{chsh}}}_\textrm{qm}| \leq 2$ up to  the Tsirelson bound $2\sqrt{2}$ (cf. (\ref{tsirelsonintro})). But this is never the case for separable states. Indeed, the correlations that separable states give rise to always allow for a LHV model. Violation of the quantum CHSH inequality  $|\av{B_{\textrm{chsh}}}_\textrm{qm}| \leq 2$ 
             is thus sufficient for entanglement detection: it allows for experimentally distinguishing separable from entangled quantum states.  This entanglement detection capability shows another interesting feature of Bell-type inequalities.  

In the multi-partite case many more different types of quantum states exist than just separable and entangled states, such as partially separable states and different kinds of entangled states. The structure of these different kinds of states as well as the correlations these states give rise to will be investigated in chapter \ref{Npartsep_entanglement} and we will give many necessary separability conditions and sufficient entanglement criteria using Bell-type inequalities. However, since the quantum body is not a polytope we do not restrict ourselves to linear Bell-type inequalities. We will therefore also look at quadratic expressions in $\av{A_1\cdots A_N}$ so as to give inequalities that  bound the quantum body. These will be called quadratic Bell-type inequalities.

We will restrict the investigation of linear and quadratic Bell-type inequalities to those cases where each of the $N$ parties measures two different dichotomous observables.  For such a scenario the following important result holds \cite{tonerverstraete}, cf. \cite{masanes05}: the maximum quantum value of any such Bell-type inequality is achieved by a quantum state of $N$-qubits (i.e., $N$ spin-$\frac{1}{2}$ particles, that each have a two dimensional Hilbert space $\H=\mathbb{C}^2$). Furthermore, one can assume  this state to have only real coefficients and that the operators corresponding to the observables are real and traceless. These are self-adjoint operators and they give a so-called projector valued measure (PVM) so one does not need consider the more general POVM operators. Such observables are in fact spin observables in some direction. These can be represented using the Pauli spin observables as follows: $A=\bm{a}\cdot \bm{\sigma}=\sum_i a_i\sigma_i$, with $\|\bm{a}\| =1$, $i=x,y,z$ and $\sigma_x, \sigma_y, \sigma_z$ the familiar Pauli spin operators on $\H=\mathbb{C}^2$.
  Accordingly, the two different types of investigation 
  mentioned above will be performed only for the case of qubits, and for the case of spin observables.    In the next section these investigations will be further introduced.

\subsubsection{On the  (non-)locality, entanglement and separability of quantum states}\label{sepentangintro}
Let us first take a closer look at  entanglement and separability of quantum states, after which we discuss  the locality properties of these states.

A bi-partite quantum state $\rho_\textrm{sep}$  on $\H=\H_1\otimes \H_2$ is separable iff \cite{werner}  it can be written as 
\begin{align}\label{sepintro}
\rho_\textrm{sep}=\sum_ip_i \rho_i^1\otimes\rho_i^2,
\end{align}
with $\sum_ip_i=1$ and $p_i\geq 0$, and where $\rho_i^1$  and $\rho_i^2$ are states for party $1$ and $2$ respectively.
A state is called entangled when it is not separable. 
Entanglement is due to the tensor product structure of a composite 
 Hilbert space and the linear superposition principle of quantum mechanics.
It  has been the focus of a lot of research over the past decade, see \citet{entanglement} for a recent very extensive overview.  

A separable state is supposed to contain only classical correlations. A physical interpretation of separability can be given in terms of the resources needed for the preparation of the state: an entangled state cannot be prepared from two previously non-interacting systems using local operations and classical communication (LOCC operations\footnote{See section \ref{qmholism} for a detailed specification of the class of LOCC operations.}). The LOCC  operations are a subset of all separable operations \cite{entanglement} that can be represented as $S(\rho) =\sum_i L^\dagger \rho L$ with $\sum_i L_i^\dagger L_i \leq \1$ and where $L_i$ is a product of local operations performed by each of the parties, i.e., $L_i=L_i^1\otimes L_i^2\otimes\cdots$, where each $L_i^j$ ($j=1,2,\ldots,N$) is some positive operator that takes states to states. The effect of such a separable operation on a state $\tilde{\rho}=\rho^1\otimes \rho^2$,  that describes the two previously non-interacting systems, is as follows
\begin{align}
S(\tilde{\rho}) &=\sum_i (L_i^1\otimes L_i^2)^\dagger( \rho^1\otimes \rho^2)  (L_i^1\otimes L_i^2)\nn\\ 
&=\sum_i{L_i^1}^\dagger \rho^1L_i^1\otimes{L_i^2}^\dagger \rho^2L_i^2 =\sum_i \tilde{\rho}_i^1\otimes\tilde{\rho}_i^2,
\end{align}
with $\tilde{\rho}_i^1={L_i^1}^\dagger \rho^1L_i^1$ and $\tilde{\rho}_i^2={L_i^2}^\dagger \rho^2L_i^2$ which are not necessarily normalized. The final state is a separable state so separable operations, and consequently LOCC operations, cannot create any entanglement. 

All correlations obtainable using a separable state can be reproduced by local correlations as defined above in \eqref{localdistr}. For completeness, let us prove this for bi-partite correlations. The $N$-partite generalization follows straightforward from the bi-partite proof. Consider the separable state (\ref{sepintro}) and the quantum correlations (\ref{quantumcorre}) this state gives rise to. These can be rewritten as
\begin{align}
P(a_1,a_2|A_1,A_2)=\textrm{Tr}[M_{a_1}^{A_1} \otimes M_{a_2}^{A_2} \rho_\textrm{sep}]&=\sum_i p_i \textrm{Tr}[M_{a_1}^{A_1}\rho_i^1]       \textrm{Tr}[M_{a_2}^{A_2}\rho_i^2]\nn\\&=\sum_i p_i P(a_1|A_1,i)P(a_2|A_2,i),
\end{align}
where $P(a_1|A_1,i)$ is the probability to find outcome $a_1$ when measuring $A_1$ on the state $\rho_i^1$ of party $1$, and analogous for $P(a_2|A_2,i)$. The bi-partite local correlations are $P(a_1,a_2|A_1,A_2)=\int_\Lambda d\lambda p(\lambda) P(a_1|A_1,\lambda)P(a_2|A_2,\lambda)$, as in  (\ref{localdistr}). If one chooses the hidden variable $\lambda$ to be the index $i$ and the distribution $p(\lambda)$ to be the discrete distribution $p_i$, one reproduces the quantum correlations  that the separable state gives rise in terms of local correlations. This ends the proof.

Let us move to non-separable, i.e., entangled states. Suppose a state is entangled, can we say how much entangled it is? This asks for the possibility of quantifying entanglement using some measure. However this appears to very difficult. Already for bi-partite systems many such measures exist. One way to see why there seems not to be a unique measure of entanglement is by noting that entanglement is not an observable. It can not be regarded a physical observable in the sense that there is no self-adjoint operator such that the value of an entanglement measure can be obtained by measuring the expectation value of the operator, for any state of the composite system. This can be seen as follows \cite{mintert}.  Entanglement is invariant under all local unitary operations. That is, for a given state $\ket{\psi_0}$ on $\H=\H_1\otimes \H_2$, all states $\ket{\psi}=U_1\otimes U_2\ket{\psi_0}$ with arbitrary unitary interaction by party $1$ and $2$  have exactly the same entanglement properties. The same holds for mixed states $\rho$.\forget{Since this holds for all states,} Hence,  any observable $E$ that is supposed to quantify entanglement  needs to have the same symmetry  $E= U_1^\dagger \otimes U_2^\dagger \, E \,U_1\otimes U_2$ for arbitrary local unitaries. However, the only operator that has this property is the identity operator $\1$ \cite{mintert}. But that is the trivial observable returning a value of $1$ for all states independent of their entanglement characteristics.

Thus in order to characterize entanglement on needs to measure more than just one observable. We will however  not be concerned with entanglement measures but with the task of determining whether a state is entangled or not. This is the so-called separability problem: Given a certain state how can one determine whether it is entangled? Of course, one may try to determine the state exactly using  full tomography and then try and see if a decomposition of the form (\ref{sepintro})  exists, but this is problematic since no general algorithm exists for such a decomposition.  Only in simple cases a necessary and sufficient criterion for entanglement exists which is the so-called positive partial transposition (PPT) criterion \cite{PPT,horodeckiPPT} that works only up to dimension six of the Hilbert space for the combined system (i.e., two qubits and a qubit and qutrit).  However, the PPT criterion is not experimentally accessible because partial transposition is not a physical operation. Furthermore, full tomography is experimentally very demanding. 
 
Other experimentally accessible necessary separability conditions have therefore been proposed whose violation is a sufficient condition for entanglement detection. These have been termed entanglement witnesses \cite{horodeckiPPT,terhal96,lewenstein,bruss02}. An entanglement witness is a self adjoint operator that upon measurement gives a sufficient criterion for the existence of entanglement. Local Bell-type inequalities are such entanglement witnesses. Indeed,  we have already seen that violation of the local CHSH inequality is sufficient for detection of bi-partite entanglement.  This is the case for all local Bell-type inequalities since, as proven above, the correlations of a separable quantum state can always be reproduced by local correlations. Thus violating the inequalities is sufficient for detecting entanglement. But unfortunately it appears not to be necessary, because not all entangled states can be made to violate a local Bell-type inequality.
 
 This already shows up in violations of the CHSH inequality: all pure entangled states can be made to violate the CHSH inequality (\ref{chshineqintro}) \cite{GisiN,POPROHR}, but for mixed states this is not the case. The latter feature is called hidden non-locality \cite{POPESCU}, because using preprocessing via local filtering techniques it can eventually be made to a violate the CHSH inequality with some non-zero probability. In the multi-partite setting the entanglement  structure already appears to be very different since there even pure entangled states exist that do not violate any of the local Bell-type inequalities from the WWZB set \cite{zukowskibrukner}, that for $N=2$ reduce to the CHSH inequality.

Since it suffices for detecting entanglement of a quantum state to show that a quantum state is non-local, it is important to know when it is non-local. But how  can one show this? One way of proving this is to show that no local model exists for all correlations the quantum state gives rise to. This has been  achieved only for quantum states with high symmetry such as the so-called Werner states \cite{werner}.  One might also ask the weaker question whether no local models exists for all correlations the state gives rise to, given a certain number of measurements and outcomes. This implies the quantum state should  violate any of the local Bell-type inequalities for this case.  But for a given state it is extremely hard to find a local Bell-type inequality and measurements such that the inequality is violated. Only for the two qubits and the CHSH inequality this has been solved analytically by  \citet{H3}.

To end this section we will show that the property of separability of quantum states depends on the chosen decomposition  $\H=\H_1\otimes \H_2$ of the total state space $\H$ (of the combined system) into a tensor product of  $\H_1$ and $\H_2$, the latter being the state spaces of the two subsystems.  Usually such a decomposition is given from the start, but what if only the state space of the combined system is known? Separability of states then might depend on the particular decomposition that is chosen.  Indeed, states exist that are separable with respect to one decomposition 
$\H=\H_1\otimes \H_2$ but that are inseparable with respect to another decomposition of $\H$ into $\H=\H_1'\otimes \H_2'$, as the following example shows. Consider a six dimensional Hilbert space $\H=\mathbb{C}^6$ and assume the following separable state $(\ket{0} +\ket{1})\otimes (\ket{0}+\ket{2})/2$ on the decomposition $\mathbb{C}^2\otimes\mathbb{C}^3$ using the basis $\{\ket{0},\ket{1}\}$ and $\{\ket{0},\ket{1},\ket{2}\}$ respectively. Surprisingly, this state is inseparable on the decomposition 
$\mathbb{C}^3\otimes\mathbb{C}^2$.  This can be verified using the positive partial transposition (PPT) criterion which is a necessary and sufficient separability criterion for these two cases.  States that are separable under any decomposition of the total state space are called `absolutely separable' \cite{kus}. In all cases to be considered the decomposition of the state space of the multi-partite system is given from the outset (we are dealing with $N$-qubits) so the question of absolute separability  will not be relevant to our investigations.  

\section[Pitfalls when using Bell-type polynomials to derive Bell-type\\ inequalities]{Pitfalls when using Bell-type polynomials to derive Bell-type inequalities}\label{pitfall}
In this section we will comment on a pitfall that lures in the background when using Bell-type polynomials to obtain Bell-type inequalities for LHV models. This exposition is inspired by an attempt to expose the flaw in the derivation of a recent Bell-type inequality for LHV models  by  \citet{chencritiq}. See \citet{seevchen} for the detailed critique.

 \citet{chencritiq} claimed ``exponential violation of local realism by separable states", in the sense
that multi-partite separable quantum states are supposed to give rise to correlations and fluctuations that violate a Bell-type
inequality that Chen claims to be obeyed by LHV models.
 However, this claim can not be true since all correlations separable quantum states give rise to have a description in terms of local correlations and thus satisfy all Bell-type inequalities for LHV models, and this holds for all number of parties. (This was explicitly  proven above for $N=2$).  We will expose the flaw in Chen's reasoning,
not merely for clarification of this issue, but perhaps even more importantly
 since it re-teaches us an old lesson J.S. Bell taught us over 40 years ago, although in a different form.
We will argue that this lesson provides us with a new morale especially relevant to modern research in Bell-type inequalities  and thus also for the research of this dissertation.
It is not important to go into the details of Chen's work, and for clarity we will not use a multi-partite but a two-partite setting. 

Consider the standard \emph{Gedankenexperiment} where one considers two systems that each are distributed to one of two parties
who measure two different dichotomous observables on the respective subsystem they have in their possession. Next consider the CHSH polynomial of (\ref{CHSHpolynomial}) that reads $B_{\textrm{chsh}}=AB +AB' +A'B -A'B'$, where for clarity we have written $A, A'$ and $B,B'$ for the observables instead of $A_1,A_1'$ and $A_2,A_2'$. Also for clarity we denote the quantum mechanical version of the CHSH polynomial by the Bell-operator $\hat{B}_\textrm{chsh}$, and the quantum observables by the operators $\hat{A},\hat{A}',\hat{B},\hat{B}'$. Consider now a separable two-qubit state $\rho_\textrm{sep}$ on $\H=\mathbb{C}^2\otimes\mathbb{C}^2$ and local orthogonal spin observables:  $\hat{A}\perp\hat{A}'$ and $\hat{B} \perp\hat{B}'$. Note that such local orthogonal spin-observables anti-commute: $\{\hat{A},\hat{A}'\}=0$ and $\{\hat{B},\hat{B}'\}=0$. Chen considered the quantities $(\hat{B}_\textrm{chsh})^2$ and $(B_{\textrm{chsh}})^2$ and calculated the bounds on these quantities as determined by separable states and local correlations respectively.

Using the fact that the operators that correspond to the local quantum mechanical observables anti-commute one obtains
\begin{align}
(\hat{B}_\textrm{chsh})^2=(\hat{A}^2+\hat{A}'^2)\otimes(\hat{B}^2 +\hat{B}'^2)-
4(\hat{A}\hat{A}'\otimes\hat{B}\hat{B}')\nonumber\\
=(\hat{A}^2+\hat{A}'^2)\otimes(\hat{B}^2 +\hat{B}'^2)+4
(\hat{A}''\otimes \hat{B}''),
\label{qmexp}
\end{align}
 with 
$\hat{A}''=[\hat{A},\hat{A}']/2i$ and $\hat{B}''=[\hat{B},\hat{B}']/2i$  spin observables orthogonal to 
both $A, A'$ and $B,B'$ respectively. However, in the local realist case where the local observables are not anti-commuting operators on a Hilbert space but some functions that take on values that represent measurement outcomes and therefore commute, one obtains
\begin{align}
(B_\textrm{chsh})^2=(A^2+A'^2)(B^2 +B'^2)+2 AA'(B^2-B'^2) +2BB'(A^2-A'^2).
\label{lhvchen}
\end{align}
Using linearity of the mean to determine the expectation values of both $(B_\textrm{chsh})^2$ and $(\hat{B}_\textrm{chsh})^2$ one obtains that local correlations give $|\av{(B_\textrm{chsh})^2}_\textrm{lhv}|\leq4$, whereas separable quantum states are able to give $|\av{(\hat{B}_\textrm{chsh})^2}_\textrm{qm}|=8$. It thus appears that separable states can give correlations that are much stronger than local correlations, hence the original claim by Chen. However, somewhere something must have gone astray since we know that all predictions separable states can give rise to can be mimicked by local correlations.

The first thing to note is that, despite the formal similarity of the expressions $(B_\textrm{chsh})^2$ and $(\hat{B}_\textrm{chsh})^2$ (i.e., when not expanded in terms of observables), expressions  (\ref{lhvchen}) and (\ref{qmexp}) cannot be considered to be counterparts of each other in case the first is supposed to be the Bell-type polynomial for LHV models and the second for quantum mechanics. The correct counterpart of $(\hat{B}_\textrm{chsh})^2$ for LHV models is obtained by translating (\ref{qmexp}) directly into
\beq
\widetilde{B}=(A^2+A'^2)(B^2 +B'^2) +4A''B'',
\eeq
with $A''$ and $B''$ some dichotomic $\pm1$ valued observables that are the local realistic counterpart of the observables that correspond to operators $\hat{A}''$ and $\hat{B}''$ respectively.  For local correlations one obtains the tight bound $|\av{\widetilde{B}}_\textrm{lhv}|\leq8$, so we see that the correlations separable states can give rise to can indeed be retrieved  using local correlations.
  The functional $\widetilde{B}$ (and not $(B_\textrm{chsh})^2$) is the Bell-polynomial that, when averaged over and using linearity of the mean, gives the Bell-type inequality which is the counterpart of the quantum mechanical inequality using $(\hat{B}_\textrm{chsh})^2$.

But what actually went wrong in considering $(\hat{B}_\textrm{chsh})^2$ and $(B_{\textrm{chsh}})^2$ to be the correct counterparts for quantum and local correlations respectively? Here we recall a lesson J.S. Bell taught us many years ago.

First we note that  it is of course not $B_{\textrm{chsh}}$ that is measured but the observables $A,A'$ and $B,B'$. However, 
the quantum mechanical counterparts of Bell-type polynomials  (i.e., in terms of the operators associated to the observables),  
such as the Bell-operator $\hat{B}_\textrm{chsh}$, can be considered to be observables themselves since a sum of self-adjoint
operators is again self-adjoint and every self-adjoint
operator is supposed to correspond to an observable.
 Furthermore, the additivity of operators gives
additivity of expectation values. 
\forget{Thus the Tsirelson inequality
expressed as  $|\av{\hat{A}\hat{B}}_\textrm{qm} +\av{\hat{A}\hat{B}'}_\textrm{qm} +\av{\hat{A}'\hat{B}}_\textrm{qm}-\av{\hat{A}'\hat{B}'}_\textrm{qm}|\leq 2\sqrt{2}$  can equally well be expressed in a shorthand notation as 
$|\av{\hat{B}_\textrm{chsh}}_\textrm{qm}|\leq 2\sqrt{2}$.}
Thus the Tsirelson inequality \begin{align}
|\av{\hat{A}\hat{B}}_\textrm{qm} +\av{\hat{A}\hat{B}'}_\textrm{qm} +\av{\hat{A}'\hat{B}}_\textrm{qm}-\av{\hat{A}'\hat{B}'}_\textrm{qm}|\leq 2\sqrt{2}
\end{align}
  can equally well be expressed in a shorthand notation as 
  \begin{align}
  |\av{\hat{A}\hat{B}+\hat{A}\hat{B}'+\hat{A}'\hat{B}-\hat{A}'\hat{B}'}_\textrm{qm}|
  =|\av{\hat{B}_\textrm{chsh}}_\textrm{qm}|\leq 2\sqrt{2}.
\end{align}

However, as noted by Bell: ``A measurement of a sum of noncommuting observables
cannot be made by combining trivially the results of separate observations
on the two terms -- it requires a quite distinct experiment. 
[\ldots] But this explanation of the nonadditivity of allowed values also established 
the non-triviality of the additivity of expectation values. 
The latter is quite a peculiar property of quantum mechanical states, 
not to be expected \emph{a priori}. There is no reason to demand 
it individually of the hypothetical dispersion free states 
[hidden-variable states $\lambda$], whose function it is to reproduce the 
\emph{measurable} peculiarities of quantum mechanics when \emph{averaged over}."  \cite{bell66}\footnote{
Note, however, that for Bell the crucial point is not that eigenvalues of self-adjoint observables do not obey the additivity rule (he gave an example using spin observables), but that the additivity rule in the case of incompatible observables cannot be justified in the light of the Bohrian point that the context of measurement plays a role in defining quantum reality: ``\forget{It will be urged that these analyses leave the real question untouched. In fact. it will be seen that these demonstrations require from the hypothetical dispersion free states, not only that appropriate ensembles thereof should have all measurable properties of quantum states, but certain other properties as well. These additional demands appear reasonable when results of measurement are loosely identified with properties of isolated systems.}They [the additivity rule] are seen to be quite unreasonable when one remembers with Bohr 'the impossibility of any sharp distinction between the behaviour of atomic objects and the interaction with the measuring instruments which serve to define the conditions under which the phenomena appear'.'' \cite{bell66}. (Bell cites N. Bohr here.) Analogously, what is important for our discussion here is not some additivity rule but that a specific inference on the hidden-variable level between incompatible observables cannot be justified in the light of the Bohrian point Bell referred to}.
 If we apply Bell's lesson to the \emph{Gedankenexperiment} considered here we realize that because 
 the CHSH polynomial $B_{\textrm{chsh}}$ contains incompatible observables it  cannot be measured by 
 combining trivially the results of separate observations
on the different terms in the polynomial -- it requires a quite distinct experiment, one that is not part of the original \emph{Gedankenexperiment}. 

The hidden variables $\lambda$ only determine the probabilities  for outcomes of 
the  individual measurements of $A,A',B,B'$ and
not probabilities for outcomes of measurement of the quantity $B_{\textrm{chsh}}$ 
since measurement of the latter would require quite a distinct experiment because  it involves incompatible observables
$A$ and $A'$  for party $1$ and $B$ and $B'$ for party $2$.  The only function 
of the CHSH polynomial  $B_{\textrm{chsh}}$  
is to provide a shorthand notation of the CHSH
inequality. Indeed, when averaged over $\lambda$  it 
gives the inequality $|\av{B_\textrm{chsh}}_\textrm{lhv}|\leq 2$, which, by using linearity of the mean,  
can be rewritten as a sum of expectation values 
in a legitimate local realistic form, namely as  the 
legitimate Bell-type inequality $|\av{AB}_\textrm{lhv}+\av{AB'}_\textrm{lhv}+\av{A'B}_\textrm{lhv}-\av{A'B'}_\textrm{lhv}|\leq 2$ that local realism must satisfy. 
Indeed, all expectation values in this Bell-type inequality  
involve only compatible quantities, and no incompatible  ones.
\forget{So although $B_{\textrm{chsh}}$ cannot be thought of as an observable specifying the sum of 
local realistic quantities determined by the hidden variable $\lambda$, and as 
such is never measured in an experiment,  when averaged over it does allow 
for a legitimate shorthand notation of the Bell-type inequality \ref{}.}
Therefore, the hidden-variable counterpart of the quantum mechanical operator $\hat{B}_\textrm{chsh}$ 
can be safely chosen to be the Bell-polynomial $B_{\textrm{chsh}}$, and vice versa.

\forget{
Furthermore, in quantum mechanics measurement of $(\hat{B}_\textrm{chsh})^2$ requires measurement of three different local measurement settings, i.e.,  $\hat{A},\hat{A}',\hat{A}''$ and $\hat{B},\hat{B}',\hat{B}''$ (see (\ref{qmexp})), of which $ \hat{A}''$ and $\hat{B}''$ are in fact not considered in the experiment. Measurement of $(B_\textrm{chsh})^2$ would thus also require three different local settings $A,A',A''$ and $B,B',B''$, of which $A''$ and $B''$ are also not considered. }

Let us now consider the expressions  $(B_\textrm{chsh})^2$  and $(\hat{B}_\textrm{chsh})^2$. 
 The reason why  $B_\textrm{chsh}$  and $\hat{B}_\textrm{chsh}$ give a legitimate shortcut formulation for a Bell-type inequality  whereas  
  $(B_\textrm{chsh})^2$  and $(\hat{B}_\textrm{chsh})^2$ do not, is that the latter two cannot be written as an expression involving expectation values of the observables $A,A',B,B'$  (or $\hat{A},\hat{A}',\hat{B},\hat{B}'$) via  a legitimate operation such as linearity of the mean, whereas the first two can. Measurement of $(B_\textrm{chsh})^2$,  and $(\hat{B}_\textrm{chsh})^2$ in the quantum case,  requires  measurement of observables  $A''$ and $B''$,  corresponding to $\hat{A}''$ and $\hat{B}''$ in the quantum case, that are not part of the \emph{Gedankenexperiment}.

Assuming that measurement of $(B_\textrm{chsh})^2$ involves measurement of only $A,A'$ and $B,B'$, as is implied by  (\ref{lhvchen}), ignores the incompatibility of the observables involved in expressions such as $AA'$, etc.  
In quantum theory, however, $\hat{A}\hat{A}'$ happens to determine another self adjoint observable $\hat{A}''$ via $[\hat{A},\hat{A}']/2i=\hat{A}''$. Measurement of the product of the incompatible observables $\hat{A}$ and $\hat{A'}$ is therefore taken care of by the quantum formalism itself\footnote{In fact, it is only the product  $\hat{A}\hat{A}'\otimes\hat{B}\hat{B}'$
of \eqref{qmexp} that is again self-adjoint and can be taken to correspond to an observable, not necessarily the terms $\hat{A}\hat{A}'$ and $\hat{B}\hat{B}'$ themselves. }. Not so for the hidden-variable formalism, where one must introduce a new observable $A''$ that on the hidden-variable level has no relationship to $A$ and $A'$. There is no reason whatsoever to presuppose an algebraic relation between the individual outcomes of measurement of  these three observables. 

We finally see where things have gone astray  in deriving that $|\av{(B_\textrm{chsh})^2}_\textrm{lhv}|\leq4$, although separable quantum states are able to give $|\av{(\hat{B}_\textrm{chsh})^2}_\textrm{qm}|=8$. It is not the strength of correlations  in separable states which
ruled out local realism, but "[i]t was the arbitrary assumption of a particular (and impossible) relation
between he results of incompatible measurements either of which \emph{might} be made on a
 given occasion but only one of which can in fact be made."\cite{bell66}

 Let us recapitulate and discuss the subtleties  that must be taken care of (also in this dissertation) when deriving Bell-type inequalities using a shorthand notation in terms of Bell-type polynomials.
  
(I) Firstly, the Bell polynomials are not to be regarded 
as observables. In general they contain incompatible observables (however, see (III) below).\forget{
 The danger is that because the operator identity in quantum mechanics
$\hat{\mathcal{B}}=\hat{A}\hat{B}+ \hat{A}\hat{B}' +
\hat{A'}\hat{B} -\hat{A}'\hat{B}'$ does indeed define a new observable $\hat{\mathcal{B}}$, one is tempted to formulate 
the hidden-variable counterpart as:
\beq\label{counter}
\mathcal{B}(\lambda)= A(\lambda)B(\lambda)+A(\lambda)B'(\lambda)
+A'(\lambda)B(\lambda)-A'(\lambda)B'(\lambda).
\eeq
However, if $\mathcal{B}(\lambda)$ is not regarded as 
merely a shorthand notation for the sum of the four terms in \Eq{counter}, but
is supposed to be the counterpart of the observable
$\hat{\mathcal{B}}$, \Eq{counter}  involves the problematic additivity of eigenvalues, 
which cannot be demanded of local realism.
Indeed, four different non-compatible setups are involved
and not just one, which the notation $\mathcal{B}(\lambda)$ 
when considered as a single observable could suggest.

Thus, when deriving local hidden-variable observables that depend on the hidden variable $\lambda$, only compatible experimental setups must be considered.}
The difficulty of measuring incompatible observables
 has to be explicitly taken into account in the hidden-variable expression.
Only in quantum mechanics this incompatibility
structure is already captured in the (non-)commutativity structure of the
operators that correspond to the observables in question.

(II) Secondly, when using a shorthand notation in terms of Bell-polynomials it must be possible (by for example 
using linearity of the mean) to translate the shorthand notation into a 
legitimate Bell-type inequality in terms of expectation values of compatible observables that are actually considered in the \emph{Gedankenexperiment}.

(III) Thirdly, suppose one would indeed regard the functionals 
$B_{\textrm{chsh}}$ to be the 
quantities of interest and regard them as observables. 
The first subtlety mentioned above shows that this is unproblematic only if they are thought of as 
being genuine irreducible observables and not to be composed out of a sum of other incompatible observables. But one then considers a different experiment. To be fair to local realism from the start
the possible values of measurement of, for example, the observable $B_{\textrm{chsh}}$ in the local hidden-variable model should
 then be equal to the eigenvalues of the quantum mechanical counterpart
$\hat{B}_{\textrm{chsh}}$.  And these eigenvalues of $\hat{B}_{\textrm{chsh}}$ are  \{$2\sqrt{2},~-2\sqrt{2},~ 0$\} respectively. The possible outcomes for the local realist quantities should equal these eigenvalues. Indeed, predictions for a single observation can always be mimicked by a local 
hidden-variable model. 

\forget{Cavalcanti: example of where things go alright (mermin, werner/wolf too)?}


\forget{\chapter{The technical results of this dissertation}

Now that we have the technical defnitions, this will be in the rest of the thesis, in technical terms: 
these techincal result will be discussed and their (foundational) relevance argued for in the respective chapters.

Part I: 
\begin{itemize}
\item Chapter: 

we chow the CHSH ineq
\item Chapter:
\item Chapter \ref{chapter_CHSHquantumorthogonal} studies the bi-partite qubit quantum correlations using the CHSH polynomial for noncommuting observables and orthogonal spin directions directions. There it is obtained that 
\end{itemize}

Part II: 
\begin{itemize}
\item Chapter:
\item Chapter:
\item Chapter:
\end{itemize}

Part III:
\begin{itemize}
\item Chapter: we conclude.
\end{itemize}

----------------------------------------

Of toch niet in een apart hoofdstuk, maar in voorgaande hfdst, of voorafgaand aan elk hoofdstuk of deel?
Of met een soort van aparte aankondiging door vorig hoofdstuk heen:

}
%
%
%
%
\clearemptydoublepage
\thispagestyle{empty}
\part{Bi-partite correlations}

\thispagestyle{empty}
\chapter[Local realism, hidden variables and correlations]{Local realism, hidden variables\\\vskip0.2cm and correlations}
\label{chapter_CHSHclassical}
\noindent
This chapter is partly based on \citet{seevnonlocal}.

\section{Introduction}\label{introCHSHclassical}
In the first part of this chapter we will mainly investigate what assumptions suffice in deriving the original CHSH inequality $|\av{B_{\textrm{chsh}}}|\leq2$ for the case of two parties and two dichotomous observables per party\forget{\footnote{In this chapter the CHSH inequality is understood to have the local upper bound of 2.}}. It is well-known that all local correlations  obey this inequality (i.e., $|\av{B_{\textrm{chsh}}}_{\textrm{lhv}}|\leq2$) but here we will show that many more correlations not of the local form also obey this inequality.  In section \ref{LRsection} we start by reviewing the fact that the doctrine of local realism that assumes free variables  and allows for local measurement contextuality in the form of measurement apparatus hidden variables must obey the CHSH inequality both in the case of deterministic and stochastic models.  In the derivation for stochastic models two conditions, first distinguished by \citet{jarrett}, are used to give a factorisability condition that enables the derivation to go through. Jarrett called them Completeness and Locality, although we will follow a different terminology.  These conditions are more general than the well-known conditions of Outcome Independence and Parameter Independence of \citet{shimony}. These latter conditions taken together give a condition called Factorisability, sometimes also referred to as Local Causality (terminology by \citet{bell76}), that also suffices to obtain the CHSH inequality. 
 The Shimony conditions follow from the Jarrett conditions when averaged over measurement apparatus hidden variables. In subsection \ref{jarrettshimony} we briefly comment on the crucial difference between these two sets of conditions and argue that they should not be conflated. 

In the following subsection, subsection \ref{shimonymaudlin} we review the fact that the Shimony conditions are not the only two conditions 
that imply Factorisability. Two different conditions, first distinguished by \citet{maudlin}, also suffice.  Although Maudlin's conditions are well-known, we have not been able to find a proof in the literature that they indeed imply Factorisability and therefore we give such a proof here (in the Appendix on p. \pageref{appendixmaudlin}).  We briefly comment on the consequences of this non-uniqueness for interpreting violations of the CHSH inequality. It has been argued that this undermines the activity called experimental metaphysics where one draws grand metaphysical conclusions based on the idea that in violations of the CHSH inequality compliance with relativity forces Outcome Independence to be violated rather than Parameter Independence.  We can only partly agree that this undermines the starting point of experimental metaphysics because Maudlin needs extra assumptions to evaluate his conditions in quantum mechanics.  We also review two other arguments against this activity and present in the next section, section \ref{non-localsection}, another difficulty for this activity.

\forget{
We next turn our attention to the difference between stochastic and deterministic local hidden-variable models. In section \ref{stochdeter} we briefly discuss their relation and comment on the fact that the assumption of perfect correlations enforces a stochastic model to be deterministic. In such a deterministic situation, and under the assumption of perfect correlation, \citet{bell64} produced the very first Bell-type inequality. We give a new derivation of this inequality thereby deriving various new previously unknown versions.  However, as first remarked by \citet{zukow1964}, the 1964 inequality by Bell is not a Bell-type inequality as we have defined them in the previous chapter. It is the extra assumption of perfect correlation for at least one pair of observables that sets this inequality apart. We are led to investigate the applicability of this inequality. Not only is the assumption of perfect correlation experimentally unachievable, it limits its usage in quantum mechanics considerably since we will show that most quantum states do not obey this extra assumption. 
}
In this section we go back to the CHSH inequality and show that both in the deterministic and stochastic case one can allow  explicit non-local setting and outcome dependence as well as dependence of the hidden variables on the settings  (i.e.,  the observables are no longer free variables), and still derive the CHSH inequality.  Violations of the CHSH inequality thus rule out a much broader class of hidden-variable models than is generally thought.\forget{ In the light of experimental violations of the CHSH inequality this shows that the question of whether it is Outcome Independence or Parameter Independence that should be abandoned is misplaced. We have no reason to expect either one of them to hold solely on the basis of the CHSH inequality. 
AANPASSEN
}
This shows that the conditions of Outcome Independence and Parameter Independence, that taken together imply the condition of Factorisability, can both be violated in deriving the inequality, i.e., they are not necessary for this inequality to obtain, but only sufficient. Therefore, we have no reason to expect either one of them to  hold solely on the basis of the CHSH inequality, i.e., satisfaction of the inequality is not sufficient for claiming that either one holds.

In section \ref{sectionleggett} we compare our findings to a recent non-local model by \citet{leggett} that violates the CHSH inequality, but which obeys a different Bell-type inequality that is violated by  quantum mechanics. The discussion of Leggett's model will show an interesting relationship between different assumptions at different hidden-variable levels. In this model parameter dependence at the deeper hidden-variable level does not show up as parameter dependence at the higher hidden-variable level (where one integrates over a deeper level hidden-variable), but only as outcome dependence, i.e., as a violation of Outcome Independence.\forget{Conversely, violations of Outcome Independence  can be a sign of a violation of deterministic Parameter Independence at a deeper hidden-variable level.} This shows 
 that which conditions are obeyed and which are not depends on the level of consideration. A conclusive picture therefore depends on which hidden-variable level is considered to be fundamental.\forget{The non-local setting and outcome dependence we allow for in deriving the CHSH inequality acts at the level of the hidden variables, i.e., on the level of what we have previously called subsurface probabilities where one conditions on the hidden variables $\lambda$. }

In section \ref{surfacesection} we will extend our investigation from the hidden-variable level (i.e., the level of subsurface probabilities where one conditions on the hidden variables $\lambda$) to the level of surface probabilities. We will present  interesting analogies between different inferences that can be made on each of these two levels. The most interesting such analogy is between, on the one hand, the subsurface inference that the condition of Parameter Independence and violation of Factorisability implies randomness at the hidden-variable level, and, on the other hand, the surface inference  that any non-local correlation that is no-signaling must be indeterministic, as was recently proven by \citet{masanes06}. An interesting corollary of this is that any deterministic hidden-variable theory that obeys no-signaling and gives non-local correlations must show randomness on the surface, i.e., the surface probabilities cannot be deterministic. The determinism thus stays beneath the surface;  the hidden variables cannot be perfectly controllable because the outcomes must show randomness at the surface. We show that Bohmian mechanics is in perfect agreement with this conclusion.

In section \ref{discerningno-signalingsection} we remain at the level of surface probabilities and further investigate the no-signaling correlations. We first show that an alleged no-signaling Bell-type inequality as proposed by \citet{roysingh} is in fact trivially true. We next derive a non-trivial no-signaling inequality in terms of expectation values. 
\forget{--- and show that it is a facet of the no-signaling polytope, i.e., it is thus both non-trivial and tight.}  In doing, so we must go beyond the analysis used in deriving the CHSH inequality, because, as has been shown in the previous chapter, this inequality is trivial for 
 no-signaling correlations. In the 4-dimensional space of product expectation values $\av{AB}$, $\av{A'B}$, etc.,  the no-signaling polytope contains only trivial facets. We therefore consider the larger space of both product and marginal expectation values (i.e., including also $\av{A}^B$, $\av{A}^{B'}$, etc.).

In the last section, section \ref{discussionchshclassical}, we discuss  a few of the most interesting open problems which have emerged from the investigations in this chapter. An investigation of quantum correlations (for both entangled and separable states)  with respect to the CHSH inequality is postponed to the next two chapters. This chapter concentrates on local, non-local and no-signaling correlations.

A list of acronyms used in this chapter can be found in section \ref{acro} on page \pageref{acro}.

\forget{, with the minor exception in section \ref{secbell64} where we consider the applicability in quantum mechanics of the original 1964 Bell inequality.}
\forget{ Note that in this chapter we will not investigate the behavior of quantum correlations (for both entangled and separable states)  with respect to the CHSH inequality. This will  be performed in the next two chapters. In this chapter we solely concentrate on local, non-local and no-signaling correlations, with the minor exception in section \ref{secbell64} where we consider the quantum applicability of the original 1964 Bell inequality.}

\section{Local realism and standard derivation of the CHSH inequality}\label{LRsection}

\subsection{Local realism and free variables}
\label{LRfree}
\noindent
It is commonly accepted that the assumptions of local realism together with the requirement that observables can be regarded as free variables ensure that deterministic and stochastic hidden-variable theories obey the  CHSH inequality. To appreciate this statement we must be precise about the notions involved in this statement. A great deal has been written about these notions; here we rely heavily on \citet{clifton} which in our opinion gives a very clear exposition. 

Realism is the idea that (i) physical systems exist independently and (ii)  posses intrinsic properties that can
be described by states. It is further assumed that these states together with the state of the measurement
apparatus completely account for outcomes of measurements and/or their statistics  (i.e., probabilities of
outcomes). The independently existing states are usually called hidden variables, and we will follow
suit for historical reasons, although there are good reasons to call them differently. Indeed, \citet{clifton} call them `existents', while \citet{bellspeakable} at some point
calls them `beables'. \forget{The motivation they use to not use the term hidden variables is to get rid of the
wrong suggestion the term could give one that these states are somehow fundamentally experimentally
 inaccessible. Indeed, this is a wrong suggestion, and the term is thus somewhat of a misnomer.}

Locality we take to be the idea that  there exists no spacelike causation \cite{clifton}, i.e., the outcomes of measurement do not depend on what happens in a spacelike separated region. Furthermore, the causal history of measurement devices is supposed to be sufficiently disentangled from other measurement devices that are spacelike separated and from the systems to be measured in the backward lightcones of the measurement events. (`sufficiently' in the sense that at least the independence conditions to be given below hold.) 

The notion of 'free variables' \cite{bellspeakable} means that the settings used to measure observables can be chosen freely, i.e., this excludes conspiracy theories such as super-determinism, as well as  retro-causal interactions. The assumption that one deals with free variables is also called the `freedom assumption', see e.g. \cite{gill} for a clear exposition.  

Models that assume local realism and that settings are free variables are  called local hidden-variable (LHV) models.
\begin{figure}[h]
\includegraphics[scale=1]{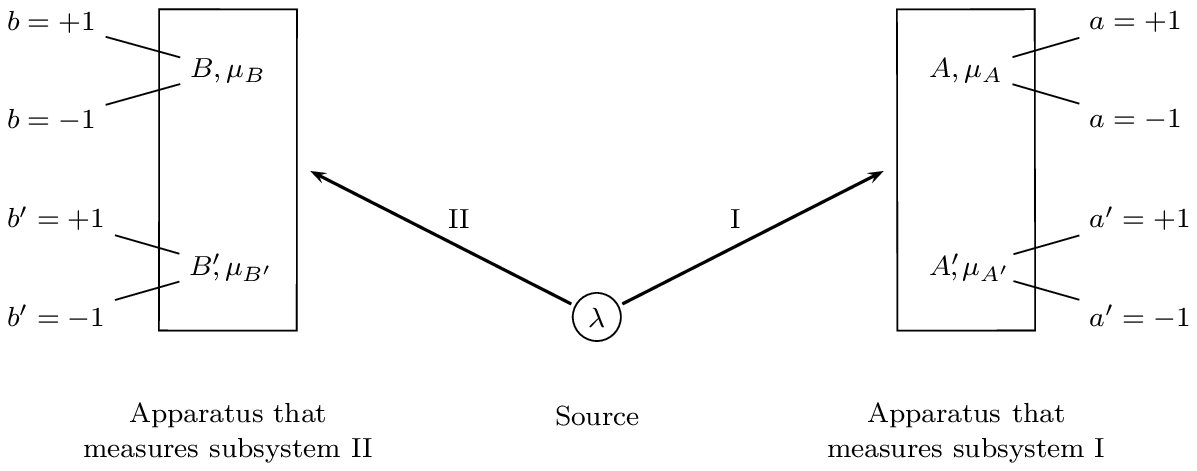}
\caption{Setup of the \emph{Gedankenexperiment}.}
\label{gedank}
\end{figure} \forget{
\begin{figure}[h]
\centerline{
\pspicture(-8,-1.5)(8,5)
\psset{linewidth=.016cm} 
\rput(-3.75,3){\rnode{A}{\footnotesize$B,\mu_B$}}
\rput(-3.75,1){\rnode{B}{\footnotesize$B'\hspace{-3pt},\mu_{B'}\hspace{-2pt}$}}
\ncbox[nodesep=.5cm,boxsize=0.70]{A}{B}
\rput[l](-6,1.5){\rnode{C}{\footnotesize $b'=+1$}}
\ncline[nodesep=.1cm]{B}{C}
\rput[l](-6,0.5){\rnode{D}{\footnotesize $b'=-1$}}
\ncline[nodesep=.1cm]{B}{D}
\rput[l](-6,3.5){\rnode{C3}{\footnotesize $b=+1$}}
\ncline[nodesep=.1cm]{A}{C3}
\rput[l](-6,2.5){\rnode{D3}{\footnotesize $b=-1$}}
\ncline[nodesep=.1cm]{A}{D3}
\rput(3.75,3){\rnode{A1}{\footnotesize${A},\mu_A$}}
\rput(3.75,1){\rnode{B1}{\footnotesize${A}'\hspace{-3pt},\mu_{A'}\hspace{-2pt}$}}
\ncbox[nodesep=.5cm,boxsize=0.70]{A1}{B1}
\rput[l](5,1.5){\rnode{C1}{\footnotesize $a'=+1$}}
\ncline[nodesep=.1cm]{B1}{C1}
\rput[l](5,0.5){\rnode{D1}{\footnotesize $a'=-1$}}
\ncline[nodesep=.1cm]{B1}{D1}
\rput[l](5,3.5){\rnode{C2}{\footnotesize $a=+1$}}
\ncline[nodesep=.1cm]{A1}{C2}
\rput[l](5,2.5){\rnode{D2}{\footnotesize $a=-1$}}
\ncline[nodesep=.1cm]{A1}{D2}
\cnodeput(0,0.5){G}{\footnotesize\sf $\lambda$}
\rput(-2.95,2){\rnode{H}{}}
\rput(2.95,2){\rnode{H1}{}}
\rput(1.4,1.5){{\footnotesize I}}
\rput(-1.4,1.5){{\footnotesize II}}
\rput(0,-0.5){{\footnotesize Source}}
\rput(3.75,-0.5){{\footnotesize Apparatus that}}
\rput(3.75,-0.85){{\footnotesize measures subsystem I}}
\rput(-3.75,-0.5){{\footnotesize Apparatus that}}
\rput(-3.75,-0.85){{\footnotesize measures subsystem II}}
\ncline[nodesep=0.04cm,linewidth=0.03cm]{<-}{H}{G}
\ncline[nodesep=0.04cm,linewidth=0.03cm]{<-}{H1}{G}
\endpspicture
}
\caption{Setup of the \emph{Gedankenexperiment}.}
\label{gedank}
\end{figure}}

Consider now the following \emph{Gedankenexperiment} (see Figure \ref{gedank}) where two different
measurements with settings $A$ (or $A'$) and $B$ (or $B'$) are performed on a certain physical
system consisting of two parts (or subsystems) called $I$ and $II$ respectively, that have originated from a common source. One can think of a spin measurement on system consisting of two spins that are created in some  decay process. 
 On part $I$ observables corresponding to settings $A$ or $A'$ are measured and on part $II$ observables corresponding to settings $B$ or $B'$. Denote the outcomes of $A$ and $B$ by $a$ and $b$ respectively (and similarly for $A'$ and $B'$).
They are assumed to be dichotomic, i.e., $a=\pm1$, $b=\pm1$. \forget{The settings $A$ and $B$ can be single
parameters, or anything more general such as whole sets of parameters, vectors or tensors.}

It is assumed that the measurement apparata are spacelike separated and that the settings are fixed just
prior to the respective measurement interactions. Let the hidden variable $\lambda$ denote the complete state (set of states) in the entire causal history of the two subsystems prior to the measurement interactions\footnote{This specification of $\lambda$ to be the complete state in the \emph{entire} causal history of the subsystems is more general than a specification of $\lambda$ as the state of the subsystems at a specific instant in time, such as the time of origination from the source. For a discussion of advantages of this specification over other specifications, see  \cite{clifton,butterfield1}.}. Note the generality of this characterization; it allows for stochastic as well as deterministic determination, the hidden variables can be anything and it allows for all possible dynamical evolutions of the system and its subsystems.
Furthermore, let the hidden variables $\mu_A$ and $\mu_{B}$ denote the states of the complete causal
history of the measurement apparata prior to any measurement of $A$ on system $I$ and $B$ on system $II$
respectively. These apparatus hidden variables of course include the (macroscopic) settings
$A$ and $B$ which are assumed to be fixed just prior to the measurement interactions. However, since these are assumed to be under control of the experimenter, they will nevertheless be explicitly notated.  The degrees of freedom that do not incorporate the setting may well be coupled to the those that characterize the setting. We  incorporate this possible dependence by letting the apparatus hidden variables  depend on the settings. We thus write $\mu_A,\, \mu_B$ instead of $\mu_I,\,\mu_{II}$, respectively.

The experiment is assumed to be performed many times on a large collection or ensemble of systems (each comprising of two subsystems $I$ and $II$), where each experiment is performed on a single system in this collection. The choice of different settings ($A$ or $A'$ and $B$ or $B'$) in each run of the experiment is assumed to be made independently. By standard sampling arguments the average of the actual data measured for a given pair of settings on a subset of the total set of measured systems, equals, for a large enough subset  (i.e., enough runs of the experiment), the average that would be obtained for all systems if they had been measured with the same pair of settings.

\subsubsection*{Assuming realism}\noindent
It is assumed that $\lambda$, together with the states of the apparata, determines the outcomes of measurement 
of any possible observable that can be measured. This is justified by the idea of realism: the intrinsic properties possessed by the system are described by a physical state (this is taken to be the hidden variable state $\lambda$) and therefore the outcomes of measurement of all possible observables (that are supposed to be reveal intrinsic properties) are dictated by $\lambda$ and perhaps also by the details of the measurement apparatus. Note that we also allow for the case where it is not the outcomes, but only their probabilities of occurring that are dictated, see below.

In order to derive Bell-type inequalities, we will consider relations among various hypothetical outcomes of a single experiment to be performed at a single time in one of several possible versions. However, in a real experiment one considers outcomes of several different versions of an experiment, all of which were actually performed at various different times.   
These actual outcomes are connected to the hypothetical outcomes in a hidden-variable model via the following conjecture which is motivated by the idea of realism. Rephrasing \citet[p. 2]{merminreply} this conjecture is that every one of the possible outcomes (or their probabilities of occurring) for every one of the possible choices one might make for the settings ($A$ or $A'$ and  $B$ or $B'$) in a single experiment performed at a single time are all predetermined by properties of the system (and measurement apparata), with one (and only one) of those predetermined outcomes actually being revealed -- namely the one associated with the particular choice of experiment actually made.\forget{\footnote{In discussions of hidden-variable theories in terms of the notion of a common cause, this relates to the assumption that different correlations between outcomes of measurement of different observables have the same common cause, a so-called common common cause, see \citet{butterfieldcommon} and references therein. 
Such an assumption has been criticized. We will here not discuss this here, but merely note that 
.}.} This allows us to consider the same hidden variable $\lambda$ when considering outcomes of two incompatible observables (e.g., $A$ and $A'$) that each require a distinct experiment in order to be measured\footnote{We do not consider the possibility where this does not suffice and where something like a common common cause is needed, cf. \citet{butterfieldcommon}.}.

The preparation of the complete set of hidden variables cannot be assumed to be perfectly controllable. Therefore, if one wants to make contact with statistics observed on ensembles of identical experiments (same settings, although the hidden variables may differ and in general will differ since they are uncontrollable), one must assume that for such an ensemble the hidden variables have some normalized distribution. Consequently, the advocate of  local realism has to use some distribution representing ignorance about which hidden variables exist in a particular experiment. This distribution is notated as
$\rho(\lambda,\mu_A,\mu_{B}|A,B)$, which is the distribution of the hidden variables given a specific setting $A$, $B$.

\forget{OUDE FORMULERING:
Note that in general the hidden variables will be different for each set of measurements $A$, $B$
since the preparation of the complete set of hidden variables cannot be assumed to be controllable.
Furthermore, if one wants to make contact with statistics observed on ensembles of identical experiments (same settings, although the hidden variables may differ and in general will differ since they are uncontrollable), one must assume that for such an ensemble the hidden variables have some normalized distribution.
Therefore, the local realist has to use some density distribution representing ignorance about which
hidden variables exist in a particular experiment. This distribution is notated as
$\rho(\lambda,\mu_A,\mu_{B}|A,B)$,
which is the distribution of the hidden variables given a specific setting $A$, $B$.
}
\forget{----------------------
  meer 
The above conditions are well-known and commonly used
(see e.g. Bell\cite{bellspeakable}, Clifton et al.\cite{clifton})  and we will not give the standard motivation.
Following \cite{clifton}, suffice it to say that
they are all plausible from the above mentioned idea of locality, freedom of the observers  and from the
assumption that the measurement apparata are spacelike separated\footnote {For
a discussion of criticism of these conditions as well as a rebuttal of these
criticism so that the assumptions remain plausible, see \cite{clifton}.
For a motivation of the conditions using the idea of screening off and the related idea of the principle of common cause, see \cite{butterfield1}. For 
the necessity of TAF in deriving constraints on probabilities of measurements in hidden-variable models, see Arntzenius \cite{arntzenius}.}. We here do not go into
whether the conditions indeed follow from the idea of local realism together with the notion of free variables. We gloss
over this issue here since it is not relevant for the results to be obtained.
[ Meer over  wat ik niet ga doen, no extensive discussion of the (alledged) sufficiency of these notion to imply the conditions  below.
----------------------}

\subsubsection*{Assuming locality and free variables}
\noindent
Assume that in the local-realist framework the settings are free variables.  We now state some conditions that we take to follow from this assumption.  Whether it is indeed the case that the doctrine of local realism supplemented with the assumption of free variables  imply these conditions is not crucial for our investigation. We could equally well assume these conditions independently, since it is the conjunction of these conditions we are concerned with, not the question what a sufficient motivation for them is.

That the settings are free variables implies that they are statistically independent from the system hidden variables. This results in the requirement of Independence of the Systems (IS):
\begin{align}\label{IS}
\mathrm{IS:}~~~~\rho(\lambda|A,B)=\rho(\lambda) ,~~\forall \lambda,A,B.
\end{align}
 This implies that $\rho(\lambda,\mu_A,\mu_{B}|A,B)=\rho(\mu_A,\mu_{B}|\lambda,A,B)\rho(\lambda),~\forall \mu_A,\mu_{B},\lambda,A,B$. Further, 
locality implies that the distribution of the apparatus hidden variables at one measurement station are independent  of what happens at the other measurement station. This results in the following conditions called Apparatus Factorisability (AF) and Apparatus Locality (AL): for all $\mu_A,\mu_B,\lambda,A,B$
\begin{align}
&\mathrm{AF:}~~
\,\,\,\,\,\rho(\mu_A,\mu_{B}|\lambda,A,B)=\rho(\mu_A |\lambda,A,B)\,\rho(\mu_B|\lambda,A,B),\\
&\mathrm{AL:}~~
\,\,\,\,\,\rho(\mu_A|\lambda,A,B)=\rho(\mu_A|\lambda,A)\,\,\mathrm{and}\,\,
\rho(\mu_B|\lambda,A,B)=\rho(\mu_B|\lambda,B),
\end{align} The probability densities
$\rho(\mu_A |\lambda,A,B,)$, etc. are defined as the marginals of the density
$\rho(\mu_A,\mu_B |\lambda,A,B,)$ and are all assumed to be positive and normalized. 

The conjunction of AF and  AL gives Total Apparatus Factorisability (TAF):
\begin{align}
&\mathrm{TAF:}\,\,\,\,\,\rho(\mu_A,\mu_B|\lambda,A,B)=\rho(\mu_A | \lambda,A)\rho(\mu_B | \lambda,B),&~~\forall \mu_A,\mu_B,\lambda,A,B.
\end{align}
If we now take the conjunction of TAF and IS we finally obtain the condition of Independence of
the Systems and Apparata (ISA):
\begin{align}\label{ISA}
&\mathrm{ISA:}\,\,\,\,\,\rho(\lambda,\mu_A,\mu_B|A,B)=\rho(\mu_A | \lambda,A)\rho(\mu_B |
\lambda,B)\rho(\lambda),&~~\forall \mu_A,\mu_B,\lambda,A,B.
\end{align}
 This condition guarantees independence of measurement apparata from distant spacelike
 apparatus hidden variables and settings, as well as independence of the (sub-) system states $\lambda$ from the
  settings.

Models of the above \emph{Gedankenexperiment}  are usually divided  into two kinds: deterministic
and stochastic. The first kind of model uses the idea that the hidden variables determine the
outcomes of measurements. Probabilities only enter as classical probability functions,
denoted by $P$, on the set $M$ of all hidden variables. Physical quantities are defined as functions on this set. Usually this
set is taken to be a phase space, cf. \cite{butterfield}.
In the second type of models, the stochastic models, the hidden variables determine the
probabilities of measurement outcomes, not the outcomes themselves. Deterministic models
are thus a special case where all probabilities are either $0$ or $1$. 

Note that the only role of the hidden variables in both kinds of models is to either fix
results or probabilities. The distinction between deterministic and stochastic  has thus
nothing to do with the issue of deterministic or indeterministic evolution of the hidden 
variables. A deterministic hidden-variable theory could thus allow for indeterministic
evolution of the systems in question, a point also made by \citet{butterfield}.

\subsection{Deterministic models}\noindent
A deterministic LHV model assumes that the outcomes of experiments are completely
determined by the hidden variables and the settings of the apparata\footnote{\citet{bell64} was the first to consider such abstract deterministic models but only for perfect
\mbox{(anti-)} correlations. The determinism was a consequence of the assumed perfect (anti-) correlation. Under this restriction he derived his famous 1964 inequality. 
\citet{chsh} relaxed this  and were able to derive a
Bell-type inequality nevertheless. \citet{bell71} generalized the result of 
\citet{chsh} by including apparatus hidden variables, thereby considering a contextual deterministic local hidden-variable model.
It is interesting to note that, because of the assumption of perfect \mbox{(anti-)} correlation, Bell's original 1964 inequality is not a Bell-type inequality as we have defined it here in \eqref{lhvbell}.}:
$a= a(A,B,\mu_A,\mu_B, \lambda)$, $b=b(A,B,\mu_A,\mu_B,\lambda)$.
The expectation value   $E(AB)$  of the product of the observables $A$ and $B$ is then determined by
\begin{align}\label{expectdeter}
E(A,B):=\int_M \,a(A,B,\mu_A,\mu_B,\lambda)\,b(A,B,\mu_A,\mu_B,\lambda)
\,\rho(\lambda,\mu_A,\mu_B|A,B)\,d\mu_A\,d\mu_B\,d \lambda,
\end{align}
where $M$ is the total set of all hidden variables and the hidden-variable
distribution is positive and normalized, i.e.,
\begin{align}
\rho(\lambda,\mu_A,\mu_B|A,B)\geq 0 ~~\textrm{and}~~
\int_M \rho(\lambda,\mu_A,\mu_B|A,B)\,d\mu_A\,d\mu_B\,d \lambda=1.
\end{align}

We now invoke to the idea of locality and  thus require
that measurement results
are not dependent on spacelike separated results, settings,
or apparatus hidden variables\footnote{No prescription is given for measurements where one of the apparata is switched off, but this can be easily accounted for by letting the variables representing states and results range over `null states' and `null results', cf. \cite{thesisclifton}. }. 
This results in the assumption of Local Determination (LD):
\begin{align}\label{LD}
\mathrm{LD:} ~~~
\left\{\begin{array}{c}
 a(A,B,\mu_{A},\mu_{B},\lambda)
=a(A,\mu_{A},\lambda), 
\\ b(A,B,\mu_{A},\mu_{B},\lambda)=b(B,\mu_{B},\lambda),
\end{array}\right.
~~~\forall\, \mu_A,\mu_B,\lambda,A,B.
\end{align}

Furthermore, without introducing any restrictions, one can assume that for
a realistic theory (i.e., obeying the idea of realism)  the set of hidden variables $M$ is the Cartesian product $\Lambda\times\Omega_{A}\times\Omega_{B}$. Here $\Lambda $ is the
set of possible values of the hidden variables associated to the two subsystems to be measured and $\Omega_{A}$
and $\Omega_{B}$ is the set of hidden variables of the two apparata that measure $A$ and $B$ respectively. The Cartesian product structure guarantees the compatibility of all the different hidden variables, cf. \cite{clifton}.

We will now use these locality conditions to obtain a non-trivial constraint. First, we
average over the apparata hidden variables to get
\begin{align}
&\overline{a}(A,\lambda):=\int_{\Omega_A} a(A,\mu_{A},\lambda)\rho(\mu_A | \lambda,A)d
\mu_{A},\nn\\ &\overline{b}(B,\lambda):=\int_{\Omega_{B}} b(B,\mu_{B},\lambda)\rho(\mu_{B} | \lambda,B)d
\mu_{B}.
\end{align}
These definitions together with the conjunction of LD and ISA now allow for rewriting  (\ref{expectdeter}) as
\begin{align}
\label{detercorr}
E(A,B)=\int_\Lambda\,\overline{a}(A,\lambda)\,\overline{b}(B,\lambda)
\,\rho(\lambda)d \lambda.
\end{align}
Finally, since $|\bar{a}(A,\lambda)|,|\bar{b}(B,\lambda)|\leq1$ this correlation has the standard form that implies CHSH inequality
\cite{chsh, bellspeakable}:
\begin{align}
|E(A,B)+E(A,B')+E(A',B)-E(A',B')|\leq2.
\end{align}
For a proof see the Intermezzo below. Thus a deterministic theory that has free variables and which obeys local realism 
has to obey this inequality\footnote{\label{alg4}If LD of (\ref{LD}) is violated we can easily get the absolute maximum of 4 for the CHSH expression. Consider the expression $a_{11}b_{11} +a_{12}b_{12}+a_{21}b_{21} -a_{22}b_{22}$. One simply chooses  $a_{11} =a_{12}=a_{21}=-a_{22}=1$ and  $b_{11} =b_{12}=b_{21}=b_{22}=1$. Here $a_{11}$ is the outcome obtained by the first party if both the first and second party choose setting $1$, $b_{12}$ the outcome obtained by the second party if the first party chooses setting $1$ and the second party setting $2$, etc.}\forget{, which, as is well-known, is violated by quantum mechanics up to the Tsirelson value of $2\sqrt{2}$.}.\forget{From a logical point of view we have obtained the following conclusion\cite{zukowski}:
\begin{align}
\neg\,( \forall E_{QM} \,\exists  E_{LR}~:~ E_{LR}=E_{QM}) \,\,
\Longleftrightarrow\,\,
\exists  E_{QM} \,\forall E_{LR} ~:~ E_{LR}\neq E_{QM},
\end{align}
where $E_{QM}$ denotes a quantum correlation and $E_{LR}$ a local correlation
as given by the local realistic model, which is a deterministic one in this case.}
For a stochastic model the same holds, as will be shown in the next subsection

\subsubsection*{Intermezzo: standard derivation of the CHSH inequality}\label{chshderivation}
Suppose $E(A,B)$ has the following form
\begin{align}\label{corr2}
E(A,B)=\int_\Lambda\,X(A,\lambda)\,Y(B,\lambda)
\,\rho(\lambda)d \lambda,
\end{align}
with $|X(A,\lambda)|\leq 1$ and $|Y(B,\lambda)|\leq1$.
Then
\begin{align}\label{chshderiv}
&|E(A,B)+E(A,B') +E(A',B)-E(A',B')|
\nn\\&~~=\int_\Lambda\,|X(A,\lambda)\,Y(B,\lambda) +
X(A,\lambda)\,Y(B',\lambda)\\&  ~~~~~~~~~~~~~~~~~~~~~~~~~~~~~~~~+
X(A',\lambda)\,Y(B,\lambda)-
X(A',\lambda)\,Y(B',\lambda)|
\,\rho(\lambda)d \lambda \nn\\&~~=\int_\Lambda\,
|X(A,\lambda)\,[Y(B,\lambda) + Y(B',\lambda)] +
X(A',\lambda)\,[Y(B,\lambda)- Y(B',\lambda)]|
\,\rho(\lambda)d \lambda \leq2,\nn
\end{align}
where in the last line it is used that $|x(y+y')+x'(y-y')|\leq 2$ for $|x|,|x'|,|y|,|y'|\leq 1$
and $x,x',y,y' \in  \mathbb{R}$, as well as that the distribution $\rho(\lambda)$ is normalized:
$ \int_{\Lambda}\,\rho(\lambda)d \lambda=1$.
Note that we use the same hidden variable $\lambda$ for all four terms in the left hand side of \eqref{chshderiv}. Above we have argued  this to be a consequence of the realism assumption.\forget{ that every one of the possible outcomes for every one of the possible choices one might make for the settings ($A$ or $A'$ and  $B$ or $B'$) in a single experiment performed at a single time are all predetermined by properties of the system (and measurement apparata), with one (and only one) of those predetermined outcomes actually being revealed -- namely the one associated with the particular choice of experiment actually made}

\subsection{Stochastic models}\noindent
We now generalize the above to stochastic models where the hidden variables only determine the probabilities\footnote{It is not necessary to take a stance on what a probability is, 
or to commit oneself to an interpretation of probability. We can be neutral as to whether probability is objective chance, a measure of partial belief, a propensity, etc.; it is sufficient to assume that it is measured by relative frequencies. However,
the idea that we are dealing with a realistic hidden-variables theory,
where the hidden variables are supposed to give a complete description of the state of
affairs favors a reading of the probabilities as objective probabilities or chances.}
for outcomes of measurement. Such as model then provides the quantity $P(a,b | A,B,\mu_{A},\mu_{B},\lambda)$ which is the joint probability that the two measurement outcomes $a,b$ are obtained for some specific settings $A$,
$B$ and hidden variables $\mu_{A}$, $\mu_{B}$ and $\lambda$. These are called subsurface probabilities, to distinguish them from the surface probabilities $P(a,b|AB)$ which are empirically accessible.

For the expectation value  $E(A,B)$ a stochastic hidden-variable theory\footnote{\citet{ch} were the first to consider stochastic hidden-variable models in 1974.
Bell followed in 1976 \cite{bell76} (However, see footnote \ref{awkward} on page \pageref{awkward}). 
  But in a footnote in 1971
Bell already mentioned that the CHSH inequality is expected to hold for stochastic hidden variables as well \cite[footnote 10]{bell71} (cf. footnote \ref{belldeter} on page \pageref{belldeter}). However, Bell's main concern was not with stochastic hidden variables  but with apparatus hidden variables
in a deterministic theory which, when averaged over, would give a condition of average locality from which the CHSH inequality would follow. See \citet[p. 145]{brown} for more on this point.}
 then yields 
\begin{align}\label{corrstoch}
E(A,B)=\int_M \sum_{a,b} ab\, P(a,b | A,B,\mu_{A},\mu_{B},\lambda)\,\rho(\lambda,\mu_A,\mu_{B}|A,B)\,
d\lambda\, d\mu_A \,d\mu_{B},
\end{align}
where we again assume  $M=\Lambda\times\Omega_{A}\times\Omega_{B}$.

Again, we invoke the idea of locality and thus require that measurement 
results are statistically independent of spacelike separated results, settings,
or values of apparatus hidden variables\footnote{\citet{timpson2} argue that the assumption of locality in a stochastic hidden-variable framework is fundamentally different from the one in the deterministic framework of the previous section, so that effectively two different notions of locality are in play. For our purposes such a difference is not important.}. Let us furthermore assume that the hidden variables
completely determine the probabilities and that\forget{ the realism assumption implies} they can serve as common causes.  These assumptions then give\footnote{Whether these assumptions are  indeed  sufficient to imply OF and OL is a matter on which opinions may differ. As mentioned before, for our purposes this is not necessary. We could equally well assume OL and OF right from the start. For a detailed motivation of the conditions using   the principle of common cause and the related idea of the idea of screening off, see \citet{clifton} and \citet{butterfield1}.} the conditions called Outcome Factorisability (OF) and
Outcome Locality (OL) (terminology from \cite{clifton}):\begin{align}
&\mathrm{OF:}\,\,\,P(a,b | A,B,\mu_{A},\mu_{B},\lambda)=
P(a | A,B,\mu_{A},\mu_{B},\lambda)P(b | A,B,\mu_{A},\mu_{B},\lambda),\nn\\&~~~~~~~~~~~~~~~~~~~~~~~~~~~~~~~~~~~~~~~~~~~~~~~~~~~~~~~~~~~~~~~~~\forall \mu_A,\mu_B,\lambda,A,B.\\
&\mathrm{OL:}\,\,\,\left\{ \begin{array}{c} P(a | A,B,\mu_{A},\mu_{B},\lambda)=P(a|A,\mu_{A},\lambda)\\
P(b |A,B,\mu_{A},\mu_{B},\lambda)=P(b|B,\mu_{B},\lambda) \end{array} \right. ,~~\forall \mu_A,\mu_B,\lambda,A,B.
\end{align}
\forget{\end{align} which is equivalent to 
\begin{align}
&\mathrm{OF:}\,\,\,\left\{ \begin{array}{c} P(a | A,B,\mu_{A},\mu_{B},\lambda)=P(a |b,A,B,\mu_{A},\mu_{B},\lambda) \\ P(b | A,B,\mu_{A},\mu_{B},\lambda)=P(b|a,A,B,\mu_{A},\mu_{B},\lambda)\end{array}\right.~,~\forall \mu_A,\mu_B,\lambda,A,B,
\end{align}
and the second reads
\begin{align}}
The probabilities $P(a | A,B,\mu_{A},\mu_{B},\lambda)$, etc. in OF are defined as
the marginals of  the joint probability $P(a,b |A,B,\mu_{A},\mu_{B},\lambda)$, and similarly
for the probabilities in OL.

OL is the condition that the probabilities for outcomes only depends on the local setting, the local apparatus hidden variable and the hidden variable that characterizes the subsystem that is locally measured.  In particular it does not depend on the faraway setting and apparatus hidden variable. OF is the condition that for given settings  and hidden variables  the distribution for outcome $a$ is independent from the outcome $b$ obtained at the other measurement station, and vice versa.

Note that OF and OL are identical to Jarrett's conditions of  `Completeness' and `Locality'  respectively \cite{jarrett}.  See section \ref{jarrettshimony} for a further discussion of Jarrett's conditions.

The conjunction of OF and OL is called Total Factorisability (TF) and gives:
\begin{align}
\mathrm{TF:}\,\,\,\,\,P(a,b | A,B,\mu_{A},\mu_{B},\lambda)=P(a | B,\mu_{A},\lambda)P(b |B,\mu_{B},\lambda),~~~\forall \mu_A,\mu_B,\lambda,A,B.
\end{align}
The conjunction of TF, and ISA allows for rewriting the product expectation value \Eq{corrstoch} as
\begin{align}\label{corr}
E(A,B)=\int_\Lambda \sum_{a,b} ab\, \overline{P}(a
|A,\lambda)\,\overline{P}(b|B,\lambda)\,\rho(\lambda)\, d\lambda,
\end{align}
where  $\overline{P}$ is a $\mu$-averaged probability,  i.e.,
\begin{align}
\overline{P}(a,b | A,B,\lambda)&:=\int_{\Omega_{A} \times\Omega_{B}}
 P(a,b | A,B,\mu_{A},\mu_{B},\lambda) \rho(\mu_{A}, \mu_{B} |\lambda,A,B) d \mu_{A}d \mu_{B},\\
 \overline{P}(a | A,B,\lambda)&:=\int_{\Omega_{A} \times\Omega_{B}}
 P(a | A,B,\mu_{A},\mu_{B},\lambda) \rho(\mu_{A}, \mu_{B} |\lambda,A,B) d \mu_{A}d \mu_{B},~\mathrm{etc}.
\end{align}
Using the definition $\overline{E}(A,\lambda):=\sum_a a\overline{P}(a|A,\lambda)$
 , and analogously for $B$, \Eq{corr} obtains the general form
 \begin{align}\label{barcorr}
E(A,B)= \int_{\Lambda} \overline{E}(A,\lambda)\overline{E}(B,\lambda)\rho(\lambda)d\lambda,
\end{align}
from which the  CHSH inequality follows in the standard way (see also the Intermezzo on p. \pageref{chshderivation}).
Thus a stochastic local realistic theory that has free variables\footnote{Models that obey local realism (i.e., they  obey TF) but that violate the freedom assumption IS of (\ref{IS}) are able to reach the absolute maximum of 4 for the CHSH expression. An example is the following model by \citet{paterek} where the hidden variable forces party $II$ to choose a specific setting depending on what outcomes $b$ and $b'$ (to be found upon measurement of either $B$ or $B'$) this hidden variable prescribes.  The choice of the setting by party $II$ consequently depends on the outcomes to be obtained that are (probabilistically) determined by the hidden variables. 
This constitutes a violation of IS.

In the model the possible results prescribed by $\lambda$ are $(a,a',b,b')=(1,1,1,1)$ with probability $1/2$ and otherwise they are $(a,a',b,b')=(1,-1,-1,1)$.  If in a specific run it will be the case that $\lambda$ prescribes that $b=b'$ the model enforces 
 that party $2$ chooses setting $B$ and otherwise $B'$ is chosen.  This scenario ensures that both $(a+a')b$ and $(a-a')b'$ are equal to $2$ in any run of the experiment, thereby ensuring that the absolute maximum of $4$ for the CHSH inequality is obtained.
 
 Another such a model by \citet{degorre} reproduces the singlet correlations of quantum mechanics as well as the corresponding marginals. Here the distribution of the hidden  variables is explicitly dependent on the setting A: $ \rho(\lambda|A,B)=|\bm{a}\cdot \bm{\lambda}|/2\pi$ (the settings and hidden variable are chosen to be vectorial quantities $\bm{a},\bm{b}, \bm{\lambda} \in \mathbb{R}^3$ respectively). If the outcomes are determined by  $a(\bm{a},\lambda)=-\textrm{sgn}(\bm{a}\cdot\bm{\lambda})$ and $b(\bm{b},\lambda)=\textrm{sgn}(\bm{b}\cdot\bm{\lambda})$, then $\av{AB}=-\bm{a}\cdot\bm{b}$ and $\av{A}=\av{B}=0$, which indeed are the singlet predictions.} has to obey the CHSH inequality.

The surface probabilities are ${P}(a,b|A,B)=\int_\Lambda d\lambda \rho(\lambda)\overline{P}(a|A,\lambda) \overline{P}(b|B,\lambda)$ 
which we defined to be local correlations in section \ref{sectionlocal}, see (\ref{localdistr}). This implies that $E(AB)$ in (\ref{corr}) is equal to $\av{AB}_\textrm{lhv}$, as previously defined in (\ref{expectation}).

This concludes the introduction of stochastic and deterministic local hidden-variable models. Stochastic hidden-variable models are more general than deterministic hidden-variable models. Indeed, the later can be obtained from the first by setting all probabilities to be either $0$ or $1$. 
In such a case the condition of OF is automatically obeyed (for a proof see, amongst others, \citet{jarrett}) and the condition of OL becomes LD, which is therefore also referred to as deterministic OL. 
\citet{suppes} showed that OF together with perfect correlation (i.e., $\av{AB}=+1$)\forget{such as for example predicted by the singlet state in quantum mechanics when $A$ and $B$ are spin measurements in opposite direction)} forces a stochastic hidden-variable model to be deterministic\footnote{This result was anticipated by Bell in 1971 who remarked that perfect correlation requires deterministic determination in a local hidden-variable theory \cite{bell71}.}. This result can be used to argue that genuinely stochastic local hidden-variables theories are a red herring \cite[p.140]{dickson} when it comes to reproducing quantum mechanics\footnote{For the point of view that for the study of non-locality in quantum mechanics stochastic hidden-variable models are beneficial over and above deterministic hidden-variable models, despite the Suppes-Zanotti result, see \cite{clifton, brown, butterfield}.}. However, the verdict on this issue is not important for our investigation, because we will later allow for the possibility that OF is violated and in such a case the Suppes-Zanotti results is no longer relevant. In the rest of this chapter we therefore unproblematically consider the more general case of stochastic hidden-variable theories.

\subsection{Jarrett vs.~Shimony. Are apparatus hidden variables necessary?}
\label{jarrettshimony}

In reply to \citet{jarrett}, who first distinguished OF and OL\footnote{\citet{jarrett} called them `completeness' and `locality'  and  \citet{ballentinejarrett} called them `predictive completeness' and `simple locality'.} and showed them to imply TF, \citet{shimony} presented the conditions of Outcome Independence (OI) and Parameter Independence (PI) which are weaker forms of OF and OL that can be considered as $\mu$- averaged versions of them\footnote{\citet{shimony84} considered these conditions already in 1984 just after Jarrett proposed his conditions, but he did not call them Outcome Independence and Parameter Independence until 1986. For early formulations of OI and PI see \cite{fraassen82} and  \cite{suppes}.}:
\begin{align}
&\mathrm{OI:}\,\,\,\,\,\overline{P}(a,b | A,B,\lambda)=
\overline{P}(a | A,B,\lambda)\overline{P}(b | A,B,\lambda),\\
&\mathrm{PI:}\,\,\,\,\,\overline{P}(a | A,B,\lambda)=\overline{P}(a|A,\lambda)\,\,\mathrm{and}\,\,
\overline{P}(b | A,B,\lambda)=\overline{P}(b|B,\lambda).
\end{align}
OI and PI give $\overline{\mathrm{TF}}$, which is the $\mu$-averaged version of TF:
\begin{align}
&\overline{\mathrm{TF}}:\,\,\,\,\,\overline{P}(a,b | A,B,\lambda)=\overline{P}(a | B,\lambda)
\overline{P}(b |B,\lambda).
\end{align}
In what follows we will call $\overline{\mathrm{TF}}$ Factorisability\footnote{\label{awkward} \citet{ch} also use Factorisability and call this Objective Locality (1974). They seem to be aware that one needs both Shimony's conditions of OI and PI to get  Factorisability, although they do not mention them explicitly. 
In 1976 Bell for the first time uses a fully probabilistic setting where he considers stochastic hidden-variable theories \cite{bell76}. He defines his notion of Local Causality and claims this assumption to give the condition of Factorisability that is subsequently used to derive the CHSH inequality. However, Bell's derivation is flawed: Factorisability does not follow from Bell's notion of Local Causality as it was give in his 1976 manuscript -- to our knowledge this has not been commented on before.

\citet{bell76} speaks of local beables A and B that are measured in regions 1 and 2 respectively. We take these beables to be the possible outcomes of measurement (which are denoted by $a,b$ in our notation). He furthermore introduces the symbol  $N$ to denote all beables in the intersection of the two backward lightcones of the two regions 1 and 2. This we denote by the hidden variable $\lambda$. He next introduces the symbols $\Lambda$ and $M$ to be the specification of some of the beables of the remainder of the backward light cone of $1$ and $2$ respectively. We denote this in our notation by  $A$ and $B$ respectively, which we take to include the settings and any other relevant local beables.
  Bell formulates Local Causality as the claim that (in our notation):
\begin{align}\label{belllc}
P(a|A,b,\lambda)=P(a|A,\lambda)\,,~~~~~ \textrm{(Eq. (2) in \cite{bell76})}. 
\end{align}
This is actually requirement P1 of Maudlin (see section \ref{shimonymaudlin} for Maudlin's assumptions). He next considers the joint probability $P(a,b|A,B,\lambda)$ which he rewrites using a standard  rule of probability into the equivalent form 
\begin{align}\label{bayes}
P(a|A,B,b,\lambda)P(b|A,B,\lambda)\,,~~~~~ \textrm{(Eq. (5) in \cite{bell76})}. 
\end{align}
Next Bell invokes Local Causality and claims that Factorisability (i.e., $P(a,b|A,B,\lambda)=P(a|A,\lambda)P(b|B,\lambda))$ follows.
However, Bell's derivation is wrong because \eqref{belllc} does not suffice to obtain Factorisability from \eqref{bayes}. One needs to assume at least one extra assumption. Indeed,  assuming Maudlin's P2 would be sufficient.

Although Bell's 1976 derivation to obtain Factorisability from Local Causality (in the form used by him in his derivation) needs a supplementary assumption to be successful, we believe that Bell in fact believed that such a supplementary assumption was not necessary. At all other later occasions he uses a different technical formulation of Local Causality and uses a one step derivation to get Factorisability from  Local Causality. Bell never used the distinctions of OI and PI (contra \citet[p. 146]{brown} and \citet[p. 5]{thesisclifton}), see also footnote \ref{roysinghfootnote} on page \pageref{roysinghfootnote}.}.
         Because Factorisability\forget{$\overline{\mathrm{TF}}$} gives the standard form of the correlation as  in \Eq{corr} Shimony's conditions together with IS  also imply the CHSH inequality. 
         
         \citet[p. 226, footnote]{shimony84} explicitly rejected the inclusion of apparatus hidden variables $\mu_A$ or $\mu_B$. \citet[p. 161]{clifton} give a critical discussion of Shimony's arguments for this rejection.  
         We will not discuss their criticism, but discard of a simple objection one might raise against the claim that Simony's conditions are weaker than  Jarrett's. The objection stems from the idea that simply including apparatus hidden variables in the hidden variable $\lambda$ would allow one to reproduce Jarrett's conditions from Shimony's. But there is a problem here. Let us take $\lambda'=(\lambda,\mu_A,\mu_B)$ and denote OI$'$, PI$'$ as the conditions OI, PI where $\lambda$ is replaced by $\lambda'$. One easily obtains that OI$'$ is equivalent to OF and that OL implies PI$'$. But now  OI$'$ and PI$'$ together are in fact more general than OF and OL, and it is the latter that are weaker, not the first. However, and this is the crucial point, the conjunction of PI$'$  with OI$'$ does not imply TF or any other similar factorisability condition where, given the hidden variables, one obtains the statistical independence between the two measurement stations. It gives $P(a,b|A,B,\lambda')=P(a|A,\mu_A,\mu_B,\lambda)P(b|B,\mu_A,\mu_B,\lambda)$, which has an unwanted non-local dependency on the apparatus hidden variables.

Let us take a closer look at Shimony's conditions before we discuss further how they relate to Jarrett's conditions. PI is the condition that the probabilities for outcomes only depend on the local setting and the hidden variable $\lambda$ that characterizes the subsystem that is locally measured.  In particular it does not depend on the faraway setting. OI is the condition that for given settings and hidden variables the distribution for outcome $a$ is independent from the outcome $b$ obtained at the other measurement station, and vice versa.

A violation of PI entails that given the value of the hidden variable $\lambda$ the statistical distribution of $A$ can be changed by changing the setting $B$ of the distant apparatus.  If the hidden variables are under control this can be used to send a spacelike signal from $I$ to $II$ or vice versa. If, however, this control is absent, there cannot be any signaling, but the non-local setting dependence remains.\forget{ and since the probability distribution of outcome $a$ depends on the instantaneous value of the setting $b$ violation of PI needs a relation of distant simultaneity and cannot be reconciled with the theory of special relativity.}  A violation of OI entails that given the settings and the value of the hidden variable $\lambda$ the statistical distribution of outcome $a$ changes if the outcome $b$ of the distant apparatus would be different.  

The two sets of conditions (Jarrett's OF and OL and Shimony's OI and PI) are often conflated, however this is faulty. The first includes apparatus hidden variables, whereas the second does not. \citet[section 4]{jonesclifton} have shown that  this difference matters and that the two sets are fundamentally different: violations of OI can be
compatible with OF.  
They also remark: ``Presumable we should take the same precaution with regard to LOC [our notation: OL] and parameter independence [PI]'' \cite[p. 310]{jonesclifton}.
\forget{
\cite{clifton},  blz 117,118 en 175: If TF and strict correlation hold in any hidden-variable state $\lambda$, then this implies $\overline{TF}$.
Furthermore, when strict anti-correlations occur (and one thus effectively has a
deterministic model) one can have TAF violating models that obey TF \cite{clifton},
blz 175 (two-step causation) for which the CHSH inequality holds.
However, not using strict correlations requires one to assume TAF to derive the
CHSH inequality.}

Although we will later not explicitly use the apparatus hidden variables, we believe it is  good practice to include them for the following three reasons. 
\begin{enumerate}
\item  Including apparatus hidden variables allows for more general hidden-variable models and we believe it is therefore to be preferred. Such an dependence on the apparatus hidden variables can be easily removed:  one simply averages over them. 
\forget{
Furthermore, we agree with \citet{clifton} that OF and OL are more directly motivated by the idea of locality and of hidden variables as common causes than their $\mu$-averaged versions OI and PI.   }
\item  Considering apparatus hidden variables allows for two different physical motivations for a stochastic model \cite{butterfield}.
The first motivation we have already encountered: a complete specification of the hidden variables determines not outcomes of measurement but only probabilities for outcomes to be obtained. Such a model incorporates some irreducible indeterminism. 
The second motivation arises when one considers averages over the apparatus hidden variables $\mu$, for example because the influence of the apparatus hidden variables cannot be accounted for precisely.  Such average values can also be interpreted, as \citet{bell71} already noted\footnote{\label{belldeter}\citet[footnote 10]{bell71}: ``We speak here [when introducing apparatus hidden variables] as if the instruments responded in a deterministic way when all variables, hidden or nonhidden are given. Clearly (6) [i.e., \eqref{barcorr} above] is appropriate also for \emph{indeterminism} with a certain local character.''}, as the predictions of an indeterministic theory. When discussing Leggett-type models in section \ref{sectionleggett} such an interpretation is explicitly spelled out. At the deeper level where one considers all hidden variables the model is deterministic but after averaging over apparatus hidden variables $\mu$ the model can be described as a stochastic hidden-variable model prescribing only probabilities for outcomes of measurement.
\item Including apparatus hidden variables  incorporates the Bohrian point of view that the total measurement context should be taken into account. 
\end{enumerate} 
In the remainder of this chapter we will only consider averages over the apparatus hidden variables, i.e., $\mu$-averages. For notational simplicity we will therefore drop the `bar' over $P$ to denote such $\mu$-averaged probabilities.   We will also deal with the $\mu$-averaged version of ISA which is the assumption of IS. This encodes the notion of `free variables' on the $\mu$-averaged level. 
\forget{ A more general analysis that includes apparatus hidden variables of what follows can presumable be carried out. }

\subsection[Shimony vs.~Maudlin: On the non-uniqueness of conditions that give Factorisability]{Shimony vs.~Maudlin: On the non-uniqueness of\\ conditions that give Factorisability}
\label{shimonymaudlin}

In the previous section we have seen that in reply to Jarrett's analysis Shimony argued for conditions where the apparatus hidden variables  are averaged over. He showed that Factorisability is the conjunction of PI and OI. In reply to Shimony's analysis, \citet[p. 95]{maudlin} has argued that this conjunction is not unique. He claims that Factorisability is  logically equivalent to the conjunction of two other conditions which he called P$1$ and P$2$. These are:
\begin{align}
&\mathrm{P}1:~~~P(a|A,b,\lambda)=P(a|A,\lambda)~~~\mathrm{and}~~~P(b|B,a,\lambda)=
P(b|B,\lambda).\\
&\mathrm{P}2:~~~P(a|A,B,b,\lambda)=P(a|A,b,\lambda)~~~\mathrm{and}~~~
P(b|A,B,a,\lambda)=
P(b|B,a,\lambda).
\end{align}
Maudlin gives no proof of his claim and \citet[p.224]{dickson} mentions that the proof is not given by Maudlin, but that it ``proceeds along lines
slightly different from the proof of Jarrett's result [\ldots]". However, the proof appears to be not straightforward at all, and since no proof was found in the literature we present one in the Appendix on page \pageref{appendixmaudlin}. This shows that P1 and P2 indeed imply Factorisability.

\subsubsection{Maudlin's and Shimony's conditions in quantum mechanics}

Quantum mechanics  can be considered as a stochastic hidden-variable theory. It then obeys PI   (also referred to as the `quantum no-signaling theorem'), but violates OI. This proven in the Appendix on page \pageref{SMQM}.
 
\citet[p. 95]{maudlin} claims that his P1 is obeyed by quantum mechanics whereas it is P2 that is violated. 
He furthermore remarks that  ``One might very well call P1 outcome independence and P2 parameter independence, since P1 concerns conditionalizing on the distant outcome and P2 on the distant setting'' \cite[p. 95]{maudlin}. Quantum mechanics is thus claimed to violate his  parameter independence but to obey his outcome independence, which -- and this is the crucial point for Maudlin -- is just the opposite from the analysis in terms of Shimony's concepts.  Indeed, quantum mechanics violates Shimony's outcome independence but obeys his parameter independence.

Before we assess the consequences of this claim for the project of understanding the violation of Factorisability in quantum mechanics and of the CHSH inequality by experiment, an important point needs to be made.  Maudlin gives no proof of his claim, but merely states that ``orthodox quantum mechanics violates P2 but not P1'' \cite[p. 95]{maudlin}. However, in order to evaluate the Maudlin distinctions in quantum mechanics  one needs to make extra assumptions not needed for evaluating Shimony's conditions. In the Appendix on page \pageref{SMQM} this is explicitly shown. The extra assumption is that one needs to provide a probability distribution $\rho(A,B)$ for what settings $A$ and $B$ are to be chosen by the two parties.  However, quantum mechanics does not prescribe anything about what observables are to be chosen. It merely gives predictions for outcomes to be obtained given an experimental context where the settings are known. 

\subsection{On experimental metaphysics} 
\label{expmeta}
Let us adopt the position that the experimentally confirmed violations of the CHSH inequality (modulo loopholes) imply that Factorisability must be violated
because the only other alternative, violation of IS, is rejected as it is too implausible. This position is adopted by the majority of philosophers and physicists (although notorious exceptions exist) because they accept that the settings can be considered free variables. Therefore, it was thought by many that in order to understand violations of the CHSH inequality one should understand failures of Factorisability. So when Jarrett and Shimony presented their two conditions that together imply respectively TF and Factorisability,  a lively debate started as to what violations of each of these two conditions entails. The focus has been on understanding violations of Factorisability rather than its counterpart TF  that uses apparatus hidden variables.


A common position in the literature is that failures of Factorisability should be understood  as a failure of OI rather than PI.  Otherwise, it is argued, the 
possibility of influencing the statistics of measurement outcomes on a system by manipulating a setting under our control on a distant system would, for a given $\lambda$, allow superluminal signaling between spacelike separate events which conflicts with relativity. It is furthermore argued that no such signaling is possible from a violation of OI because the outcomes are only constrained stochastically and are not under our control. 
Thus violations of OI are supposedly consistent with relativity and no-signaling, whereas violations of PI are not, and it is furthermore believed that  correlations that violate OI do not exhibit spacelike causation.   \citet[p. 226]{shimony84} referred to this position as one of `peaceful coexistence' between quantum mechanics and relativity: action at a distance (violation of PI) is avoided but we must accept a sort of a new sort of non-causal connection called  `passion at a distance' (violation of OI) \cite[p. 227]{shimony84}, cf. \cite{redhead}. Such an activity has been called `experimental metaphysics'\footnote{\citet[p. 296]{jonesclifton} characterize this activity as follows: ``First we demonstrate that any empirically adequate model of the Bell-type correlations which does not contain any superluminal signaling will have a particular formal feature. [\ldots] Then  we adduce an argument which purports to show that the formal feature in question is evidence of a certain metaphysical state of affairs. It this works we have powerful argument from weak and general premises (namely, empirical adequacy and a ban on superluminal signaling) to a rich an momentum conclusion about the structure of the world''.}. Others have suggested different interpretations of the alleged violation of OI that have resulted in equally startling metaphysical conclusions\footnote{For example: the existence of holism of some stripe \cite{teller86,teller89, healey91};  incompleteness as a property of nature \cite[p. 700]{ballentinejarrett}; the necessity of broadening the classical concept of a localized event \cite[p. 30]{shimony1989}; adopting relative identity for physical individuals \cite[p. 250]{howard89}.}.

But the view that it is OI that must be abandoned whereas it is PI that must be retained has been challenged by several authors. We will not discuss this issue very extensively.  We merely review three arguments that exist in the literature against this position and therefore against the specific form of experimental metaphysics that takes this position as its starting point. Then in the next section two other difficulties for advocating this position will be provided, which we briefly discuss here already.

{\bf (I)} \citet[p. 95]{maudlin} comments on the discussion about whether it is OI or PI that should be abandoned, that ``the entire analysis is somewhat arbitrary'' because of the non-uniqueness of carving up the condition of Factorisation. Why focus on OI and PI rather than on  P1 and P2? Furthermore, in a quantum context, where OI is violated and PI is obeyed,  focusing on P1 and P2 shows a different picture: P2 is violated by quantum mechanics thereby indicating a form of setting dependence instead of outcome dependence.  The `passion at a distance' has become `action at a distance'.  Because we have no reason to favor the distinction between OI and PI over the one between P1 and P2 the starting point for experimental metaphysics is blocked.

However, we note that this argument glosses over an important difference between the Shimony and Maudlin distinctions.  
We have just seen that Maudlin needs to make additional assumptions -- not needed by Shimony --  in order to evaluate his conditions in quantum mechanics.  The non-uniqueness is thus only on a formal level, when actually applied to quantum mechanics the Maudlin distinctions becomes unnatural because of the supplementary assumptions. Maudlin's argument that in evaluating the condition of Factorisability in quantum mechanics one can equally well use his way of carving up this condition instead of Shimony's thus breaks down. 

Despite the failure of Maudlin's argument, the argument against the importance of the Shimony way of carving up Factorisability can be somewhat saved  by noting that it has not been shown that Shimony's way is unique. OI and PI are sufficient to get Factorisability\forget{, but perhaps not necessary}. But  Maudlin's conditions suffice too, although, as we have shown above, they can arguably be dismissed as unnatural.  In any case, we have no necessary and sufficient set of conditions, therefore we cannot say what a violation of Factorisability amounts to. This point will return in the next section, where we show that  they are not necessary for deriving the CHSH inequality.


{\bf (II)} The relationships between on the one hand OI and PI  and on the other hand spacelike causation, signaling and relativity are much more subtle than is acknowledged in most of the experimental metaphysics projects\footnote{For an extensive discussion of these relationships, see \cite{clifton}, \cite{maudlin}, \cite{dickson}, \cite{butterfield} and \cite{berkovitz1998a,berkovitz1998b}.
}.
\begin{description}
  \item{}(a) First of all,  PI may be violated without there being any signaling.  It could be that the hidden variables $\lambda$ may not be controllable, thereby blocking the route to changing the faraway outcome using the local setting under control, cf. (III) below.

  \item{}(b) In the literature it is argued that relativity plays identical roles in justifying completeness  and locality, but for both not a decisive role. For example, \citet[p. 76]{butterfield} claims that ``[t]he prohibition of superluminal causation plays the same role in justifying completeness and locality`'. In justifying both conditions extra assumptions over and above relativity are needed, such as  e.g., Reichenbach's common cause principle, cf. \citet[especially chapter 4]{maudlin}.

\item{}(c)  It is furthermore argued that peaceful coexistence with relativity does not favor giving up OI instead of  PI, i.e.,  violations of OI and PI are equally at odds with a ban on superluminal signaling, and, furthermore, that there is nothing intrinsically non-causal about correlations that violate OI. \citet{jonesclifton} make this point very convincingly for the conditions OF and OL that include the apparatus hidden variables $\mu$  (recall that the $\mu$-averaged conditions of OF and OL are OI and PI respectively). They show that violation of OF can be used to signal superluminally if the right conditions were satisfied. And the conditions that would have to be satisfied are just the same (mutatis mutandis) as the conditions that would have to be satisfied for us to be able to put violations of OL to  use in signaling superluminally -- roughly speaking, in both cases we need to assume that we would be able to control (or at least influence) the values of all the variables that appear in  the relevant conditional probabilities\footnote{A violation of OF implies indeterminism, i.e., probabilistic determination of the outcomes. But this does not imply that one cannot signal superluminally: ``The possibility remains open that the experimenter might use some controllable feature of the experimental situation as a ``trigger'' which operates \emph{stochastically}  on the outcomes at her own end of the experiment. The signaler could then influence, without complete controlling, the result in the individual case, and could thus signal superluminally by employing an array of identically prepared experiments---just as in Jarrett's own argument for the claim that a failure of OL for stochastic theories makes superluminal signaling possible.'' \cite[p. 301]{jonesclifton}.}. So no important asymmetry has been established between OL and OF with regard to superluminal signaling.  The same analysis holds for the $\mu$-averaged conditions OI and PI. Indeed, \citet{kronz} proved that under certain circumstances one can use violations of OI to signal superluminally.

\forget{There is nothing intrinsically non-causal about correlations that violate OF.  This can be separated into two points. First, a purely comparative point: As violations of OF are just as closely connected to the possibility of signaling, so to speak, as violations of OL (and given that no other relevant considerations have been presented), we have no reason to think that violations of OF are any more likely to have an non-causal explanation than violations of OL. Second, and furthermore, given that violations of OF clearly could be exploited for the purposes of signaling if the appropriate conditions were satisfied, there is no particular reason to think that violations of OF could not have a causal explanation. They might have an non-causal explanation, but there's no good reason to think that they would have to have, or even that they would be most naturally interpreted as non-causal.}
 
\citet[p. 77]{butterfield} is of the same opinion: `To sum up: relativity's lack of superluminal causation does not favor giving up OI over giving up PI. It leaves the issue open.', cf. \cite[pp. 131-135]{butterfield1}.\footnote{Points (b) and (c) go against the idea that OI is solely a condition of completeness or sufficiency which has nothing to do with locality and superluminal causation.  For example, Uffink (private communication) defends such a view: OI can be given an interpretation  in terms of Bayesian statistics where $\lambda$ and the settings are a sufficient statistic. 
We do not comment further on whether locality and/or spacelike causation need to be invoked to argue for OI.  What seems to matter most is 
not what is sufficient for justifying OI but what violations of it amount to. The fact that violations of OI can, under the right circumstances, lead to signaling  is enough to block the starting point of this form of experimental metaphysics, which is that violations of OI peacefully coexist with relativity.
 \forget{
But let's us recall Maudlin's distinction. A interesting question would then be if the violation by QM of Maudlin's condition, which is P2 can also be given such a motivation. And this is very unlikely, since it is the conditional dependence on a  setting that is relevant, not on an outcome.
Koren op de molen van Maudlin. de conditie die geschonden wordt bij Maudlin kan moeilijk een interpretatie in termen van sufficient statistics gegeven worden.}} 
\end{description}

{\bf (III)} The mere existence of Bohmian mechanics undermines the starting point of experimental metaphysics:  violations of PI do not  have to lead to signaling.  This theory violates PI, obeys OI, and is empirically equivalent to quantum mechanics and thus has no-signaling for the surface probabilities. Of course, there is a problem with reconciling Bohmian mechanics with relativity (Lorentz invariance of the dynamics is a problem), but this is a different point.  The comparison should be made to non-relativistic quantum mechanics, so Lorentz invariance is not the issue.

{\bf (IV)} In the next section we show that the CHSH inequality can be obeyed by hidden-variable models that violate the conditions OI, PI and IS. These models are setting and outcome dependent in a specific way. This shows explicitly that none of these three conditions are necessary for this inequality to obtain, i.e, they are only sufficient. Therefore, we have no reason to expect either one of them to hold solely on the basis of the CHSH inequality.
Of course, when confronted with  experimental violations of the CHSH inequality one must still give up on at least one of the conditions OI, PI or IS. However, the crucial point is that giving up only one might not be sufficient. The CHSH inequality does not allow one to infer which of the conditions in fact holds.  The results of the next section show that even satisfaction of the inequality is not sufficient for claiming that either one holds. It could be that all three conditions are violated in such a situation.
\forget{
The results of the next section show that even adopting a theory for which the inequality is obeyed is not sufficient for claiming that either one holds. It could be that all three are violated in such a theory.

The results of the next section show that even satisfaction of the inequality  is not sufficient for claiming that either one  holds. It could be that all three are violated in such a situation.
}

{\bf (V)} Another difficulty for the project of experimental metaphysics, which is related to what was remarked about Bohm's theory in (III) above, is that  with respect to violations of the CHSH inequality it will be shown in subsection \ref{sectionleggett} that which conditions are obeyed and which are not depends on the level of consideration.  The verdict whether it is OI or PI that is to blame in violations of the CHSH inequality  may thus change depending on the level of consideration. A conclusive picture therefore depends on which hidden-variable level is considered to be fundamental. 
But since it is impossible to know whether one has obtained the fundamental level, one cannot argue that it is OI that has to be abandoned when being confronted with violations of the CHSH inequality. It might be that a deeper hidden variable level exists at which it is deterministic PI that is violated and not OI. This again undercuts the starting point of any form of experimental metaphysics that takes the failure of OI to be responsible for violations of the CHSH inequality.

\forget{
\subsection{On spacelike causation and signaling}\label{spacelike}
imperfect match between causal structure from probabilistic constraints.
===============

Why Jarrett calls this completeness. If one has complete information about the experiments and the
subsystems that are measured (i.e., of one knows  $A,B,\mu_A,\mu_{B},\lambda$) then finding out
outcome $b$ cannot tell you anything about the probability of outcome $a$. 

It is tempting to interpret this as a condition that has nothing to do with non-locality, for a non-local observer would adopt the Jarrett completeness condition as long as he has complete information. 

================

Onderscheid in Jones/Clifton twee zaken
1) hoe de Jarrett-condities met Shimony gerelateerd zijn. sectie 4 v/h artikel. ovr werk van Kronz
2) hoe de Jarrett condities met signallign en spacelike causation gerelateerd zijn. Sectie 1-3 v/h artikel.

Punt 1) behandel ik in andere sectie, punt 2 behandel ik hier. Bub vat dit goed samen.

ad 2):Jones and Clifton:
 some forms of violation of parameter dependence (jarrett locality)
implies signalling
(when full control over hidden apparatus variables)
(perhaps only stochastically)
conversely: no- signaling sometimes (?) implies parameter independence
But not the other way around. some parameter dependent models do not allow
signalling.
\\\\
Jones and Clifton:
 determinism implies Jarrett completeness. Thus when Jarrett completeness
fails we have some form of indeterminism (the  hidden variables determine the results only
stochastically). However, this does not exclude
signaling. There could be signaling stochastically. This what they set
out to prove, using one additional assumption called MC. Against experimental
 metaphysics that we have to accept that violations of the Bell ineqaulity tell us that
completeness does not hold and  give a metaphysical account of it.

CONCLUSION: Violations of OI and PI are equally at odds with  a ban on spacelike causation. Butterfield agrees.

OL en PI hebben niets met signaling te maken, maar met causation:
Signaling requires particular abilities to control physical states. The hidden variables are not assumed to be perfectly controllable, it is not assumed that the experimenters can perfectly control any disered ensemble of perfectly specified physical states.

But one can even allow for signaling at the hidden-variable level and still be no-signaling at the surface level. Bohmian mechanics is an example of this.  This theory violates subsurface signaling (it violates 
)  but has surface signaling since the hidden variables are uncontrolable and the hidden-variable distributions are such that they prevent the subsurface signaling from appearing at the surface.

A distinction between (spacelike) causation and (spacelike) signaling is important. \cite{maudlin}

OL and PI forbid spacelike(superluminal) causation, not spacelike(superluminal) signaling. 
For example, if one takes PI to capture the idea of the impossibility of spacelike causation, then Bohm has spacelike causation (violates PI) 
but has no spacelike signalling. it obeys the no-signalling principle.

For a discussion see:  see \citet{clifton}, \citet{maudlin}, \citet{dickson} and \cite{berkovitz1998a,berkovitz1998b}
Whether the dynamics is Lorentz invariant: this is a onother issue

ALS VOETNOOT??

Ballentine \& Jarrett claim that in a deterministic setting
parameter independence and Factorisability are equivalent, and
furthermore that determinism implies outcome independence. They then reason as follows:
determinism and parameter independence imply Factorisability.
Quantum mechanics violates Factorisability, but obeys parameter independence, and thus it must
violate determinism. They take the violation of the Bell inequalities as an
indication that a deterministic substratum does not exist.

However, they admit in footnote 12 that there is another possibility, which they however
dismiss on personal grounds. This possibility is that deterministic hidden variables exist
and violate parameter independence, although we are not able to control them
and thus cannot send signals superluminally.
  Bohmian mechanics is an example of such a HV theory. It is obeys no signaling for the expectation values (when averaged over the beables)
on the observational level, however, it doesn't obey parameter independence on the subsurface level; these are non-locally determined.

}
\section[Non-local hidden-variable models obeying the CHSH inequality]{Non-local hidden-variable models obeying the\\ CHSH inequality}\label{non-localsection}
\forget{
In  section \ref{LRsection} we have seen that in the derivation of the CHSH
inequality we had to assume a  locality assumption for both deterministic and  stochastic models\footnote{
For a deterministic model the conjunction of the assumptions of LD and ISA
(or their $\mu$-averaged versions)
resulted in the product expectation value to be of the standard form from which the CHSH inequality follows,
and to get LD we needed locality. For a stochastic
model it was the conjunction of TF and ISA (or again their $\mu$-averaged versions)
that led to the desired expression for the product expectation value,  and to get TF we needed to assume OL which
required locality.}.
However, it was not shown that this assumption of locality is necessary in order to arrive at an expression for the product expectation value that leads to the CHSH inequality can be derived.
This section, which investigates
  possibilities of weakening the assumptions of the previous section, will show that this is indeed not necessary.
\forget{And indeed, it is not necessary, as will be shown in this section, 
which investigates
  possibilities of weakening the assumptions of the previous section. }
  }

  In this section we will derive the CHSH inequality firstly, by relaxing the condition LD (in the deterministic case) or OI  and PI (in the stochastic case), and secondly, we will allow for specific hidden-variable distributions that violate IS (i.e., we will not assume  the hidden variables to be free variables).  Although this weakens the assumptions of the previous section considerably, we nevertheless show that   the derivation of the CHSH inequality, both for deterministic as well as for stochastic models, still goes through.
This analysis generalizes investigations by \citet{fahmi1} and \citet{fahmi,fahmi2}.

We perform the analysis for $\mu$-averaged assumptions. No apparatus
hidden variables besides the settings are taken into account.
The generalization to models that include apparatus hidden variables
\forget{are here omitted, since they}gives no new interesting results,
but can easily be done. Alternatively, we could think of the
apparatus hidden variables to be included in the settings for
notational simplicity, cf.~\citet{butterfield}, but
note that this inclusion glosses over the differences between
the two types of models (i.e., with or without apparatus hidden variables), as discussed in section \ref{jarrettshimony}, although these differences do not matter here.

\subsection{Deterministic case}\noindent

Consider the \emph{Gedankenexperiment} of Figure \ref{gedank}. Recall from section \ref{LRfree} that a deterministic local hidden-variables model which obeys the assumptions of LD  \Eq{LD} and ISA  \Eq{ISA} must obey the CHSH inequality.
However, we will now weaken these two requirements and show that they are nevertheless sufficient in order to derive the CHSH inequality. 

 Let us first weaken LD by explicitly allowing for non-locality. Assume a deeper hidden-variable level which is represented by hidden variables $\omega$ for system $1$ and $\nu$ for system $2$, with corresponding distribution functions
$k(\omega)$ and $l(\nu)$. The outcomes of measurement  $a(A,B,\lambda)$ and $b(A,B,\lambda)$
are now assumed to be determined by the deeper level hidden variables in the following way \cite{fahmi1}:
\begin{align}
a(A,B,\lambda)&= \int  f_1(A,\lambda,\omega) \,g_1(B,\lambda,\omega)\,k(\omega)\, d\omega , \label{non-localeq}\\
b(A,B,\lambda)&=\int f_2(A,\lambda,\nu) \,g_2(B,\lambda,\nu)\, l(\nu)\, d\nu ,\label{non-localeq2}
\end{align}
with
\begin{align}\label{relaties}
&-1\leq f_1(A,\lambda,\omega) ,\,g_1(B,\lambda,\omega),\,
f_2(A,\lambda,\nu) ,\,g_2(B,\lambda,\nu)\leq 1 ,\nonumber \\
& \int k(\omega)d\omega\,=\,\int l(\nu) d\nu =1\,, ~~\textrm{and}~~k(\omega)\geq 0,~ l(\nu) \geq \,0.
\end{align}
The functions $f$ and $g$ represent response functions that encode the way the
non-local hidden-variables theory determines the measurement outcomes. For example, in \Eq{non-localeq}
the outcome experimenter $1$ (who measures system $1$) will obtain
when $A$ on system $1$ is being measured and $B$ on
system $2$ (by experimenter $2$) for a given hidden variable $\lambda$ is determined firstly by averaging over
the deeper level hidden variable $\omega$ and secondly by
the local response function
$f_1(A,\lambda,\omega)$ and the non-local response function $g_1(B,\lambda,\omega)$ (note that both response functions are for system $1$).\\\\
Note that for the special case of $k(\omega)=\delta(\omega -\omega_0)$ and $l(\nu)=\delta(\nu -\nu_0)$ we get
 the non-local relations of \citet{fahmi}): $a(A,B,\lambda)=f_1(A,\lambda) \,g_1(B,\lambda)$ and $b(A,B,\lambda)=f_2(A,\lambda) \,g_2(B,\lambda)$.

The non-local relations \Eq{non-localeq} and \Eq{non-localeq2} lead to
\begin{align}\label{correlation}
&E(A,B)=\int a(A,B,\lambda)\,b(A,B,\lambda)
\,\rho(\lambda)\,d\lambda\\
&=\int \Big( \int  f_1(A,\lambda,\omega) \,g_1(B,\lambda,\omega)\,k(\omega)\,
d\omega \int f_2(A,\lambda,\nu) \,g_2(B,\lambda,\nu)\, l(\nu)\, d\nu \Big) \rho(\lambda)\,d\lambda\nn\\
&=\int\int l(\nu) k(\omega) d\omega d\nu\Big( \int f_1(A,\lambda,\omega)\,g_1(B,\lambda,\omega)
f_2(A,\lambda,\nu) \,g_2(B,\lambda,\nu)\,\rho(\lambda)\,d\lambda \Big).\nn
\end{align}
Now define:
\begin{align}
U(A,\lambda,\omega,\nu)&:= f_1(A,\lambda,\omega)\, f_2(A,\lambda,\nu),\\
W(B,\lambda,\omega,\nu)&:= g_1(B,\lambda,\omega)\,g_2(B,\lambda,\nu).
\end{align}
It follows that $|U(A,\lambda,\omega,\nu)|\leq 1$, and $|W(B,\lambda,\omega,\nu)|\leq1$.
Let
\begin{align} E_{(\omega,\nu)}(A,B):=\int U(A,\lambda,\omega,\nu)\,W(B,\lambda,\omega,\nu)
\,\rho(\lambda)\, d\lambda.
\end{align}
This quantity is in the standard factorisable form, and thus a CHSH inequality holds for
$E_{\omega,\nu}(A,B)$ (see the Intermezzo on p. \pageref{chshderivation}). 

After averaging over $\mu, \nu$ we finally obtain the expectation value $E(A,B)$:
\begin{align}\label{eab}
E(A,B)=\int \int E_{(\omega,\nu)}(A,B)l(\nu) k(\omega) d\omega d\nu.
\end{align}
Thus $E(A,B)$ is a ($\omega,\nu$)-average of
$E_{(\omega,\nu)}(A,B)$,
and therefore a CHSH inequality also holds for $E(A,B)$,
since averaging cannot increase expectation values.   
\forget{
An explicit proof of the CHSH inequality
for the results $a(A,B,\lambda)$ and $b(A,B,\lambda)$ as determined
in \Eq{non-localeq} and \Eq{non-localeq2}
 is the following \cite{fahmi1}. Suppose that on system $1$ experimenter $1$ can also
 measure observable $A'$ and that
on system $2$ experimenter $2$ can also measure the observable $B'$.
Using \Eq{relaties} the following relation now holds for the response functions $f$ and $g$:
\begin{align}\label{rel}
&|f_1(A,\lambda,\omega) g_1(B,\lambda,\omega) f_2(A,\lambda,\nu)
g_2(B,\lambda,\nu)+\nonumber\\&
f_1(A,\lambda,\omega) g_1(B',\lambda,\omega) f_2(A,\lambda,\nu)
g_2(B',\lambda,\nu) +\nonumber\\&
f_1(A',\lambda,\omega) g_1(B,\lambda,\omega) f_2(A',\lambda,\nu)
g_2(B,\lambda,\nu) -\nonumber\\&
f_1(A',\lambda,\omega) g_1(B',\lambda,\omega) f_2(A',\lambda,\nu)
g_2(B',\lambda,\nu) |\leq2.
\end{align}
This follows because in general $|x_1y_1+x_1y_2+x_2y_1-x_2y_2|=|x_1(y_1 +y_2)+x_2(y_1-y_2)|\leq 2$,
for all $|x_i|\leq1$, $|y_i|\leq1$, $i=1,2$.
Averaging over the deeper level hidden variables gives:
\begin{align}
|a(A,B,\lambda)b(A,B,\lambda)+&
a(A',B,\lambda)b(A',B,\lambda)\nn\\&+a(A,B',\lambda)b(A,B',\lambda)-
a(A',B',\lambda)b(A',B',\lambda)|\leq 2.
\end{align}
Averaging over the hidden variable $\lambda$  with distribution $\rho(\lambda)$
now gives the CHSH inequality for the non-local hidden-variable theory:
\begin{align}
\label{chsh}
 |E(A,B)+E(A,B')+E(A',B) -E(A',B')|\leq 2,
\end{align}
with the expectation value $E(A,B):=
\int a(A,B,\lambda)b(A,B,\lambda)\rho(\lambda)d\lambda$.
}
Note that one could have started out with only the deeper level hidden variables
and thus eliminate the hidden variable $\lambda$. However, for ease of
comparison to the standard CHSH inequality derivation this has not been performed.
 \\\\
Before providing a generalization of the above to the stochastic case,
we show that we can  weaken the assumption IS which was part of the assumption ISA that we previously used to derive the CHSH inequality.  We thus no longer assume that we deal with free variables, i.e., the freedom assumption that gives IS  of (\ref{IS}) need not be made: The distribution
$\rho(\lambda)$ of the hidden variable $\lambda$ does not  have to be obey  $\rho(\lambda|A,B)=\rho(\lambda)$.
Indeed, a normalized distribution of the form \cite{fahmi1}
\begin{align}\label{non-localdistr}
\rho(\lambda|A,B)=\int\tilde{\rho}(\lambda|A,\gamma)\tilde{\tilde{\rho}}(\lambda|B,\gamma) m(\gamma) d \gamma,
\end{align}
also suffices to derive the CHSH inequality\footnote{
$E(A,B)$ of \Eq{correlation} is then defined as
$E(A,B)=\int a(A,B,\lambda)\,b(A,B,\lambda)
\,\rho(\lambda|A,B)\,d\lambda$.}. Here $\gamma$ is a deeper level hidden variable
with distribution $m(\gamma)$, and where $0\leq\tilde{\rho}(\lambda|A,\gamma)\leq 1, ~
0\leq \tilde{\tilde{\rho}}(\lambda|B,\gamma)\leq 1$ and $\int m(\gamma) d\gamma =1$.
 Note that if $m(\gamma)=\delta(\gamma -\gamma_0)$ we get for $\rho(\lambda)$:
\begin{align}\label{deltaIS}
\rho(\lambda|A,B)=\tilde{\rho}(\lambda|A,\gamma_0)\tilde{\tilde{\rho}}(\lambda|B,\gamma_0).
\end{align}
The distribution of the hidden variables of the system
 explicitly depends on the settings of both measurement apparata\footnote{Here we have assumed that $\rho(\lambda|A,B)$ is normalized. If this is not the case, the distribution must have the following factorising form: $\rho(\lambda|A,B)=
\tilde{\rho}(\lambda|A)\tilde{\tilde{\rho}}(\lambda|B)/\int \tilde{\rho}(\lambda|A)d\lambda \int \tilde{\tilde{\rho}}(\lambda|B) d\lambda.$
}.

The proof that a setting dependent hidden-variable distribution of the form (\ref{non-localdistr}) suffices to obtain the CHSH inequality 
goes exactly analogously to the above proof that establishes that $E(A,B)$ of \eqref{correlation}  with  $\rho(\lambda|A,B)=\rho(\lambda)$ has to obey the CHSH inequality\footnote{More explicitly, one inserts \eqref{non-localdistr} in  \eqref{correlation}  in place of $\rho(\lambda)$ and redefines 
$U$ and $W$ to be $U(A,\lambda,\omega,\nu,\gamma):= f_1(A,\lambda,\omega)\, f_2(A,\lambda,\nu)\rho(\lambda|A,\gamma)$ and $
W(B,\lambda,\omega,\nu):= g_1(B,\lambda,\omega)\,g_2(B,\lambda,\nu)\rho(\lambda|B,\gamma)$ respectively.
The expression $ E_{(\omega,\nu,\gamma)}(A,B):=\int U(A,\lambda,\omega,\nu,\gamma)\,W(B,\lambda,\omega,\nu,\gamma)
\,\rho(\lambda)\, d\lambda$ then has the standard form to give the CHSH inequality, and therefore $E(A,B)=\int \int\int E_{(\omega,\nu,\gamma)}(A,B) l(\nu) k(\omega) m(\gamma)d\omega d\nu d\gamma$ does so as well.}.
\forget{  as the same as follows. From  \Eq{non-localdistr} the outcomes at both
 measurement setups  become (i.e., \Eq{non-localeq} and \Eq{non-localeq2}
 after averaging over $\lambda$ become),
 \begin{align}
 a(A,B)&=& \int \int \int  f_1(A,\lambda,\omega)\,\tilde{\rho}(\lambda|A,\gamma)
 \,g_1(B,\lambda,\omega)\,
 \tilde{\tilde{\rho}}(\lambda|B,\gamma)\,k(\omega)\,m(\gamma)\,d\gamma d\omega d\lambda, \\
b(A,B)&=&\int\int\int f_2(A,\lambda,\nu) \,\tilde{\rho}(\lambda|A,\gamma)
\,g_2(B,\lambda,\nu)\,
 \tilde{\tilde{\rho}}(\lambda|B,\gamma)\,l(\nu)\,m(\gamma)\,d\gamma d\nu d\lambda,
 \end{align}
which after rewriting \Eq{rel} accordingly\footnote{The first term of \Eq{rel}
would be $f_1(A,\lambda,\omega)\tilde{\rho}(\lambda|A,\gamma)
g_1(B,\lambda,\omega)\tilde{\tilde{\rho}}(\lambda|B,\gamma) f_2(A,\lambda,\nu)\tilde{\rho}(\lambda|A,\gamma)
g_2(B,\lambda,\nu)\times$ $\tilde{\tilde{\rho}}(\lambda|B,\gamma)$ which can be rewritten as
$X_1(A,\lambda,\omega,\gamma)Y_1(B,\lambda,\omega,\gamma)X_2(A,\lambda,\nu,\gamma)
Y_2(B,\lambda,\nu,\gamma)$, whose absolute value is less than or equal to one.
Since the other terms follow analogously we see that the proof goes through.}
 and averaging this over the deeper level hidden variables $\gamma,~\omega,~\nu$
gives the CHSH expression of \Eq{chsh}.}

Note that the non-local distribution \Eq{non-localdistr} again has a form of `factorisability'
of the settings $A,B$
(or product form with respect to dependency on $A$ and $B$), just as in
\Eq{non-localeq} and \Eq{non-localeq2}. It is this fact which is
responsible for the derivation to go through.
This realization tells us that the following form of extreme non-locality still 
suffices to derive the CHSH inequality:
\begin{align}
\label{crazylocal}
a(A,B,\lambda)=a(B,\lambda)~~ \textrm{and}~~
b(A,B,\lambda)=b(A,\lambda).
\end{align}
The outcomes at one setup now depend not on the local
parameter but only on the non-local parameter (and the hidden variable $\lambda$, of course).
Since the expectation value $E(A,B)$ obtains the following product form
\begin{align}
  E(A,B)=\int a(B,\lambda)b(A,\lambda)\rho(\lambda)d\lambda,
\end{align}
with `factorisation' of the settings $A,B$ the derivation of the CHSH inequality goes through.
 The same holds true if we use the
non-local hidden-variable distribution of \Eq{non-localdistr} which does not obey IS.
\\\\
Thus a hidden-variable theory in which the outcome of one measurement is allowed to depend
on the setting of the measurement apparatus of the other particle as in
\Eq{non-localeq} and \Eq{non-localeq2} for all possible response functions $f_1,f_2,g_1,g_2$ must still
obey the CHSH inequality. However, quantum mechanics violates this inequality. Thus neither a
local nor a non-local hidden-variable theory of the form here considered can reproduce the
predictions of quantum mechanics.

So surely it cannot be non-locality per se that
is the cause of the violation of the Bell inequalities. What can be the cause?
Three candidates seem to remain: one or more of the following assumptions do not in fact obtain:

 (i) The assumption of realism as used here, i.e., that
outcomes of measurement are determined by hidden variables and
deterministic (though perhaps even contextual) response functions. 

(ii) The form of `factorisability'
of \Eq{non-localeq} and \Eq{non-localeq2}, i.e. the assumption of a \emph{product} of
 response functions  $f$ and $g$. Note that Bell's non-local hidden-variable model that reproduces
 quantum mechanical predictions of the singlet state \cite{bell64} can thus not be of
 the form of \Eq{non-localeq} and \Eq{non-localeq2}. Indeed, it is not\footnote{\label{bellmodel}
 Bell's model has
 $ E(\vec{a},\vec{b})=\frac{1}{2 \pi}\int\, \mathrm{sgn}(\vec{a}'\cdot\vec{\lambda})\,
 \mathrm{sgn}(\vec{b}\cdot\vec{\lambda})\, d\vec{\lambda}
 $
 with setting $\vec{a}'$ non-locally determined by: $\vec{a}\cdot\vec{b}=1-\frac{2}{\pi}
\mathrm{arccos}(\vec{a}'\cdot\vec{b})$.}.

 (iii) The weakened version of IS as given in (\ref{non-localdistr}) that has a specific  dependence on the settings in the hidden-variable distribution.

 We will come back to the issue of what to make of the
violation of the CHSH inequality in section \ref{remarks} and in section \ref{discussionchshclassical}. In the next subsection we will generalize the  previous results to the case of stochastic hidden-variable theories.

\subsection{Stochastic case}\label{non-localstochastic}
\noindent\forget{
\emph{The following intuition lies behind exploring the stochastic case:
In the deterministic setting a certain form of parameter dependence (of the local outcomes and
of the hidden-variable distribution) was allowable to nevertheless still
derive the CHSH inequality. Could it be that in the stochastic case one also can
allow for some parameter dependence and still derive the CHSH inequality, i.e.
that some weaker set of assumptions than the conjunction of
OF, OL, and ISA (or of OI, PI and IS, when the apparatus hidden variables are not considered) would
be sufficient for deriving the CHSH inequality? We will show that this is indeed the case.}\\\\}
As a first remark, and as a  warming-up exercise, note that the previous analysis using the extreme form of non-locality
as in \Eq{crazylocal} indeed generalizes to the stochastic setting.
For if the probabilities obey
\begin{align} \label{crazy}
P(a|A,B,\lambda)=P(a|B,\lambda)~\textrm{and}~~ P(b|A,B,\lambda)=P(a|B,\lambda),
\end{align}
then the joint probability $P(a,b|A,B,\lambda)$ has
such a form of non-local `factorisation',
\begin{align}
 P(a,b|A,B,\lambda)=P(a|B,\lambda)\,P(b|A,\lambda),
\end{align}
from which we get a CHSH inequality using the standard derivation.
Thus parameter independence (understood as no dependence on the distant
parameter) is not necessary in order to derive the CHSH inequality for a
stochastic hidden-variable
theory. However, one could argue that this example has a form of parameter
independence, although not of the distant parameter
but of the local parameter. \forget{Indeed, but in any case, some form of non-locality
(non-local dependence) can be allowed for 
in the stochastic hidden-variable theory (either of the form in \Eq{non-localdistr} or
in \Eq{crazy})  to obtain the CHSH inequality.
\\\\}

Let us now continue with a less contrived approach.
Consider a stochastic hidden-variable model for the \emph{Gedankenexperiment} of Figure \ref{gedank}. 
We will  derive the CHSH inequality under specific violations of outcome independence (OI),  of parameter independence (PI)
\forget{, i.e., we do not assume $P(a|A,B,b,\lambda)=
P(a|A,B,\lambda),~P(b|A,B,a,\lambda)=
P(b|A,B,\lambda) $ and  $P(a|A,B,\lambda)=P(a|A,\lambda), P(b|A,B,\lambda)=P(b|A,\lambda)$ respectively.} 
and of Independence of the Source (IS).\forget{ i.e., we do not assume that the distribution of the hidden variables is independent of the settings, thus we do not assume $\rho(\lambda|A,B)=\rho(\lambda)$} We allow that the probability that a certain local outcome is obtained can be dependent on the local setting, the hidden variable as well as on the distant setting and outcome in the following way:
\begin{subequations}\label{non-localstoch}\begin{align}
P(a|A,B,b,\lambda)&=f(a,A,\lambda)\,x(b,B,\lambda),\label{nlstoch1}\\
P(a|A,B,\lambda)&=\overline{f}(a,A,\lambda)\,\overline{x}(B,\lambda),\label{nlstoch2}
\\
P(b|A,B,a,\lambda)&=g(b,B,\lambda)\,y(a,A,\lambda),\label{nlstoch3}\\
	P(b|A,B,\lambda)&=\overline{g}(b,B,\lambda)\,\overline{y}(A,\lambda).\label{nlstoch4}
\end{align}
\end{subequations}
Here the response functions $f$, $\overline{f}$, $g$, $\overline{g}$, $x$, $\overline{x}$, $y$ and $\overline{y}$ have their range in the interval $[0,1]$ and are  possibly further restricted by normalization conditions. We have now explicitly incorporated some non-local setting and outcome dependence, i.e., OI and PI are not assumed.  
 Furthermore, the distribution of the hidden variables is allowed to depend on the settings, and thereby to violate IS, as in \eqref{deltaIS}: 
\begin{align}\label{rholambda}
\rho(\lambda |A,B)=\tilde{\rho}(\lambda |A)\tilde{\tilde{\rho}}(\lambda |B).
\end{align} 

The identity $P(a,b|A,B,\lambda)=P(a|A,B,b,\lambda)\,P(b|A,B,\lambda)$ together with  the assumptions \eqref{non-localstoch}  and \eqref{rholambda} allows for rewriting the expectation value $E(A,B)$ as follows:
\begin{align}
E(A,B)&:=\int \sum_{a,b} ab\, P(a,b|A,B,\lambda)\,\rho(\lambda |A,B)\,d\lambda,\nn\\
&=\int \sum_{a,b} ab\, P(a|A,B,b,\lambda)\,P(b|A,B,\lambda)\,\rho(\lambda |A,B)\,d\lambda,\nn\\
&=     \int \sum_{a,b} ab\,   f(a,A,\lambda)\,x(b,B,\lambda)\, \overline{g}(b,B,\lambda)\,\overline{y}(A,\lambda) \,             \tilde{\rho}(\lambda |A)\tilde{\tilde{\rho}}(\lambda |B)\,d\lambda,\nn\\
&=     \int \sum_{a} a\,   f(a,A,\lambda)\,\overline{y}(A,\lambda)\sum_b b\,x(b,B,\lambda)\, \overline{g}(b,B,\lambda)\, \tilde{\rho}(\lambda |A)\tilde{\tilde{\rho}}(\lambda |B)\,d\lambda,\nn\\
&= \int F(A,\lambda)\, G(B,\lambda) \, \tilde{\rho}(\lambda |A)\,\tilde{\tilde{\rho}}(\lambda |B)\, d\lambda,\nn\\
&= \int \underbrace{ F(A,\lambda)\,  \tilde{\rho}(\lambda |A)}_{A}\,\underbrace{G(B,\lambda)   \,\tilde{\tilde{\rho}}(\lambda |B)}_{B}\, d\lambda,\label{standard}
\end{align}
with
\begin{align}
F(A,\lambda) :=  \sum_a a f(a,A,\lambda)\,\overline{y}(A,\lambda) ~    ,~\textrm{and}~
G(B,\lambda):=   \sum_b b\, \overline{g}(b,B,\lambda)\,x(b,B,\lambda)  .
\end{align}
The expectation value $E(A,B)$ in \Eq{standard} thus  has obtained a product form in terms of the settings $A$ and $B$. Furthermore, since $|F(A,\lambda)|\leq1$, $|G(B,\lambda)|\leq1$, $\int\tilde{\rho}(\lambda |A)d\lambda= 1$, and $\int\tilde{\tilde{\rho}}(\lambda |B)d\lambda= 1$, it follows that $E(A,B)$ has obtained the standard form from which one derives the CHSH inequality.
This  proof is not symmetric with respect to $a$ and $b$, but this is not important. Starting with the identity  $P(a,b|A,B,\lambda)=P(b|A,B,a,\lambda)\,P(a|A,B,\lambda)$ gives the same result.

\forget{
Let us assume outcome independence (OI) (but not parameter independence (PI) or at
this stage independence of the source (IS)):
\begin{align}
P(a,b|A,B,\lambda)=P(a|A,B,b,\lambda)\,P(b|A,B,\lambda)\stackrel{OI}{=}
P(a|A,B,\lambda)\,P(b|A,B,\lambda)
\end{align}
We allow that the probability for a local outcome is depends on the local setting, the hidden 
variable as well as on the distant setting in the following way:
\begin{align}\label{non-localstoch}
P(a|A,B,\lambda)&=f_a(A,\lambda)\,x(B,\lambda),
\\
P(b|A,B,\lambda)&=f_b(B,\lambda)\,y(A,\lambda).\label{non-localstoch2}
\end{align}
Here the response functions $f_a$, $f_b$, $x$ and $y$ have their range in the interval $[0,1]$ and are
 possibly further restricted by normalization conditions. We have now explicitly incorporated some non-local dependence. 
Since we have possible outcomes $a,b=\pm1$, the expectation value $E(A,B)$ becomes:
\begin{align}
E(A,B)&=\int \sum_{a,b} ab\, P(a,b|A,B,\lambda)\,\rho(\lambda)\,d\lambda,\nn\\
&= \int \Big[\Big(f_{(a=1)}(A,\lambda) -f_{(a=-1)}(A,\lambda) \Big)\,y(A,\lambda)\Big]\times\nn\\
&~~~~~~~~~~~~~~~~~~\Big[\Big( f_{(b=1)}(B,\lambda) -f_{(b=-1)}(B,\lambda) \Big) \,x(B,\lambda) \Big] \,\rho(\lambda)\, d\lambda,\nn\\
&=\int F(A,\lambda)\, G(B,\lambda)                   \, \rho(\lambda)\, d\lambda,\label{standard}
\end{align}
with
\begin{align}
F(A,\lambda)  &:=  (f_{(a=1)}(A,\lambda) -f_{(a=-1)}(A,\lambda) )\,y(A,\lambda)      ,\\
G(B,\lambda)  &:=   (f_{(b=1)}(B,\lambda) -f_{(b=-1)}(B,\lambda) )\, x(B,\lambda)     .
\end{align}
Since $|F(A,\lambda)|\leq1$, $|G(B,\lambda)|\leq1$
we see that the correlation of  \Eq{standard}
again has the standard Bell form and thus the CHSH inequality follow s(see  the Intermezzo on p.~\pageref{chshderivation}). 

We have used the assumptions of outcome
independence (OI) and the
non-local determination of the probabilities $P(a|A,B,\lambda)$, $P(b|A,B,\lambda)$
as displayed in \Eq{non-localstoch}, as well as the assumption of  independence of the source (IS).
}

\forget{
The assumptions can be further relaxed by introducing the deeper level hidden variables $\gamma$, $\omega$, and $\nu$   and averaging over them as was done in the previous section.  Firstly, the conditions in \eqref{non-localstocha} then become
\begin{subequations}\label{non-localstocha}\begin{align}
P(a|A,B,b,\lambda)&=\int f(a,A,\lambda,\omega)\,x(B,b,\lambda,\omega) \, k(\omega)\,d \omega +\alpha_1
,\label{nlstoch1a}\\
P(a|A,B,\lambda)&=\int \overline{f}(a,A,\lambda,\omega)\,\overline{x}(B,\lambda,\omega) \, k(\omega)\,d \omega +\alpha_2,\label{nlstoch2a}
\\
P(b|A,B,a,\lambda)&=\int g(b,B,\lambda,\nu)\,y(a,A,\lambda,nu)\,l(\nu) \,d \nu+\beta_1,\label{nlstoch3a}\\
	P(b|A,B,\lambda)&=\int \overline{g}(b,B,\lambda,\nu)\,\overline{y}(A,\lambda,\nu)\,l(\nu) \,d \nu+\beta_2,
\label{nlstoch4a}
\end{align}
\end{subequations}
where $\omega$, $\nu$ are deeper hidden variables for particles $1$ and $2$ respectively,
and $\alpha_i, \beta_i \in \mathbb{R}$, $i=1,2,$ are some constants solely restricted by normalization conditions.
Secondly,  the distribution of the hidden variable $\lambda$  as in \eqref{rholambda} can weakened  to be \Eq{non-localdistr}.
}

Our weakest set of assumptions leading up to the CHSH inequality\footnote{As mentioned before, the analysis here is performed without explicitly mentioning apparatus hidden variables since this complicates the notation and it has (as far as we can see) no advantage to be included. It can be easily seen
however that if one would include the apparatus hidden variables the more general
non-$\mu$-averaged conclusion would be obtained: The CHSH inequality can be derived while weakening OL, OF and ISA in the
appropriate way so as to allow explicit setting and outcome dependence.} are thus:
\begin{description}
\item{(i)} the non-local dependence of the
distribution of the hidden variable $\lambda$ on the settings $A$, $B$ as in \Eq{non-localdistr}, and
\item{(ii)} the setting and outcome dependent determination of the conditional marginal probabilities 
 as displayed in \Eq{non-localstoch}.
 \forget{
 \footnote{Let us give an explicit example where PI and OI are violated,
while the conditions of \Eq{non-localstoch} are obeyed.
Suppose the settings $A$ and $B$ are some vectorial quantities $\bm{a}$ and $\bm{b}$ respectively, just as $\bm{\lambda}$ is. We now choose
\begin{align}
P(a=\pm1|\bm{a},\bm{b},b,{\lambda})&=
\frac{1}{2} \pm b \,\mathrm{sgn}(\bm{a}\cdot\bm{\lambda})\,
\mathrm{sgn}(\bm{b}\cdot\bm{\lambda}),\nonumber \\
P(a=\pm1|\bm{a},\bm{b},\bm{\lambda})&=\frac{1}{2} 
\pm \alpha ,\nonumber\\
P(b=\pm1|\bm{a},\bm{b},a,{\lambda})&=
\frac{1}{2} \pm a \,\mathrm{sgn}(\bm{a}\cdot\bm{\lambda})\,
\mathrm{sgn}(\bm{b}\cdot\bm{\lambda}),\nonumber \\
P(b=\pm1|\bm{a},\bm{b},\bm{\lambda})&=\frac{1}{2} \pm \alpha \,\mathrm{sgn}(\bm{b}\cdot\bm{\lambda})\,
\mathrm{sgn}(\bm{a}\cdot\bm{\lambda}),
\label{modelnon-local}
\end{align}
where $-1/2\leq \alpha\leq 1/2$, and sgn$(\phi)$ is equal to $1$ if $\phi\geq0$ and equal to $-1$ if $\phi <0$.  This distribution gives $E(\bm{a},\bm{b})=1$, from which it is easily seen that the CHSH inequality must be obeyed. These relations violate PI (e.g., $P(a|\bm{a},\bm{b},\bm{\lambda})\neq
P(a|\bm{a},-\bm{b},\bm{\lambda}$)$\,$)  and OI  (e.g., $P(a|\bm{a},\bm{b},b,\bm{\lambda})\neq
P(a|\bm{a},\bm{b},-b,\bm{\lambda}$)$\,$) but obey \Eq{non-localstocha}.
}
 \forget{Let us give an explicit example where PI is violated,
while the conditions of \Eq{non-localstocha} are obeyed.
Suppose the settings $A$ and $B$ are some vectorial quantities $\bm{a}$ and $\bm{b}$ respectively, just as $\bm{\lambda}$ is. We now choose
\begin{align}
P(a=\pm1|\bm{a},\bm{b},\bm{\lambda})&=
\frac{1}{2} +\Big(\pm \alpha \,\mathrm{sgn}(\bm{a}\cdot\bm{\lambda})\Big)\,
\mathrm{sgn}(\bm{b}\cdot\bm{\lambda}),\nonumber \\
P(b=\pm1|\bm{a},\bm{b},\bm{\lambda})&=\frac{1}{2} +
\Big(\pm \beta \,\mathrm{sgn}(\bm{b}\cdot\bm{\lambda})\Big)\,
\mathrm{sgn}(\bm{a}\cdot\bm{\lambda}),
\label{modelnon-local}
\end{align}
where sgn$(\phi)$ is equal to $1$ if $\phi\geq0$ and equal to $-1$ if $\phi <0$.
Furthermore $-1/2\leq \alpha,\beta\leq 1/2$. Suppose OI holds, then this gives $E(\bm{a},\bm{b})=4\alpha\beta$, from which it is easily seen that the CHSH inequality must be obeyed. These relations violate PI (e.g., $P(a|\bm{a},\bm{b},\bm{\lambda})\neq
P(a|\bm{a},-\bm{b},\bm{\lambda}$)$\,$) but obey \Eq{non-localstocha}.}}
 \end{description}

Finally, note that the extreme non-local dependence as in \Eq{crazy} can be
written in the general form of \Eq{non-localstoch}.
Indeed, choosing $\overline{f}(a,A,\lambda)$ independent of $a,\, A$ and
$\overline{g}(b,B,\lambda)$ independent of $b,\, B$ will suffice.

\subsection{Remarks}\label{remarks}
\noindent
{\bf (0)} Assuming local realism and that observables are free variables is sufficient for deriving the CHSH inequality, but not necessary.  Indeed, 
the above results show that the assumptions OI, PI and IS can be relaxed considerably while still implying the CHSH inequality. 
Violations of the CHSH inequality thus not only exclude models in which OI, PI and IS hold, but also some models in which none of these three  assumptions hold. Thus, a larger class of models than previously considered is ruled out by quantum theory, and modulo some loopholes also by experiment.  \forget{Because the CHSH inequality is compatible with violations of both OI and PI, an attempt to understand experimental violations of the CHSH inequality no longer justifies the question whether it is OI or PI that should be abandoned. 
We believe this blocks any experimental metaphysics of the form described in section \ref{expmeta}.}

Note that the assumptions that are used to give the CHSH inequality are not directly experimentally testable since they involve the hidden variable $\lambda$, i.e., the assumptions are at the subsurface level. It is only the surface probabilities not the subsurface probabilities that are determined via measurement of relative frequencies in experiment.  Therefore, experiment cannot tell us which of the assumptions are violated and which ones are not. 
\forget{
with respect to the business of doing experimental metaphysics: 

All experiment tells us is that the CHSH inequality is violated, not which of the subsurface conditions is to be rejected. The latter are experimentally inaccessible. This undercuts the significance of Factorisability and the conditions of  OI or PI. The question should be rephrased: how to understand  violation of the setting and outcome dependent conditions used.

In the next section we show that the CHSH inequality can be obeyed by models that violate OI, PI and IS. These models are setting and outcome dependent in a specific way. 
We have no reason to expect either one of them to hold solely on the basis of the CHSH inequality.
}
\\\\
{\bf (1)}
The crucial point that is responsible for the derivation of  the CHSH inequality, is that after incorporating all assumptions and averaging over all deeper level hidden variables a form of factorisability in the expression for the product expectation value was obtained. When this expression has the 
form $E(A,B)=\int_\Lambda\,X(A,\lambda)\,Y(B,\lambda)\,\rho(\lambda)d \lambda,
$ with $|X(A,\lambda)|\leq 1$ and $|Y(B,\lambda)|\leq1$ the CHSH inequality follows.
And to get such a form it was not necessary to assume IS, or, in the case of stochastic local realistic models, the conditions of PI and OI whose conjunction gives Factorisability. Nor was it needed to assume the independence of the local outcomes on the distant settings (i.e., $a(A,B,\lambda)=a(A,\lambda)$, etc.) in the case of deterministic local hidden-variable models.
\\\\ 
\forget{
to assume in the case of stochastic local realistic models the conditions of PI and OI whose conjunction gives Factorisability (i.e., $P(a,b|A,B,\lambda)=P(a|A,\lambda)P(|B,\lambda)$. Nor was it needed to assume the independence of the local outcomes on the distant settings  (i.e., LD) 
as was the case in the discussion of deterministic local hidden-variable models. 
}
\forget{However,  in the stochastic case we still had to assume OI.  We have tried to obtain the CHSH inequality by weakening OI, but have not succeeded. It would be interesting to determine if OI is necessary tor derive the CHSH inequality, and if indeed so, why that is the case. In this regard it is interesting to remark that although OI could be a necessary assumption for deriving the CHSH inequality, it 
		is not the case that any model that obeys OI has to obey the CHSH inequality. For example, Bohmian mechanics obeys  OI since it is deterministic but violates the CHSH inequality. Furthermore one can obey OI but still give the absolute maximum of 4 for the CHSH expression.  An example of such a model was shown in footnote \ref{alg4} (this model obeys OI since it is deterministic).}
\forget{
ON BOHM: violates IS and PI, but obeys OI and gives QM.
Dickson gives explicit contextualism in Bohmian mechanics. 117,228, en 201 laat
expliciet zien hoe TAF noodzakelijk is in de afleidng van de bell ineq, en waarom
bohmian mechanics, die TAF schendt, dus de bell ineq kan schenden.
 de schending van TF in bohmian mechaincs geeft geen niet-localiteit. Het is
een vorm van superdeterminisme dat de afleiding van de Bell ineq. blokkeert.
see Butterfield, Bell's theorem on what it takes. veel opmerkingen!
Dewdney et al .results? zij geven expliciet aan hoe PI en SI worden geschonden.
}
%
\forget{
{\bf (2)} The deeper hidden variables $\omega,\nu,\gamma$ we introduced are not new, (e.g., Bell already introduced
them \cite{bell71}), nor are they essential. Indeed, they can be assumed to be
absent as was shown by giving them delta distributions.
\\\
JOS SNAPTE (2) NIET.
\\\\
}
{\bf (2)} From a mathematical point of view it is no surprise that
the contrived dependence as in (\ref{crazylocal}) and (\ref{crazy}), where the
outcomes depended solely upon the non-local settings (and not on
 the local ones), imply  the CHSH inequality.
Compared to the standard assumptions, the settings $A$ and $B$
were merely interchanged, but what
 was important about the condition, the factorisability or product form
of the expressions, was retained. This situation has striking
similarity to Maudlin's assumptions, where after interchanging
outcome $a$ and setting $A$ (analogous for $b$ and $B$) in Jarrett's
 or Shimony's assumptions one could still obtain Factorisability (see section \ref{shimonymaudlin}).

But are these possibilities merely mathematical and nothing else? Perhaps, since
the newly obtained conditions might not be easily given a physical motivation, but their mere
possibility, even if only mathematical, shows that one should be
 careful in analyzing what the experimentally confirmed violation of the
 CHSH inequality means. \forget{This complex issue will be picked up again
 in section () in the light of all results of this paper, including these ones. }

The above remark raises the question
whether it is possible to derive the CHSH inequality by weakening Maudlin's assumptions P1 and P2 whose conjunction also gives Factorisability.
We have tried but did not succeed in doing  so. 
 If it is indeed impossible to weaken P1 or P2 to get the CHSH inequality, then this would show an interesting and novel  difference between the Shimony-assumptions OI and PI and the Maudlin-assumptions P1 and P2. Such a difference could possibly be used to argue for a foundational difference between the Shimony factorisation or the Maudlin factorisation.  
 \\\\
{\bf (3)}
\citet{jones} have studied so-called `inseparable hidden-variable
 models' for three and more subsystems and have shown that such models have to
 obey generalized CHSH inequalities (so called Svetlichny inequalities, see chapter \ref{chapter_svetlichny}), which quantum mechanics violates. The inseparable models
they have studied are non-local setting dependent models. For three or more systems
 they thus showed that quantum correlations are stronger than
 the correlations of some such models.

 We have complemented this analysis to the case of two subsystems
and showed that, because quantum mechanics violates the CHSH inequality for bi-partite systems, quantum correlations are stronger than
a large class of non-local correlations. 
\\\\
{\bf (4)}
Non-local hidden-variable models have been constructed that reproduce some
of the quantum correlations that violate the CHSH inequality. These are thus necessarily not of the non-local setting and outcome dependent 
 forms considered above. Indeed, they are not. For example, Bell's hidden-variable model \cite{bell64})
  is not of any of these forms. In fact, it is not even analytic, cf. footnote \ref{bellmodel}.
   But non-analytic models exist that violate PI but nevertheless obey the CHSH inequality\footnote{An example of such a model is the following. Suppose the settings $A$ and $B$ are some vectorial quantities $\bm{a}$ and $\bm{b}$ respectively, just as $\bm{\lambda}$ is. We now choose
  $P(a|\bm{a},\bm{b},\bm{\lambda})=
\frac{1}{2}+ a\, \alpha \,\mathrm{sgn}(\bm{a}\cdot\bm{\lambda})\,
\mathrm{sgn}(\bm{b}\cdot\bm{\lambda})$, and 
$P(b|\bm{a},\bm{b},\bm{\lambda})=\frac{1}{2} +
b\,\beta \,\mathrm{sgn}(\bm{b}\cdot\bm{\lambda})\,
\mathrm{sgn}(\bm{a}\cdot\bm{\lambda})$, 
\forget{
\begin{align}
P(a|\bm{a},\bm{b},\bm{\lambda})&=
\frac{1}{2}+ a \alpha \,\mathrm{sgn}(\bm{a}\cdot\bm{\lambda})\,
\mathrm{sgn}(\bm{b}\cdot\bm{\lambda}),\nonumber \\
P(b|\bm{a},\bm{b},\bm{\lambda})&=\frac{1}{2} +
b\beta \,\mathrm{sgn}(\bm{b}\cdot\bm{\lambda})\,
\mathrm{sgn}(\bm{a}\cdot\bm{\lambda}),
\label{modelnon-local}
\end{align}}
where $-1/2\leq \alpha,\beta\leq 1/2$, and sgn$(\phi)$ is equal to $1$ if $\phi\geq0$ and equal to $-1$ if $\phi <0$. Suppose OI and IS obtains, then one obtains that  $E(\bm{a},\bm{b})=4\alpha\beta$.
This model violates PI (e.g., $P(a|\bm{a},\bm{b},\bm{\lambda})\neq
P(a|\bm{a},-\bm{b},\bm{\lambda}$)$\,$) but obeys \Eq{non-localstoch}, and therefore obeys the CHSH inequality.}. Thus non-analyticity of the non-locality
is not sufficient to violate the CHSH inequality, but it is in many
cases necessary \cite{socolovsky}.  It is an open question
 what form of non-locality is necessary and sufficient to imply the CHSH inequality.

\subsection{Comparison to Leggett's non-local model}
\label{sectionleggett}

In the previous section we have derived the CHSH inequality while explicitly allowing for some non-local setting and parameter dependence that violated the assumptions PI, OI and IS. \citet{leggett} recently derived a different inequality than the CHSH inequality while also allowing a form of non-locality. Both the CHSH and Leggett's inequality are violated by quantum mechanics, but satisfaction of Leggett's inequality allows for correlations that violate the CHSH inequality. It is therefore interesting to compare the two different forms of non-locality involved. This is the goal of this subsection.

Leggett considers two parties, $I$ and $II$ respectively, that each hold a subsystem on which they measure different dichotomous observables which  are indicated by the settings $A$ and $B$ and that have outcomes $a=\pm1, b=\pm1$ respectively. 
He furthermore considers a deterministic hidden-variable model that is supposed to give the outcomes of measurement. The model assumes three hidden variables $\lambda, \vec{u},\vec{v}$. For these hidden variables he assumes that IS holds, i.e.,  their distribution is independent of the settings $A,B$: $\rho(\lambda, \vec{u},\vec{v}|A,B) =\rho(\lambda, \vec{u},\vec{v})$.  The hidden variable $\lambda$ specifies the total system and the vectors $\vec{u}$ and $\vec{v}$ are further specifications of the subsystems held by party $I$ and $II$ respectively. The outcomes $a,b$ are deterministically determined by the settings $A,B$ and the hidden variables, i.e.,   $a=f(A,B,\lambda, \vec{u},\vec{v})$ and  $b=g(A,B,\lambda, \vec{u},\vec{v})$.

Since the model is deterministic the assumption OI is automatically obeyed\footnote{
Leggett, remarks ``\ldots I shall rather arbitrarily assert assumption (4) (outcome independence). The reason for doing  so is not so much  that it is particularly ``natural'' [\ldots] but it is a purely practical one.; if one relaxes (4) [i.e., OI] it appears quite unlikely (though I have no rigorous proof) that one can prove anything useful at all, and in particular it appears very likely that one can reproduce quantum-mechanical results for an arbitrary experiment.''\citep[p.1475]{leggett}. Leggett does not seem to realize that his starting point, a deterministic theory, automatically enforces OI to be obeyed. In order to allow for the possibility of a violation of OI he should consider indeterministic hidden-variable theories from the start. However, as we will argue below, after averaging over $\lambda$, i.e., on the level $\vec{u}$ and $\vec{v}$,  OI is violated in Leggett's model.}. However, in order to obtain a non-trivial result Leggett does not assume the locality assumption LD, i.e., he does not require that $a=f(A,\lambda, \vec{u})$ and  $b=g(B,\lambda,\vec{v})$.  He thus allows for a possible non-local setting dependence of the local outcomes, which Leggett interprets as a violation of the assumption of PI\footnote{Using our definitions this is a violation of LD. 
But it is of course possible to view this as a violation of PI for the deterministic case where all probabilities are $0$ and $1$. We call such a situation deterministic PI.}.

Leggett now introduces some further assumptions from which he derives his inequality. These assumptions are not at the level of the three hidden variables $\lambda, \vec{u},\vec{v}$, but at the level of the two hidden variables $\vec{u},\vec{v}$ where one has averaged over $\lambda$ using some distribution of the hidden variables $\rho(\lambda,\vec{u},\vec{v})$ that only has to obey normalization constraints.  These assumptions (to be given below) are thus imposed on the following $\lambda$-averaged quantities (where we follow the notation of \citet{branciard}): 
\begin{subequations}\label{leggettaverages}
\begin{align}
M^I_\xi(A,B)&=\int \, d\lambda \, \rho(\lambda,\xi) \,f(A,B,\lambda, \xi), \label{LA1} \\
M^{II}_\xi(A,B)&=\int \, d\lambda \, \rho(\lambda,\xi) \,g(A,B,\lambda, \xi),      \label{LA2}  \\
C_\xi(A,B)&=  \int \, d\lambda \, \rho(\lambda,\xi) \,f(A,B,\lambda, \xi)\,g(A,B,\lambda,\xi). \label{LA3}      
\end{align}
\end{subequations}
 These are equations (2.9a), (2.9b) and (2.11) of Leggett respectively \cite[p. 1477]{leggett}. Here we have introduced the notation $\xi$ for the pair $(\vec{u},\vec{v})$.  For a given value of $\xi$, $M^I_\xi(A,B)$ and $M^{II}_\xi(A,B)$ are the marginal expectation values for party $I$ and $II$ respectively, and $C_\xi(A,B)$ is the product expectation value. The expectation value of measuring $A$ and $B$ jointly is then given by $\av{AB}=\int \, d\xi \, \tilde{\rho}(\xi) \,C_\xi(A,B)$, with $\tilde{\rho}(\xi)=\int \rho(\lambda,\xi) d \lambda$.

Introduction of $\lambda$ is not necessary for the derivation of the Leggett-inequality because 
the physical assumptions Leggett uses (to be shown below) are imposed only on the quantities in the left hand side of \eqref{leggettaverages} and these   depend only on the hidden variable $\xi$. However, including $\lambda$ gives the hidden-variable model a radically different character. By including this extra hidden variable Leggett  is able to propose a deterministic hidden-variable model. But the average values over $\lambda$ in  \eqref{leggettaverages} can also be interpreted as the predictions of a stochastic hidden-variable model (see section \ref{jarrettshimony} were this has been also discussed).  Below we will choose this option and interpret these average values as predictions of an indeterministic model. This model gives the subsurface probabilities $P(a,b|A,B,\xi)$, i.e., the probabilities to obtain the outcomes $a,b$ when measuring $A,B$ on a
system in state $\xi$. Accordingly, the quantities in \eqref{leggettaverages}  can be assumed to be determined by the subsurface correlations $P(a,b|A,B,\xi)$ in the following way:
\begin{subequations}\label{leggettprobs}
\begin{align}
M^I_\xi(A,B)&=\sum_{a,b}a\, P(a,b|A,B,\xi) \label{LP1},\\
M^{II}_\xi(A,B)&=\sum_{a,b}b\, P(a,b|A,B,\xi)\label{LP2},\\
C_\xi(A,B)&=\sum_{a,b}ab \,P(a,b|A,B,\xi) \label{LP3}.
 \end{align}
 \end{subequations}
Note that in general the subsurface probabilities can be written as:
\begin{align}\label{defcorre}
P(a,b|A,B,\xi)= \frac{1}{4}\big(1+ a\, M^I_\xi(A,B) +b\, M^{II}_\xi(A,B) +ab\, C_\xi(A,B)\big).
\end{align}
Because the probabilities on the left hand side are non-negative  the marginals  $M^I_\xi(A,B)$ and  $M^{II}_\xi(A,B)$ are not completely independent of the product expectation value $C_\xi(A,B)$, and vice versa.

Before we discuss Leggett's assumptions that allow him to derive his inequality, we first determine what the assumptions of OI and PI imply for
the quantities in \eqref{leggettprobs} and \eqref{defcorre}.  For a given $\xi$, PI requires the marginal expectation value for party $I$ ($II$)  to be independent of the setting chosen by $II$ ($I$)\footnote{The presentation of \citet{branciard} also discusses Leggett-type models at the $\xi$-level. They 
formulate this condition as a no-signaling condition at the level of the hidden variables $\xi$. We call this PI and reserve the notion of no-signaling to the surface probabilities  $P(a,b|A,B)$ only.}, i.e., $M^I_\xi(A,B)=M^I_\xi(A)$, and $M^{II}_\xi(A,B)=M^{II}_\xi(B)$, whereas 
OI requires that  $C_\xi(A,B)$ must have the product form $C_\xi(A,B)=M^{I}_\xi(A) M^{II}_\xi(B)$\footnote{That PI implies that $M^I_\xi(A,B)$ must be independent of $B$ can be easily seen:
\begin{align}
M^I_\xi(A,B)=\sum_{a,b}a\, P(a,b|A,B,\xi) & =\sum_{a}a \, P(a|A,B,\xi) \nn\\&  \overset{\textrm{PI}}{=}\sum_{a}a\, P(a|A,B',\xi) =\sum_{a,b}a\, P(a,b|A,B',\xi)= M^I_\xi(A,B'),
 \end{align}
 and  analogous for $M^{II}_\xi(A,B)$  independent of $A$. Likewise, it is easy to see that  OI  implies that $C_\xi(A,B)=M^{I}_\xi(A) M^{II}_\xi(B)$:
 \begin{align}
C_\xi(A,B)=  \sum_{a,b}ab \,P(a,b|A,B,\xi) & \overset{\textrm{OI}}{=}\sum_{a,b}ab P(a|A,B,\xi)P(b|A,B,\xi)=    \sum_{a}a P(a|A,B,\xi)\sum_b bP (b|A,B,\xi) \nn\\
&=\sum_{a,b}a P(a,b|A,B,\xi)\sum_{a,b} bP(a,b|A,B,\xi)=M^{I}_\xi(A) M^{II}_\xi(B).
\end{align}
 }. 
Inserting this into \eqref{defcorre} gives: 
\begin{align}
\textrm{OI}~~~\Longrightarrow~~~& P(a,b|A,B,\xi)= \frac{1}{4} \big(1+ a\, M^I_\xi(A,B)\big) \big(1+b\, M^{II}_\xi(A,B)\big)\label{LOI}\\
\textrm{PI}~~~\Longrightarrow ~~~&P(a,b|A,B,\xi)= \frac{1}{4} \big(1+ a\, M^I_\xi(A) +b\, M^{II}_\xi(B) +ab\, C_\xi(A,B)\big)\label{LPI}
\end{align}

Leggett's model has a particular non-trivial form of the local marginal expectation values $M^I_\xi(A,B)$ and $M^{II}_\xi(A,B)$ that enforces PI so as to give  \eqref{LPI}, but puts no explicit constraints on $C_\xi(A,B)$. The latter  is  only constrained by the fact that  $P(a,b|A,B,\xi)$ must be give a valid probability distribution over the outcomes $a,b$ for all choices of $A,B$  \cite[cf.][]{paterek}.  It is thus explicitly not required that  $C_\xi(A,B)=M^{I}_\xi(A) M^{II}_\xi(B)$ which is equivalent to OI. Because Leggett allows PI to be violated, he does not want to require this condition OI, for he would then in fact require Factorisability (because PI is already assumed) from which one trivially obtains the CHSH inequality\footnote{``It is immediately clear that a necessary (but by no means sufficient) condition for a [Leggett-type model] to be nontrivial is that the subensemble averages fail to satisfy $\overline{AB}=\overline{A}\cdot\overline{B}$ [in our notation: $C_\xi(A,B)=M^{I}_\xi(A) M^{II}_\xi(B)$].''\cite[p. 1485]{leggett}}. We conclude that Leggett allows for violations of OI\footnote{\citet[p. 1.]{branciard} formulate this as:  ``\ldots, only the correlation coefficient [\ldots] $C_\xi(A,B)$ can be non-local, \ldots''  This we believe to be a confusing way of putting things.  
 Furthermore, they remark, that ``Leggett's assumption concerns only the local part of the probability  distributions $P_\xi$; it is thus somewhat confusing to name it [Leggett's model] a nonlocal model, though it is clearly nonlocal  in the sense  of not satisfying Bell's locality assumption" \cite[p. 2.]{branciard}.  However, this statement itself is a bit confusing. We can state things more clearly: at the $\xi$-level Leggett's model obeys PI, has specific constraints on the local marginal expectation values for a given $\xi$ [to be specified below in \eqref{marginal}], but allows for violations of OI. Therefore, being the conjunction of OI and PI,  Factorisability need not hold.}.  It is interesting to contrast this with what happens on the deterministic $(\lambda, \bm{u},\bm{v})$-level described above, and which Leggett originally considered. On this level Leggett's model obeys OI, but allows for violations of deterministic PI.  This difference will be further discussed below.  But we first present Leggett's further assumptions explicitly.
 
Leggett's fundamental assumption is that locally the systems party $I$ and $II$ possess behave as if they were in a pure qubit quantum state, i.e., each local system when analyzed individually is in a pure quantum state. This is encoded in the formalism  presented above  in the following way: 
 $\xi$ describes the hypothetical pure states of the qubits held by parties $I$ and $II$, and these are denoted by normalized vectors $\vec{u}, \vec{v}$ on the Poincar\'e sphere, so as to give: $\xi=\ket{\vec{u}}\otimes\ket{\vec{v}}$. As a consequence of Leggett's assumption the local marginal expectation values are the ones predicted 
by quantum mechanics:
\begin{align}
M^I_{\vec{u},\vec{v}}(\vec{a})&=\bra{\vec{u}}\vec{a}\cdot \bm{\sigma}\ket{\vec{u}}=\vec{u}\cdot \vec{a},\nn\\ 
M^{II}_{\vec{u},\vec{v}}(\vec{b})&=\bra{\vec{v}}\vec{b}\cdot \bm{\sigma}\ket{\vec{v}}=\vec{v}\cdot \vec{b}.\label{marginal}
\end{align}
Here the measurement settings are represented as unit-vectors on the Poincar\'e sphere: $A\rightarrow \vec{a}$, $B\rightarrow\vec{b}$.
In Leggett's  model the qubits are encoded in polarization degrees of freedom of photons. Each photon is assumed to have a definite polarization in directions $\vec{u}$ and $\vec{v}$ respectively and the local marginal expectation values should obey Malus' law.

If we now consider the correlations \eqref{defcorre}, where $\xi =\ket{\vec{u}}\otimes\ket{\vec{v}}$ and $\rho(\xi)$ some distribution of the polarizations $\vec{u},\vec{v}$, we see that Leggett's model requires that 
\begin{align}\label{propleggett}
P(a,b|\vec{a},\vec{b},\vec{u},\vec{v})= \frac{1}{4}\big(1+ a\,\vec{u}\cdot \vec{a} +b\, \vec{v}\cdot \vec{b} + ab\, C_{\vec{u},\vec{v}}(\vec{a},\vec{b})\big).
\end{align}
This explicitly incorporates  Leggett's assumption  that the local marginal expectation values should obey \eqref{marginal}.\forget{The product expectation value $C_{\vec{u},\vec{v}}(\vec{a},\vec{b})$ is constrained only 
by the requirement that \eqref{propleggett} must give a probability distribution over the outcomes $a,b$ for all choices of $\vec{a},\vec{b}$ \cite{paterek}.} 
Leggett showed that this constraint leads to an inequality which is violated by the singlet state correlations of quantum mechanics. Because Leggett's original inequality was not amenable to experimental testing other Leggett-type inequalities have been derived which are violated in recent experiments (see e.g. \cite{branciard}
and references therein).

Let us present the strongest known Leggett-type inequality of \citet{branciard}. 
This inequality uses three triplets of settings  $(\vec{a}_i,\vec{b}_i,\vec{b}_i')$  where party 1 thus chooses 3 settings and party 2 chooses 6 settings. Party 2  chooses the same angle $\phi$  between all pairs $(\vec{b}_i,\vec{b}_i')$ and such that 
  $\vec{b}_i-\vec{b}'_i=2 \sin \frac{\phi}{2} \vec{e}_i$, where $\{\vec{e}_1,\vec{e}_2,\vec{e}_3\}$ form an orthogonal basis.  
   The Leggett-type inequality of \citet{branciard} reads
\begin{align}
\frac{1}{3}\sum_{i=1}^{3}|\av{\vec{a}_i\vec{b}_i} +\av{\vec{a}_i\vec{b}_i'}|\leq 2-\frac{2}{3}|\sin\frac{\phi}{2}|,
\end{align}
where $\av{\vec{a}_i\vec{b}_i}=\int \,d\vec{u} \,d\vec{v} \, \rho(\vec{u},\vec{v})\, \sum_{ab}\, ab\,P(a,b|\vec{a}_i,\vec{b}_i,\vec{u},\vec{v})$.
The singlet state gives a value of $2|\cos \frac{\phi}{2}|$ which violates this inequality for a large range of values of $\phi$.
\forget{
Lastly, we mention an interesting result by \citet{branciard} which 
 can be rephrased as follows.  Consider a certain hidden variable $\xi$ which needs not be of the product form $\ket{\bm{u}}\otimes\ket{\bm{v}}$. Any model that obeys PI at this level and that wants to give the singlet state quantum predictions when averaged over the hidden variable at this level, i.e., $\av{\vec{a}\vec{b}}=\int d\xi \,\rho(\xi)\, C_\xi(\bm{a},\bm{b})=-\bm{a} \cdot \bm{b}$, needs to have vanishing marginal expectation values at this level, i.e.  $M^I_\xi(A)=M^{II}_\xi(B)=0$. \citet{branciard} remark that Leggett's model is not of this form (see \eqref{marginal}) and that this explains why the model cannot reproduce the singlet state predictions.
}
\subsubsection*{Discussion\footnote{A more profound discussion as well as a deeper analysis can be found in \cite{seevdeep}.}}

The above exposition of Leggett's model has presented us with an interesting relationship between the way different assumptions at the two different hidden-variable levels are related. At the $(\bm{u},\bm{v})$-level Leggett's model obeys PI, but allows for violations of OI.  But this was shown to be a consequence of opposite behavior on the deeper deterministic $(\lambda, \bm{u},\bm{v})$-level: on this level OI is obeyed, but deterministic PI is allowed to be violated.  We thus see that which conditions are obeyed and which are not depends on the level of consideration.
A conclusive picture therefore depends on which hidden-variable level is considered to be fundamental.

This can be nicely illustrated in a different hidden-variable model. Consider Bohmian mechanics where the deeper hidden-variable level is the description that contains the positions of the particles involved as well as the quantum state of these particles\footnote{\citet[p. 120]{bohm} take as the hidden variables of Bohmian mechanics ``the overall wave function together with the coordinates  of the particles''.}. At this level Bohmian mechanics is deterministic and thus obeys OI, whereas it is well known that deterministic PI (and IS) is violated \cite{dewdney}. However, at the level of the quantum state that is obtained by averaging over the positions of the particles we retrieve the quantum mechanical situation, as discussed in section \ref{shimonymaudlin},  where OI is violated, but PI is obeyed.
  
This shows explicitly that parameter dependence at the deeper deterministic hidden-variable level does not always show up as parameter dependence at the higher hidden-variable level, but sometimes as outcome dependence, i.e., as a violation of OI. In other words, violation of OI could be a sign of a violation of deterministic PI at a deeper hidden-variable level.  

It is known that any stochastic hidden-variable model can be made deterministic by adding additional variables. Here we should note that mathematically this always works\footnote{\label{mathematical}Mathematically introduce a deeper level hidden variable $\zeta$  with a distribution $\rho(\zeta)$ and a deterministic response function $\chi_{A,B}(a,b,(\lambda,\zeta))$ such that  $P(a,b|A,B,\lambda)= \int \chi_{A,B}(a,b,(\lambda,\zeta))\rho(\zeta) d\zeta$. This is always possible, for example choose $\zeta$ uniformly distributed and set  $\chi_{A,B}(a,b,(\lambda,\zeta))=1$  if $\zeta\leq P(a,b|A,B,\lambda)$ and $\chi_{A,B}(a,b,(\lambda,\zeta))=0$ otherwise (Cf. \citet{werwolf} and \citet{jones}).}, but only if physically one assumes that the stochastic model is incomplete since a deeper hidden-variable description is assumed to exist.  In such a case the feature above is generic: a violation of OI implies a violation of deterministic PI at the deeper hidden-variable level where the model is deterministic. The reason being that determinism implies OI (see next section) thus any violation of Factorisability must be because of violation of PI at this deeper level.

\subsubsection*{Comparison}

We will compare the Leggett-type model to the models of the previous section. These latter were shown to violate OI, PI and IS in a specific way but to nevertheless obey the CHSH inequality. Because we have considered stochastic models such a comparison must be performed at the at the 
$(\bm{u},\bm{v})$-level, and not at the deterministic $(\lambda,\bm{u},\bm{v})$-level.  We again write $\xi$ for the pair $(\bm{u},\bm{v})$. Let us recall both sets of assumptions involved.
\begin{enumerate}
\item {\bf Models of section \ref{non-localstochastic}}: We allow  $P(a,b|A,B,\xi)$  to be of the form \Eq{non-localstoch}. This violates both OI an PI. 
 We furthermore allow the distribution of $\xi$ to depend on the settings $A$, $B$ as in \Eq{non-localdistr}, thereby violating IS.
 \item {\bf  Leggett-type models}: Both at the $(\lambda,\xi)$-level and at the $\xi$-level  IS must be obeyed\footnote{With respect to possible violations of IS Leggett remarks: ``It might, for example, be thought at least plausible \emph{a priori} to reject the second postulate [IS], and in particular to allow the hidden-variable distribution $\rho(\lambda)$ to depend on the settings $\bm{a}$ and $\bm{b}$ of the polarizers. Whether any non-trivial results could be obtained under this assumption is a question I have not so far investigated''. \citep[p. 1492]{leggett}. But we have investigated this, and have found a non-trivial result.} (assumption 2 of \citet[p. 1473]{leggett}), i.e., both hidden-variable distributions $\rho(\lambda|\xi)$ and $\rho(\xi)$ must be independent of $A,B$. At the $\xi$-level Leggett-type models obey PI and furthermore require the marginal expectation values for a given value of $\xi$ to be equal to Malus' law at this level. OI is not assumed and possible violations of it are only constrained by consistency requirements, not by any other restrictions.
\end{enumerate}
 Comparing these assumptions reveals the following.  At the $\xi$-level where the physical assumptions are made Leggett-type models obey PI and IS and therefore also our weakened version of PI and IS, but they must allow for violations of our weakened version of OI.  The latter must be the case because the Leggett-type assumptions taken together are mathematically weaker than ours since a Leggett-type model has been given that violates the CHSH inequality \cite{paterek}. 
We conclude that although Leggett-type models impose a lot of structure (i.e., locally Malus' law needs to be obeyed at the $\xi$-level) the fact that violations of OI are not physically constrained in any way gives Leggett-type models greater correlative power than our models that allow for restricted violations of PI, OI as well as IS.

 \forget{
3) An open question remains: what forms of parameter and setting dependence are sufficient and/or necessary for reproducing QM?  This is an active field of study.e Seevinck and Leggett forms of non-locality do not suffice. One thus needs models that have more non-locality.

The non-local setting and outcome dependence of the distribution of the hidden variable $\xi$ on the settings $A$, $B$ as in \Eq{non-localdistr}, 

also for experiment since they have been shown to be violated (modulo some well-known loopholes).
}

\section[Subsurface vs.~surface probabilities: determinism and randomness]{Subsurface vs.~surface probabilities:\\ determinism and randomness}\label{surfacesection}
The previous section considered only subsurface probabilities and relationships between different kinds of assumptions at different hidden-variable (i.e. subsurface) levels. In this section we investigate relationships between surface and subsurface  probabilities and various constraints that can  be imposed on both types of probabilities. 

Let us first consider subsurface probabilities. These are conditioned on the hidden variable $\lambda$, which we take to be completely specified\footnote{In case the hidden variables are not completely specified (i.e., extra relevant information exists)  the trivial inference follows that outcomes will be probabilistically determined, i.e., deterministic determination is excluded from the start.  We are interested in non-trivial inferences and therefore assume that the hidden variables are completely specified.}. Suppose the hidden variables
are deterministic. This implies that OI is always obeyed, because if the outcomes are determined completely by the settings and the hidden variable, additionally specifying the outcome that was obtained by some distant party, cannot change any probabilities (for a formal proof see, amongst others, \citet{jarrett}). Thus,  if OI  is violated there must be some randomness at the hidden-variable level.

Let us consider what this implies for a situation where Factorisability is violated  (i.e.,  $P(a,b|A,B,\lambda)\neq P(a|A,\lambda)P(b|B,\lambda)$) so as to give non-local correlations that violate the CHSH inequality.  Recalling that Factorisability follows from the conjunction of both OI and PI we obtain the following inferences:
\begin{description}
\item (i) Deterministic hidden variables and violation of Factorisability implies violation of PI.
\item (ii) PI and violation of Factorisability implies  randomness at the hidden-variable level. 
\end{description}

Thus (i) says that any theory that gives violation of Factorisability but that obeys PI must have non-deterministic determination of the outcomes.
Quantum mechanics, where one takes the quantum state to be the hidden variable, is an example of such a theory:  it obeys PI and  the outcomes of measurement are probabilistically determined by the quantum state. However, not all hidden-variable theories that violate Factorisability  have this feature.
Indeed, as (ii) says, one can allow for a deterministic substratum at the hidden-variable level, but at the price of violating PI. 
 Bohmian mechanics is an example of such a latter theory. But we know that  it reproduces the predictions of quantum mechanics 
and therefore obeys no-signaling for the surface probabilities it predicts. Such a no-signaling requirement is quite constraining.
To see this we must look for analogs for the case of surface probabilities of the inferences (i) and (ii)  stated above.  

There is no straightforward surface analog of (i), since violations of PI not necessarily imply parameter dependence at the surface level (which would imply a violation of the no-signaling constraint), because the hidden variables need not be under control of the experimenter. However, (ii) does have a surface analog: No-signaling correlations that are non-local, but which are given by a hidden-variable model that obeys PI must be indeterministic, i.e., it  must show randomness in determining the outcomes.  However, this does not apply to Bohmian mechanics since it violates PI, so it would be interesting to see if such an inference can be made for any no-signaling correlation, independent of whether they violate PI or not. Surprisingly, this is indeed the case as was recently shown by  \citet{masanes06} using the following proof.

Consider a deterministic surface probability distribution $P_\textrm{det}(a,b|AB)$.  The outcomes $a$ and $b$ are deterministic functions of  $A$ and $B$: $a=a[A,B]$ and $b=b[A,B]$. Suppose it is a no-signaling distribution, then
\begin{align}
P_\textrm{det}(a,b|AB)=&\delta_{(a,b),(a[(A,B],b[A,B])}=\delta_{a,a[A,B]}\delta_{b,b[A,B]}\nn\\&~~~~~~~~~~~~~~~~~~~=P(a|A,B)P(b|A,B)=P(a|A)P(b|B).
\end{align}
The right hand side is a local distribution (i.e., it is of the form (\ref{localdistr})) and therefore any deterministic no-signaling correlation must be local. This results implies the following inferences for the correlations that are defined in terms of 
the surface probabilities:
\begin{description}
\item (iii) Any non-local correlation that is deterministic must be signaling.
\item (iv) Any non-local correlation that is no-signaling must be indeterministic, i.e., it determines the outcomes only probabilistically.
\end{description}
The inference (iii) and (iv) are the surface analogs of (i) and (ii).
 
If we now again consider Bohmian mechanics, we see that because it obeys no-signaling and gives rise to non-local correlations (since it violates the CHSH inequality) it must determine the outcomes only probabilistically. So although this theory  has deterministic hidden variables, this determinism must stay beneath the surface:  the hidden variables cannot be perfectly controllable because the outcomes must show randomness at the surface. In other words, although fundamentally deterministic it must necessarily be predictively indeterministic. Thus no Bohmian demon can have perfect control over the hidden variables and still be non-local and no-signaling at the surface.
This is not specific to Bohmian mechanics: any deterministic hidden-variable theory that obeys no-signaling and gives non-local correlations at the surface must have the same feature:  it must determine the outcomes of measurement indeterministically.

The inferences (iii) and (iv) show that requiring no-signaling  in conjunction with some other constraint has strong consequences. But what if we solely require no-signaling? 
\forget{The question we therefore put ourselves to answer is:  can we discern no-signaling correlations from more general ones using some experimentally accessible condition?   
In the next section we show this question has an affirmative answer.}In the next section we derive non-trivial constraints using only this condition and that are solely in terms of expectation values.

\section{Discerning no-signaling correlations}\label{discerningno-signalingsection}

\forget{
As mentioned in the previous chapter, the facets of the convex no-signaling polytope follow from the defining 
conditions \eqref{nosignalingdistr} on the the space of correlations  $P(a,b|A,B)$.  But we have seen that in terms of expectation values these conditions become trivial. They give no explicit restrictions on either the marginals $\av{A}$, $\av{B}$ or the product expectation values $\av{AB}$. They merely state that $\av{A}\leq\av{A}$ and $\av{A}\geq\av{A}$, see WEGGEHAALD.  Of course, it is the case that  non-negativity of  the joint probabilities  $P(a,b|A,B)$ implies that the marginals  $\av{A}$ and $\av{B}$  are not independent of the product expectation value $\av{AB}$, and vice versa, but this is a trivial constraint that cannot be violated by any correlation  (see e.g. \eqref{roy1}, \eqref{roy2} for such constraints).  
}
\forget{
As mentioned in the previous chapter, the facets of the convex no-signaling polytope follow from the defining 
conditions \eqref{nosignalingdistr} on the the space of correlations  $P(a_1,\ldots, a_N|A_1,\ldots,A_N)$.  But we have seen that in terms of expectation values these conditions become trivial. They give no explicit restrictions on either the marginals $\av{A_i}$ ($i=1,2,\ldots, N$) or the product expectation values $\av{A_1\cdots A_N}$. They merely state that $\av{A_i}\leq\av{A_i}$ and $\av{A_i}\geq\av{A_i}$.
Of course, it is the case that  non-negativity of  the joint probabilities    $P(a_1,\ldots, a_N|A_1,\ldots,A_N)$ implies that the marginals  $\av{A_1}$, etc.,  are not independent of the product expectation value $\av{A_1\cdots A_N}$, and vice versa, but this is a trivial constraint that cannot be violated by any correlation  (see e.g. \eqref{roy1}, \eqref{roy2} for such constraints).   
}
In this section we search for non-trivial constraints on the expectation values that are a consequence of no-signaling. We derive a non-trivial  Bell-type inequality for the no-signaling correlations in terms of both product and marginal expectation values. It thus discerns such correlations from more general correlations. Although the inequalities do not indicate facets of the no-signaling polytope we show that they can provide interesting results nevertheless. They provide  constraints on no-signaling correlations that are required to reproduce the perfectly correlated and anti-correlated quantum predictions of the singlet state.

Before we present our new inequalities, we first take a look at a previous attempt to formulate such a non-trivial inequality which we show to be flawed.

\subsection[The Roy-Singh no-signaling Bell-type inequality is trivially true]{The Roy-Singh no-signaling Bell-type inequality is\\ trivially true}

\citet{roysingh} claimed to have obtained a non-trivial no-signaling Bell-type inequality in terms of expectation values. They assumed no-signaling  by requiring that the expectation value of the observable corresponding to setting $A$ only depends on this setting and not on the faraway setting $B$, and vice versa. Thus $\av{A}_\textrm{ns}=f(A)$ and  $\av{B}_\textrm{ns}=g(B)$ where $f$ and $g$ are some functions\footnote{This notation by Roy and Singh is awkward since it suggests that the expectation value solely depends on the setting and not also on the state of the system one is measuring. However, this is not the case since they in fact use the definition $\av{A}_{\textrm{ns}}:=\int d \lambda \rho(\lambda,A,B)a(\lambda,A,B)$, that incorporates the hidden-variable distribution of the system under consideration, and where the dependence on $B$ on the left hand side is left out because of no-signaling.}. The inequalities of Roy and Singh \cite{roysingh} read:
\begin{align}
|\,\av{AB}_\textrm{ns} \pm \av{A}_\textrm{ns}\,|&\leq 1\pm \av{B}_\textrm{ns}, \label{roy1}\\
|\,\av{AB}_\textrm{ns} \pm \av{B}_\textrm{ns}\,|&\leq 1\pm \av{A}_\textrm{ns} \label{roy2}.
\end{align}
Roy and Singh interpret their inequalities as testing theories that obey no-signaling against more general signaling theories, i.e., their inequalities are  supposed to give a non-trivial bound for no-signaling correlations.

We mention two points of criticism; the first minor, the second major: First, one should  include the far-away setting in the marginals expectation values (i.e., use $\av{A}_\textrm{ns}^{B}$ and $\av{B}_\textrm{ns}^{A}$) as was argued in footnote \ref{wrongsignaling} on page \pageref{wrongsignaling}.  Secondly, no correlation whatsoever can violate these inequalities, whether they are signaling or not.  The inequalities are trivially true and are therefore irrelevant. The reason for this is that they follow from the trivial constraint that the probabilities $P(a,b|A,B)$ are non-negative. Let us show why this is the case.

\forget{
In the following we do not suppose any restrictions on the probabilities for outcomes. Consider first the identities that hold for general correlations and for the case of dichotomous outcomes $a,b=\pm1$:
\begin{subequations}
\label{identityroy}
\begin{align}
\av{A}^B&:= \sum_{a,b}aP(a,b|A,B)=\sum_a a P(a|A[B])=P(+|A[B])-P(-|A[B]), \\
\av{B}^A&:= \sum_{a,b}bP(a,b|A,B)=\sum_b b P(b|B[A])=P(+|B[A])-P(-|B[A]), \\
\av{AB}&=4P(+,+|A,B)-2P(+|A[B])-2P(+|B[A]) +1,\\
1&=P(+|B[A])+P(-|B[A]),
\end{align}
\end{subequations}
where the two outcomes are denoted $+$ and $-$ respectively, and  $P(a|A[B]):=\sum_{b}P(a,b|A,B)$ and $P(b|B[A]):=\sum_{a}P(a,b|A,B)$. Here we have explicitly taken the possibility of signaling into account by letting $\av{A}^B$  ($\av{B}^A$) depend on the setting $B$ ($A$) because the marginal probabilities can depend on the setting at the other side. Thus $P(\pm|A[B])$ indicates the probability for obtaining outcome $\pm$ when measuring $A$ on the first party and when the setting for the second  party is $B$ (analogously for  $P(\pm|B[A])$). 

Consider the first inequality (\ref{roy1}) of Roy and Singh\forget{ where we first choose the $+$ sign and next the $-$ sign respectively}. Using the identities (\ref{identityroy}) we rewrite this as 
\begin{align}
&|4P(+,+|A,B)-2P(+|B[A]) | \leq   2P(+|B[A]),
 \end{align} when choosing $+$ in (\ref{roy1}), and 
 \begin{align}
&|4P(+,+|A,B)-4P(+|A[B])-2P(+|B[A])+2| \leq   2P(-|B[A]),
\end{align}
when choosing $-$ in (\ref{roy1}).
Suppose terms in between $|\cdot|$ in the left hand side of  these two inequalities are positive, respectively negative, we then obtain:
\begin{align}
&P(+|B[A])\geq P(+,+|A,B)~~ \textrm{and}~~P(+,+|A,B)\geq 0. \label{roy3}\\
&P(+|A[B])\geq P(+,+|A,B)~~ \textrm{and}~~P(+,+|A,B)+1\geq P(+|A[B]) +P(+|B[A]).\label{roy4}
\end{align}
This can never be violated because the second inequality in (\ref{roy3}) is true because of non-negativity and the first inequality in (\ref{roy3}) and both  the inequalities in (\ref{roy4}) are trivially true since $P(+,+|A,B)= P(+|B[A])-P(-+|A,B])=P(+|A[B])-P(+-|A,B])$ for all probability distributions.
Thus the first inequality (\ref{roy1}) of Roy and Singh is true for all possible correlations, i.e., it excludes no possibilities. By symmetry the same conclusion of course holds for the second inequality (\ref{roy2}). 

}

The Roy-Singh inequalities \eqref{roy1} and \eqref{roy2} are in fact equivalent to the set of inequalities 
\begin{align}\label{seteq}
-1+|\av{A}^B+\av{B}^A|\leq \av{AB}\leq 1-|\av{A}^B-\av{B}^A|
\end{align}
that can be easily shown to hold for any possible correlation. Note that we leave out the subscript `ns', but include in the marginal expectation values $\av{A}^B$, $\av{B}^A$ the setting at the other side because there might be a dependency on the far-away setting as we are no longer restricting ourselves to no-signaling correlations.\forget{ cf. footnote \ref{trivnontrivmarginal} on page \pageref{chsintrotech}}

 The inequality \eqref{seteq} was first derived by \citet{leggett} in the following way (cf. \cite{paterek,branciard}).
For quantities $A,B$ that can take outcomes $a=\pm1$ and $b=\pm1$  the following identity holds:
\begin{align}\label{outcomesid}
-1+|a+b|=ab=1-|a-b| .
\end{align}
Let the outcome $a$ be determined\footnote{Without any further constraints, it is mathematically always possible to let the outcomes be determined by a deterministic hidden variable model. This was explicitly shown in footnote \ref{mathematical} on page \pageref{mathematical}.} by some hidden variable $\lambda$ and by the settings $A,B$: \mbox{$a:=a(\lambda,A,B)$}.  Furthermore, let  $\av{A}^B:= \int_\Lambda d\lambda \mu(\lambda|A,B) a(\lambda,A,B)$ be the average of quantity $A$ with respect to some positive normalized weight function $ \mu(\lambda|A,B)$ over the hidden variables. This function can contain any non-local or signaling dependencies on the setting $A$ and $B$. Define similarly the quantity $\av{B}^A$ and the average of the product $AB$ denoted by $\av{AB}$. Taking the average of the expression in \eqref{outcomesid} and using the fact that the average of the modulus is greater or equal to the modulus of the averages  one obtains the set of inequalities \eqref{seteq}.

Although the Roy-Singh inequalities indicate that the marginals  $\av{A}^B$ and  $\av{B}^A$ are not independent of the product expectation value $\av{AB}$, and vice versa, this is only a consequence of non-negativity of joint probabilities and not of the requirement of no-signaling. 
In conclusion, the Roy-Singh inequalities fail to show what they were supposed to do\footnote{\label{roysinghfootnote}Roy and Singh remark that J.S. Bell gave their manuscript a critical reading and that he commented upon some aspects of their manuscript. But apparently Bell did not comment on the fact that the inequalities are trivially true. What is interesting though is that Roy and Singh mention that Bell informed them of the manuscript by \citet{ballentinejarrett} in which the distinction between OL and OF is made (there called weak locality and predictive completeness) whose conjunction gives the condition of Factorisability used in deriving the Bell theorem. 
This is the only reference we know that indicates that Bell was aware of this distinction by Jarrett. We therefore cannot agree with \citet[p. 146]{brown} that \citet{bell81} was aware of any such distinction by 1981.  On all occasions where Bell argues for Factorisability 
 [i.e., in \cite{bell76,bell77,bell80,bell81,bell90}] this is performed using only a single step that is motivated by his condition of Local Causality. For Bell Factorisability is a package deal. Indeed, he nowhere uses a two step derivation that makes use of the conditions of OL and OF or some variants such as Shimony's PI and  OI.  (However, see footnote \ref{awkward} for the awkward derivation Bell uses to obtain Factorisability in his \cite{bell76}.)
It seems that Bell regarded local outcomes and settings on equal footing, i.e., both as local beables, and therefore it did not make sense for him to conditionalize on one but not on the other, a point also advocated by Hans Westman (private communication).
}. However, we next present a derivation that does meet this task of providing a non-trivial no-signaling Bell-type inequality in terms of both product and marginal expectation values.

\subsection{Non-trivial  
 no-signaling Bell-type inequalities}\label{nontrivnosignal}

Recall that the CHSH inequality does not suffice for discerning no-signaling correlations from general correlations because no-signaling correlations can reach the absolute maximum of this inequality. Indeed, using only product expectation values it was shown that the no-signaling polytope in the corresponding 4-dimensional space of vectors with components  $\av{A,B}, \av{A,B'},\av{A',B},\av{A',B'}$ is the trivial unit-cube (cf. section (\ref{techbellineq})). Our analysis must thus be performed in a larger space, and we consider the vectors that have as components  in addition to the product expectation values the marginal ones, i.e., we also consider the quantities $\av{A}^B, \av{A}^{B'},\av{A'}^B$, etc. In this space we obtain a set of non-trivial no-signaling Bell-type inequalities that discerns the no-signaling correlations from more general correlations.

\forget{
. The no-signaling polytope is 8-dimensional and we  therefore consider the larger space of vectors that have as components  in addition to the product expectation values the marginal ones, i.e., we also consider the quantities $\av{A}^B, \av{A'},\av{B},\av{B'}$. In this space we obtain a set of non-trivial no-signaling Bell-type inequalities that discerns the no-signaling correlations from more general correlations. }

The trick we use to obtain the new set of inequalities is to combine two different Roy-Singh inequalities where the no-signaling constraint is invoked to
set  $\av{A}_\textrm{ns}^B=\av{A}_\textrm{ns}^{B'}:=\av{A}_{\textrm{ns}}$, etc. 

For example, consider the following two Roy-Singh inequalities that hold for all correlations: 
\begin{align}
|\,\av{AB} \pm \av{A}^B\,|&\leq 1\pm \av{B}^A, \\
|\,\av{A'B} \pm \av{A'}^B\,|&\leq 1\pm \av{B}^{A'}.
\end{align}
Using  the inequality $|x+y|\leq |x|+|y|$ ($x,y\in\mathbb{R}$) we obtain
\begin{align}
|\av{AB} +\av{A}^B +\av{A'B} -\av{A'}^B|&\leq |\av{AB} +\av{A}^B| +|\av{A'B} -\av{A'}^B|\nn\\&
\leq 2+\av{B}^A-\av{B}^{A'}.
\end{align}
Assuming no-signaling (i.e., we set  $\av{B}_\textrm{ns}^A=\av{B}_\textrm{ns}^{A'}:=\av{B}_{\textrm{ns}}$) gives\footnote{That this is non-trivial can be shown by  giving an example of a signaling correlation that violates \eqref{nosignontriveq}. Consider a deterministic protocol where if $A$ and $B$ are measured jointly party 1 obtains outcome  $a_{11}$ and party 2 obtains outcome $b_{11}$, and, alternatively, if $A'$ and $B$ are measured jointly party 1 obtains outcome  $a_{21}$ and party 2 obtains outcome $b_{21}$, where $b_{11}\neq b_{21}$.
Then $\av{AB}=a_{11}b_{11}$, $\av{A'B}=a_{21}b_{21}$, $\av{A}^B=a_{11}$,  $\av{A'}^B=a_{21},\av{B}^A=b_{11},\av{B}^{A'}=b_{21}$. This is a one-way signaling protocol because $\av{B}^A\neq \av{B}^{A'}$.  If one chooses $a_{11}=b_{11}=1$ and $a_{21}=b_{21}=-1$ a value of 4 is obtained for the left hand-side of \eqref{nosignontriveq} clearly violating this inequality.}:
\begin{align}\label{nosignontriveq}
|\av{AB}_{\textrm{ns}} +\av{A'B}_{\textrm{ns}} +\av{A}_{\textrm{ns}}^B -\av{A'}_{\textrm{ns}}^B|\leq 2.
\end{align}
A total of 32 different such inequalities can be obtained that we can write as 
\begin{subequations}\label{nontrivnonsig2}
\begin{align}
(-1)^\gamma \av{AB}_{\textrm{ns}} +(-1)^{\beta +\gamma}\av{A'B}_{\textrm{ns}} +(-1)^{\alpha +\gamma} \av{A}_{\textrm{ns}}^B +(-1)^{\alpha+\beta+\gamma+1}\av{A'}_{\textrm{ns}}^B\leq 2,\\
(-1)^\gamma \av{AB}_{\textrm{ns}} +(-1)^{\beta +\gamma}\av{AB'}_{\textrm{ns}} +(-1)^{\alpha +\gamma} \av{B}_{\textrm{ns}}^A +(-1)^{\alpha+\beta+\gamma+1}\av{B'}_{\textrm{ns}}^A\leq 2,\\
(-1)^\gamma \av{A'B'}_{\textrm{ns}} +(-1)^{\beta +\gamma}\av{A'B}_{\textrm{ns}} +(-1)^{\alpha +\gamma} \av{B}_{\textrm{ns}}^{A'} +(-1)^{\alpha+\beta+\gamma+1}\av{B'}_{\textrm{ns}}^{A'}\leq 2,\\
(-1)^\gamma \av{A'B'}_{\textrm{ns}} +(-1)^{\beta +\gamma}\av{AB'}_{\textrm{ns}} +(-1)^{\alpha +\gamma} \av{A}_{\textrm{ns}}^{B'} +(-1)^{\alpha+\beta+\gamma+1}\av{A'}_{\textrm{ns}}^{B'}\leq 2,
\end{align}
\end{subequations}
with $\alpha,\beta,\gamma \in \{0,1\}$.

If we compare these inequalities to the CHSH inequality  $|\av{AB}_{\textrm{lhv}} +\av{A'B}_{\textrm{lhv}} +\av{AB'}_{\textrm{lhv}} -\av{A'B'}_{\textrm{lhv}}|\leq 2$ for local correlations we see a structural similarity; we only have to replace two product expectation values by some  specific marginal expectation values.

Adding two different Roy-Singh inequalities  and assuming no-signaling gives a slightly different  inequality that contains six terms\footnote{This is indeed non-trivial. The deterministic signaling protocol where the outcomes are $a_{11}=a_{22}=-1$ and $a_{12}=b_{12}=a_{21}=b_{21}=b_{11}=b_{22}=1$ 
gives $\av{AB}=a_{11}b_{11}=-1$, $\av{A'B'}=a_{22}b_{22}=-1$, and  $\av{A}^{B'}=a_{12}=1,\av{A'}^{B}=a_{21}=1,\av{B}^{A'}=b_{21}=1,\av{B'}^A=b_{12}=1$ so as to give a value of 6 on the left hand side of \eqref{nsres333} and which violates this inequality.}:
 \begin{align}\label{nsres333}
 -\av{AB}_\textrm{ns}-\av{A'B'}_\textrm{ns}+\av{A}_\textrm{ns}^{B'}+\av{B}_\textrm{ns}^{A'}+\av{A'}_\textrm{ns}^{B}+\av{B'}_\textrm{ns}^{A}\leq 2
\end{align}
Using permutations of observables and outcomes in \eqref{nsres333}\footnote{There are 6 different permutations that are of two types:  3 different permutations of the outcomes:  for party 1, for party 2 and for both parties; and 3 different permutations for the observables: permute $A$ with $A'$, $B$ with $B'$ or perform both permutations at once. All different combinations of these six give 64 possibilities of which only 14 give distinct non-trivial inequalities.}
  a total of 14 different non-trivial inequalities  can be obtained. These  can be compactly written as
\begin{subequations}\label{14ineq}
\begin{align}\label{14ineqa}
&-\av{AB}_\textrm{ns}-\av{A'B'}_\textrm{ns}-(-1)^\alpha\av{A}_\textrm{ns}^{B'}-(-1)^\alpha\av{B}_\textrm{ns}^{A'}-
\nn\\&~~~~~~~~~~~~~~~~~~~~~~~~~~~~~~~~~~~~~~~~~
(-1)^\beta\av{A'}_\textrm{ns}^{B}-(-1)^\beta\av{B'}_\textrm{ns}^A\leq 2,\\
&-(-1)^\gamma\av{AB}_\textrm{ns}-(-1)^{\gamma+1}\av{A'B'}_\textrm{ns}-(-1)^{1+\gamma\delta}\av{A}_\textrm{ns}^{B'}-(-1)^{1-\gamma(\delta+1)}\av{B}_\textrm{ns}^{A'}\nn\\&~~~~~~~~~~~~~~~~~~~~~~~-(-1)^{(\delta+1)(1-\gamma)+1}\av{A'}_\textrm{ns}^B-(-1)^{1+\delta(1-\gamma)}\av{B'}_\textrm{ns}^A\leq 2,\label{14ineqb}
\end{align}
\end{subequations}
where $\alpha,\beta,\gamma,\delta\in \{0,1\}$ except for the case $\alpha=\beta=0$ which is excluded since it gives a trivial inequality (see \eqref{result2}). This specifies 7 inequalities and the other 7 are obtained by interchanging $A$ by $A'$.

None of the above no-signaling inequalities are facets. They are saturated by only 7 affinely independent extreme points instead of the required 8 which is necessary for a facet.

\forget{

 \subsection{OUDE AFLEIDING}
Although all 16 vertices of the eight-dimensional no-signaling polytope are known for the case of two parties and two dichotomous observables per party (see (\ref{extremens})), we know of no non-trivial Bell-type inequalities for this polytope or of any non-trivial description of its no-signaling facets. Recall that the CHSH inequality does not suffice for discerning no-signaling correlations from general correlations because no-signaling correlations can reach the absolute maximum of this inequality. Recall furthermore that using only product expectation values the no-signaling polytope in the corresponding 4-dimensional space of vectors with components  $\av{A_1,A_2}, \av{A_1,A_2'},\av{A_1',A_2},\av{A_1',A_2'}$ is the trivial unit-cube (cf. section (\ref{techbellineq})). We will therefore consider the larger  $16$-dimensional space of vectors  with components 
$P(a_1,a_2|A_1,A_2),$ $P(a_1',a_2|A_1,A_2), \ldots, P(a_1',a_2'|A_1',A_2')$ to obtain a non-trivial no-signaling Bell-type inequality that discerns the no-signaling correlations from more general correlations. 

Consider the following two inequalities that hold under the assumption of no-signaling 
\begin{align}\label{ns1}
P(-|A'[B']) +P(++|A'B)\geq P(+|B[A]),
 \\\label{ns2}
P(-|B'[A']) +P(++|AB')\geq P(+|A[B]).
\end{align}
To see that they follow from no-signaling, let us proof the first of these inequalities. By symmetry the other one follows immediately. Consider the following inequality, where in its derivation we have explicitly indicated what role the no-signaling assumption plays\footnote{Here we have used the identities $P(+|A[B])+P(-|A[B])=1$, 
$P(+|A[B])=P(++|AB) +P(+-|AB)$,  $P(+|B[A])=P(++|AB) +P(-+|AB)$ and $P(--|AB)=1+P(++|AB)-P(+|A[B])-P(+|B[A])$, and analogous for the other choices of observables.}
\begin{align}
 P(-|A'[B']) +P(++|A'B)&\overset{\textrm{ns}}{=}P(-|A'[B]) +P(++|A'B)\nn\\&=P(-+|A'B)+P(--|A'B)+P(++|A'B)\nn\\&=P(--|A'B)+P(+|B[A'])
\nn\\&\geq P(+|B[A'])\overset{\textrm{ns}}{=} P(+|B[A]),
\end{align}
which indeed gives (\ref{ns1}).  Here the symbol  $\overset{\textrm{ns}}{=}$  means that the equality holds under the assumption of no-signaling, i.e., the marginal probabilities do not depend on the observable chosen by the other party.
If we now add (\ref{ns1}) and  (\ref{ns2}) we obtain 
\begin{align}\label{nonsigeq}
P(++|A'B)+ P(++|AB')+P(-|A'[B'])+
P(-|B'[A']) \geq P(+|A[B])+P(+|B[A])
\end{align}
which in  terms of the joint probability distributions becomes:
\begin{align}\label{nsres2}
P(++|A'B)+P(++|AB') +P(--|A'B')-P(++|A'B') + P(--|AB)-P(++|AB)\geq0
\end{align}
Note that since we assume no-signaling we may write $P(+|A)$ instead of $P(+|A[B])$, etc., so that \eqref{nonsigeq} can be rewritten as 
\begin{align}\label{nsres1}
P(++|A'B)+ P(++|AB')+P(-|A')+P(-|B')- P(+|A)-P(+|B)\geq0,
\end{align}
In terms of single and product expectation values this becomes\footnote{\label{proprelations}Here we use that $P(++|AB)=(\av{AB}_\textrm{ns}+\av{A}_\textrm{ns}+\av{B}_\textrm{ns} +1)/4$ and $P(--|AB)=(\av{AB}_\textrm{ns}-\av{A}_\textrm{ns}-\av{B}_\textrm{ns} +1)/4$, etc. \forget{Recall that since we assume no-signaling we may here write $P(+|A)$ instead of $P(+|A[B])$, etc.} Here  $\av{A}_\textrm{ns}:=P(+|A)-P(-|A)$ and analogous for $\av{B}_\textrm{ns}$, etc.
}
\begin{align}\label{nsres3}
 \av{AB'}_\textrm{ns}+\av{A'B}_\textrm{ns}-\av{A'}_\textrm{ns}-\av{B'}_\textrm{ns}-\av{A}_\textrm{ns}-\av{B}_\textrm{ns}\geq -2
\end{align}
So we see that not only expectation values of products of observables, but also of single observables need to be taken into account to get a nontrivial no-signaling inequality on the level of expectation values. 

Using permutations of observables and outcomes in \eqref{nsres3}\footnote{There are 6 different permutations that are of two types:  3 different permutations of the outcomes:  for party 1, for party 2 and for both parties; and 3 different permutations for the observables: permute $A$ with $A'$, $B$ with $B'$ or perform both permutations at once. All different combinations of these six give 64 possibilities of which only 14 give distinct non-trivial inequalities.}  a total of 14 different non-trivial inequalities that can be obtained.  These  can be compactly written as
\begin{align}\label{14ineqa}
&\av{AB}_\textrm{ns}+\av{A'B'}_\textrm{ns}+(-1)^\alpha\av{A}_\textrm{ns}+(-1)^\alpha\av{B}_\textrm{ns}+(-1)^\beta\av{A'}_\textrm{ns}+(-1)^\beta\av{B'}_\textrm{ns}\geq -2,\\
&(-1)^\gamma\av{AB}_\textrm{ns}+(-1)^{\gamma+1}\av{A'B'}_\textrm{ns}+(-1)^{1+\gamma\delta}\av{A}_\textrm{ns}+\nn\\&~~~~~~~~~~~~~~~~~~~(-1)^{1-\gamma(\delta+1)}\av{B}_\textrm{ns}+(-1)^{(\delta+1)(1-\gamma)+1}\av{A'}_\textrm{ns}+(-1)^{1+\delta(1-\gamma)}\av{B'}_\textrm{ns}\geq -2,\label{14ineqb}
\end{align}
where $\alpha,\beta,\gamma,\delta\in \{0,1\}$ except for the case $\alpha=\beta=0$ which is excluded since it gives a trivial inequality (see \eqref{result2}). This specifies 7 inequalities and the other 7 are obtained by interchanging $A$ by $A'$.

Inequality (\ref{nsres2}) is \forget{the sought-after} a non-trivial constraint on no-signaling correlations. To show that it is non-trivial it must be the case that signaling correlations exist that violate the inequality.  For example, choose  $P(++|AB') =P(++|A'B) =P(--|A'B')= P(--|AB)=0$ and $P(++|A'B')=P(++|AB)=1$. This signaling correlation violates (\ref{nsres2})\footnote{This correlation of course also allows for a violation of (\ref{nsres3}). It gives $\av{A}_\textrm{ns}=\av{B}_\textrm{ns}=\av{A'}_\textrm{ns}=\av{B'}_\textrm{ns}=1$, and allows for  $ \av{AB'}_\textrm{ns}=\av{A'B}_\textrm{ns}=-1$, when choosing $P(--|AB') =P(--|A'B)=0$ as well. This gives a value of $-6$ for the left hand side of (\ref{nsres3}), clearly violating the inequality.}.

It is easily verified that $11$ extreme points of the no-signaling polytope are able to give equality in (\ref{nsres2}). Of these $11$ extreme points  a total of $9$ are local extreme points (i.e., local deterministic distributions), and two are non-local as given in (\ref{extremens})\footnote{\label{2nosig}The two non-local no-signaling correlations are given by the following correlations:  (i) the $AB$ probabilities are $1/2$ for equal outcomes and zero for different outcomes, whereas  the probabilities for $AB'$, $A'B$ and $A'B'$ are zero for equal outcomes and  $1/2$ for different outcomes, (ii) the same as in (i) but with $AB$ and $A'B'$ interchanged. Note that these correlations  give $\av{AB'}_\textrm{ns}=\av{A'B}_\textrm{ns}=-1$ and $\av{A}_\textrm{ns}=\av{B}_\textrm{ns}=\av{A'}_\textrm{ns}=\av{B'}_\textrm{ns}=0$ and thus also give equality in (\ref{nsres3}), as was expected.}. 
Considered as vectors the 11 extreme points  determine a set of only 7 linearly independent vectors\footnote{To verify this, we considered the matrix that contains the $11$ extreme points that give equality in (\ref{nsres2}) and determined its rank to be $7$.}.  Since the null vector is not among the 11 original vectors  this number should be 8 in order for the inequalities to determine a facet because the no-signaling polytope is $8$-dimensional  \cite{masanes02}.   Therefore, equality in (\ref{nsres2}) does not describe a facet of the no-signaling polytope.  For the other 13 equivalent inequalities the same analysis goes through, so these are neither facets of the no-signaling polytope. 


\subsubsection{Comparison to CHSH and CH inequality}
Comparing (\ref{nsres3}) to the CHSH inequality $| \av{A'B'}_\textrm{lhv}+\av{AB'}_\textrm{lhv}+\av{A'B}_\textrm{lhv}-\av{AB}_\textrm{lhv}|\leq2$  shows similarities, but also a crucial difference since expectation values of single observables instead of solely of products of observables appear.  The polynomial one would associate to such an inequality is therefore not of the standard form (\ref{bellpoly}) as mentioned in chapter \ref{definitionchapter} because the latter has only product expectation values.

 It therefore seems to be more instructive to compare the no-signaling inequality (\ref{nsres2})  to the Clauser-Horne inequalities \cite{ch} that are equivalent to the CHSH inequalities (see \citet{redhead} for a proof of this equivalence). One of the CH inequalities reads:
\begin{align}\label{chineq}
P(++|AB') +P(++|A'B) +&P(++|A'B')\nn\\&- P(++|AB)- P(+|A') -P(+|B')\geq -1.
\end{align}
For no-signaling correlations it is the case  that 
\begin{align}
P(--|A'B')=1+P(++|A'B')-P(+|A')-P(+|B'),
\end{align} so that we can rewrite the CH inequality  (\ref{chineq}) as
\begin{align}\label{chineq2}
         P(--|A'B') +P(++|A'B) +P(++|AB')-P(++|AB)\geq 0.
\end{align}
If we compare this to the no-signaling inequality  (\ref{nsres2}) we see that the expression that is bounded from below has a two extra terms (adding $P(--|AB)$, but subtracting $P(++|A'B')$) as compared to the original CH inequality (\ref{chineq}).

}

\subsubsection{Reproducing perfect (anti-) correlations}\forget{ constrains admissible no-signaling models}

The set of non-trivial inequalities \eqref{14ineq} shows an interesting constraint on no-signaling correlations that are required to reproduce the perfectly correlated and anti-correlated quantum predictions of the two-qubit singlet state $(\ket{01}-\ket{10})/\sqrt{2}$. Consider spin measurements in directions $\vec{a}$ and $\vec{b}$  on each of the two qubits.  It is well known that the singlet  state gives perfect anti-correlated predictions when the measurements are in the same direction, and perfect correlated predictions when they are in opposite directions:  
\begin{align}
\forall \,\vec{a},\,\vec{b}:~~\av{\vec{a}\vec{b}}=-1,~~\textrm{when} ~~\vec{a}=\vec{b}, \label{perfectanticorre}\\
\forall \,\vec{a},\,\vec{b}:~~\av{\vec{a}\vec{b}}=1,~~\textrm{when} ~~\vec{a}=-\vec{b}.\label{perfectcorre}
\end{align} 
Suppose one wants to reproduce these correlations using a no-signaling correlation, i.e., for all $\vec{a},\vec{a}',\vec{b},\vec{b}'$ inequalities \eqref{14ineqa} and \eqref{14ineqb} for all admissible $\alpha,\beta,\gamma,\delta$ must hold, where the settings $A,B,A,B'$ have been denoted by  
the vectors  $\vec{a},\vec{a}',\vec{b},\vec{b}'$ respectively.   Because of no-signaling the dependence of the marginals on far-away settings is dropped,  i.e., $\av{\vec{a}}^{\vec{b}}=\av{\vec{a}}^{\vec{b}'}:=\av{\vec{a}}$, etc.

\forget{We derive two non-trivial constraints from this set of assumptions and the .}

In the case where $\vec{a}'=\vec{b}=\vec{b}'=\vec{a}$ the assumption \eqref{perfectanticorre}  together with the constraint \eqref{14ineqa} for $\alpha=\beta=1$ implies, for all $\vec{a}$:
\begin{align}\label{result1}
-\av{\vec{a}}^I_\textrm{ns}-\av{\vec{a}}^{II}_\textrm{ns}\geq 0.
\end{align}
where the two different parties $I$ and $II$ are explicitly indicated, i.e., $\av{\vec{a}}^I_\textrm{ns}$ for party $I$ and $\av{\vec{a}}^{II}_\textrm{ns}$ for party $II$.

 Furthermore, non-negativity gives  $4P(++|\vec{a}\vec{b})+4P(++|\vec{a}'\vec{b}')\geq0$, which is identical to
  \begin{align}\label{result2}
 \av{\vec{a}\vec{b}}_\textrm{ns}+ \av{\vec{a}'\vec{b}'}_\textrm{ns}+\av{\vec{a}}_\textrm{ns}+\av{\vec{b}}_\textrm{ns}+\av{\vec{a}'}_\textrm{ns}+\av{\vec{b}'}_\textrm{ns}+2 \geq0.
 \end{align}  
 In the case where $\vec{a}'=\vec{b}=\vec{b}'=\vec{a}$  assumption \eqref{perfectanticorre}  and the constraint \eqref{result2} imply, for all 
 $\vec{a}$: $\av{\vec{a}}^I_\textrm{ns}+\av{\vec{a}}^{II}_\textrm{ns}\geq 0$.
 Together with \eqref{result1} we thus obtain, for all $\vec{a}$:
\begin{align}\label{result4}
\av{\vec{a}}^I_\textrm{ns}+\av{\vec{a}}^{II}_\textrm{ns}= 0.
\end{align}
This is the first non-trivial constraint.

The second constraint follows from the case where $-\vec{a}'=\vec{b}=\vec{b}'=\vec{a}$. In this case the assumption \eqref{perfectcorre}  together with the constraints of \eqref{14ineqb} for $\gamma=\delta=0$ and $\gamma=0,\delta=1$ implies, for all $\vec{a}$: $\av{\vec{a}}^I_\textrm{ns}=\av{-\vec{a}}^{II}_\textrm{ns}$. Together with \eqref{result4} this implies, for all $\vec{a}$:
\begin{align}\label{result5}
\av{-\vec{a}}^I_\textrm{ns}=-\av{\vec{a}}^I_\textrm{ns}.
\end{align}
By symmetry the same holds for party $II$. 

Thus \eqref{result4} and \eqref{result5} are necessary conditions for any no-signaling model to reproduce the singlet state perfect (anti-)correlations. These conditions state that the marginal expectation values for party $I$ and $II$ must add up to zero  for measurements in the same direction, and the individual marginal expectation values must be odd functions of the settings.
Consequently, any model reproducing the singlet state perfect (anti-) correlations and  which does not obey either one (or both) of these conditions must be signaling.

In case the no-signaling model treats the systems held by party $I$ and $II$ the same, i.e., $\av{\vec{a}}^I_\textrm{ns}=\av{\vec{a}}^{II}_\textrm{ns}$, it must have vanishing marginal expectation values: $\av{\vec{a}}^I_\textrm{ns}=\av{\vec{a}}^{II}_\textrm{ns}=0$.  All marginal probabilities then must be uniformly distributed: $P(+|\vec{a})=P(-|\vec{a})=1/2$, etc. 
 \forget{
 We conjecture the following: A necessary condition for a no-signaling model to reproduce the conditions of perfect (anti-)correlation \eqref{perfectanticorre} and \eqref{perfectcorre} is that its marginal expectation values must vanish. If true, any such model reproducing the singlet perfect (anti-) correlations and  which has non-vanishing marginal expectation values must be signaling. }
\forget{
The no-signaling inequalities we used do not give facets of the no-signaling polytope and thus stronger such inequalities exist. 
}

In case one requires not only the perfect (anti-) correlations for parallel and anti-parallel settings but the full singlet state correlation $\av{\vec{a}\vec{b}}=-\vec{a}\cdot\vec{b}$, $\forall\,\vec{a},\vec{b}$, the requirement of vanishing marginal expectation values must indeed obtain. \citet{branciard} established this for hidden-variable models  of the Leggett type (see section \ref{sectionleggett}), but it holds also for general no-signaling models.

 \section{Discussion}\label{discussionchshclassical}

Many of the investigations in this chapter are not final. We will discuss four interesting and important open problems. The first three are more of a technical nature, the fourth has a foundational character.

(1) We have shown that a large class of hidden-variable models must obey the CHSH inequality despite the fact that the probabilities for outcomes and  the hidden-variable distributions are non-locally setting and outcome dependent. Such a form of setting and outcome dependence at the subsurface level is thus sufficient to derive the CHSH inequality. An open question remains what forms of setting and outcome dependence  would be necessary and sufficient.  

In view of the comparison of our result to Leggett's model, a related question, not investigated here, arises. What forms of non-local setting and outcome dependence are necessary and sufficient for reproducing quantum mechanical predictions for bi-partite quantum systems? Despite interesting progress, see e.g. \citet{brunner0803}, even in the most simple case of two dichotomous observables per party this is an open question.

(2)  The analysis of Leggett-type models has presented us with interesting relationships between the way different assumptions at the two different hidden-variable levels of such models are related. It is the case that in such models parameter dependence at the deeper hidden-variable level does not show up as parameter dependence at the higher hidden-variable level, but only as setting dependence, i.e., as a violation of OI.  Conversely, for such models, and in fact for any hidden-variable model for which there is a deeper deterministic level,  a violation of OI can be regarded as a sign of a violation of deterministic PI at a deeper hidden-variable level.   We thus see that which conditions are obeyed and which are not depends on the level of consideration and on which hidden-variable level is considered to be fundamental. An interesting avenue for future research would be to search for other such relationships. 
\forget{ We leave as an interesting open problem the investigation of the relationship between the way different assumptions at different hidden-variable levels are related in more general models than Leggett-type models.}

(3)  The surprising result obtained in section \ref{surfacesection} that any hidden-variable model that is deterministic at the subsurface level but which has no-signaling non-local correlations at the surface must show randomness in the distribution of the outcomes, asks for a further investigation of the relationship between inferences and results that exist at the levels of surface and subsurface probabilities.  

\forget{

(4) We have presented a set of 14 non-trivial no-signaling inequalities that were unfortunately not tight inequalities that describe
bound the no-signaling polytope. It would be very interesting to find the full set of non-trivial tight facet inequalities.  This set would specify the necessary and sufficient conditions for a correlation to be reproducible using no-signaling correlations. Experimental investigations of this set would allow for tests whether nature is no-signaling.  

}

(4)  Although the investigations are not final, we can nevertheless already claim that a foundational question should be asked.
Given the arguments against experimental metaphysics that were reviewed in section \ref{expmeta},  and the novel one presented here where  a class of non-local setting and outcome dependent hidden-variable models that violate OI, PI and IS was shown to nevertheless obey the CHSH inequality, we are led to ask the following question: how should we understand violations of the CHSH inequality?  This is a difficult question to answer, since we only have rather trivial necessary conditions and some sufficient conditions for when a hidden-variable model obeys this inequality.  But no necessary and sufficient condition has been found. Therefore, we do not know precisely what a violation amounts to.

  We think we can say at least this: violation of the CHSH inequality shows that we must give up on one (or more) of the following:
  \begin{enumerate}
\item The non-local outcome dependent versions of OI (as given in \eqref{nlstoch1} and \eqref{nlstoch3}.   Giving up this forces us to include even more non-local outcome dependence. 
\item The non-local setting dependent versions of PI (as  in \eqref{nlstoch2} and \eqref{nlstoch4}).\forget{ which we have shown to be sufficient to derive  the CHSH inequality.} Giving up this forces us to include even more non-local setting dependence. 
\item  The setting dependent version of IS (as given in (\ref{non-localdistr})).\forget{which was also shown to be sufficient to derive the CHSH inequality.} Opting to give up this assumption forces us to give up on even more freedom of the observers to choose settings.
\item One of Maudlin's conditions  P1 or P2 (or both). But note that  in section \ref{shimonymaudlin} it was argued that Maudlin's conditions are rather unnatural 
because of the extra assumptions that are needed to evaluate them in quantum mechanics.
\end{enumerate}

If one not carefully takes these findings into account, and acknowledges that perhaps more may be found,  experimental metaphysics becomes a very dangerous field, full of perhaps metaphysically interesting, but non-instantiated conclusions. 
It would be too much to ask for a ban on interpreting violations of the CHSH inequality until a final technical investigation of the issue is available, but we believe we ask not too much to acknowledge the limitations of the technical results upon which one bases its philosophical endeavors.
It is important to recognize this if we are to have a proper appreciation of the epistemological situation we are in when we attempt to glean metaphysical implications  of the failure of the CHSH inequality.
\label{listalternative}
\forget{
(5)  Although the investigations are not final, we can nevertheless already claim that a foundational question should be asked.
Given the facts that (i) the condition of Factorisability is not uniquely determined by the Shimony conditions (we have proven that the Maudlin conditions suffice as well), and (ii) that a class of non-local setting and outcome dependent hidden-variable models that violate OI, PI and IS nevertheless obey the CHSH inequality, we are led to ask the following question: how should we understand violations of the CHSH inequality? 
This is a difficult question to answer, since we only have rather trivial necessary conditions and some sufficient conditions for when a hidden-variable model obeys this inequality.  But no necessary and sufficient set of conditions has been found. Therefore, we do not know precisely what a violation amounts to.

  We think we can say at least this: violation of the CHSH inequality shows that we must give up on one (or more) of the following:
  \begin{enumerate}
\item The non-local outcome dependent versions of OI (as given in \eqref{nlstoch1a} and \eqref{nlstoch3a}.   Giving up this forces us to include even more non-local outcome dependence. 
\item The non-local setting dependent versions of PI (as given in \eqref{nlstoch2a} and \eqref{nlstoch4a}).\forget{ which we have shown to be sufficient to derive  the CHSH inequality.} Giving up this forces us to include even more non-local setting dependence. 
\item  The setting dependent version of IS (as given in (\ref{non-localdistr})).\forget{which was also shown to be sufficient to derive the CHSH inequality.} Opting to give up this assumption forces us to give up on even more freedom of the observers to choose settings.
\item One of Maudlin's conditions  P1 or P2 (or both). 
\end{enumerate}

If one not carefully takes these findings into account, and acknowledges that perhaps more may be found,  experimental metaphysics becomes a very dangerous field, full of perhaps metaphysically interesting, but non-instantiated conclusions. 
It would be too much to ask for a ban on interpreting violations of the CHSH inequality until a final technical investigation of the issue is available, but we believe we ask not too much to acknowledge the limitations of the technical results upon which one bases its philosophical endeavors.
It is important to recognize this if we are to have a proper appreciation of the epistemological situation we are in when we attempt to glean metaphysical implications  of the failure of the CHSH inequality.
}	

\section{Appendices}
\subsection{On Shimony and Maudlin factorisation}
\noindent
In this Appendix we will prove\footnote{We thank Sven Aerts for crucial help in establishing the proof.}~that the conjunction of Maudlin's assumptions implies Factorisability, just as the conjunction of Shimony's assumption does. Their interrelationship will also be investigated. For completeness, we state Shimony's and Maudlin's assumptions again:
\begin{description}
\item{{ Shimony: }}
\begin{align}\label{appendixmaudlin}
\textrm{OI}:&~~~P(a|A,B,b,\lambda)=P(a|A,B,\lambda)~~~\mathrm{and}~~~
P(b|A,B,a,\lambda)=P(b|A,B,\lambda).\\
\textrm{PI}:&~~~P(a|A,B,\lambda)=P(a|A,\lambda)~~~\mathrm{and}~~~
P(b|A,B,\lambda)=P(b|B,\lambda).
\end{align}
\item{{ Maudlin:}}
\begin{align}
&\mathrm{P}1:~~~P(a|A,b,\lambda)=P(a|A,\lambda)~~~\mathrm{and}~~~P(b|B,a,\lambda)=
P(b|B,\lambda).\\
&\mathrm{P}2:~~~P(a|A,B,b,\lambda)=P(a|A,b,\lambda)~~~\mathrm{and}~~~
P(b|A,B,a,\lambda)=
P(b|B,a,\lambda).
\end{align}
\end{description}
OI and PI together imply Factorisability, i.e., $P(a,b|A,B,\lambda)=P(a|A,\lambda)P(b|B,\lambda)$. The conjunction of P1 and P2 also implies this. We prove this as follows.
Consider the general result from the law of conditional probability that:
\begin{align}\label{general}
P(a,b|A,B,\lambda)&=P(a|A,B,b,\lambda)\,P(b|A,B,\lambda)=P(b|A,B,a,\lambda)\,P(a|A,B,\lambda).
\end{align}
Applying P1 and P2 we get:
\begin{align}\label{maudlin1}
P(a,b|A,B,\lambda)&=P(a|A,\lambda)\,P(b|A,B,\lambda)=P(b|B,\lambda)\,P(a|A,B,\lambda).
\end{align}
Consider now the second equality. Supposing that $P(b|B,\lambda)$ and $P(a|A,\lambda)$
are non-zero, we can write this as:
\begin{align}\label{breuk}
\frac{P(a|A,b,\lambda)}{P(a|A,\lambda)}=\frac{P(b|A,B,\lambda)}{P(b|B,\lambda)}.
\end{align}
Maudlin's claim that the conjunction of P1 and P2 give Factorisability
will hold if it is the case that \Eq{breuk} equals the numerical constant 1.
%
Note that this effectively states the condition of parameter independence.
This indeed follows from the conjunction of Maudlin's assumptions P1 and P2.
\\\\
{\bf Proof}: Suppose we would have taken another outcome $a'$ in \Eq{general},
then applying P1 and P2 again we would obtain
in the same way as which gave us \Eq{breuk} the following
\begin{align}\label{breuk3}
\frac{P(a'|A,B,\lambda)}{P(a'|A,\lambda)}=\frac{P(b|A,B,\lambda)}{P(b|B,\lambda)}.
\end{align}
Combining this with \Eq{breuk} we get
\begin{align}\label{breuk4}
\frac{P(a|A,B,\lambda)}{P(a|A,\lambda)}=\frac{P(a'|A,B,\lambda)}{P(a'|A,\lambda)}.
\end{align}
We now suppose we are dealing with a standard Bell experiment
where all measurements have dichotomous outcomes.
The possible outcomes of measuring $A$ are  thus $a$, $a'$. We therefore have
\begin{align}
P(a|A,B,\lambda)+P(a'|A,B,\lambda) =1~~~\textrm{and}~~~ P(a|A,\lambda)+P(a'|A,\lambda)=1.
\end{align}
If we substitute this into \Eq{breuk4} we get $P(a|A,B,\lambda)=P(a|A,\lambda)$.
We thus have parameter independence and if we use this in \Eq{maudlin1} we get
Factorisability, having assumed only P1 and P2 and the
requirement of dichotomous observables:
\begin{align}\label{maudlin2}
P(a,b|A,B,\lambda)&=P(a|A,\lambda)\,P(b|B,\lambda).
\end{align}
\subsection*{The requirement of dichotomous observables is not necessary}\noindent
Suppose $a$ (and $b$) are possible outcome variables which 
have more than two possible real-valued outcomes.
Divide the domain of $a$ into two
measurable subsets such that the intersection is zero
and the union equal the domain of $a$. Call them $S$ and $S^{c}$,
where the latter is the complement of the first. We thus obtain a two valued
observable with outcomes $S$ and $S^{c}$, which, for convenience, can be given the
values $+1$ and $-1$ if one wants to.
Next define the probability for obtaining one of these two values as:
\begin{align}
P(S)=\int_S\, P(a)\, da~,~~~~P(S^c)=\int_{S^c}\, P(a)\, da,
\end{align}
and analogously for the conditional probabilities.
\\\\
One would then get according to P$1$:
\begin{align}
P(S|A,b,\lambda)=\int_S P(a|A,b,\lambda) da
=\int_S P(a|A,\lambda)da=P(S|A,\lambda).
\end{align}
And according to P$2$:
\begin{align}
P(S|A,B,b,\lambda)=\int_S P(a|A,B,b,\lambda) da
=\int_S P(a|A,B,\lambda)da=P(S|A,b,\lambda).
\end{align}
The same holds for $P(S^c|\ldots)$.

Thus all functional dependencies of the $a$ probabilities are reproduced
on the level of the $S$ probabilities. Suppose that we divide the set
of outcomes of the observable $b$ into two subsets $T$ and $T^c$,
where the latter is again the complement of the first.
From the above proof that P$1$ and P$2$ imply Factorisability
for the case of dichotomous observables we  can conclude that since $S$, $S^c$ and $T$, $T^c$ represent
dichotomous observables, that $P(S,T|a,b,\lambda)$ factorises. Since this has
to hold for all possible choices of measurable subsets
$S$, $S^c$ and $T$, $T^c$ Factorisability must also hold for
  $P(a,b|A,B,\lambda)$.
\subsection*{On the conjunction of Maudlin's P1 and P2}\noindent
From the above proof  we see that
the following requirement also implies Factorisability
\begin{align}
\mathrm{P}3:~~~P(a|A,B,b,\lambda)=P(a|A,\lambda)~~~\mathrm{and}~~~
P(b|A,B,a,\lambda)=P(b|B,\lambda).
\end{align}
P$1$ and P$2$ imply P$3$ as can be seen by combining (\ref{general}) and (\ref{maudlin1}). But could it be that P3 is weaker than P1 and P2 in conjunction? That is,
is it possible that
$P(a|A,B,b,\lambda)=P(a|A,\lambda)$, and either $\neg\,\mathrm{P}1$ (i.e., 
$P(a|A,b,\lambda)\neq P(a|A,\lambda)$) or $\neg\,\mathrm{P}2$ (i.e., $P(a|A,B,B,\lambda)
\neq P(a|A,B,\lambda)$)?
Since P1, P2 and P3 have to hold for all possible outcomes this is not possible.\\\\
{\bf Proof:}

(A) P3$\Longrightarrow$ P1: We have that
\begin{align}\label{xyz}
P(a|A,b,\lambda)\,=\,P(a|A,B,b,\lambda)\,P(b|A,B,\lambda) +P(a|A,B, \neg\, b,\lambda)\,(1-P(b|A,B,\lambda)),
\end{align}
because $P(X|Y)=P(X|YZ)\,P(Y|Z)+P(X| \neg\, Y Z)\,(1-P(Y|Z))$ for all $X,Y,Z$. 
If we assume P3, i.e., if $P(a|A,B,b,\lambda)=P(a|A,\lambda)$ and
$P(a|A,B,\neg \,b,\lambda)=P(a|A,\lambda)$, then from
\Eq{xyz} we see that $P(a|A,b,\lambda)=P(a|A,\lambda)$.

(B) P3$\Longrightarrow$ P2: Since P3 implies P1, it follows from
P3 that P3 $\land$ P1. Thus:
\begin{align}
\big[ P(a|A,B,b,\lambda)=P(a|A,\lambda)\big] &\land \big[P(a|A,b,\lambda)=P(a|A,\lambda)\big]\nn\\
&\Longrightarrow P(a|A,B,b,\lambda)= P(a|A,b,\lambda),
\end{align}
which is P2.
\subsection*{Conclusion}\noindent
We have shown that the conjunction of P1 and P2  is equivalent to P3 which is equivalent to Factorisability.
Shimony has already shown that  the conjunction of  OI and PI is equivalent to Factorisability.  This gives us the
following logical relations:
\begin{align}
\mathrm{OI}\land \mathrm{PI}&\Longleftrightarrow \mathrm{P}1\land
\mathrm{P}2\Longleftrightarrow \mathrm{P}3\Longleftrightarrow\mathrm{Factorisability}
\end{align}

\subsection[Shimony's and Maudlin's conditions in quantum mechanics]{Shimony's and Maudlin's conditions in quantum\\ mechanics}\label{SMQM}

Quantum mechanics  can be considered as a stochastic hidden-variable theory where the hidden variable $\lambda$ is effectively the quantum state $\ket{\psi}$, i.e., $\rho(\lambda)= \delta(\lambda-\lambda_0)$ with $\lambda_0=\ket{\psi}$ (actually, the quantum state need not be a pure state).  The joint probabilities\footnote{These probabilities conditional on quantum states denote probabilities prescribed by those states. Although this commits us to probabilities for certain quantum states to be prepared, this can be easily removed by a reformulation where states are treated as parameters and not as random variables, cf. \cite[p. 5]{clifton}.} are then obtained via 
\begin{align}\label{qm1}
P(a_i, b_j|A,B,\lambda_0)=\textrm{Tr}[\hat{A}_i\otimes\hat{B}_j\ket{\psi}\bra{\psi}],
\end{align}
 and the marginals are 
\begin{align}\label{qm2}
P(a_i|A,\lambda_0)=\textrm{Tr}[\hat{A}_i\rho^I], ~~~\textrm{and}~~~ P(b_j|B,\lambda_0)=\textrm{Tr}[\hat{B}_j\rho^{II}],
\end{align}
 with $\rho^I$ and $\rho^{II}$ the reduced density matrices for subsystems $I$ and $II$ respectively,  and we consider POVM's  $\{ \hat{A}_i\}$, and $\{\hat{B}_j\}$ with $\sum_i \hat{A}_i=\sum_j\hat{B}_j=\1$.
\\\\
{\bf (I)} Shimony's OI and PI: Considered as a stochastic hidden-variable theory quantum mechanics obeys PI but violates OI. Proof: 
\begin{itemize}
\item
PI is obeyed, because $\forall i,j$:   
$P(a_i|A,B,\lambda_0)=\sum_j \textrm{Tr}[\hat{A}_i\otimes\hat{B}_j\ket{\psi}\bra{\psi}]=\textrm{Tr}[\hat{A}_i\otimes\1\ket{\psi}\bra{\psi}]=
\textrm{Tr}[\hat{A}_i\rho^I]=P(a_i|A,\lambda_0)$, and analogously we find $P(b_j|A,B,\lambda_0)= P(b_j|B,\lambda_0)$. 
\item OI is violated. For example, take $\ket{\psi}$ to be the singlet state $(\ket{01}-\ket{10})/\sqrt{2}$. In case $A$ and $B$ are chosen to be equal this state predicts: \begin{align}
& P(a_+,b_+|A,B,\lambda_0)=\textrm{Tr}[\hat{A}_+\otimes\hat{B}_+\ket{\psi}\bra{\psi}]=0 ,\nn\\
& P(a_+|A,\lambda_0)=\textrm{Tr}[\hat{A}_+\rho^I]=P(b_+|B,\lambda_0)=\textrm{Tr}[\hat{B}_+\rho^{II}]=\frac{1}{2}.
 \end{align} Here $a_+$ and $b_+$ denote the outcomes $+1$ and $\hat{A}_+$, $\hat{B}_+$ the POVM element associated to these outcomes respectively.
The predictions of the singlet state violate OI, since $ P(a_+,b_+|A,B,\lambda_0)\neq P(a_+|A,B,\lambda_0)P(b_+|A,B,\lambda_0)$. 
 Indeed, 
$P(a_+,b_+|A,B,\lambda_0)=\textrm{Tr}[\hat{A}_+\otimes\hat{B}_+\ket{\psi}\bra{\psi}]=0$, whereas
\begin{align}  
P(a_+|A&,B,\lambda_0)P(b_+|A,B,\lambda_0)=\sum_b P(a_+,b|A,B,\lambda_0) 
\sum_a P(a,b_+|A,B,\lambda_0)\nn\\&=\sum_j \textrm{Tr}[\hat{A}_+\otimes\hat{B}_j\ket{\psi}\bra{\psi}] \,
\sum_i \textrm{Tr}[\hat{A}_i\otimes\hat{B}_+\ket{\psi}\bra{\psi}]\nn\\&=
\textrm{Tr}[\hat{A}_+\otimes\1\ket{\psi}\bra{\psi}] \,
 \textrm{Tr}[\1\otimes\hat{B}_j\ket{\psi}\bra{\psi}]=
\textrm{Tr}[\hat{A}_+\rho^I]\textrm{Tr}[\hat{B}_+\rho^{II}]=\frac{1}{4}.
\end{align}
\end{itemize}
{\bf (II)} Maudlin's P1 and P2:\\
In order to evaluate P1 and P2 we need  to evaluate the probabilities  
$P(a|A,\lambda)$, $P(b|B,\lambda)$ via \eqref{qm2} and $P(a|A,B,b,\lambda)$, $P(b|A,B,a,\lambda)$ via \eqref{qm1}; but we need also evaluate $P(a|A,b,\lambda)$ and $P(b|B,a,\lambda)$. However, quantum mechanics, when considered as a stochastic hidden-variable theory, does not specify such latter probabilities. Quantum mechanics only specifies probabilities for outcomes given that one has chosen certain settings, i.e., it only allows one to calculate \eqref{qm1} and \eqref{qm2}. The theory does not specify probabilities for settings to be chosen, and we need these  to evaluate Maudlin's conditional probabilities $P(a|A,b,\lambda)$ and $P(b|B,a,\lambda)$, as we will now show.

Consider the big joint probability $P(a,b,A,B,\lambda)$. Note that this assumes the settings $A$ and $B$ to be random variables (ranging over some set $\Lambda_A$ and $\Lambda_B$ respectively). The relation
\begin{align}
\int_{\Lambda_B} dB \,P(a,b,A,B, \lambda)= &P(a,b,A,\lambda)= P(a|A,b,\lambda)P(b,A,\lambda)=\nn\\&p(a|A,b,\lambda)
\int_{\Lambda_B} dB \int_{\Lambda_a}da \,P(a,b,A,B, \lambda),
\end{align}
gives the sought after conditional probability
\begin{align}\label{cond}
P(a|A,b,\lambda)= \frac{\int_{\Lambda_B} dB \,P(a,b,A,B, \lambda)}{\int_{\Lambda_B} dB \int_{\Lambda_a}da \,P(a,b,A,B, \lambda)},
\end{align}
and analogous for $P(b|B,a,\lambda)$. 
We can furthermore write the joint probability $P(a,b,A,B,\lambda)$ as $P(a,b,A,B,\lambda)=P(a, b|A,B,\lambda)\rho(\lambda,A,B)$ by the law of conditional probability, where $\rho(\lambda,A,B)$ is a joint probability distribution of the hidden-variable $\lambda$ and settings $A,B$. 

Now we invoke quantum mechanics. This theory obeys IS, i.e., $\rho(\lambda,A,B)=\rho(\lambda)\rho(A,B)$ because the quantum state is independent of the settings chosen. The hidden variable $\lambda$ is again chosen to be the quantum state $\ket{\psi}$, i.e. $\rho(\lambda)= \delta(\lambda-\lambda_0)$ with $\lambda_0=\ket{\psi}$. In this way quantum mechanics gives \eqref{qm1}, \eqref{qm2} (where, without restriction, the outcomes are taken to be discrete so that they can be labeled by $i,j$ respectively). But this is not sufficient to evaluate \eqref{cond}. Quantum mechanics does not specify how to proceed any further, and, in order to do so we have to make some extra assumption about the probabilities $\rho(A,B)$ for settings to be chosen.

The extra assumption we adopt is that the observables can be chosen freely. We take this to imply two things. Firstly, that the observables measured on each subsystem are independent, i.e.,  $\rho(A,B)=\tilde{\rho}(A)\tilde{\tilde{\rho}}(B)$.  Secondly, the specific way outcomes $a_i$ are related to POVM elements  $\hat{A}_{i}$ is asymmetric: once a POVM is chosen the relationship between an outcome that can be obtained and its associated POVM element is uniquely determined, but if only an outcome is given, many POVM's can be associated to this outcome, as well as many POVM elements. All that matters is some unique labeling between POVM elements and outcomes. This can be chosen freely. But after it is chosen one should stick to it for consistency.

Let us now label the  POVM's by $x$ and its POVM elements by $j^x$, so a POVM is given by $\{\hat{B}_{j^x}\}$, with $\sum_{j^x} \hat{B}_{j^x}=\1$, $\forall x$. The distribution $\tilde{\tilde{\rho}}(B)$ gives a POVM $\{\hat{B}_{j^x}\}$ a weight $\gamma_x$, with  $\sum_x \gamma_x=1$.  
Since the outcomes are discrete, $\int_{\Lambda_a}da$ is a sum over $i$: $\sum_i$.  Also, since we only consider a given outcome $b_j$, and not some  particular observable,  we are free to chose which POVM is going to be measured and which POVM element we associate to this outcome, and thus  $\int_{\Lambda_B}dB$ is a sum over both $x$ and $j^x$: $\sum_x\sum_{j^x}$. 

This finally allows for rewriting \eqref{cond} as: 
\begin{align}\label{10}
P(a_i|b_j,A,\lambda)= \frac{\sum_x\sum_{j^x} \gamma_x\textrm{Tr}[\hat{A}_i\otimes\hat{B}_{j^x}\ket{\psi}\bra{\psi}]}{\sum_x\sum_{j^x} \sum_i \gamma_x
\textrm{Tr}[\hat{A}_i\otimes\hat{B}_{j^x}\ket{\psi}\bra{\psi}]}
\end{align}
Performing the summations\forget{(first over $i, j^x$ and then over $x$)}  gives:
\begin{align}
P(a_i|b_j,A,\lambda)= \frac{\sum_x \gamma_x\textrm{Tr}[\hat{A}_i\otimes\1\ket{\psi}\bra{\psi}]}{ \sum_x\gamma_x\textrm{Tr}[\1\otimes\1\ket{\psi}\bra{\psi}]}&=\textrm{Tr}[\hat{A}_i\otimes\1\ket{\psi}\bra{\psi}]\nn\\&=\textrm{Tr}[\hat{A}_i\rho^I]=P(a_i|A,\lambda)
\end{align}
\noindent
 This implies that P1 is obeyed: $P(a_i|A,b_j,\lambda)= P(a_i|A,\lambda)$. And, of course, by symmetry we obtain $P(b_j|B,a_i,\lambda_0)= P(b_j|B,\lambda)$.
 
 P2 is violated. Proof: Consider again the singlet state $\lambda_0=\ket{\psi}$. This state gives  $P(a_+|A,B,b_+,\lambda_0)=P(a_+,b_+|A,B,\lambda_0)/P(a_+|A,B,\lambda_0)=0$ whereas it is the case that\\ $P(a_+|A,b_+,\lambda_0)\overset{\textrm{P}1}{=}P(a_+|A,\lambda_0)=1/2$  so that $P(a_+|A,B,b_+,\lambda_0)\neq P(a_+|A,\lambda_0)$. Here we have had to use P1. 
 
 Analogously we obtain that $P(b_+|A,B,a_+,\lambda_0)\neq P(b_+|B,a_+,\lambda_0)$.

\newpage
\section{List of acronyms for this chapter}\label{acro}
\hskip0.5cm
\begin{tabbing}
\noindent
\medskip
\medskip
IS: Independence of the Source
\hskip0.6cm \=$\rho(\lambda|AB)=\rho(\lambda)$\\\\
AF: Apparatus Factorisability \>$\rho(\mu_A,\mu_{B}|\lambda,A,B)=\rho(\mu_A |\lambda,A,B)\,\rho(\mu_B|\lambda,A,B)$\\\\
AL: Apparatus Locality\>$\begin{array}{l}\negmedspace\negthickspace\rho(\mu_A|\lambda,A,B)=\rho(\mu_A|\lambda,A) \\
\negmedspace\negthickspace\rho(\mu_B|\lambda,A,B)=\rho(\mu_B|\lambda,B)
\end{array}$\\\\
TAF: Total Apparatus Factorisability\\\>
$ \rho(\mu_A,\mu_B|\lambda,A,B)=\rho(\mu_A | \lambda,A)\,\rho(\mu_B | \lambda,B)$\\\\
ISA: Independence of the Source and Apparata\\\> 
$\rho(\lambda,\mu_A,\mu_B|A,B)=\rho(\mu_A | \lambda,A)\,\rho(\mu_B |
\lambda,B)\rho(\lambda)$\\\\
%
LD: Local Determination\>$\begin{array}{l}\negmedspace\negthickspace a(A,B,\mu_{A},\mu_{B},\lambda)=a(A,\mu_{A},\lambda)\\
\negmedspace\negthickspace b(A,B,\mu_{A},\mu_{B},\lambda)=b(B,\mu_{B},\lambda)
\end{array}$\\\\
OL: Outcome Locality\>$\begin{array}{l}\negmedspace\negthickspace P(a | A,B,\mu_{A},\mu_{B},\lambda)=P(a|A,\mu_{A},\lambda)\\
\negmedspace\negthickspace P(b |A,B,\mu_{A},\mu_{B},\lambda)=P(b|B,\mu_{B},\lambda)\end{array}$
\\\\
OF: Outcome Factorisability\>$P(a,b | A,B,\mu_{A},\mu_{B},\lambda)=$\\
\>~~~~~~~~~~~~$P(a | A,B,\mu_{A},\mu_{B},\lambda)\,P(b | A,B,\mu_{A},\mu_{B},\lambda)$\\\\
TF: Total Factorisability\>$P(a,b | A,B,\mu_{A},\mu_{B},\lambda)=P(a | B,\mu_{A},\lambda)\,P(b |B,\mu_{B},\lambda)$ 
\\\\
%
PI: Parameter Independence\>${P}(a,b | A,B,\lambda)={P}(a | A,B,\lambda)\,{P}(b | A,B,\lambda)$\\\\
OI: Outcome Independence\>$\begin{array}{l}\negmedspace\negthickspace {P}(a | A,B,\lambda)={P}(a|A,\lambda)\\
\negmedspace\negthickspace {P}(b | A,B,\lambda)={P}(b|B,\lambda)\end{array}$\\\\
P1: Maudlin's OI\>$\begin{array}{l}\negmedspace\negthickspace P(a|A,b,\lambda)=P(a|A,\lambda)\\
\negmedspace\negthickspace P(b|B,a,\lambda)=P(b|B,\lambda)\end{array}$\\\\
P2: Maudlin's PI\>$\begin{array}{l}\negmedspace\negthickspace P(a|A,B,b,\lambda)=P(a|A,b,\lambda)\\
\negmedspace\negthickspace P(b|A,B,a,\lambda)=P(b|B,a,\lambda)\end{array}$\\\\
P3: \>$\begin{array}{l}\negmedspace\negthickspace P(a|A,B,b,\lambda)=P(a|A,\lambda)\\
\negmedspace\negthickspace P(b|A,B,a,\lambda)=P(b|B,\lambda)\end{array}$\\\\
Factorisability:\> ${P}(a,b | A,B,\lambda)={P}(a | A,\lambda)\, {P}(b |B,\lambda)$  
\end{tabbing}


\forget{OUDE STUK VOORDAT IK BEWIJS P1 EN P2 AANGEPAST HEB.

\subsubsection{Maudlin's and Shimony's conditions in quantum mechanics}

Quantum mechanics  can be considered a stochastic hidden-variable theory where the hidden-variable $\lambda$ is effectively the quantum state $\ket{\psi}$, i.e. $\rho(\lambda)= \delta(\lambda-\lambda_0)$ with $\lambda_0=\ket{\psi}$ (here the quantum state need not be a pure state).  The joint probabilities\footnote{These probabilities conditional on quantum states denote probabilities prescribed by those states. Although this commits us to probabilities for certain quantum states to be prepared, this can be easily removed by a reformulation where states are treated as parameters and not as random variables, cf. \cite[p. 5]{clifton}.} are then obtained via 
$P(a_i, b_j|A,B,\lambda_0)=\textrm{Tr}[\hat{A}_i\otimes\hat{B}_j\ket{\psi}\bra{\psi}]$, and the marginals are $P(a_i|A,\lambda_0)=\textrm{Tr}[\hat{A}_i\rho^I]$ and $P(b_j|B,\lambda_0)=\textrm{Tr}[\hat{B}_j\rho^{II}]$, with $\rho^I$ and $\rho^{II}$ the reduced density matrices for subsystems $I$ and $II$ respectively,  and we consider POVM's  $\{ \hat{A}_i\}$, and $\{\hat{B}_j\}$ with $\sum_i \hat{A}_i=\sum_j\hat{B}_j=\1$.

Considered as a stochastic hidden-variable theory quantum mechanics obeys\footnote{Proof: PI is obeyed since $\forall i,j$:   
$P(a_i|A,B,\lambda_0)=\sum_j \textrm{Tr}[\hat{A}_i\otimes\hat{B}_j\ket{\psi}\bra{\psi}]=\textrm{Tr}[\hat{A}_i\otimes\1\ket{\psi}\bra{\psi}]=
\textrm{Tr}[\hat{A}_i\rho^I]=P(a_i|A,\lambda_0)$, and analogously we find $P(b_j|A,B,\lambda_0)= P(b_j|B,\lambda_0)$.  P1 is obeyed since $\forall i,j $: $P(a_i|A,b_j,\lambda_0)= ?=P(a_i|A,\lambda_0)$  and $P(B_j|B,a_i,\lambda_0)= ?=P(b_j|B,\lambda)$  BEWIJS NOG GEVEN.

}  PI (also referred to as the 'quantum no-signaling theorem') as well as P1, but violates\footnote{Proof:
 Take $\ket{\psi}$ to be the singlet state $(\ket{+-}-\ket{-+})/\sqrt{2}$. In case $A$ and $B$ are chosen to be equal the singlet state predicts:  
$P(a_+,b_+|A,B,\lambda_0)=\textrm{Tr}[\hat{A}_+\otimes\hat{B}_+\ket{\psi}\bra{\psi}]=0$ and $P(a_+|A,\lambda_0)=\textrm{Tr}[\hat{A}_+\rho^I]=1/2$ and $P(b_+|B,\lambda_0)=\textrm{Tr}[\hat{B}_+\rho^{II}]=1/2$. Here $a_+$ and $b_+$ denote the outcomes $+1$ and $\hat{A}_+$  and $\hat{B}_+$ the POVM element associated to these outcomes respectively.

OI is violated since $ P(a_+,b_+|A,B,\lambda_0)\neq P(a_+|A,B,\lambda_0)P(b_+|A,B,\lambda_0)$. This follows because 
$P(a_+,b_+|A,B,\lambda_0)=\textrm{Tr}[\hat{A}_+\otimes\hat{B}_+\ket{\psi}\bra{\psi}]=0$ whereas
\begin{align} 
P(a_+|A&,B,\lambda_0)P(b_+|A,B,\lambda_0)=\sum_b P(a_+,b|A,B,\lambda_0) 
\sum_a P(a,b_+|A,B,\lambda_0)\nn\\&=\sum_j \textrm{Tr}[\hat{A}_+\otimes\hat{B}_j\ket{\psi}\bra{\psi}] \,
\sum_i \textrm{Tr}[\hat{A}_i\otimes\hat{B}_+\ket{\psi}\bra{\psi}]=
\textrm{Tr}[\hat{A}_+\otimes\1\ket{\psi}\bra{\psi}] \,
 \textrm{Tr}[\1\otimes\hat{B}_j\ket{\psi}\bra{\psi}]\nn\\&=
\textrm{Tr}[\hat{A}_+\rho^I]\textrm{Tr}[\hat{B}_+\rho^{II}]=1/4.
\end{align}
\forget{
\begin{align}
&P(a_+,b_+|A,B,\lambda_0)=\textrm{Tr}[\hat{A}_+\otimes\hat{B}_+\ket{\psi}\bra{\psi}]=0\\
&P(a_+|A,B,\lambda_0)P(b_+|A,B,\lambda_0)=\sum_b P(a_+,b|A,B,\lambda_0) 
\sum_a P(a,b_+|A,B,\lambda_0)\nn\\&~~~~~~~~~=\sum_j \textrm{Tr}[\hat{A}_+\otimes\hat{B}_j\ket{\psi}\bra{\psi}] \,
\sum_i \textrm{Tr}[\hat{A}_i\otimes\hat{B}_+\ket{\psi}\bra{\psi}]\\&~~~~~~~~~=
\textrm{Tr}[\hat{A}_+\otimes\1\ket{\psi}\bra{\psi}] \,
 \textrm{Tr}[\1\otimes\hat{B}_j\ket{\psi}\bra{\psi}]=
\textrm{Tr}[\hat{A}_+\rho^I]\textrm{Tr}[\hat{B}_+\rho^{II}]=1/4\nn.
\end{align}
}
 P2 is violated since  $P(a_+|A,B,b_+,\lambda_0)=P(a_+,b_+|A,B,\lambda_0)/P(a_+|A,B,\lambda_0)=0$ whereas $P(a_+|A,b_+,\lambda_0)\overset{\textrm{P}1}{=}P(a_+|A,\lambda_0)=1/2$  so that $P(a_+|A,B,b_+,\lambda_0)\neq P(a_+|A,\lambda_0)$, and analogously we obtain $P(b_+|A,B,a_+,\lambda_0)\neq P(b_+|B,a_+,\lambda_0)$. Here we use P1 which is indeed obeyed in quantum mechanics as was proven in the previous footnote.} OI and P2.
 
\citet[p. 95]{maudlin} remarks that  ``One might very well call P1 outcome independence and P2 parameter independence, since P1 concerns conditionalizing on the distant outcome and P2 on the distant setting''.
In Maudlin's terminology quantum mechanics violates his  parameter independence but obeys his outcome independence, which
is just the opposite from the analysis in terms of Shimony's concepts. Indeed, quantum mechanics violates
Shimony's outcome independence but obeys his parameter independence.

\subsubsection{Experimental metaphysics and consequences of the non-uniqueness}\label{expmeta}
Ever since Jarrett and Shimony  came up with their two conditions that together imply respectively TF and Factorisability, there has been a lively debate as to what violations of each of these two conditions entails. 
A common position in the literature is that failures of Factorisability should be understood  as a failure of OI rather than PI.  Otherwise, it is argued, the 
possibility of influencing the statistics of measurement outcomes on a system by manipulating a setting under our control on a distant system would, for a given $\lambda$, allow superluminal signaling between spacelike separate events which conflicts with relativity. It is furthermore argued that no such signaling is possible from a violation of OI because the outcomes are only constrained stochastically and are not under our control. 
Thus violations of OI are supposedly consistent with relativity and no-signaling, whereas violations of PI are not, and it is furthermore believed that  correlations that violate OI do not exhibit spacelike causation.   \citet[p. 226]{shimony84} referred to this position as one of `peaceful coexistence' between quantum mechanics and relativity: action at a distance (violation of PI) is avoided but we must accept a sort of a new sort of non-causal connection called  `passion at a distance' (violation of OI) \cite[p. 227]{shimony84}, cf. \cite{redhead}. Such an endeavor has been called `experimental metaphysics'\footnote{\citet[p. 296]{jonesclifton} characterize this endeavor as follows: 'First we demonstrate that any empirically adequate model of the Bell-type correlations which does not contain any superluminal signaling will have a particular formal feature. [\ldots] Then  we adduce an argument which purports to show that the formal feature in question is evidence of a certain metaphysical state of affairs. If this works we have a powerful argument from weak and general premises (namely, empirical adequacy and a ban on superluminal signaling) to a rich an momentous conclusion about the structure of the world'.}. Other have suggested different interpretations of the alleged violation of OI that have resulted in equally startling metaphysical conclusions\footnote{For example: the existence of holism of some stripe \cite{teller86,teller89, healey91};  incompleteness as a property of nature \cite[p. 700]{ballentinejarrett}; the necessity of broadening the classical concept of a localized event \cite[p. 30]{shimony1989}; adopting relative identity for physical individuals \cite[p. 250]{howard89}.}.

But the view that it is OI that must be abandoned whereas it is PI that must be retained has been challenged by several authors. We will not discuss this issue very extensively.  We merely review three arguments that exist in the literature against this position and therefore against the specific form of experimental metaphysics that takes this position as its starting point. Further, we will provide another in the next section.

{(I)} \citet[p. 95]{maudlin} comments on the discussion about whether it is OI or PI that should be abandoned, that ``the entire analysis is somewhat arbitrary'' because of the non-uniqueness of the conditions that imply Factorisation. Why focus on OI and PI and not on P1 and P2? Furthermore, focusing on P1 and P2 shows a different picture: P2 is violated by quantum mechanics thereby indicating a form of setting dependence instead of outcome dependence which  was the case with  violation of OI.  The `passion at a distance' has become `action at a distance'.  Because we have no reason to favor the distinction between OI and PI over the one between P1 and P2 the experimental metaphysics arguments are blocked.


}

\clearemptydoublepage
\thispagestyle{empty}
\chapter[Strengthened CHSH separability inequalities]{Strengthened CHSH separability\\\vskip0.2cm inequalities}
\label{chapter_CHSHquantumorthogonal}
\noindent
This chapter is largely based on \citet{uffseev}.

\section{Introduction}

The current interest in the study of entangled quantum states derives from two
sources: their role in the foundations of quantum mechanics \cite{entanglement} and
their applicability in practical problems of information processing such as quantum communication and computation \cite{nielsen}.

Bell-type inequalities likewise serve a dual purpose. Originally, they
were designed in order to answer a foundational question dealt with in the previous chapter: to test
the predictions of quantum mechanics against those of a local
hidden-variable (LHV) theory. However, these
inequalities also provide a test to distinguish entangled from
separable (unentangled) quantum states \cite{GISIN91,H3}. Indeed, experimenters
routinely use violations of a CHSH inequality to check whether
they have succeeded in producing bi-partite entangled states. This problem of
entanglement detection is crucial in all experimental applications
of quantum information processing.\forget{\\\\
Benadruk dat het een studie van de CHSH ineq. is voor quantum systemen, en dat we CHSH versterken.\\\\
}

It is the goal of this chapter to report that in the case of the standard CHSH
inequality experiment, i.e., for two distant spin-${1}/{2}$ particles,
significantly stronger inequalities hold for separable states in the case of
locally orthogonal observables. These inequalities provide sharper tools for
entanglement detection, and are readily applicable to recent experiments such as performed by \citet{volz} and \citet{stevenson}.  In fact,
if they hold for all sets of locally orthogonal observables they are necessary and sufficient for
separability, so the violation of these separability inequalities is not only a sufficient but also
a necessary condition for entanglement.
 They furthermore advance upon the necessary and sufficient sepa-\newpage\noindent rability inequalities of
\citet{hefei}, since, in contrast to these, the inequalities
presented here do not need to assume that the orientations of the
measurement basis for each qubit are the same, so no shared reference frame is necessary.

We show the strength and efficiency of the separability criteria by showing that they are stronger than other sufficient and experimentally accessible criteria for
two-qubit entanglement while using the same measurement settings. These are (i) the so-called fidelity
criterion \cite{sackett,seevuff}, and (ii) recent linear and non-linear
entanglement witnesses \cite{yu,nonlinear,zhang}.
However, in order to implement all of the above criteria successfully, the observables have to be chosen
in a specific way which depends on the state to be detected. So in general one needs some prior knowledge
about this state. In order to circumvent this experimental drawback we discuss the problem of whether
a finite subset of the separability inequalities could already provide a necessary and sufficient condition for
separability. For mixed states we have not been able to resolve this, but for pure states a set of six inequalities using only
three sets of orthogonal observables is shown to be already necessary and sufficient for separability.

The inequalities, however, are not applicable to the original purpose of
testing LHV theories. This shows that the purpose of testing
 entanglement within quantum theory, and the purpose of testing quantum
mechanics against LHV theories  are not equivalent, a point already
demonstrated by \citet{werner}. Our analysis follows up on Werner's
observation by showing that the correlations achievable by all separable
two-qubit states in a standard Bell experiment are tied to a bound strictly less
than those achievable for LHV models. In other words, quantum theory needs
entangled two-qubit states even to produce the latter type of correlations. As an
illustration, we exhibit a class of entangled two-qubit states, including the Werner
states, whose correlations in the standard Bell experiment possess a
reconstruction in terms of a local hidden-variable model.

This chapter is organized as follows.  In section 2, we rehearse the CHSH
inequalities for separable two-qubit states in the standard setting and derive a
stronger bound for orthogonal observables. In section 3, we compare this result
with that of LHV theories and argue that the stronger bound
does not hold in that case. In section 4, we return to quantum theory and
derive an even stronger bound than in section 2 which provides a necessary and sufficient criterion for
separability of all quantum two-qubit states, pure or mixed. Furthermore, it is shown
that the orientation of the measurement basis is not relevant for the criterion
to be valid. Section 5 compares the strength of these inequalities with some
other criteria for separability of two-qubit states, not based on Bell-type inequalities. Also, it is investigated whether
a finite subset of the inequalities of section 4 could already provide a necessary and  sufficient separability condition.
Section 6 summarizes our conclusions.

\section[Bell-type inequalities as a test for entanglement]{Bell-type inequalities as a test for\\ entanglement}
\noindent

 Consider a bipartite quantum system in the familiar setting of a standard Bell experiment: Two experimenters at distant sites each receive one subsystem and choose to measure  one of  two dichotomous observables: $A$ or $A'$ at the first site, and $B$ or $B'$ at the second.
 We assume that all  observables have the spectrum $\{-1,1\}$.  Let us consider the so-called CHSH operator \cite{braunstein} 
  \begin{align}\label{operator1}
  \mathcal{B}:= AB +AB' +A'B-A'B'.
 \end{align}
  We write $AB$ etc., as shorthand for $A\otimes B$ and $\av{AB}_\rho :=
\mbox{Tr}[\rho A\otimes B]$ or $\av{A B}_\Psi = \bra{\Psi}A \otimes B
\ket{\Psi}$ for the expectations\footnote{In this chapter, as well as in chapters \ref{chapter_CHSHquantumtradeoff} and \ref{Npartsep_entanglement},  we will leave out the subscript `qm'  in $\av{A}_\textrm{qm}$, etc., for ease of notation.} of $AB$ in the mixed state $\rho$ or pure
state $\ket{\Psi}$.

 Since $\av{\mathcal{B}}_\rho:=\mathrm{Tr}[\mathcal{B}\rho]$ is a convex function of the quantum state $\rho$ for the system, its maximum is attained for pure states.
In fact, \citet{cirelson} already proved that
$\max_{\rho} |\av{\mathcal{B}}_\rho|$  can be attained in a pure
two-qubit state (with associated Hilbert space
$\H=\mathbb{C}^2\otimes \mathbb{C}^2$) and  for spin observables.

In the following it  will thus suffice to consider only qubits
(spin-${1}/{2}$ particles) and the usual traceless spin
observables, e.g.\ $A=\bm{a}\cdot \bm{\sigma}=\sum_i
a_i\sigma_i$, with $\|\bm{a}\| =1$, $i=x,y,z$ and $\sigma_x,
\sigma_y, \sigma_z$ the familiar Pauli spin operators on the state space 
$\H=\mathbb{C}^2$, which has as a standard basis the set $\{\ket{0}, \ket{1}\}$ which are the spin-states for ``up'' and ``down'' in the $z$-direction of a single qubit.

\forget{
Consider a system composed of a pair of spin-${1}/{2}$
particles (qubits) on the Hilbert space $\mathcal{H}=\mathbb{C}^2 \otimes \mathbb{C}^2$  in the familiar setting of a standard Bell
experiment consisting of two distant sites, each receiving one of
the two particles, and where, at each site, a choice of measuring
either of two spin observables is made.
 Let $A,
A'$ denote the  two spin observables on the first particle, and $B, B'$ on the
second.
}

It is well known that for all such observables and all separable
states, i.e., states of the form $\rho= \rho_1\otimes\rho_2$ or
convex mixtures of such states (to be denoted as $\rho\in \mathcal{D}_{\textrm{sep}}$), the Bell
 inequality in the form derived by Clauser, Horne, Shimony and Holt (CHSH) \cite{chsh} 
holds:
 \begin{align}|\av{\mathcal{B}}_\rho|= |\langle AB + AB' + A'B -
A'B'\rangle_\rho|  \leq 2, ~~~\forall\rho\in\mathcal{D}_{\textrm{sep}}.   
\label{1} 
\end{align} The maximal
violation of (\ref{1})
 follows from an inequality by \citet{cirelson} 
 (cf. \cite{landau}) that holds for all quantum states (denoted as $\rho\in \mathcal{D}$):
 \begin{align} |\av{\mathcal{B}}_\rho|=|\langle AB + AB' +
A'B - A'B'\rangle_\rho | \leq 2 \sqrt{2}, ~~~\forall\rho\in\mathcal{D}. 
\label{2}
\end{align}
Equality in \eqref{2} --and thus the maximal violation of inequality
(\ref{1}) allowed in quantum mechanics-- is attained by e.g. the pure
entangled  states
$\ket{\phi^{\pm}}=\frac{1}{\sqrt{2}}(\ket{00} \pm
\ket{11})$ and
$\ket{\psi^{\pm}}=\frac{1}{\sqrt{2}}(\ket{01} \pm
\ket{10})$.

\citet{uffink} furthermore showed that for all such observables and for all states $\rho$ the following inequality must be obeyed:
\begin{align}
\av{AB' + A'B }_\rho^2 + \av{A B - A' B'}_\rho^2 \leq 4,~~~\forall\rho\in\mathcal{D},
\label{3}
\end{align}
which strengthens the Tsirelson inequality (\ref{2}). This quadratic inequality (\ref{3}) is
likewise saturated for maximally entangled states like $\ket{\psi^\pm}$ and
$\ket{\phi^\pm}$.
 Unfortunately, no smaller
bound  on the left-hand side of (\ref{3}) exists for separable
states, as long as the choice of observables is kept  general. (To
verify this, take $\ket{\Psi} = \ket{00}$ and $A =A'=B=B' =
\sigma_z$). Thus, the quadratic inequality (\ref{3}) does not
distinguish entangled and separable states.  We now show that a
much more stringent bound can be found  on the left-hand side of
(\ref{3}) for separable two-qubit states when a suitable choice of observables is
made, exploiting an idea used in a different context by \citet{tothseev}.

For the case of the singlet state $\ket{\psi^-}$, it has long been
known that an optimal choice of the spin observables for the
purpose of finding violations of the Bell inequality requires that
$A,A'$ and $B,B'$ are pairwise orthogonal, and many experiments
have chosen this setting. And for general states, it is only in
such locally orthogonal configurations that one can hope to attain
equality in inequality (\ref{2}) \cite{popescu169,wernerwolf,tradeoff}. It is  not true,
however, that for any given state $\rho$ the maximum of the left
hand side of the CHSH inequality always requires orthogonal
settings \cite{GISIN91,H3,POPROHR}.

However this may be, we will from now on assume local orthogonality,
 i.e., $A\perp A'$  and $B\perp
B'$ (for the case of two qubits this amounts to the local observables
anti-commuting with each other: $\{A,A'\}=0=\{B,B'\}$). Furthermore, assume for
the moment that the two-particle state is pure and separable. We may thus write
$\rho = \ket{\Psi}\bra{\Psi}$, where $\ket{\Psi} = \ket{\psi} \ket{\phi}$, to
obtain:
\begin{align} \av{AB' + A'B }_\Psi^2 + &\av{A B - A' B'}_\Psi^2 \nn\\&=
 \big(\av{A}_\psi\av{B'}_\phi + \av{A'}_\psi\av{B}_\phi \big)^2  +  \big(
\av{A}_\psi\av{ B}_\phi - \av{ A'}_\psi \av{ B'}_\phi\big)^2 \nn \\&
=\big( \av{A}_\psi^2 + \av{A'}_\psi^2 \big) \big(\av{B}_\phi^2 + \av{B'}_\phi^2\big). 
\label{4q}
 \end{align}
 Now, for any spin-$\half$  state $\rho$ on $\mathcal{H}=\mathbb{C}^2$, and any orthogonal triple of
spin components $A, A'$ and $A''$, one has
\begin{align}\label{quadratic1}
\av{A}_\rho^2 +\av{A'}_\rho^2 +\av{A''}_\rho^2\leq 1,
\end{align}
with equality for pure states only. 
Thus we have for any pure state $\ket{\psi}$ that  $\av{A}_\psi^2 +
\av{A'}_\psi^2 + \av{A''}_\psi^2 =1$, and similarly $\av{B}_\phi^2
+ \av{B'}_\phi^2 + \av{B''}_\phi^2=1$. Therefore, we can write (\ref{4q})
as: \begin{align} \av{AB' + A'B }_\Psi^2 + \av{A B - A' B'}_\Psi^2  =
\left(1 - \av{A''}_\psi^2\right)\left(1 - \av{B''}_\phi^2\right) .
\label{newa} \end{align} \forget{As a result, the left hand side of
(\ref{3}) is bounded by $1$ for any pure product state.} This
result for pure separable states can be extended to any  mixed
separable state by noting that the density operator of any such
state is a convex combination of the density operators for pure
product-states, i.e. $\rho = \sum_i p_i \ket{\Psi_i}\bra{\Psi_i}$,
with $\ket{\Psi_i} =\ket{\psi_i} \ket{\phi_i}$,  $p_i\geq 0$ and
$\sum_i p_i=1$.  We may thus write for such states:
\begin{align}   & \av{AB' + A'B }_\rho^2 + \av{A B - A' B'}_\rho^2 \leq\big(\sum_{i} p_i  \sqrt{\av{AB' + A'B }^2_i  +  \av{A B - A'B'}^2_i } \big)^2
\nn\\  & =   \big(\sum_i p_i \sqrt{ \big( 1 - \av{A''}^2_i\big) \big( 1 -
\av{B''}^2_i\big)}\big)^2   \leq \big( 1 - \av{A''}^2_ \rho\big) \big( 1 -
\av{B''}^2_\rho\big),~~~\forall\rho\in\mathcal{D}_{\textrm{sep}}.
\label{6q}
\end{align}
Here, $\av{\cdot}_i$ \  an expectation value with
respect to $\ket{\Psi_i}$ (e.g.,
$\av{A''}_i:=\bra{\Psi_i} A''\otimes\1\ket{\Psi_i}$) and
$\av{A''}_\rho:=\av{A''\otimes\1}_\rho$, etc.

The first inequality follows because
  $\sqrt{\av{AB' + A'B }^2_\rho  +  \av{A B - A'B'}^2_\rho }$ is a
convex function of $\rho$  and the second
because $\sqrt{ \big( 1 - \av{A''}^2_ \rho\big) \big( 1 - \av{B''}^2_\rho\big)}$ is concave
in $\rho$. (To verify this, it is helpful to use the general lemma that for all
positive concave functions $f$ and $g$, the function $\sqrt{fg}$ is concave.)

 Thus, we obtain for all two-qubit separable states and
locally orthogonal triples $A\perp A'\perp A''$, $B\perp B'\perp
B''$: \begin{align} \av{AB' + A'B }_\rho^2 + \av{A B - A' B'}_\rho^2 \leq
\left(1 - \av{A''}_\rho^2\right)\left(1 - \av{B''}_\rho^2\right),~~~\forall\rho\in\mathcal{D}_{\textrm{sep}}
. \label{New} \end{align}
Clearly, the right-hand side of this inequality is bounded by 1. However, as
noted\\ before, entangled states can saturate inequality (\ref{3}) -- even for
orthogonal observables -- and attain the value of 4  for the left-hand side and
thus clearly violate the bound (\ref{New}). In contrast to (\ref{3}), inequality
(\ref{New}) thus does provide a criterion for testing entanglement.    The
strength of this bound for entanglement detection in comparison with the
traditional CHSH inequality (\ref{1}) may be illustrated by considering
the region of values it allows  in the $(\av{X}_\rho,\av{Y}_\rho)$-plane, where
$\av{X}_\rho = \av{AB - A'B' }_\rho$ and $\av{Y}_\rho = \av{AB' + A'B }_\rho$,
cf.\ Fig. \ref{figxy}. Note that even in the weakest case, (i.e., if $\av{A''}_\rho=
\av{B''}_\rho=0$), it wipes out just over 60\% of the area allowed by
inequality (\ref{1}). This quadratic inequality even implies
 a strengthening of the CHSH inequality (\ref{1}) by a factor
$\sqrt{2}$:
\begin{align} |\av{\mathcal{B}}_\rho|=|\langle AB + AB' + A'B -
A'B'\rangle_\rho|  \leq \sqrt{2},~~~\forall\rho\in\mathcal{D}_{\textrm{sep}},   \label{sqrt2}
\end{align} recently
obtained by \citet{roy}. In fact, even if one chooses only one pair (say
$B,B'$) orthogonal, and let $A,A'$ be arbitrary, one would obtain an upper
bound of 2 in (\ref{New}), and still improve upon the CHSH inequality.
Another pleasant feature of inequality (\ref{New}) is that for pure states it's
violation, for all sets of local orthogonal observables, is a necessary and sufficient condition for entanglement (see
Appendix A on p. \pageref{appendix_orthoA}). Also, for future purposes we note that the expression in left-hand
side is invariant under rotations of $ A, A'$ around the axis $A''$ and
rotations of $B,B'$ around $B''$.

\begin{figure}[h]
\includegraphics[scale=1]{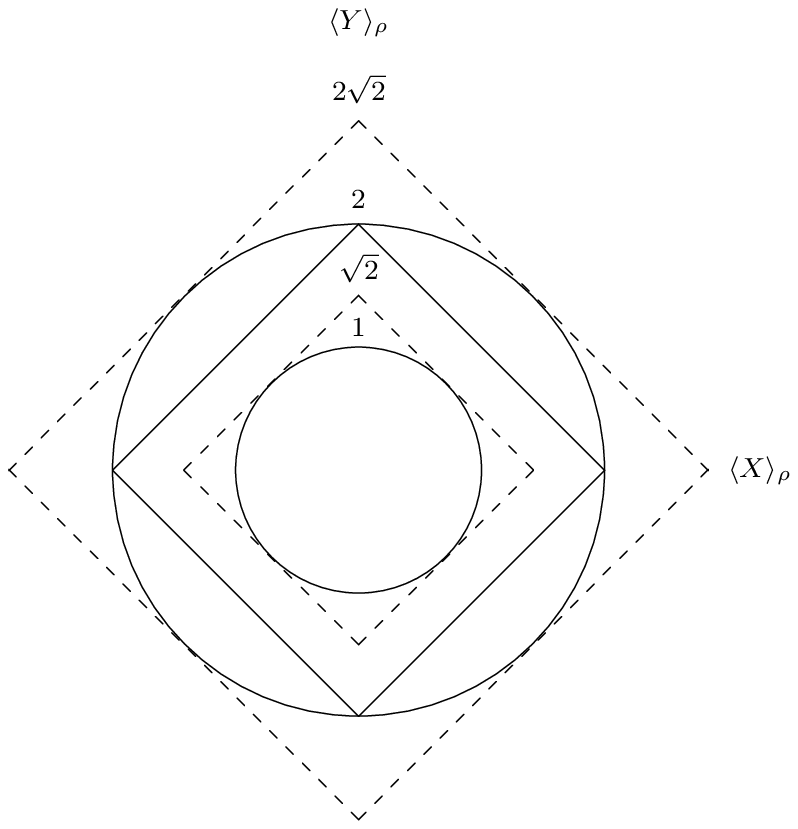}
\caption{
Comparing the regions in the
$(\av{X},\av{Y})$-plane allowed (i) by the tight bound  (\ref{3}) for all quantum states (inside the largest circle); (ii) 
by the CHSH bound (\ref{1}) for all separable states (inside the second largest tilted square); (iii)
by the stronger tight bound (\ref{New}) for separable states in case of usage of locally orthogonal observables  (inside the circle with radius $1$). The quadratic bounds give rise to the familiar Tsirelson bound (\ref{2}) (inside the largest tilted square; interrupted line);  and the linear bound (\ref{sqrt2}) (inside the smallest tilted square; interrupted line).}
\label{figxy}
\vspace{\baselineskip}
\end{figure} \forget{
 \setlength{\unitlength}{0.25 mm}
 \begin{figure}[h]
 \begin{center}
\begin{picture}(200,310)(-100,-125)
\linethickness{0.1mm} 
\multiput(0,-71)(6.8,6.8){11}{\line(1,1){3}}
\multiput(0,-71)(-6.8,6.8){11}{\line(-1,1){3}}
\multiput(0,71)(6.8,-6.8){11}{\line(1,-1){3}}
\multiput(0,71)(-6.8,-6.8){11}{\line(-1,-1){3}}
  \put(0,-100){\line(1,1){100}}
 \put(0,-100){\line(-1,1){100}}
  \put(0,100){\line(1,-1){100}}
 \put(0,100){\line(-1,-1){100}}
\multiput(0,-142)(6.63,6.63){22}{\line(1,1){3}}
\multiput(0,-142)(-6.63,6.63){22}{\line(-1,1){3}}
\multiput(0,142)(6.63,-6.63){22}{\line(1,-1){3}}
\multiput(0,142)(-6.63,-6.63){22}{\line(-1,-1){3}}
 \put(0,0){\circle{100}}
 \put(0,0){\circle{200}}
 \put(163,0){\makebox(0,0){\footnotesize{$\av{X}_\rho$}}}
 \put(0,182){\makebox(0,0){\footnotesize{$\av{Y}_\rho$}}}
 \put(0,82){\makebox(0,0){\footnotesize{$\sqrt{2}$}}}
  \put(0,58){\makebox(0,0){\footnotesize{$1$}}}
 \put(0,110){\makebox(0,0){\footnotesize{$2$}}}
\put(0,155){\makebox(0,0){\footnotesize{$2\sqrt{2}$}}}
\end{picture}\end{center}
\caption{
Comparing the regions in the
$(\av{X},\av{Y})$-plane allowed (i) by the tight bound  (\ref{3}) for all quantum states (inside the largest circle); (ii) 
by the CHSH bound (\ref{1}) for all separable states (inside the second largest tilted square); (iii)
by the stronger tight bound (\ref{New}) for separable states in case of usage of locally orthogonal observables  (inside the circle with radius $1$). The quadratic bounds give rise to the familiar Tsirelson bound (\ref{2}) (inside the largest tilted square; interrupted line);  and the linear bound (\ref{sqrt2}) (inside the smallest tilted square; interrupted line).}
\label{figxy}
\vspace{\baselineskip}
\end{figure}}

The inequalities (\ref{New}) present a necessary criterion for a
quantum state to be separable\footnote{
Note that using only two correlation terms (instead of four as in
(\ref{sqrt2})) one can already find a
separability criterion for the case of two qubits (also noted
by \citet{tothguhne1}), namely (\ref{New}) implies that $\forall\rho \in {\cal
D}^{2\textrm{-sep}}_2\,:\, | \av{AB - A'B' }_\rho|\leq 1/2$.
Violation of this inequality thus gives an entanglement criterion,
i.e., if $| \av{AB - A'B' }_\rho|> 1/2$ then $\rho$ is entangled. In
fact, a maximally entangled state can give rise to
$| \av{AB - A'B' }_\rho|=1$ a factor two higher than for separable
states. The same of course holds for the choice $\av{AB' + A'B }_\rho$.} --and its violation thus a
sufficient criterion for entanglement--, but in contrast to pure
states,  they are  clearly not sufficient for separability  of mixed
states. In section  \ref{necsuf} we shall present an even stronger
set of inequalities that is necessary and sufficient for mixed states as
well,  but we will first present an alternative form of Figure \ref{figxy} as well as discuss in section \ref{comparisonlhvN2} the results obtained so far  in the light of
LHV theories.

\subsection*{Figure \ref{figxy} in terms of the CHSH inequality}

We give another geometrical representation of the inequalities obtained so far. We believe it is easier to relate to than the representation in Figure \ref{figxy} because it is not in terms of the rather unfamiliar quantities $\av{X}_\rho$ and $\av{Y}_\rho$ but in terms of the expectation values of the more familiar CHSH operators  $\mathcal{B}$ and $\mathcal{B}'$, where $\mathcal{B}'=A'B'+AB'+A'B-AB$ (i.e., compared to $\mathcal{B}$ the primed and unprimed observables are interchanged). The alternative representation follows form the identity
\begin{align}\label{equiv}
\av{\mathcal{B}}_\rho^2+\av{\mathcal{B}'}_\rho^2= 2\av{A'B+AB'}_\rho^2+
2\av{AB-A'B'}_\rho^2\end{align}
 that  allows us to reformulate the inequalities of this section as follows.

\noindent
All states obey the quadratic bound  \eqref{3} which reads in terms of our reformulation\footnote{\citet{pitowsky08} has recently obtained an interesting similar inequality using a geometrical analysis:\\
 $\max_{A,A',B,B'}\,|-\av{\mathcal{B}}_\rho^2+\av{\mathcal{B}'}_\rho^2+\av{\mathcal{B}''}_\rho^2+\av{\mathcal{B}'''}_\rho^2|\leq 8,$ $\forall\rho\in\mathcal{D}$, with $\mathcal{B}''=AB -AB'+A'B+A'B' $  and $\mathcal{B}'''=AB+AB'-A'B+A'B'$.}:
\begin{align} \label{ufffig}
 \max_{A,A',B,B'}\,\av{\mathcal{B}}_\rho^2+\av{\mathcal{B}'}_\rho^2\leq 8,~~~\forall\rho\in\mathcal{D},
\end{align}
This implies the Tsirelson inequality of \eqref{2}:
\begin{align}\label{cirelsonfig}
\max_{A,A',B,B'}\,|\av{\mathcal{B}}_\rho|,|\av{\mathcal{B}'}_\rho|\leq 2\sqrt{2},~~~\forall\rho\in\mathcal{D}.
\end{align}
Separable states must obey the more stringent bound of \eqref{1}:
\begin{align}\label{chshfig}
\max_{A,A',B,B'}\,|\av{\mathcal{B}}_\rho|,|\av{\mathcal{B}'}_\rho|\leq 2,~~~\forall\rho\in\mathcal{D}_{\textrm{sep}}.
\end{align}
For orthogonal measurements we get the sharper quadratic inequality of \eqref{New}:
\begin{align}
\max_{A\perp A',B\perp B'}\av{\mathcal{B}}_\rho^2+\av{\mathcal{B}'}_\rho^2
\leq2,~~~\forall\rho\in\mathcal{D}_{\textrm{sep}},
\label{sharpsep}
\end{align} 
which in turn gives the linear inequalities \eqref{sqrt2}:
\begin{align}\label{sharpseplinear}
\max_{A\perp A',B\perp B'}\,|\av{\mathcal{B}}_\rho|,|\av{\mathcal{B}'}_\rho|\leq \sqrt{2},~~~\forall\rho\in\mathcal{D}_{\textrm{sep}}.
\end{align}
All these bounds are plotted in Figure \ref{chshfigure}. 

\begin{figure}[h]
\includegraphics[scale=1]{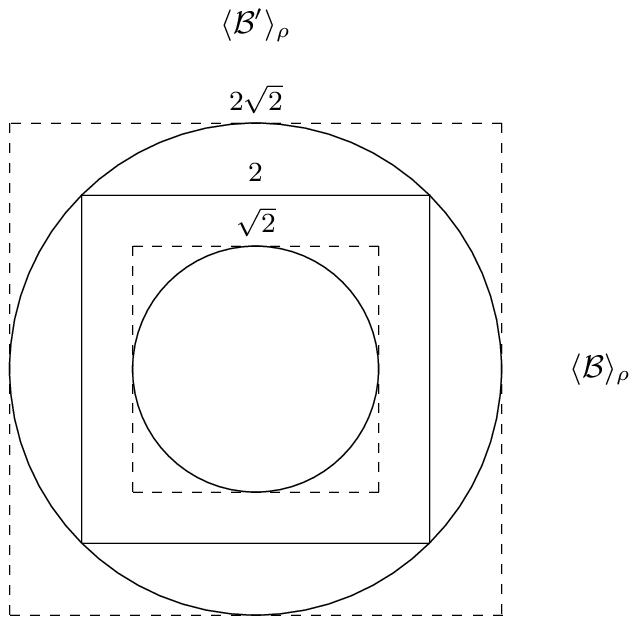}
\caption{Comparing the regions  in the $(\av{\mathcal{B}}_\rho,
\av{\mathcal{B}'}_\rho)$-plane. (i) by the tight bound  (\ref{ufffig}) for all quantum states (inside the largest circle); (ii) 
by the CHSH bound (\ref{chshfig}) for all separable states (inside the second largest square); (iii)
by the stronger tight bound (\ref{sharpsep}) for separable states in case of usage of locally orthogonal observables  (inside the circle with radius $1$). The quadratic bounds give rise to the familiar Tsirelson bound (\ref{cirelsonfig}) (inside the largest square; interrupted line);  and the linear bound (\ref{sharpseplinear}) (inside  the smallest square; interrupted line).}
\label{chshfigure}
\vspace{\baselineskip}
\end{figure} \forget{

 \setlength{\unitlength}{0.25 mm}
\begin{figure}[h]\begin{center}
\begin{picture}(200,310)(-100,-125)
\linethickness{0.1mm} 
\multiput(-100,-100)(0,8.17){25}{\line(0,1){4}}
\multiput(-100,-100)(8.18,0){25}{\line(1,0){4}}
\multiput(100,100)(0,-8.15){25}{\line(0,-1){4}}
\multiput(100,100)(-8.15,0){25}{\line(-1,0){4}}
 \put(-70.7,-70.7){\line(0,1){141.5}}
 \put(-70.7,-70.7){\line(1,0){141.2}}
 \put(70.7,70.7){\line(0,-1){141.2}}
\put(70.7,70.7){\line(-1,0){141.5}}
\multiput(-50,-50)(0,8){13}{\line(0,1){4}}
\multiput(-50,-50)(8,0){13}{\line(1,0){4}}
\multiput(50,50)(0,-8){13}{\line(0,-1){4}}
\multiput(50,50)(-8,0){13}{\line(-1,0){4}}
 \put(0,0){\circle{200}}
 \put(0,0){\circle{100}}
 \put(140,0){\makebox(0,0){$\av{\mathcal{B}}_\rho$}}
 \put(0,140){\makebox(0,0){$\av{\mathcal{B}'}_\rho$}}
 \put(0,110){\makebox(0,0){\footnotesize{$2\sqrt{2}$}}}
 \put(0,80){\makebox(0,0){\footnotesize{$2$}}}
   \put(0,60){\makebox(0,0){\footnotesize{$\sqrt{2}$}}}
 \end{picture}\end{center}
\caption{Comparing the regions  in the $(\av{\mathcal{B}}_\rho,
\av{\mathcal{B}'}_\rho)$-plane. (i) by the tight bound  (\ref{ufffig}) for all quantum states (inside the largest circle); (ii) 
by the CHSH bound (\ref{chshfig}) for all separable states (inside the second largest square); (iii)
by the stronger tight bound (\ref{sharpsep}) for separable states in case of usage of locally orthogonal observables  (inside the circle with radius $1$). The quadratic bounds give rise to the familiar Tsirelson bound (\ref{cirelsonfig}) (inside the largest square; interrupted line);  and the linear bound (\ref{sharpseplinear}) (inside  the smallest square; interrupted line).}
\label{chshfigure}
\vspace{\baselineskip}
\end{figure}}

\section{Comparison to local hidden-variable theories}\label{comparisonlhvN2}
\noindent
It is interesting to ask whether one can obtain a similar stronger
inequality as  (\ref{New}) in the context of local hidden-variable
theories. It is well known that inequality (\ref{1})  holds also
for any such theory in which dichotomous outcomes $a,b \in \{+, -\}$
are subjected to a probability distribution \begin{align} p(a,b) =
\int_\Lambda \!d\lambda\, \rho(\lambda) P_{\bm{a}}(a|\lambda)
P_{\bm{b}} (b|\lambda).\end{align} Here, $\lambda\in \Lambda$ denotes the
``hidden variable'', $\rho(\lambda)$ denotes a probability density
over $\Lambda$, $\bm{a}$ and $\bm{b}$ denote the `parameter settings', i.e., the directions of
the spin components measured, and $P_{\bm{a}}(a|\lambda)$, $P_{\bm{b}}(b|\lambda)$
are the probabilities (given $\lambda$) to obtain outcomes $a$ and $b$ when measuring the settings
$\bm{a}$ and $\bm{b}$ respectively. The locality condition is expressed by the
factorization condition $P_{\bm{a}, \bm{b} } (a,b|\lambda)
=P_{\bm{a}}(a|\lambda) P_{\bm{b}} (b|\lambda)$.

The assumption to be added to such an LHV theory in order to obtain
the strengthened inequality (\ref{New}) is the requirement that for any
orthogonal choice of $A, A'$ and $A''$ and for every given $\lambda$
we have the analog of (\ref{quadratic1}) which is \begin{align}
\av{A}_\textrm{lhv}^2 + \av{A'}_\textrm{lhv}^2  + \av{A''}_\textrm{lhv}^2 = 1,
\label{reqa}\end{align} or at least \begin{align} \av{A}_\textrm{lhv}^2 +
\av{A'}_\textrm{lhv}^2 \leq 1 \label{req},\end{align}
where $\av{A}_\textrm{lhv}
= \sum_{a=\pm1}  a\, P_{\bm{a}}\,(a|\lambda)$, \forget{$\av{A'}_{\rm hv}
= \int d \lambda \rho(\lambda) p_\bm{a'}(a'= +|\lambda -
p_\bm{a'}(a'= -|\lambda$; and similar for $\av{B}_{\rm
hv},\av{B'}_{\rm hv}$.} etc.

But a requirement like (\ref{reqa}) or (\ref{req}) is by no means
obvious for a local hidden-variable theory. Indeed, as has often
been pointed out, such a theory may employ a mathematical
framework which is completely different from quantum theory. There
is no \emph{a priori} reason why the orthogonality of spin
directions should have any particular significance in the
hidden-variable theory, and  why such a theory should confirm to
quantum mechanics in reproducing (\ref{req})  if one
conditionalizes on a given hidden-variable state. (One is reminded
here of Bell's critique \cite{bell66,bell71} on von Neumann's `no-go
theorem', cf. chapter \ref{definitionchapter}, section \ref{pitfall}.) Indeed, (\ref{req}) is violated by Bell's own example of
an LHV model \cite{bell64} and in fact it must fail in every
deterministic LHV theory
(where all probabilities $P_{\bm{a}}(a|\lambda)$, $P_{\bm{b}}(b|\lambda)$ are either $0$ or $1$), since for those theories
$\av{A}_\textrm{lhv}^2 =\av{A'}^2_\textrm{lhv} = \av{A''}^2_\textrm{lhv} =1$.
 Thus, the additional requirement (\ref{req})
would appear entirely  unmotivated  within an LHV theory.

It thus appears that testing for entanglement within quantum
theory and testing quantum mechanics against the class of all LHV
theories are not equivalent issues. Of course, this conclusion is
not new: \citet{werner} already constructed an
explicit LHV model for a specific two-qubit entangled state. Consider the
so-called Werner states: $ \rho_W = \frac{1-p}{4} \1   + p
\ket{\psi^-}\bra{\psi^-}$,   $p\in[0,1]$. \citet{werner} showed that
these states are entangled if \mbox{$p> {1}/{3}$}, but
nevertheless possess an LHV model for \mbox{$p= {1}/{2}$}.  The
above inequality (\ref{New}) suggests that the phenomenon
exhibited by this Werner state is much more ubiquitous, i.e., that
many more entangled two-qubit states have an LHV model. We will show that
this is indeed the case.

 It is not easy to find the general set of quantum states that
 possess an LHV model \cite{werner, acingisintoner}. Certainly, the question cannot be decided by considering orthogonal observables only.
However, as shown in Appendix B on p. \pageref{appendix_orthoB}, it is possible to determine the class of two-qubit states for which
 \begin{align}  \max_{A\perp A',
B\perp B'} \av{AB' + A'B }_\rho^2 + \av{A B - A' B'}_\rho^2 >1
\end{align} holds (they are thus entangled), and which in addition
satisfy the CHSH inequalities of \mbox{Eq. (\ref{1})} for
\emph{all} choices of observables, i.e., not restricted to
orthogonal directions.

Since the latter are known \cite{wernerwolf2,zukowskibrukner,fine,pitowsky}
to form a necessary and sufficient set of conditions for the existence
of an LHV model for all standard Bell experiments on spin-${1}/{2}$ particles,
 we conclude that all
correlations obtained from such entangled two-qubit states can be
 reconstructed by an LHV model\footnote{Note that experiments with more general measurement scenarios
(e.g., collective, sequential
or postselected measurements) might still produce
correlations incompatible with any LHV model.
 However, we will not discuss this issue.}.
It  follows from Appendix B on p.~\pageref{appendix_orthoB} that  this class of states includes
the Werner states for the region ${1}/{2} < p \leq {1}/{\sqrt2}$,
which complements results obtained by \citet{H3} in which the
non-existence of an LHV model is demonstrated for ${1}/{\sqrt2}< p
\leq1$.

It is important to realize that the above only holds for the case of qubits. The crucial relation (\ref{quadratic1}) can be violated for 
systems whose state space is a larger Hilbert space than the single qubit state space $\H=\mathbb{C}^2$.  The observables $A$, $A'$, $A''$ that are locally orthogonal spin observables  correspond to pairwise anti-commuting operators only in the case of a qubit. For systems with a large enough Hilbert space they can be commuting.  Simply choose this Hilbert space to be the direct sum of the eigenspaces of the three spin observables so that they do not have any overlap. Thus by choosing the Hilbert space of the systems under consideration to be large enough one can obtain commuting observables for any choice of observables. Using separable states   of a system consisting of two such systems one can, after all, reproduce the predictions of all LHV models. 

Thus one may take an experimental violation of the separability inequality (\ref{New}) (and its strengthened version, see the next section) using two-qubits to mean two things: (i)  either one can conclude that the state of the two-qubits is entangled, or (ii) the state might be separable but then one is not dealing with qubits after all and some degrees of freedom must have been overlooked.

\section[A necessary and sufficient condition for separability]{A necessary and sufficient condition for\\ separability \label{necsuf}}\noindent The inequalities
(\ref{New}) can be strengthened even further. To see this it is
useful to introduce, for some given pair of locally orthogonal
triples $(A, A', A'')$ and $(B,B',B'')$, eight new two qubit operators on $\mathcal{H}=\mathbb{C}^2\otimes\mathbb{C}^2$:
\begin{align}
I &:= \half (\1 + A'' B'') &
 \tilde{I} &:= \half (\1 - A'' B'' )
 \nn\\
X&:=   \half (A B - A' B' )
& \tilde{X}&:=   \half (A B  + A' B' )\nn\\
Y&:=   \half (A'B  +   AB'  )&
\tilde{Y}&:=   \half (A' B  -  AB' ) \nn\\
 Z &:=  \half (A'' + B'' ) & \tilde{Z}& :=  \half (  A'' -  B'' )\, ,
  \label{2operators}
   \end{align}
where   $ \half (A'' + B'' )$ is shorthand for $\half(A''\otimes\1 +\1\otimes B'')$, etc.
Note that $X^2 = Y^2 = Z^2 = I^2 =I$ and similar for their tilde versions,
  and that all eight operators mutually anti-commute.
Furthermore, if the orientations of the two triples is the same (e.g., $[A, A']=2iA''$ and $[B, B']=2iB''$), they form
   two representations of the generalized
Pauli-group, i.e. they have the same commutation relations as the
Pauli matrices on $\mathbb{C}^2$, i.e.: $ [X,Y]=2iZ$, etc., and $\av{X}^2+\av{Y}^2+\av{Z}^2=\av{I}^2$ (analogous for the tilde version).  Note that
these two sets transform in each other by replacing $B' \goesto -
B'$ and $B'' \goesto - B''$.

Now we can repeat the argument of section 2. Let us first
temporarily assume the state to be pure and separable, $\ket{\Psi}
=\ket{\psi}\ket{\phi}$.  We then obtain:
\begin{align}
\av{X}_\Psi^2 +
\av{Y}_\Psi^2 &= \frac{1}{4} \left( \av{AB -A'B'}_\Psi^2 +
\av{A'B + A B'}_\Psi^2\right) \nn\\
&=\frac{1}{4} \left( \av{A}_\psi^2 + \av{A'}_\psi^2 \right) \left(\av{B}_\phi^2 + \av{B'}_\phi^2\right) 
= \av{\tilde X}_\Psi^2 + \av{\tilde Y}_\Psi^2 \label{XY}
\end{align}
and similarly: \begin{align}
\av{I}_\Psi^2 - \av{Z}_\Psi^2& = \frac{1}{4} \left( \av{1 + A'' B''}_\Psi^2 - \av{A'' + B''}_\Psi^2 \right)\nn\\
&= \frac{1}{4} \left( 1 -\av{A''}_\psi^2 \right) \left( 1 - \av{B''}_\phi^2 \right)= \av{\tilde I}_\Psi^2  -\av{\tilde Z}_\Psi^2  \label{IZ}.\end{align}
In view of (\ref{newa}) we conclude that for all pure separable
states all expressions in the equations (\ref{XY}) and (\ref{IZ})
are equal to each other.   Of course, this conclusion does not
hold for mixed separable states.

 However,  $\sqrt{\av{ X}_\rho^2 +
\av{ Y}_\rho^2}$ and   $ \sqrt{\av{\tilde X}_\Psi^2 + \av{\tilde
Y}_\rho^2}$ are convex functions of $\rho$ whereas
 the three expressions $\sqrt{\av{I}_\rho^2 - \av{Z}_\rho^2}$, $\frac{1}{4}\sqrt{ \left(
1 -\av{A''}_\rho^2 \right)
 \left( 1 - \av{B''}_\rho^2\right)}$ and $\sqrt{\av{\tilde I}_\rho^2 - \av{\tilde Z}_\rho^2}$
 are all concave in $\rho$.
 Therefore we can repeat
 a similar
 chain of reasoning as in (\ref{6q}) to  obtain  the following inequalities, which are valid
for all mixed two-qubit separable states: \begin{align}\max \left\{
\begin{array}{c}
\av{X}_\rho^2 + \av{Y}_\rho^2 \\
\av{\tilde{X}}_\rho^2 + \av{\tilde{Y}}_\rho^2 \end{array} \right\}
\leq\min
\left\{ \begin{array}{c} \av{\tilde I}_\rho^2 - \av{\tilde Z}_\rho^2 \\
\frac{1}{4} \left( 1 -\av{A''}_\rho^2 \right)
 \left( 1 - \av{B''}_\rho^2\right)\\
 \av{I}_\rho^2 - \av{Z}_\rho^2 \end{array}
 \right\} ,~~~\forall\rho\in\mathcal{D}_{\textrm{sep}}.\label{mixsep} \end{align}

This result extends the previous inequality  (\ref{New}).
  The next obvious  question is then which of the three  right-hand
sides in (\ref{mixsep}) provides  the lowest  upper bound. It is
not difficult to show that the ordering of these three expressions
depends on the correlation coefficient $C_\rho = \av{A''B''}_\rho
- \av{A''}_\rho\av{B''}_\rho$. A straightforward calculation shows
that if $C_\rho \geq 0$, \begin{align} \av{I}^2_\rho - \av{Z}^2_\rho \leq \frac{1}{4}
\left( 1 -\av{A''}^2_\rho \right)
 \left( 1 - \av{B''}^2_\rho\right)  \leq \av{\tilde I}^2_\rho - \av{\tilde Z}^2_\rho \end{align}
 while the above inequalities are inverted  when $C_\rho \leq 0$.  Hence, depending on the sign of $C_\rho$, either $\av{I}^2_\rho
-\av{Z}^2_\rho$ or $\av{\tilde I}^2_\rho - \av{\tilde{Z }}^2_\rho$ yields the
sharper upper bound. In other words, for all separable two-qubit quantum
states one has:
\begin{align} 
\max\left\{\begin{array}{c}
\av{X}_\rho^2 + \av{Y}_\rho^2 \\
\av{\tilde{X}}_\rho^2 + \av{\tilde{Y}}_\rho^2 \end{array} \right\}
\leq\min
\left\{ \begin{array}{c} \av{\tilde I}_\rho^2 - \av{\tilde Z}_\rho^2 \\
 \av{I}_\rho^2 - \av{Z}_\rho^2 \end{array}
 \right\},~~~\forall\rho\in\mathcal{D}_{\textrm{sep}}.\label{mixsep2} \end{align}
For completeness we mention that the right hand side has an upper bound of $1/4$. This set of inequalities provides the announced strengthening of
(\ref{New}). This improvement pays off: in contrast to
(\ref{New}), the validity of the inequalities (\ref{mixsep2}) for all orthogonal triples $A,A',A''$ and $B,B',B''$ provides a necessary
and sufficient condition for separability for all two-qubit states, pure or
mixed. (See Appendix C on  p. \pageref{appendix_orthoC} for a proof). 

Furthermore, \eqref{mixsep2} detects entanglement of the Werner states 
$\rho= (1-p)\ket{\psi^-}\bra{\psi^-} + p\1/4$ (i.e, the singlet state mixed with a fraction $p$ of white noise) for $p<2/3$. Since the PPT criterion \cite{PPT,horodeckiPPT} gives the same bound and because it is necessary and sufficient for entanglement of two qubits, our criterion  \eqref{mixsep2} thus detects all entangled Werner states. It furthermore detects entanglement also if the singlet state is replaced by any other maximally entangled
state, a feature which is not possible using linear entanglement witnesses. 

We note that a special case of the inequalities (\ref{mixsep2}), to wit
\begin{align} \av{\tilde{X}}_\rho^2 + \av{\tilde{Y}}_\rho^2
\leq \av{I}_\rho^2 - \av{Z}_\rho^2  \label{hef} \end{align} was already found by \citet{hefei}, by a rather different argument. These authors stressed that
the orientation of the locally orthogonal observables play a crucial role in
this inequality: if one chooses both triples to have a \emph{different}
orientation (i.e., $A = i [A', A'']/2$ and $B = -i [B', B'']/2$ or $A =
-i [A', A'']/2$ and $B = i [B', B'']/2$) the inequality (\ref{hef}) holds
trivially for all quantum states $\rho$, whether entangled or not. It is only
when the orientation between those two triples is \emph{the same} that
inequality (\ref{hef}) can be violated by entangled  quantum states.

 The present result (\ref{mixsep2})
 complements their findings by showing that  the
relative orientation of the two triples is not a crucial factor in
entanglement detection. Instead, if the orientations are the same,
both of the following inequalities contained in (\ref{mixsep2}) 
\begin{subequations}
\begin{align} \av{X}_\rho^2 +
\av{Y}_\rho^2 &\leq \av{\tilde{I}}_\rho^2 -
\av{\tilde{Z}}_\rho^2 \label{21} \\
\av{\tilde X}_\rho^2 + \av{\tilde Y}_\rho^2  &\leq
\av{{I}}_\rho^2 - \av{{Z}}_\rho^2 \label{22} 
\end{align} 
\end{subequations}
are useful
tests for entanglement, while the remaining two become trivial. If on the other hand, the orientations are opposite,
their role is taken over by
\begin{subequations}
 \begin{align} \av{X}_\rho^2 + \av{Y}_\rho^2
&\leq \av{{I}}^2_\rho -
\av{{Z}}_\rho^2 \\
\av{\tilde X}_\rho^2 + \av{\tilde Y}_\rho^2 & \leq
\av{\tilde{I}}_\rho^2 - \av{\tilde{Z}}_\rho^2
 \end{align} 
 \end{subequations}
 while
(\ref{21}) and (\ref{22}) hold trivially.
\forget{\\\\
Hier in termen van Pauli-Operatoren en in dichtheidsmatrix formulatie?
}

\section{Experimental strength of the new inequalities}
\label{exp2ortho}
 In this section we compare
the strength of the inequalities (\ref{mixsep2}) to some other experimentally feasible
 conditions to distinguish separable and entangled two-qubit states that are not based on Bell-type inequalities.
 Also, we discuss the problem of whether a
finite set of triples for the inequalities (\ref{mixsep2}) could
be necessary and sufficient for separability.

A well-known alternative condition for separability of two qubit states is the
fidelity condition,  which says that for all separable states the
fidelity $F$ (i.e., the overlap with a Bell state
$\ket{\phi_\alpha^+} = \frac{1}{\sqrt{2}}(\ket{00} +
e^{i\alpha}\ket{11}$), $\alpha \in \mathbb{R}$) is bounded
as \begin{align}
 F (\rho) :=\max_\alpha
 \bra{\phi_\alpha^+}\rho\ket{\phi_\alpha^{+}}
=\frac{1}{2}(\rho_{1,1} +\rho_{4,4}) + |\rho_{1,4} |  \leq \frac{1}{2}. \label{fid}
\end{align}  Here, $\rho_{1,1}=\bra{00} \rho \ket{00}$, $\rho_{4,4}=\bra{11} \rho \ket{11}$  and 
$\rho_{1,4}$ denotes the extreme anti-diagonal element of $\rho$, i.e., $\rho_{1,4}=\bra{00} \rho \ket{11}$. 
For a proof, see \cite{sackett,seevuff}. An equivalent formulation  of (\ref{fid}), using
$\mbox{Tr}\rho =1$ is
 \begin{align}  2 | \rho_{1,4}|
\leq\rho_{2,2}+\rho_{3,3} \label{fid2}.\end{align}
here $\rho_{2,2}=\bra{01} \rho \ket{01}$ and $\rho_{3,3}=\bra{10} \rho \ket{10}$.   A second alternative condition, the Laskowski-\.Zukowski condition  \cite{laskowzukow}, states that separable states must obey $|\rho_{1,4}|\leq 1/4$.

However, choosing the Pauli
matrices for  both triples, i.e., $(A, A', A'') =\\( B,B', B'') =
(\sigma_x, \sigma_y , \sigma_z)$ we obtain  from (\ref{mixsep2})
 \begin{align} 
 \av{X}^2 + \av{Y}^2 \leq  \av{\tilde I}_\rho^2 -
\av{\tilde Z}_\rho^2 ~~~ \equivto ~~~|\rho_{1,4}|^2\leq \rho_{2,2}\rho_{3,3},~~~\forall\rho\in\mathcal{D}_{\textrm{sep}},
\label{matrixtwo}
\end{align} which strengthens both these alternative conditions as we will now show.

We use the trivial inequalities that hold for every state:  $(\sqrt{\rho_{2,2}} - \sqrt{\rho_{3,3}})^2\geq0\Longleftrightarrow2\sqrt{\rho_{2,2}\rho_{3,3}}\leq \rho_{2,2} +\rho_{3,3}$  and $| \rho_{1,4}|^2\leq\rho_{1,1} \rho_{4,4}$ (the latter follows from the semi-definiteness of every state).  Let us denote by the symbols $\overset{A}{\leq}$ and $\overset{\textrm{sep}}{\leq}$ inequalities that either hold for all two-qubit states or for states that are separable.  Then we obtain from (\ref{matrixtwo}) and the trivial inequalities that hold for all states that \begin{align}
4|\rho_{1,4}|- (\rho_{1,1}+\rho_{4,4}) \overset{A}{\leq}2 |\rho_{1,4}| \overset{\textrm{sep}}{\leq} 2\sqrt{\rho_{2,2}\rho_{3,3}}  \overset{A}{\leq}\rho_{2,2}+ \rho_{3,3}. 
\label{matrixtwocompare}
\end{align}
The strongest inequality is that between the second and third term which is (\ref{matrixtwo}). This inequality implies the inequalities that use the second and fourth term and the first and fourth term, which are the fidelity condition and the Laskowski-\.Zukowski condition respectively. Note that (\ref{matrixtwocompare}) also shows that the fidelity condition implies the Laskowski-\.Zukowski condition. Lastly, using the first and third term gives a new separability condition not mentioned before, but which is also weaker than (\ref{matrixtwo}). Violation of  (\ref{matrixtwo}) is thus the strongest entanglement criterion, i.e., it will detect more entangled
states than these other criteria.

As another application, consider
the following entanglement witnesses\footnote{An entanglement witness \cite{horodeckiPPT,terhal96,lewenstein,bruss02} is a self-adjoint operator $W$ that has (i) positive expectation value for all separable states, i.e,  $\av{W}_\rho\geq0,~ \rho\in \mathcal{D}_\textrm{sep}$, (ii) but that has
at least one negative eigenvalue. Thus if it is the case that $\av{W}_\rho<0$ then $\rho$ is entangled. Property (ii) ensures that every entanglement witness detects some entanglement, i.e., it detects the states in the eigenspace corresponding to the negative eigenvalue of $W$.} for so-called local orthogonal
observables (LOOs) $\{G_k^A\}_{k=1}^4$ and $\{G_k^B\}_{k=1}^4$: a linear one
presented by  \citet{yu}: \begin{align} \av{\mathcal{W}}_\rho = 1- \sum_{k=1}^4  \av{G_k^A
\otimes G_k^B}_\rho, \end{align} and  a nonlinear witness from \citet{nonlinear}
given by \begin{align} \mathcal{F}(\rho)=1-\sum_{k=1}^4\langle G_{k}^{A}\otimes
G_{k}^{B}\rangle_{\rho}-\frac{1}{2}\sum_{k=1}^4\langle
G_{k}^{A}\otimes \1-\1\otimes
G_{k}^{B}\rangle^{2}_\rho.\end{align}
Here, the set $\{ G_k^A\}_{k=1}^4$ is a set of four observables that form a basis
for all operators in the Hilbert space of a single qubit and which satisfy
orthogonality relations Tr$[G_kG_{k'}]=\delta_{kk'}$ ($k,k'=1,\dots 4$).
A typical complete set of LOOs is formed by any orthogonal triple of spin
directions  conjoined with the identity operator, i.e., in the notation of this paper,
 $\{ G_k^A\}_{k=1}^4 = \{ \1 ,  A, A', A''\}/\sqrt{2} $ and similarly    for $\{ G_k^B\}_{k=1}^4 $.

   These witnesses provide tests for two-qubit entanglement in the sense that for all separable two-qubit states
$   \av{\mathcal{W}}_\rho \geq  0 ,~ \mathcal{F}(\rho) \geq 0$ must hold
     and a violation of either of these inequalities is thus a sufficient condition for entanglement.
 An optimization procedure for the choice of LOOs in
these two witnesses is given by \citet{zhang}.

The strength of these two criteria has been studied for the noisy singlet state
introduced by \citet{nonlinear}: 
\begin{align}
\rho=p \ket{\psi^{-}} \bra{\psi^-}+(1-p)\rho_{sep}\,,
\end{align}  with
$\ket{\psi^{-}}=$\mbox{$(\ket{10}-\ket{01})/\sqrt{2}$} the singlet
state and the separable noise is $\rho_{sep}=(2 \ket{00} \bra{00} + \ket{01}\bra{01})/3$. The
Peres-Horodecki criterion \cite{PPT,horodeckiPPT} gives that this state $\rho$ is entangled for any
$p>0$. Under the complete set of LOOs
$\{-\sigma_{x},-\sigma_{y},-\sigma_{z},\1\}^{A}/\sqrt{2}$,
$\{\sigma_{x},\sigma_{y},\sigma_{z},\1\}^{B}/\sqrt{2}$, the linear
witness given above can detect the entanglement for all $p>0.4$
\cite{zhang}, and the nonlinear one detects the entanglement for
$p>0.25$ \cite{nonlinear}.
 Using the optimization procedure of  \citet{zhang} the
optimal choice of LOOs for the linear witness can detect the
entanglement for all $p>0.292$, whereas the nonlinear witness
appears to be already optimal.

Using the same set of LOOs as above, the quadratic separability inequality (\ref{mixsep2})
detects the entanglement already for $p>0$ (i.e., every entangled state is detected), and it is thus stronger
than these two witnesses for this particular state.

As a final topic, we wish to point out that, in spite of the
strength of the inequalities (\ref{mixsep2}), they also have an
important drawback from an experimental point of view as a
necessary and sufficient condition for separability.  In order to
check their validity or violation one would have to measure for
\emph{all} locally orthogonal triples of observables, a task which
is obviously unfeasible since there are uncountably many of those.
Because of this one must generally gather some prior knowledge about
the state whose entanglement is to be detected, so that one can choose settings that give a violation.
It is therefore highly interesting to ask whether a finite collection of orthogonal triples could be found for which the
satisfaction of these inequalities would already provide a
necessary and sufficient condition for separability, since then such prior knowledge would no longer be necessary. Measuring
the finite collection of settings would then be always sufficient for entanglement detection, independent of the state to be detected.

We have performed an (unsystematic) survey of this problem. A
first natural attempt would be to consider the triples obtained
by permutations of the basis vectors. Thus, consider the set of three
 inequalities obtained by taking for both  triples $(A, A',
A'')$ and $(B, B' , B'')$  the  choices $\alpha=(\sigma_x, \sigma_y
, \sigma_z)$, $\beta=(\sigma_z, \sigma_y, \sigma_x)$ and $\gamma=(\sigma_z, \sigma_x, \sigma_y)$. (Other permutations do not
contribute independent inequalities.)

Under this choice,  (\ref{mixsep2}) leads to
 the six inequalities 
 \begin{subequations}
 \label{6ineq}
 \begin{align} \av{X_k}_\rho^2 + \av{Y_k}_\rho^2
&\leq
\av{\tilde{I}_k}_\rho^2 - \av{\tilde{Z_k}}_\rho^2 , \\
\av{\tilde{X}_k}_\rho^2 + \av{\tilde{Y}_k}_\rho^2 &\leq
 \av{{I}_k}_\rho^2 -
\av{{Z_k}}_\rho^2 , 
\end{align}
\end{subequations}
for $k= \alpha,\beta,\gamma$.

For a general pure state $\ket{\Psi} = a \ket{00} + b \ket{01} + c
\ket{10} + d\ket{11}$, the satisfaction of these inequalities (\ref{6ineq}) 
boils down to three equations:
\begin{subequations}
\begin{align}  
 |ad|   &= |bc| , \\
|   (a+d)^2-  (b+c)^2 |   &=   |(a-d)^2   - (b-c)^2 |,\\
|(b+c)^2   +  (a-d)^2 | &= | (b-c)^2 +  (a+d)^2|. 
\end{align}
\end{subequations}
 However,  these equations are satisfied if $a =c =i$, $ -b= d =1$, i.e.
for an entangled pure state. This shows that the  choice
$\alpha$, $\beta$, $\gamma$ above does not produce a
 sufficient condition for separability.

However, let us make an amended choice $\beta'$: take the observables
$\beta$ and apply a rotation $U$ for the observables of particle 1 around
the $y$-axis over 45 degrees, i.e.  take $(A, A', A'')_{\beta'} = (
U\sigma_zU^\dagger , \sigma_y , U \sigma_x U^\dagger)$ and
$(B,B',B'')_{\beta'} = (\sigma_z, \sigma_y, \sigma_x)$
; and $\gamma'$: take the observables of choice $\gamma$ and
apply rotation $U$  on the observables for particle 1   (i.e., over 45 degrees around the
$y$-axis) followed up by rotation $V$ over 45 degrees around the $z$-axis on the same observables,   in other
words:  $(A,A',A''))_{\gamma'} = (VU\sigma_zU^\dagger V^\dagger,VU
\sigma_xU^\dagger V^\dagger ,VU \sigma_yU^\dagger V^\dagger)$ and
$(B,B',B'')_{\gamma'} = (\sigma_z, \sigma_x, \sigma_y)$.

  The choice $\alpha$, $\beta'$ and $\gamma'$  gives for the above arbitrary  pure
state $\ket{\Psi}$:
\begin{subequations}
\begin{align} |ad|   &= |bc| , \\
|   (a+c)  (b-d) |   &=   |(a-c)  (b+d) |,\\
 | (a+ i c) (b-id) | &= | (a -ic)(b+id)|. 
 \end{align}
 \end{subequations}
A tedious but straightforward calculation shows that these
equations are fulfilled \emph{only} if $ad =bc$, i.e., if
$\ket{\Psi}$ is separable.  Hence, by  measuring the observables
in the directions indicated by  the choice $\alpha$, $\beta'$ and $\gamma'$,
the inequalities  (\ref{mixsep2}) do
 provide a necessary and sufficient criterion for separability for
 pure two-qubit states.  We have  not been able to check whether this result extends
 to mixed states.

\section{Discussion}
It has been shown that for two spin-${1}/{2}$ particles (qubits)
and orthogonal spin components  quadratic separability inequalities hold
that impose much tighter bounds on the correlations in separable
states than the traditional CHSH inequality.
\forget{ This result provides
a quadratic entanglement witness that improves on all
existing witnesses derived from Bell inequalities.}
In fact, the quadratic inequalities (\ref{mixsep2}) are so strong that
their validity for all orthogonal bases is a necessary and sufficient condition
for separability of all states, pure or mixed, and a subset of these
inequalities for just three orthogonal bases (giving six inequalities) is a necessary and sufficient
condition for the separability of all pure states.
Furthermore, the orientation of the measurement basis is shown to be irrelevant,
which ensures that no shared reference frames needs to be established
between the measurement apparata for each qubit.

The quadratic inequalities
 (\ref{mixsep2}) have been shown to be stronger
 than both the fidelity criterion and the linear and non-linear entanglement witnesses based on LOOs as given by \citet{yu} and \citet{nonlinear}.
Experimental tests for entangled states using orthogonal directions can
therefore be considerably strengthened by means of the quadratic
inequalities (\ref{mixsep2}). As we will discuss
in chapter \ref{Npartsep_entanglement}, these inequalities provide tests of
entanglement that are much more robust against noise than many
alternative criteria. There we will also extend the analysis
to the $N$-qubit case by generalizing the method of
section \ref{necsuf} to more than two qubits.

Furthermore, we have argued that these quadratic Bell-type inequalities do
not hold in LHV theories.  This provides a more general example of
the fact first discovered by Werner, i.e.,  that some entangled
two-qubit states do allow an LHV reconstruction for all correlations in a
standard Bell experiment.
What is more, there appears to be a  `gap' between the
correlations that can be obtained by separable two qubit quantum states and
those obtainable by LHV models. This non-equivalence between the
correlations obtainable from separable two-qubit quantum states and from LHV
theories means that, apart from the question raised and answered
by Bell (can the predictions of quantum mechanics be reproduced by
an LHV theory?)
  it is also interesting to ask whether separable two-qubit quantum states
can reproduce the predictions of an LHV theory. The answer, as we
have seen, is negative:  quantum theory generally needs entangled two-qubit 
states even in order to reproduce the classical correlations of
such an LHV theory. In fact, as we will show in chapter \ref{Npartsep_entanglement}, the gap between the correlations allowed for by local
hidden-variable theories and those achievable by separable qubit
states increases exponentially with the number of particles.


\section*{Appendices}

\noindent
{\bf  Appendix A} \label{appendix_orthoA}---
Here we prove that any pure two-qubit state
satisfying (\ref{New}), for all sets of local orthogonal observables, must be separable.  By the bi-orthogonal
decomposition theorem, and following \citet{GISIN91}, any pure
state can be written in the form $ \ket{\Psi} =$ \mbox{$r\ket{10} -
s\ket{01}$}, with $r,s \geq 0$, $r^2 +s^2 =1$. For this
state $\av{\bm{a}\cdot \bm{\sigma} \otimes \bm{b}\cdot
\bm{\sigma}}_{\Psi} =-a_z b_z - 2 rs \left(a_x b_x +
a_yb_y\right)$, etc.
 Using this and choosing $\bm{a} =(0, \,0,\, 1),~\bm{a'}=(1, \,0,\, 0)$ and $
\bm{b} = (\sin\beta,\,0,\,\cos\beta),~\bm{b'} = (-\cos
\beta,\,0 ,\,\sin \beta)
$ we obtain $\av{AB' + A'B}^2 + \av{AB - A' B'}^2
= (1 + 2rs)^2$. This is the maximum value that can be obtained for any set of locally orthogonal observables. If (\ref{New}) holds, this
expression is smaller than or equal to 1, and
 it follows that $rs=0$, i.e., the state $\ket{\Psi}$  is not
entangled.
\\\\
\noindent
{\bf Appendix B} \label{appendix_orthoB}--- Here we provide further examples of entangled states that satisfy
the CHSH inequalities (\ref{1}) for all observables in the standard
Bell experiment. First note \cite{H3} that any two-qubit state can
be written in the form $\rho = \frac{1}{4} ( \1 \otimes \1 +
\bm{r}\cdot \bm{\sigma} \otimes \1 + \1\otimes \bm{s}\cdot
\bm{\sigma} + \sum_{ij=1}^3 t_{ij}\, \sigma_i \otimes \sigma_j )
$, where $ \bm{r} = \Tr\rho (\bm{\sigma}\otimes \1)$, $\bm{s} =
\Tr \rho (\1\otimes \bm{\sigma}) $  and $t_{ij} = \Tr \rho
(\sigma_i\otimes \sigma_j)$. By employing the freedom of choosing
local coordinate frames at both sites separately, we can bring the
matrix $(t_{ij})$ to diagonal form \cite{H2}, i.e., $t=
\mbox{diag~}(t_{11}, t_{22}, t_{33})$, and arrange that $t_{ii}
\geq 0$. Furthermore, since the labeling of the coordinate axes
is arbitrary, we can also pick an ordering such that $t_{11} \geq
t_{22}\geq t_{33}$.

Now let $\alpha, \alpha', \beta, \beta'$ denote two pairs of
arbitrary spin observables, for particle 1 and 2 respectively,
$\alpha = \bm{\alpha}\cdot \bm{\sigma} \otimes \1$, $ \beta =
\1 \otimes \bm{\beta} \cdot \bm{\sigma}$ and similar for the
primed observables. It is easy to see that
 the maximum of
 $|\av{\cal \alpha\beta + \alpha\beta' + \alpha'\beta - \alpha'\beta'}_\rho|$
 for all choices of observables  will
 be attained  by taking the vectors $\bm{\alpha}, \bm{\alpha'},
\bm{\beta},\bm{\beta'}$ coplanar \footnote{Here `coplanar' refers
to a single plane in the local frames of reference. Since these
frames may have a different orientation, this does not
necessarily refer to a single plane in real space.}, and in
fact, in the plane spanned by the two eigenvectors of $t$ with the
largest eigenvalues, i.e., $t_{11}$ and $t_{22}$. As shown by
\citet{H3}, this maximum is $
\max_{\alpha,\beta,\alpha',\beta'}|\av{ \alpha\beta +
\alpha\beta' + \alpha'\beta - \alpha'\beta'}_\rho|= 2
\sqrt{t_{11}^2 + t_{22}^2 }$. Thus $\rho$ will satisfy all
CHSH inequalities if $t^2_{11} + t^2_{22} \leq 1$, which is
the necessary and sufficient condition for the existence of an LHV
model \cite{zukowskibrukner}.

Now  consider  the maximum of $\av{ AB - A'B'}_{\rho}^2  + \av{AB' +
A'B}_\rho^2$, with $A\perp A'$ and $B \perp B'$. Clearly, these
spin observables  should be chosen in the same plane as before,
spanned by the eigenvectors corresponding to $t_{11}$ and
$t_{22}$. As mentioned in the text, the expression is invariant
under rotations of $A,A'$ or $B,B'$ in this plane. Choosing
$A = B =\sigma_x $, $A'= - B' = \sigma_y$ the maximum is equal to
$\max_{A\perp A', B\perp B'} \av{ AB - A'B'}^2 + \av{AB' + A'B}^2
 = (t_{11} + t_{22})^2. $
Clearly,  state $\rho$ will be  both entangled and satisfy
all CHSH inequalities for all observables
(and thus have an LHV description)  if $   t_{11} +
t_{22}>1$ and $t^2_{11} + t^2_{22} \leq 1$.
\\\\

\noindent
{\bf Appendix C} \label{appendix_orthoC}---
Here we will prove that any state $\rho$ that satisfies the inequalities (\ref{mixsep2}) for
all orthogonal triples $A, A', A''$, and  $B, B', B''$ must be separable (the converse has already been proven above).

We proceed from the well-known Peres-Horodecki lemma  \cite{PPT,horodeckiPPT}
that a state of two qubits is separable iff $\rho^{\rm PT}\geq 0 \label{PT}$ where 'PT' denotes partial transposition.
Equivalently, the state is entangled iff, for all pure states
$\ket{\Psi}$: \begin{align} \bra{\Psi} \rho^{\rm PT} \ket{\Psi} =\Tr
\rho^{\rm PT} \ket{\Psi}\bra{\Psi} = \Tr \rho
(\ket{\Psi}\bra{\Psi})^{PT} \geq 0 \label{PT'}.\end{align}
We shall show that (\ref{PT'}) holds whenever $\rho$ obeys (\ref{hef}). Indeed, according to the bi-orthonormal decomposition theorem (cf. \citet{GISIN91}), we can find bases
$\ket{0}, \ket{1}$ on ${\cal H}_1$ and $\ket{0},\ket{1}$ on ${\cal
H}_2$ such that \mbox{$\ket{\psi} =   \sqrt{p} \ket{01}  + \sqrt{1-p} \ket{10}$}.
Choosing these bases to be the eigenvectors of $A''$ and $B''$ respectively, we
thus find \begin{align}
 \ket{\Psi}\bra{\Psi} &= \half \tilde{I} + (p-\half)\tilde{ Z} +
 \sqrt{p(1-p)} \tilde{X}, \nn\\
\ket{\Psi}\bra{\Psi}^{PT} &= \half \tilde{I} + (p-\half)\tilde{
Z} +
 \sqrt{p(1-p)} X .\end{align}
Hence \begin{align} \bra{\Psi}\rho^{PT} \ket{\Psi} = \half \av{\tilde{I}} +
(p-\half) \av{\tilde{Z}} + \sqrt{p(1-p)} \av{X} ,\end{align} where the
last two terms can be bounded by a  Schwartz inequality to yield
\begin{align}  | (p-\half) \av{\tilde{Z}} + \sqrt{p(1-p)} \av{X}| \leq
\half \sqrt{\av{\tilde{Z}}^2  + \av{X}^2} \end{align} and we find $ \bra{\Psi}\rho^{PT} \ket{\Psi} \geq  \half \av{\tilde{I}}  -
\half \sqrt{\av{\tilde{Z}}^2  + \av{X}^2}$.  But (\ref{hef}) demands $ \av{X}^2_\rho + \av{\tilde{Z}}_\rho^2  \leq
 \av{\tilde{I}}^2_\rho $ from which it follows that \begin{align} \bra{\Psi}\rho^{PT} \ket{\Psi} \geq 0\end{align} so
that the state $\rho$ is separable.

\clearemptydoublepage
\thispagestyle{empty}
\chapter[Local commutativity and CHSH inequality violation]{Local commutativity and CHSH\\\vskip0.2cm inequality violation}
\label{chapter_CHSHquantumtradeoff}
\forget{\section{Abstract}
By introducing a quantitative `degree of commutativity' in terms
of the angle between spin-observables  we present  two tight
quantitative  trade-off relations in the case of two qubits:
First, for entangled states, between the degree of commutativity
of local observables  and the maximal amount of violation of the
Bell inequality:   if both local angles increase from zero to
$\pi/2$ (i.e., the degree of local commutativity decreases), the
maximum violation of the Bell inequality increases. Secondly, a
converse trade-off relation holds for separable states: if both
local angles approach  $\pi/2$  the maximal value obtainable for
the correlations in the Bell inequality decreases and thus the
non-violation increases.
  As expected, the extremes of these
relations are found in the case of anti-commuting local
observables where respectively the bounds of $2\sqrt{2}$ and
$\sqrt{2}$ hold for the expectation value of the Bell operator. The
trade-off relations show that non-commutativity gives ``a more
than classical result" for entangled states, whereas ``a less than
classical result" is obtained for separable states. The
experimental relevance of the trade-off relation for separable
states   is that it provides an experimental test for two-qubit
entanglement. Its advantages are twofold: in comparison to
violations of Bell inequalities it is a stronger criterion and in
comparison to entanglement witnesses
 it needs to make less strong assumptions about the  observables implemented in the
experiment.
}
This chapter is largely based on \citet{tradeoff}.
\section{Introduction}
\forget{
[Benadrukken dat dit de eis uit het vorige hoofdstuk van locale orthogonaliteit loslaat. Ook een studie van de CHSH ongelijkheid voor quantum systemen.]\\\\
}
\noindent
The previous chapter considered a strengthening of the CHSH inequality as a separability condition for the choice of locally orthogonal spin observables. In the case of qubits such a choice amounts to choosing anti-commuting observables.  In this chapter we relax this condition of anti-commutation of the local observables and study the bound on the CHSH inequality for the full spectrum of non-commuting observables, i.e., ranging from commuting to anti-commuting observables.  We provide analytic expressions for the bounds for both entangled and separable qubit states.

The CHSH inequality is satisfied for every separable quantum state, but may be violated by any pure entangled
state \cite{GisiN,GISIN91,POPROHR}. It is well-known that in order to achieve such a violation one
must make measurements of pairs of non-commuting spin-observables
for both particles, which we can take to be qubits. It is also well-known (thanks to the work of
 \citet{cirelson}) that in order to achieve the maximum
violation allowed for by quantum theory, one must choose both pairs of
these local observables to be anti-commuting. It is tempting to
introduce a quantitative `degree of commutativity' by means of the
angle between two spin-observables: if their angle is zero, the
observables commute; if their angle is $\pi/2$ they anti-commute,
which may thought of as the extreme case of non-commutativity.
Thus one may expect that there is a trade-off relation between the
degrees of local commutativity and the degree of CHSH inequality
violation, in the sense that if both local angles increase from 0
towards $\pi/2$ (i.e., the degree of local commutativity
decreases),  the maximum violation of the CHSH inequality
increases. It is one of the purposes of this chapter to provide a
quantitative tight expression of  this relation for arbitrary
angles.

It is less well-known that there is also a converse trade-off
relation for separable two-qubit states. For these states, the bound implied
by the CHSH inequality may be reached, but only if at least one of
the pairs of local observables commute, i.e., if at least one of
the angles is zero. It was shown in the previous chapter 
 that if both pairs anti-commute (i.e, are locally orthogonal), such states can
only reach a bound which is considerably smaller than the
bound set by the CHSH inequality, namely $\sqrt{2}$ instead  of
$2$. Thus, for separable two-qubit states there appears to be a trade-off
between local commutativity and CHSH inequality
\emph{non}-violation. The quantitative expression  of this
separability inequality was already investigated by
\citet{roy} for the special case when the local angles between
the spin observables are equal. It is a second purpose of this
chapter to report an improvement of this result and extend it to
the general case of unequal angles. As in the case of entangled
states mentioned above, the quantitative expression reported will
be tight.

Apart from the purely theoretical interest of these two trade-off
relations, we will show that the last one also has experimental
relevance. This latter trade-off relation is a separability
condition, i.e., it must be obeyed by all separable two-qubit states, and
consequently, a violation of this trade-off relation is a
sufficient condition for the presence of two-qubit entanglement.  Indeed,
this separability condition is strictly stronger as a test for
entanglement than the ordinary CHSH inequality whenever both
pairs of local observables are non-commuting (i.e., for
non-parallel settings).

Furthermore, since the relation is linear in the state $\rho$ it
can be easily formulated as an entanglement witness \cite{horodeckiPPT,terhal96,lewenstein,bruss02}
for two qubits in terms of locally measurable observables
\cite{guhnewitness,guhnewitness1}. It has the advantage, not shared by ordinary
entanglement witnesses \cite{horodeckiPPT,terhal96,lewenstein,bruss02,nonlinear,yu,zhang,guhnewitness,guhnewitness1}, that it
is not necessary that one has exact knowledge about the
observables one is implementing in the experimental procedure.
Thus,  even in the presence of some uncertainty about the
observables measured, the trade-off relation of this chapter allows
one to use an explicit entanglement criterion nevertheless.

The structure of this chapter is as follows. Before presenting the
trade-off relations in section \ref{tradeoffsection} we will
review some requisite background  in  section \ref{review}. In
section \ref{discussion} we will discuss the import of the
relations obtained.

\section{CHSH inequality and local commutativity}\label{review} 
\noindent
Consider again a bi-partite quantum system in the familiar setting of a standard Bell experiment.   Let us further recall the CHSH operator
 \begin{align}\label{operator}
  \mathcal{B}:= A(B +B') +A'(B-B'),
 \end{align} and that in order to obtain the quantum bounds of $\av{\mathcal{B}}_\rho$ it suffices to consider only qubits
and the usual traceless spin
observables, e.g.\ $A=\bm{a}\cdot \bm{\sigma}=\sum_i
a_i\sigma_i$, with $\|\bm{a}\| =1$, $i=x,y,z$ and $\sigma_x,
\sigma_y, \sigma_z$ the familiar Pauli spin operators on
$\H=\mathbb{C}^2$.

In the previous chapter the bounds on the CHSH operator that hold for separable and entangled states have been shown. 
For convenience they will be repeated here.  For the set $\mathcal{D}_{\textrm{sep}}$ of all separable states the bound is 
 \begin{align}  \label{chshineq}
     | \langle\mathcal{B}\rangle_\rho|  \leq 2.
 \end{align}
However, for the set $\mathcal{D}$ of all (possibly entangled)
quantum states \citet{cirelson} (cf. \citet{landau}) showed that  \begin{align} \label{Tsi}
|\av{\mathcal{B}}_\rho|\leq \sqrt{4+|\av{[A, A'] \otimes
[B,B']}_\rho|}\,\,, \end{align}
which, has a numerical upper bound of $2\sqrt{2}$ (cf. (\ref{2})).

\subsection{Maximal violation requires local anti-commutativity}\label{IIA}\noindent
The Tsirelson inequality  \Eq{Tsi} tells us that the only way to
get a violation of the CHSH inequality  (\ref{chshineq}) is when both
pairs of local observables are non-commuting: If one of the two
commutators in \Eq{Tsi} is zero  there will be no violation of
\eqref{chshineq}.  Furthermore, we see from \Eq{Tsi} that in order to maximally
violate inequality \Eq{chshineq} (i.e., to get
$|\av{\mathcal{B}}_\rho|= 2\sqrt{2}$) the following condition must
hold \cite{cirelson, tonerverstraete}: \begin{align}
|\av{[A,A']\otimes[B,B']}_\rho|=4.
\label{condition2} 
\end{align} The  local observables $i[A,A']/2$ and
$i[B,B']/2$ (which are both dichotomous and have their spectra
within $[-1,1]$) must thus be maximally correlated.

However, the condition \Eq{condition2} is only necessary for a
maximal violation, but not sufficient.
 Separable states are also able to obey this condition while  such states never violate the CHSH inequality.
For example, choose $A=B=\sigma_y$, $A'=B'=\sigma_x$. This gives
$[A,A'] \otimes [B,B'] = - 4 \sigma_z \otimes \sigma_z$. The
condition \Eq{condition2} is then satisfied  in the separable two-qubit 
state $(\ket{00}\bra{00}+
\ket{11}\bra{11})/2$ in the $z$-basis.

Nevertheless, we can infer from \Eq{condition2}  that for maximal
violation the local observables must anti-commute, i.e.,
$\{A,A'\}=\{B,B'\}=0$ (a result already obtained in a different
way by \citet{popescu169}).
  To see this, consider local qubit observables, which are not necessarily anti-commuting and note that
  $ i [A, A']/2 = - ( \bm{a} \times
\bm{a'})\cdot \bm{\sigma} $  and analogously $i[B,B]/2= -  (
\bm{b} \times \bm{b'})\cdot \bm{\sigma} $. We thus get
\begin{align}\label{angleres} |\av{[A, A'] \otimes [B',B]}_\rho|=
4|\av{(\bm{a} \times \bm{a'})\cdot\bm{\sigma}\otimes(\bm{b}
\times \bm{b'})\cdot\bm{\sigma}}_\rho       |. \end{align} This can
equal 4 only if
$||\,\bm{a}\times\bm{a'}||=||\,\bm{b}\times\bm{b'}||=1$, which
implies that  $\bm{a}\cdot\bm{a'}=0$ and
$\bm{b}\cdot\bm{b'}=0$, since  $\bm{a}$, $\bm{a'}$, $\bm{b}$
and $\bm{b'}$ are unit vectors.

If we denote by $\theta_A$ the angle between observables $A$ and
$A'$ (i.e., $\cos\theta_A=\bm{a}\cdot\bm{a'}$) and analogously
for $\theta_B$, we see that the local observables must thus be
orthogonal: $\theta_A=\theta_B=\pi/2$ (mod $\pi$), or
equivalently, they must anti-commute. Thus the condition
\Eq{condition2}   implies that we need locally anti-commuting
observables to obtain a maximal violation of the CHSH
inequality.

 As mentioned in the introduction, local commutativity  (i.e.,  $[A,A']=[B,B']=0$) corresponds to the observables  being  parallel or anti-parallel, i.e., $\theta_A =\theta_B =0$ (mod.\ $\pi$), and  local  anti-commutativity (i.e.,  $\{A,A'\}=\{B,B'\}=0$) corresponds to the observables  being orthogonal, i.e., $\theta_A, \theta_B=  \pm\pi/2$.
Therefore, in order to obtain any violation at all it is necessary
that the local observables are at some angle to each other, i.e.,
$\theta_A\neq0, ~\theta_B\neq0$, whereas maximal violation is only
possible if the local observables are orthogonal.

This suggests that there exists a quantitative trade-off relation
that expresses exactly how the amount of violation depends on the
local angles $\theta_A,~\theta_B$ between the spin observables. In
other words, we are interested in determining the form of \begin{align}
C(\theta_A, \theta_B) :=  \max_{\rho \in \cal{D}}
|\av{\mathcal{B}}_\rho |\end{align}
 In the next section we
 will present such a relation.

However, before doing so, we continue our review for the case of
separable two-qubit states.  In this case, a more stringent bound on
the  expectation value of the CHSH operator is obtained than the
usual bound  of 2.

\subsection{Local anti-commutativity and separable states}\label{IIB}\noindent
 Using the quadratic separability inequality (\ref{mixsep}) of the previous chapter for anti-commuting observables
 ($\{A,A'\}=\{B,B'\}=0$),  the identity (\ref{equiv})  and the definitions of  (\ref{2operators}) we get for all two-qubit states in  $
\mathcal{D}_{\textrm{sep}}$:
\begin{align}\label{quadr}
\av{\mathcal{B}}_\rho^2 +\av{\mathcal{B}'}_\rho^2\leq
2[\av{\1\otimes \1 -A''\otimes B''}^2_\rho -
\av{A''\otimes \1 -\1\otimes
B''}^2_\rho], \end{align} where $\mathcal{B}'$ is the same as
$\mathcal{B}$ but with the local observables interchanged (i.e.,
$A \leftrightarrow A'$ , $B \leftrightarrow B'$), and where we
have also used the shorthand notation $A''=i[A,A']/2$ and
$B''=i[B,B']/2$. Note that the triple $A,A',A''$ are mutually
anti-commuting and can thus be easily extended to form a set of local orthogonal
observables for $\mathbb{C}^2$ (so-called LOO's
\cite{nonlinear,yu,zhang}).

The separability inequality  (\ref{quadr}) provides a very strong
entanglement criterion, as was shown in the previous chapter, but it is here used to
derive a (weaker) separability inequality in terms of the Bell
operator $\mathcal{B}$ for all two-qubit states in $
\mathcal{D}_{\textrm{sep}}$: \begin{align}\label{Tsisep}
|\av{\mathcal{B}}_\rho| \leq
\sqrt{2(1-\frac{1}{4}|\av{[A,A']}_{\rho_1}|^2)(1-\frac{1}{4}|\av{[B,B']}_{\rho_2}|^2)}.
\end{align} Here $\rho_1$ and $\rho_2$ are the reduced single qubit
states that are obtained from $\rho$ by partial tracing over the
other qubit. The inequality \Eq{Tsisep} is the separability
analogue for anti-commuting observables of the Tsirelson
inequality \Eq{Tsi}. Note that even in the weakest case  ($
\av{[A,A']}_{\rho_1} = \av{[B,B]}_{\rho_2} =0 $) it implies
$|\av{\mathcal{B}}_\rho|\leq \sqrt{2}$, which is the strengthening of the
original CHSH inequality (\ref{chshineq}) already shown in the previous chapter. Thus, for
separable states, a reversed effect of the requirement of local
anti-commutativity appears than for entangled quantum states.
Indeed, for locally anti-commuting observables we deduce from
\Eq{Tsisep} that the maximum value of $\av{\mathcal{B}}_\rho$ is
considerably less than the maximum value of $2$ attainable using
commuting observables. In contrast to entangled states, the
requirement of anti-commutativity, which, as we have seen, is
equivalent to local orthogonality of the spin observables,  thus
decreases the maximum expectation value of the CHSH operator
$\mathcal{B}$ for separable two-qubit states.

An interesting question is now: what happens to the maximum
attainable by separable two-qubit states for locally non-commuting
observables that are not precisely anti-commuting? Or put
equivalently, how does this bound depend on the angles between the
local spin observables when the observables are neither parallel
nor orthogonal? From the above one would expect the bound to drop
below the standard bound of $2$ as soon as  the settings are not
parallel or anti-parallel. Just as in the case of general quantum
states it would thus be interesting to get a quantitative
trade-off relation that expresses exactly how the maximum bound
for $\av{\mathcal{B}}_{\rho}$ depends on the local angles of the
spin observables. In other words, we need to establish  \begin{align}
D(\theta_A, \theta_B) :=  \max_{\rho \in
\cal{D}_{\textrm{sep}}} |\av{\mathcal{B}}_\rho|, \end{align}
from which we obtain the separability inequality
\begin{align}\label{sepineq} |\av{\mathcal{B}}_\rho|\leq D(\theta_A,
\theta_B), ~~~\forall\rho \in \cal{D}_{\textrm{sep}}. \end{align}
 In the following we present such a tight trade-off relation.

\section{Trade-off relations}\label{tradeoffsection} \subsection{General qubit states}\noindent

 \noindent
It was already pointed out by \citet{landau} that inequality
(\ref{Tsi}) is tight, i.e.,  for all choices of the observables,
there exists a two-qubit state $\rho$ such that : \begin{align}
  \max_{\rho\in \mathcal{D} }
|\langle\mathcal{B}_\rho\rangle|=
 \sqrt{4+|\av{[A, A'] \otimes
[B',B]}_\rho|}. \label{tight}\end{align} This maximum is invariant under
local unitary transformations $U\otimes U'$, since Tr$[(U\otimes
U')^\dagger\mathcal{B}(U\otimes U')\rho]=$
Tr$[\mathcal{B}\tilde{\rho}]$ with $\tilde{\rho}=(U\otimes
U')\rho(U\otimes U')^\dagger$. This invariance amounts to a
freedom in the choice of the local reference frames.

Hence, without loss of generality, we can  choose \begin{align} \bm{a}&
=(1,0,0),~\bm{a'}=( \cos \theta_A ,  \sin \theta_A ,
0),\nonumber\\ 
\bm{b}& =(1,0,0),~\bm{b'}=( \cos \theta_B ,  \sin
\theta_B , 0).\label{choiseobs} \end{align}  This  choice  
 gives  $ i [A, A']/2 =  - \sin \theta_A
~\sigma_z $
   and, analogously,
$i[B,B']/2= - \sin \theta_B  ~\sigma_z $. Hence,   we immediately
obtain \begin{align}\label{Tsiangles}
  \max_{\rho\in \mathcal{D} }
|\langle\mathcal{B}_\rho\rangle| = \sqrt{4+ 4 |\sin \theta_A \sin
\theta_B \av{\sigma_z\otimes\sigma_{z}}_\rho|}. \end{align} To obtain a
state independent bound, it remains to be shown that we can choose
$\rho$ such that $|\av{\sigma_z\otimes\sigma_{z}}_\rho|=1$ in
order to conclude  that \begin{align} C(\theta_A , \theta_B)  = \sqrt{4+ 4
|\sin \theta_A \sin \theta_B|}.   \label{C} \end{align}

To see that  (\ref{C}) holds,  note that the CHSH operator for the
above choice (\ref{choiseobs}) of observables becomes:
\begin{align}\mathcal{B}=
\alpha\ket{00}\bra{11}
+\beta\ket{01}\bra{10} +
\alpha^*\ket{10}\bra{01}
+\beta^*\ket{11}\bra{00}, \end{align}
with \begin{align} \alpha&= 1+e^{-i \theta_A}+e^{-i \theta_B}-e^{-i
(\theta_A+\theta_B)}\label{A},\\ \beta&=1+e^{-i \theta_A}+e^{i
\theta_B}-e^{-i (\theta_A-\theta_B)}\label{BB}. \end{align}
 We distinguish two cases: (i) when
$\sin \theta_A \sin \theta_B \geq 0 $  (i.e. when
  $0\leq \theta_A, \theta_B\leq \pi$
 or $\pi \leq \theta_A,\theta_B\leq 2 \pi$),  choose the pure  state $\ket{\phi_{\tau}^+}=\frac{1}{\sqrt{2}}(\ket{00}
+e^{i\tau}\ket{11})$. Then:
\begin{align}
\max_{\tau}\,\mathrm{Tr}[\mathcal{B}\ket{\phi_{\tau}^+}\bra{\phi_{\tau}^+}]=
 \max_{\tau}[\mathrm{Re}(\alpha)
\cos\tau+\mathrm{Im}(\alpha)\sin\tau]
= |\alpha|=
\sqrt{4+4\sin\theta_A\sin\theta_B}.
\end{align}
Similarly,  (ii)  for   $\sin \theta_A \sin\theta_B \leq 0$ (i.e.,
$0\leq \theta_A \leq \pi$, $\pi \leq \theta_B\leq 2 \pi$ or
$\pi\leq \theta_A \leq 2 \pi$, $0 \leq \theta_B \leq \pi$), and
the pure state $
\ket{\psi_\tau^+}=\frac{1}{\sqrt{2}}(\ket{01}
+e^{i\tau}\ket{10})$ we find
\begin{align}
\max_{\tau}\,\mathrm{Tr}[\mathcal{B}\ket{\psi_\tau^+}\bra{\psi_\tau^+}]=
 \max_{\tau}[\mathrm{Re}(\beta)
\cos\tau+\mathrm{Im}(\beta)\sin\tau] =
|\beta|=\sqrt{4-4\sin\theta_A\sin\theta_B}.
\end{align}
Since
$|\av{\sigma_z\otimes\sigma_{z}}_{\phi_{\tau}^+}|=|\av{\sigma_z\otimes\sigma_{z}}_{\psi_\tau^+}|=1$
we see that the bound in \Eq{C} is saturated. The  shape of the
function $C(\theta_A, \theta_B)$ as determined in (\ref{C})  is
plotted in Figure~\ref{plaatje_general}.

We thus see that  $C(\theta_A, \theta_B)$  becomes greater and
greater when the angles approach orthogonality.  Obviously, for
the extreme cases of parallel and completely orthogonal settings
(i.e., $\theta_A=\theta_B=0$  or $ \pi/2$) we retrieve the results
mentioned in section \ref{IIA}.

\begin{figure}[h!]
\includegraphics[scale=0.9]{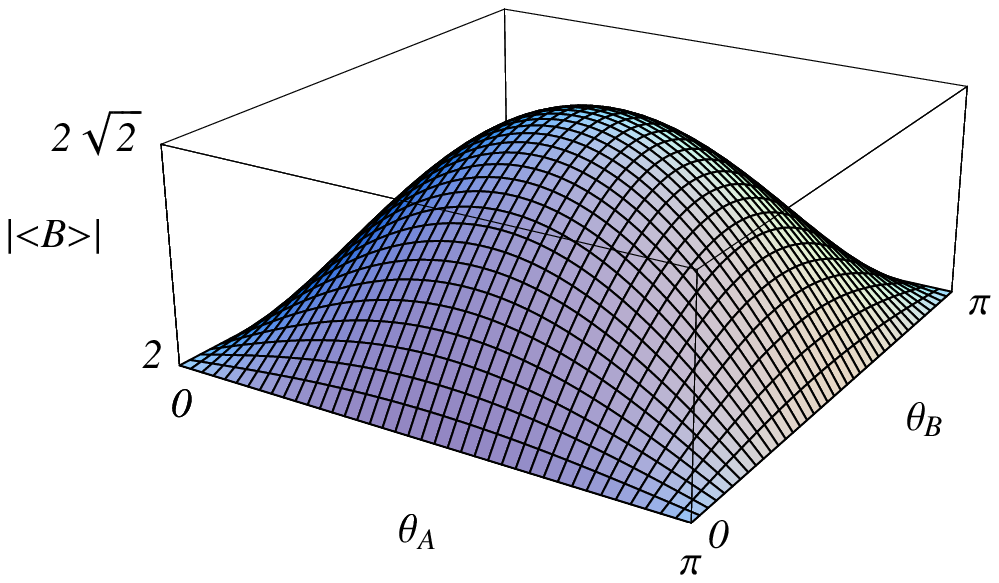}
\caption{Plot of $C(\theta_A,\theta_B)=  \max_{\rho \in \cal{D}}
|\av{\mathcal{B}}_\rho| $ as given in \Eq{C} for $0\leq
\theta_A,\theta_B\leq \pi$.} \label{plaatje_general} \end{figure}

If both angles are chosen the same, i.e.,
$\theta_A=\theta_B:=\theta$, \Eq{C} simplifies to
\begin{align}\label{Tsiequalangles}  C(\theta, \theta) = \sqrt{4+ 4 \sin^2
\theta }, \end{align} which is plotted in Figure~\ref{grafiek1}.

\subsection{Separable qubit states}
\noindent The set $ \mathcal{D}_{\textrm{sep}}$ of
separable two-qubit states is closed under local unitary transformations.
Therefore, to find $\max_{\rho \in  \mathcal{D}_{\textrm{sep}}} |\av{\mathcal{B}}_\rho|$, we may consider the same choice
of observables as before in \Eq{choiseobs} without loss of
generality. Further, we only have to consider pure states and can
take the state \mbox{$\ket{\Psi}=\ket{\psi_1}\ket{\psi_2}$} with
$\ket{\psi_1} =
  \cos \gamma_1 e^{-i \phi_1/2} \ket{0} +  \sin \gamma_1 e^{i \phi_1 /2}\ket{1} $
  and $\ket{\psi_2} =  \cos \gamma_2 e^{-i \phi_2/2} \ket{0} +
  \sin \gamma_2 e^{i \phi_2 /2} \ket{1}$.
We then obtain \begin{align}
&\av{A}_{\psi_1} =
 \sin 2\gamma_1  \cos\phi_1, ~~~ \av{A'}_{\psi_1} = \sin 2\gamma_1 \cos (\phi_1 -\theta_A),\nn\\
 &\av{B}_{{\psi_2}} = \sin 2\gamma_2   \cos\phi_2, ~~~ \av{B'}_{{\psi_2}}= \sin 2\gamma_2 \cos (\phi_2 -\theta_B).
 \end{align}

Since  $\ket{\Psi}$ is separable, we get $\av{A\otimes B}_{\Psi}=\av{A}_{{\psi_1}}\av{B}_{{\psi_2}}$, etc., and the maximal expectation  value of the CHSH operator becomes 
\begin{align} D(\theta_A, \theta_B) &= \max_{{\Psi}}\, \av{\mathcal{B}}_{\Psi } \nn\\&=\max_{\gamma_1, \gamma_2, \phi_1, \phi_2}\sin 2\gamma_1 \sin 2\gamma_2 [\cos \phi_1 (\cos\phi_2 + \cos(\phi_2 -\theta_B))\nonumber\\
 &~~~~~~~~~~~~~~~~~~~~~~~~+\cos(\phi_1 -\theta_A)(\cos\phi_2 - \cos(\phi_2 -\theta_B) )].\label{sepA} \end{align} This  maximum is attained for $\gamma_1=\gamma_2= \pi/4$ and \Eq{sepA} reduces to:
 \begin{align}
 D(\theta_A, \theta_B)=\max_{\phi_1, \phi_2} &\,\cos \phi_1 (\cos\phi_2 + \cos(\phi_2 -\theta_B))
\nn\\&+\cos(\phi_1 -\theta_A)(\cos\phi_2 - \cos(\phi_2 -\theta_B) ).
\end{align}
A tedious but straightforward calculation yields that the maximum
over $\phi_1$ and $ \phi_2$ is given by 
\begin{align}
D(\theta_A,\theta_B)=\sqrt{2\big(1 + \sqrt{1 - \sin^2\theta_A\sin^2\theta_B}\,\big)}
\label{maxSep}
\end{align}
\forget{
\begin{align}\nonumber
D(\theta_A,\theta_B)=& \Big | \mathrm{W}_+ (1 +
\mathrm{X}_{\pm}^{2})^{-1/2} \\ & +
\mathrm{cos}(\mathrm{arctan}(\mathrm{X}_{\pm}) - \theta_A)\,
 \mathrm{W}_-
\Big|,\label{maxSep}
\end{align}
\noindent with
\begin{align}
\mathrm{W}_{\pm}:&=\nonumber (1 + \frac {\mathrm{Z}^{2}}
{\mathrm{sin}^{2}\theta_B\,\mathrm{Y}^{2}})^{-1/2}
    \\&\pm \mathrm{cos}(\mathrm{arctan}(
        {
                \frac {\mathrm{Z}}
            {\mathrm{sin}\theta_B\,\mathrm{Y}}
        } ) + \theta_B),    \\
\mathrm{X}_{\pm} :&=(
\sin\theta_A\cos^2\theta_A\sin^2\theta_B)^{-1}\big(-\cos\theta_A(\cos\theta_B
\nonumber\\&+\cos^2\theta_B +\cos^2\theta_A\sin^2\theta_B)
\pm(\cos^2\theta_A\\
&\times(1+\cos^2\theta_B) (\cos^2\theta_B
+\cos^2\theta_A\sin^2\theta_B))^{1/2}
 \big)\nonumber\\
\mathrm{Y} :&= \mathrm{X}_{\pm}(1- \cos\theta_A + \sin\theta_A),
\\ \mathrm{Z} :&=\mathrm{X}_{\pm}(1 +\cos\theta_B +\cos\theta_A-
\cos\theta_B\,\cos\theta_A) + \nonumber\\&\cos\theta_B\sin\theta_A
-\sin\theta_A, \end{align} \noindent where in $\mathrm{X}_{\pm}$
the $+$ sign is chosen for $-\pi/2\leq\theta_A\leq \pi/2$  and the
$-$ sign is chosen for $\pi/2\leq\theta_A\leq 3\pi/2$ (both modulo
$2\pi$). 
}
The function (\ref{maxSep}) is plotted in
Figure~\ref{plaatje_separable}.

From this figure we conclude that the maximum of
$|\av{\mathcal{B}}_\rho|$ for separable two-qubit states  becomes smaller
and smaller when the angles approach orthogonality. For parallel
and completely orthogonal settings we again retrieve the results
of section \ref{IIB}. 
\noindent \begin{figure}[h!]
\includegraphics[scale=0.9]{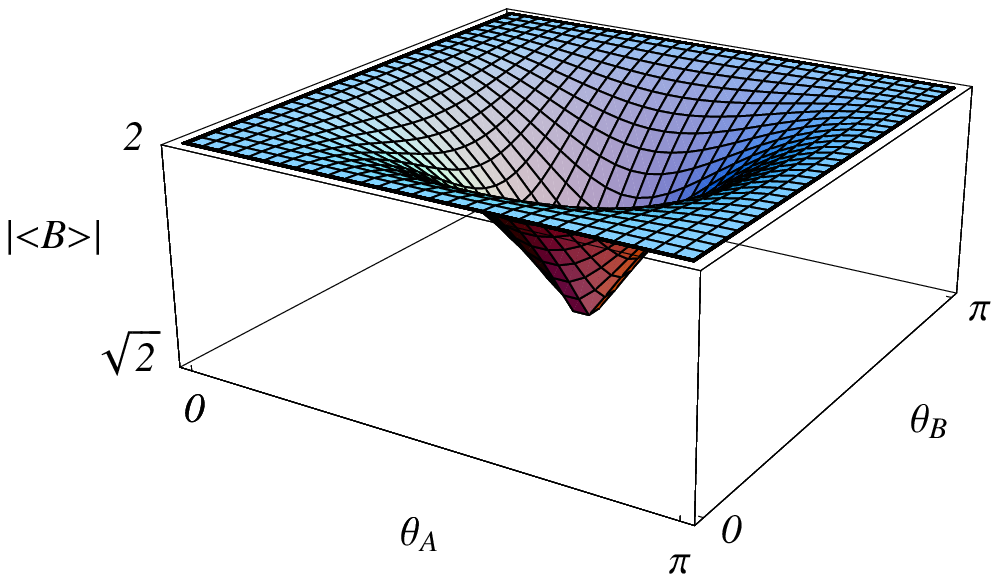} \caption{Plot of
$D(\theta_A,\theta_B) :=\max_{\rho \in \cal{D}_{\textrm{sep}}} |\av{\mathcal{B}}_\rho|$ as given in \Eq{maxSep} for $0\leq
\theta_A,\theta_B\leq \pi$.} \label{plaatje_separable}
\end{figure} 
\noindent As a special case, suppose we choose
$\theta_A=\theta_B:=\theta$. Then, (\ref{maxSep}) reduces to the
much simpler expression \begin{align} D(\theta, \theta)=
|\cos\theta| +\sqrt{1+\sin^2\theta}. \label{equalsep}
\end{align}
This result strengthens the bound obtained previously by \citet{roy} for this special case, which is: \begin{align}
D(\theta, \theta)\leq \left \{ \begin{array}{ll} \sqrt{2}(|\cos\theta| +1),& |\cos \theta|\leq3-2\sqrt{2},\\
1+2\sqrt{|\cos \theta|}-|\cos\theta|,& ~~\mathrm{otherwise}.
\end{array} \right. \label{royeq}
\end{align} Both functions are shown in Figure~\ref{grafiek1}.

\begin{figure}[!h]
\includegraphics[scale=0.9]{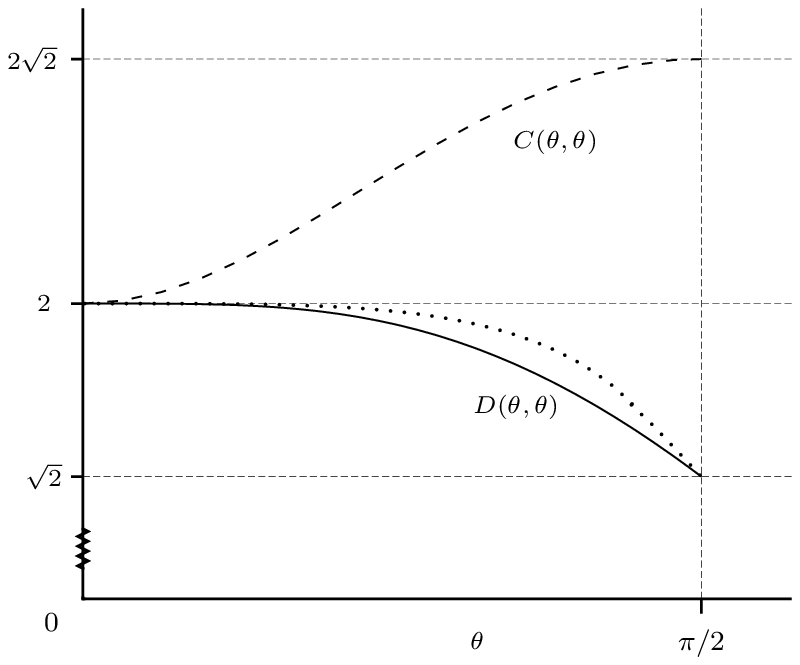}
\caption{Plot of the results  \Eq{Tsiequalangles} (dashed line)
and \Eq{equalsep} (uninterrupted line),  and of the bound by \citet{roy} given in (\ref{royeq}) (dotted line).} \label{grafiek1}
\end{figure}

\section{Discussion}\label{discussion}\noindent
In this chapter we have given tight quantitative expressions for
two trade-off relations. Firstly, between the degrees of local
commutativity, as measured by the local angles $\theta_A$ and
$\theta_B$,  and the maximal degree of CHSH inequality
violation, in the sense that if both local angles increase towards
$\pi/2$ (i.e., the degree of local commutativity decreases),  the
maximum violation of the CHSH inequality increases. Secondly,
a  converse trade-off relation holds for separable two-qubit states: if both
local angles increase towards  $\pi/2$, the value attainable for
the expectation of the CHSH operator decreases and thus the
\emph{non}-violation of the CHSH inequality increases. The
extreme cases of these relations are obtained for anti-commuting
local observables where the bounds of $2\sqrt{2}$ and $\sqrt{2}$
hold, which reproduces these results of the previous chapter. For the case of equal angles the trade-off relation for
separable states strengthens a previous result of \citet{roy}.

Our results are complementary to the well studied question what
the maximum of the expectation value of the CHSH operator is when
evaluated in a certain state (see e.g., \cite{GisiN,GISIN91,POPROHR}). Here we have
not focused on a certain given state, but instead on the
observables chosen, i.e., we asked, independent of the specific
state of the system, what the maximum of the expectation value of
the CHSH operator is when using certain local observables. The
answer found shows a diverging trade-off relation for the two
classes of separable and non-separable two-qubit states.

Indeed,
these two trade-off relations show that local non-commutativity has
two diametrically opposed features: On the one hand, the choice of
locally non-commuting observables is necessary to allow for any
violation of the CHSH inequality in entangled states (a
``more than classical'' result). On the other hand, this  very
same choice of non-commuting observables  implies  a ``less than
classical'' result  for separable two-qubit  states: For such states the
correlations (in terms of $\av{\mathcal{B}}_\rho$) obey a more
stringent bound than  allowed for by  local hidden-variable
theories, cf. the CHSH inequality (\ref{chshineq}).

These trade-off relations are useful for experiments aiming to
detect entangled states. They have an experimental advantage above
both Bell-type inequalities and entanglement witnesses as tests for two-qubit
entanglement.
 This will be discussed next.

\begin{figure}[h]
\includegraphics[scale=1]{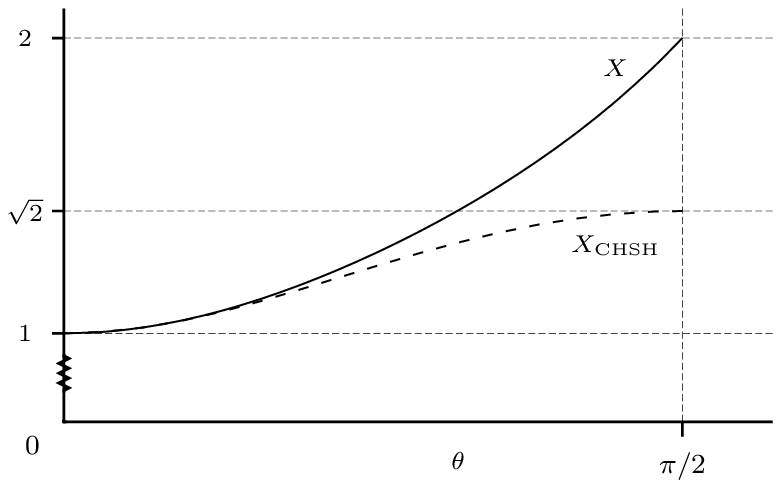}
\caption{Violation factor $X$ (uninterrupted line)  and
$X_{\mathrm{CHSH}}$ (dashed line) for
$\theta_A=\theta_B:=\theta$.} \label{grafiek2} \end{figure}
\noindent

 For comparison to the CHSH inequality as a test of entanglement,
let us define the 'violation factor' $X$ as the ratio $C(\theta_A,
\theta_B)/D(\theta_A, \theta_B) $, i.e. the maximum correlation
attained by entangled states divided by the maximum correlation
attainable for separable states. In Figure (\ref{grafiek2}) we
have plotted this violation factor $X$  for the special case of
equal angles, cf.\ \Eq{Tsiequalangles}  and \Eq{equalsep} and
compared it to the ratio by which these maximal correlations
violate the  CHSH inequality \Eq{chshineq}, i.e.
$X_{\mathrm{CHSH}} := C(\theta, \theta)/2$. Figure \ref{grafiek2}
shows  that the  violation factor $X$ is always higher than
$X_{\rm CHSH}$ except when $\theta =0$.  For angles
$\theta\lesssim\pi/4$ these two factors   differ only slightly,
but the violation factor $X_{\mathrm{}}$ increases to $\sqrt{2}$
times the original factor $X_{\mathrm{CHSH}}$ when $\theta$
approaches $\pi/2$. Furthermore note that the factor
$X_{\mathrm{}}$ increases more and more steeply, whereas
$X_{\mathrm{CHSH}}$ increases less and less steeply. For the case
of unequal angles the same features occur, as is evident from
comparing Figures \ref{plaatje_general} and
\ref{plaatje_separable}.

Therefore, the comparison of the correlation in entangled qubit states to the maximum
correlation obtainable in separable states yields a stronger test for
entanglement than violations of the CHSH inequality. Indeed, the violation
factor may reach 2 instead of $\sqrt{2}$. This means that the separability
inequality \Eq{sepineq} detects more entangled states and tolerates 
greater noise robustness in detecting entanglement (cf. section \ref{exp2ortho}).
Clearly, the optimal case of this relation obtains when the local observables
are exactly orthogonal to each other. On the other hand, in the case where at
least one of the local pairs of observables are parallel, no improvement upon
the CHSH inequality  is obtained. But that case is trivial, i.e., no
entangled state can violate  either \Eq{chshineq} or \Eq{sepineq} in that case.

Other criteria for the detection of qubit entanglement than the CHSH inequality
 have been developed in the
form of entanglement witnesses. In general,  these criteria have
two experimental drawbacks\footnote{See also \cite{enk} where the assumptions  needed in various entanglement
verification procedures are extensively  discussed.}: (i) they are usually
designed for the detection of a particular entangled state  and
hence require some a priori knowledge about the state, and  (ii)
they require the implementation of a specific  set of local
observables (e.g., locally orthogonal ones \cite{nonlinear,yu,zhang}). The
separability inequality \Eq{sepineq} compares favorably on these
two points, as we will discuss next.

In real experimental situations one might not be
completely sure about which observables are being measured. For
example, one might not be sure that the local angles are
\emph{exactly} orthogonal in the optimal setup.  However,
  even in such cases, one might be reasonably sure
that the angles are close to 90 degrees, e.g., that
these angles certainly lie within some finite-sized interval $\epsilon$ around 90
degrees. In that case, the bound \Eq{sepineq}  for
separable states would of course be higher than the optimal value
of $\sqrt{2}$ and the increase depends on the size of the
interval specified.  But the trade-off relation presented in this letter
tells us exactly how much higher the bound becomes as a function
of the angles (e.g., $\theta= \pi/2\pm \epsilon$), so one can still obtain a relevant  bound on
$|\av{\mathcal{B}}|$. One can thus still use it as a criterion for
testing entanglement in the presence of some ignorance about  the measured observables.
Entanglement witnesses do not share this
feature since no other  trade-off relations have been obtained (at
least to our knowledge) that quantify how the performance of the witness is changed
when one allows for uncertainty in the observables that feature in
the witness.

Note that for two qubits this result answers the question raised by \citet{nagataPRL} where it was asked how separability inequalities for orthogonal observables could allow for some uncertainty $\epsilon$ in the orthogonality, i.e., allowing for  $|\{A,A'\}|\leq \epsilon$ (analogous for $B$, $B'$). 
A further advantage of the separability inequalities  \Eq{sepineq}
is that they are not state-dependent and are formulated in terms
of locally measurable observables from the start, whereas it is
usually the case (apart from a few simple cases) that
constructions of entanglement witnesses involve some extremization
procedure and are state-dependent. Furthermore, finding the
decomposition of witnesses in terms of a few locally measurable
observables is not always easy \cite{guhnewitness,guhnewitness1}. However, it must
be said that  choosing the optimal set of observables in the
separability inequalities for detecting a specific state of
course also requires some prior knowledge of this state.

The results presented here only concern the case of two qubits\footnote{For the case of quantum systems that have a larger Hilbert space than $\mathbb{C}^2$ as their state space, see the discussion at the end of section \ref{comparisonlhvN2} that deals with the question of what happens to the trade-off relation for separable states in that case.} and the bi-partite linear
Bell-type inequality. It might prove useful to look for similar
trade-off relations for nonlinear separability inequalities as
well as for entanglement witnesses. Furthermore, it would be
interesting to extend this analysis to the multi-partite Bell-type
inequalities involving two dichotomous observables per party such
as the Werner-Wolff-\.Zukowski-Brukner inequalities \cite{wernerwolf2,zukowskibrukner}
or the Mermin-type inequalities \cite{mermin}. For the latter the
situation for local anti-commutativity has already been investigated
\cite{roy, nagataPRL, partsep}, but for non-commuting observables that are
not anti-commuting no results have yet been obtained.

%
%
%
%
\clearemptydoublepage
\thispagestyle{empty}
\part{Multi-partite  correlations}     

\thispagestyle{empty}
\chapter[Partial separability and multi-partite entanglement]{Partial separability and\\ \vskip-0.1cm entanglement criteria for\\\vskip0.2cm multi-qubit quantum states}
\label{Npartsep_entanglement}
\noindent 
This chapter is largely based on (i) \citet{partsep}, (ii) \citet*{tothseev}, and (iii) \citet{seevuff}.

\section{Introduction}\noindent
The problem of characterizing entanglement for multi-partite
quantum systems has recently drawn much attention. An important
issue in this study is that, apart from the extreme cases of full
separability and full entanglement of all particles in the system,
one also has to face the intermediate situations
 in which only some subsets of particles are
entangled and others not.  The latter states are usually called
`partially separable with respect to a specific partition' 
or, more precisely, $k$-separable with respect to a specific partition if the
$N$-partite system is separable into a specific partition of $k$ subsystems ($k \leq N$)
\cite{duer2,duer22,duer,duer22,nagataPRL} .  The partial
separability structure of multi-qubit states has been classified by
\citet{duer2,duer22}. This classification consists of   a hierarchy
of levels corresponding to the $k$-separable states for $k=1,
\ldots N$, and within each level different classes are
distinguished by specifying  under which partitions of the system
the state is $k$-separable or $k$-inseparable. As we shall argue,
however, it is useful to extend  this classification with one
more class at each level $k$, since the notion of $k$-separability with respect to a specific partition does not exhaust all partial separability properties.

Several experimentally accessible conditions to characterize
$k$-separable multi-qubit states have already been proposed, e.g., Bell-type inequalities  \cite{laskowzukow,nagataPRL,roy, uffink,seevsvet,collins,gisin} and, more generally, in terms of entanglement witnesses
\cite{tothguhne2}. However, these conditions address only part of
the full classification since they do not distinguish between the
various classes within a level. Here we derive separability
conditions that do address the full classification of partial
separability.  This will be performed by generalizing the derivation of the two-qubit separability conditions of chapter \ref{chapter_CHSHquantumorthogonal} to the multi-qubit setting.

These new conditions take the form of inequalities that provide bounds on
experimentally accessible correlations for the standard
Bell-type experiments (involving at each site measurement of two
dichotomic observables). These inequalities form a hierarchy with  strong state-dependent bounds and 
numerical bounds that decrease by a factor of four for each level
in the partial separability hierarchy.  For the classes within a
given level, the inequalities give state-dependent bounds,
differing for each class. Violations of the inequalities provide strong
sufficient criteria for various forms of non-separability and
multi-qubit entanglement.

We next demonstrate the strength of these conditions in two ways:
Firstly, by showing that they imply several other general
experimentally accessible entanglement criteria, namely the
fidelity criterion \cite{sackett, seevuff,fidelity}, the Laskowski-\.Zukowski
condition \cite{laskowzukow} (with a strict improvement
for $k=2,N$),  and the D\"ur-Cirac
criterion \cite{duer2,duer22}. The first two are conditions for
separability in general  and the third is a condition for
separability under specific partitions. We furthermore show that
the new conditions imply the \forget{linear and quadratic}
Mermin-type separability inequalities of \cite{nagataPRL,roy,uffink, seevsvet,collins,gisin}.  We also show
that the latter are equivalent to the Laskowski-\.Zukowski
separability condition.

 Secondly, we  compare the conditions to other state-specific
multi-qubit entanglement criteria \cite{tothguhne2,guhne2007,chen}
both for their white noise robustness and for the number of
measurement settings required in their implementation. In particular, we show (i) detection of bound
entanglement for $N\geq 3$ with noise robustness for detecting the
bound entangled states of \citet{duer22} that goes to $1$ for
large $N$ (i.e., maximal noise robustness), (ii) detection of the 
 four qubit Dicke state with noise robustness $0.84$ and $0.36$ for detecting it as entangled and fully entangled respectively, (iii) great noise and decoherence robustness
\cite{noise2,noise} in detecting entanglement of the $N$-qubit GHZ state where for colored noise and for decoherence due to dephasing 
the robustness for detecting full entanglement goes to $1$ for large $N$, and lastly,
(iv) better white noise robustness than the stabilizer witness criteria of
\citet{tothguhne2} for detecting the $N$-qubit GHZ states. In all
these cases it is shown that only $N+1$ settings are needed.

Choosing the familiar Pauli matrices as the local orthogonal
observables  yields  a convenient matrix  element representation
of the partial separability conditions. In  this representation,
the inequalities give specific bounds on the anti-diagonal matrix
elements in terms of the diagonal ones. Further, some comments
will be made along the way on how these results relate to the
original purpose \cite{bell64} of Bell-type inequalities to test
local hidden-variable models against  quantum
mechanics. Most notably,  when the number of parties is increased,
there is not only
 an exponentially increasing factor that separates the correlations
 allowed in maximally entangled states in comparison to those of local
 hidden-variable theories,
 but, surprisingly, also an exponentially increasing factor between the
 correlations allowed by LHV models and those allowed by
 non-entangled qubit states.

This chapter is structured as follows. In section \ref{partsep} we
define the relevant partial separability notions and extend the hierarchic partial separability classification of 
\citet{duer2,duer22}. There we also introduce the notions of $k$-separable
entanglement  and of $m$-partite entanglement. Using these notions
we investigate the relation between partial separability and
multi-partite entanglement and show it to be non-trivial. The four experimentally accessible partial separability conditions
are presented that are to be strengthened in the next section.
 In section  \ref{criteria}  we derive the announced partial separability
 criteria for $N$ qubits in terms of experimentally accessible
quantities.  They provide the desired
necessary conditions for the full hierarchic separability
classification. From these we obtain the sufficient non-separability
and entanglement criteria. In section \ref{strengthsection} the experimental strength of these criteria is discussed. We end in section
\ref{discussionn} with a discussion of the results obtained.

\section[Partial separability and multi-partite entanglement]{Partial separability and multi-partite\\ entanglement}\label{partsep}
In this section we introduce terminology and definitions
to be used in later sections. We define the notions of
$k$-separability, $\alpha_k$-separability, $k$-separable entanglement and $m$-partite
entanglement and use these notions to capture aspects
of the separability and entanglement structure in multi-partite
states.   We review  the separability hierarchy
introduced by \citet{duer2,duer22} and extend their classification. We also
discuss four partial separability conditions known in the literature \forget{and whose violations are sufficient
criteria for forms of entanglement.}
 These conditions will be strengthened in \ref{criteria}.

\subsection{Partial separability and the separability hierarchy}\label{generalksep}
 Consider an $N$-qubit system\footnote{The definitions and results of this subsection are not limited to qubits only. The dimension of the Hilbert space can be any finite number. However, since we restrict ourselves in all other sections to qubits we adopt qubits in this section too.}   with Hilbert space
 $\H=\mathbb{C}^{2}\otimes\ldots\otimes\mathbb{C}^{2}$.  Let
$ \alpha_k= (S_1, \ldots, S_k)$ denote a partition of $\{ 1,
\ldots ,N\}$ into $k$ disjoint nonempty subsets
  \mbox{($k\leq N$)}. Such a partition  corresponds to a division of
the system into $k$ distinct subsystems, also called a $k$-partite
split \cite{duer2}. A quantum state $\rho$ of this $N$-qubit
system is $k$-separable under a specific $k$-partite split
$\alpha_k$ \cite{duer2,duer22,duer,duer22,nagataPRL} iff it is fully separable in terms of the $k$ subsystems in
this split, i.e., iff
  \beq\label{kseprel}
\rho=\sum_i p_i
  \otimes_{n=1}^{k} \rho_i^{S_n } \,,~~~~~p_i\geq 0,~~\sum_i
p_i =1, \eeq where
 $\rho^{S_n }$ is  a state of the subsystem  corresponding to $S_n$ in the split $\alpha_k$.
We denote such states as  $\rho \in {\cal D}_N^{\alpha_k}$ and
also call them $\alpha_k$-separable, for short.  Clearly, ${\cal
D}_N^{\alpha_k}$  is a convex set. A state of the $N$-qubit
system outside this set is
 called $\alpha_k$-inseparable.

 More generally, a state  $\rho$ is called
$k$-separable \cite{laskowzukow,guhnetothbriegel,tothguhne1,acin, haffner}, denoted
as $\rho \in {\cal D}_N^{k\textrm{-sep}}$, iff there exists a convex decomposition \beq\label{ksep}
\rho=\sum_j p_j
  \otimes_{n=1}^{k}\rho^{S^{(j)}_n } \,,~~~~~p_j\geq 0,~~\sum_j
p_j =1, \eeq where each state $\otimes_{n=1}^{k}\rho^{S^{(j)}_n }$
 is a tensor product of $k$
density matrices of the subsystems corresponding to some such
partition $\alpha^{(j)}_k$, i.e., it factorizes under this split
$\alpha_k^{(j)}$. In this definition, the partition may vary for
each $j$, as long as it is a $k$-partite split, i.e., contains $k$
disjoint non-empty sets. Clearly ${\cal D}_N^{k\textrm{-sep}}$ is
also convex; it is the convex hull of the union  of all $\ {\cal
D}_N^{\alpha_k}$ for fixed values of $k$ and $N$.   States that
are not $k$-separable will be called $k$-inseparable. Note that  a
$k$-separable state need not be $\alpha_k$-separable for any
particular split $\alpha_k$\forget{For example,  consider the
    following two three-qubit states with the three qubits denoted by
    $a$, $b$ and $c$: $\ket{\psi_1}=\ket{0}\otimes(\ket{01}
    -\ket{10})/\sqrt{2}$, $\ket{\psi_2}=\ket{01}
    -\ket{10})/\sqrt{2}\otimes\ket{0}$. The state $\ket{\psi_1}$ is
    bi-separable under split $a$-$(bc)$, i.e., $\ket{\psi_1}\in {\cal
    D}_3^{a\textrm{-}(bc)}$, and $\ket{\psi_2}$ is bi-separable under
    split $c$-$(ab)$, i.e., $\ket{\psi_2}\in {\cal
    D}_3^{c\textrm{-}(ab)}$. Now form a convex mixture of these two
    states: $\rho=\frac{1}{2}(\ket{\psi_1}\bra{\psi_1}
+\ket{\psi_2}\bra{\psi_2})$. This state $\rho$ is not bi-separable
under any split, yet it is by construction bi-separable, i.e,
$\rho\in {\cal D}_3^{2\textrm{-sep}}$, and is thus not fully
inseparable \footnote{ Furthermore, using an eigenvalue analysis
of the partial transposed state under each bi-partite split, we
obtain that $\rho$ has negative partial transposition (NPT) under
all bi-partite splits.  Although NPT under a split is sufficient
for inseparability under this split, NPT under all bi-partite
splits is  not sufficient for full inseparability. This is in some
sense analogous to the fact that positive partial transposition
(PPT) with respect all bi-partite splits is not sufficient for
full separability.}.}\footnote{
\label{example2partiteentang}
For example, consider the following construction (which was inspired by \citet{tothguhne1}) where we use 
the $N$-qubit states $\ket{\psi_1}=\ket{e}_{12}\ket{0}_3\ket{0}_4
\cdots \ket{0}_N$; $\ket{\psi_2}=\ket{0}_1\ket{e}_{23}\ket{0}_4
\cdots\ket{0}_N$; \ldots,
$\ket{\psi_N}=\ket{0}_2\ket{0}_3 \cdots \ket{e}_{N,1}$, where $\ket{e}_{ij}$ is any entangled pure state of the two
parties $i$ and $j$ (mod $N$). Then the state $\rho =\sum_{i=1}
^N\ket{\psi_i}\bra{\psi_i}/N$  is inseparable under all splits, yet  by
construction $(N-1)$-separable.}.
     And even the converse  implication need not hold:
  If a state is bi-separable under every bipartition, it does not have
  to be fully separable, as shown
  by the three-partite examples of
  \cite{bennett,eggeling,acin} that give states that are bi-separable with respect to all
 bi-partite partitions, yet are not fully separable, i.e.,  they are
 three-inseparable. Similar observations (using different terminology) were obtained by \citet{guhnetothbriegel} and \citet{tothguhne1}, but below we
 will present a more systematic investigation.

The notion of $k$-separability naturally induces a hierarchic
ordering of the $N$-qubit states. Indeed, the sequence of sets
${\cal D}_N^{k\textrm{-sep}}$ is nested:
 ${\cal D}_N^{N\textrm{-sep}} \subset {\cal D}_N^{(N-1)\textrm{-sep}}
 \subset \cdots \subset
  {\cal D}_N^{1\textrm{-sep}}$.  In other words, $k$-separability implies $\ell$-separability for all $\ell \leq k$.
  We call a  $k$-separable state that is not $(k +1)$-separable  ``$k$-separable entangled''. Thus,  each $N$-qubit state can be characterized by the level  $k$ for which it is $k$-separable entangled, and these levels provide a hierarchical ranking: at one extreme end are the $1$-separable entangled states which are fully  entangled (e.g., the GHZ states), at the other end are the $N$-separable or fully separable states (e.g. product states or   the ``white noise state''  $\openone/ 2^N$).

  Often, it is interesting to know  how many qubits
 are entangled in a $k$-separable entangled state. However, this question does not have a
unique answer.
For example, take $N=4$ and  \mbox{$k=2$} (bi-separability). In
this case two types of states may occur in the decomposition
(\ref{ksep}), namely $\rho^{\{ij\}}\otimes\rho^{\{kl\}}$ and
$\rho^{\{i\}}\otimes\rho^{\{jkl\}}$ ($i,j,k,l=1,2,3,4$). A
$2$-separable entangled four-partite state might thus be two- or three-partite
entangled.

In general, an $N$-qubit state $\rho$ will be called $m$-partite
entangled iff  a decomposition of the state such as in
(\ref{ksep}) exists such that each subset $S^{(i)}$ contains at
most $m$ parties, but no such decomposition is possible when all
the $k$ subsets are required to contain less than $m$ parties
 \cite{seevuff}. (\citet{guhnetothbriegel} and \citet{tothguhne1} call this 
 `not producible by $(m-1)$-partite entanglement').
 It follows that a $k$-separable entangled state
is also $m$-partite entangled, with $\integer{N/k}\leq m\leq N-k+1$.
Here $\integer{N/k}$ denotes the smallest integer which is not less
than $N/k$. Thus, a state that is
$k$-separably entangled ($k< N$) is at least
$\integer{N/k}$-partite entangled and might be up to
\mbox{$(N-k+1)$}-partite entangled.
Therefore,
conditions that
 distinguish $k$-separability from
\mbox{$(k+1)$}-separability also provide conditions for
 $m$-partite entanglement, but generally allowing a wide range of values of $m$.
For example, for  $N=100$ and $k=2$, $m$ might lie anywhere between 50 and 99.

Of course, a much tighter conclusion about $m$-partite
entanglement can be drawn if we know exactly under which splits
the state is separable. This is why the notion of
$\alpha_k$-separability is helpful, since it provides these finer
distinctions. For example, suppose that a 100-qubit state is
separable under the bi-partite split $(\{1\}, \{ 2, \ldots 100\})$
but under no other bi-partite split. This state would then  be
$2$-separable (bi-separable) but now we could also infer that
$m=99$. On the other hand, if the state were only separable under
the split $\{ 1, \ldots 50\}, \{51, \ldots 100 \}$, it would still
be bi-separable, but only $m$-partite entangled for  $m= 50$.

\citet{duer2} provided such a fine-grained
classification of $N$-qubit states  by considering their
   separability or inseparability under  all $k$-partite splits.
Let us introduce this classification (with a slight extension) by means of the example of
three qubits,  labeled as $a,b,c$.

\emph{Class 3.} Starting with the lowest level $k=3$, there is
only one 3-partite split, $a$-$b$-$c$,  and consequently only  one
class to be distinguished at this level , i.e. ${\cal
D}_3^{a\textrm{-}b\textrm{-}c}$. This set coincides with ${\cal
D}^{3\textrm{-sep}}_3$.

\emph{Classes 2.1---2.8} Next, at level $k=2$, there are three
bi-partite splits: $a$-$(bc)$, $b$-$(ac)$ and $c$-$(ab)$ which
define the sets ${\cal D}_3^{a\textrm{-}(bc)}$, ${\cal
D}_3^{b\textrm{-}(ac)}$, and ${\cal D}_3^{c\textrm{-}(ab)}$.  One
can further distinguish classes defined by all logical combinations of
separability and inseparability under these  splits, i.e. all the
set-theoretical intersections and complements shown in Figure 1.
This leads to classes 2.2 -- 2.8. \citet{duer2,duer22}  showed that all
these classes are non-empty.  To these, we add one more class
2.1: the set of bi-separable states that are not separable under
any split. As we have seen, this set is non-empty too.

\emph{Class 1.} Finally,  at level $k=1$ there is again only one
(trivial) split $(abc)$,
 and thus only one class, consisting of all the fully entangled states,  i.e., ${\cal
D}_3^{1\textrm{-sep}} \setminus {\cal D}_3^{2\textrm{-sep}}$.

\begin{figure}[h]
\includegraphics[scale=0.45]{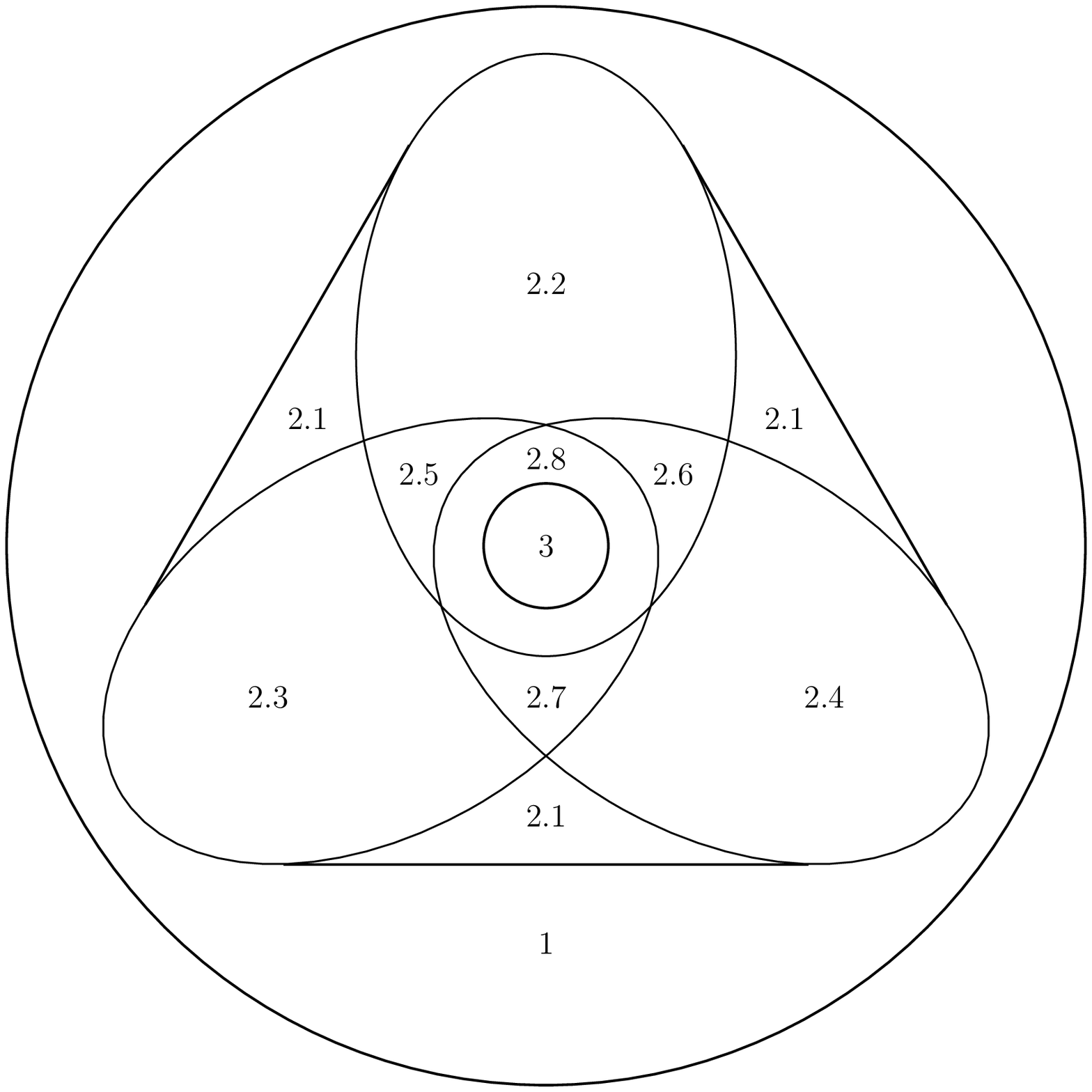}
\caption{Schematic representation of the 10 partial separability
classes of three-qubit states} \label{figclass}
\end{figure}

We feel that the above extension is desirable since otherwise  the
D\"ur-Cirac classification would not
 distinguish between class 2.1 and class 1. However,  states in class 2.1 are
simply convex combinations of states that are bi-separable under
different bi-partite splits. Such states can be realized by mixing
the bi-separable states, and are conceptually different from the
fully inseparable states of class 1.

This three-partite example serves to illustrate how the D\"ur-Cirac separability
classification works for general $N$. Level $k$ ($1\leq k \leq N$)
of the separability hierarchy consists of all $k$-separable
entangled states.  Each level is further divided into distinct
classes by considering all logically possible combinations of
separability and inseparability under the various $k$-partite
splits. The number of such classes increases rapidly with $N$, and
therefore we will not attempt to list them. In general, all such
classes may be non-empty.   As an extension of the D\"ur-Cirac
classification, we distinguish at each level $1< k<N$ one further
class, consisting of $k$-separable entangled states that are not
separable under any $k$-partite split.

In order to find relations between these classes, the notion of a
\emph{contained split} is useful \cite{duer2}. A $k$-partite split
$\alpha_k$ is contained in a $l$-partite split $\alpha_l$, denoted
as $\alpha_k \prec \alpha_\ell$,  if $\alpha_l$  can be obtained
from $\alpha_k$ by joining some of the subsets of $\alpha_k$. The
relation $\prec$ defines a partial order between splits at
different levels. This
partial order is  helpful because $\alpha_k$-separability implies
$\alpha_\ell$-separability of all splits $\alpha_\ell$ containing
$\alpha_k$. \forget{and $\alpha_\ell$-inseparability implies
inseparability under all splits that are contained in $\alpha_l$.}
We will use this implication below to obtain conditions for
separability of a $k$-partite split at level $k$ from such
conditions on all  $(k-1)$-partite splits at level $k-1$ this $k$-partite
split is contained in. One can thus construct separability
conditions for all classes at higher levels from the separability
conditions for classes at level $k=2$. Conditions at a lower level thus imply conditions at a higher level.

The multi-partite entanglement properties of  $k$-separable  or
$\alpha_k$-separable  states are subtle, as can be seen from the
following examples.

(i) mixing states does not conserve $m$-partite entanglement. Take
$N=3$, then mixing the $2$-partite entangled $2$-separable states
$\ket{0}\otimes(\ket{00}+\ket{11})/\sqrt{2}$ and
$\ket{0}\otimes(\ket{00} -\ket{11})/\sqrt{2}$ with equal weights
gives  a $3$-separable state $(\ket{000}\bra{000}
+\ket{011}\bra{011})/2$.

(ii) an $N$-partite state can be $m$-partite entangled ($m<N$)
even if it has no $m$-partite subsystem whose (reduced) state is
$m$-partite entangled \cite{seevuff,guhnetothbriegel}. Such states are said to have
irreducible $m$-partite entanglement\footnote{Note that \citet{walck} use the same notion of `irreducible $m$-partite entanglement', but with a different meaning. Their notion is used to denote multi-partite states whose set of reduced states does not suffice to uniquely determine the state. This we believe is better referred to as `underdetermination by the set of reduced states'.}. Thus, a state
of which some reduced state is $m$-partite entangled is itself at
least $m$-partite entangled,
 but the converse need not be true.

(iii)  consider a bi-separable entangled state that is only
separable under the bi-partite split $(\{1\},\{2,\ldots,N\})$. One cannot infer
that the subsystem $\{2,\ldots,N\}$ is \mbox{$(N-1)$}-partite
entangled. A counterexample is
 the three-qubit state $\rho = (
\ket{0}\bra{0}\otimes P^{(bc)}_{-} + \ket{1}\bra{1}\otimes
P^{(bc)}_{+})/2
 $
which is bi-separable only under the partition $a\textrm{-}(bc)$,
and thus bi-partite entangled, but has no bi-partite subsystem
whose reduced state is entangled. Here $P^{(bc)}_{+}$ and
$P^{(bc)}_{-}$ denote projectors on the Bell states
$\ket{\psi^{\pm}}=\frac{1}{\sqrt{2}}(\ket{01} \pm \ket{10})$ for
parties $b$ and $ c$, respectively.

(iv)  a state that is inseparable under all splits but which is
not fully inseparable (i.e., $\rho \in  {\cal
D}^{k\textrm{-sep}}_N$ with $k>1$ and  $\rho\notin\cup_{\alpha_k}
{\cal D}^{\alpha_k}_N,~\forall \,\alpha_k,k$) might still have all forms of $m$-partite
entanglement apart from full entanglement, i.e., it could be
$m$-partite entangled with $2\leq m\leq N-1$. Thus the state could
even have $m$-partite entanglement as low as $2$-partite
entanglement, although it is inseparable under all splits.  For
example, \cite{tothguhne1} consider a mixture
of two $N$-partite states where each of them is $(\lceil N/2
\rfloor)$- separable according to different splits. This mixed
state is by construction $(\lceil N/2 \rfloor)$- separable, not
bi-separable under any split, yet only $2$-partite entangled. See
also the example in footnote~\ref{example2partiteentang} which is $(N-1)$-separable and only $2$-partite entangled.

(v) Lastly, $N$-partite fully entangled states exist where no
$m$-partite reduced state is entangled (such as
 $N$-qubit GHZ state) and also where all $m$-partite reduced states
 are entangled (such as the $N$-qubit W-states) \cite{durR}.

These  examples serve to emphasize that one should be very
cautious in inferring the existence of entanglement in subsystems
of a larger system which is known to be $m$-partite entangled or $k$-separable entangled for some specific value of $m$ and $k$.

\subsection{Separability Conditions}\label{introsepconditions}
We now review four separability conditions for qubits,  which will
all be strengthened in the next section. These are necessary
conditions for  states  to be $k$-separable, $2$-separable, and
$\alpha_k$-separable respectively.

(I) \citet{laskowzukow} showed that for
any $k$-separable $N$-qubit state  $\rho$ the anti-diagonal matrix
elements (denoted by $\rho_{j,\bar{\jmath}}$, where
$\bar{\jmath}=d+1-j$, $d =2^N$ ) must satisfy
 \beq\label{anti-diagonal}  \max_j | \rho_{j,\bar{
 \jmath}}|  \forget{, \rho_{2, (d-1)},
 \ldots , \rho_{d, 1}\}
|\rho_{\nearrow}^{}|}
\leq
\big(\frac{1}{2}\big{)}^{k},~~~\forall\rho_{} \in {\cal
D}^{k\textrm{-sep}}_N. \eeq
 This condition can be
 easily proven by the observation that
for any density matrix to be physically meaningful its
anti-diagonal matrix elements must not exceed $1/2$. Therefore,
anti-diagonal elements of a product of $k$ density matrices cannot
be greater than $(1/2)^{k}$. For pure states this can be
checked directly\footnote{
The proof that the modulus of an anti-diagonal matrix element in a physically
allowable state must be less than
$1/2$ runs as follows. Consider a $d$-dimensional system with orthonormal
basis $\ket{1}, \ket{2},\ldots,\ket{d}$.
Next consider a general pure state of the form
$\ket{\psi}= \alpha\ket{1}+\ldots+\beta\ket{d}$. Normalization gives
$|\alpha|^2 +|\beta|^2\leq1$ ($\alpha,\beta \in \mathbb{C}$).
The anti-diagonal matrix element $\rho_{1,d}=\bra{1}\rho \ket{d}$
 is equal to
$\alpha\beta^*$. Since we are interested in the maximum absolute value of this
element we choose all other coefficients zero, to obtain  $|\alpha|^2
+|\beta|^2=1$, and hence $|\rho_{1,d}|\leq 1/2.$ The proof for all other
anti-diagonal matrix elements is analogous.} and by convexity, this results then holds
all $k$-separable states. Note that this condition is not basis
dependent.

 It follows  from (\ref{anti-diagonal}) that if the anti-diagonal matrix elements
of state $\rho$ obey
 \beq\label{anti-diagonalentang}
 \big(\frac{1}{2}\big)^{k} \geq  \max_j
 |\rho_{j,\bar{\jmath}}|>   \big(\frac{1}{2}\big)^{k+1},
\eeq then  $\rho$ is at most $k$-separable, i.e.,  $k$-separable
entangled, and thus at least $m$-partite entangled, with
$m\geq\integer{N/k}$.

 \forget{
Thus,
$\max |\rho_{\nearrow}|>{1}/{4}$
 is a sufficient condition for full $N$-partite entanglement,
  $ {1}/{4} \geq\max |\rho_{\nearrow}|>{1}/{8}$   is sufficient for  $2$-separable entanglement  (and at least $m$-partite
entangled with $m\geq\integer{N/2}$), and so on. If furthermore
the number of parties per subset in the partition is specified,
i.e., if a specific $k$-partite split is known under which the
state is separable, we get sufficient $m$-partite entanglement
criteria for a \emph{definite} number $m$. To choose an example,
suppose a state is bi-separable under the bipartition of the form
$\{1\}\{2,\ldots,N\}$. Then the state is \mbox{$(N-1)$}-partite
entangled, whereas if it is separable under a bipartition  that
splits the set of parties in half (i.e., $\{1,\ldots,
\integer{N/2}\}\{\integer{N/2}+1,\ldots ,N\}$) the state is only
$\integer{N/2}$-partite entangled.
}
The partial separability condition (\ref{anti-diagonal}) does not
yet explicitly refer to directly experimentally accessible
quantities. However, in the next section we will rewrite this
condition in terms of expectation values of local observables, and
show that they are  equivalent to Mermin-type separability
inequalities. \forget{ We strengthen this condition for $k=2$ and
$k=N$ considerably. This yields stronger sufficient criteria for
when states are entangled or fully entangled respectively. For
other $k$ it is unknown if the new conditions yield stronger
sufficient criteria for $k$-inseparability.}

(II) Mermin-type separability inequalities  \cite{nagataPRL,roy, uffink,seevsvet,gisin,collins}. Consider the familiar CHSH operator for two qubits (labeled as $a$ and $b$) which is defined by: 
\beq
M^{(2)} :=
X^{}_a\otimes X^{}_b + X^{}_a\otimes Y^{}_b + Y^{}_a\otimes
X^{}_b- Y^{}_a\otimes Y^{}_b. \label{111} 
\eeq  Here, $X^{}_a$ and $Y^{}_a$ denote two spin observables on the Hilbert spaces ${\cal H}_a$ and ${\cal  H}_b$ of qubit $a$, and $b$. The so-called
Mermin operator \cite{mermin} is  a generalization of this operator to $N$
qubits (labeled as $(a, b, \ldots n)$), defined by the recursive relation: \beq \label{merminN}
M^{(N)}:= \frac{1}{2}M^{(N-1)} \otimes(X^{}_n +Y^{}_n) +
 \frac{1}{2}M'^{(N-1)}\otimes(X^{}_n -Y^{}_n),
\eeq where $M'$ is the same operator as $M$ but with
all  $X$'s and $Y$'s  interchanged.

In the special case where, for each qubit, the spin observables $X$ and $Y$ are orthogonal, i.e. $ \{ X_i, Y_i\} =0$ for $i\in \{ a, \ldots n\}$,  \citet{nagataPRL} obtained the following $k$-separability
 conditions:
 \beq \label{quadraticN} \av{M^{(N)}}^2
+\av{M'^{(N)}}^2
\leq2^{(N+3)}\big(\frac{1}{4}\big)^k,~~ \forall \rho
\in {\cal D}_N^{k\textrm{-sep}}. \eeq
As just mentioned,  the next section will show that  these inequalities are equivalent to the
Laskowski-\.Zukowski inequalities.  The quadratic inequalities  (\ref{quadraticN}) also imply the following sharp linear Mermin-type
inequality  for $k$-separability: \beq\label{linearN}
|\av{M^{(N)}}|\leq 2^{(\frac{N+3}{2})}\big(\frac{1}{2}\big)^k,~~
\forall \rho \in {\cal D}_N^{k\textrm{-sep}}. \eeq \forget{This
inequality  is sharp: $\sup_{\rho \in {\cal D}_N^{k\textrm{-sep}}}
|\av{M_N} | = 2^{(N+3-2k)/2}$. } \forget{If (\ref{linearN}) is
violated the $N$ qubit state $\rho$ is $(k-1)$-separably
 entangled and it has at least $m$-partite entanglement,
 with $m\geq \integer{N/(k-1)}$.} For $k=N$ inequality (\ref{linearN})  reproduces a
 result obtained by  \citet{roy}.

(III). The fidelity $F(\rho)$ of a $N$-qubit state $\rho$ with
respect to the generalized $N$-qubit GHZ state $\ket{\Psi_{{\rm
GHZ},\alpha}^N}:= ( \ket{0}^{\otimes N}
+e^{i\alpha}\ket{1}^{\otimes N})/\sqrt{2}$ ($\alpha \in
\mathbb{R}$) is defined as
\begin{equation}
 F(\rho) := \max_\alpha \langle \Psi_{{\rm GHZ},\alpha}^N |\rho|\Psi_{{\rm GHZ},\alpha}^N\rangle=
 \frac{1}{2} (\rho_{1,1}+\rho_{d,d})
 +|\rho_{1,d}|,
 \label{Fidelity}
\end{equation} The fidelity
condition \cite{sackett,seevuff,fidelity} (also known as the
projection-based witness \cite{tothguhne2}) says that for all
bi-separable $\rho$:
  \beq  F(\rho) \leq 1/2, ~~~ \forall \rho \in {\cal D}_N^{2\textrm{-sep}}. \label{fidelitycriterion}  \enq
In other words,  $ F(\rho) > 1/2 $
  is a sufficient condition for full $N$-partite entanglement.
An equivalent formulation of (\ref{fidelitycriterion})  is:
\beq\label{equivfidelity} 2|\rho_{1,d}| \leq \sum_{j\neq 1, d}
\rho_{j,j},~~~ \forall\rho_{} \in {\cal D}^{2\textrm{-sep}}_N
.\eeq

 Of course, analogous conditions may be obtained by
replacing $\ket{\Psi_{{\rm GHZ},\alpha}^N}$ in the definition
(\ref{Fidelity}) by any other maximally entangled state
\cite{fidelity,nagata2002}.  Exploiting this feature,  one can
reformulate (\ref{equivfidelity}) in a basis-independent form:
\beq\label{equivfidelity2} 2 \max_{j} |\rho_{j,\bar{\jmath}}| \leq
\sum_{i\neq j,\bar{\jmath} } \rho_{i,i},~~~ \forall\rho_{} \in
{\cal D}^{2\textrm{-sep}}_N .\eeq

Note that in contrast to the Laskowski-\.Zukowski condition and
the Mermin-type separability inequalities, the fidelity condition does not
distinguish bi-separability and other forms of $k$-separability.
Indeed, a fully separable state (e.g. $\ket{0^{\otimes N}}$ can
already attain the value $F(\rho)=1/2$. Thus, the fidelity
condition only distinguishes full inseparability \mbox{(i.e.,
$k=1$)} from other types of separability ($k\geq 2$). However, as
will be shown in the next section, violation of the fidelity
condition yields a stronger test for full entanglement than
violation of the Laskowski-\.Zukowski condition.

(IV) The D\"ur-Cirac depolarization method \cite{duer,duer2} gives
necessary conditions for partial separability under specific
bi-partite splits. It uses a two-step procedure in which a general
state $\rho$ is first depolarized to become a member of a special
family of states, called $\rho_N$, after which this depolarized
state is tested for $\alpha_2$-separability under a  bi-partite
split $\alpha_2$. If the depolarized state $\rho_N$ is not
separable under  $\alpha_2$, then neither is the original state
$\rho$, but not necessarily vice versa since the depolarization
process can decrease inseparability.

 The special family of states $\rho_N$ is  given by
\beq\label{DuerStates}
\rho_N= \lambda_0^+\ket{\psi_0^+}\bra{\psi_0^+} +\lambda_0^-\ket{\psi_0^-}\bra{\psi_0^-}+
\sum_{j=1}^{2^{N-1}-1}\lambda_j(\ket{\psi_j^+} \bra{\psi_j^+}+\ket{\psi_j^-}\bra{\psi_j^-}),
\eeq
with the so-called orthonormal GHZ-basis $\ket{\psi_j^{\pm}}=
\frac{1}{\sqrt{2}}\ket{j0} \pm\ket{j'1})$,
  where $ j=j_1j_2\ldots j_{N-1}$ is in binary notation (i.e., a string of
$N-1$ bits),  and $j'$ means a bit-flip of $j$:  $
j'=j'_1j'_2\ldots j'_{N-1}$, with $j'_i=1,0$ if $j_i=0,1$.
\forget{Note that $\ket{\psi_0^+}$ and $\ket{\psi_0^-}$ are the
states $\ket{\Psi_{{\rm GHZ},\alpha}^N}$ with $\alpha =0,\pi$
respectively.} The depolarization process does not  alter the
values  of
$\lambda_0^{\pm}=\bra{\psi_0^{\pm}}\rho\ket{\psi_0^{\pm}}$ and of
$\lambda_j=(\bra{\psi_j^{+}}\rho\ket{\psi_j^{+}}+\bra{\psi_j^{-}}\rho\ket{\psi_j^{-}})/2$
of the original state $\rho$. The values of $j=j_1j_2\ldots
j_{N-1}$ can be used to label the various bi-partite splits by
stipulating that $j=j_1j_2\ldots j_{N-1}$,  $j_n=0,(1)$ corresponds
to the $n$-th qubit belonging (not belonging) to the same subset as
the last qubit. For example, the splits $a$-$(bc)$, $b$-$(ac)$,
$c$-$(ab)$ have labels $j=10,01,11$ respectively.

The D\"ur-Cirac condition \cite{duer2} says that a state $\rho$ is
separable under a specific bi-partite split $j$ if
\beq\label{dccondition} |\lambda_0^+-\lambda_0^-|\leq2
\lambda_j~~~\Longleftrightarrow~~~ 2|\rho_{1,d}|\leq \rho_{l,l}
+\rho_{\bar{l},\bar{l}},~~~~~ \forall\rho_{} \in {\cal D}^{j}_N,
~~~ \bar{l} = d+1 -l,  \eeq For the states (\ref{DuerStates}) this
condition is in fact necessary and sufficient. In the right-hand
side of the second inequality of (\ref{dccondition}) $l$ is
determined from $j$ using Tr$[\rho \ket{\psi_j^+}
\bra{\psi_j^+}+\ket{\psi_j^-}\bra{\psi_j^-}]= \rho_{l,l}
+\rho_{\bar{l},\bar{l}}$.

Separability conditions for multi-partite splits are  constructed
from the conditions (\ref{dccondition}) by means of the partial
order $\prec$ of containment. As mentioned above, if a state is
$\alpha_k$-separable, then it is also $\alpha_2$-separable for all
bi-partite splits  $\alpha_k \prec \alpha_2$. Therefore,  the
conjunction of all $\alpha_2$-separability conditions  must hold
for such a state.

 \forget{ Thus if $
|\lambda_0^+-\lambda_0^-|>2 \lambda_j$ the state is inseparable
under the split $j$.}
 Note that if
$|\lambda_0^+-\lambda_0^-|>2\max_{j} \lambda_j$,  the state is
inseparable under all bi-partite splits, but this does not imply
that it is fully inseparable (cf.\ footnote~\ref{example2partiteentang}).
 Indeed,  this feature also exists for states  of the form (\ref{DuerStates}) as the following example shows.
 Take the following two members of the  family (\ref{DuerStates}) for $N=3$:
for
 $\rho_3^i$ we choose $\lambda_0^+=1/2$, $\lambda_0^- =0$, $\lambda_{01}=0$, $\lambda_{10}=1/4$, $\lambda_{11}=0$,
 and for
 $\rho_3^{ii}$ :  $\lambda_0^+=1/2$, $\lambda_0^- =0$, $\lambda_{01}=0$, $\lambda_{10}=0$, $\lambda_{11}=1/4$.
\forget{From the results in   \cite{duer2} it follows that a
state $\rho_3$ has PPT and is also separable under the split
$a$-$(bc)$ iff  $2\lambda_{10}\geq |\lambda_0^+ - \lambda_0^-| $,
has PPT and is separable under the split  $b$-$(ac)$ iff
$2\lambda_{01}\geq |\lambda_0^+ - \lambda_0^-| $, has PPT and is
separable under the split  $c$-$(ab)$ iff $2\lambda_{11}\geq
|\lambda_0^+ - \lambda_0^-| $. Furthermore the state is
$3$-separable iff all bi-partite splits are separable. Applying
these results we obtain that}  It follows from condition
(\ref{dccondition}) that $\rho_3^i$ is separable under split
$a$-$(bc)$ and inseparable under  other  splits, while
$\rho_3^{ii}$ is separable under the split $c$-$(ab)$ and
inseparable under any other split. Now form a convex mixture of
these two states: $\tilde{\rho}_3 =\alpha\rho_3^i +\beta
\rho_3^{ii}$ with $\alpha +\beta =1$ and $\alpha,~ \beta \in
(0,1)$.  This state $\tilde{\rho}_3$ is still of the form
(\ref{DuerStates})\forget{with values $\lambda_0^+=1/2$,
$\lambda_0^- =0$, $\lambda_{01}=0$, $\lambda_{10}=\alpha/4$,
$\lambda_{11}=\beta/4$}, so that we can again  apply condition
(\ref{dccondition}) to conclude that $\tilde{\rho}_3$ is not
separable under any bi-partite split, yet bi-separable by
construction.

 In
the next section we give necessary conditions for $k$-separability
and $\alpha_k$-separability   that are stronger than the
Laskowski-\.Zukowski condition (for $k=2,N$), the fidelity
condition and the D\"ur-Cirac condition. \forget{ so that violation of these new
conditions yields stronger sufficient criteria for inseparability
under splits.}

\section{Deriving new partial separability conditions} \label{criteria}

In this section we present $k$-separability conditions for all
levels and classes in the hierarchic classification of $ N$-qubit
states. Violations of the conditions give strong criteria for
specific forms of non-separability and $m$-partite entanglement.  The starting point will be the two-qubit results of chapter \ref{chapter_CHSHquantumorthogonal}, whose result we rehearse here for both convenience and for introducing the notation to be used in this chapter.
We next move on to the slightly
more complicated case of three qubits, for which explicit
separability conditions are given for each of the 10 classes in
the separability hierarchy that were depicted in Figure \ref{figclass}.  Finally, the case of $N$-qubits is
treated by a straightforward generalization.

\subsection{Two-qubit case: setting the stage}\label{twoqubitsection}
 For  two-qubit systems the separability
hierarchy is very simple: there is only one possible split, and
consequently just one class at each of the two levels $k=1$ and
$k=2$, i.e., states are either inseparable (entangled) or
separable.

Consider a system composed of a pair of qubits  \forget{on the
Hilbert space $\mathcal{H}=\mathbb{C}^2 \otimes \mathbb{C}^2$} in
the familiar setting of two  distant sites, each receiving one of
the two qubits, and where, at each site, a  measurement of either
of two spin observables is made. We will  focus on the special
case that these local spin observables are mutually orthogonal.
 Let $(X^{(1)}_a, Y^{(1)}_a, Z^{(1)}_a)$ denote three orthogonal spin observables on  qubit  $a$,
 and    $(X^{(1)}_b, Y^{(1)}_b, Z^{(1)}_b)$ on qubit $b$.
(The superscript $1$ denotes that we are
 dealing with  single-qubit operators.)
 A familiar choice  for the orthogonal
triples $\{ X^{(1)},Y^{(1)},Z^{(1)} \}$  are the Pauli matrices
$\{\sigma_x,\sigma_y,\sigma_z\}$.  But note that the choice of the two sets need not coincide.
We further define $I^{(1)}_{a,b}:=\1$. For all single-qubit pure states $\ket{\psi}$  we have
\beq\label{spin}\av{X^{(1)}_j}_{\psi}^2+\av{Y^{(1)}_j}_{\psi}^2+
\av{Z^{(1)}_j}_{\psi}^2=\av{I^{(1)}_j}_{\psi}^2, ~~~j=a,b, \eeq and for mixed
states $\rho$ \beq
\label{spin2}\av{X^{(1)}_j}^2+\av{Y^{(1)}_j}^2+\av{Z^{(1)}_j}^2\leq
\av{I^{(1)}_j}^2, ~~~ j=a,b. \eeq

  We write $X_a X_b$  or even $XX$ etc.\, as shorthand for $X_a\otimes X_b$ and
$\av{X X}:= \mathrm{Tr}[\rho X_a\otimes X_b]$ for the
expectation value  in  a general state $\rho$, and $\av{XX}_\Psi := \bra{\Psi}X_a \otimes X_b \ket{\Psi}$  for the
expectation in  a pure state $\ket{\Psi}$.

\forget{ We also assume local orthogonality of the spin
observables
 i.e., $A\perp A'\perp A''$  (where we have included a third orthogonal observable $A''$
 orthogonal to the first two), and $B\perp B'\perp B''$.
\forget{(for the case of two qubits this amount to the local
observables anti-commuting with each other: $\{A,A'\}=0=\{B,B'\}$).
}From now on, we denote such sets of single-qubit orthogonal spin
observables by
 $\{ X^{(1)},Y^{(1)},Z^{(1)}\}$,
 with the choice  of $X^{(1)},~Y^{(1)},~Z^{(1)}$  orthogonal but further
 arbitrary,}
\forget{ It is well known that for all such observables and all
separable states, the Bell
 inequality in the form derived by Clauser, Horne, Shimony and Holt (CHSH) \cite{chsh}
holds:
 \begin{equation}
  |\langle XX + XY + YX -
YY\rangle|  \leq 2,~~\forall\rho_{} \in {\cal
D}^{2\textrm{-sep}}_2~.
 \label{1}
 \end{equation}
}

So, let two triples of locally orthogonal observables be given, denoted as  $\{
X^{(1)}_a,Y^{(1)}_a,Z^{(1)}_a\}$ and $\{
X^{(1)}_b,Y^{(1)}_b,Z^{(1)}_b\}$, where $a,b$ label the different
qubits. We further introduce two sets of four two-qubit operators on
$\mathcal{H}=\mathbb{C}^2\otimes\mathbb{C}^2$, labeled by the
subscript  $x= 0,1$:
\begin{align}
X_0^{(2)}&:=   \half (X^{(1)}X^{(1)} - Y^{(1)}Y^{(1)})
 & X_1^{(2)}&:=   \half (X^{(1)}X^{(1)} + Y^{(1)}Y^{(1)})\nn\\
Y_0^{(2)}&:=   \half (Y^{(1)}X^{(1)} + X^{(1)}Y^{(1)} )
&{Y}_1^{(2)}&:=   \half (Y^{(1)}X^{(1)} - X^{(1)}Y^{(1)} ) \nn\\
 Z_0^{(2)} &:=  \half (Z^{(1)}I^{(1)} + I^{(1)}Z^{(1)})
 &{Z}_1^{(2)}& :=  \half ( Z^{(1)}I^{(1)} - I^{(1)}Z^{(1)} ) \nn\\
I_0^{(2)} &:= \half (I^{(1)}I^{(1)} + Z^{(1)}Z^{(1)})
 &{I}_1^{(2)}& := \half (I^{(1)}I^{(1)} - Z^{(1)}Z^{(1)}).
 \label{set2}
   \end{align}
 Here,  the superscript label
indicates that we are dealing with two-qubit operators. Later on,
$X_x^{(2)}$
 will sometimes be notated as $X_{x,ab}^{(2)}$,  and similarly for $Y_x^{(2)}$,
 $Z_x^{(2)}$ and $I_x^{(2)}$. This more extensive labeling will prove
 convenient for the multi-qubit generalization.
  Note that $(X^{(2)}_x)^2 = (Y^{(2)}_x)^2 = (Z^{(2)}_x)^2 = (I^{(2)}_x)^2 =I^{(2)}_x$
 for $x=0,1$,  and that all eight operators mutually anti-commute.
Furthermore, if the orientations of the two triples is the same,
these two sets  form
 representations of the generalized
Pauli group, i.e., they have the same commutation relations as the
Pauli matrices on $\C^2$, i.e.:
$[X_x^{(2)},Y_x^{(2)}]=2iZ_x^{(2)}$, etc.\, and \beq
\av{X_x^{(2)}}^2+\av{Y_x^{(2)}}^2+\av{Z_x^{(2)}}^2\leq\av{I_x^{(2)}}^2,~~~
x\in \{0,1\}, \label{altijd} \enq with equality only for pure
states.   Note that we can rewrite the CHSH inequality
(\ref{1}) in terms of these observables as: $|\av{X_0^{(2)} +
Y_0^{(2)}}|\leq 1, ~ \forall\rho_{} \in {\cal
D}^{2\textrm{-sep}}_2$.

Let us now consider the separability inequalities of chapter \ref{chapter_CHSHquantumorthogonal}. In terms of the observables of (\ref{set2}) the separability inequality of (\ref{New}) becomes:
\begin{align} \av{X_x^{(2)}}^2 + \av{Y_x^{(2)}}^2   \leq
\frac{1}{4}(\av{I_a^{(1)}} - \av{Z^{(1)}_a}^2)(\av{I_b^{(1)}} - \av{Z^{(1)}_b}^2)
,~~~\forall\rho_{} \in {\cal
D}^{2\textrm{-sep}}_2 .\label{NEW}
\end{align}
\forget{
Clearly, the right-hand side of this inequality is bounded by 1/4.
However, for  entangled states (e.g., for the Bell states
$\ket{\phi^{\pm}}=(\ket{00} \pm
\ket{11})/\sqrt{2}$ and
$\ket{\psi^{\pm}}=(\ket{01} \pm \ket{10})/\sqrt{2}$) the left-hand side can attain the
value of 1 and thus violate the bound (\ref{NEW}).  Hence,
inequality (\ref{NEW}) provides a non-trivial bound for separable
states, and thus a criterion for testing entanglement.
}
\forget{It is also instructive to relate the results obtained so far to the original Bell-CHSH inequality (\ref{chsheq}).} 
Since $(\av{X_0^{(2)}+ Y_0^{(2)}})^2 +(\av{X_0^{(2)} - Y_0^{(2)}})^2=2 (\av{X_0^{(2)}}^2 + \av{Y_0^{(2)}}^2)$ we obtain from (\ref{NEW}) that $|\av{X_0^{(2)} + Y_0^{(2)}}|\leq \sqrt{1/2}$, thereby reproducing (\ref{sqrt2}) which shows the strengthening of the CHSH separability inequality by a factor $\sqrt{2}$.

 \forget{
 Usually Bell inequalities are considered
interesting only if there exist a certain quantum state that
violates it for certain measurement settings. However, we see here
that it is also very interesting to ask not what quantum states
violate a certain Bell inequality, but what quantum states cannot
violate such a Bell inequality, and by what factor.  This is
especially so if this factor is lower than the bound for LHV models since  then LHV correlations would
exist which cannot be reproduced by separable quantum states,  cf.\  \cite{uffseev}. In fact,
section \ref{Nqubitsection} shows that this
divergence between  the correlations obtainable by LHV models and by
separable quantum states generalizes to the $N$-qubit case and
increases exponentially with $N$.
}

The separability condition of (\ref{NEW}) can be strengthened even further as was done in section \ref{necsuf} to produce (\ref{mixsep2}). In terms of the notation of this chapter this separability condition is
\forget{. Let us first
temporarily assume the state to be pure and separable, $\ket{\Psi}
=\ket{\psi}\ket{\phi}$.  Then we obtain from (\ref{onep})
\begin{align}
\av{X_0^{(2)}}_{{\Psi}}^2 + \av{Y_0^{(2)}}_{\Psi}^2 =\av{X_1^{(2)}}_{{\Psi}}^2
+ \av{Y_1^{(2)}}_{{\Psi}}^2  \label{XYxy},
\end{align}
and similarly: \begin{align}
\av{I_0^{(2)}}_{{\Psi}}^2 - \av{Z_0^{(2)}}_{{\Psi}}^2 =
\frac{1}{4} \left( \av{I_a^{(1)}}_\psi -\av{Z_a^{(1)}}_\psi^2
\right) \left( \av{I_b^{(1)}}_\phi - \av{Z_b^{(1)}}_\phi^2 \right)
= \av{I_1^{(2)}}_{{\Psi}}^2 - \av{Z_1^{(2)}}_{{\Psi}}^2  \label{IZiz}.\end{align}
In view of (\ref{NEWA}) we conclude that for all pure separable
states all expressions in the equations (\ref{XYxy}) and (\ref{IZiz})
are equal to each other.   Of course, this conclusion does not
hold for mixed separable states. However,  $\sqrt{\av{X_0^{(2)}}^2 + \av{Y_0^{(2)}}^2}$
and   $\sqrt{\av{X_1^{(2)}}^2 + \av{Y_1^{(2)}}^2}$ are convex functions of $\rho$ whereas
 the three expressions
 $\sqrt{\av{I_0^{(2)}}^2 - \av{Z_0^{(2)}}^2}$,
 $\sqrt{\frac{1}{4} ( \av{I_a^{(1)}} -\av{Z_a^{(1)}}^2 )
(\av{I_b^{(1)}} - \av{I_a^{(1)}}^2 )}$
 and $\sqrt{\av{I_1^{(2)}}^2 - \av{Z_1^{(2)}}^2}$
 are all concave in $\rho$.
 Therefore we can repeat
 a similar
 chain of reasoning as in (\ref{convexA})  to  obtain  the following inequalities, which are valid
for all mixed separable states:
\forget{\begin{align} \left.
\begin{array}{c}
\av{X_0^{(2)}}^2 + \av{Y_0^{(2)}}^2 \\
\av{X_1^{(2)}}^2
+ \av{Y_1^{(2)}}^2  \end{array} \right\}
\leq
\left\{ \begin{array}{c}
\av{I_0^{(2)}}^2 - \av{Z_0^{(2)}}^2 \\
\frac{1}{4} \left( \av{I_a^{(1)}} -\av{Z_a^{(1)}}^2 \right)
\left(\av{I_b^{(1)}} - \av{Z_b^{(1)}}^2 \right)\\
\av{I_1^{(2)}}^2 - \av{Z_1^{(2)}}^2
\end{array}
\right\} \leq\frac{1}{4}, ~~~\forall\rho_{} \in {\cal
D}^{2\textrm{-sep}}_2.\label{Mixsep} \end{align}

{\bf Alternative formulation:}}
\begin{align} \max\left \{
\begin{array}{c}
\av{X_0^{(2)}}^2 + \av{Y_0^{(2)}}^2 \\
\av{X_1^{(2)}}^2
+ \av{Y_1^{(2)}}^2  \end{array} \right\}
\leq \min
\left\{ \begin{array}{c}
\av{I_0^{(2)}}^2 - \av{Z_0^{(2)}}^2 \\
\frac{1}{4} \left( \av{I_a^{(1)}} -\av{Z_a^{(1)}}^2 \right)
\left(\av{I_b^{(1)}} - \av{Z_b^{(1)}}^2 \right)\\
\av{I_1^{(2)}}^2 - \av{Z_1^{(2)}}^2
\end{array}
\right\} \leq\frac{1}{4}, ~~~\forall\rho_{} \in {\cal
D}^{2\textrm{-sep}}_2.
\label{Mixsep}\end{align}

This result extends the previous inequality  (\ref{NEW}).
  The next obvious  question is then which of the three  right-hand
sides in (\ref{Mixsep}) provides  the lowest  upper bound. It is
not difficult to show that the ordering of these three expressions
depends on the correlation coefficient $C = \av{Z_a^{(1)}Z_b^{(1)}}
- \av{Z_a^{(1)}}\av{Z_b^{(1)}}$. A straightforward calculation shows
that if $C \geq 0$, we get
\begin{align}
\av{I_0^{(2)}}^2 - \av{Z_0^{(2)}}^2 \leq
\frac{1}{4} \left( \av{I_a^{(1)}} -\av{Z_a^{(1)}}^2 \right) \left(
\av{I_b^{(1)}} - \av{Z_b^{(1)}}^2 \right)
\leq \av{I_1^{(2)}}^2 - \av{Z_1^{(2)}}^2,
 \end{align}
 while these inequalities are inverted  when $C \leq 0$.  Hence,
 depending on the sign of $C$,
 either $\av{I_0^{(2)}}^2 - \av{Z_0^{(2)}}^2$ or
 $\av{I_1^{(2)}}^2 - \av{Z_1^{(2)}}^2 $ yields the
sharper upper bound. In other words, for all separable quantum
states one has:
\forget{
\begin{align} \left.
\begin{array}{c}\av{X_0^{(2)}}^2 + \av{Y_0^{(2)}}\\
\av{X_1^{(2)}}^2 + \av{Y_1^{(2)}}\end{array} \right\}
\leq
\left\{ \begin{array}{c}
\av{I_0^{(2)}}^2 - \av{Z_0^{(2)}}^2\\
\av{I_1^{(2)}}^2 - \av{Z_1^{(2)}}^2  \end{array}
 \right\}\leq \frac{1}{4}, ~~\forall\rho_{} \in {\cal
D}^{2\textrm{-sep}}_2.\label{Mixsep2} \end{align}
{\bf Alternative formulation:}}
}

\begin{align} 
\max \left\{ \begin{array}{c}\av{X_0^{(2)}}^2 + \av{Y_0^{(2)}}\\
\av{X_1^{(2)}}^2 + \av{Y_1^{(2)}}\end{array} \right\}
\leq
\min \left\{ \begin{array}{c}
\av{I_0^{(2)}}^2 - \av{Z_0^{(2)}}^2\\
\av{I_1^{(2)}}^2 - \av{Z_1^{(2)}}^2  \end{array}
 \right\}\leq \frac{1}{4}, ~~\forall\rho_{} \in {\cal
D}^{2\textrm{-sep}}_2. \label{Mixsep2} 
\end{align}
\forget{This set of inequalities provides the announced strengthening of (\ref{NEW}).
This improvement pays off: in contrast to (\ref{NEW}), the validity of the
inequalities (\ref{Mixsep2}) for all orthogonal triples $\{ X^{(1)}_a,Y^{(1)}_a,Z^{(1)}_a\}$ and
$\{ X^{(1)}_b,Y^{(1)}_b,Z^{(1)}_b\}$ provides a necessary and sufficient condition for separability for
all two-qubit states, pure or mixed. (See \cite{uffseev} for a proof).}

If we leave out the upper bound of $1/4$ in (\ref{Mixsep2}), then of
the four inequalities in (\ref{Mixsep2}) two are trivially true for
all states, whether separable or not, and the other two in fact
provide the separability criteria. Which two of the four depends on
the orientation of the local orthogonal observables.\forget{, for by changing
the relative orientation the correlation coefficient $C$ changes
sign.} Let us choose the orientations for both parties to be the
same, then the non-trivial inequalities are $\av{X_0^{(2)}}^2 +
\av{Y_0^{(2)}}\leq \av{I_1^{(2)}}^2 - \av{Z_1^{(2)}}^2$ and
$\av{X_1^{(2)}}^2 + \av{Y_1^{(2)}} \leq \av{I_0^{(2)}}^2 -
\av{Z_0^{(2)}}^2$.
 Choosing the orientations to be different gives the other two non-trivial inequalities.

\forget{
To conclude this section we give an explicit example of the
separability inequalities of (\ref{Mixsep2}) by choosing the
ordinary Pauli matrices $\{\sigma_x,\sigma_y,\sigma_z\}$ for the
triples $\{ X^{(1)}_a,Y^{(1)}_a,Z^{(1)}_a\}$ and $\{
X^{(1)}_b,Y^{(1)}_b,Z^{(1)}_b\}$.
This choice allows us to write the inequalities in terms of the
density matrix elements on the standard two qubit $z$-basis
$\{\ket{00},\ket{01},\ket{10},\ket{11}\}$: if $\rho_{} \in {\cal
D}^{2\textrm{-sep}}_2$ then we must have\footnote{Here we use that this choice of observables give
the following relations on the $z$-basis:
$X_0^{(2)}= \ket{00}\bra{11} +\ket{11}\bra{00}$, $\av{ X_0^{(2)}}=2\mathrm{Re}\, \rho_{14}$,
$Y_0^{(2)}= -i\ket{00}\bra{11} +i\ket{11}\bra{00}$, $\av{ Y_0^{(2)}}=-2\mathrm{Im}\, \rho_{14}$,
$I_0^{(2)}=  \ket{00}\bra{00} +\ket{11}\bra{11}$, $\av{ I_0^{(2)}}=\rho_{11}+ \rho_{44}$,
$Z_0^{(2)}= \ket{00}\bra{00} -\ket{11}\bra{11}$, $\av{ Z_0^{(2)}}=\rho_{11} -\rho_{44}$.
$X_1^{(2)}= \ket{01}\bra{10} +\ket{10}\bra{01}$, $\av{ X_1^{(2)}}=2\mathrm{Re}\, \rho_{23}$,
$Y_1^{(2)}= -i\ket{01}\bra{10} +i\ket{10}\bra{01}$, $\av{ Y_1^{(2)}}=-2\mathrm{Im}\, \rho_{23}$,
$I_1^{(2)}=  \ket{01}\bra{01} +\ket{10}\bra{10}$, $\av{ I_1^{(2)}}=\rho_{22}+ \rho_{33}$,
$Z_1^{(2)}= \ket{01}\bra{01} -\ket{10}\bra{10}$, $\av{ Z_1^{(2)}}=\rho_{22} -\rho_{33}$.}:
$\max\{ |\rho_{1,4}|^2,\, |\rho_{2,3}|^2 \} \leq \min\{
\rho_{1,1}\rho_{4,4},\,\rho_{2,2}\rho_{3,3}\}\leq \frac{1}{16}$,
(where of course $|\rho_{1,4}|=|\rho_{4,1}|$ and
$|\rho_{2,3}|=|\rho_{3,2}|$). 

As was shown in chapter \ref{chapter_CHSHquantumorthogonal} this strengthens the Laskowski-\.Zukowski 
condition (\ref{anti-diagonal}) as well as the fidelity condition (\ref{Fidelity}) in the $2$ qubit case.
}

To conclude this section we give an explicit form of the
separability inequalities (\ref{Mixsep2}) by choosing the
Pauli matrices $\{\sigma_x,\sigma_y,\sigma_z\}$ for both triples
$\{ X^{(1)}_a\!,Y^{(1)}_a\!,Z^{(1)}_a\}$ and $\{
X^{(1)}_b\!,Y^{(1)}_b\!,Z^{(1)}_b\}$. This choice enables us to write
the inequalities (\ref{Mixsep2}) in terms of the density matrix
elements on the standard  $z$-basis
$\{\ket{00},\ket{01},\ket{10},\ket{11}\}$, labeled here as  $\{\ket{1},\ket{2},\ket{3},\ket{4}\}$.
  This choice of observables yields  $\av{
X_0^{(2)}}=2\mathrm{Re}\, \rho_{1,4}$,  $\av{
Y_0^{(2)}}=-2\mathrm{Im}\, \rho_{1,4}$, $\av{ I_0^{(2)}}=\rho_{1,1}+
\rho_{4,4}$,  $\av{ Z_0^{(2)}}=\rho_{1,1} -\rho_{4,4}$,
 $\av{X_1^{(2)}}=2\mathrm{Re}\, \rho_{2,3}$,
 \forget{ $Y_1^{(2)}=
-i\ket{01}\bra{10} +i\ket{10}\bra{01}$,}
 $\av{ Y_1^{(2)}}=-2\mathrm{Im}\, \rho_{2,3}$, \forget{, $I_1^{(2)}=
\ket{01}\bra{01} +\ket{10}\bra{10}$,} $\av{ I_1^{(2)}}=\rho_{2,2}+
\rho_{3,3}$, \forget{ $Z_1^{(2)}= \ket{01}\bra{01}
-\ket{10}\bra{10}$,} $\av{ Z_1^{(2)}}=\rho_{2,2} -\rho_{3,3}$.
So, in this choice, we can write  (\ref{Mixsep2}) as:
\begin{align}
\max\{ |\rho_{1,4}|^2,\, |\rho_{2,3}|^2 \} \leq \min\{
\rho_{1,1}\rho_{4,4},\, \rho_{2,2} \rho_{3,3} \}\leq \frac{1}{16}, ~~~~   \rho_{} \in {\cal
D}^{2\textrm{-sep}}_2.
 \label{2sepmatrix} 
\end{align}

In the form (\ref{2sepmatrix}), it is easy to compare the result
to the separability conditions reviewed in subsection II.B\footnote{ This comparison has already been performed in chapter 
 \ref{chapter_CHSHquantumorthogonal}  but will be repeated here for completeness}.  
Assume for simplicity that $|\rho_{1,4}|$ is the largest of all
the anti-diagonal elements  $|\rho_{j\bar{\jmath}}|$. Then, for
$\rho_{} \in {\cal D}^{2\textrm{-sep}}_2$, and using $\av{M^{(2)}}^2 +\av{M'^{(2)}}^2=8(\av{X_0^{(2)}}^2+\av{Y_0^{(2)}}^2)$ the Mermin-type separability inequality \eqref{quadraticN} becomes $|\rho_{1,4}|^2\leq 1/16$, which is equivalent to the Laskowski-\.Zukowski condition $|\rho_{1,4}|\leq 1/4$; the
fidelity/D\"ur-Cirac conditions read: $2|\rho_{1,4}|\leq
\rho_{2,2} +\rho_{3,3}$; and the condition (\ref{2sepmatrix}):
 $|\rho_{1,4}|^2\leq \rho_{2,2}\rho_{3,3}$. Using the trivial inequality $(\sqrt{\rho_{22}} -\sqrt{\rho_{33}})^2\geq0\Longleftrightarrow  2\sqrt{ \rho_{2,2}\rho_{3,3}}\leq \rho_{2,2} +\rho_{3,3}$, we can then write the following chain of inequalities
\beq
  4|\rho_{1,4}|- (\rho_{1,1}+\rho_{4,4}) \overset{A}{\leq}2 |\rho_{1,4}| \overset{\textrm{sep}}{\leq} 2\sqrt{\rho_{2,2}\rho_{3,3}}  \overset{A}{\leq}\rho_{2,2} + \rho_{3,3}\,,
  \label{inequ2}
  \eeq
where we used the symbols  $\overset{A}{\leq}$
and $\overset{\textrm{sep}}{\leq}$ to denote inequalities that hold for all states, and for the separability condition (\ref{2sepmatrix}) respectively.

The Laskowski-\.Zukowski condition is then recovered by comparing the first and fourth expressions in this chain, the fidelity/ D\"ur-Cirac conditions by comparing the second and fourth expression, and a new condition -- not previously mentioned -- can be obtained by
comparing the first and third term, whereas condition (\ref{2sepmatrix}), i.e.
the comparison between the second and third expression in (\ref{inequ2}), is the strongest inequality in this chain, and thus implies and strengthens all of these other conditions.

\subsection{Three-qubit case}\label{N3}

 We now derive separability conditions that distinguish
the $10$ classes in the $3$-qubit classification of section
\ref{generalksep} by generalizing the method of section
\ref{twoqubitsection}. To begin with, define four sets of
three-qubit observables from the two-qubit operators  (\ref{set2}):
\begin{align}
X_0^{(3)} &:=\frac{1}{2}\,(X^{(1)}   X^{(2)}_0 -Y^{(1)}  Y^{(2)}_0)
&{X}_1^{(3)}
&:=\frac{1}{2}\,(X^{(1)}   {X}^{(2)}_0 +Y^{(1)}  Y^{(2)}_0)\nn
\\
Y_0^{(3)}  &:= \frac{1}{2}\,(Y^{(1)}  X^{(2)}_0+X^{(1)}   Y^{(2)}_0) &
{Y}_1^{(3)}
&:=\frac{1}{2}\,(Y^{(1)}   X^{(2)}_0-X^{(1)}  Y^{(2)}_0 )\nn
\\
Z_0^{(3)}  &:= \frac{1}{2}\,(Z^{(1)}   I^{(2)}_0+I^{(1)}  Z^{(2)}_0) &{Z}_1^{(3)}
&:=\frac{1}{2}\,(Z^{(1)}  I^{(2)}_0-I^{(1)}   Z^{(2)}_0 )\nn
\\
I_0^{(3)}&:=\frac{1}{2}\,(I^{(1)}   I^{(2)}_0 +Z^{(1)}  Z^{(2)}_0) &{I}_1^{(3)}
&:=\frac{1}{2}\,(I^{(1)}   {I}^{(2)}_0 -Z^{(1)}  Z^{(2)}_0)\nn
\\\nn\\
X_2^{(3)}  &:=\frac{1}{2}\,(X^{(1)}   X^{(2)}_1 -Y^{(1)}
Y^{(2)}_1) &
X_3^{(3)} &:=\frac{1}{2}\,(X^{(1)}
 X^{(2)}_1 +Y^{(1)} Y^{(2)}_1)\nn
\\
Y^{(3)}_2 &:=\frac{1}{2}\,(Y^{(1)}   {X}^{(2)}_1 +X^{(1)}
{Y}^{(2)}_1 ) &Y_3^{(3)} &:=\frac{1}{2}\,(Y^{(1)}   {X}^{(2)}_1-X^{(1)}
{Y}^{(2)}_1 )\nn
\\
Z_2^{(3)} &:=\frac{1}{2}\,(Z^{(1)}  {I}^{(2)}_1 +I^{(1)}
{Z}^{(2)}_1) &Z_3^{(3)} &:=\frac{1}{2}\,(Z^{(1)}
  {I}^{(2)}_1-I^{(1)}  {Z}^{(2)}_1 )\nn
\\
I_2^{(3)} &:=\frac{1}{2}\,(I^{(1)}   {I}^{(2)}_1 +Z^{(1)}
 Z^{(2)}_1)
 &I^{(3)}_3 &:=\frac{1}{2}\,(I^{(1)}   {I}^{(2)}_1
 -Z^{(1)}  {Z}^{(2)}_1)
,\label{N3operators}
\end{align}
where $X^{(1)}X^{(2)}_{0}= X^{(1)}_a \otimes X^{(2)}_{0,bc}$,
etc.,  $a,b,c$ label the three qubits. In analogy to
 the two-qubit case, we note that all these operators anticommute and that if the orientations of the triples  for each qubit
are the same, the operators in (\ref{N3operators}) yield
representations of the generalized Pauli group:
$[X_x^{(3)},Y_x^{(3)}]=2iZ_x^{(3)}$, for    $x=0,1,2,3$. For
convenience, we will indeed assume these orientations to be the
same, unless noted otherwise. Choosing orientations differently
would yield similar separability
 conditions, in the same vein as in the previous section.
Under this choice we have, for all $k$,  \beq
\av{X_x^{(3)}}^2+\av{Y_x^{(3)}}^2+\av{Z_x^{(3)}}^2\leq\av{I_x^{(3)}}^2,~~~\forall
\rho \in {\cal D}_N^{k\textrm{-sep}} \label{eenheid} \eeq
 with equality only for pure states.

We now derive conditions for the different levels and classes of
the partial separability classification. Most of the proofs are by
straightforward generalization of the method of the previous
section and these will be omitted.

Suppose first that the three-qubit state  is pure and separable
under split $a$-$(bc)$. From the definitions  (\ref{N3operators})
we obtain: 
\begin{alignat}{2}
 \av{X_0^{(3)} }^2 +\av{Y_0^{(3)} }^2&=
\frac{1}{4}\,(\,\av{X^{(1)}_a}^2+\av{Y^{(1)}_a}^2\,)\,
(\,\av{X^{(2)}_{0,bc}}^2 +\av{Y^{(2)}_{0,bc}}^2\,)&&=
 \av{X_1^{(3)} }^2 +\av{Y_1^{(3)} }^2 \nn\\&= \label{2_sep_a_bcX}
\\
\av{I_0^{(3)} }^2 -\av{Z_0^{(3)}
}^2&=\frac{1}{4}\,(\,\av{I^{(1)}_a}^2
-\av{Z^{(1)}_a}^2\,)\,(\,\av{I^{(2)}_{0,bc}}^2
-\av{Z^{(2)}_{0,bc}}^2\,)&&=
 \av{I_1^{(3)} }^2 -\av{Z_1^{(3)} }^2,
 \nn
\\
\av{X_2^{(3)} }^2 +\av{Y_2^{(3)}
}^2&=\frac{1}{4}\,(\,\av{X^{(1)}_a}^2
+\av{Y^{(1)}_a}^2\,)\,(\,\av{X^{(2)}_{1,bc}}^2
+\av{Y^{(2)}_{1,bc}}^2\,) &&= \av{X_3^{(3)} }^2 +\av{Y_3^{(3)} }^2\nn\\
&= \label{2_sep_a_bcI}
\\
\av{I_2^{(3)} }^2 -\av{Z_2^{(3)} }^2&=\frac{1}{4}\,(\,\av{I^{(1)}_a}^2
-\av{Z^{(1)}_a}^2\,)\,(\,\av{I^{(2)}_{1,bc}}^2 -\av{Z^{(2)}_{1,bc}}^2\,)&&=
 \av{I_3^{(3)} }^2 -\av{Z_3^{(3)} }^2.\nn
  \end{alignat}
  Similarly, for pure states  that are separable under
split $b$-$(ac)$, we obtain  analogous equalities by interchanging
the labels  $x=1$ and $x=3$ (denoted as $1 \leftrightarrow
3$); and for split $c$-$(ab)$ by $1\leftrightarrow 2$.

Of course,  these equalities  hold for pure states only, but by
 the convex analysis of section \ref{twoqubitsection} we obtain from \eqref{2_sep_a_bcX}, \eqref{2_sep_a_bcI} inequalities for all mixed states that are bi-separable under
the split $a$-$(bc)$:
\begin{align}\begin{array}{clcl}
\max\limits_{x \in \{0,1\}} \av{X_x^{(3)} }^2 +\av{Y_x^{(3)} }^2 \leq
\min\limits_{x \in \{ 0,1\}} \av{I_x^{(3)} }^2 -\av{Z_x^{(3)} }^2
\leq \frac{1}{4}
\\
\max\limits_{x \in \{2,3\}} \av{X_x^{(3)} }^2 +\av{Y_x^{(3)} }^2 \leq
\min\limits_{x \in \{ 2,3\}} \av{I_x^{(3)} }^2 -\av{Z_x^{(3)} }^2 \leq
\frac{1}{4}
\end{array}, ~~ ~\forall\rho \in {\cal D}_3^{a\textrm{-}(bc)}.
  \label{ineq_2sep}
  \end{align}
For states that are bi-separable under split $b$-$(ac)$ the
analogous inequalities with $1\leftrightarrow 3$ hold, i.e.,
\begin{align}
\begin{array}{clcl}\max\limits_{x \in \{0,3\}} \av{X_x^{(3)} }^2 +\av{Y_x^{(3)} }^2 \leq
\min\limits_{x \in \{ 0,3\}} \av{I_x^{(3)} }^2 -\av{Z_x^{(3)} }^2
\leq \frac{1}{4}\\
\max\limits_{x \in \{1,2\}} \av{X_x^{(3)} }^2 +\av{Y_x^{(3)} }^2 \leq
\min\limits_{x \in \{ 1,2\}} \av{I_x^{(3)} }^2 -\av{Z_x^{(3)} }^2 \leq
\frac{1}{4}
\end{array},  ~~ ~\forall\rho \in {\cal D}_3^{b\textrm{-}(ac)}.
  \label{ineq_2sepb}
  \end{align}
 and for
the split $c$-$(ab)$ we need to replace $1 \leftrightarrow 2$:
\begin{align}
\begin{array}{clcl}\max\limits_{x \in \{0,2\}} \av{X_x^{(3)} }^2 +\av{Y_x^{(3)} }^2 \leq
\min\limits_{x \in \{ 0,2\}} \av{I_x^{(3)} }^2 -\av{Z_x^{(3)} }^2
\leq \frac{1}{4}
\\
\max\limits_{x \in \{1,3\}} \av{X_x^{(3)} }^2 +\av{Y_x^{(3)} }^2 \leq
\min\limits_{x \in \{ 1,3\}} \av{I_x^{(3)} }^2 -\av{Z_x^{(3)} }^2 \leq
\frac{1}{4}
\end{array},
 ~~ ~\forall\rho \in {\cal D}_3^{c\textrm{-}(ab)}.
  \label{ineq_2sepc}
  \end{align}

A general bi-separable state $\rho_{} \in {\cal
D}^{2\textrm{-sep}}_3$  is a  convex mixture of states that are
separable under some bi-partite split, i.e.,
 $ \rho=p_1 \rho_{a\textrm{-}(bc)}+p_2 \rho_{b\textrm{-}(ac)}+p_3 \rho_{c\textrm{-}(ab)}$ with
$\sum_{j=1}^3 p_j=1$.
 Since $\sqrt{\av{{X^{(3)}_0}}^2 +\av{{Y^{(3)}_0}}^2}$ is convex in $\rho$ we get from (\ref{ineq_2sep}- \ref{ineq_2sepc})
 for such a state:
 \begin{align}\label{convex2sep3}
\sqrt{\av{{X^{(3)}_0}}^2 +\av{{Y^{(3)}_0}}^2} = p_1\sqrt{
\av{{X^{(3)}_0}}^2_{ \rho_{a\textrm{-}(bc)}} \!\!+
\av{{Y^{(3)}_0}}^2_{ \rho_{a\textrm{-}(bc)}}}\!&+ p_2\sqrt{
\av{X^{(3)}_0}^2_{ \rho_{b\textrm{-}(ac)}} \!\!+\av{Y^{(3)}_0}^2_{ \rho_{b\textrm{-}(ac)}}}\! \nn\\ 
& +
p_3\sqrt{\av{X^{(3)}_0}^2_{ \rho_{c\textrm{-}(ab)}}\!\! +\av{Y^{(3)}_0}^2_{ \rho_{c\textrm{-}(ab)}}}\nn\\
\leq p_1 \sqrt{\av{I^{(3)}_1}^2_{ \rho_{a\textrm{-}(bc)}}
\!\! -  \av{Z^{(3)}_1}^2_{ \rho_{a\textrm{-}(bc)}}}\! &+ p_2\sqrt{
\av{I^{(3)}_3}^2_{ \rho_{b\textrm{-}(ac)}}\!\! - \av{Z^{(3)}_3}^2_{ \rho_{b\textrm{-}(ac)}}}\! 
\nn\\& + p_3 \sqrt{\av{I^{(3)}_2}^2_{ \rho_{c\textrm{-}(ab)}}\!\! -
\av{Z^{(3)}_2}^2_{ \rho_{c\textrm{-}(ab)}}}.
\end{align}
Here $\av{\cdot}_{ \rho_{a\textrm{-}(bc)}}$ means taking the
expectation value in the state $ \rho_{a\textrm{-}(bc)}$, etc.
Analogous bounds hold for the expressions $\sqrt{\av{X_x^{(3)} }^2
+\av{Y_x^{(3)} }^2}$ for  $x=1,2,3$.

From the numerical upper bounds in the conditions
(\ref{ineq_2sep}- \ref{ineq_2sepc}) it is easy to obtain a first
bi-separability  condition: \beq\label{first3} \av{X_x^{(3)} }^2
+\av{Y_x^{(3)} }^2\leq 1/4, ~~~ \forall\rho_{} \in {\cal
D}^{2\textrm{-sep}}_3, ~~~x\in\{0,1,2,3\}. \eeq This is
equivalent to the Laskowski-\.Zukowski condition
(\ref{anti-diagonal}) for $k=2$, as will be shown below.
 However, a stronger condition can be obtained  by noting that $\sqrt{\av{I^{(3)}_y}^2
-\av{Z^{(3)}_y}^2}$ is concave in $\rho$ so that
\begin{align} \label{concave}
p_1\sqrt{
\av{{I^{(3)}_y}}^2_{ \rho_{a\textrm{-}(bc)}} -
\av{{Z^{(3)}_y}}^2_{ \rho_{a\textrm{-}(bc)}}}&+ p_2\sqrt{
\av{I^{(3)}_y}^2_{ \rho_{b\textrm{-}(ac)}} -\av{Z^{(3)}_y}^2_{ \rho_{b\textrm{-}(ac)}}}\\&
\hskip-1cm+
p_3\sqrt{\av{I^{(3)}_y}^2_{ \rho_{c\textrm{-}(ab)}} -\av{Z^{(3)}_y}^2_{ \rho_{c\textrm{-}(ab)}}}\leq
\sqrt{\av{I^{(3)}_y}^2 -\av{Z^{(3)}_y}^2}.\nn
\end{align}
After taking a sum over $y\neq x$ in (\ref{concave}),  the left hand
side of (\ref{concave}) is larger than the right hand side of
(\ref{convex2sep3}). This yields a stronger condition for
bi-separability  of  $3$-qubit states 
\begin{align}
\sqrt{\av{X^{(3)}_{x}}^2 +\av{Y^{(3)}_{x}}^2} \leq\sum_{y\neq x}
\sqrt{ \av{I^{(3)}_{y}}^2 -\av{Z^{(3)}_{y}}^2}, ~~~ \forall\rho
\in {\cal D}^{2\textrm{-sep}}_3, ~~~x,y\in\{0,1,2,3\}. 
\label{2sep3}
\end{align}

 That
(\ref{2sep3}) is indeed a stronger than (\ref{first3}) will be
shown below using the density matrix representation of this
condition.
 If one
would alter the orientation of the orthogonal triple of
observables for a certain qubit, then the right-hand side of
(\ref{2sep3}) changes by adding either $1$, $2$ or $3$ (modulo
$3$) to $x$  in the sum on the right hand side, depending on for
which qubit the orientation was changed.

Next, consider the case  of a  $3$-separable state, $\rho \in
{\cal D}_3^{3\textrm{-sep}}$.  One might then use the fact that
this split is contained in all three bi-partite splits $a$-$(bc)$,
$b$-$(ac)$ and $c$-$(ab)$ to conclude that the  inequalities
(\ref{ineq_2sep}, \ref{ineq_2sepb}, \ref{ineq_2sepc}) must hold
simultaneously. Thus,  3-separable states must obey: \beq \max_{x}
\{ \av{X^{(3)}_{x}}^2 +\av{Y^{(3)}_{x}}^2 \} \leq \min_{x} \{
\av{I^{(3)}_{x}}^2 -\av{Z^{(3)}_{x}}^2\} \leq
\frac{1}{4},~~~\forall \rho \in {\cal D}_3^{3\textrm{-sep}}
\label{3sep3}.
 \eeq
However, a more stringent condition
 holds by virtue of
the following equalities  for  pure $3$-separable states: 
\begin{align}
\av{{X^{(3)}_0}}^2 +\av{{Y^{(3)}_0}}^2 &=
\frac{1}{16}\,(\,\av{X^{(1)}_a}^2+\av{Y^{(1)}_a}^2\,)
\,(\,\av{X^{(1)}_b}^2+\av{Y^{(1)}_b}^2\,)\,(\,\av{X^{(1)}_c}^2+\av{Y^{(1)}_c}^2\,)
\nn\\
&=\av{X^{(3)}_1}^2 +\av{Y^{(3)}_1}^2=\av{ X^{(3)}_2}^2 +
\av{Y^{(3)}_2}^2=\av{ X^{(3)}_3}^2 +
\av{Y^{(3)}_3}^2,
\\
\av{{I^{(3)}_0}}^2 -\av{{Z^{(3)}_0}}^2 &=
\frac{1}{16}\,(\,\av{I^{(1)}_a}^2-\av{Z^{(1)}_a}^2\,)
\,(\,\av{I^{(1)}_b}^2-\av{Z^{(1)}_b}^2\,)\,(\,\av{I^{(1)}_c}^2-\av{Z^{(1)}_c}^2\,)
\nn\\
&=\av{I^{(3)}_1}^2 -\av{Z^{(3)}_1}^2=\av{ I^{(3)}_2}^2 -
\av{Z^{(3)}_2}^2=\av{ I^{(3)}_3}^2 -
\av{Z^{(3)}_3}^2.\label{completesep} 
\end{align} From these equalities
for pure states it is easy to obtain, by a convexity argument
similar to previous cases, an upper bound of $1/16$
instead of $1/4$ in (\ref{3sep3}):
  \beq \max_{x}
\{ \av{X^{(3)}_{x}}^2 +\av{Y^{(3)}_{x}}^2 \} \leq \min_{x} \{
\av{I^{(3)}_{x}}^2 -\av{Z^{(3)}_{x}}^2\} \leq
\frac{1}{16},~~~\forall \rho \in {\cal D}_3^{3\textrm{-sep}}
\label{3sep3b}.
 \eeq\forget{Note that by the same convex
analysis of section \ref{twoqubitsection} these equalities allow
for direct derivation of the inequality (\ref{3sep3}) without
using the bi-separability inequalities (\ref{ineq_2sep}).}

In summary, the states at the different separability levels $k=1,2,3$ have state independent bounds for the quantities $\av{X^{(3)}_{x}}^2 +\av{Y^{(3)}_{x}}^2$, $x\in\{0,1,2,3\}$, that differ a factor $4$ for each level. They are respectively $1$, $1/4$ and  $1/16$. These bounds can be strengthened by state-dependent bounds that use the quantities  $\av{I^{(3)}_{x}}^2 -\av{Z^{(3)}_{x}}^2$.
This gives the separability inequalities (\ref{2sep3}), (\ref{3sep3}) and \eqref{3sep3b}.   Separability with respect to specific splits results in strong  state-dependent bounds such  as for example given in (\ref{ineq_2sep}) for the split $a$-$(bc)$.  The conditions obtained give 
different conditions for each of the 10 classes
in the full separability classification of three qubits,
summarized in table \ref{table1}.\vskip0.3cm\noindent

\begin{table}[!h]
\begin{center}{
\begin{tabular}{|l|l|} \hline
Class &  Separability conditions \\ \hline
1 &  (\ref{eenheid}) \\
2.1& (\ref{2sep3})\\
 2.2 & (\ref{ineq_2sep})  \\
 2.3 & (\ref{ineq_2sepb})  \\
 2.4 & (\ref{ineq_2sepc}) \\
 2.5  &  (\ref{ineq_2sep}) \&  (\ref{ineq_2sepb}) but not (\ref{ineq_2sepc})\\
2.6  &   (\ref{ineq_2sep}) \&  (\ref{ineq_2sepc}) but not
(\ref{ineq_2sepb})
 \\
2.7 &  (\ref{ineq_2sepb}) \& (\ref{ineq_2sepc}) but not (\ref{ineq_2sep}) \\
2.8 &  ((\ref{ineq_2sep}) \& (\ref{ineq_2sepb})\& (\ref{ineq_2sepc}))$ \equivto$ (\ref{3sep3}) \\
 3 &   (\ref{3sep3b})\\ \hline
\end{tabular}
\caption{Separability conditions for the 10 classes in the
separability classification of three-qubit states.} \label{table1}
}\end{center}
\end{table}
Violations of these partial separability conditions
  give sufficient conditions for particular types of entanglement. For example,  if inequality (\ref{3sep3b}) is violated, then the state  must be in one of the
bi-separable classes $2.1$ to $2.8$ or in class $1$, which implies
that the state is at least $2$-partite entangled;  if (\ref{2sep3})
violated it is in class $1$  and thus fully inseparable (fully
entangled), and so on.

In order to gain further familiarity with the above separability
inequalities, we choose the ordinary Pauli matrices $\{\sigma_x,
\sigma_y, \sigma_z\}$ for the locally orthogonal observables $ \{
X^{(1)}, Y^{(1)}, Z^{(1)}\}$, and formulate them in terms of
density matrix elements in the standard $z$-basis.
Inequalities (\ref{ineq_2sep},\ref{ineq_2sepb},\ref{ineq_2sepc}) now read successively:\\
\begin{align}
\begin{array}{l}
\max\{
 |\rho_{1,8}|^2,
|\rho_{4,5}|^2
 \}
\leq
\min \{\rho_{4,4}\rho_{5,5}
,
\rho_{1,1}\rho_{8,8}
  \}\leq 1/16\\
 \max\{ |\rho_{2,7}|^2
,
|\rho_{3,6}|^2
 \}
\leq
\min\{
 \rho_{2,2}\rho_{7,7}
, \rho_{3,3}\rho_{6,6} \} \leq 1/16
\end{array},~~~\forall\rho \in {\cal
D}_3^{a\textrm{-}(bc)}, \label{a-bc3sep}
 \end{align}\forget{
while bi-separability under split $b$-$(ac)$ (i.e., $\rho \in
{\cal D}_3^{b\textrm{-}(ac)}$) gives:}
    \begin{align}
    \begin{array}{l}
    \max\{
 |\rho_{1,8}|^2,
|\rho_{3,6}|^2
 \}
\leq
\min \{\rho_{3,3}\rho_{6,6}
,
\rho_{1,1}\rho_{8,8}
  \}\leq 1/16\\
 \max\{ |\rho_{2,7}|^2
,
|\rho_{4,5}|^2
 \}
\leq
\min\{
 \rho_{2,2}\rho_{7,7}
, \rho_{4,4}\rho_{5,5} \}\leq 1/16\end{array}, ~~~   \forall \rho \in {\cal
D}_3^{b\textrm{-}(ac)}, \label{b-ac3sep}
 \end{align} \forget{
     and under split $c$-$(ab)$ (i.e., $\rho \in {\cal
     D}_3^{c\textrm{-}(ab)}$):}
      \begin{align}\begin{array}{l}
\max\{
 |\rho_{1,8}|^2,
|\rho_{2,7}|^2
 \}
\leq
\min \{\rho_{2,2}\rho_{7,7}
,
\rho_{1,1}\rho_{8,8}
  \}\leq 1/16 \\
 \max\{ |\rho_{3,6}|^2
,
|\rho_{4,5}|^2
 \}
\leq
\min\{
 \rho_{3,3}\rho_{6,6}
, \rho_{4,4}\rho_{5,5} \}\leq 1/16\end{array}, ~~~ \forall \rho \in {\cal
     D}_3^{c\textrm{-}(ab)}. \label{c-ab3sep}
 \end{align}
       For a general bi-separable state  we can rewrite
       (\ref{first3}) as:
 \beq
 \max\{|\rho_{1,8}, |\rho_{2,7}|, |\rho_{3,6}|,|\rho_{4,5}| \} \leq 1/4
 ~~~ \forall \rho \in {\cal D}_3^{2\textrm{-sep}}.\eeq It can easily
 be seen that
 this is equivalent to Laskowski-\.Zukowski's condition
(\ref{anti-diagonal}) for $k=2$. The condition (\ref{2sep3}) for
bi-separability yields: 
\begin{align}
\begin{array}{l}
|\rho_{1,8}| \leq
\sqrt{\rho_{2,2}\rho_{7,7}}+\sqrt{\rho_{3,3}\rho_{6,6}}+\sqrt{\rho_{4,4}\rho_{5,5}}\\
 |\rho_{2,7}| \leq \sqrt{\rho_{1,1}\rho_{8,8}}+
\sqrt{\rho_{3,3}\rho_{6,6}}+\sqrt{\rho_{4,4}\rho_{5,5}}\\
  |\rho_{3,6}|  \leq \sqrt{\rho_{1,1}\rho_{8,8}}+\sqrt{\rho_{2,2}\rho_{7,7}}+\sqrt{\rho_{4,4}\rho_{5,5}}\\
|\rho_{4,5}|\leq\sqrt{ \rho_{1,1}\rho_{8,8}}+
\sqrt{\rho_{2,2}\rho_{7,7}}+\sqrt{\rho_{3,3}\rho_{6,6}}\end{array},~~~\forall\rho \in {\cal
D}_3^{2\textrm{-sep}}\label{matrixbisep3}.\end{align}
Finally,  condition
(\ref{3sep3}) for general $3$-separable states becomes:  $\forall\rho \in {\cal
D}_3^{3\textrm{-sep}}$: \begin{align}
\max\{ |\rho_{1,8}|^2, |\rho_{2,7}|^2, |\rho_{3,6}|^2,
|\rho_{4,5}|^2\}\leq \min  \{ \rho_{1,1}\rho_{8,8},
\rho_{2,2}\rho_{7,7}, \rho_{3,3}\rho_{6,6},
\rho_{4,4}\rho_{5,5}\}\leq \frac{1}{64}. \label{matrix3sep3}\end{align}  Note  that the
separability inequalities (\ref{a-bc3sep})-(\ref{matrix3sep3}) all
give bounds on anti-diagonal elements in terms of diagonal
elements.

 We will now show that these bounds  improve upon the  separability conditions  discussed  in section \ref{introsepconditions}.
We focus on the anti-diagonal element $\rho_{1,8}$ (i.e., we
suppose that this is the largest anti-diagonal matrix element)
since this is easiest for comparison. However, the same argument
holds for any other anti-diagonal matrix element.

The D\"ur-Cirac conditions in terms of $|\rho_{1,8}|$ read as
follows. For partial separability under  the split $a$-$(bc)$: 
$2|\rho_{1,8}| \leq \rho_{4,4} +\rho_{5,5}$, under the split
$b$-$(ac)$: $2|\rho_{1,8}| \leq \rho_{3,3} +\rho_{6,6}$, and
lastly under the split $c$-$(ab)$: $2|\rho_{1,8}| \leq \rho_{2,2}
+\rho_{7,7}$. Next, the
Laskowski-\.Zukowski condition (\ref{anti-diagonal})  gives for
$\rho \in  {\cal D}_3^{2\textrm{-sep}}$  that $|\rho_{1,8}|\leq
1/4$ and for $\rho \in {\cal D}_3^{3\textrm{-sep}} $ that
$|\rho_{1,8}|\leq1/8$. The fidelity condition (\ref{Fidelity})
gives that if $\rho\in {\cal D}_3^{2\textrm{-sep}}$  then
$2|\rho_{1,8}|\leq \rho_{2,2} +\ldots+ \rho_{7,7}$.

In order to show that all these conditions are implied  by our
separability conditions, we employ some inequalities which hold
for all states $\rho$: $|\rho_{1,8}|^2\leq \rho_{1,1}\rho_{8,8}$
(this follows from (\ref{eenheid})), and $(\sqrt{\rho_{4,4}} -
\sqrt{\rho_{5,5}})^2\geq0\Longleftrightarrow2\sqrt{\rho_{4,4}\rho_{5,5}}\leq
\rho_{4,4} +\rho_{5,5}$, and similarly
$2\sqrt{\rho_{3,3}\rho_{6,6}}\leq \rho_{2,2} +\rho_{6,6}$ and
$2\sqrt{\rho_{2,2}\rho_{7,7}}\leq \rho_{2,2} +\rho_{7,7}$. Using
these trivial inequalities  one easily sees  that the conditions
(\ref{a-bc3sep})-(\ref{c-ab3sep}) imply the D\"ur-Cirac conditions
for separability under the three bi-partite splits.  It is also
easy to see that the condition for 3-separability
(\ref{matrix3sep3}) strengthens the Laskowski-\.Zukowski condition
(\ref{anti-diagonal}) for $k=3$. However, it is not so easy to see
 that (\ref{matrixbisep3}) strengthens both the fidelity and
Laskowski-\.Zukowski condition for $k=2$. We will nevertheless
show that this is indeed the case.

Let us use the symbols $\overset{A}{\leq}$ and
$\overset{2\textrm{-sep}}{\leq}$ to denote inequalities that  hold
for all states or for bi-separable  states respectively. Combining
the above  trivial inequalities  with  condition
(\ref{matrixbisep3}) yields the following sequence of
inequalities: \begin{align}
  4|\rho_{1,8}|- (\rho_{1,1}+\rho_{8,8}) \overset{A}{\leq}2 |\rho_{1,8}| \overset{2\textrm{-sep}}{\leq} 2\sqrt{\rho_{4,4}\rho_{5,5}} +&2\sqrt{\rho_{3,3}\rho_{6,6}} +2\sqrt{\rho_{2,2}\rho_{7,7}} \nn\\& \overset{A}{\leq}\rho_{2,2} + \cdots + \rho_{7,7}.
  \label{inequ3}
  \end{align}
The inequality between the second and third expression is
(\ref{matrixbisep3}). It implies the other inequalities that
follow from (\ref{inequ3}). Comparing the first and fourth
expression of (\ref{inequ3}) one obtains the Laskowski-\.Zukowski
condition (\ref{anti-diagonal}), while a comparison of the second
and fourth yields the fidelity criterion (\ref{Fidelity}). Comparing the first and third term gives a 
new condition which was not previously mentioned. All these are implied by
condition (\ref{matrixbisep3}).

To end this section we show that the separability inequalities for $x=0$ give
Mermin-type separability inequalities. Consider the
Mermin operator \cite{mermin} for three qubits:
 \beq
  M^{(3)}:= X^{(1)}_a X^{(1)}_bY^{(1)}_c+ Y^{(1)}_aX^{(1)}_bX^{(1)}_c+
  X^{(1)}_aY^{(1)}_bX^{(1)}_c- Y^{(1)}_aY^{(1)}_bY^{(1)}_c,  \label{M3}
 \eeq
and define $M'^{(3)}$  in the same way, but with all
$X$ and $Y$ interchanged.   We can now use the
identity  $16( \av{{X^{(3)}_0}}^2 +\av{{Y^{(3)}_0}}^2)
=\av{M^{(3)}}^2+\av{M'^{(3)}}^2$ to obtain  from the  separability
conditions  (\ref{first3}) and (\ref{3sep3b})    the following quadratic
inequality for $k$-separability:
\begin{align}
   \label{quadratic3}
  16( \av{{X^{(3)}_0}}^2 +\av{{Y^{(3)}_0}}^2) =\av{M^{(3)}}^2+\av{M'^{(3)}}^2
   \leq 64\big(\frac{1}{4}\big)^k, ~~ \forall \rho \in {\cal
D}_3^{k\textrm{-sep}}.
   \end{align}
Of course, a similar bound holds when  $\av{X_0}^2 + \av{Y_0}^2$
in the left-hand side is replaced by   $\av{X_x}^2 + \av{Y_x}^2$
for $x=1,2,3$.
   This reproduces, for $N=3$,  the result  (\ref{quadraticN}) of \citet{nagataPRL}.    From the density matrix representation, we
see that these Mermin-type separability conditions are in fact
equivalent to the Laskowski-\.Zukowski condition
(\ref{anti-diagonal}).\forget{ when choosing the Pauli observables
for the set of local observables (for then one in effect obtains
the Laskowski-\.Zukowski condition for $k$-separability
$|\rho_{1,8}|^2\leq (1/4)^k$ which was shown to
be implied by our separability conditions). However, this holds
for any choice of orthogonal local observables since using a
suitable local unitary transformation one can transform the chosen
set of observables into the Pauli set.}\forget{These
 quadratic inequalities imply the following linear Mermin-type Bell-inequalities
 for partial separability: \beq |\av{M_3} | \leq 2^{3-k}, ~~ \forall \rho \in {\cal
D}_3^{k\textrm{-sep}}. \eeq These inequalities are sharp, i.e.\
$\sup_{\rho \in {\cal D}_3^{k\textrm{-sep}}} |\av{M_3} | =
2^{3-k}$. For full separability ($k=3$) this reproduces a result
obtained by \citet{roy}, and for bi-separability  ($k=2$) this
reproduces the result of \cite{tothseev}.} 
These
 quadratic inequalities imply the following linear Mermin-type Bell-inequalities
 for partial separability:
\begin{align} |\av{M_3} | \leq 2^{3-k}, ~~ \forall \rho \in {\cal
D}_3^{k\textrm{-sep}}. 
\label{merminsharpen}
\end{align} These inequalities are sharp so that
$\sup_{\rho \in {\cal D}_3^{k\textrm{-sep}}} |\av{M_3} | =
2^{3-k}$. For full separability ($k=3$) this reproduces a result
obtained by \citet{roy}, and for bi-separability ($k=2$) this reproduces the result of \citet{tothseev}.
Note that these conditions do not distinguish the different classes
within level $k=2$, as was the case in \eqref{a-bc3sep}-\eqref{c-ab3sep}.

Lastly, we note that  (\ref{merminsharpen}) for $k=2$, which holds for locally orthogonal observables, strengthens the following Mermin-type bi-separability inequality that holds for general observables $A,A'$, $B,B'$ and $C,C'$ \cite{gisin,seevuff}:
\begin{align}
|\av{ABC' +AB'C +A'BC -A'B'C'}|\leq 2^{3/2},~~~\forall \rho \in {\cal D}_3^{2\textrm{-sep}}.
\label{merminoriginal}
\end{align}
Again we see that the restriction to orthogonal observables gives stronger conditions.

\subsubsection{Analysis of experiment producing full three-qubit entanglement}
In an early work by \citet{seevuff} several experiments were discussed and it was
investigated whether three-qubit entanglement was present in
these experiments. It was there concluded that the experiments did not provide conclusive evidence for such full three-qubit entanglement. However, in that analysis the Mermin-type bi-separability inequality (\ref{merminoriginal}) was used (amongst others) to test for full entanglement. But as shown above, this bi-separability condition can be sharpened for the case of orthogonal observables to (\ref{merminsharpen}), i.e., to $|\av{M_3}|\leq 2$. 

The experiment described by \citet{PAN}\footnote{See also \citet[p.~209]{BOU2000}
.} aimed to create a GHZ state with three photons  and indeed used such orthogonal observables. They obtained a value of $\av{M_3}=2.83\pm0.09$.  We can thus conclude that although this experiment did not violate the original Mermin-type separability inequality (\ref{merminoriginal}) it does violate the condition (\ref{merminsharpen}) that holds for orthogonal observables  thereby indicating full three-qubit entanglement. This sheds new light on old experimental data in \cite{PAN} and shows that
genuine three-qubit entanglement has already been realized experimentally.

Note that \citet{duer2,duer22} also claimed that this experiment created full three-qubit entanglement. However, their analysis in fact only tested for inseparability with respects to all possible bi-partite splits, and not for full inseparability, i.e., for full entanglement. Thus they can indeed claim that the experiment has shown the existence of a state that is inseparable with respect to all splits, but not that it has shown full three-qubit entanglement.

 Finally, we note that using some idealization assumptions about the data 
of the experiment of  \citet{PAN} \citet{nagata2002} showed that the experiment contained three-particle
entanglement but without using a Mermin-type separability inequality.

\subsection{$N$-qubit case}\label{Nqubitsection}

In this section we generalize the analysis of the previous section
to $N$ qubits to obtain conditions for $k$-separability  and
$\alpha_k$-separability. The proofs are analogous to the previous
cases, and will be omitted. Explicit conditions for
$k$-separability are presented for all levels $k = 1, \ldots, N$.
Further, we give a recursive procedure to derive
$\alpha_k$-separability conditions for each $k$-partite split
$\alpha_k$  at all level $k$. From these, one can easily construct
the conditions that distinguish all the classes in $N$-partite
separability classification by enumerating all possible logical
combinations of separability or inseparability under each of these
splits at a given level. We will however not attempt to write down
these latter  conditions explicitly since the number of classes
grows exponentially with the number of qubits.  We  start by
considering  bi-partite splits, and bi-separable states (level
$k=2$), and then move upwards to obtain separability conditions
for splits on higher levels.

We define $2^{(N-1)}$ sets of four observables
$\{X^{(N)}_x,~Y^{(N)}_x,~Z^{(N)}_x,~I^{(N)}_x \}$ , with  $x\in
\{0,1, \ldots,\\ 2^{(N-1)}-1\}$
 recursively from the
$(N-1)$-qubit observables:
\begin{align}
X^{(N)}_y &:=\frac{1}{2}\,(X^{(1)} \otimes X_{y/2}^{(N-1)} -Y^{(1)}\otimes Y_{y/2}^{(N-1)})
\nn\\X^{(N)}_{y+1} &:=\frac{1}{2}\,(X^{(1)} \otimes X_{y/2}^{(N-1)} +Y^{(1)}\otimes Y_{y/2}^{(N-1)})\nn
\\
Y^{(N)}_y  &:= \frac{1}{2}\,(Y^{(1)}\otimes X_{y/2}^{(N-1)}+X^{(1)} \otimes Y_{y/2}^{(N-1)} )
\nn\\Y^{(N)}_{y+1}  &:= \frac{1}{2}\,(Y^{(1)}\otimes X_{y/2}^{(N-1)}-X^{(1)} \otimes Y_{y/2}^{(N-1)} )\nn\\
Z^{(N)}_y  &:= \frac{1}{2}\,(Z^{(1)}\otimes I_{y/2}^{(N-1)}+I^{(1)} \otimes Z_{y/2}^{(N-1)} )
\nn\\Z^{(N)}_{y+1}  &:= \frac{1}{2}\,(Z^{(1)}\otimes I_{y/2}^{(N-1)}-I^{(1)} \otimes Z_{y/2}^{(N-1)} )\nn\\
  I^{(N)}_y  &:= \frac{1}{2}\,(I^{(1)} \otimes I_{y/2}^{(N-1)} +Z^{(1)}\otimes Z_{y/2}^{(N-1)})
    \nn\\I^{(N)}_{y+1}  &:= \frac{1}{2}\,(I^{(1)} \otimes I_{y/2}^{(N-1)}-Z^{(1)}\otimes Z_{y/2}^{(N-1)}),
  \label{Noperators}
  \end{align}
 with $y~\mathrm{even}$, i.e., $y\in \{0,2,4,\ldots\}$.   Analogous relations between these observables hold as those between the
  observables
  (\ref{set2})  and  (\ref{N3operators}). In particular, if the orientations of each triple  of local orthogonal
  observables is the same,  these sets form
  representations of the generalized
  Pauli group, and  every $N$-qubit state  obeys $\av{X^{(N)}_x}^2
+\av{Y^{(N)}_x}^2
  \leq\av{I^{(N)}_x}^2 -\av{Z^{(N)}_x}^2$,
  with equality only for pure states.

\subsubsection{Bi-separability}

Consider a state that is separable under some bi-partite split
$\alpha_2$ of the $N$ qubits. For each such split we get
$2^{(N-1)}$ separability inequalities in terms of the sets
$\{X^{(N)}_x,~Y^{(N)}_x,~Z^{(N)}_x,~I^{(N)}_x \}$ labeled by
$x\in\{0,1\ldots ,2^{(N-1)}-1\}$. These separability inequalities
provide necessary conditions for the $N$-qubit state to be
separable under the split under consideration. In order to find
these inequalities, we  first determine  the $N$-qubit  analogs of
the three-qubit pure state equalities (\ref{2_sep_a_bcX}) and
(\ref{2_sep_a_bcI}) corresponding to this bi-partite split.  We
have not found a generic  expression  that lists them all for each
possible split and all $x$. However, for the split where the first
qubit is separated from the $(N-1)$ other qubits, i.e. $\alpha_2 =
a\textrm{-}(bc \ldots n)$ a generic form can be given:
\begin{align}
\av{X_x^{(N)} }^2 +\av{Y_x^{(N)} }^2&=
\frac{1}{4}\,(\,\av{X^{(1)}_a}^2+\av{Y^{(1)}_a}^2\,)\,
(\,\av{X^{(N-1)}_{x/2}}^2 +\av{Y^{(N-1)}_{x/2}}^2\,)\nn\\&=
 \av{X_{x+1}^{(N)} }^2 +\av{Y_{x+1}^{(N)} }^2  =\nn\\
 \av{I_x^{(N)} }^2 -\av{Z_x^{(N)} }^2&=
\frac{1}{4}\,(\,\av{I^{(1)}_a}^2-\av{Z^{(1)}_a}^2\,)\,
(\,\av{I^{(N-1)}_{x/2}}^2 -\av{Z^{(N-1)}_{x/2}}^2\,)\nn\\&=
 \av{I_{x+1}^{(N)} }^2 -\av{Z_{x+1}^{(N)} }^2,
 \label{partialeq}
\end{align}
where, without loss of generality, $x$ is chosen to be even, i.e.
$x\in\{0,2,4,\ldots\}$. For  other bi-partite splits the sets of
observables labeled by $x$ are permuted, in a way depending on
the particular split.

For example,  for $N=4$  where $x\in \{0,1,\ldots,7\}$ the
equalities  (\ref{partialeq}) give the result for the split
$a$-$(bcd)$.  The corresponding equalities for  other bi-partite splits are
obtained by the following permutations of $x$: for split $b$-$(acd)$:
$1 \leftrightarrow 3$ and $5 \leftrightarrow 7$; for split
$c$-$(abd)$: $1 \leftrightarrow 6$ and $3 \leftrightarrow
4$; and for split $d$-$(abc)$: $1 \leftrightarrow 4$ and $3
\leftrightarrow 6$. For the split $(ab)$-$(cd)$: $1
\leftrightarrow 2$ and $5 \leftrightarrow 6$; for  $(ac)$-$(bd)$:  $1 \leftrightarrow 7$ and $3
\leftrightarrow 5$; and lastly, for $(ad)$-$(bc)$:
$1 \leftrightarrow 5$ and $3 \leftrightarrow 7$.

For mixed states that are separable under a given bi-partite split
the equalities (\ref{partialeq}) (and  their analogs obtained via
suitable permutations) become inequalities. We again state them
for
 the split $a$-$(bc\ldots n)$:
\begin{align} 
\max \left\{ \begin{array}{c}
\av{X_x^{(N)} }^2 +\av{Y_x^{(N)} }^2
\\
\av{X_{x+1}^{(N)} }^2 +\av{Y_{x+1}^{(N)} }^2
\end{array} \right\}
\leq&
\min \left\{ \begin{array}{c}
\av{I_{x}^{(N)} }^2 -\av{Z_{x}^{(N)} }^2
\\
\av{I_{x+1}^{(N)} }^2 -\av{Z_{x+1}^{(N)} }^2
 \end{array}
 \right\}\leq \frac{1}{4}, ~~ \forall \rho \in {\cal
D}_N^{a\textrm{-}(bc\ldots n)}, 
 \label{partialeqM}
  \end{align}
with $x\in\{0,2,4,\ldots\}$. 
The proof of (\ref{partialeqM}) is a straightforward
generalization of the convex analysis in section
\ref{twoqubitsection}. Again, for the other bi-partite splits, the
 labels  $x$ are permuted in a way depending on the particular
split.

For a general bi-separable state  $ \rho \in {\cal
D}_N^{2\textrm{-sep}}$, we thus obtain the following bi-separability
conditions: \beq\label{Nk2first} \av{{X}_x^{(N)}}^2
+\av{{Y}_x^{(N)}}^2 \leq 1/4,~ \forall x, ~~~\forall \rho \in
{\cal D}_N^{2\textrm{-sep}} , \eeq which is equivalent to the
Laskowski-\.Zukowski condition for $k=2$ (as will be shown below).  And
just as in the three-qubit case, we also obtain a stronger condition
\beq\label{Nk2} \sqrt{\av{{X}_x^{(N)}}^2 +\av{{Y}_x^{(N)}}^2 }\leq
\sum_{y \neq x} \sqrt{\av{I_{y}^{(N)}}^2 -\av{Z_{y}^{(N)}}^2},
~~~\forall \rho \in {\cal D}_N^{2\textrm{-sep}},
 \eeq
 with $x,y=0,1,\ldots, 2^{(N-1)}-1$.
 Violation of this inequality is a sufficient
condition for  full inseparability, i.e., for full $N$-partite
entanglement.

The inequalities (\ref{Nk2}) are stronger than the fidelity
criterion (\ref{Fidelity}) and the Laskowski-\.Zukowski criterion
(\ref{anti-diagonal}) for $k=2$,  and inequalities
(\ref{partialeqM}) are  stronger than the D\"ur-Cirac condition
(\ref{dccondition}) for separability under bi-partite splits. This
will be shown below in subsection \ref{Nmatrix}. 

\subsubsection{Partial separability criteria for levels $2<k\leq N$}
For levels  $k>2$ we  sketch  a procedure to find
$\alpha_{k+1}$-separability inequalities recursively from
inequalities at the preceding level. Suppose that at level $k$ the
inequalities are given for separability under each $k$-partite
split $\alpha_k$ of the $N$ qubits, and  that these
$\alpha_k$-separability inequalities take the form:
 \beq
 \max_{ x \in z_i^{\alpha_k}}  \av{{X}_x^{(N)}}^2 +\av{{Y}_x^{(N)}}^2 \leq
 \min_{x \in z_i^{\alpha_k}} \av{I_x}^2 - \av {Z_x}^2
 \leq
\frac{1}{4^{(k-1)}},\forget{~~i=1,2,\ldots, 2^{(N-k)},} ~~ \forall
\rho \in {\cal D}_N^{\alpha_k},
\label{recursivek} 
\eeq
 with $i\in\{1,2,\dots,2^{(N-k)}\}$.\forget{with
$\mathcal{Q}_{z_i^{\alpha_k}}$ the maximum of some set of
expressions $\av{{X}_y^{(N)}}^2 +\av{{Y}_y^{(N)}}^2$  and
$\mathcal{P}_{z_i^{\alpha_k}}$ the minimum of some set  of
expressions  $\av{I_{y}^{(N)}}^2 -\av{Z_{y}^{(N)}}^2$} where
$z_i^{\alpha_k}$ denote `solution sets' for the specific
$k$-partite split $\alpha_k$. \forget{This gives a total of
$2^{(N-k)}$ inequalities.   A solution set $z_i^{\alpha_k}$ is a
set of values for $x$ that are grouped together in a separability
inequality for the specific split $\alpha_k$.} For example, in the
case of three qubits,  the solution sets   for  the bi-partite
split $a$-$(bc)$ are $z_1^{a\textrm{-}(bc)}=\{0,1\}$ and
$z_2^{a\textrm{-}(bc)}=\{2,3\}$, as can be seen from
(\ref{ineq_2sep}). The solution sets for other bi-partite splits can be read off (\ref{ineq_2sepb}) and
(\ref{ineq_2sepc}) so as to give: $z_1^{b\textrm{-}(ac)}=\{0,3\}$, 
$z_2^{b\textrm{-}(ac)}=\{1,2\}$, and  $z_1^{c\textrm{-}(ab)}=\{0,2\}$, 
$z_2^{c\textrm{-}(ab)}=\{1,3\}$. And for future purposes we list them for the case of four qubits in table \ref{tabel1} below. These were obtained by  determining \eqref{partialeqM} for $N=4$ and for all bi-partite splits $\alpha_2$.
 \begin{table}[h]
\begin{centerline}{
\begin{tabular}{|c||c|c|c|c|c|c|c|}
\hline
split $\alpha_2$&$a$-$(bcd)$&$b$-$(acd)$&$c$-$(abd)$&$d$-$(abc)$&$(ab)$-$(cd)$&$(ac)$-$(bd)$&$(ad)$-$(bc)$\\
\hline\hline
$z_1^{\alpha_2}$&$\{ 0,1  \}$&$\{ 0 ,3  \}$&$\{ 0 ,6  \}$&$\{ 0 ,4  \}$&$\{0  ,2  \}$&$\{ 0 ,7  \}$&$\{  0, 5 \}$\\
$z_2^{\alpha_2}$&$\{  2,3  \}$&$\{ 1 ,2  \}$&$\{ 1 , 7 \}$&$\{1  ,5  \}$&$\{ 1 ,3  \}$&$\{  1,6  \}$&$\{ 1 ,4  \}$\\
$z_3^{\alpha_2}$&$\{  4,5  \}$&$\{  5,6  \}$&$\{  2,4  \}$&$\{ 2 ,6  \}$&$\{ 4 ,6  \}$&$\{ 2 ,5  \}$&$\{ 2 ,7  \}$\\
$z_4^{\alpha_2}$&$\{  6,7  \}$&$\{  4, 7 \}$&$\{  3, 5 \}$&$\{ 3 ,7  \}$&$\{ 5 ,7  \}$&$\{ 3 ,4  \}$&$\{ 3 ,6  \}$\\
\hline
\end{tabular}}
\end{centerline}
\caption{Solution sets for the seven different bi-partite splits of four qubits.} \label{tabel1}
\end{table}

\forget{Let us denote the total number of possible splits for an
$N$-partite state on level $k$ by $\sharp_k^N$
\cite{explicit_expression}.} Now move one level higher and
consider a given $(k+1)$-partite split $\alpha_{(k+1)}$. This
split  is contained in a total number of $\binom{k+1}{2}=k(k+1)/2$~ 
$k$-partite splits $\alpha_k$. Call the collection of these
$k$-partite splits $\mathcal{S}_{\alpha_{(k+1)}}$. We then obtain
preliminary  separability inequalities for the  split
$\alpha_{k+1}$ from the conjunction of all separability
inequalities for the splits $\alpha_k$ in the set
$\mathcal{S}_{\alpha_{(k+1)}}$.
 To be
specific, this yields:
 \begin{align}\max_{ \alpha_k \in \mathcal{S}_{\alpha_{k+1}}}
\max_{ x \in z_i^{\alpha_k}}
 \av{{X}_x^{(N)}}^2 +\av{{Y}_x^{(N)}}^2\leq \forget{\min_{ x\in z_i^{\alpha_k}, \alpha_k\in\mathcal{S}_{\alpha_{(k+1)}}}}
\min_{ \alpha_k \in\mathcal{S}_{\alpha_{(k+1)}}} &\min_{x \in
z_i^{\alpha_k}}   \av{I^{(N)}_x}^2 - \av{Z^{(N)}_x}
^2
   \nn\\ &\leq
\frac{1}{4^{k-1}}, ~~ \forall \rho \in {\cal
D}_N^{\alpha_{(k+1)}}, \label{recursivek+1}
 \end{align}
This may be written more compactly as \begin{align}
  \max_{ x \in
z_i^{\alpha_{k+1}}}
 \av{{X}_x^{(N)}}^2 +\av{{Y}_x^{(N)}}^2\leq \min_{x \in
z_i^{\alpha_{k+1}}}   \av{I^{(N)}_x}^2 - \av{Z^{(N)}_x} ^2
   \leq
\frac{1}{4^{k-1}}, ~~ \forall \rho \in {\cal D}_N^{\alpha_{(k+1)}}
\label{recursivek+1compact},
 \end{align} 
with $i\in\{1,2,\dots,2^{(N-k-1)}\}$.  (In fact, this can be regarded as an implicit definition of the
 solution sets  $z^{\alpha_{k+1}}_i$.)
 More importantly,   by
an argument similar to that leading from (\ref{3sep3}) to
(\ref{3sep3b}) one finds a stronger numerical bound in the utmost
right-hand side of these inequalities, namely $4^{-k}$ instead
of $4^{-(k-1)}$. Thus, the final result is: \begin{align}  \max_{ x \in
z_i^{\alpha_{k+1}}}
 \av{{X}_x^{(N)}}^2 +\av{{Y}_x^{(N)}}^2\leq \min_{x \in
z_i^{\alpha_{k+1}}}   \av{I^{(N)}_x}^2 - \av{Z^{(N)}_x} ^2
   \leq
\frac{1}{4^{k}}, ~~ \forall \rho \in {\cal
D}_N^{\alpha_{(k+1)}},
\label{recursivek+1final} 
\end{align} 
with $i\in\{1,2,\dots,2^{(N-k-1)}\}.$ This shows that the
$\alpha_k$-separability  inequalities  indeed take the same  form
as (\ref{recursivek}) at all levels.

As an example of this recursive procedure, take  $N=4$, set $k=3$,
and choose the split $a$-$b$-$(cd)$. This split is contained in
three $2$-partite splits $a$-$(bcd)$, $b$-$(acd)$ and
$(ab)$-$(cd)$.\forget{The separability inequalities for these splits have
the following solution sets (four per split):  $z_i^{a\textrm{-}(bcd)}$:
$\{0,1\},\{2,3\},\{4,5\},\{6,7\}$, $z_i^{b\textrm{-}(acd)}$:
$\{0,3\},\{1,2\}\{4,7\},\{5,6\}$, and $z_i^{(ab)\textrm{-}(cd)}$:
$\{0,2\},\{1,3\}\{4,6\},\{5,7\}$ respectively.} 
Using (\ref{recursivek+1})  and the first, second and fifth column of table \ref{tabel1}  
 one obtains the following two solutions sets  for the split  $a$-$b$-$(cd)$:    $z_1^{a\textrm{-}b\textrm{-}(cd)}=\{0,1,2,3\}$ and $z_2^{a\textrm{-}b\textrm{-}(cd)}=\{4,5,6,7\}$. This leads to the separability
inequalities:
\begin{align}\begin{array}{clcl}
 \max\limits_{x\in \{ 0,1,2,3\}} \av{X_{x}^{(4)} }^2 +\av{Y_{x}^{(4)}
}^2  &\leq&
 \min\limits_{x\in \{ 0,1,2,3\}} \av{I_{x}^{(4)} }^2 -\av{Z_{x}^{(4)}
}^2&
 \leq \frac{1}{16} \\
 \max\limits_{x\in \{ 4,5,6,7\}} \av{X_{x}^{(4)} }^2 +\av{Y_{x}^{(4)}
}^2 &\leq&
 \min\limits_{x\in \{ 4,5,6,7\}} \av{I_{x}^{(4)} }^2 -\av{Z_{x}^{(4)}
}^2&
 \leq \frac{1}{16}
 \end{array},   \forall \rho \in {\cal
D}_4^{a\textrm{-}b\textrm{-}(cd)}.
\label{43}
 \end{align}
For  other $3$-partite splits\forget{besides  the split
$a$-$b$-$(cd)$} the inequalities can be obtained in a similar way so as to give table \ref{tabel2} below.

\begin{table}[h]
\begin{centerline}{
\begin{tabular}{|c||c|c|c|c|c|c|} \hline
split $\alpha_3$ &  $a$-$b$-$(cd)$ & $(ab)$-$c$-$d$& $a$-$b$-$(cd)$&  $(ac)$-$b$-$d$& $(ad)$-$b$-$c$ & $(bd)$-$a$-$c$\\
\hline\hline $z_1^{\alpha_3}$ &  \{0,1,2,3\} &\{0,2,4,6\} &\{0,1,4,5\}
&\{0,3,4,7\} &\{0,3,5,6\} &\{0,1,6,7\} \\ $z_2^{\alpha_3}$ &
 \{4,5,6,7\} &\{1,3,5,7\} &\{2,3,6,7\}
&\{1,2,5,6\} &\{1,2,4,7\} &\{2,3,4,5\}\\\hline
\end{tabular}}
\end{centerline}
\caption{Solution sets for the six different $3$-partite splits of four qubits.} \label{tabel2}
\end{table}

As a special case, we mention the result for full separability,
i.e., for $k=N$. There is only one $N$-partite split, namely where
all qubits end up in a different set.  Further, there is only one
solution set $z_i^{\alpha_N}$ and it contains all $x \in \{0,1,
\ldots,2^{(N-1)}-1\}$. States $\rho$ that are separable under this
split thus obey: \begin{align} \max_{x} \av{{X}_x^{(N)}}^2
+\av{{Y}_x^{(N)}}^2 \leq \min_{x} \av{I_{x}^{(N)}}^2
-\av{Z_{x}^{(N)}}^2 \leq \frac{1}{4^{(N-1)}}, ~~~\forall \rho \in
{\cal D}_N^{N\textrm{-sep}}. \label{NNsep} \end{align}
 Violation of this inequality is a sufficient condition for
some entanglement to be present in the $N$-qubit state. The
condition (\ref{NNsep}) strengthens the Laskowski- \.Zukowski
condition  (\ref{anti-diagonal}) for $k=N$ (to be shown below).

For an $N$-qubit $k$-separable state $ \rho \in {\cal
D}_N^{k\textrm{-sep}}$, i.e., a state that is a convex mixture of
states that are separable under some $k$-partite split,
 we  obtain from (\ref{recursivek+1final}) the following
$k$-separability  conditions: \beq\label{Nksepfirst}
\av{{X}_x^{(N)}}^2 +\av{{Y}_x^{(N)}}^2\leq
\frac{1}{4^{(k-1)}},~\forall x, ~~~\forall \rho \in {\cal
D}_N^{k\textrm{-sep}}, \eeq which is equivalent to the
Laskowski-\.Zukowski condition (\ref{anti-diagonal}) for all $N$
and $k$ (this will be shown below using the density matrix
formulation of these conditions).  However, in analogy to
(\ref{2sep3}) we also obtain  the stronger condition: \beq
\sqrt{\av{{X}_x^{(N)}}^2 +\av{{Y}_x^{(N)}}^2}\leq \min_l  \sum_{y
\in \mathcal{T}_{k,l}^{N,x} }\sqrt {\av{{I}_y^{(N)}}^2
-\av{{Z}_y^{(N)}}^2 } ,~~~\forall \rho \in {\cal
D}_N^{k\textrm{-sep}}, \label{Nksep} \eeq   where, for given $N,
k$ and $x$, $\mathcal{T}_{k,l}^{N,x}$ denotes  a tuple  of values
of $y\neq x$, each one being picked  from each of the solutions
sets $z_i^{\alpha_{k}}$ that contain $x$, where $\alpha_k$ ranges
over all the $k$-partite splits of the $N$ qubits. In general,
there will be many ways of picking such values, and we use $l$ as
an index to label such tuples.

For example, in the case $N=3$, there are a total of 6 solution
sets (two for each of the three bi-partite splits): $\{ 0,1\},
\{2,3\}, \{ 0,2\}, \{1,3\} ,  \{ 0,3\}, \{1,2\}$. If we set $x=0$
and pick a member different  from  0 from each of those sets that
contain $0$, we find: $\mathcal{T}_{2,1}^3= \{ 1,2,3\}$. This is
in fact the only such choice and thus $l=1$.  Thus, in this
example condition (\ref{Nksep}) reproduces the result
(\ref{2sep3}).

As a more complicated example, take $N=4$, $k=3$, and choose again
$x=0$. In this case there are six 3-partite splits each of which has two solution sets, as given in table \ref{tabel2}. 
 The solution sets that contain 0 are all on the top row of this
table. There are now many ways of constructing a tuple by picking
elements  that differ from 0 from  each of these sets , for
example ${\cal T}^{4,0}_{3,1}= \{ 1,2,1,3,3, 1\}$, ${\cal
T}^{4,0}_{3,2} = \{ 1,2,1,3,3,6\}$, etc. In this case one  has to
take a minimum in (\ref{Nksep}) over all these $l=1, \ldots, 3^6$
tuples.

\forget{
 The minimum is taken over all distinct sets
$l$. In each set $l$ the distinct occurrences of $y$ are picked
from a total number of sets that is equal to the total number
$\sharp_k^N$ \cite{explicit_expression} of possible splits
$\gamma_{k}$ that exist  for an $N$-partite state on this level
$k$.} \forget{ Violation of condition (\ref{Nksep}) is a
sufficient criterion for $(k)$-inseparability, i.e., for
$(k-1)$-separable entanglement (and thus for at least $
\integer{N/(k-1)}$-partite entanglement). {\bf niet
k-separabiliteit?}} For $k=2$, condition (\ref{Nksep}) reduces
to (\ref{Nk2}) and for $k=N$ to (\ref{NNsep}). For these values of
$k$, the condition is stronger than (\ref{Nksepfirst}) (see the
next section). For $k\neq 2,N$, this is still an open question.

To conclude this subsection, let us recapitulate.   We have found
separability conditions in terms of local orthogonal observables
for each of the $N$ parties that are  necessary for
$k$-separability  and for separability under  splits $\alpha_k$ at
each level on the hierarchic separability classification.
\forget{The criteria can distinguish all different levels from
each other by a bound that decrease by a factor of four for each
level. Furthermore, they allow for distinguishing the classes
within each level, using the finer characterization made possible
by the interconnections between splits at different levels as
captured by the notion of contained splits.} Violations of these
separability conditions give sufficient criteria for $k$-separable
entanglement and $m$-partite entanglement with $\integer{N/k}\leq
m\leq N-k+1$. The separability conditions are stronger than the
D\"ur-Cirac condition for separability under specific splits, and
stronger than the fidelity condition and the Laskowski-\.Zukowski
condition for bi-separability. The latter condition is also
strengthened for $k=N$. These implications are shown in the next
section.\forget{The strength of the conditions will be further
commented on in section \ref{strengthsection}. We first take a
further look at the derived separability conditions.}

\subsubsection{The conditions in terms of matrix elements}\label{Nmatrix}
Choosing the Pauli matrices $\{\sigma_x^{(j)},
\sigma_y^{(j)},\sigma_z^{(j)}\}$ as local orthogonal observables,
with the same orientation at each qubit,    allows one to
formulate the separability conditions in terms of the density
matrix elements $\rho_{i,j}$ on the standard $z$-basis\footnote{In the standard $z$-basis, $\rho_{i,j}=\bra{i'}\rho\ket{j'}$ with $i'=i-1$, $j'=j-1$
and where $i'=i_1i_2\ldots i_N$ and $j'=j_1j_2\ldots j_N$ are  in
binary notation. For example, for $N=4$: $\rho_{1,16}
=\bra{0000}\rho\ket{1111}$   and
$\rho_{9,12}=\bra{1000}\rho\ket{1011}$.}.
 For these choices we obtain:\begin{alignat}{2}
X_0^{(N)}&= \ket{0}\bra{1}^{\otimes N} +\ket{1}\bra{0}^{\otimes N},
&&\av{ X_0^{(N)}}=2\mathrm{Re}\, \rho_{1,d}, \nn\\
Y_0^{(N)}&= -i\ket{0}\bra{1}^{\otimes N} +i\ket{1}\bra{0}^{\otimes N}, \qquad
&&\av{Y_0^{(N)}}=-2\mathrm{Im}\, \rho_{1,d}, \nn\\
I_0^{(N)}&=  \ket{0}\bra{0}^{\otimes N} +\ket{1}\bra{1}^{\otimes N},
&&\av{ I_0^{(N)}}=\rho_{1,1}+ \rho_{d,d},\nn\\
Z_0^{(N)}&= \ket{0}\bra{0}^{\otimes N} -\ket{1}\bra{1}^{\otimes N},
&&\av{ Z_0^{(N)}}=\rho_{1,1} -\rho_{d,d}, \label{zbasisrel}
\end{alignat}
where $d= 2^N$.  Analogous relations hold for $X_x^{(N)},
~Y_x^{(N)},~ Z_x^{(N)},~I_x^{(N)}$ for $x\neq 0$.

Let us treat the case  $N=4$ in detail. First, consider the level
$k=2$. Bi-separability under the split $a$-$(bcd)$ gives the
following inequalities for the anti-diagonal matrix elements:
\begin{align}
\label{n4k2}
\begin{array}{clcl}
 \max \{ |\rho_{1,16}|^2, |\rho_{8,9}|^2\} &\leq& \min\{ \rho_{1,1}\rho_{16,16}, \rho_{8,8}\rho_{9,9} \} &\leq1/16 \\
 \max \{|\rho_{2,15}|^2, |\rho_{7,10}|^2\} &\leq& \min\{\rho_{2,2}\rho_{15,15}, \rho_{7,7}\rho_{10,10} \} &\leq1/16\\
\max \{|\rho_{3,14}|^2, |\rho_{6,11}|^2\} &\leq& \min\{\rho_{3,3}\rho_{14,14}, \rho_{6,6}\rho_{11,11} \} &\leq1/16\\
 \max \{|\rho_{5,12}|^2, |\rho_{4,13}|^2\} &\leq&\min\{\rho_{5,5}\rho_{12,12}, \rho_{4,4}\rho_{13,13} \}&\leq1/16
 \end{array},~~~~ \mbox{ $ \forall \rho \in {\cal D}_4^{a\textrm{-}(bcd)}$}
\end{align}
 The analogous inequalities for separability under  other
bi-partite splits are obtained  by suitable permutations on the
labels. Indeed,  for split $b$-$(acd)$ labels 8 and 5, 9 and 12, 2 and 3, 5 and 14 are permuted, which we denote as: $(8,9, 2,15)
\leftrightarrow ( 5, 12,3,14)$; and for split $c$-$(abd)$:
$(8,9,2,15)\leftrightarrow (3,14,5,12)$; for split
$d$-$(abc)$: $(8,9, 3,14) \leftrightarrow (2,15, 5,12)$; for
the split $(ab)$-$(cd)$: $(8,9,3,14 ) \leftrightarrow (4,13,
7,10)$; for  $(ac)$-$(bd)$:\,$(8,9, 5,12)  \leftrightarrow
(6,11, 7,10)$; and lastly, for the split $(ad)$-$(bc)$:\,$(8,9,
5,12) \leftrightarrow (7,10, 6,11)$. For a general bi-separable
state we obtain \beq
 |\rho_{1,16}| \leq \sqrt{\rho_{2,2}\rho_{15,15}} +
\sqrt{\rho_{3,3}\rho_{14,14}}+ \ldots+\sqrt{\rho_{8,8}\rho_{9,9}},
 ~~~ \forall \rho \in {\cal D}_4^{2\textrm{-sep}}
,\eeq
and analogous for the other anti-diagonal elements.

Next, consider one level higher, i.e., $k=3$.  There are six
different $3$-partite splits  for a system consisting of four
qubits. For separability under each such split a different set of
inequalities can be obtained from (\ref{recursivek+1}). To be more
precise,  such a set consists of the conjunction of all the
separability inequalities for the bi-partite splits at level $k=2$
this particular $3$-partite split is contained in. For $N=4$ each
$3$-partite split is contained in three bi-partite splits. For
example, for separability under split $a$-$b$-$(cd)$
 we obtain:
 \begin{align}\max 
&\left\{ \begin{array}{l}  |\rho_{1,16}|^2, 
|\rho_{8,9}|^2,  
|\rho_{4,13}|^2, 
|\rho_{5,12}|^2
        \end{array}\right\}\nn\\
      & ~~~~~~~~~~~~~~~~~~~~ \leq\min 
        \left\{ \begin{array}{l} \rho_{1,1}\rho_{16,16},~\rho_{8,8}\rho_{9,9},~
                    \rho_{4,4}\rho_{13,13},~
                    \rho_{5,5}\rho_{12,12}
        \end{array}\right\}\leq1/64,\nn
       \\
      \max 
&\left\{ \begin{array}{l}  |\rho_{2,15}|^2, 
|\rho_{3,14}|^2,
|\rho_{6,11}|^2, 
|\rho_{7,10}|^2
        \end{array}\right\}\nn\\
       &~~~~~~~~~~~~~~~~~~~~\leq\min 
        \left\{ \begin{array}{l} \rho_{2,2}\rho_{15,15},~\rho_{3,3}\rho_{14,14},~
                    \rho_{6,6}\rho_{11,11},~
                    \rho_{7,7}\rho_{10,10}
        \end{array}\right\}\leq1/64.        
        \end{align}
              This is the density matrix formulation of (\ref{43}).

A general $3$-separable state  $\rho \in {\cal
D}_4^{3\textrm{-sep}}$ is a convex mixture of states that each are
separable under some such $3$-partite split. The separability
condition follows from (\ref{Nksep}): \beq\label{MelementsN4k3}
 |\rho_{1,16}|\leq \min_{l}
(\sum_{j\in\tilde{\mathcal{T}}_{3,l}^{4,0}}
\sqrt{\rho_{j,j}\rho_{17-j,17-j} }), ~~~\forall \rho\in {\cal
D}_4^{3\textrm{-sep}}, \eeq
 where $\tilde{\mathcal{T}}_{3,l}^{4,0}$  is the tuple of indices $j\in\{1,16\}$ that label the anti-diagonal density matrix elements $\rho_{j,17-j}$ corresponding to the density matrix formulation of the set of operators  $\av{{X}_y^{(4)}}^2 +\av{{Y}_y^{(4)}}$ with $y$ determined by 
 $\mathcal{T}_{3,l}^{4,0}$. Here we have used that the anti-diagonal element $\rho_{1,16}$ corresponds to   $\av{{X}_0^{(4)}}^2 +\av{{Y}_0^{(4)}}^2$. 
              For  $N=4$, $k=3$ there are six possible splits, so for each $l$, $j$ is picked from a total of six sets.
                For the case under consideration the sets are $\{1,4,5,8 \}, \{1,2,3,4 \}, \{1,3,5,7 \}, \{1,2,5,6 \},\\ \{1,2,7,8 \}$, and $\{1,3,6,8 \}$. For each $l$ one chooses  a tuple  of values of $j$ where one value is picked from each of these six sets, except for the value $1$ which is excluded.    Analogous inequalities are obtained for the other anti-diagonal matrix elements.
Finally for full separability ($k=4$) we get: \begin{align}
\max\{|\rho_{1,16}|^2, |\rho_{2,15}|^2,\ldots,
|\rho_{8,9}|^2\}\leq \min \{  \rho_{1,1}\rho_{16,16},
\rho_{2,2}\rho_{15,15}, 
\ldots,\rho_{8,8}\rho_{9,9}   \}\leq 1/256,      \end{align}
with $\forall \rho \in {\cal D}_4^{4\textrm{-sep}}$. 

 For general $N$, it is easy to see that (\ref{Nk2first}) yields
the Laskowski-\.Zukowski condition (\ref{anti-diagonal}).
 It is instructive to look
at the extremes of bi-separability  and full separability, since
for them explicit forms can be  given. For $k=2$ condition
(\ref{Nk2}) reads: \begin{align} |\rho_{{l},\bar{l}}|\leq \sum_{n \neq
l,\bar{l}}\sqrt{\rho_{n,n}\rho_{\bar{n},\bar{n}}}/2 ,~~ \forall \rho \in
{\cal D}_N^{2\textrm{-sep}}   ~~~\mbox{where $\bar{l}=d+1-l$,
$\bar{n}=d+1-n$}, \label{N2gen} \end{align}
with $l,n\in \{1,\dots,d\}$. 
\forget{
 For example, for $l=1$ this gives
 \begin{align}
 \label{Nmatrixelements}
|\rho_{1,d}|\leq \sqrt{\rho_{2,2}\rho_{d-1,d-1}}+
\sqrt{\rho_{3,3}\rho_{d-2,d-2}}+\ldots+\sqrt{\rho_{d/2,d/2}\rho_{d/2+1,d/2+1}},~~
\forall \rho \in {\cal D}_N^{2\textrm{-sep}}. 
\end{align}} For  $k=N$, we
can reformulate condition (\ref{NNsep}) as \begin{align}
 \max\{|\rho_{1,d}|^2, |\rho_{2,d-1}|^2\ldots         \}
 \leq\min\{  \rho_{1,1}\rho_{d,d}, \rho_{2,2}\rho_{d-1,d-1}, \ldots
        \}\leq1/4^{N},~~ \forall \rho \in {\cal D}_N^{N\textrm{-sep}} .\label{matrixfull}
\end{align}  It is easily seen that the condition (\ref{matrixfull}) is
stronger than the Laskowski-\.Zukowski condition
(\ref{anti-diagonal}) for this case.

Again, these inequalities  give bounds on anti-diagonal matrix
elements in  terms of diagonal ones on the $z$-basis.
 These density matrix representations depend on the choice of the Pauli matrices as the local observables.
 However, every other triple of  locally orthogonal observables with the same orientation can
 be obtained from the Pauli matrices by suitable local basis transformations, and therefore this matrix
 representation does not loose generality.
 Choosing different orientations of the triples  one obtains the corresponding inequalities by
  suitable permutations of anti-diagonal matrix elements.

We will now show that (\ref{N2gen}) is indeed stronger than the
fidelity condition (\ref{Fidelity}) and the Laskowski-\.Zukowski
condition (\ref{anti-diagonal}) for $k=2$ by following the same
analysis as in the three-qubit case. We again assume, for
convenience, that the anti-diagonal element $\rho_{1,d}$ is the
largest of all anti-diagonal elements. Using some inequalities that
hold for all states together with the condition (\ref{N2gen}) for
bi-separability  we get the following sequence of inequalities for
 $\rho_{1,d}$: 
  \begin{align}
  4|\rho_{1,d}|- (\rho_{1,1}+\rho_{d,d}) \overset{A}{\leq}2 |\rho_{1,d}| \overset{2\textrm{sep}}{\leq}& 2\sqrt{\rho_{2,2}\rho_{d-1,d-1}} + \cdots+ 2\sqrt{\rho_{d/2,d/2}\rho_{d/2+1,d/2+1}} \nn\\& \overset{A}{\leq}\rho_{22} + \cdots + \rho_{d-1,d-1}. 
  \label{stronginequ}
  \end{align}
The inequality in the middle is (\ref{N2gen}). It implies all
other inequalities in the sequence (\ref{stronginequ}). The
inequality between the first and fourth term yields the
Laskowski-\.Zukowski condition for $k=2$, and between the second
and fourth gives the fidelity criterion in the formulation
(\ref{equivfidelity}). One also sees that the fidelity criterion
is stronger than the Laskowski-\.Zukowski condition for $k=2$.

We finally discuss two examples showing that the bi-separability
 condition (\ref{N2gen}) is stronger in detecting full
entanglement than other methods.
 First, consider the family of $N$-qubit
states \begin{align}\label{states} \rho_N'=
\lambda_0^+\ket{\psi_0^+}\bra{\psi_0^+}
+\lambda_0^-\ket{\psi_0^-}\bra{\psi_0^-}+
\sum_{j=1}^{2^{N-1}-1}\lambda_j(\ket{\psi_k^+}+\ket{\psi_j^-})(\bra{\psi_j^+}+\bra{\psi_j^-}).
\end{align} The states (\ref{states})  violate (\ref{N2gen}) for all
$|\lambda_0^+ - \lambda_0^-|\neq 0$ and are thus detected as fully
entangled by that condition.   In that case they are also
inseparable under any split. The fidelity criterion
(\ref{equivfidelity}), however, detects these states as
fully entangled only for
 $|\lambda_0^+ - \lambda_0^-|\geq \sum_j \lambda_j$.
Violation of (\ref{N2gen}) thus allows for detecting more  states
of the form $\rho_N'$ as fully entangled than violation of the
fidelity criterion. Further, the D\"ur-Cirac criteria detect these
states  as inseparable under any split  for $|\lambda_0^+ -
\lambda_0^-|> 2\lambda_j$, $\forall j$, which includes less states
than a violation of (\ref{N2gen}).
 This generalizes the observation of \citet{ota} from two qubits to the $N$-qubit case.

Secondly, consider the $N$-qubit GHZ-like states
$\ket{\theta}=\cos{\theta}\ket{0}^{\otimes
N}+\sin{\theta}\ket{1}^{\otimes N}$    with density matrix
 \beq\label{thetastate}
\ket{\theta}\bra{\theta}=
 \left (
\begin{array}{lll}
\cos^2{\theta} &\cdots&\cos{\theta}\sin{\theta}\\
~~\vdots&&~~\vdots\\
\cos{\theta}\sin{\theta} &\cdots&\sin^2{\theta}\\
\end{array}\right ).
 \eeq
 We can easily read off from the density matrix $\ket{\theta}\bra{\theta}$
 that the far off-anti-diagonal matrix elements $\rho_{1,d}=\rho_{d,1}$ is equal
 to $\cos{\theta}\sin{\theta}$  and that the diagonal  matrix elements
 $\rho_{2,2}, \ldots, \rho_{d-1,d-1}$ are all equal to zero.
  Using (\ref{N2gen}) we see that these states are
fully $N$-partite entangled for $\rho_{1,d}=\cos{\theta}\sin{\theta}\neq 0$,
  i.e., for all $\theta\neq 0,\pi/2$ (mod $\pi$). Thus, all fully
entangled states of this form are detected by condition
(\ref{N2gen}), including those not  detectable by any standard
multi-partite Bell inequality \cite{zukow2002}.

\subsubsection{Relationship to Mermin-type inequalities for partial separability and to LHV models}\label{LHVmermin}
We will now show that the separability inequalities of the
previous section imply already known Mermin-type inequalities
for partial separability.

Using the identity $ 2^{(N+1)}(\av{X^{(N)}_0}^2
+\av{Y^{(N)}_0}^2)=\av{M^{(N)}}^2 +\av{M'^{(N)}}^2$, for the Mermin
operators (\ref{merminN})  together with the upper bound for the
separability inequality  of (\ref{Nksepfirst}) for $x=0$ gives the
following sharp quadratic inequality: 
\beq \label{quadraticN2}
\av{M^{(N)}}^2 +\av{M'^{(N)}}^2\leq2^{(N+3)}\big(\frac{1}{4}\big)^k,~~
\forall \rho \in {\cal D}_N^{k\textrm{-sep}}. 
\eeq 
which
reproduces the  result (\ref{quadraticN}) found by
\citet{nagataPRL}. Since (\ref{Nk2first}) is equivalent to
(\ref{anti-diagonal}) we see that the Mermin type separability
condition is in fact one of Laskowski-\.Zukowski conditions
written in terms of local observables $X$ and $Y$. \forget{ These
quadratic inequalities
  straightforwardly give rise to sufficient conditions for
  $k$-separable  entanglement, i.e., if
 $\av{M^{(N)}}^2 +\av{M'^{(N)}}^2 >2^{(N+3)}({1}/{4})^{(k+1)}$,
  then the $N$-partite state $\rho$ is $k$-separable entangled and not
 \mbox{$(k+1)$}-separable entangled,
 and at least $m$-partite entangled, with $m\geq \integer{N/k}$.
If the state is separable according to some single split where the largest split contains exactly $m'$
qubits then it is a sufficient condition for
$m'$-partite entanglement.

 Furthermore, the quadratic inequalities  (\ref{quadraticN}) imply the following sharp linear Mermin-type
inequality  for $k$-separability: \beq 
|\av{M^{(N)}}|\leq 2^{(\frac{N+3}{2})}\big(\frac{1}{2}\big)^k,~~
\forall \rho \in {\cal D}_N^{k\textrm{-sep}}. \eeq \forget{This
inequality  is sharp: $\sup_{\rho \in {\cal D}_N^{k\textrm{-sep}}}
|\av{M_N} | = 2^{(N+3-2k)/2}$. }If (\ref{linearN}) is violated the
$N$ qubit state $\rho$ is $(k-1)$-separably
 entangled and it has at least $m$-partite entanglement,
 with $m\geq \integer{N/(k-1)}$. For $k=N$ inequality (\ref{linearN})  reproduces a
 result obtained by \citet{roy}.}

As a special case we consider a split of the form
$\{1\},\ldots,\{\kappa\},\{\kappa+1,\ldots,n\}$). Any state that
is separable under this  split is $(\kappa +1)$-separable so we
get the condition $\av{M^{(N)}}^2 +\av{M'^{(N)}}^2\leq
2^{(N-2\kappa+1)}$, and hence  $|\av{M^{(N)}}| \leq
2^{(N-2\kappa+1)/2}$. This strengthens the result of \citet{gisin} by a factor $2^{\kappa/2}$ for
these specific Mermin operators (\ref{merminN}).

As another special case of the inequalities (\ref{quadraticN2}),
consider $k=N$. In this case, the inequalities express a condition
for full separability of $\rho$. These inequalities are maximally
violated by fully entangled states by an exponentially increasing
factor of $2^{N-1}$, since the maximal value of $|\av{M^{(N)}}|$ for
any quantum state $\rho$ is $2^{(N+1)/2}$ \cite{wernerwolf}.
  Furthermore,  LHV models violate them also by an exponentially increasing factor of $2^{(N-1)/2}$, since
for all $N$, LHV models allow a maximal value for $|\av{M^{(N)}}|$  of
$2$ \cite{gisin,seevuff}, which is  a factor $2^{(N-1)/2}$ smaller
than the quantum maximum using entangled states. This bound for LHV models is sharp since
the maximum is attained by  choosing  the LHV expectation
values $\av{\sigma_x^i}=\av{\sigma_y^i}=1$ for all $i \in \{1,
\ldots, N\}$.   This shows that there are exponentially increasing
gaps between the values of  $|\av{M^{(N)}}|$ attainable by fully separable states,
fully entangled states and LHV models. This is shown in Figure \ref{grafiek11}.

That the maximum violation of multi-partite Bell inequalities allowed
by quantum mechanics grows exponentially with $N$ with respect to
the value obtainable by LHV models has been known for quite some
years \cite{mermin,wernerwolf}. However, it is
equally remarkable that the maximum value obtainable by separable
quantum states \emph{exponentially decreases} in comparison  to the
maximum value obtainable by LHV models, cf.\  Fig.\ \ref{grafiek11}.
 We thus see exponential divergence between separable quantum
states and LHV theories:  as $N$ grows,  the latter are able to give
correlations that need more and more entanglement in order to be
reproducible in quantum mechanics.

But why does quantum mechanics  have correlations larger than those
obtainable by a LHV model? Here we give an argument showing that
it is not the degree of entanglement but the degree of
inseparability that is responsible. The degree of entanglement of
a state may be quantified by the value $m$ that indicates the
$m$-partite entanglement of the state, and the degree of
inseparability by the value of $k$ that indicates the
$k$-separability of the state. Now suppose we have $100$ qubits.
For partial separability of $k\geq 51$ no state of these $100$
qubits can violate the Mermin inequality (\ref{linearN}) above the
LHV bound, although the state could be up to $50$-partite
entangled ($m\leq50$). However, for $k=2$, a state is possible
that is also $50$-partite entangled, but which allows for violation of  the
Mermin inequality by an exponentially large factor of $2^{97/2}$.
For $k<N$, a $k$-separable state is always entangled in some way,
so we see that it is the degree of partial separability, not the
amount of entanglement  in a multi-qubit state that determines the
possibility of a violation of the Mermin inequality. Of course,
some entanglement must be present, but the inseparability aspect
of the state determines the possibility of a violation. This is
also reflected in the fact that for a given $N$ it is the value of
$k$, and not that of $m$, which determines the sharp upper bounds
of the Mermin inequalities.
\begin{figure}[!h]
\includegraphics[scale=0.9]{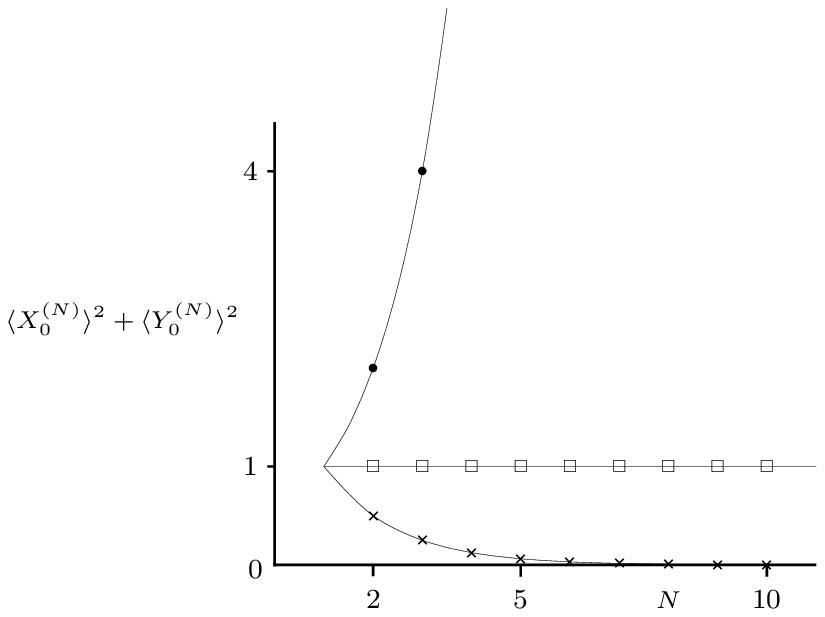}
\caption{The maximum value for $\av{X_0}^2 + \av{Y_0}^2$
obtainable by entangled quantum states (dots), by separable
quantum states (crosses) and by LHV models (squares), plotted as a
function of the number of qubits $N$.  Note the exponential
divergence between both the maxima obtained for entangled states
as well as for separable states compared to the LHV value, where
the former maximum is exponentially increasing and the latter
maximum is exponentially decreasing.} \label{grafiek11}
\end{figure}

\section[Experimental strength of the $k$-separable entanglement criteria]{Experimental strength of the $k$-separable\\ entanglement criteria}\label{strengthsection}
  Violations of the above
conditions for partial separability  provide sufficient
criteria for detecting $k$-separable entanglement (and $m$-partite
entanglement with $\integer{N/k}\leq m\leq N-k+1$). It has already
been shown that these criteria are stronger than the Laskowski-\.Zukowski
criterion for  $k$-inseparability  for $k=2,N$ (i.e., detecting
some and full entanglement), the fidelity criterion for full
inseparability (i.e., full entanglement) and the D\"ur-Cirac
criterion for inseparability under splits. In this section we will
elaborate further on the experimental usefulness and strength of
these entanglement criteria, when focusing on specific $N$-qubit states. The strength of an
entanglement criterion to detect a given entangled state may be
assessed by determining how well it copes with two desiderata
\cite{tothguhne2}:  the noise robustness of the criterion for this given state should be high, and the number of local measurements
settings needed for its implementation should be small.

In this section we will first take a closer look at the issue of
noise robustness and at the number of required settings for
implementation of the separability criteria, both in the general
state-independent case and in the case of detecting target states.
We then show the strength of the criteria for a variety of
specific $N$-qubit states.\forget{ In particular, we show (i)
detection of bound entanglement for $N\geq 3$, (ii) detection of
$N$-qubit $W$-states with great noise robustness, (iii) detection
of a six qubit cluster state and four qubit Dicke state, (iv)
great noise and decoherence robustness in detecting the $N$-qubit
GHZ state and better noise robustness than the stabilizer witness
criterion for detecting these latter states.}

\subsection[Noise robustness and the number of  measurement settings]{Noise robustness and the number of  measurement\\ settings}

White noise robustness  of an entanglement criterion for a given
entangled state is the maximal fraction $p_0$ of  white noise which may be admixed to this state  so that the state can no longer be detected as entangled by the
criterion. Thus, for a given entangled state $\rho$, the
noise robustness of a criterion is the threshold value $p_0$ for which the state
$\rho=p\,\1/2^N +(1-p) \rho$, with $p\geq p_0$ can
no longer be detected by that criterion.

So, for the criterion for detecting full entanglement
(\ref{N2gen}), the white noise robustness is found by solving the
threshold equation for $p_0$: \beq |(1-p_0) \rho_{l,\bar{l}}| =
\sum_{j\neq l} \sqrt{(\frac{p_0}{2^N}
+(1-p_0)\rho_{j,j})(\frac{p_0}{2^N}
+(1-p_0)\rho_{\bar{\jmath},\bar{\jmath}})}, \label{noiseeq} \eeq
\forget{This equation is quadratic in $p_0$ and can be easily solved. For
$l$ that gives the maximum of $|\rho_{l,\bar{l}}|- \sum_j
\sqrt{\rho_{j,j}\rho_{\bar{\jmath},\bar{\jmath}}},  j\neq l$.} The
state is fully entangled for $p<p_0$.

For the criterion (\ref{matrixfull}), for detecting some entanglement,
  one finds a similar threshold equation: \beq \max_l \{|(1-p_0)
\rho_{l,\bar{l}}|^2\} = \min_j \{(\frac{p_0}{2^N}
+(1-p_0)\rho_{j,j})(\frac{p_0}{2^N}
+(1-p_0)\rho_{\bar{\jmath},\bar{\jmath}})\}.
 \label{noiseeqsep}
\eeq This equation is quadratic and easily solved. Again, the state
is entangled for $p<p_0$.

A local measurement setting
\cite{setting,terhalComputSci,guhnehyllus}  is an observable such
as $\mathcal{M}=\sigma_1\otimes\sigma_l \ldots\otimes \sigma_N$,
where $\sigma_l$ denote single qubit  observables for each of the
$N$ qubits. Measuring such a setting (determining all coincidence
probabilities of the $2^N$ outcomes) also enables one to determine
the probabilities for observables like $\1\otimes\sigma_2
\ldots\otimes \sigma_N$, etc.\ \cite{guhne2007}.  Now consider the
observables $X_x^{(N)}$ and $Y_x^{(N)}$ that appear in the
separability criteria of (\ref{partialeq})-(\ref{Nksep}). As is
easily seen from their definitions in (\ref{Noperators}), one can
measure such an observable   using $2^N$  local settings. However,
  these  same $2^N$ settings then suffice to measure the observables $X_x^{(N)}$ and
$Y_x^{(N)}$ for all other $x$ since these are linear combinations
of the same settings. Thus,  $2^N$ measurement settings are
sufficient to determine $\av{{X}_x^{(N)}}$ and $\av{{Y}_x^{(N)}}$
for all $x$. It remains to determine the number of settings needed
for the terms $\av{I_{x}^{(N)}}$ and $\av{Z_{x}^{(N)}}$. For all
$x$ these terms contain only two single-qubit observables:
$Z^{(1)}$ and $I^{(1)}= \1$. They can thus be measured by a single
setting, i.e.,  $\left (Z^{(1)}\right)^{\otimes N}$.

 Thus, in total $2^N+1$ settings are needed in order to test the separability conditions. This number
grows exponentially  with the number of qubits. However, this is
the price we pay for being so general, i.e., for having criteria
that work for all states.  If we apply the criteria to detecting
forms of inseparability and entanglement of specific entangled
$N$-qubit states, this number can be greatly reduced. Knowledge of
the target state enables one to select a single separability
inequality for an optimal value of $x$ in
(\ref{partialeq})-(\ref{Nksep}). Violation of this single
inequality is then sufficient for detecting the entanglement in
this state, and, as we will now show, the required number of
settings then grows only linear in $N$, with $N+1$ being the
optimum for many states of interest.

For simplicity, assume  that the local observables featuring in
the criteria are the Pauli spin observables with the same
orientation for each qubit.
  We can then readily use the density matrix representations of the separability
criteria given at the end of each subsection in the previous
section. Choosing the local observables  differently amounts to
performing suitable bases changes to the density matrix
representations and would not affect the argument.

The  matrix representations  of the conditions show that only some
anti-diagonal matrix elements and the values of some diagonal
matrix elements have to be determined in order to test whether
these inequalities are violated. Indeed, observe that for all $x$~
$\av{I_{x}^{(N)}}^2-\av{Z_{x}^{(N)}}^2=4\rho_{j,j}\rho_{
\bar{\jmath}, \bar{\jmath}}$ with $\bar{\jmath}=d+1-j$ for some $j
\in \{1,2,\ldots,d\}$ and $\av{X_{x}^{(N)}}^2 -\av{Y_{x}^{(N)}}
^2=4|\rho_{j ,\bar{\jmath}}|^2$ denotes some
anti-diagonal matrix element.
 It suffices to consider $x=0$ since conditions for other values of $x$
are obtained by some local unitary basis changes that
will be explicitly given later on. We now want to rewrite the
density matrix representation for this single separability
inequality with $x=0$ in terms of less than $2^N+1$ settings.

Determining the diagonal matrix elements requires only a single
setting, namely $\sigma_z^{\otimes N}$.  Next,  we should
determine the modulus of the far-off anti-diagonal element $\rho_{1,d}$
($d=2^N$) by measuring
  $X_0^{(N)}$
   and
$Y_0^{(N)}$, since $\av{X_0^{(N)}} =2$Re$\rho_{1,d}$ and
$\av{Y_0^{(N)}} =2$Im$\rho_{1,d}$ (cf. (\ref{zbasisrel})).
Following the method of  \citet{guhne2007}, these matrix elements
can be obtained from two settings $\mathscr{M}_l$  and
$\tilde{\mathscr{M}}_l$, given by
\begin{align}\label{real}
\mathscr{M}_l&=\big(   \cos(\frac{l \pi}{N})\sigma_x + \sin(\frac{l
\pi}{N})\sigma_y \big)^{\otimes N},~~~
l=1,2,\ldots,N~,\\
\label{imaginary0} \tilde{\mathscr{M}}_l&= \big( \cos(\frac{l
\pi+\pi/2}{N})\sigma_x +\sin( \frac{l \pi+\pi/2}{N})\sigma_y
          \big)^{\otimes N},~~~ l=1,2,\ldots,N.
\end{align}
These  operators obey:
\begin{align}
\sum_{l=1}^N (-1)^l\, \mathscr{M}_l&=   N\, X_0^{(N)}, \label{real2}\\
 \sum_{l=1}^N (-1)^l\,\tilde{\mathscr{M}}_l&= N\, Y_0^{(N)}. \label{imaginary}
\end{align}
The proof of (\ref{real2}) is given by \cite{guhne2007} and
(\ref{imaginary}) can be proven in the same way.

These relations show that the imaginary and the real part of an
anti-diagonal element can  be determined by the $N$ settings
$\mathscr{M}_l $ and $\tilde{\mathscr{M}}_l$ respectively.  This
implies that the bi-separability  condition (\ref{N2gen}) needs
only $2N+1$ measurement settings. However, if each anti-diagonal
term is  real valued (which is often the case for states of
interest) it can be determined by the $N$ settings $\mathscr{M}_l
$, so  that in total $N+1$ settings suffice.

Implementation of the criteria for other $x$ involves determining
the modulus of  some other anti-diagonal matrix element instead of
the far-off anti-diagonal element $\rho_{1,d}$. The settings that
allow for this  determination can be obtained from  a local
unitary rotation on the settings $\mathscr{M}_l$ and
$\tilde{\mathscr{M}}_l$ needed to measure $|\rho_{1,d}|$. This can
be done as follows.

Suppose we want to determine the modulus of the matrix element
$\rho_{j,\bar{\jmath}}$. The unitary rotation to be applied is
given by $
U_j=\sigma_{j_1}\otimes\sigma_{j_2}\otimes\ldots\otimes\sigma_{j_N}$
with $j=j_1 j_2\ldots j_N$ in binary notation, with $\sigma_0=\1$
and $\sigma_1=\sigma_x$.  The settings that suffice are then given
by $\mathscr{M}_{j,l}=U_j \,\mathscr{M}_l\,U_j^{\dagger}$  and
$\tilde{\mathscr{M}}_{j,l}
=U_j\,\tilde{\mathscr{M}}_l\,U_j^{\dagger}$  ($l=1,2,\ldots, N$).
For example,  take  $N=4$ and suppose we want to determine
$\rho_{5,4}$. We obtain the required settings by applying the
local unitary $U_5=\1\otimes\sigma_x\otimes \1\otimes\sigma_x$
(since the binary notation of $5$ on four bits is $0101$) to the
two settings $\mathscr{M}_l $ and $\tilde{\mathscr{M}_l}$  given
in (\ref{real}) and (\ref{imaginary}) respectively that for $N=4$
allow for determining $|\rho_{1,16}|$. In conclusion, using the
above procedure the modulus of each anti-diagonal element can be
determined using $2N$ settings, and in case they are  real (or
imaginary) $N$ settings suffice.

Since the strongest separability inequality for the specific
target state  under consideration is chosen, this reduction in the
number of settings does not reduce the noise robustness for
detecting forms of entanglement  as compared to that obtained
using the entanglement criteria  in terms of the usual settings
$X_x^{(N)}$, etc.  \forget{In fact, the white noise robustness
$p_0$ for detecting full entanglement is determined via
(\ref{N2gen}) by solving the following bi-separability
\emph{simpliciter} threshold equation for $p_0$: \beq |(1-p_0)
\rho_{l,\bar{l}}| = \sum_{j\neq l} \sqrt{(\frac{p_0}{2^N}
+(1-p_0)\rho_{j,j})(\frac{p_0}{2^N}
+(1-p_0)\rho_{\bar{\jmath},\bar{\jmath}})}, \label{noiseeq} \eeq
\forget{This equation is quadratic in $p_0$ and can be easily
solved.} for $l$ that gives the maximum of $|\rho_{l,\bar{l}}|-
\sum_j \sqrt{\rho_{j,j}\rho_{\bar{\jmath},\bar{\jmath}}},  j\neq
l$. The state is fully entangled for $p<p_0$.

For detecting some entanglement  the white noise robustness $p_0$ is
determined via (\ref{matrixfull}) by solving the following $N$-separability \emph{simpliciter} threshold
equation for $p_0$:
\beq
\max_l \{|(1-p_0) \rho_{l,\bar{l}}|^2\} =
\min_j \{(\frac{p_0}{2^N} +(1-p_0)\rho_{j,j})(\frac{p_0}{2^N} +(1-p_0)\rho_{\bar{\jmath},\bar{\jmath}})\}.
 \label{noiseeqsep}
\eeq This equation is quadratic in $p_0$ and can be easily solved.
The state is entangled for $p<p_0$. }

 In conclusion, if the state
to be detected is known, the $2N$ settings of  (\ref{real}) and
(\ref{imaginary0})  together with the single setting
$\sigma_z^{\otimes N}$ suffice, and in case this state has solely
real or imaginary anti-diagonal matrix elements only $N +1$
settings are needed. The white noise robustness using these
settings  is just as great as using the general condition that use
the observables $X_x^{(N)}$ and $Y_x^{(N)}$, and is found by
solving (\ref{noiseeq}) or (\ref{noiseeqsep}) for detecting full
and some entanglement respectively.

As a final note, we observe that in order to determine the modulus
of not just one but of all anti-diagonal matrix elements it is
more efficient to use the observables $X_x^{(N)}$, $Y_x^{(N)}$
than the observables of (\ref{real}) and (\ref{imaginary0}). The
first method needs $2^N$ settings to do this and the second needs
$2^NN/2$ settings (since  there are $2^N/2$ independent
anti-diagonal elements), i.e.,  the latter needs more settings
than the former for  all $N$.

\forget{\subsubsection{examples}
 We will next apply the above procedure to two examples.
 The celebrated $N$-qubit GHZ state $\ket{\Psi_{\textrm{GHZ},0}^N}=
 (\ket{0^{\otimes N}}+\ket{1^{\otimes N}})/\sqrt{2}$
has only two non-zero off diagonal matrix elements $\rho_{1,d}$ and $\rho_{d,1}$
which are real valued and thus equal to each other. This state can be detected as
fully entangled using the criterion
$(\rho_{1,d})^2>\max\{ \rho_{j,j}\rho_{\bar{\jmath},\bar{\jmath}} \}$,  $j\in \{2,\dots,d-1\}$.  As
discussed above, for its implementation it  needs only
$N+1$  measurement settings ($N$ settings $\mathscr{M}_l$ given in (\ref{real}) plus the setting $
\sigma_z^{\otimes N}$).  The criterion has noise robustness of $p_0=1/(1+2^{(1-N)})$,
which goes to $1$ for large $N$. (In section \ref{GHZexample} this state is further analyzed
and other criteria for detecting it as entangled are also considered).
}

Let us apply the above procedure to an example, taken from  \citet{guhne2007},  the so-called four-qubit singlet state, which
is\forget{a state in a decoherence free subspace (i.e., invariant
under a simultaneous unitary rotation on all qubits). It is} given
by: \beq\label{4singlet} \ket{\Phi_4}=(\ket{0011} +\ket{1100}
-\frac{1}{2}(\ket{01} +\ket{10}) \otimes(\ket{01}+\ket{10}       )
)/\sqrt{3}. \eeq For detecting it as fully entangled
(\ref{noiseeq}) gives a noise robustness $p_0=12/29\approx0.41$,
and for detecting it as entangled (\ref{noiseeqsep}) gives a noise
robustness of $16/19\approx 0.84$. The implementation needs
$16+1=17$ settings.

 This number of settings can be reduced by using the fact that this state has
only real anti-diagonal matrix elements and that we need only look
at the largest anti-diagonal element. As shown above, this matrix
element can be measured in $4$ settings. Thus the total number of
settings required is reduced to only $5$.  The off-diagonal matrix
element to be determined is $\ket{0011} \bra{1100}$. The four
settings that allow for this determination are obtained from the
four settings given in (\ref{real}) by applying the unitary
operator $U_3=\1\otimes\1\otimes\sigma_x\otimes \sigma_x$ to these
settings.

For comparison, note that \citet{guhne2007} showed
that the projector-based witness for the state
($\ref{4singlet}$) detects full entanglement with a white noise robustness $p_0=0.267$ and
uses $15$ settings, whereas  the optimal witness from
\citet{guhne2007} uses only  $3$ settings and has $p_0=0.317$. Here
we obtain $p_0\approx0.41$ using $5$ settings, implying a
significant  increase in  white noise robustness using only two
settings more.

This  example gives the largest noise robustness when the
conditions are measured in the standard $z$-basis.  However, sometimes one obtains larger noise
robustness when the state is first rotated so as to be expressed
in a different basis before it is analyzed.  For example, consider
the four qubit Dicke state $\ket{2,4}$, where
$\ket{l,N}=\binom{N}{l}^{-1/2}\sum_k
\pi_k(\ket{1_1,\ldots,1_l,0_{l+1},\ldots,0_N})$ are the symmetric
Dicke states \cite{dicke}  (with $\{ \pi_k (\cdot)\}$  the set of
all distinct permutations of the $N$ qubits).  In the standard
basis this state does not violate any of the separability
conditions we have discussed above. However, if each qubit is
rotated around the $x$-axis by $90$ degrees all of the
separability conditions can be violated with quite high noise
robustness. Indeed, it is detected as inseparable under all splits
through violation of conditions (\ref{partialeqM}) for
$p<p_0=16/19\approx0.84$ and as fully entangled through violation
of condition (\ref{Nk2}) for $p<p_0=4/11\approx0.36$ using $5$
settings. For comparison,  \citet{chen}  used
specially constructed entanglement witnesses for detection of full
entanglement in these states, and they obtained  as noise
robustness $p_0=2/9\approx 0.22$ using only $2$ settings.  We have not
performed an optimization procedure, so it is unclear whether or
not the values obtained for $p_0$ can be improved.

\subsection{Noise and decoherence robustness for the $N$-qubit GHZ state}\label{GHZexample}
In this subsection  we  determine the robustness of our
separability criteria for detecting the $N$-qubit GHZ state in
five kinds of noise processes (admixing white and colored noise,
and three types of decoherence: depolarization, dephasing and
dissipation of single qubits). We give the noise robustness as a function
of $N$ for detecting some entanglement, inseparability with respect to all splits and  full entanglement. We  compare the results for white noise robustness of the criteria for full entanglement to that of the fidelity criterion (\ref{fidelitycriterion}) and to that of the so called stabilizer criteria of \citet{tothguhne2,tothguhne3}.

 The $N$-qubit GHZ state
$\ket{\Psi_{\mathrm{GHZ},0}^N}
=\frac{1}{\sqrt{2}}(\ket{0}^{\otimes N} +\ket{1}^{\otimes N}$) can
be transformed into a mixed state $\rho_N$ by  admixing noise to
this state or by decoherence. Let us consider the following five
such processes.

(i) Mixing in a fraction $p$ of white noise gives:
\beq\label{whitenoiseghz}
\rho_N^{\textrm{(i)}}=(1-p)\ket{\Psi_{\mathrm{GHZ},0}^N}\bra{\Psi_{\mathrm{GHZ},0}^N} +p\frac{\1}{2^N}.
\eeq

(ii) Mixing in a fraction $p$ of colored noise \cite{noise2} gives:
\beq
\rho_N^{\textrm{(ii)}}= (1-p)\ket{\Psi_{\mathrm{GHZ},0}^N}\bra{\Psi_{\mathrm{GHZ},0}^N} +\frac{p}{2}
(\ket{0\ldots0}\bra{0\ldots0} +\ket{1\ldots1}\bra{1\ldots1}).
\eeq

(iii) A depolarization process \cite{noise} with a
depolarization degree $p$ of a single qubit gives: 
\begin{alignat}{2}
&\ket{i}\bra{j}\longrightarrow (1-p)\ket{i}\bra{j}
+p\,\delta_{ij} \frac{\1}{2},\nn \\
&\rho_N^{\textrm{(iii)}}=\frac{1}{2}\big{[}  \big{(}
(1-\frac{p}{2}) \ket{0}\bra{0} + \frac{p}{2}\ket{1}\bra{1}
\big{)}^{\otimes N}
+\big{(}&&\frac{p}{2}\ket{0}\bra{0} +  (1-\frac{p}{2})\ket{1}\bra{1}\big{)}^{\otimes N}\nn\\
&&&+(1-p)^N\big{(}\ket{0}\bra{1}^{\otimes N}+\ket{1}\bra{0}^{\otimes N}\big{)}\big{]}.
\end{alignat}

(iv) A dephasing process \cite{noise} with a dephasing
degree $p$ of a single qubit gives:  \begin{align}
&\ket{i}\bra{j}\longrightarrow (1-p)\ket{i}\bra{j}
+p\,\delta_{ij} \ket{i}\bra{j}, \nn\\&
 \rho_N^{\textrm{(iv)}}  =
 \frac{1}{2} \big{[} \ket{0}\bra{0}^{\otimes
N}+\ket{1}\bra{1}^{\otimes N}+ (1-p)^N(\ket{0}\bra{1}^{\otimes
N}+\ket{1}\bra{0}^{\otimes N}) \big{]}. \end{align}

 (v) A dissipation process \cite{noise} with a dissipation degree $p$ of a single qubit (where
the ground state is taken to be $\ket{0}$) gives: 
\begin{align}
&\ket{i}\bra{i}  \longrightarrow  (1-p)\ket{i}\bra{i} +  p\ket{0}\bra{0},  \nn\\
&\ket{i}\bra{j}  \longrightarrow  (1-p)^{1/2}\ket{i}\bra{j},
~~~~i\neq j ,    \nn \\
 &\rho_N^{\textrm{(v)}}  =  \frac{1}{2}
\big{[} \ket{0}\bra{0}^{\otimes N}+(p\ket{0}\bra{0}+
(1-p)\ket{1}\bra{1})^{\otimes N}+\nn\\&~~~~~~~~~~~~~~~~~~~~~~~~~~~~~~~~~~~~~~~~~~~~
(1-p)^{N/2}(\ket{0}\bra{1}^{\otimes N}+\ket{1}\bra{0}^{\otimes N})
\big{]}. 
\end{align}

We now consider the question for what values of $p$ these states
$\rho_N^{(\textrm{i})}$ to $\rho_N^{(\textrm{v})}$ are detected, firstly, as containing some entanglement (using the condition \eqref{matrixfull}) and, secondly, as inseparable under any split (using the conditions of the form (\ref{partialeqM}) for all bi-partite splits). In other words, we determine the noise (or
decoherence) robustness of violations of all these conditions for $\rho_N^{(\textrm{i})}$ to $\rho_N^{(\textrm{v})}$.
We find the following threshold values $p_0$.
 \begin{alignat}{2}
 \label{GHZsomeN}
&\textrm{(i)}&& p_0= \frac{1}{1+2^{(1-N)}}, \nn
\\
&\textrm{(ii)}&& p_0 =1,~~~~ \forall N,\nn
\\
&\textrm{(iii)}~&& (1-p_0)^N= 
(1-\frac{p_0}{2})^{\alpha}(\frac{p_0}{2})^{(N-\alpha)} +(1-\frac{p_0}{2})^{(N-\alpha)}(\frac{p_0}{2})^{\alpha},\\
&\textrm{(iv)}&&p_0 =1,~~~~\forall N,
\nn\\
&\textrm{(v)}&& p_0= 1,~~~~\forall N.\nn 
\end{alignat} 
For cases (i), (ii), (iv) and (v) the threshold values $p_0$ for detecting some entanglement and inseparability with respect to all splits are the same because for these cases the product of the diagonal matrix elements  $\rho_{j,j}\rho_{\bar{j},\bar{j}}$ is the same for all $j\neq 1,d$.
 Only in case (iii) is this product different for different $j$.
We then have to take the minimum and maximum value, respectively, from which it follows that 
 $\alpha$ is to be set to $[N/2]$ for detecting some entanglement and to $1$ for detecting inseparability with respect to all splits. Here $[N/2]$ is the largest integer smaller or equal to $N/2$.

 The result in case (i)
is in accordance with the results of \cite{duer2,duer},
where it is furthermore shown that the opposite holds as well,
i.e., iff $p<1/(1+2^{(1-N)})$ then $\rho_N^{\textrm{(i)}}$ is
inseparable under any split and otherwise it is fully separable. Thus all states of the form
(\ref{whitenoiseghz}) that are inseparable under any split are
detected by violations of  the conditions of the form (\ref{partialeqM}) for all bi-partite splits. The same holds for cases (ii),
(iv) and (v), since all states $\rho_N^{\textrm{(ii)}}$,
$\rho_N^{\textrm{(iv)}}$ and $\rho_N^{\textrm{(v)}}$ are
inseparable under any split for all $p<1$. In other words, as soon
as a fraction of the GHZ state is present, these states
 are inseparable under any split. In case
 (i) $p_0$  increases  monotonically from $p_0=2/3$ for $N=2$ to
$p_0=1$ for large $N$. For process (iii) these  limiting values
are not so straightforward: $p_0=(3-\sqrt{3})/3\approx0.42$ for
$N=2$, and $p_0=(5-\sqrt{5})/5\approx0.55$ for  large $N$. In
conclusion, the noise and decoherence robustness is high for all
$N$, except maybe for case (iii).

Next, consider the noise robustness for detecting full entanglement
by means of  the bi-separability condition (\ref{Nk2}). The result is the
following: 
\begin{alignat}{2}
\label{GHZfullN} 
&\textrm{(i)}&& p_0=1/(2(1-2^{-N})),
\nn
\\
&\textrm{(ii)}&& p_0 =1,~~~~ \forall N,\nn
\\
&\textrm{(iii)}~~&& p_0 \approx 0.42,0.28,0.22,0.18,~~ N=2,3,4,5.
\\
&\textrm{(iv)}&&p_0=1,~~~~\forall N,\nn
\\
&\textrm{(v)}&& p_0\approx 1,0.48,0.39,0.35, ~~N=2,3,4,5.\nn 
\end{alignat}

For case (i) the noise robustness is equivalent to the fidelity
criterion (\ref{fidelitycriterion}). For large $N$  $p_0$ decreases  to the limit value $p_0=1/2$.
Case (ii) and (iv) have $p_0=1$, thus as soon as the states
$\rho_N^{\textrm{(ii)}}$ and $\rho_N^{\textrm{(iv)}}$ are
entangled they are fully entangled. For cases  (iii) and (v) we listed the
noise robustness found numerically for $N=2$ to $N=5$. These values decrease for increasing $N$.

Let us  compare the results for white noise robustness (case (i))
to the results  obtained from the so-called stabilizer formalism \cite{gottesman}.
This formalism   is used by \citet{tothguhne2,tothguhne3} to
derive entanglement witnesses  that are especially useful for minimizing the number of  settings
required to detect either full or some entanglement. Here we will
only consider the criteria formulated for detecting entanglement
of the $N$-qubit GHZ states. 
 The stabilizer witness by T\'oth \& G\"uhne that detects some
entanglement has $p_0=2/3$, independent of  $N$, and requires only
three settings (cf.\  Eq. (13) in \cite{tothguhne2}). The
strongest witness
 for full entanglement  of  T\'oth \& G\"uhne has a robustness
 $p_{0}=1/(3-2^{(2-N)})$ and requires only two settings  (cf.\ Eq. (23) in \cite{tothguhne2}).

Figure \ref{grafiek4} shows these threshold noise ratios for
detecting full entanglement for these three criteria. Note that
the criterion of  \cite{tothguhne2} needs only
two measurement settings, whereas our criteria need $N+1$
settings. So although the former are less robust against white
noise admixture, they compare favorably with respect to minimizing
the number of measurement settings.

Although we give a criterion for full entanglement that is
generally stronger than the fidelity criterion, for the
$N$-partite GHZ state this does not lead to  better noise
robustness. It appears that for large $N$ the noise threshold
$p_0=1/2$ is the best one can do. However, in the limit of large
$N$ the GHZ state is inseparable under all splits for all $p_0<1$,
as was shown in (i) in (\ref{GHZsomeN}). See also Figure
\ref{grafiek4}.

We have seen that if the state $\rho_N^{\textrm{(i)}}$ is entangled it is also inseparable under any split. 
 Because of the high symmetry of both the GHZ state and white noise, one might
 conjecture that if this state is entangled it is also fully entangled.
 At present, however it is unknown whether this is indeed true.
Detecting the states $\rho_N^{\textrm{(i)}}$ as fully entangled
appears to be a much more demanding task than detecting them as
inseparable under all splits. In the first case, for large $N$,
only a fraction of $50 \%$ noise is permitted,  in the second case
one can permit any noise fraction  (less than $100 \%$). Note that we have given explicit examples of states that are diagonal in the GHZ basis (cf. \eqref{dccondition} of section \ref{introsepconditions}), and that are inseparable under any split, but not fully entangled. But these are not of the form  $\rho_N^{\textrm{(i)}}$.

Lastly, we mention that  our criteria detect the various forms of entanglement and inseparability also if the state $\ket{\Psi_{\mathrm{GHZ},0}^N}$ is replaced by any other maximally entangled state (i.e., any state of the GHZ basis, cf. \eqref{DuerStates}), a feature which is not possible using linear entanglement witnesses.  There is no single linear witness  that detects entanglement of all maximally entangled states.

\begin{figure}[!h]
\includegraphics[scale=1.0]{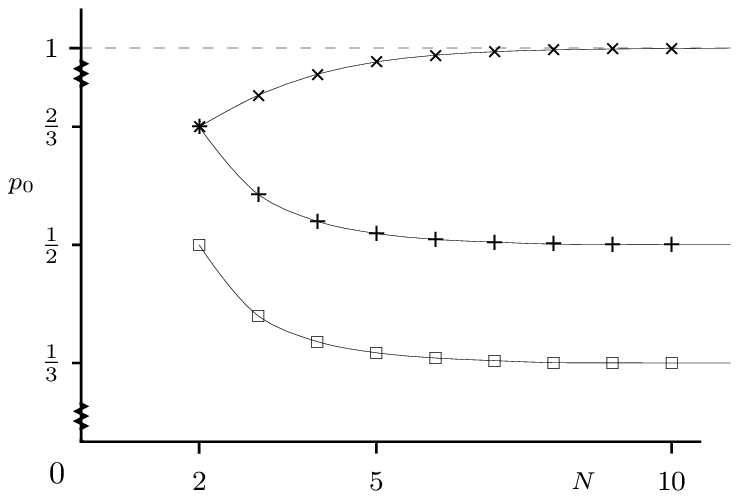}
\caption{ The threshold noise ratios $p_0$ for detection of full
$N$-qubit entanglement when admixing white noise to the $N$-qubit
GHZ state for the criterion (\ref{Nk2}) derived here (plus-signs)
and  for the stabilizer witness of \citet{tothguhne2}
 (squares). The noise robustness for
detecting  inseparability  under all splits as given in (i) in
(\ref{GHZsomeN}) is
 also plotted (crosses).
}
\label{grafiek4}
\end{figure}

\subsection{Detecting bound entanglement for $N\geq 3$}
 Violation of the separability inequality (\ref{NNsep}) allows for detecting all bound
 entangled states of  \citet{dur}. These states have the form
\beq \label{boundstate}
\rho_B=\frac{1}{N+1} \left( \ket{\Psi_{{\rm
GHZ},\alpha}^N}\bra{\Psi_{{\rm GHZ},\alpha}^N}
+\frac{1}{2}\sum_{l=1}^{N} P_l +\bar{P}_l \right),
 \eeq with $P_l$ the projector on the state
$\ket{0}_1\ldots\ket{1}_l \ldots \ket{0}_N$, and where $\bar{P}_l$
is obtained from $P_l$ by replacing all zeros by ones and vice
versa. For $N\geq4$ these states are entangled and have positive
partial transposition (PPT) with respect to transposition of any
qubit. This means they are  bound entangled \cite{bound}. Note
that they are detected as entangled  by  the $N$-partite Mermin
inequality $|M_N|\leq2$ of section \ref{Nqubitsection} only for
$N\geq8$ \cite{dur}. However, the condition (\ref{NNsep}) detects
them as entangled  for $N\geq 4$.  Thus all bound entangled states
of this form are detected as entangled by this latter condition.
\forget{, cf.\cite{nagata2007}. }The white noise robustness for
this purpose is $p_0=2^N/(2+2 N+ 2^N)$, which for $N=4$ gives
$p_0=8/13\approx0.615$ and  goes to $1$ for large $N$.
 Note that  for $N=4$, this state violates  the condition for
$4$-separability, and the condition for $3$-separability
(\ref{Nksep}), but not  the condition for $2$-separability. It is
thus at least $2$-separable entangled. It is not detected as fully
entangled by these criteria. (Of course, it could still be fully
entangled since these criteria are only sufficient and not
necessary for entanglement).  For general $N$ we have not
investigated the $k$-separable entanglement of the states
(\ref{boundstate}), although this can be readily performed using
the criteria of (\ref{Nksep}).

Another interesting bound entangled state is the so-called four
qubit Smolin state \cite{smolin} \beq \rho_S=
\frac{1}{4}\sum_{j=1}^4
\ket{\Psi^j_{ab}}\bra{\Psi^j_{ab}}\otimes\ket{\Psi^j_{cd}}\bra{\Psi^j_{cd}},
\eeq where $\{ \ket{\Psi^j} \}$ is the set of four Bell states $\{
\ket{\phi^{\pm}}, \ket{\psi^{\pm}} \}$, and $a,b,c,d$ label the
four qubits. This state is also detected as entangled by the
criterion (\ref{NNsep}), and with white noise robustness  $p_0=2/3$.
The Smolin state violates the separability conditions
(\ref{partialeqM}) for bi-separability under the splits
$a$-$(bcd)$, $b$-$(acd)$, $c$-$(abd)$, $d$-$(abc)$. However, it is
separable under the splits $(ab)$-$(cd)$, $(ac)$-$(bd)$,
$(ad)$-$(bc)$ (cf.\ \cite{smolin}). This state is thus inseparable under
splits that partition the system into two subsets with one and
three qubits, but it is separable when each subset contains two
qubits.

So far we have detected bound entanglement for $N\geq 4$.  What
about $N=3$? Consider the three-qubit bound entangled state of
\citet{duer22}: \begin{align} \rho=
\frac{1}{3}\ket{\Psi_{GHZ,0}^3}\bra{\Psi_{GHZ,0}^3}
+\hskip-0.025cm\frac{1}{6}(\ket{001}\bra{001} +\hskip-0.025cm\ket{010}\bra{010}
+\hskip-0.025cm\ket{101}\bra{101} +\hskip-0.025cm\ket{110}\bra{110} ). \end{align} This state is
detected as entangled by the criterion (\ref{3sep3}), with white
noise robustness  $p_0= 4/7\approx0.57$. It violates the
bi-separability condition (\ref{ineq_2sep}) for the split
$a$-$(bc)$ so it is at least bi-separable entangled, but does not
violate the condition (\ref{2sep3}) for bi-separability
 i.e., it is not detected as fully entangled.
In fact, it can be shown using the results of \citet{duer}
that this state is separable under the splits $b$-$(ac)$ and
$c$-$(ab)$.

\section{Discussion}
\label{discussionn}
We have discussed partial separability of quantum states by
distinguishing $k$-separa-bility from $\alpha_k$-separability and used
these distinctions to extend the classification proposed by D\"ur
and Cirac. We discussed the relationship of partial separability
to multi-partite entanglement and distinguished the notions of a
$k$-separable entangled state and a $m$-partite entangled state
and indicated the interrelations of these kinds of entanglement.

 Next, we have presented necessary conditions for partial separability
  in the  hierarchic separability classification. These are  formulated
     in terms of experimentally accessible correlation inequalities for operators
     defined by products of local orthogonal observables.
   Violations of these inequalities provide, for all $N$-qubit states, criteria for
   the entire hierarchy of $k$-separable
entanglement, ranging from the levels $k$=1 (full or genuine
$N$-particle entanglement) to $k=N$ (full separability, no
entanglement), as well as for specific classes within each level.
Choosing the  Pauli matrices as the locally orthogonal observables
provided matrix representations of the criteria that bound
anti-diagonal matrix elements in terms of diagonal ones.

Further, the $N$-qubit Mermin-type separability inequalities for
partial separability were shown to follow from the partial
separability conditions derived in this chapter. The bi-separability
 conditions are stronger than the
fidelity criterion and the Laskowski-\.Zukowski criterion, and the
latter criterion is also shown to be strengthened for full
separability and biseparability. For separability under splits the conditions are
stronger than the D\"ur-Cirac conditions. Violation of these
conditions thus give entanglement criteria that  detect more
entangled states than violations of these three other separability conditions. \forget{
We therefore believe these state-independent entanglement criteria
to be the strongest experimentally accessible conditions for
multi-qubit entanglement applicable to all multi-qubit states.}

We have furthermore shown that the required number of measurement
settings for implementation of these criteria, which is $2^N +1$
in  general, can be drastically reduced if entanglement of a given
target state is to be detected. In  that case, it may be reduced
to  $2N+1$, and for multi-qubit states with either real or
imaginary anti-diagonal matrix elements, only $N+1$ settings are
needed.

When comparing the entanglement criteria to other state-specific
multi-qubit entanglement criteria it was found that the white noise
robustness was high for a great variety of interesting multi-qubit
states, whereas  the number of required settings was only $N+1$.
However, these other state-specific entanglement criteria need
less settings although for the states analyzed here they give
lower noise robustness. Analyzing some specific target states
shows that the entanglement criteria detect bound entanglement for
$N\geq3$.

Furthermore, we applied  the entanglement criteria  for some and
full entanglement to  the $N$-qubit GHZ state subjected to two
different kinds of noise and three different kinds of decoherence.
The robustness against colored noise and against dephasing turns out
to be maximal (i.e., $p_0=1$) both for detecting some and full
entanglement.  It is remarkable that for large $N$ the GHZ state
allows for maximal white noise robustness for the state to remain
inseparable under all possible splits, whereas for detecting full
entanglement  the best known result -- to our best knowledge --
only allows for a white noise robustness of $p_0=1/2$. It would be very
interesting to search for full entanglement criteria that  can
close this gap, or if this is shown to be impossible to understand
why this is the case.

Orthogonality of the local observables is  crucial in the above
derivation of separability conditions. It is due to this
assumption that the multi-qubit operators  form representations of
the generalized Pauli group. It would be interesting to analyze
the role of orthogonality in deriving the inequalities. For two
qubits it has been shown by \citet{tradeoff}, see chapter \ref{chapter_CHSHquantumtradeoff}, that when  orthogonality
is relaxed the separability conditions become less strong, and we
conjecture the same holds for their multi-qubit analogs. Relaxing
the requirement of orthogonality has the advantage that some
uncertainty in the angles may be accommodated, which is desirable
since in real experiments it may be hard to measure perfectly
orthogonal observables.

It is also interesting that the separability inequalities are
equivalent to bounds on anti-diagonal matrix elements in terms of
products of diagonal ones.  We thus gain a new perspective on why
they allow for entanglement detection: they probe the values of
anti-diagonal matrix elements, which encode entanglement
information about the state; and if these elements are large
enough, this entanglement is detected.
 Note furthermore that compared to the Mermin-type separability inequalities
we need not do much more to obtain our stronger inequalities. We
must solely determine some diagonal matrix elements, and this can
be easily performed using the single extra setting
$\sigma_z^{\otimes N}$. 
\forget{It is also noteworthy that the comparison to the Mermin-type
separability inequalities shows that the strength of the
correlations allowed for by separable states is exponentially
decreasing when compared to the strength of the correlations
allowed for by LHV models.}

Our recursive definition of the multi-partite correlation operators
 (see (\ref{Noperators})) is by no means unique. One can generate
many new inequalities by choosing the locally orthogonal
observables differently, e.g.,  by permuting their order  in each
triple of local observables. It could well be that combining such
new inequalities with those presented here yield even stronger
separability conditions, as is indeed the case for pure two-qubit
states \cite{uffseev}, see chapter \ref{chapter_CHSHquantumorthogonal}. Unfortunately, we have no conclusive
answers  for this open question.

We end by suggesting three further lines of future research.
Firstly, it would be interesting to apply the entanglement
criteria to an even larger variety of $N$-qubit states than
analyzed here, including for example all $N$-qubit graph and Dicke
states. Secondly, the generalization from qubits to qudits (i.e.,
$d$-dimensional quantum systems)
 would, if indeed possible, prove very useful since strong partial separability criteria for $N$ qudits
have -- to our knowledge -- not yet been obtained. And finally, it would be beneficial to have optimization
procedures for choosing the set of local orthogonal observables featuring in the
entanglement criteria that gives the highest noise robustness for a given set of states.
We believe we have chosen such optimal sets for the variety of states analyzed here, but since no rigorous optimization was performed, our choices could perhaps be improved.

To end this chapter we  comment on the noteworthy result that  the Mermin-type separability inequalities show that the strength of the correlations in  separable qubit states is exponentially decreasing when compared to the strength of the correlations  allowed for by LHV models. From a more fundamental point of view it is quite remarkable  that the strengthened Bell-type inequalities which were shown to hold for
separable qubit states, do not hold  for LHV theories, for which
the Bell-type inequalities were originally designed.  This  shows that the latter theories are able to give
correlations for which quantum mechanics, in order to reproduce them using qubit states, 
needs recourse to entangled states; and even more and more so when
the number of particles increases.

 Assuming that the LHV doctrine is a necessary ingredient
 for the notion of classicality, the idea that it is
  the separable qubit states which are the classical states  among all qubit states needs revision.
  Although \citet{werner} was the first to point to this, we here show a
 much more radical and general departure, especially when the number of qubits grows. Of course,
 if more general measurement scenarios than the standard Bell experiment setup are allowed things might change.
 Given the surprising results found here between separability of
qubit states and local hidden-variable structures, the question
what is exactly the classical part of quantum mechanics  seems to
still be not fully answered and open for new investigations.

\forget{

This `non-classicality' of separable qubit states raises
interesting questions. Is it the case that quantum separability and
LHV separability (locality) are fundamentally different and not
equally strong requirements? What is the relationship between the
independence notions derived from the principle of locality and from
a state being separable? In what way is mathematical separability of
states to be understandable as an expression of physical
separability or of independence? Given the surprising results found here between separability of
qubit states and local hidden-variable structures, the question
what is exactly the classical part of quantum mechanics  seems to
still be not fully answered and open for new investigations.
}

However, it should be mentioned that, just as was the case in chapter \ref{chapter_CHSHquantumorthogonal} for the two-qubit separability inequalities,  these findings hold only  for the case of qubits.\forget{
It is important to realize that the above only holds for the case of qubits. For the crucial relation (\ref{quadratic1}) can be violated for 
systems whose state space is a larger Hilbert space than the single qubit state space $\H=\mathbb{C}^2$.  The observables $A$, $A'$, $A''$ that are locally orthogonal spin observables  correspond to pairwise anti-commuting operators only in the case of a qubit. For systems with a large enough Hilbert space they can be commuting.  Simply choose this Hilbert space to be the direct sum of the eigenspaces of the three spin observables so that they do not have any overlap. Thus} By choosing the Hilbert space of the systems under consideration to be large enough  any choice of observables can be made commuting\footnote{This is  is easily obtained by generalizing the argument given in section \ref{comparisonlhvN2}  from two to $N$ qubits.}. Using separable states  of a system consisting of  such systems one can, after all, reproduce the predictions of all LHV models. 

Thus one may take an experimental violation of the Mermin-type separability inequalities  by $N$-qubits to mean two things: (i)  either one can conclude that the state of the $N$-qubits is entangled, or (ii) the state might be separable but then one is not dealing with qubits after all and some degrees of freedom must have been overlooked.

\clearemptydoublepage
\thispagestyle{empty}
\chapter{Monogamy of correlations}
\forget{\chapter{Classification and monogamy of three-qubit bi-separable correlations}}
\label{chapter_monogamy}
\noindent
This chapter is in part based on \citet{seevmon}.

\section[Introduction]{Introduction to the monogamy of entanglement\\ and of correlations}
If a pure quantum state of two systems is entangled, then none of the two systems can be entangled with a third system. This can be easily seen.  Suppose that systems $a$ and $b$ are in a pure entangled state. Then when the system $ab$ is considered as part of a larger system, the reduced density operator for $ab$ must by assumption be a pure state. However, for the composite system $ab$ (or for any of its subsystems $a$ or $b$) to be entangled with another system, the reduced density operator of $ab$ must be a mixed state.  But since it is by assumption pure, no entanglement between $ab$ and any other system can exist. This feature is referred to as the monogamy of pure state entanglement\footnote{This is sometimes confusingly referred to as the claim that in quantum theory a system can be pure state entangled with only one other system \cite{spekkens}. But what about the GHZ state $(\ket{000}+\ket{111})/\sqrt{2}\,$? All three parties are entangled to each other in this pure state, so this seems to be a counterexample to the claim. What is actually meant is that if a pure state of two systems is entangled, then none of the two systems can be entangled with a third system. This is the formulation we will use.}. 

This monogamy can also be understood as a consequence of the linearity of quantum mechanics that is also responsible for the no-cloning theorem. For suppose that party\footnote{For ease of notation we will use the same symbols to refer to parties and the systems they possess, e.g., party $a$ possesses system $a$.}  $a$ has a qubit which is maximally pure state entangled to both a qubit held by party $b$ and a qubit held by party $c$. Party $a$ thus has a single qubit coupled to two perfect entangled quantum channels, which this party could exploit  to teleport two perfect copies of an unknown input state, thereby violating the no-cloning theorem, and thus the linearity of quantum mechanics \cite{terhal}.

If the state of two systems is not a pure entangled state but a mixed entangled state,  then it is possible that both of the two systems  are entangled to a third system. For example, the $W$-state $\ket{\psi}=(\ket{001}+\ket{010}+\ket{100})/\sqrt{3}$\forget{(cf. \ref{WState})} has bi-partite reduced states that are all identical and entangled. This feature is called `sharing of mixed state entanglement',  or `promiscuity of entanglement'. So we see that entanglement is strictly speaking only monogamous in the case of pure entangled states. In the case of mixed entangled states  it can be promiscuous.  But this promiscuity is not unbounded:  although some entangled bi-partite states may be shareable with some finite number of parties, no entangled bi-partite state can be shared with an infinite number of parties\footnote{This is also referred to as `monogamy in an asymptotic sense'  by \citet{terhal}, but we believe that this feature is better captured by the term `no unbounded promiscuity'}.
 Here  a bi-partite state $\rho_{ab}$ is said to be $N$-shareable  when it is possible to find a quantum state $\rho_{a{b_1}{b_2}\ldots{b_N}}$ such that $\rho_{ab}=\rho_{{a{b_1}}}=\rho_{a{b_2}}=\ldots =\rho_{a{b_N}}$, where  $\rho_{a{b_k}}$ is the reduced state for parties $a$ and $b_k$. Consider the following theorem \cite{fannes,raggio}:  A bi-partite state is $N$-shareable for all $N$ (also called $\infty$-shareable \cite{masanes06}) iff it is separable. Thus no bi-partite entangled state, pure or mixed, is $N$-shareable for all $N$.

The monogamy of entanglement was first quantified by \citet*{ckw}  who gave a trade-off relation
between how entangled $a$ is with $b$, and how entangled $a$ is with $c$ in a three-qubit system $abc$ that is in a pure state, using the measure of bi-partite entanglement called the tangle \cite{osborne}. It states that $\tau(\rho_{ab})+\tau(\rho_{ac})\leq \tau(\rho_{a(bc)})$ where $\tau(\rho_{ab})$ is the tangle\footnote{The tangle $\tau(\rho_{ab})$ is the square of the concurrence $\mathcal{C}(\rho_{ab}):=\max\{0,\sqrt{\lambda_1}-\sqrt{\lambda_2}-\sqrt{\lambda_3}-\sqrt{\lambda_4}\}$, where the $\lambda_i$ are the eigenvalues of the matrix $\rho_{ab}(\sigma_y\otimes\sigma_y)\rho_{ab}^*(\sigma_y\otimes\sigma_y)$ in non-decreasing order, with $\sigma_y$ the Pauli-spin matrix for the $y$-direction.} between $A$ and $B$, analogous for $\tau(\rho_{ac})$ and $\tau(\rho_{a(bc)})$ is the bi-partite entanglement\footnote{In case of three qubits the tangle $\tau(\rho_{a(bc)})$ is equal to $4\,\textrm{det}\rho_a$, with $\rho_a=\textrm{Tr}_{bc}[\ket{\psi}\bra{\psi}]$ and $\ket{\psi}$ the pure three-qubit state.}
 across the split $a$-$(bc)$. The multi-partite generalization has been recently proven by \citet{osborne}. In general, $\tau$ can vary between $0$ and $1$, but monogamy constrains the entanglement (as quantified by $\tau$) that  party $a$ can have with each of parties $b$ and $c$.

Classically one does not have such a trade-off. All classical probability distributions can be shared \cite{toner2}. If parties $a$, $b$ and $c$ have bits instead of quantum bits (qubits) and if $a$'s bit is always the same as $b$'s bit then there is no restriction on how $a$'s bit is correlated to $c$'s bit.

Let us however not just look at entanglement but also at correlations that result from making local measurements on quantum systems. As we have seen many times already,  these correlations can violate Bell-type inequalities that hold for local hidden-variable models. Such Bell-type inequality violating correlations turn out to be monogamous. This is termed `monogamy of quantum non-locality' or `non-local correlations are monogamous'. 

Let us show this in a setup where each party implements two possible dichotomous observables. The CHSH inequality is the only non-trivial local Bell-type inequality for this setup. All quantum correlations that violate this inequality are monogamous as follows from the following tight trade-off inequality for a three-partite system $abc$ proven by \citet{tonerverstraete}:
\begin{align}
\av{\mathcal{B}_{ab}}_{\textrm{qm}}^2 +\av{\mathcal{B}_{ac}}_{\textrm{qm}}^2\leq8,
\label{tonerverstraeteineq}
\end{align}  
where $\mathcal{B}_{ab}$ is the CHSH operator (\ref{operator1}) for parties $a$ and $b$, and analogous for $\mathcal{B}_{ac}$. 

\begin{figure}[!hb]
\includegraphics[scale=1]{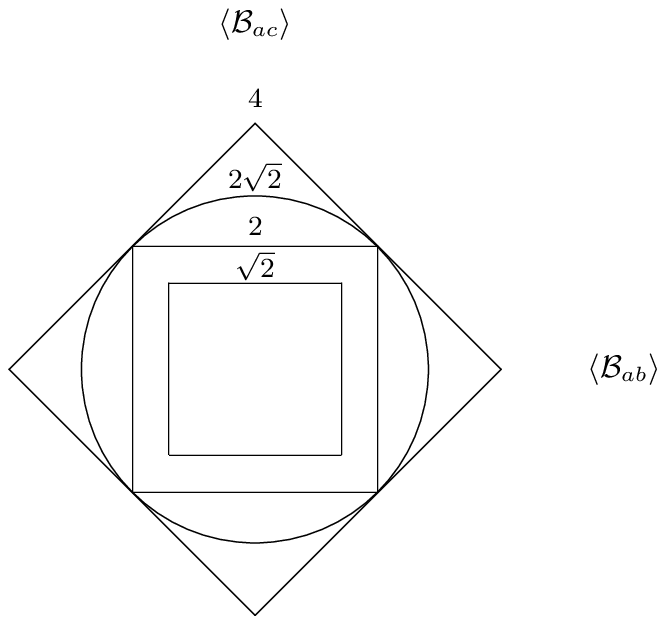}
\caption{Monogamy of quantum and no-signaling correlations. All quantum correlations lie within the circle, and all no-signaling correlations lie within the tilted larger square. For comparison the classical correlations are also shown. These lie within the square with edge length $2$. The correlations obtainable by orthogonal measurements on separable two-qubit states lie within the smallest square.}\label{figmonogamy}
\end{figure} 
\forget{
 \setlength{\unitlength}{0.25 mm}
 \begin{figure}[h]\begin{center}
\begin{picture}(200,270)(-100,-100)
\linethickness{0.1mm} 
\put(-100,0){\line(1,1){100}}
\put(100,0){\line(-1,1){100}}
\put(-100,0){\line(1,-1){100}}
\put(100,0){\line(-1,-1){100}}
 \put(-50,-50){\line(0,1){100}}
 \put(-50,-50){\line(1,0){100}}
 \put(50,50){\line(0,-1){100}}
\put(50,50){\line(-1,0){100}}
\put(-35,-35){\line(0,1){70}}
\put(-35,-35){\line(1,0){70}}
\put(35,35){\line(0,-1){70}}
\put(35,35){\line(-1,0){70}}
 \put(0,0){\circle{141}}
 \put(150,0){\makebox(0,0){$\av{\mathcal{B}_{ab}}$}}
 \put(0,140){\makebox(0,0){$\av{\mathcal{B}_{ac}}$}}
 \put(0,78){\makebox(0,0){\footnotesize{$2\sqrt{2}$}}}
 \put(0,58){\makebox(0,0){\footnotesize{$2$}}}
  \put(0,42){\makebox(0,0){\footnotesize{$\sqrt{2}$}}}
   \put(0,110){\makebox(0,0){\footnotesize{$4$}}}
\end{picture}\end{center}
\caption{Monogamy of quantum and no-signaling correlations. All quantum correlations lie within the circle, and all no-signaling correlations lie within the tilted larger square. For comparison the classical correlations are also shown. These lie within the square with edge length $2$. The correlations obtainable by orthogonal measurements on separable two-qubit states lie within the smallest square.}\label{figmonogamy}
\end{figure}
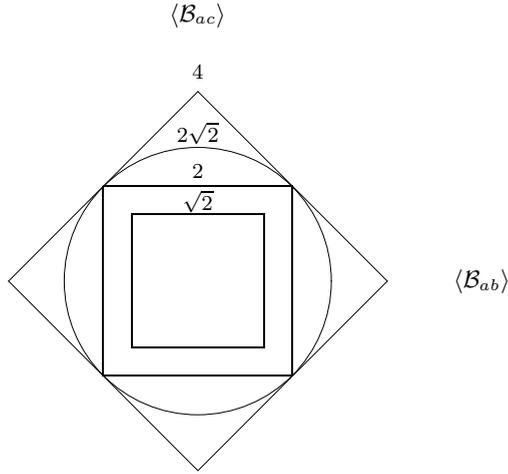 }\forget{
=============}
Quantum correlations thus show an interesting trade-off relationship: In case the correlations between party $a$ and $b$ are non-local (i.e., when $|\av{\mathcal{B}_{ab}}_{\textrm{qm}}|>2$) the correlations between parties $a$ and $c$ cannot be non-local (i.e., necessarily $|\av{\mathcal{B}_{ac}}_{\textrm{qm}}|\leq 2$), and vice versa (cf. \citet{scarani}). These non-local quantum correlations can thus not be shared. Furthermore, in case they are maximally non-local, i.e., $|\av{\mathcal{B}_{ab}}_{\textrm{qm}}|=2\sqrt{2}$ the other must be uncorrelated, i.e., it must be that  
$|\av{\mathcal{B}_{ac}}_{\textrm{qm}}|=0$, and vice versa.
Here it is crucial that the measurements performed by party $a$ are the same in both expressions. 

This trade-off relation is plotted in Figure \ref{figmonogamy}. It provides a non-trivial bound on the set of quantum correlations because a three-partite system $abc$ cannot simultaneously violate the CHSH inequality for correlations between $ab$ (summing over $c$'s outcomes) and between $ac$ (summing over $c$'s outcomes). Both general unrestricted and local correlations do not obey such a monogamy trade-off. The first type can reach the absolute maxima $|\mathcal{B}_{ab}|_\textrm{max}=|\mathcal{B}_{ac}|_\textrm{max}=4$, and the second type can attain the maximal value for local correlations, i.e,   
$\av{\mathcal{B}_{ab}}_\textrm{lhv}=\av{\mathcal{B}_{ac}}_\textrm{lhv}=2$.

The reason for this is that  general unrestricted correlations and local correlations can be shared. The latter fact is proven by  \citet{masanes06} and the first we will prove here. However, first we need the relevant definitions. Shareability of a general unrestricted probability distribution is defined as follows (where for simplicity we restrict ourselves to shareability of bi-partite distributions). A bi-partite distribution $P(a,b_1|A,B_1,\ldots, B_N)$ is $N$-shareable with respect to the second party if an $(N+1)$-partite distribution $P(a,b_1,\ldots, b_{N}|A,B_1,\ldots, B_N)$ \mbox{exists} that is symmetric with respect to $(b_1,B_1), \,(b_2,B_2),\,\ldots,\,(b_N,B_N)$ and with marginals $P(a,b_i|A,B_1,\ldots, B_N)$ equal to the original distribution $P(a,b_1|A,B_1,\ldots, B_N)$, for all $i$. For notational clarity we use $b_i$ and $B_i$ (instead of $a_i$ and $A_i$) to denote outcomes and observables respectively for the parties other than the first party. If a distribution is shareable for all $N$ it is called  $\infty$-shareable.

Shareability of a no-signaling probability distribution is defined analogously: A no-signaling distribution $P(a,b_1|A,B_1)$ is $N$-shareable with respect to the second party if there exist an $(N+1)$-partite distribution $P(a,b_1,\ldots, b_{N}|A,B_1,\ldots, B_N)$ being symmetric with respect to $(b_1,B_1), \,(b_2,B_2),\,\ldots,\,(b_N,B_N)$ with marginals $P(a,b_i|A,B_i)$ equal to the original distribution $P(a,b_1|A,B_1)$, for all $i$. The difference between shareability of unrestricted correlations and of no-signaling correlations is that in the first case the marginals depend on all $N+1$ settings, whereas in the latter case they only depend on the two settings $A$ and $B_i$.

Suppose we are given a general unrestricted correlation $P(a,b_1|A,B_1,\ldots, B_N)$. We can then construct  
\begin{align} P(a,b_1,\ldots, b_{N}|A,B_1,\ldots, B_N)=P(a,b_1|A,B_1,\ldots, B_N)\delta_{b_1,b_2}\cdots\delta_{b_1,b_N},
\end{align}  which has by construction the same marginals $P(a,b_i|A,B_1,\ldots, B_N)$ equal to the original distribution $P(a,b_1|A,B_1,\ldots, B_N)$. This holds for all $i$, thereby proving the $\infty$-shareability. Thus an unrestricted correlation can be shared for all $N$.  If we restrict the distributions to be no-signaling, \citet{masanes06} proved that $\infty$-shareability  implies that the distribution is local, i.e., it can be written as  \begin{align}
P(a,b_1,\ldots, b_N)|A,B_1\ldots, B_N)=\int_\Lambda d\lambda p(\lambda)P(a|A,\lambda)P(b_1|B_1,\lambda)\cdots P(b_N|B_N,\lambda), 
\end{align} for some  local distributions  $P(a|A,\lambda),P(b_1|B_1,\lambda),\ldots, P(b_N|B_N,\lambda)$ and hidden-variable distribution $p(\lambda)$.

Because general unrestricted correlations and local ones can be shared they both will not show any monogamy. This implies that partially-local correlations also do not show any monogamy, since these are combinations of local and general unrestricted correlations between subsystems of the $N$-systems.

The above result by \citet{masanes06} shows that quantum and no-signaling correlations can not be $\infty$-shareable and they must therefore show monogamy effects. Monogamy of quantum correlations has already been shown above via the trade-off relation (\ref{tonerverstraeteineq}), so let us move to no-signaling correlations. First consider a very strong monogamy property for extremal no-signaling correlations, already mentioned by \citet{barrett05}. Suppose one has some no-signaling three-party probability distribution $P(a_1,a_2,a_3|A_1,A_2,A_3)$ for parties $a$, $b$ and $c$. In case the marginal distribution $P(a_1,a_2|A_1,A_2)$ of system $ab$ is extremal then it cannot be correlated to the third system $c$,
 as the following proof by \citet{barrett05} shows.

\forget{
=============

Here it is crucial that the measurements for party $A$ are the same in both expressions. This trade-off relation is plotted in Figure \ref{figmonogamy}. It provides a non-trivial bound on the set of quantum correlations since one cannot simultaneously have CHSH inequality violating correlations between $ab$ (summing over $c$'s outcomes)
and between $ac$ (summing over $c$'s outcomes). Both general and local correlations do not have such a monogamy trade-off. General correlations can reach the absolute maximum bound for both CHSH expressions since they are unrestricted, and local correlations can give the maximal local value of $2$ for both expressions. The reason for this latter fact is that local correlations are derivable from local probability distributions which can all be shared. In fact, locality of a no-signaling distribution is equivalent to $N$-shareability for all $N$ ($\infty$-shareability) \cite{masanes06}. Here shareability of a probability distribution is defined as follows (where for simplicity we restrict ourselves to shareability of bi-partite distributions). A distribution $P(a_1,b_2|A_1,B_2)$ is shareable with respect to party $2$ if there exist an $(N+1)$-partite distribution $P(a_1,b_2,\ldots, b_{N}|A_1,B_2,\ldots, B_N)$ being symmetric with respect to $(b_1,B_1),\ldots(b_N,B_N)$ with marginals $P(a_1,b_i|A_1,B_i)$ equal to the original distribution $P(a_1,b_2|A_1,B_2)$. For notational clarity we used $b_i$ and $B_i$ (instead of $a_i$ and $A_i$) to denote outcomes and observables respectively for the parties other than $1$.

Since general correlations are unrestricted and local ones can be shared which ensures they both do not show any monogamy, this implies that the partially-local correlations also will not 
show any monogamy.  This follows from the fact that the partially-local correlations 
are combinations of local and general correlations between subsystems of the $N$-systems, each of which do not show monogamy.

=================

No-signaling correlations, however, can in general not be shared and they thus show monogamy just as some quantum correlations do. Let us first consider a very strong monogamy property for extremal no-signaling correlations. Suppose one has some no-signaling three-party probability distribution $P(a_1,a_2,a_3|A_1,A_2,A_3)$ for parties $a$, $b$ and $c$. Then in case the marginal distribution $P(a_1,a_2|A_1,A_2)$ or $ab$ is extremal then cannot be correlated to the third system $c$,
 as the following proof by \cite{barrett05} shows.
 }
Bayes' rule and no-signaling give  
\begin{align}
P(a_1,a_2,a_3|A_1,A_2,A_3)&= P(a_1,a_2,|A_1,A_2,A_3,a_3)P(a_3|A_1,A_2,A_3)\nn\\
&= P(a_1,a_2|A_1,A_2,A_3,a_3)P(a_3|A_3).
\end{align}
Therefore the marginal $P(a_1,a_2|A_1,A_2)$  can be rewritten as 
\begin{align}
P(a_1,a_2|A_1,A_2)&= \sum_{a_3} P(a_1,a_2,a_3|A_1,A_2,A_3)\nn\\&=\sum_{a_3} P(a_1,a_2|A_1,A_2,A_3,a_3)P(a_3|A_3), ~\forall A_3.
\end{align} 
Since by supposition $P(a_1,a_2|A_1,A_2)$ is extremal the decomposition is unique, this gives\\ \mbox{$P(a_1,a_2|A_1,A_2,A_3,a_3)=P(a_1,a_2|A_1,A_2), \forall a_3,A_3$}. Then combining all this gives:
\begin{align}
P(a_1,a_2,a_3|A_1,A_2,A_3)=P(a_1,a_2|A_1,A_2)P(a_3|A_3),
\end{align}
which implies that party $c$ is completely uncorrelated with party $ab$: the extremal correlation $P(a_1,a_2|A_1,A_2)$ is completely monogamous.  Note that this implies that all local Bell-type inequalities for which the maximal violation consistent with no-signaling is attained by a unique correlation  have monogamy constraints. This follows because all Bell-type inequalities are linear in the correlations, therefore, if the maximal violation is produced by a unique correlation, it can only be produced by an extreme point of the no-signaling polytope. Otherwise the correlation that produces maximal violation would not be unique. An example is the CHSH inequality, as will be shown below.

Extremal no-signaling correlations thus show monogamy, but what about non-extremal no-signaling correlations?  Just as was the case for quantum states where non-extremal (mixed state) entanglement can be shared, non-extremal no-signaling correlations can be shared as well. This can be seen from the fact that no-signaling correlations obey the following tight trade-off relation in terms of the CHSH operators \cite{toner2}: 
\begin{align}\label{monogamynosignaling}
|\av{\mathcal{B}_{ab}}_{\textrm{ns}}| +|\av{\mathcal{B}_{ac}}_{\textrm{ns}}|\leq 4.
\end{align} 
This is also depicted in Figure \ref{figmonogamy}.  Extremal no-signaling correlations can attain $|\av{\mathcal{B}_{ab}}_{\textrm{ns}}|=4$ so that necessarily $|\av{\mathcal{B}_{ac}}_{\textrm{ns}}|=0$, and vice versa (this is monogamy of extremal no-signaling correlations), whereas non-extremal ones are shareable since the correlation terms $|\av{\mathcal{B}_{ab}}_{\textrm{ns}}|$ and  $|\av{\mathcal{B}_{ac}}_{\textrm{ns}}|$ can both be non-zero at the same time. But note that in case the no-signaling correlations are non-local they can not be shared, i.e., it is not possible that 
$|\av{\mathcal{B}_{ab}}_{\textrm{ns}}|\geq2$ and $|\av{\mathcal{B}_{ac}}_{\textrm{ns}}|\geq 2$. This shows that if these non-local correlations can be shared they must be signaling.

For general unrestricted correlations no monogamy holds, i.e., $|\av{\mathcal{B}_{ab}}|$ and $|\av{\mathcal{B}_{ac}}|$ are not mutually constrained and can each obtain a value of $4$ so as to give the absolute maximum of the left hand side of (\ref{monogamynosignaling}) which is the value $8$. The monogamy bound (\ref{monogamynosignaling}) therefore gives a way of discriminating no-signaling from general correlations:  if it is violated the correlations cannot be no-signaling (i.e., they must be signaling).  This discerning inequality uses product expectation values only, in contrast to the facets of the no-signaling polytope that only give non-trivial constrains on the marginal expectation values, as was discussed in chapter \ref{definitionchapter}, section \ref{chsintrotech}.

For classical correlations no such trade-off  as in (\ref{tonerverstraeteineq}) or as in (\ref{monogamynosignaling}) holds. Indeed, it is possible to have both $|\av{\mathcal{B}_{ab}}_{\textrm{lhv}}|=2$ and $|\av{\mathcal{B}_{ac}}_{\textrm{lhv}}|=2$, see also Figure \ref{figmonogamy}. \forget{Thus only classical correlations are shareable, whereas quantum and no-signaling ones need not be shareable.} This reflects the fact that classical correlations are always shareable. 
The correlations  that separable quantum states allow for are also shareable. Indeed, in the $\av{\mathcal{B}_{ab}}_{\textrm{qm}}$-$\av{\mathcal{B}_{ac}}_{\textrm{qm}}$ plane of Figure \ref{figmonogamy} such correlations can reach the full square with edge length $2$. 
 Analogous to what we have seen in chapter \ref{chapter_CHSHquantumtradeoff}, it is the case that when considering qubits and measurements that are restricted to orthogonal ones only one obtains tighter bounds. These restrict the possible values of $\av{\mathcal{B}_{ab}}_{\textrm{qm}}$ and $\av{\mathcal{B}_{ac}}_{\textrm{qm}}$ to the smallest square of Figure \ref{figmonogamy}:  $|\av{\mathcal{B}_{ab}}_{\textrm{qm}}|, |\av{\mathcal{B}_{ac}}_{\textrm{qm}}|\leq \sqrt{2}$, $\rho \in \mathcal{D}_{\textrm{sep}}$.
But again there is no monogamy for separable states in this case since this full square can be reached.

\subsection[A stronger monogamy relation for the non-locality of bi-partite quantum correlations]{A stronger monogamy relation for the non-locality of\\  bi-partite quantum correlations}

We will now give an alternative simpler proof of  the inequality (\ref{tonerverstraeteineq}) that also allows us to strengthen it as well.
The proof uses the idea that (\ref{tonerverstraeteineq}), which describes the interior of a circle in the  $\av{\mathcal{B}_{ab}}$-$\av{\mathcal{B}_{ac}}$ plane,  is equivalent to the interior of the set of tangents to this circle. It is thus 
a compact way of writing the following infinite set of linear equalities
\begin{align}\label{linmon}
\mathcal{S}=\max_\theta \av{\mathcal{S}_\theta}_{\textrm{qm}}\leq 2\sqrt{2},
\end{align}
where we have used $\sqrt{x^2 +y^2} =\max_\theta (\cos\theta\, x +\sin\theta\, y)$, and where 
$\mathcal{S}_{\theta}=\cos\theta\, \mathcal{B}_{ab}+\sin\theta \,\mathcal{B}_{ac}$. 

We will now prove this by showing that $|\av{\mathcal{B}_{ab} \cos\theta + \mathcal{B}_{ac} \sin\theta }_{\textrm{qm}}|\leq 2\sqrt{2}$ for all $\theta$, using  a method presented by \citet{dieks} in a different context.  In this proof we only consider quantum correlations  so for brevity we drop the subscript `qm' from the expectation values. Let us first write 
\begin{align}
\mathcal{B}_{ab} \cos\theta + \mathcal{B}_{ac} \sin\theta =(A+A')B\cos\theta +&(A-A')B'\cos\theta+
\nn\\ &(A+A')C\sin\theta +(A-A')C\sin\theta.
\label{tussen}
\end{align}
Next we express $A$ and $A'$  in terms of orthogonal Pauli observables in some basis using the geometry of  Figure \ref{hoek}: $A= \cos\gamma \sigma_x +\sin\gamma\sigma_z$ and $A'= \cos\gamma \sigma_x -\sin\gamma\sigma_z$. This gives $A+A'=2\cos\gamma\sigma_x,~ A-A'=2\sin\gamma \sigma_z$.
\begin{figure}
\includegraphics[scale=1]{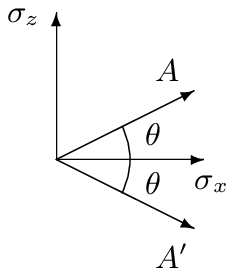}
\caption{Expressing $A$ and $A'$ in terms of orthogonal Pauli spin observables in some basis.}
\label{hoek}
\end{figure} \forget{
\begin{figure}
{ \setlength{\unitlength}{1mm}
\begin{picture}(30,35)(-55,-15)
\linethickness{0.1mm} 
 \put(0,0){\vector(0,1){15}}
 \put(0,0){\vector(1,0){15}}
\put(0,0){\vector(2,1){14}}
\put(0,0){\vector(2,-1){14}}
  \put(0,0){\arc{15}{5.82}{0.47}}
 \put(-5,14){\mbox{$\sigma_z$}}
\put(14,-3){\mbox{$\sigma_{x}$}}
 \put(9,1.5){\mbox{$\theta$}}
  \put(9,-3.5){\mbox{$\theta$}}
 \put(10,8){\mbox{$A$}} 
\put(10,-11){\mbox{$A'$}} 
 \end{picture}
}
\caption{Expressing $A$ and $A'$ in terms of orthogonal Pauli spin observables in some basis.}
\label{hoek}
\end{figure}
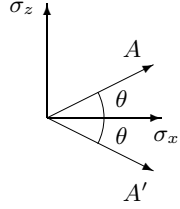}
Taking the expectation value of (\ref{tussen}) gives
\begin{align}
|\av{\mathcal{B}_{ab} \cos\theta}_{ab} + \av{\mathcal{B}_{ac} \sin\theta}_{ac}|= ~&2| \av{\sigma_x B}_{ab}\cos\gamma\cos\theta +\av{\sigma_z B'}_{ab}\sin\gamma\cos\theta\nn\\& +\av{\sigma_x C}_{ac}\cos\gamma\sin\theta  +\av{\sigma_z C'}_{ac}\sin\gamma\sin\theta |
\end{align}
The right hand side can be considered to be twice the absolute value of the inproduct of the two four-dimensional vectors $\bm{a}=(\av{\sigma_x B}_{ab}, \av{\sigma_z B'}_{ab},\av{\sigma_x C}_{ac},\av{\sigma_z C'}_{ac})$ and $\bm{b}=
(\cos\gamma\cos\theta,$ $ \sin\gamma\cos\theta,$$ \cos\gamma\sin\theta,$$\sin\gamma\sin\theta)$. If we now apply the Cauchy-Schwartz  inequality  $|(\bm{a},\bm{b})|\leq ||\bm{a}||\,||\bm{b}||$ we find 
\begin{align}
|\av{\mathcal{B}_{ab} \cos\theta}_{ab} &+ \av{\mathcal{B}_{ac} \sin\theta}_{ac}|\nn\\
&\leq2\sqrt{\av{\sigma_xB}_{ab}^2+\av{\sigma_zB'}_{ab}^2+ \av{\sigma_xC}_{ac}^2+\av{\sigma_zC'}_{ac}^2}~\times\nn\\&\qquad\qquad\qquad \qquad\sqrt{\cos^2\gamma(\cos^2\theta+ \sin^2\theta)+\sin^2\gamma(\cos^2\theta+ \sin^2\theta)}\nn\\
&\leq 2\sqrt{2(\av{\sigma_x}_{a}^2+\av{\sigma_z}_{a}^2)}\nn\\
&\leq2 \sqrt{2}\sqrt{1-\av{\sigma_y}_{a}^2}\label{proof1}\\
&\leq 2\sqrt{2}\label{proof2}
\end{align}
This proves (\ref{linmon}). Here we have used that $\av{\sigma_x}_{\textrm{qm}}^2+\av{\sigma_y}_{\textrm{qm}}^2 +\av{\sigma_z}_{\textrm{qm}}^2\leq1$ for all single qubit quantum states, and for clarity we have used the subscripts $ab$, $ac$ and $a$ to indicate with respect to which subsystems the quantum expectation values are taken. 
Using (\ref{proof1}) we obtain 
\begin{align}
\av{\mathcal{B}_{ab}}_{\textrm{qm}}^2 +\av{\mathcal{B}_{ac}}_{\textrm{qm}}^2\leq8(1-\av{\sigma_y}_{a}^2),
\label{strongmon}
\end{align} which strengthens the original monogamy inequality (\ref{tonerverstraeteineq}). An alternative strengthening  of  (\ref{tonerverstraeteineq}) 
was already found by \citet{tonerverstraete}:\,$\av{\mathcal{B}_{ab}}_{\textrm{qm}}^2 +\av{\mathcal{B}_{ac}}_{\textrm{qm}}^2\leq8(1-\av{\sigma_y\sigma_y}_{bc}^2)$.

So far we have only focused on subsystems $ab$ and $ac$, and not on the subsystem $bc$. One could thus also consider the quantity 
$\av{\mathcal{B}_{bc}}_{\textrm{qm}}$.  The above method would give the intersection of the three cylinders $\av{\mathcal{B}_{ab}}_{\textrm{qm}}^2+\av{\mathcal{B}_{ac}}_{\textrm{qm}}^2\leq8$, $\av{\mathcal{B}_{ab}}_{\textrm{qm}}^2+\av{\mathcal{B}_{bc}}_{\textrm{qm}}^2\leq8$, $\av{\mathcal{B}_{ac}}_{\textrm{qm}}^2+\av{\mathcal{B}_{bc}}_{\textrm{qm}}^2\leq8$. It is known \cite{tonerverstraete} that this bound is not tight.

It might be tempting to think that because of these results we could have the following even stronger inequality than  (\ref{tonerverstraeteineq}):
\beq\label{mono1}
\av{\mathcal{B}_{ab}}_{\textrm{qm}}^2+\av{\mathcal{B}_{ac}}_{\textrm{qm}}^2+\av{\mathcal{B}_{bc}}_{\textrm{qm}}^2\leq8.
\eeq
However, this is not true. For a pure separable state (e.g. $\ket{000}$) the left hand side has a maximum of $12$, which violates \Eq{mono1}. But inequality (\ref{mono1}) is true for the exceptional case that we have maximal violation for one pair, say $ab$, since we know from (\ref{tonerverstraeteineq}) that $\av{\mathcal{B}_{ac}}_{\textrm{qm}}$ and  $\av{\mathcal{B}_{bc}}_{\textrm{qm}}$ for the  other two pairs must then be zero.
We can see the monogamy trade-off at work: in case of maximal violation of the CHSH inequality (i.e., for maximal entanglement) the left hand side of \Eq{mono1} has a maximum of 8, whereas in case of no violation of the CHSH inequality it allows for a maximum value of 12, which can be obtained by pure separable states. Thus we see the opposite behavior from what is happening in the ordinary CHSH inequality: for the expression considered here  separability gives higher values, and entanglement necessarily lower values.

A correct bound is obtained from (\ref{proof1}) and the two similar ones for the other two expressions $\av{\mathcal{B}_{ab}}_{\textrm{qm}}^2+\av{\mathcal{B}_{bc}}_{\textrm{qm}}^2$ and  $\av{\mathcal{B}_{ac}}_{\textrm{qm}}^2+\av{\mathcal{B}_{bc}}_{\textrm{qm}}^2$. This gives:
\beq
\av{\mathcal{B}_{ab}}_{\textrm{qm}}^2+\av{\mathcal{B}_{ac}}_{\textrm{qm}}^2+\av{\mathcal{B}_{bc}}_{\textrm{qm}}^2\leq12- 4(\av{\sigma_y}_a^2+\av{\sigma_y}_b^2+\av{\sigma_y}_c^2).
\eeq
\forget{$a,b,c$ staan voor elk van de drie deelsystemen. Hierbij is de basis voor de Pauli operatoren zo dat de de $z$-richting voor $a$ en $b$ bepaald worden door loodrecht te staan op de spinobservabelen die voorkomen in $\mathcal{B}_{ab}$ voor $a$ en $b$ respectievelijk. En zo ook voor de $z$-richting van deeltje $c$ die vastgelegd worden door of $ \mathcal{B}_{ac}$ of $\mathcal{B}_{bc}$ (beide geven hetzelfde).} However, it is unknown if this inequality is tight.

\subsection[Monogamy of non-local quantum correlations vs. monogamy\\ of entanglement]{Monogamy of non-local quantum correlations vs.\\  monogamy of entanglement}\label{comparingmonogamies}

Two types of monogamy and shareability have been discussed: of entanglement and of correlations. These are different in principle, although sometimes they go hand in hand. Monogamy (shareability) of entanglement is a property of a quantum state, whereas monogamy (shareability) of correlations
 is not solely determined by the state of the system under consideration, but it is also dependent on 
  the specific setup used to determine the correlations. That is, it is crucial to also know the number of observables per party and the number of outcomes per observable.  It is thus possible that  a quantum state can give non-local correlations that are monogamous when obtained in one setup, but which are shareable when obtained in another setup. An example of this will be given below. This example also shows that shareability of non-local quantum correlations and shareability of entanglement are related in a non-trivial way.

\citet{masanes06} already remarked (and as was discussed above) that, if we consider an unlimited number of parties, locality and $\infty$-shareability of bi-partite correlations are identical properties. This is analogous to the fact that quantum separability and $\infty$-shareability of a quantum state are identical in the case of an unlimited number of parties. But if we consider shareability with respect to only one other party the analogy between locality, separability and shareability breaks down. Instead we have the following result: Shareability of non-local quantum correlations implies shareability of entanglement of mixed states, but not vice versa. The proof runs as follows. Because by assumption the correlations are shareable they are identical for parties $a$ and $b$ and $a$ and $c$. Furthermore, because the correlations are non-local, the quantum states for $ab$ and $ac$ that are supposed to give rise to these correlations must be entangled. They furthermore must be non-pure, i.e., mixed, because entanglement of pure states can not be shared. This concludes the proof. Below we give an example of this and show that the converse implication does not hold. In order to do so we will first discuss methods that reveal the shareability of non-local correlations. 


In general a bi-partite quantum state can be investigated using different setups that each have a different number of observables per party and outcomes per observable. In each such a setup the monogamy of the correlations  that are obtainable via measurements on the state can be investigated. This is performed via a Bell-type inequality that distinguishes local from non-local correlations in the setup used.

Let us first assume the case of two parties that each measure two dichotomous observables. For this case the only relevant local Bell-type inequality is the CHSH inequality for which we have seen that the Toner-Verstraete trade-off \eqref{tonerverstraeteineq} implies that all quantum non-local correlations must be monogamous: it is not possible to have correlations between party $a$ and $b$  of subsystem $ab$ and between $a$ and $c$ of subsystem $ac$ such that both $|\av{\mathcal{B}_{ab}}_{\textrm{qm}}|$ and $|\av{\mathcal{B}_{ac}}_{\textrm{qm}}|$ violate the LHV bound. 
 
It is tempting to think that\forget{because a violation of the CHSH inequality requires the (reduced) bi-partite states of subsystems $ab$ and $ac$ to be entangled, that} those entangled states that show monogamy of non-local quantum correlations will also show monogamy of entanglement. This, however, is not the case. We have seen that in general entanglement of mixed states can be shared to another party, and for our particular case considered here three-party pure entangled states exist whose reduced bi-partite states are identical, entangled and able to violate the CHSH inequality (e.g., the W-state of \eqref{WState} has such reduced bi-partite states). These reduced bi-partite states are mixed and their entanglement is shareable, yet they show monogamy of the non-local correlations obtainable from these states in a setup that has two dichotomous observables per party.  Thus we cannot infer from the monogamy of non-local correlations that quantum states responsible for such correlations have monogamy of entanglement; some of them have shareable mixed state entanglement. Consequently,  the study of the non-locality of correlations in a setup that has two dichotomous observables per party, and consequently the CHSH inequality, does not reveal  shareability of the entanglement of bi-partite mixed states. 

It is possible to reveal shareability of entanglement of bi-partite mixed states using a Bell-type inequality. But for that it is necessary that the non-local correlations which are obtained from the state in question are not monogamous, i.e., a setup must be used in which some non-local quantum correlations turn out to be shareable.  The case of two dichotomous observables per party was shown not to suffice. 
However, adding one observable per party does suffice. Consider the setup where each of the two parties measures three dichotomic observables, which will be denoted by $A,A',A''$ and $B,B,B''$ respectively. \citet{collinsgisin} have shown that for this setup only one relevant new inequality besides the CHSH inequality can be obtained (modulo permutations of observables and outcomes). This inequality reads:  
\begin{align}\av{\mathcal{C}}_{\textrm{lhv}}:=~&\av{AB}_{\textrm{lhv}}+
\av{A'B}_{\textrm{lhv}}+
\av{A''B}_{\textrm{lhv}}+
\av{AB'}_{\textrm{lhv}}+
\av{A'B'}_{\textrm{lhv}}+
\av{AB''}_{\textrm{lhv}}\nn\\
&-\av{A''B'}_{\textrm{lhv}}
-\av{A'B''}_{\textrm{lhv}}
+\av{A}_{\textrm{lhv}}
+\av{A'}_{\textrm{lhv}}
-\av{B}_{\textrm{lhv}}
-\av{B'}_{\textrm{lhv}}
\leq4\label{collgisineq}
\end{align}
\citet{collinsgisin} show that the fully entangled pure three-qubit state $\ket{\phi}= \mu \ket{000} +\sqrt{(1-\mu^2)/2}(\ket{110}+\ket{101})$ 
\forget{with $\mu=0.852$} gives for some values of $\mu$ correlations between party $a$ and $b$ of subsystem $ab$ and between $a$ and $c$ of subsystem $ac$ such that the inequality is violated:  
$\av{\mathcal{C}_{ab}}_{\textrm{qm}}\geq4$  and $\av{\mathcal{C}_{ac}}_{\textrm{qm}}\geq4$. 
Some of the non-local correlations between party $a$ and $b$ can thus be shared with party $a$ and $c$. 

Since $\ket{\phi}$ is a pure entangled three-qubit state the two-qubit reduced states $\rho_{ab}$ and $\rho_{ac}$ of subsystem $ab$ and $ac$ respectively are mixed. Furthermore, since the state $\ket{\phi}$ is symmetric with respect to qubit $b$ and $c$ these reduced states are identical. They must also be entangled because they violate the two-party inequality \eqref{collgisineq}. Therefore, the two-qubit mixed entangled state $\rho_{ab}$ is shareable to at least one other qubit. 
This shows that the inequality \eqref{collgisineq} is  suitable to reveal shareability of entanglement of mixed states. \forget{, something which was not possible using the CHSH inequality. }

It would be interesting to investigate the multi-partite extension of these results. Does monogamy exist for quantum correlations that violate a $N$-qubit Bell-type inequality, such as the $N$-partite Mermin-type inequalities? Are these inequalities also suitable for revealing shareability of entanglement of mixed $N$-qubit states for some definite number $N$?  In the next section, section \ref{montribi}, such an investigation is performed for $N=3$: we study the monogamy of bi-separable three-partite quantum correlations that violate a three-qubit Bell-type inequality that has two dichotomic measurement per party. For this specific Bell-type inequality we find that maximal violation by the bi-separable three-partite quantum correlations is monogamous. This is to be expected because maximal quantum correlations are obtained from pure state entanglement which is monogamous,  but we non-trivially find that the correlations that give non-maximal violations can be shared.

\section[Monogamy of three-qubit bi-separable  quantum correlations]{Monogamy of three-qubit bi-separable\\  quantum correlations}\label{montribi}
\noindent
\forget{Although Bell inequalities have been originally proposed to test quantum
mechanics against local realism, nowadays  they also serve another purpose, namely investigating
quantum entanglement. Indeed, Bell inequalities were  used to give detailed characterizations
of multi-partite entangled states by giving bounds on the correlations present in these states\cite{gisin,uffink,seevsvet,yu03}.
}
Recently a set of Bell-type inequalities was presented by \citet{sun} that gives a finer
classification for entanglement in three-partite systems than was previously known. 
The inequalities
distinguish three different types of bi-partite entanglement that may exist in three-partite systems. 
They not only determine if one of the three parties is separable with respect to the
other two, but also which one. It was shown that the three inequalities give a bound that 
can be thought of as
tracing out a sphere in the space of expectations of the three Bell operators that were used in
the inequalities.  Here we strengthen this bound 
by showing that all states are confined within the interior of the intersection of three cylinders and
the already mentioned sphere.

Furthermore, in chapters  \ref{chapter_CHSHquantumorthogonal} and \ref{Npartsep_entanglement} it was shown that considerably stronger separability
inequalities for the expectation of Bell operators can be obtained if one restricts oneself
 to local orthogonal spin observables (so-called LOO's \cite{nonlinear,yu}).
We will show that the same is the case for the Bell operators considered here by strengthening
all above mentioned three-partite inequalities under the restriction of orthogonal observables.

The relevant three-partite inequalities are included in the $N$-partite inequalities derived by
\citet{chenalbeveriofei}. It was shown that these $N$-partite inequalities can be violated maximally by the
$N$-partite maximally entangled GHZ states \cite{chenalbeveriofei}, but, as will be shown here, they can
also be maximally violated by states that contain only $(N-1)$-partite entanglement.  Although
these inequalities thus give a further classification of multi-partite entanglement (besides
some other interesting properties), they can not be used to distinguish
full $N$-partite entanglement from $(N-1)$-partite entanglement in $N$-partite states. It is
shown that this is neither the case for the stronger bounds that are derived for the case of
LOO's.

In subsection \ref{unrestricted} the case of  unrestricted spin observables is analyzed and 
subsection \ref{restricted} is devoted to  the restriction to LOO's. Lastly, in the discussion of subsection \ref{discusss} we will interpret the presented quadratic inequalities 
as indicating a type of monogamy 
of maximal bi-separable three-party quantum correlations. 
Non-maximal correlations can however be shared. This is contrasted to the Toner-Verstraete monogamy inequalities 
(\ref{tonerverstraeteineq}).

\subsection{Analysis for unrestricted observables}\label{unrestricted}\noindent 
\citet{chenalbeveriofei} consider $N$-parties that each have two alternative dichotomic measurements
denoted by $A_j$ and $A_j'$  (outcomes $\pm1$) and show that local hidden-variable models (LHV) require that
\begin{align}\label{ineqN}
|\av{D_N^{(i)}}_{\textrm{lhv}}|:= \frac{1}{2}|\av{B_{N-1}^{(i)}(A_i +A_i ') +(A_i -A_i ')}_{\textrm{lhv}} |\leq1,
\end{align}
for $i=1,2,\ldots, N$, where $B_{N-1}^{(i)}$  is the Bell polynomial of the Werner-Wolf-\.Zukowski-Brukner (WWZB)
inequalities \cite{wernerwolf2,zukowskibrukner} for the $N-1$ parties, except for party $i$.
These Bell-type inequalities have only two different local settings and  are contained in the general inequalities for $N>2$ parties that have more than two alternative measurement settings derived by \citet{laskowski}.
Indeed, they follow from the latter when choosing certain settings equal.  Note furthermore that the WWZB
inequalities are contained in the inequalities of \Eq{ineqN} by choosing $A_N=A_N'$.

The quantum mechanical counterpart of the Bell-type inequality of \Eq{ineqN}  
is obtained by introducing operators $A_k$, $A_k'$ for each party $k$ that represent the 
dichotomic observables in question. Let us define analogously \citet{sun} the operator
\begin{align}\label{qmcounter}
\mathcal{D}_N ^{(i)}:=\mathcal{B}_{N-1}^{(i)}\otimes(A_i +A_i')/2 + \1_{N-1}^{(i)}\otimes(A_i -A_i')/2,
\end{align}
for $i=1,2,\ldots, N$. Here $\mathcal{B}_{N-1}^{(i)}$
and $ \1_{N-1}^{(i)}$ are respectively
the Bell operator of the WWZB inequalities and the identity operator both for the $N-1$ qubits not involving qubit $i$.

Quantum mechanical counterparts of the local realism inequalities of \Eq{ineqN} are obtained by deducing relevant bounds on the expression $\av{\mathcal{D}_N ^{(i)}}_{\textrm{qm}}:=\mathrm{Tr}[\mathcal{D}_N ^{(i)}\rho]$, where
$\rho$ is a $N$-party quantum state.
For example, separable states must obey \begin{align}\label{ineqNqm}
|\av{\mathcal{D}_N ^{(i)}}_{\textrm{qm}}|\leq 1.
\end{align}
\forget{
obtained  considering the expression using 
\begin{align}\label{ineqNqm}
|\av{\mathcal{D}_N ^{(i)}}_{\textrm{qm}}|\leq 1,
\end{align}
$\av{\mathcal{D}_N ^{(i)}}_{\textrm{qm}}:=\mathrm{Tr}[\mathcal{D}_N ^{(i)}\rho]$, where
$\rho$ is a $N$-party quantum state.}
 In the remainder we only consider quantum correlations  so we drop the subscript `qm' from the expectation value expressions.

Since the Bell inequality of \Eq{ineqNqm} uses only two alternative dichotomic observables for each party the maximum violation of this Bell inequality is obtained for an $N$-party pure qubit state and furthermore for projective observables \cite{masanes,masanes05,tonerverstraete}. 
In the following we will thus consider qubits only and the observables will be represented by the spin operators
$A_k=\bm{a}_k\cdot \bm{\sigma}$ and $A_k'=\bm{a}'_k\cdot \bm{\sigma}$ with $\bm{a}_k$ and $\bm{a}_k '$
unit vectors that denote the measurement settings and
$\bm{a}\cdot \bm{\sigma}=\sum_l a_l\sigma_l$ where $\sigma_l$ are the familiar Pauli spin observables for $l=x,y,z$ on $\H=\mathbb{C}^2$. In fact, it suffices \cite{tonerverstraete} to consider only real and traceless observables, so we can set $a_y=0$ for all observables.

An interesting feature of the inequalities in \Eq{ineqNqm} is that all
generalized GHZ states $\ket{\psi_\alpha^N}=\cos \alpha \ket{0}^{\otimes N} +\sin
\alpha \ket{1}^{\otimes N}$ can be made to violate them 
for all $\alpha$, which is not the case for the WWZB inequalities  \cite{chenalbeveriofei,laskowski}. Furthermore,
the maximum is given by
\begin{align}\label{maxn}
\max_{A_i,A_i'}|\av{\mathcal{D}_N ^{(i)}}|=
 2^{(N-2)/2},\end{align}  as was
proven by \citet{chenalbeveriofei}. They also noted that this maximum is
attained for the maximally entangled $N$-party GHZ state $\ket{GHZ_N}$
(i.e., $\alpha=\pi/4$) and for all local unitary transformations of this state.
However, not noted by \citet{chenalbeveriofei} is the fact that the maximum is also
obtainable by $N$-partite states that only have $(N-1)$-partite entanglement,
which is the content of the following theorem.

\emph{Theorem 1.} Not only can the maximum value of $2^{(N-2)/2}$ for
$\av{\mathcal{D}_{N}^{(i)}}$ be reached by fully $N$-partite entangled states
(proven by \citet{chenalbeveriofei}) but also by $N$-partite states that only have
$(N-1)$-partite entanglement.

\emph{Proof}: Firstly, $(\mathcal{B}_{N-1}^{(i)})^2\leq2^{(N-2)} \1_{N-1}^{(i)}$ (as
proven in \cite {wernerwolf2}). Here $X\leq Y$ means that $Y-X$ is semi-positive definite. Thus the
maximum possible eigenvalue of $\mathcal{B}_{N-1}^{(i)}$
is $2^{(N-2)/2}$. Consider a state $|
\Psi_{N-1}^{(i)}\rangle$ for which
$\av{\mathcal{B}_{N-1}^{(i)}}_{\ket{\Psi_{N-1}^{(i)}}}=2^{(N-2)/2}$. This must
be \cite{wernerwolf2} a maximally entangled $(N-1)$-partite state (for the $N$
parties except for party $i$), such as the state $\ket{GHZ_{N-1}}$.
Next consider the state
$\ket{\xi^{(i)}}=\ket{\Psi_{N-1}^{(i)}}\otimes\ket{0_i}$, with $\ket{0_i}$ an
eigenstate of the observable $A_i$ with eigenvalue $1$. This is an $N$-partite
state that only has $(N-1)$-partite entanglement. Furthermore choose
$A_i=A'_i$ in \Eq{qmcounter}. We then obtain
$\av{\mathcal{D}_{N}^{(i)}}_{\ket{\xi^{(i)}}}=
\av{\mathcal{B}_{N-1}^{(i)}}_{\ket{\Psi_{N-1}^{(i)}}}\av{A_i}_{\ket{0_{i}}}=2^{(N-2)/2}$,
which was to be proved. $\square$

This theorem thus shows that the Bell inequalities of \Eq{ineqNqm} can not
distinguish between full $N$-partite entanglement  and $(N-1)$-partite
entanglement, and thus can not serve as full $N$-partite entanglement
witnesses.

Let us now concentrate on the three-partite case ($N=3$ and $i=1,2,3$). \citet{sun} obtain that fully separable three-partite states  satisfy
 $|\av{\mathcal{D}_3
^{(i)}}|\leq 1$, which does not violate the local
realistic bound of \Eq{ineqN}.  General states give
$|\av{\mathcal{D}_3^{(i)}}| \leq \sqrt{2}$, which follows from \Eq{maxn}. As
follows from Theorem 1  this can be saturated
by both fully entangled states as well as for bi-separable
entangled states (e.g., two-partite entangled three-partite
states).

\citet{sun} have furthermore presented a set of Bell-type inequalities that
distinguish three possible forms of bi-separable entanglement. They consider
bi-separable states that are separable with respect to partitions $1-23$, $2-13$ and $3-12$ respectively, 
where the set of states in these partitions is denoted as $S_{1-23}, S_{2-13},
S_{3-12}$ and which we label by $j=1,2,3$ respectively. These sets contain
states such as $\rho_{1}\otimes\rho_{23},~\rho_{2}\otimes\rho_{13}$, and
$\rho_{3}\otimes\rho_{12}$ respectively. We call the correlations obtained from a state that is bi-separable with respect to one of the three partitions  `bi-separable three-partite correlations'. 
 
For states in partition $j$  (and for $i=1,2,3$) \citet{sun} obtained
 \begin{align}
|\av{\mathcal{D}_3^{(i)}}|\leq \chi_{i,j}\,,
 \end{align}
 with $\chi_{i,j}=\sqrt{2}$ for $i=j$  and $\chi_{i,j}=1$ otherwise.

 They furthermore proved that for all three qubit states
   \begin{align}\label{4}
 \av{\mathcal{D}_3^{(1)}} ^2+\av{\mathcal{D}_3^{(2)}}^2+
 \av{\mathcal{D}_3^{(3)}}^2 \leq 3,~~~\forall \rho.
 \end{align}
   Although this inequality is stronger than the set above (for details see Figure 1 in \cite{sun}),
   it can be saturated by fully separable states. For example, choose the state $\ket{000}$ and 
all observables to be projections 
   onto this state. Then we get
   $\av{\mathcal{D}_3^{(1)}}_{\ket{000}} ^2+\av{\mathcal{D}_3^{(2)}}_{\ket{000}}^2+
   \av{\mathcal{D}_3^{(3)}}_{\ket{000}}^2 =3$.

   Let us consider $\mathcal{D}_3^{(i)}$ (for $i=1,2,3$) to be three coordinates of a space
   in the same spirit as \citet{sun} did.
They showed that the  fully separable states are confined to a cube with edge length $2$ and the
bi-separable states in partition $j=1,2,3$ are confined to cuboids with size
either $ 2\sqrt{2}\times2\times2 ,~ 2\times  2\sqrt{2}\times2$, or
$2\times2\times2\sqrt{2}$.  Note that states exist that are bi-separable with respect to all three partitions (and thus must lie within the cube with edge length $2$), but which are not fully separable \cite{bennett}. Furthermore, all three-qubit states are in the
intersection of the cube with size $2\sqrt{2}$ and of the sphere with radius
$\sqrt{3}$. \citet{sun} note that this sphere is just the external sphere of the
cube with edge $2$, which is consistent with the above observation that fully
separable states can lie on this sphere. If we look at the
$\mathcal{D}_3^{(i)}-\mathcal{D}_3^{(i+1)}$ plane we get Figure \ref{fig1}. The
fully separable states are in region I; region II belongs to  the bi-separable
states of partition $j=i+1$; and region III belongs to states of partition
$j=i$. Other bi-separable states and fully entangled states are outside these
regions but within the circle with radius $\sqrt{3}$. 
   However, in the following theorem we show a quadratic inequality even stronger than
   \Eq{4} which thus strengthens the bound in Figure \ref{fig1} given by the circle of radius $\sqrt{3}$
and which forces the bi-separable states just mentioned into the black regions.
\begin{figure}[!h]
\includegraphics[scale=1]{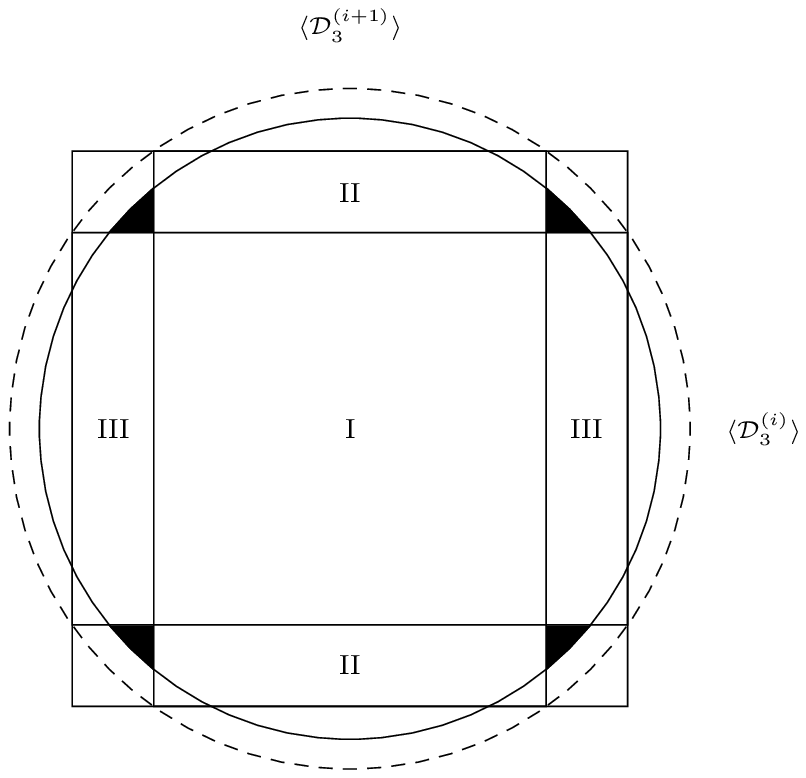}
\caption{$\mathcal{D}_3^{(i)}-\mathcal{D}_3^{(i+1)}$ plane with the stronger bound given by
the circle with radius $\sqrt{5/2}$ which strengthens the less strong bound with radius $\sqrt{3}$ that is
given by the dashed circle.}
\label{fig1}
\end{figure}

\emph{Theorem 2}.
  For the case where each observer chooses between two settings all three-qubit states
  obey the following inequality:
  \begin{align}\label{new}
  \av{\mathcal{D}_3^{(i)}} ^2+\av{\mathcal{D}_3^{(i+1)}}^2\leq  \frac{5}{2},~~~\forall \rho,
 \end{align}
 for $i=1,2,3$ and where $i$ and $i+1$ are both modulo 3.

 \emph{Proof}: The proof uses the exact same steps as the proof of \Eq{4} by \citet[proof of Theorem 2]{sun} and can be easily performed, although the left hand side of \Eq{new} contains only two terms instead of the three terms in the left hand side of
 \Eq{4}.  This results in only  a minor change in calculations\footnote{
 In further detail,
steps (1) to (4) of the proof by \citet{sun} become (using the terminology of their proof): 
 (1):  $\omega = 2(\bm{s}_1\otimes\bm{s}_2\otimes\bm{s}_3\cdot \bm{Q})^2=2\bra{\Psi}C_1C_2C_3\ket{\Psi}^2\leq 2$,  (2):  $\omega = 2(\bm{s}_1\otimes\bm{s}_2\otimes\bm{s}_3\cdot \bm{Q}+\bm{s}_1\otimes\bm{s}_2\otimes\bm{t}_3\cdot \bm{Q})^2=2\bra{\Psi}C_1C_2(C_3+D_3)\ket{\Psi}^2\leq 2$,  
 (3): $\omega=(5/4)(\cos(\theta_1+\theta_2+\theta_3) -\sin(\theta_1+\theta_2+\theta_3))^2\leq5/2$, 
 (4):  $\omega=(\cos(\theta_1+\theta_2+\theta_3) -\sin(\theta_1+\theta_2+\theta_3))^2\leq2$.
Here $\omega=\av{\mathcal{D}_3^{(i)}} ^2+\av{\mathcal{D}_3^{(i+1)}}^2$ (i.e., the l.h.s. of \Eq{new}), where  we have chosen $i =1$. Note that by symmetry the proof goes analogous for $i=2,3$. It follows that step (3) has the highest bound of $5/2$.}.  Case (3) in this proof then has the highest bound of $5/2$, whereas  the other three cases give a lower bound equal to $2$.   
 $\square$
 
Note that in contrast to \Eq{4} the inequality of \Eq{new} can not be saturated by separable
states, since the latter have a maximum of 2 for the left hand expression in \Eq{new}.

If we again look at the space  given by the coordinates $\mathcal{D}_3^{(i)}$ (for $i=1,2,3$),
we have thus found that all states are, firstly, confined within the intersection of the three
orthogonal cylinders
$\av{\mathcal{D}_3^{(i)}} ^2+\av{\mathcal{D}_3^{(i+1)}}^2\leq 5/2$ (with $i+1$
and $i+2$  both modulo $3$)
each with radius $\sqrt{5/2}$ and, secondly, they must furthermore still lie within the cube
of edge length $2\sqrt{2}$, and thirdly they must also lie within the sphere with radius $\sqrt{3}$.
In Figure \ref{fig1} we see the strengthened bound of \Eq{new} as compared to the bound of \citet{sun}. 
However,  we see from this figure that
neither the intersection of the three cylinders, nor the sphere,
nor the cube give tight bounds.

\noindent
The black areas in Figure \ref{fig1} are non-empty.
For the case of \Eq{new} states thus exist that have both $|\av{D_{3}^{(i)}}|>1$ and $|\av{D_{3}^{(i+1)}}|>1$ 
(for some $i$). For example, the so-called $W$-state
\begin{align} \label{WState}
\ket{W}=(\ket{001} +\ket{010} +\ket{100})/\sqrt{3},
\end{align}
 gives  $|\av{D_{3}^{(i)}}|=1.022$ for all $i$  when the observables are chosen as follows:  $A_i= \cos\alpha_i \, \sigma_z +\sin\alpha_i\,\sigma_x$  with $ \alpha_i= -0.133 $  and $A_i' = \cos\beta_i\,  \sigma_z +\sin\beta_i\,\sigma_x$
with $ \beta_i =0.460$.
 
\subsection{Restriction to local orthogonal spin observables}\label{restricted}
\noindent \citet{roy} and \citet{uffseev} have shown that considerably stronger separability
inequalities for the expectation of the bi-partite Bell operator
$\mathcal{B}_2$ can be obtained if one restricts oneself to local orthogonal observables (LOO's). See chapter \ref{chapter_CHSHquantumorthogonal}. 
We will now show that the same is the case for the Bell operator $\mathcal{D}_3^{(i)}$. 
The following theorem strengthens all previous bounds of section {II} for general observables.

\emph{Theorem 3}.
Suppose all local observables are orthogonal, i.e., $\bm{a}_i\cdot\bm{a}_i'=0$,
then the following inequalities hold:
\begin{description}
\item (i) For all states: $|\av{\mathcal{D}_{3}^{(i)}} |\leq \sqrt{3/2}$.
\item (ii) For fully separable states:  $|\av{\mathcal{D}_{3}^{(i)}} |\leq\sqrt{3/4}$. 
\item (iii) For bi-separable states in partition $j=1,2,3$:
 \begin{align}
|\av{\mathcal{D}_3^{(i)}}|\leq \chi_{i,j}\,,
 \end{align}
 with $\chi_{i,j}=\sqrt{3/2}$ 	
 for $i=j$  and $\chi_{i,j}=\sqrt{3/4}$ 
  otherwise.
\item (iv) Lastly,
 for all states:
\begin{align}\label{new2}
\av{D_{3}^{(i)}}^2 +\av{D_{3}^{(i+1)}}^2  \leq 2.\end{align}
\end{description}

\emph{Proof}:
(i) The square of $\mathcal{D}_3 ^{(i)}$ is given by
\begin{align}\label{square}
(\mathcal{D}_3 ^{(i)})^2=(\mathcal{B}_{2} ^{(i)})^2
\otimes\frac{1}{2}(1+\bm{a}_i\cdot\bm{a}'_i) \1_i
+ \1_{2}^{(i)}\otimes\frac{1}{2}(1-\bm{a}_i\cdot\bm{a}'_i) \1_i,
\end{align}
where $ \1_{2}^{(i)}$ is the identity operator for the $2$ qubits not
including qubit $i$. For orthogonal observables we get $\bm{a}_i\cdot\bm{a}'_i=0$, and
$(\mathcal{B}_{2} ^{(i)})^2\leq 2 \1_2^{(i)}$ (as proven in \cite{roy,uffseev}). 
The maximum eigenvalue of $(\mathcal{D}_3 ^{(i)})^2$ is thus $3/2$, which implies that
$|\av{\mathcal{D}_{3} ^{(i)}}|\leq \sqrt{3/2}$.

(ii) For fully separable states we have from \Eq{qmcounter} that
\begin{align}\label{proofi}
\av{\mathcal{D}_3 ^{(i)}}=\frac{1}{2}(\av{\mathcal{B}_{2}^{(i)}} \av{(A_i +A_i')} +\av{(A_i -A_i')}).
\end{align}
Furthermore for the case of orthogonal observables $|\av{\mathcal{B}_{2}^{(i)}}| \leq 1/\sqrt{2}$
\cite{roy, uffseev}. Thus $|\av{\mathcal{D}_3 ^{(i)}}|\leq|( \av{(A_i +A_i')}/\sqrt{2} +\av{(A_i -A_i')})/2|.$
Since the averages are linear in the state $\rho$ the maximum is obtained for a pure state of qubit $i$.
This state can be represented as $1/2( \1 +\bm{o}\cdot\bm{\sigma})$, with $|\bm{o}|=1$ and
$\bm{o}\cdot\bm{\sigma}=\sum_k o_k \sigma_k$ ($k=x,y,z$).
Take $C=(A_i +A_i')$, $D=(A_i -A_i')$ and
$\bm{s}=\bm{a}_i+\bm{a}_i'$, $\bm{t}=\bm{a}_i-\bm{a}_i'$. We get $|\bm{s}|=|\bm{t}|=\sqrt{2}$.
Choose now without losing generality \cite{tonerverstraete}
 $\bm{s}=\sqrt{2}(\cos\theta,0,\sin\theta)$ and $\bm{t}=\sqrt{2}(-\sin\theta,0,\cos\theta)$.
 Then
 \begin{align}
| \av{\mathcal{D}_3 ^{(i)}}|&\leq |( \bm{s}\cdot\bm{o}/\sqrt{2} +  \bm{t}\cdot\bm{o} )/2|\nonumber\\
&= |\frac{1}{2}\big((o_z-\sqrt{2}o_x)\sin\theta +(o_x+\sqrt{2}o_z)\cos\theta|\big).\nonumber
  \end{align}
 Maximizing over $\theta$ (i.e., $\max_\theta (X\cos\theta +Y\sin\theta) =\sqrt{X^2 +Y^2}$) and
 using $o_x^2 +o_y^2+o_z^2=1$
 we finally get
  \begin{align}
| \av{\mathcal{D}_3 ^{(i)}}|&\leq |\sqrt{3/4(o_x^2 +o_z^2)}|\leq\sqrt{ 3/4}
.
  \end{align}

(iii)  For bi-separable states in partition $j=i$ we get the same as in \Eq{proofi}, but now
  $|\av{\mathcal{B}_{2}^{(i)}}| \leq \sqrt{2}$. Using the method of (ii) we get
  \begin{align}
| \av{\mathcal{D}_3 ^{(i)}}|&\leq| ( \sqrt{2}\,\bm{s}\cdot\bm{o} +  \bm{t}\cdot\bm{o} )/2|
\leq\sqrt{3/2}
.
  \end{align}
For bi-separable states in partition $i+1$ and $i+2$ a somewhat more elaborate proof is needed.
Let us set $i=1$ and $j=3$ for convenience (for the other partition $j=2$ we get the same result).
The maximum is again obtained for pure states. Every pure state in partition $j=3$ can be written as
$\ket{\psi}=\ket{\psi}_{12}\otimes\ket{\psi}_3$. Then
\begin{align}
|\av{\mathcal{D}_3 ^{(i)}}|=\,|      &  \frac{1}{4}\av{(A_1+A_1')(A_2+
A_2')} _{\ket{\psi_{12}}}\av{A_3}_{\ket{\psi_3}}\nonumber\\&
     +\frac{1}{4} \av{(A_1+A_1')(A_2-A_2')} _{\ket{\psi_{12}}}\av{A_3'}_{\ket{\psi_3}}   + \frac{1}{2}\av{(A_1-A_1')\otimes \1_2}_{\ket{\psi_{12}}}|
  \end{align}
  Using the technique in (ii) above it is found that the maximum over $\ket{\psi_3}$ gives
\begin{align}
|\av{\mathcal{D}_3 ^{(i)}}|  \leq\, 
| \frac{\sqrt{2}}{4} \, \big( \,
 \av{(A_1+A_1')\,A_2} _{\ket{\psi_{12}}}^2
 &+\av{(A_1+A_1')\,A_2'} _{\ket{\psi_{12}}}^2\,
 \big)^{1/2}\nn\\& + \frac{1}{2}\av{(A_1-A_1')\otimes \1_2}_{\ket{\psi_{12}}}  |.
\end{align}
Without losing generality we choose $A_i$, $A_i'$ in the $x-z$ plane \cite{tonerverstraete} and  $\ket{\psi}_{12}=\cos\theta \ket{01}+\sin\theta\ket{10}$.  We can use the symmetry to set $A_1=A_2=A$ and $A_1'=A_2'=A'$. This gives
\begin{align}
|\av{\mathcal{D}_3 ^{(i)}}|  \leq \,|\frac{1}{2}(a_z -a_z') \cos( 2\theta) +
\frac{\sqrt{2}}{4}\big( (a_z+a_z')^2+((a_x+a_x')^2\sin(2\theta))^2 \big)^{1/2}|.
\end{align}
Since the observables $A$ and $A'$ must be orthogonal (i.e., $\bm{a}\cdot\bm{a}'=0$),
 this expression obtains its maximum for $a_x=a_x'=1/\sqrt{2}$ and $a_z=-a_z'=1/\sqrt{2}$. We finally get:
\begin{align}
|\av{\mathcal{D}_3 ^{(i)}}|  \leq \frac{\sqrt{2}}{2}\cos( 2\theta) +\frac{1}{2}\sin( 2\theta)\leq \sqrt{3/4}.
\end{align}

(iv)   We use the exact same steps of the proof of Sun \& Fei of \Eq{4}  (i.e., 
\citet[proof of Theorem 2]{sun}) but since the observables are orthogonal only case (4) of that proof needs
to be evaluated. This can be easily performed for the left hand side of \Eq{new2} that contains only two
terms instead of the three terms on the right hand side of \Eq{4}, thereby resulting in only a minor modification of the calculations \footnote{In further detail,
 the proof by \citet{sun} for the case of orthogonal observables amounts to  (using the terminology of their proof)  $|\bm{s}_i|=|\bm{t}_i|=\sqrt{2}/2$. Thus only step (4) needs to be evaluated and this gives
  $\omega=(\cos(\theta_1+\theta_2+\theta_3) -\sin(\theta_1+\theta_2+\theta_3))^2\leq2$.
As in the proof of Theorem 2 we have  $\omega=\av{\mathcal{D}_3^{(i)}} ^2+\av{\mathcal{D}_3^{(i+1)}}^2$ (i.e., the l.h.s. of \Eq{new2}),where again we have chosen $i =1$, but by symmetry the proof goes analogous for $i=2,3$. }  giving the result  $\av{D_{3}^{(i)}}^2 +\av{D_{3}^{(i+1)}}^2  \leq 2$. $\square$

 \begin{figure}[!h]
\includegraphics[scale=1]{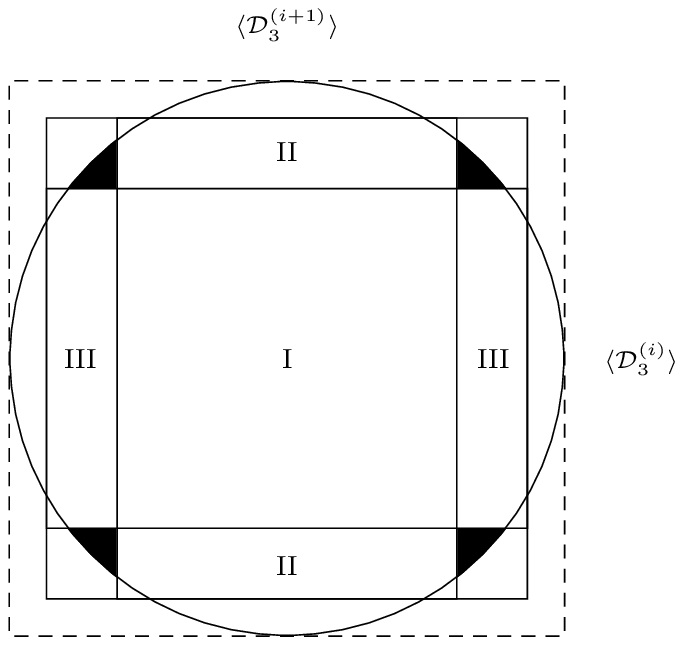}
\caption{The results of Theorem 3 for orthogonal observables. For comparison to the case
where the observables were not restricted to be orthogonal, the dashed square is included
that has edge length $2\sqrt{2}$ and which is the largest square in Figure \ref{fig1}. }
\label{fig2}
\end{figure}

These results for orthogonal observables can again be interpreted in terms of  the space given by the
coordinates $\mathcal{D}_3^{(i)}$ (for $i=1,2,3$). The same structure as in Figure \ref{fig1} then arises
but with the different numerical bounds of Theorem 2.
The fully separable states are confined to a cube with edge length $\sqrt{3}$ and the bi-separable
states in partition $j=1,2,3$ are confined to cuboids with size either $\sqrt{6}\times\sqrt{3}\times\sqrt{3} ,~
\sqrt{3}\times \sqrt{6}\times\sqrt{3}$, or $\sqrt{3}\times\sqrt{3}\times\sqrt{6}$. Furthermore, all three-qubit
states are in the intersection of firstly the cube with edge length $\sqrt{6}$, secondly of  the three orthogonal
cylinders with radius $\sqrt{2}$, and thirdly of the sphere with radius $\sqrt{3}$.

The corresponding $\mathcal{D}_3^{(i)}-\mathcal{D}_3^{(i+1)}$ plane  is drawn in Figure \ref{fig2}.
Compared to the case where no restriction was made to orthogonal observables  (cf. Figure \ref{fig1}) we see that we can
still distinguish the different kinds of bi-separable states, but they can still
not be distinguished from
fully three-partite entangled states since both types of states still have the same maximum for
$\av{\mathcal{D}_{3}^{(i)}}$. Furthermore, the ratio of the different maxima of
$\av{\mathcal{D}_3^{(i)}}$ for fully separable  and bi-separable states is still the same,
i.e., the ratio is $\sqrt{2}/1=(\sqrt{3/2})/(\sqrt{3/4})=\sqrt{2}$.

The black areas in Figure \ref{fig2} are again non-empty since states exist that have both $|\av{D_{3}^{(i)}}|>\sqrt{3/4}$ and $|\av{D_{3}^{(i+1)}}|>\sqrt{3/4}$ for the case of orthogonal observables. 
For example, the $W$-state of \Eq{WState} gives $|\av{D_{3}^{(i)}}|=0.906$ for all $i$, for the local angles $\alpha_i= 0.54=\beta_i -\pi/2$ in the $x$-$z$ plane.

\subsection{Discussion of the monogamy aspects}\label{discusss}\noindent
Let us take another look at the quadratic inequalities $\av{D_{3}^{(i)}}^2 +\av{D_{3}^{(i+1)}}^2  \leq 5/2$ of
\Eq{new} for general observables and $\av{D_{3}^{(i)}}^2 +\av{D_{3}^{(i+1)}}^2  \leq 2$  of \Eq{new2}
for orthogonal observables. These can be interpreted as monogamy inequalities for maximal bi-separable
three-qubit quantum correlations (i.e., bi-separable correlations that saturate the inequalities), since the inequalities show that a state that has maximal
bi-separable correlations for a certain partition can not have it maximally for another partition.
Indeed, when
partition $i$ gives  $|\av{D_{3}^{(i)}}|= \sqrt{2}$ it must be the case according to  \Eq{new} that
for the other two partitions both $|\av{D_{3}^{(i+1)}}|\leq  \sqrt{1/2}$ and
 $|\av{D_{3}^{(i+2)}}|\leq \sqrt{1/2}$ must hold. The latter two must thus be non-maximal as soon as the first type of
 bi-separable correlation is maximal. And for the second inequality of \Eq{new2} using orthogonal observables we get that when
 $|\av{D_{3}^{(i)}}|= \sqrt{3/2}$ (this is maximal) it must be the case that both
 $|\av{D_{3}^{(i+1)}}|\leq  \sqrt{1/2}$ and
 $|\av{D_{3}^{(i+2)}}|\leq \sqrt{1/2}$, which is non-maximal.

 From this we see that the first (i.e., \Eq{new} for general observables)
 is a stronger monogamy relationship than the second (i.e., \Eq{new2} for 
 orthogonal observables) in the sense that the trade-off
 between how much the maximal value for $|\av{D_{3}^{(i)}}|$ for one 
 partition $i$ restricts
 the value of $|\av{D_{3}^{(i+1)}}|$, $|\av{D_{3}^{(i+2)}}|$ 
 for the other two partitions below the maximal value  
 is larger in the first case than in the second case.
 
 Let us see how this compares to the Toner-Verstraete monogamy inequality      
     $ \av{\mathcal{B}^{(i)}_2}^2 +\av{\mathcal{B}^{(i+1)}_2}^2\leq 2$ of  (\ref{tonerverstraeteineq}). Here $|\av{\mathcal{B}^{(i)}_2}_\textrm{lhv} |\leq 1$ is the ordinary CHSH inequality (scaled down by a factor of $2$) for the local correlations of the two qubits other than qubit $i$ (cf. \eqref{tonerverstraeteineq}). The Toner-Verstraete monogamy inequality  is even stronger than the ones presented here, because  when  $|\av{\mathcal{B}^{(i)}_2}|$ obtains its maximal value of $\sqrt{2}$ it must be that     $|\av{\mathcal{B}^{(i+1)}_2}|$=$|\av{\mathcal{B}^{(i+2)}_2}|=0$.

Furthermore, in section \ref{comparingmonogamies} we have seen that the Toner-Verstraete monogamy relationship  shows that the non-locality indicated by correlations that violate the CHSH inequality cannot be shared: as soon as for some $i$ one has 
$|\av{\mathcal{B}^{(i)}_2}| > 1$, it must be that  both $|\av{\mathcal{B}^{(i+1 )}_2} |< 1$ and $|\av{\mathcal{B}^{(i+2 )}_2} |< 1$.  But we have also seen that  \citet{collinsgisin} have nevertheless shown that the quantum non-locality indicated by a violation of the bi-partite Bell-type inequality \eqref{collgisineq} indicates can be shared. Since $|\av{D_{3}^{(i)}}_\textrm{lhv}|\leq1$ are local  Bell-type inequalities (see \Eq{ineqN}) whose violation can be seen to indicate some non-locality, the inequalities considered here could possibly also indicate some quantum non-locality sharing. 
Indeed, this is the case since it was shown that the black areas in Figure \ref{fig1} are non-empty.
Violation of the Bell-type inequalities given here thus indicate shareability of the non-locality of bi-separable
three-qubit quantum correlations.
 
In conclusion, we have presented stronger bounds for bi-separable correlations in three-partite
systems than were given by \citet{sun} and extended this analysis to the case of the restriction to orthogonal observables
which gave even stronger bounds. The quadratic inequalities for bi-separable correlations give
a monogamy relationship for correlations that violate these inequalities maximally (i.e., such correlations cannot be shared), but they indicate shareability of the non-maximally violating correlations.  

We hope that future research will reveal more of 
the monogamy of multi-partite quantum correlations.  It could therefore be fruitful to generalize this work from three to a larger number of parties. Even more interesting would be including also no-signaling correlations besides correlations that come from quantum states.

\section{Discussion}

In this chapter we have seen that, apart from using Bell-type inequalities in terms of all parties involved, another fruitful way of studying the different kinds of correlations  is via the question whether the correlations can be shared.  Here one focuses on subsets of the parties and whether their correlations can be extended to parties not in the original subsets. This can be done either directly in terms of joint probability distributions or in terms of relations between Bell-type inequalities that hold for different, but overlapping subsets of the parties involved.

 We have proven that unrestricted general correlations can be shared to any number of parties (called $\infty$-shareable). In the case of no-signaling correlations it was already known that  such correlations can be  $\infty$-shareable  iff the correlations are local. We have shown that this implies, firstly, that partially-local correlations are also $\infty$-shareable, since they are combinations of local and unrestricted correlations between subsets of the parties. Secondly, it  implies that both quantum and no-signaling correlations  that are non-local are not $\infty$-shareable and we have shown monogamy constraints for such correlations.
 
We have investigated the relationship between sharing non-local quantum correlations and sharing mixed entangled states, and already for the simplest bi-partite correlations this was shown to be non-trivial. The Collins-Gisin Bell-type inequality indicates that non-local  quantum correlations can be shared and it thus indicates sharing of entanglement of mixed states. The CHSH inequality was shown not to indicate this. This shows that non-local bi-partite correlations in a setup with two-dichotomous observables per party cannot be shared, whereas this is possible in a setup with one observable per party more.
 
We have given a  simpler proof of the monogamy relation $\av{\mathcal{B}_{ab}}_{\textrm{qm}}^2 +\av{\mathcal{B}_{ac}}_{\textrm{qm}}^2\leq8$ of \citet{tonerverstraete}. We have furthermore provided a different strengthening  of this constraint than the one given by Toner and Verstraete.
For no-signaling correlations we have argued that the monogamy constraint $|\av{\mathcal{B}_{ab}}_{\textrm{ns}}| +|\av{\mathcal{B}_{ac}}_{\textrm{ns}}|\leq 4$ of \citet{toner2} can be interpreted as  a non-trivial bound on the set of three-partite no-signaling correlations. 
This discerning inequality uses product expectation values only. We know of no other such non-trivial bounds for no-signaling correlations of three or more parties (in the next chapter this will be further discussed).

Lastly, we have derived monogamy constraints for three-qubit bi-separable quantum correlations, which is a first example of investigating monogamy of quantum correlations using a three-partite Bell-type inequality.

%
%
%
%
\clearemptydoublepage
\thispagestyle{empty}
\chapter[Discerning multi-partite correlations]{Discerning multi-partite partially-\\\vskip-0.1cm local, quantum mechanical and\\\vskip0.2cm no-signaling correlations}

\label{chapter_svetlichny}

\noindent  
This chapter is in part based on \citet{seevsvet}.

\section{Introduction}
    
    In the previous chapters we have seen that quadratic and linear Bell-type inequalities 
    distinguish the correlations of various types of multi-partite quantum states. We have also we seen that Mermin-type inequalities discern LHV models from quantum mechanics, i.e., they discern local correlations from quantum correlations.
  In this chapter we will construct new Bell-type inequalities to discern partially-local from quantum mechanical correlations and also 
  discuss the issue of discerning multi-partite no-signaling correlations. Unfortunately, the Mermin-type inequalities do not suffice for either purpose.

Let us recall the notion of partial locality by reviewing some of the definitions that were given in chapter \ref{definitionchapter}. For $N=2$ locality and partial locality coincide so we start our investigation at $N=3$.  Here we consider three-partite models where arbitrary correlations (e.g., a signaling correlation) are allowed between two of the three parties but only local correlations between these two and the third party. The two parties that are non-locally correlated need not be fixed in advance, but can be chosen with probability $p_i$. The correlations (joint probability distributions) are thus of the form
\begin{align}
P(a_1,a_2,a_3|A_1,A_2,A_3)=\int_\Lambda d\lambda \,
[\,&p_1\,\rho_1(\lambda) P_1(a_1|A_1,\lambda)P_1(a_2,a_3|A_2,A_3,\lambda)\nn\\
&+p_2\,\rho_2(\lambda) P_2(a_2|A_2,\lambda)P_2(a_1,a_3|A_1,A_3,\lambda)\nn\\
&+p_3\,\rho_3(\lambda) P_3(a_3|A_3,\lambda)P_3(a_1,a_2|A_1,A_2,\lambda)
\,].
\label{localnonlocal3}
\end{align}
with $A_i$ observables and $a_i$ outcomes and where  $P_1(a_2,a_3|A_2,A_3,\lambda)$ can be any probability distribution; it need not factorise into $P_1(a_2|A_2,\lambda)P_1(a_3|A_3,\lambda)$. Analogously for the other two joint probability terms. 
The $\rho_i(\lambda)$ are the hidden-variable distributions. Models that allow for correlations of the form (\ref{localnonlocal3}) are called partially-local hidden-variable models (PLHV) models. Models whose correlations cannot be written in this form are fully non-local, i.e., they are said to contain full non-locality.

For the three-partite case \citet{svetlichny} derived a non-trivial Bell-type inequality for partially-local correlations of the form (\ref{localnonlocal3}). This inequality can thus  distinguish between full three-partite non-locality and two-partite non-locality in a three-partite system.
A priori it is not clear if the correlations of the form (\ref{localnonlocal3}) are stronger than quantum mechanical correlations.
However,  Svetlichny showed that quantum states exist that give correlations that violate the inequality, thereby proving that these correlations are fully non-local. Furthermore, no-signaling correlations were shown to violate the inequality maximally \cite{jones,barrett05}. Thus three-partite quantum and no-signaling correlations exist that cannot be reproduced by any three-partite PLHV model, despite the fact that PLHV models allow for arbitrary strong signaling correlations between any two of the three parties.

In this chapter we generalize Svetlichny's inequalities to the multi-partite case and we call them Svetlichny inequalities. Quantum mechanics violates these inequalities for some fully entangled multi-qubit states and these thus contain fully non-local correlations.  In a recent four particle experiment such a violation was observed, so full non-locality occurs in nature.
  It is an open question whether all fully entangled states  imply full non-locality.  If they do, this cannot always be shown by violations of the Svetlichny inequalities, because we will show that fully entangled states exist that do not violate any of them.

 After we announced the multi-partite generalization of Svetlichny's three-partite inequality, as published in \cite{seevsvet}, \citet{collins} independently also presented such a generalization. \citet{CER} commented upon the original three-partite case, and \citet{jones} performed an extension of the generalization and furthermore showed that no-signaling correlations can give maximal violation of the Svetlichny  inequalities.

The outline of this chapter is as follows. In section \ref{preliminaries}  some preliminary results and \mbox{notations} are presented. In section  \ref{3svet} the three-partite case is treated as a stepping stone to the multi-partite generalization of section \ref{Nsvet}.
In presenting the three-partite case we use the presentation as in \citet{collins} and \citet{CER}. For the multi-partite generalization we use the original proof given by us, and present some further multi-partite results by \citet{jones}. In section \ref{moreQM} we look at some further aspects of quantum mechanical violations of the generalized Svetlichny inequalities. In section \ref{nosignalsvet} we comment on the fact that although the Svetlichny inequalities discern partially-local and quantum correlations from the most general correlations, they cannot do so for no-signaling correlations.
 What set of inequalities that bound some linear sum of product expectation values (possibly including some marginal expectation values) and that would discern the multi-partite no-signaling correlations, we pose as an interesting open problem. Lastly, in section \ref{concsvet} we give a conclusion and discussion of the results obtained.

\section{Preliminaries}\label{preliminaries}
In order to introduce the Svetlichny inequality and to give its multi-partite generalization in the next section it is helpful to introduce the so-called Mermin polynomials, whose quantum counterpart we have already encountered in chapter \ref{Npartsep_entanglement} and that were used to give the Mermin-type separability inequalities (\ref{linearN}). Let $A_j$ and $A_j'$ be dichotomic observables for parties $j=1,\ldots N$. 
The Mermin polynomials  are defined in the following way:
Let $M_2=A_1A_2+A_1'A_2+A_1A_2'-A_1'A_2'$ (i.e., analogous to the 
Bell operator $\mathcal{B}$), and define recursively
\begin{align}
M_j:=\frac{1}{2}(M_{j-1} (A_j +A_j')+ M_{j-1}' (A_j -A_j')),
\label{recursiveM}
\end{align}
where for $M_j'$ all primed and non-primed observables are exchanged.
For $N=3$ we get
\begin{align}
M_3:=&A_1'A_2A_3+A_1A_2'A_3+A_1A_2A_3'-A_1'A_2'A_3',\\
M_3':=&A_1A_2'A_3'+A_1'A_2A'_3+A_1'A'_2A_3-A_1A_2A_3.
\end{align}

\forget{We will consider expectation values of these polynomials, e.g.,
$\av{M_2}= \av{A_1A_2}+ \av{A_1'A_2}+ \av{A_1A_2'}-\av{A_1'A_2'}$, which is the CHSH expression.}
  In the following we consider expectation values  of these polynomials as predicted by the different types of correlations as distinguished in chapter \ref{definitionchapter}. That is, we consider the expectation values
 $\av{M_N}_\textrm{plhv}$,  $\av{M_N}_\textrm{qm}$ and $\av{M_N}_\textrm{ns}$. \forget{That is we consider the expectation values of the (i) by LHV models, denoted by $\av{M_N}_\textrm{lhv}$ where $M_N$ is evaluated using a fully factorisable probability distribution, (ii) by PLHV models , as denoted by $\av{M_N}_\textrm{plhv}$, where $M_N$ is evaluated using  the partial factorisable probability distribution that for $N=3$ is given in (\ref{localnonlocal3}), and (iii) by quantum mechanics denoted by $\av{M_N}_\textrm{qm}=\textrm{Tr}[M_N\rho]$ where the observables in $M_N$ are given their quantum mechanical 
counterparts\forget{\footnote{We will consider only projective dichotomic measurements on qubits. Each observable $A_j$ 
can thus be written as some spin observable $\vec{a}_j\cdot\bm{\sigma}$ with $\vec{a}_j$ some unit vector and $\bm{\sigma}$ the Pauli matrices.}} (i.e., some self-adjoint operator).
}
 Furthermore, the absolute maximum on $\av{M_N}$ is denoted by $|M_N|_\textrm{max}$ and is equal to the number of terms in the Mermin polynomial. It is always possible to find a fully non-local model that is able to give this absolute maximum.  

 \citet{gisin} were the first to derive that $ | \av{M_N}_\textrm{lhv} | \leq 2 $,  $|\av{M_N}_\textrm{qm}|\leq 2^{(N+1)/2}$,  $|M_N|_{\textrm{max}}=2^{(N+1)/2}$ for $N$ = odd, and $|M_N|_{\textrm{max}}=2^{N/2}$ for $N$ = even.
 \forget{  $\max|\av{M_N}_\textrm{qm}| =|M_N|_{\textrm{max}}$ for $N$ = odd and $\max|\av{M_N}_\textrm{qm}| = |M_N|_{\textrm{max}}/\sqrt{2}$ for $N$ = even.}
It was shown in chapter \ref{Npartsep_entanglement} that the tight quantum bounds on Mermin polynomials distinguish various forms of entanglement including full entanglement. An interesting question now is if these polynomials are also suitable for detecting full multi-partite non-locality. For this purpose one needs to find the bounds on $\av{M_N}_\textrm{plhv}$ so as to answer if one can use some Mermin-type inequality to distinguish a partially-local from a fully non-local model. It will be shown that for $N$ = even that this is indeed the case, but not for $N$ = odd.  Consequently, for $N$ = odd new inequalities need to be found.  Svetlichny provided the case $N=3$, which we will review in the next section. Later we will generalize his inequality to all $N$. 
 
\section{Three-partite partial locality}\label{3svet}
Consider the Mermin polynomial $M_3$. For this case $\max|\av{M_3}_\textrm{qm}| =|M_3|_{\textrm{max}}=4$.  We want to determine  $\max|\av{M_3}_\textrm{plhv}|$. \citet{collins} obtained this as follows.  Consider the recursive relation (\ref{recursiveM}) so as to obtain  $M_3=(M_2 (A_3 +A_3') +M_2'(A_3-A_3'))/2$. Now assume partial factorisability in the sense that party $3$ factorises from party $1$ and $2$. We note that this is not a limiting restriction because the same results follows for the two other choices or convex combinations of these three possibilities. The desired maximum becomes: $\max|\av{M_3}_\textrm{plhv}|=\max | \,|M_2|_{\textrm{max}}\av{A_3 +A_3'}+|M_2|_{\textrm{max}}\av{A_3 -A_3'}|/2$.  Here the absolute maxima for the expectation values of $M_2$ and $M_2'$ can be attained since arbitrary strong correlations between party $1$ and $2$ are allowed.  
 Since we are dealing with dichotomic observables with outcomes $\pm1$, the maximum of $|\av{M_3}_\textrm{plhv}|$ is obtained if $|\av{A_3}|=|\av{A_3'}|=1$. Without loss of generality we choose $\av{A_3}=\av{A_3'}=1$ so that  
  $\max| \av{M_3}_\textrm{plhv}|= |M_2|_{\textrm{max}}=|M_3|_{\textrm{max}}=4$.   In conclusion, we have obtained the tight bound
  \begin{align}
|\av{M_3}_\textrm{plhv}|,| \av{M_3}_\textrm{qm}|\leq |M_3|_{\textrm{max}}=4,
  \end{align}
  from which it follows that $M_3$ does not distinguish between PLHV models, quantum mechanics and models that allow for full unrestricted non-locality between all parties.
 
 The problem lies in the fact that $M_3$ only has four correlation terms.  \citet{CER} showed 
 that a PLHV model  with correlations as in (\ref{localnonlocal3}) can reproduce whatever values are assumed for the four expectation values in $M_3$.\forget{i.e., in the set $E_1=\{ \av{A_1'A_2A_3}$,$\av{A_1A_2'A_3}$, $\av{A_1A_2A_3'}$,$\av{A_1'A_2'A_3'}\}$\forget{\footnote{\citet{mitchell} show by explicit construction  that the measurements on a three-partite GHZ state of the set $E_1$ with $A_i=\sigma_x$ and $A_i'=\sigma_y $, $ i=1,2,3$,  can always be reproduced by a PLHV model with correlations as in (\ref{localnonlocal3}).}}.} Likewise another such PLHV model can be found that reproduces the expectation values in $M_3'$. Thus, in order to give a non-trivial bound for PLHV models one needs to consider at  least more than four product expectation values.  Svetlichny considered all eight possible terms using the following two polynomials:
\begin{align} 
S^\pm_3:= M_3\pm M_3'.
\end{align}
We call these Svetlichny polynomials. Both polynomials have eight terms from which one obtains $|S_3^{\pm}|_{\textrm{max}}=8$.
Using the recursive relation (\ref{recursiveM}) we see that $S_3^\pm$ is equal to $M_2A_3' \pm M_2'A_3$. Then for the case of partial factorisability where party $3$ factorises from party $1$ and $2$ one obtains that the maximum of $|\av{S_3^\pm}_\textrm{plhv}|$
is given by   $|M_2 \pm M_2'|_\textrm{max}= 2 |A_1A_2' \pm A_1'A_2|_\textrm{max}=4$, which is half the absolute maximum $|S_3^\pm|_\textrm{max}=8$. This finally gives a non-trivial inequality:  all PLHV models that allow correlations of the form (\ref{localnonlocal3}) must obey the following non-trivial bound 
\begin{align}
|\av{S^\pm_3}_\textrm{plhv}|\leq 4.
\label{Sv3}
\end{align}
Explicitly the inequalities read:
  \begin{align}
| \av{S_3^{\pm}}|=&| \av{A_1A_2A_3}\pm\av{A_1A_2A_3'}\pm\av{A_1A_2'A_3}\pm\av{A_1'A_2A_3}\nn\\
 &-\av{A_1A_2'A_3'} -\av{A_1'A_2A_3'}-\av{A_1'A_2'A_3}\pm\av{A_1'A_2'A_3'}|\leq 4,\label{SVETA_3H2}
\forget{| \av{S_3^-}|=& |\av{A_1A_2A_3}-\av{A_1A_2A_3'}-\av{A_1A_2'A_3}-\av{A_1'A_2A_3}\nn\\
 &-\av{A_1A_2'A_3'} -\av{A_1'A_2A_3'}-\av{A_1'A_2'A_3}+\av{A_1'A_2'A_3'}|\leq 4.
  \label{SVETA_3H2a}}
  \end{align}
  \forget{These inequalities were first obtained by \citet{svetlichny}\footnote{There is a sign error in front of the \(E(A_1A_2A_3')\) term of equation (6) in \cite{svetlichny}.} }These are necessary conditions to be obeyed by all three-partite PLHV models.

Since $\av{S_3^+}^2+\av{S_3^-}^2=2(\av{M_3}^2+\av{M_3'}^2)$ we obtain from the Mermin-type separability inequalities (\ref{linearN}) that the maximum value attainable by quantum mechanics is $\max|\av{S^\pm_3}_\textrm{qm}|=4\sqrt{2}$. This is attained for a GHZ state, as first shown by \citet{svetlichny}.  Quantum mechanics thus contains states that have full non-locality. 

In conclusion, Svetlichny obtained the following bounds: 
\begin{align}
2\max|\av{S^\pm_3}_\textrm{plhv}|=\sqrt{2}\max|\av{S^\pm_3}_\textrm{qm}|=|S^\pm_3|_\textrm{max}=8.
\end{align}
The three bounds are each time increased with a factor $\sqrt{2}$. 
In the next section we give the multi-partite generalization of these bounds.
It is noteworthy that no-signaling correlations can reach the absolute maximum \cite{jones}: $\max|\av{S^\pm_3}_\textrm{ns}|=|S^\pm_3|_\textrm{max}$. This means that the Svetlichny polynomials cannot be used to distinguish three-partite no-signaling correlations from more general correlations that are  signaling. This will be further commented upon in section \ref{nosignalsvet}.

To end this section we mention that it is an open question what is the minimum number of correlation terms one should consider in a Svetlichny-like polynomial in order to distinguish between bi-partite non-locality and full three-partite  non-locality. We have seen that four terms does not suffice, whereas eight terms does suffice, but perhaps one can do with less.

 \section{Generalization to $N$-partite partial locality}\label{Nsvet}
  For four parties  the strategy of the previous section gives \begin{align}
  \max |\av{M_4}_\textrm{plhv}|=4,~\max |\av{M_4}_\textrm{qm}|=4\sqrt{2},~\textrm{whereas}~|M_4|_\textrm{max}=8,
  \end{align}
   and analogously for $M_4'$.  This shows that for four parties the Mermin polynomial gives non-trivial bounds on PLHV models. Let us now generalize this by showing that (up to a numerical factor) the Mermin polynomials $M_N$ and $M_N'$ for $N$ = even give valid Svetlichny inequalities that test partial factorisability.  However, just as was the case for $N=3$  it will be shown that for $N$ = odd one should take a linear combination of the two Mermin polynomials $M_N $ and $M_N'$.
   
  Consider an $N$-partite system and let us now make the
    following partial factorisability assumption (we recall this from chapter \ref{definitionchapter}): An ensemble of such
 systems consists of subensembles in which each one of the
    subsets of the \(N\) parties form extended systems, whose subsystems can be correlated in any way (e.g., entangled, fully non-local)  which however are uncorrelated to each other.   Let us for the time 
    being focus our attention on one of these subensembles, formed by
    a system consisting of two subsystems of \(k<N\) and \(N-k<N\)
    parties which are uncorrelated to each other. Assume also for the time being that
    the
    first subsystem is formed by parties \(1,2,\dots,k\) and the other by the
    remaining. We
    express our partial factorisability hypothesis by assuming a factorisable
    expression for the probability \( p (a_1,a_2,\ldots,a_N|A_1,A_2,\cdots,A_N)\) for
    observing the results \(a_i\), for the observables \( A_i\):
    \begin{align}\nonumber
    \lefteqn{p (a_1,a_2,\ldots, a_N|A_1,A_2,\ldots,A_N)=} \\  \label{EQ1}
    & &  \int P(a_1,\ldots,a_k|A_1,\ldots,A_k,\lambda)
    P(a_{k+1},\ldots,a_N|A_{k+1},\ldots,A_N,\lambda)\,\rho(\lambda)\, d\lambda,
    \end{align}
    where the probabilities are conditioned to the hidden variable \(\lambda\) with probability measure \(d\rho\). Formulas
    similar to (\ref{EQ1}) with different choices of the composing
    parties and different value of \(k\) describe the other subensembles.
    We need not consider decomposition into more than two subsystems
    as then any two can be considered jointly as parts of one
    subsystem still uncorrelated with respect to the others.   
\forget{  Though (\ref{EQ1}) expresses a hidden-variable model for the local (i.e.
    uncorrelated)  
    behavior of the two subsystems in relation to each other, we shall show that
    the same inequality derived below can be used to distinguish, in the quantum
    mechanical case, between fully entangles states and those only partially
    entangled. }

    Consider the expectation value of the product of the observables in the
    original ensemble        
    \begin{align}\nonumber
     \langle A_1A_2 \cdots A_N\rangle_{\textrm{plhv}} =
    \sum_J(-1)^{n(J)}p(J),
    \end{align}
    where \(J\) stands for an \(N\)-tuple \(j_1,\dots,j_N\) with \(j_k=\pm 1\),
    \(n(J)\) is the number of \(-1\) values in \(J\) and \(p(J)\) is the
    probability of achieving the indicated values of the observables. 
    Using the hypothesis of Eq.\ (\ref{EQ1}) as a constraint we now  
    derive non-trivial inequalities satisfied by the numbers
    \(\av{A_1A_2\cdots A_N}_{\textrm{plhv}}\) when introducing two alternative dichotomic 
    observables \(A_i^1,A_i^2\), \(i=1,2,\dots ,N\)  (here we write $A^1_i$ and $A^2_i$ instead of $A_i$ and $A'_i$ for the dichotomic observables for party $i$). To simplify the notation we write \(\av{i_1i_2\cdots i_N}_{\textrm{plhv}}\) for
    \(\av{A_1^{i_1}A_2^{i_2}\cdots A_N^{i_N}}_{\textrm{plhv}}\), where $i_1=1,2$ denotes which of the two dichotomic observables is chosen for party $1$, etc.   For any value of \(k\)
    and any choice of these \(k\) parties to comprise one of the subsystems we
    obtain (proof in the Appendix on page \pageref{appsvet}) the following inequalities:   
    \begin{align}\label{EQ3}
    \sum_I\nu^\pm_{t(I)}\av{i_1i_2\cdots i_N}_{\textrm{plhv}}\le 2^{N-1} ,
    \end{align}
    where \(I=(i_1,i_2,\dots,i_N)\), \(t(I)\) is the number of
    times index \(2\) appears in \(I\), and \(\nu^\pm_k\) is a sequence 
    of signs given by 
    \begin{align}\label{eq:nupm}
    \nu^{\pm}_k = (-1)^{\frac{k(k\pm 1)}{2}}.
    \end{align}
    These sequences have
    period four with cycles \((1,-1,-1,1)\) and \((1,1,-1,-1)\)
    respectively.  We call these inequalities alternating. They are
    direct generalizations of the three-partite inequalities  by \cite{svetlichny}.
    The
    alternating inequalities are satisfied by a system with any form of
    partial 
    factorisability, so their violation is a sufficient indication of full
    non-factorisability.  
              
    Introduce now the operator
    \begin{align}\label{eq:sop}
    S_N^\pm = \sum_I\nu_{t(I)}^\pm A_1^{i_1}\cdots A_N^{i_N}.
    \end{align}
    Using Eq. (\ref{EQ3}) the $N$-partite  alternating
    inequalities can be expressed as 
    \begin{align}\label{NPartite}
    |\langle S^\pm_N\rangle_{\textrm{plhv}} |\le 2^{N-1}.
    \end{align}
    For even \(N\) the two inequalities are interchanged by a global
    change of labels $1$ and $2$ and are thus equivalent. However for odd \(N\) this
    is not the case and thus they must be considered a-priori independent. To see
    this consider the effect of such a change upon the cycle $(1,-1,-1,1)$. 
    If $N$ is even, we get $(-1)^{N/2}(1 ,1,-1,-1)$ which gives the second
    alternating inequality. For $N$ = odd, we get  $\pm(1 ,-1,-1,1)$, which results in
    the same inequality. Similar results hold for the other cycle.
    The inequalities (\ref{NPartite}) are necessary conditions for a PLHV model to exist. It is not known what a necessary and sufficient set would be.

The bound (\ref{NPartite}) for PLHV models is sharp
since it  can be obtained by considering for example the
bi-separable partition
$\{1, \ldots, N-1\}\,,\, \{N\} $ and choosing
the absolute maximum for $S_{N-1}^{\pm}$, which is $2^{N-1}$ since
there are just that many terms in the operator $S_{N-1}^{\pm}$,
and choosing $\av{A_N}=\av{A_N'}=1$ for party  $N$.

 Let us consider the Svetlichny polynomials at closer scrutiny. The following recursive relation holds:
    \begin{align}\label{eq:rec}
    S_N^\pm = S_{N-1}^\pm A_N \mp S_{N-1}^\mp
    A_N',\end{align}
 with $S_2^+=-M_2$ and $S_2^-=  M_2'$ . Consider the term \( S_{N-1}^\pm A_N \). The maximum of  \(|\langle S_{N-1}^\pm A_N\rangle |\) is equal to the maximum of 
    \(|\langle S_{N-1}^\pm\rangle|\) since $\max|\av{A_N}|=1$. Similarly for the other term. Thus one can take the \(N\)-partite
    bound as twice the \((N-1)\)-partite bound.

The Svetlichny polynomials  $S_N$  are related to the Mermin polynomials  $M_N$
by the following linear recursive relations \cite{uffink}:  \begin{alignat} {2}
&S^{\pm}_{N} = 2^{l-1}
\left( (-1)^{l(l\pm1)/2} M_{N} \mp (-1)^{l(l\mp1)/2} M'_{N}\right), ~~~~&&\textrm{for}~N\textrm{ = odd, and }N=2\,l+1,
\nn\\ 
& S^{\pm}_{N} =  2^{l-1} (-1)^{l(l\pm1)/2} M_N^{\pm}, &&\textrm{for}~N\textrm{ = even, and }N=2\,l,
\label{evenodd}
\end{alignat} 
where $M_N^{+}:= M_N$  and $M_N^{-} := M_N'$.
            
            The expectation values are thus related as:
            \begin{align}
                   | \av{S_N^{\pm}}|=\left\{\begin{array}{cc}      2^{(N-2)/2} |\av{M_N^{\pm}} | , &\textrm{if}~   N\textrm{ = even},\\
2^{(N-3)/2}   |\av{M_N^\pm \pm M_N^\mp}| ,~~~&\textrm{if}~         N\textrm{ = odd}.
            \end{array}\right.
            \end{align}
 Note that from the above relations we get the following identity:
\begin{align}\label{Ident}
\av{S_N^{+}}^2+\av{S_N^{-}}^2=2^{N-2}(\av{M_N}^2+\av{M_N'}^2).
\end{align}
Hence, the quadratic separability inequalities of \Eq{quadraticN} for multi-partite quantum states can be equally expressed in terms of operators $S_N$. The maximal quantum mechanical violation the left-hand side of the
    \(N\)-partite alternating inequalities of  (\ref{NPartite}) is thus equal to \(2^{N-1}\sqrt2\). 
This upper bound is in fact achieved for the GHZ states for appropriate values of
    the polarizer angles of the relevant spin observables\footnote{The settings are obtained as follows. 
    Let 
    \(A_k^i=\cos\alpha_k^i \sigma_x +
    \sin\alpha_k^i \sigma_y\) denote spin observables with angle
    \(\alpha_k^i\) in the \(x\)-\(y\) plane. A simple calculation
    shows 
    \begin{align}\label{eq:eghz} \av{i_1\cdots
    i_N}_{\textrm{qm}}=\pm\cos(\alpha_1^{i_1}+\cdots+\alpha_N^{i_N}) ,
    \end{align}
    where the sign is the sign chosen in the GHZ state.
    We now note that for \(k=0,1,2,\dots\) one has:
    \(\cos\left(\pm\frac{\pi}{4}+k\frac{\pi}{2}\right)=\nu^{\pm}_k\frac{\sqrt
    2}{2}\) where \(\nu^{\pm}_k\) is given by (\ref{eq:nupm}). This means that by a
    proper choice of angles, we can match, up to an overall sign, the sign of the
    cosine in (\ref{eq:eghz}) with the  sign in  front of \(\av{i_1\cdots i_N}_{\textrm{qm}}\) as
    it appears in the inequality, forcing the left-hand side of the inequality to be
    equal to
    \(2^{N-1}\sqrt 2\). This can be easily done if each time an index \(i_j\)
    changes from \(1\) to \(2\), the argument of the cosine is increased by
    \(\frac{\pi}{2}\).  Choose therefore
    $\left(\alpha^1_1,\alpha^1_2,\dots,\alpha^1_N \right)
    =\left(\pm\frac{\pi}{4},0,\dots,0\right),
$ and $\left(\alpha^2_1,\alpha^2_2,\dots,\alpha^2_N \right)=   
    \left(\pm\frac{\pi}{4}+\frac{\pi}{2},\frac{\pi}{2},\dots,\frac{\pi}{2}\right)$, 
    where the sign indicates which of the two $S^{\pm}_N$ inequalities is used.}.

In conclusion, fully entangled quantum states can violate the Svetlichny  inequalities by a factor as large as $\sqrt{2}$ \cite{seevsvet,collins}, thereby proving that quantum correlations contain full multi-partite non-locality. The absolute maximum is a factor $\sqrt{2}$ larger than the maximum quantum bound. Thus 
\begin{align}\label{svetresult}
2\max| \av{S_N}_\textrm{plhv}|=\sqrt{2}\max |\av{S_N}_\textrm{qm}|= |S_N|_\textrm{max}=2^N.
\end{align}

Note that quantum mechanics can never attain the absolute maximum for the expectation value of the Svetlichny polynomials $S^\pm_N$. This is in contradistinction to what was the case for the Mermin polynomials $M_N$ and $M_N'$, where for $N$ = odd quantum mechanics is able to give the absolute maximum on $|\av{M_N}_{\textrm{qm}}|$ and $|\av{M_N'}_{\textrm{qm}}|$. It is thus the quantum bound on the Svetlichny polynomials, and not on the Mermin polynomials that distinguishes quantum correlations from more general correlations for all $N$.\forget{, since these are able to violate the quantum bound on the Svetlichny polynomials, but not this bound on the Mermin polynomials for $N$ = odd.}

  The two alternating solutions for \(N=2\) are the usual CHSH inequalities, i.e, $|\av{M_2}_{\textrm{lhv}}|\leq2$, and $|\av{M_2}_{\textrm{lhv}}'|\leq 2$, where for $N=2$ there is of course no distinction between a LHV and PLHV model.
    The  ones for \(N=3\) give rise to the two inequalities found in Svetlichny
    \cite{svetlichny} that are also given in (\ref{Sv3}),  and for \(N=4\) we have $|\av{S_4^+}_{\textrm{plhv}}|=2|\av{M_4}_{\textrm{plhv}}|\leq 8$ and\forget{        \begin{align} M_4
    & |E(1111)-E(2111)-E(1211)-E(1121)-
     E(1112)-E(2211)-E(2121)-E(2112)- \nonumber \\
     & E(1221)-E(1212)-E(1122)+E(2221)+ 
     E(2212)+E(2122)+E(1222)+E(2222)|\le 8, 
    \end{align}}
    where the second inequality is $|\av{S_4^-}_{\textrm{plhv}}|=2|\av{M_4'}_{\textrm{plhv}}|\leq 8$.

\subsection{Alternative formulation}
After \citet{seevsvet} announced their generalized Svetlichny inequalities \citet{collins} independently announced similar inequalities. They define  the following Svetlichny polynomials
            \begin{align}\label{alt}
                   \widetilde{S}_N=\left\{\begin{array}{cc}     M_N , &\textrm{if}~   N\textrm{ = even},\\
 \frac{1}{2}(M_N + M_N') ,~~~&\textrm{if}~         N\textrm{ = odd},
            \end{array}\right.
            \end{align}
 and  prove\forget{the following bounds}
for $N$ = odd: $|\av{\widetilde{S}_N}_{\textrm{plhv}}|\leq 2^{(N-1)/2}$, and for $N$ = even: $|\av{\widetilde{S}_N}_{\textrm{plhv}}|\leq 2^{(N-2)/2}$. The quantum bounds are a factor $\sqrt{2}$ higher and the absolute maxima $|\widetilde{S}_N|_{\textrm{max}}$ are a factor $2$ higher.  These bounds give the same structure as  in (\ref{svetresult})  which was obtained using the Svetlichny polynomial $S^{\pm}_N$ used here. This formulation (\ref{alt}) is used by \citet{jones} and \citet{marcovitch}.

Although using (\ref{alt}) gives a simpler recursive relation in terms of the Mermin polynomials than was the case for $S^\pm_N$ as given in (\ref{evenodd}), the bounds for $\widetilde{S}_N$ now depend on whether $N$ is  even or odd, which we regard to be an unwelcome feature.

\section[Further remarks on quantum mechanical violations]{Further remarks on quantum mechanical\\ violations}\label{moreQM}

We  have seen that using $N$-partite GHZ states the Svetlichny inequalities can be violated by as large as a factor $\sqrt{2}$. The GHZ states are fully entangled and this full entanglement is a necessary feature to give a violation. This follows from the fact  that (\ref{quadraticN}) of chapter \ref{Npartsep_entanglement} shows (using the identity (\ref{Ident})) that any bi-separable state (i.e., $k=2$) has a maximal value for $\av{S_N^{\pm}}_{\textrm{qm}}$ equal to $2^{(N-3/2)}$, which is a factor $\sqrt{2}$ below the PLHV bound as given in (\ref{NPartite}).
Thus a `gap' appears between the correlations
that can be obtained by bi-separable quantum states and those obtainable by
PLHV models. It thus takes fully entangled states to obtain all the correlations
obtainable by a PLHV model. This is analogous to the results found in section
\ref{LHVmermin} for the LHV case, where it was shown that one needs entangled states to give all the correlations that are producible by LHV models.

 These results imply
that the mere requirement of locality (factorisability) between just
two subsets, although within each subset full blown non-locality is
admissible, already forces the correlations to be less strong than
some of the quantum mechanical correlations, although they are
nevertheless still stronger than those obtainable from bi-separable
quantum states.

An interesting question to ask next is whether the $N$-partite non-locality that is found in the fully entangled GHZ states is generic or whether it can only be found in some specific states. That is, can we generalize the observation that $N$-party entangled  pure states contain $2$-partite non-locality \cite{GisiN,GISIN91,POPROHR} (any such state can be made to violate the CHSH inequality for some set of observables) so as to be able to claim that all fully entangled pure states are fully non-local? \forget{Only a partial answer exists.  All three-partite fully entangled pure three-partite quantum states contain full $3$-partite non-locality, as they can be made to violate the Svetlichny ineq. (\ref{NPartite}) for $N=3$ \cite{thesiscollins} [KLOPT DIT?]. The question whether a similar result holds for $N\geq 4$ is still open.}  This question is still open. However, for mixed states this probably does not hold. Indeed, fully entangled mixed states exist that cannot be made to violate the Svetlichny inequality, as we will now show. 

\subsection{Hidden full non-locality?}\label{hiddenfullnonlocal}

Let us consider the GHZ states mixed with white noise, notated as: 
\begin{align}\rho_N=(1-p)\ket{\psi_{{\rm \sc
GHZ},\alpha}^N}\bra{\psi_{{\rm \sc GHZ},\alpha}^N} + p\,\1/2^N,
\end{align} with
$0\leq p\leq1$.  The white noise robustness of the GHZ states so as to exhibit full non-locality is obtaining by determining for which value of $p$ the Svetlichny inequalities can be violated. It is easily found that this gives $p<1-1/\sqrt{2}\approx 0.29$. We already know that for $p< 1/(2(1-2^{-N}))$ this set is fully $N$-partite entangled (it violates the
sufficient criterion of (\ref{Nk2}), cf. (\ref{GHZfullN})). For $N=3$ this gives $p<4/7$ and for large $N$ this goes to $p<1/2$. Consequently, for $1-1/\sqrt{2} <p< 1/(2(1-2^{-N}))$ the set $\rho_N$ is fully $N$-partite entangled, but nevertheless cannot be made to violate the Svetlichny inequality (\ref{NPartite}). 

Note however, that does result not prove that these states are not fully non-local since the Svetlichny inequalities are not sufficient for a PLHV model to exist, i.e., they are only necessary requirements.  What it does show is that in case these states are fully non-local  (to be shown by some other method), this non-locality cannot be revealed using a Svetlichny inequality. If such states indeed exist, they contain what we propose to call `hidden                              full non-locality'. This terminology is motivated by an analogous feature for the two-partite case: bi-partite states exist that are entangled and which have a local model for all measurements using two dichotomic observables per party (and thus cannot violate the CHSH inequality) whose  non-locality can nevertheless be revealed using a local filtering process. \citet{POPESCU} called these `hidden nonlocal'. \forget{
Contrast this to noise robustness for mere non-locality not full non-locality:  For $p<1- 2^{(1-N)/2}$  the state violates the Mermin inequality $\av{M_N}\leq 2$ that holds for LHV models. This goes to one for large $N$.  }

\subsection*{Experiments indicating full non-locality in a quantum system}

Although the experiment by \citet{PAN} did create full three-qubit entanglement, as was argued for in section \ref{N3}, it is unclear if full non-locality was experimentally  produced, since no violation of a three-partite Svetlichny inequality has been tested. However for $N=4$, such a violation is reported by \citet{zhao} since there the Svetlichny inequality using $S_4^+=M_4$ was violated using a GHZ state, which confirms full four-partite non-locality.

 \section[On discriminating no-signaling correlations]{On discriminating no-signaling correlations\\ using expectation values only} \label{nosignalsvet}
    
We have seen that quantum mechanics cannot maximally violate the Svetlichny inequalities, i.e., $\max |\av{S_N^\pm}_\textrm{qm}|= |S_N^\pm|_\textrm{max}/\sqrt{2}$. Thus the inequalities obtained from the Svetlichny polynomials  allow for distinguishing quantum correlations from more general correlations, something the Mermin polynomials were unable to do for odd $N$. However, the Svetlichny polynomials unfortunately do not 
distinguish no-signaling correlations from the most general correlations, for it is the case that $\max |\av{S_N^\pm}_\textrm{ns}|= |S_N^\pm|_\textrm{max}$, as proven by \citet{jones}. Thus no non-trivial bound for the no-signaling correlations is obtained.  

Of course, the defining conditions of no-signaling themselves give the facets of the no-signaling polytope. However, we believe it is interesting to ask for non-trivial inequalities in terms of product expectation values $\av{A_1\cdots A_N}$ despite the fact that these cannot be facets of the no-signaling polytope. 
For $N=2$ we were able to find such a set of Bell-type inequalities\forget{with a non-trivial bound on product and marginal expectation values for bi-partite no-signaling correlations}  (in section \ref{nontrivnosignal}) and for $N=3$ it was argued in the previous chapter that the monogamy inequality (\ref{monogamynosignaling}) is able to discriminate no-signaling from general three-partite correlations.  But for $N>3$ no such Bell-type inequalities or monogamy inequalities are known to exist.\forget{
The condition of no-signaling is by no means a trivial condition. A good case can be made that it must be obeyed by any theory that respects special relativity. It would be very interesting to be able to find non-trivial inequalities for this condition, such that the maximum bound on a certain correlation polynomial is strictly smaller than the absolute maximum. Such an inequality would then be able to discriminate between no-signaling correlations and arbitrary correlations.
} We thus leave as an open question the search for non-trivial no-signaling Bell-type inequalities in terms of product expectation values for $N>3$.  The Svetlichny polynomials use all possible combinations of the  products $A_1A_2 \cdots A_N$  for all $A_1,\ldots, A_N$.  It does not seem likely that, when compared to $S_N^\pm$, using different linear combinations of these terms with coefficients $\pm1$ will help. One must thus probably resort to including 
marginal expectation values\forget{also expectation values of products of observables that do not contain all parties} that  have less than $N$ terms, just as was the case in for example the bi-partite inequalities (\ref{nontrivnonsig2})\forget{ that gave a non-trivial bound for bi-partite no-signaling correlations}.  It might furthermore be necessary  to allow  for more than just two local settings or for more than just two possible outcomes per observable.
We expect that the method used in the bi-partite case that gave the non-trivial no-signaling inequalities  \eqref{nontrivnonsig2} and \eqref{14ineq} generalizes to the multi-partite case, but we have not performed such a generalization.  

\forget{the extreme no-signaling point are known (barrett et al.), but are the facets known? are the given a bell-type expression?}

          \section{Discussion}\label{concsvet}
          The multi-partite investigation of discriminating partially-local, quantum mechanical and no-signaling correlations has given us many results, but some interesting questions remain unsolved, as we will now discuss.
          
                           In this chapter we have derived Bell-type inequalities -- which we have called Svetlichny inequalities -- that discriminate partially-local correlations from quantum correlations, and also quantum correlations from no-signaling correlations.  It is however unknown if these inequalities are tight, i.e., if they give facets of the partially-local polytope. It would be interesting to try and find the full set of tight Svetlichny inequalities for $N$ parties, although this might be a computationally hard problem. For three parties, however, it is likely that this problem is computationally tractable.
                                                      
                           The Svetlichny inequalities do not discriminate no-signaling correlations from general unrestricted correlations.                                                    
                           For no-signaling correlations no non-trivial bound exists on the expectation value of the Svetlichny polynomials. Providing such discriminating conditions for multi-partite no-signaling correlations ($N>3$) in terms of product expectation values  (possibly including some marginal expectation values)  is left as an open problem.

        Fully entangled quantum states were shown to violate the Svetlichny inequality, thereby showing that they are fully non-local: no PLHV model can give rise to these quantum correlations.  However, we showed that fully entangled mixed states exist that cannot be made to violate the inequalities.  Their full non-locality, if indeed present, thus needs to be shown in a different, yet hitherto unknown way.                                     
                                   
                               \forget{       In chapter \ref{chapter_monogamy} we have seen that local correlations, partially-local and fully general correlations are not monogamous, whereas bi-partite quantum and no-signaling correlations exist that are monogamous. We conjecture that  fully non-local correlations, i.e. those that violate the Svetlichny inequalities, are also monogamous. If indeed so, this provides a means of discriminating them from general unrestricted correlations.  However, we leave such an investigation for future research.}
                               
                                    \forget{OUDE FORMULERING
                                      In chapter \ref{chapter_monogamy} we have seen that local correlations and fully general correlations are not monogamous, whereas bi-partite quantum and no-signaling correlations exist that are monogamous. What about partially-local ones? We conjecture that they are not monogamous, but that  fully non-local correlations, i.e. those that violate the Svetlichny inequalities, are monogamous. If indeed so, this provides a means of discriminating them from general unrestricted correlations.   However, we leave such an investigation for future research. }
                                 \forget{   In chapter \ref{chapter_monogamy} we have seen that local correlations and fully general correlations are not monogamous, whereas bi-partite quantum and no-signaling correlations exist that are monogamous. What about partially-local ones? We conjecture that they are not monogamous, but also leave also this as an open question for future research. However, it seems plausible that fully non-local correlations, i.e. those that violate the Svetlichny inequalities, are monogamous. If indeed so, this provides a means of discriminating them from general unrestricted correlations. 
                                   }                     
                                                        \forget{It could well be that investigating the monogamy aspects of correlations that violate the Svetlichny inequalities will prove monogamy of multi-partite no-signaling correlations, thereby providing a means of discriminating them from general unrestricted correlations. This was indeed the case in the three-partite case for non-local correlations between two of the three parties that violate the CHSH inequality.                                                   }        
                                      
          Lastly, we note that \citet{jones} consider a class of models more general than we have considered here, and they showed that these models must still obey the Svetlichny inequalities. The models they considered did not impose partial locality on the correlations, but allowed for specific fully non-local correlations that follow from a so-called partially paired communication graph which represents a specific signaling pattern between all of the parties. It was shown that these models can not be made to violate the Svetlichny inequalities. Thus the non-locality needed in obtaining a violation of the Svetlichny inequalities must be stronger than the non-locality of these partially paired communication graphs. Because quantum mechanics violates the Svetlichny inequalities, \citet{jones}  interpret their result as indicating that quantum correlations are much more non-local than previously thought.

      \forget{     Violation of the Svetlichny ineq. shows that to simulate $N$-partite quantum correlations using signaling (communication) requires communication that links all the parties. Thus any such classical model that reproduces such correlations, must use signaling that links all parties. But note that the quantum correlations themselves are non-signaling.
           
        We will not investigate the precise simulation of quantum correlations, and what nonlocal resources are needed for that. This is being investigated at the moment.}

    \section*{Appendix: Proof of inequality (\ref{EQ3})}\label{appsvet}
    We seek inequalities of the form
    \begin{align}
    \sum_I\sigma_I \av{i_1i_2\cdots i_N}_{\textrm{plhv}} \le M ,    \end{align}
    where \(\sigma_I\) is a sign and $M$ non-trivial.
    Following almost verbatim the analysis in \cite{svetlichny}, one must look for
    \(\sigma_I\) which solve the minimax problem
    \begin{align}\label{eq:minmax}
    m=\min_\sigma m_\sigma =\min_\sigma\max_{\xi,\eta}
    \sum_{I}\sigma_I\xi_{i_1\cdots i_k}\eta_{i_{k+1}\cdots i_N},
    \end{align}
    where \(\xi_{i_1\cdots i_k}=\pm 1\) and \(\eta_{i_{k+1}\cdots i_N}=\pm 1\)
    are also signs.  Without loss of generality we can take  \(k\ge N-k\).

    One can derive some useful
    upper bound on \(m\). Toward this end, we choose to set $\eta_{i_{k+1}\cdots i_{N-1}2}=
    \zeta_{i_{k+1}\cdots i_{N-1}}\eta_{i_{k+1}\cdots i_{N-1}1}$ for some sign
    \(\zeta_{i_{k+1}\cdots i_{N-1}}\), using the fact that $i_N=1,2$. Taking into
    account that \(\sigma_I^2=1\), and denoting by \(\hat I\) the \((N-1)\)-tuple
    \((i_1,\dots,i_{N-1})\) we have:
    \begin{align}
   m_\sigma = \max \sum_{\hat I} \sigma_{\hat I1}\eta_{i_{k+1}\cdots i_{N-1}1}
    \xi_{i_1\cdots i_k}(1  + \sigma_{\hat I1}\sigma_{\hat
    I2}\zeta_{i_{k+1}\cdots
    i_{N-1}}).
    \end{align}
    The maximum being over  \(\xi_{i_1\cdots i_k}\), \(\eta_{i_{k+1}\cdots
    i_{N-1}1}\), and \(\zeta_{i_{k+1}\cdots i_{N-1}}\).

    Now certainly one has
    \begin{align}\label{eq:rm}
    m_\sigma \le \hat m_\sigma=\max\sum_{\hat I}|1  + \sigma_{\hat
    I1}\sigma_{\hat I2}\zeta_{i_{k+1}\cdots i_{N-1}}|,
    \end{align}
    the maximum taken over \(\zeta_{i_{k+1}\cdots i_{N-1}}\).

    If we define \( \hat m=\min_\sigma\hat m_\sigma\) one easily sees that
    \(\hat m = 2^{N-1}\). This can only be achieved under the following condition:
    \begin{align}\label{eq:mincon} \begin{array}{l} \hbox{\sl For each fixed\ }
    (i_{k+1},\dots, i_{N-1}) \hbox{\sl\ exactly\ } 2^{k-1} \\
    \hbox{\sl of the quantities\ } \sigma_{\hat I1}\sigma_{\hat I2} \hbox{\sl\
    are\ } +1 \hbox{\sl\ and\ } 2^{k-1} \hbox{\sl\ are\ } -1.\end{array}
    \end{align}
    Although it may be that \(m<\hat m\) we have proven that \(m=\hat m=8\)
    in all cases for \(N=4\), and \(m = \hat m\) for \(k=N-1\) for any \(N\).

    We shall call any choice of the \(\sigma_I\) satisfying this condition
    a minimal solution.

    What immediately follows from the above is that any
    solution of (\ref{eq:mincon}) for a given value of \(P\) is a
    solution for all greater values of \(k\le N-1\).
    A violation of an inequality so obtained for the smallest possible
    value of \(k\ge N/2\) precludes then any PLHV
    model of the \(N\)-partite correlations.

Assume provisionally that only bi-partitions $\{1,\ldots,k\},\{N-k, \ldots,N\}$ occurs.\forget{    Assume provisionally that the only bi-partite partitions  are those in which an
    $\{1,\ldots,k\},\{N-k, \ldots,N\}$ split occurs.} The whole ensemble  consists of subensembles corresponding to different choices of the    \(k\) parties.  We do not know in any particular system  to which of the subensembles the system belongs. 
    To take account of this, our inequality must be one that would arise under any
    choice of the \(k\) parties.  Call a minimal solution \(\sigma_I\) 
    admissible if \(\sigma_{\pi(I)}\) is also a minimal
    solution for any permutation \(\pi\). An inequality that follows
    from an admissible solution will therefore be one that must be
    satisfied by any subensemble of systems with split $\{1,\ldots,k\},\{N-k, \ldots,N\}$.

    The set of admissible solutions  breaks up into orbits by the
    action of the permutation group. The overall sign of \(\sigma_I\)
    is not significant and two solutions that differ by a sign are
    considered equivalent. The set of these equivalence classes also
    breaks up into orbits by the action of the permutation group. It
    is remarkable that there are orbits consisting of one equivalence
    class only. For such, one must have \(\sigma_{\pi(I)} = \pm
    \sigma_{I}\). The sign in front of the right-hand side must be a
    one-dimensional representation of the permutation group, so one
    must have either \(\sigma_{\pi(I)} =  \sigma_{I}\) or
    \(\sigma_{\pi(I)} =  (-1)^{s(\pi)}\sigma_{I}\), where \(s(\pi)\)
    is the parity of \(\pi\). The second case is impossible since one
    then would have \(\sigma_{11\cdots 1}=-\sigma_{11\cdots 1}\) as a
    result of a flip permutation. Since an overall sign is not
    significant one can now fix \(\sigma_{11\cdots 1}=1\).  As the
    only permutation invariant of \(I\) is \(t(I)\), the number of
    times index \(2\) appears in \(I\), we must have \(\sigma_I =
    \nu_{t(I)}\) for some \((N+1)\)-tuple (\(\nu_0=1\) by convention)
    \(\nu=(1,\nu_1,\nu_2,\dots,\nu_N)\). We must now solve for the
    possible values of \(\nu\).

    Let \(a=t(i_{k+1}\cdots i_{N-1})\) and \(b=t(i_1\cdots i_k)\), then
    condition  (\ref{eq:mincon}) for our choice of \(\sigma_I\), is equivalent
    to
    \(\nu\) satisfying
    \begin{align}\label{eq:nucond}
    \sum_{b=0}^k\left(\begin{array}{c}k \\ b\end{array}\right)
    \nu_{a+b}\nu_{a+b+1} = 0,\quad a=0,1,\dots N-k-1.
    \end{align}
    Let $\mu_k=\nu_k\nu_{k+1}$. Eq. (\ref{eq:nucond}) then becomes
    \begin{align}\label{eq:mucond}
    \sum_{b=0}^k\left(\begin{array}{c}k \\ b\end{array}\right) \mu_{a+b} =
    0,\quad a=0,1,\dots N-k-1.
    \end{align}

    Now it is obvious that there are at least two solutions of (\ref{eq:mucond})
    valid for all \(k\), to wit \(\mu_k = \pm (-1)^{k}\) since then
    (\ref{eq:mucond}) is just the
    expansion of \((1-1)^k\) or \((-1+1)^k\). Call these solutions the alternating solutions.  Finally we get from $\mu_k=\nu_k\nu_{k+1}$ the two
    solutions (\ref{eq:nupm}) once we've chosen the overall sign to set 
    \(\nu_0=1\).

%
%
%
%
\clearemptydoublepage
\thispagestyle{empty}
\part{Quantum philosophy}

\clearemptydoublepage
\thispagestyle{empty}
\chapter[The quantum world and correlations]{The quantum world is not built up\\\vskip0.2cm from correlations}
\label{chapter_quantumworldcorrelations}
\forget{\section{Abstract}
It is known that the global state of a composite quantum system can be
completely determined by specifying correlations between measurements performed on subsystems only.
Despite the fact that the quantum correlations thus suffice to reconstruct the quantum state,
we show, using a Bell inequality argument, that they cannot be regarded as
objective local properties of the composite system in question.
It is well known since the work of
 J.S. Bell, that one cannot have locally determinate values for all physical quantities,
 irrespective of  whether they are determined in a deterministic or stochastic way.
 The Bell inequality argument we present here shows this is also impossible for
correlations among subsystems of an individual isolated
composite system.
Neither of them can be used to build up a world consisting of some
local realistic structure.  The argument has an important advantage over others because it
 does not need perfect correlations but only statistical correlations.
 It can therefore easily be tested in currently
 feasible experiments using four particle entanglement.
 As a corollary to the result we argue that
entanglement cannot be considered ontologically robust. We present four conditions  that arguably can be regarded as necessary conditions for ontological robustness of entanglement and show that they are all four violated by quantum mechanics.
 }
\noindent
This chapter is a slightly adapted version of \citet{seevcor}.
\section{Introduction}\label{intro}
  What is quantum mechanics about? This question has haunted the physics community
ever since the conception of the theory in the 1920's. Since the
work of J.S. Bell we know at least that quantum mechanics is not
about a local realistic structure built up out of values of
physical quantities \cite{bell64}. This is because of the well known
fact, that if one considers the values of physical quantities to
be locally real (i.e., if they are to obey the doctrine of local realism), then they must obey a local Bell-type inequality, which
quantum mechanics violates. The paradigmatic example of a quantum
system that gives such a violation is the singlet state of two
spin-${\frac{1}{2}}$ particles. This state describes two particles that
are anti-correlated in spin. Bell's result shows that the two particles in the singlet state cannot be regarded to possess local realistic\footnote{The adjective `local realistic' is to be understood as obeying the doctrine of local realism, 
 cf. chapter \ref{chapter_CHSHclassical}.}  values for all their (single particle) physical quantities, values  which do not vary depending on what one does to another spatially separated system.  Instead, the
singlet state tells us that upon measurement the spin values, if
measured in the same direction on each particle, will always be
found anti-parallel. Because this (anti-) correlation is found in
all such measurements, an obvious question to ask is whether
or not we can think of this (anti-) correlation as a real property
of the two-particle system independent of measurement.

Could it be that what is real about two systems in the singlet
state are
 not the local spin values,
 but merely the correlations between the two systems? Is quantum
mechanics about a physical world consisting not of systems that have objective local realistic values of
quantities but solely of objective local realistic  correlations, of which
some are non-contextually revealed in experiment? In other words, is there a
fundamental difference according to quantum mechanics as regards
the physical status of values of quantities and of correlations,
as for example Mermin\footnote{N.D. Mermin, in a series of papers \cite{merminithaca1,merminithaca2,merminithaca3}, tried to defend this
 fundamental interpretational difference between values of quantities and correlations. He used the phrase `correlations without correlata' for this position.} seems to suggest?

 There is good reason to think that these questions should be
 answered in the positive, since a non-trivial theorem (which is true
 in quantum mechanics) points into this direction.
  The theorem (to be treated in the next section) shows that the global state of a composite quantum system
 can be completely determined by specifying
 correlations (i.e., joint probability distributions) when sufficient
 local measurements are performed on each subsystem.
 It thus suffices to consider only  correlations  between measurements performed on subsystems only in order to completely specify
 the state of the composite system.  
  But can one also think of these correlations to be
  objective properties that pertain local realistically to the composite quantum system in question?
  As mentioned before in the case of the anti-correlation
  of the singlet state, one is tempted to think that this is indeed the case.
  However in this chapter we will demonstrate that, however tempting,
  no such interpretation is possible and that
  these questions (as well as the questions
mentioned earlier) can thus not be answered in the positive.  This is shown using a Bell-type inequality argument which shows that 
 the correlations cannot be regarded as objective properties constrained by local realism that somehow pertain to the composite system in question. Our strategy is analogous to the one Bell adopted  when he showed that in quantum mechanics one cannot have values for all physical quantities that are determined via deterministic or stochastic local hidden variables.  We extend Bell's analysis by showing that this is also impossible for correlations among subsystems of an individual isolated composite system. Neither of them can be used to build up a world consisting of some local realistic structure\footnote{This is not to be understood as a claim which is supposed to show the impossibility of defining local elements of reality, but as one that shows the impossibility of these elements of reality to obey the doctrine of local realism when required to reproduce the quantum predictions.}.

\citet{cabello} and \citet{jordan} give the same answer to similar questions using a Kochen-Specker-type \cite{kochenspecker}, Greenberger-Horne-Zeilinger (GHZ) -type 
 \cite{ghz,GHZ} or Hardy-type \cite{hardy}
argument.
Besides giving the Bell-type inequality version of the argument
(which in a sense completes the discussion because it was still lacking),
the advantage of the argument given in this chapter above these previous arguments,
is that it is more easily experimentally accessible using current
technology.
For this purpose, we explicitly present
a quantum state and the measurements that are to be performed in
order to test the inequality.

The structure of this chapter
 is as follows.
In sec. \ref{qmcorr}
we will present an argument to the effect that quantum correlations
are objective local properties that pertain to composite quantum systems and that do not vary depending on what one does to another, spatially separated system.
In the next two sections we will however show that this
  line of thought is in conflict with quantum mechanics itself.
To get such a conclusive result we need to be very formal and rigorous. In sec. \ref{bell} we will
therefore define our notion of correlation and derive a Bell-type inequality for correlations using a stochastic
hidden-variable model under the assumption of local realism. This formalizes the idea of correlations as objective local realistic properties.  In sec. \ref{qmbell} we show that this inequality, when turned into it's
quantum mechanical form, is violated by quantum correlations. We
present a quantum state and a set of measurements that allow for
such a violation and furthermore show that it is the maximum
possible violation.

In sec. \ref{ontol} we apply this result to argue that 
entanglement cannot be considered ontologically robust
 when the quantum state is taken to be a complete description of the system in question.
 We present four conditions  that arguably can be regarded as necessary conditions for ontological robustness of entanglement and show that they are all four violated by quantum mechanics.
However, we argue that it nevertheless can be considered a resource
in quantum information theory to perform computational and
information-theoretic tasks.
In the last section, sec. \ref{conclu},  we briefly
discuss the implications of our results, compare our argument
 to the ones given by \citet{cabello} and by \citet{jordan} and return to the questions stated in the beginning of this introduction.

\section[Does the quantum world consist of correlations?]{Does the quantum world consist of\\ correlations?}\label{qmcorr}

In many important instances a system can be regarded as composed out
of separate subsystems.
In a physical theory that describes such composite
systems it can be asked whether one can assume that
the global state of the system
 can be completely determined by specifying
 correlations (joint probability distributions)
 when a sufficient number of local\footnote{Note that here (and in the rest of the chapter)
`local' is taken to be opposed to `global'
and thus not in the sense of spatial localization. Local
thus refers to being confined to a subsystem of a larger system, without requiring the
subsystem itself to be localized (it can thus itself exist of
spatially separated parts).} measurements
 are performed on each subsystem.
\citet{barrett} calls this the global state assumption.
 Perhaps not surprisingly, the assumption holds for classical
 probability theory and for quantum mechanics on a complex Hilbert space.
 However, it need not be satisfied in an arbitrary theory,
 which shows that the theorem is non-trivial. For example, \citet{wootters}
 has shown that for quantum mechanics on a real Hilbert
 space the assumption does not hold because the correlations between subsystems do not suffice
 to build up the total state. By counting available degrees of freedom of
 the state of a composite system and of the states of its subsystems one
 can easily convince oneself that this is the case\footnote{J. Barrett (private communication) gives the following counting argument.
  A density matrix on a real Hilbert space with dimension $d$ has
  $M= (d^2-d)/2 +d=(1/2)d(d+1)$ parameters (without normalization),
  and a density matrix on a $d\otimes d$-dimensional real Hilbert space
  has $(1/2)d^2(d^2 +1)$ parameters, which is too many because it is more than $M^2$.
      On the other hand, for complex Hilbert spaces we have that a density 
matrix has  $N = d^2$ real parameters. So a density matrix on a $d\otimes d$-dimensional complex Hilbert space has 
$d^4$ real parameters  which is indeed $N^2$.}.

\citet{merminithaca1} has called the fact that in quantum mechanics the global state assumption holds
sufficiency of subsystems correlations, or the SSC theorem. He phrases it as follows.  Given a system $\mathcal{S}=\mathcal{S}_1+\mathcal{S}_2$
with density matrix $\rho$,
then $\rho$ is completely determined by correlations $P(a,b|A,B)$ (joint probability distributions conditioned on the settings chosen, see section \ref{qmcorrsection})  that determine the mean values $\av{A\otimes B}_{\textrm{qm}}=\textrm{Tr}[\rho(A\otimes B)]=\sum_{a,b} ab\,P(ab|AB)$, for an appropriate set
of observable pairs $\{A\},\{B\}$. The proof\footnote{\citet{wootters} has also independently proven this.} relies on three facts:
Firstly, the mean values of \emph{all} observables for the entire system determine its state.
  Secondly, the set of all products over subsystems of subsystem observables
  (i.e., the set $\{ A\otimes B \}$) contains
 a basis for the algebra of \emph{all} such system-wide observables.
  Thirdly, the algorithm that supplies observables with their mean value is
linear on the algebra of observables.

As an example of the theorem, consider the well known singlet state $\ket{\psi^-}=(\ket{01}-\ket{10})/\sqrt{2}$  of two qubits (spin-$\frac{1}{2}$ particles) written
as the one-dimensional projection operator
 \begin{align}
\hat{P}_{s}=\ket{\psi^-}\bra{\psi^-}= \frac{1}{4}(\1 -\sigma_z \otimes\sigma_z
-\sigma_x \otimes\sigma_x-\sigma_y \otimes\sigma_y).
\end{align}
The mean value of $\hat{P}_{s}$ is determined by the mean values
of the products of the  $x$, $y$ and $z$ components of the individual spins:
\begin{align}
\av{\hat{P}_s}_{\textrm{qm}}=\frac{1}{4}(1-\av{\sigma_z\otimes\sigma_z}_{\textrm{qm}}-\av{\sigma_x\otimes\sigma_x}_{\textrm{qm}}-\av{\sigma_y\otimes\sigma_y}_{\textrm{qm}}).
\end{align}
Since the mean value of this projector is 1 for the singlet,
the singlet state is thus determined by
the spin correlations in $x$, $y$ and $z$ direction having the value $-1$ for 
$\av{\sigma_z\otimes\sigma_z}_{\textrm{qm}}$, $\av{\sigma_x\otimes\sigma_x}_{\textrm{qm}}$ and $\av{\sigma_y\otimes\sigma_y}_{\textrm{qm}})$ which is perfect anti-correlation in all these three directions. 
Because of rotational invariance of the singlet state one can choose any three orthogonal $x$, $y$ and $z$ directions.   Perfect anti-correlation of any three orthogonal components is
thus enough to ensure that the global state is the singlet state.
Thus correlations among all subsystems completely determine
the density matrix for the composite system they make up,
or in Mermin's words \cite{merminithaca2}:
``anything you can say in terms of quantum states can be translated into
a statement about subsystem correlations, i.e., about joint distributions." Note that while these correlations are relational properties of the two individual systems (i.e., qubits), they are taken to be intrinsic properties of the joint system composed of the two qubits, as the possession of the correlation properties by the joint system does not, we may suppose, depend on any relation the joint system has to any further (possibly spatially separated) systems.

It is tempting to think that because of
this SSC theorem and because of the fact that Bell has shown that
a quantum state is not a prescription of local realistic values of
physical quantities,
that we can take a quantum state to
be nothing but the encapsulation of all the quantum correlations
present in the quantum system.
Indeed, the SSC theorem was used by  \citet{merminithaca1,merminithaca2,merminithaca3}
to argue for the idea that correlations are physically real and give a local realistic underpinning of quantum mechanics, whereas
values of quantities do not 
(although by now he has set these ideas aside\footnote{N.D. Mermin, personal communication.}).
Without wanting to claim that Mermin is committed to the issue we address next,
we explore if correlations between subsystems of an individual isolated composed system,
although determining the state of the total composite system,
can also be be considered to be real objective local properties of such a system.
That is, can one consider quantum correlations to be properties obeying the doctrine of local realism that
somehow (pre-)exist in the quantum state? Are correlations somehow atomic local realistic building blocks of the (quantum) world?

In the next two sections we will show that none of these questions
can be answered in the positive.  The supposition we made above that the correlations of the composite two-qubit system 
in the singlet state are intrinsic to this two-qubit system and thus do not depend on the relation this system has to a further possibly spatially separated system turns out to be false. Therefore, and arguably surprisingly, they cannot be regarded to be local realistic properties. We will be formal and rigorous and
follow the road paved by Bell for us, but enlarge it
to not only include values of quantities but also correlations.

\section{A Bell-type inequality for correlations between correlations}\label{bell}
\noindent Consider two spatially separated parties $I$ and $II$ which each have a
bi-partite system. Furthermore,
assume that each party determines the correlations of the
bi-partite system at his side. By correlations we here mean the conditional 
joint probability distributions $P^I(ab|AB)$ and $P^{II}(cd|CD)$,
where $A$ and $B$ are physical quantities each associated
to one of the subsystems in the bi-partite
system that party $I$ has, and where $a$ and $b$ denote the possible
values these quantities can obtain. The same holds for quantities
$C$, $D$ and possible outcomes $c$, $d$
but then for party $II$.
We now assume local realism for these correlations
in the following well-known way. The correlations party $I$ finds are
determined by some hidden variable $\lambda \in \Lambda$ (with distribution
 $\rho(\lambda)$ and hidden-variable space $\Lambda$).
 The same of course holds for $II$.
We next look at the relationship correlations within each of $I$ and $II$ have with the correlations possessed by the total system composed of $I$ and $II$, i.e., we consider correlations between the correlations $P^I(ab|AB)$ and $P^{II}(cd|CD)$.
Because of locality the correlations one party will obtain
 are for a given $\lambda$ statistically independent of the correlations
that the other party will find.
 Under these assumptions the joint
probability distribution that encodes the correlations between the correlations
 factorises, i.e., 
 \begin{align}
 P(ab,cd|AB,CD,\lambda)=P^I(ab|AB,\lambda)\,
P^{II}(cd|CD,\lambda),
\end{align}
 so as to give\footnote{For clarity we group the outcomes and observables for both parties together in the  probability $P(ab,cd|AB,CD)$, etc.}
 \begin{align} \label{Factorisability}
P(ab,cd|AB,CD)= \int_{\Lambda}P^I(ab|AB,\lambda)\,
P^{II}(cd|CD,\lambda)\, \mu(\lambda)\,d\lambda.
 \end{align}
Here we assume a so-called stochastic hidden-variable model where the
hidden-variables $\lambda$ determine only the correlations 
$P^I(ab|AB,\lambda)$, $P^{II}(cd|CD,\lambda)$, and not the
values $a$,$b$,$c$,$d$ of the quantities $A$,$B$,$C$,$D$. Neither does it determine the probabilities for these values to be found. Thus the correlations $P^I(ab|AB,\lambda)$ and $P^{II}(cd|CD,\lambda)$
itself need not factorise (if they would factorise one obtains the familiar situation of local realism for values of quantities). 

Suppose now that we deal with dichotomic quantities $A,B,C,D$
with possible outcomes  $a,b,c,d \in \{-1,1\}$.
We denote the mean value of the product of the tuples $AB$ and $CD$  by 
\begin{align}
\av{AB,CD}_\textrm{lc}= \sum_{a,b,c,d } abcd \,P(ab,cd|AB,CD)
\end{align}
 where the subscript `lc' stands for `local correlations', indicating that the joint distribution $P(ab,cd|AB,CD)$ is given by (\ref{Factorisability}) that encodes the idea of local realism for correlations.

Then because of the factorisability in (\ref{Factorisability}) we get the following
Bell-type inequality, in familiar CHSH form,
\begin{align}\label{bellineq}
|\av{AB,CD}_\textrm{lc}+
     \av{AB,(CD)'}_\textrm{lc}+
     \av{(AB)',CD}_\textrm{lc}-\av{(AB)',(CD)'}_\textrm{lc}
   |\leq 2.
\end{align}
Here $AB,\,(AB)'$ denote two sets of quantities that give rise to
two different joint probabilities (i.e., correlations) at party
$I$. Similarly for the set $CD$ and $(CD)'$ at party $II$. 

This is a Bell-type inequality which relates the correlations between $I$ and $II$ to the correlations within each of $I$ and $II$. In the next section we show that quantum mechanics violates it by  a suitably chosen entangled state of the composite system comprising both $I$ and  $II$. Despite the resemblance between our inequality and the usual CHSH inequality,
they are fundamentally different because the latter is in terms of correlations between values of subsystem quantities whereas the former is in terms of correlations between correlations and does not assume anything about the values of subsystem quantities.

\section{Quantum correlations are not local elements of reality}
\label{qmbell}
Consider a four-partite quantum system $\mathfrak{O}$ that consists of two
pairs of qubits (spin-$\frac{1}{2}$ particles) where parties $I$ and $II$ each receive a single pair.
In this section we will provide an entangled state of the four-qubit
quantum system $\mathfrak{O}$ and specific sets of two-qubit observables each performed
by parties $I$ and $II$ (that each have a pair of  qubits)
with the following property: These observables give rise to
correlations in the
two-qubit subsystems, which violate the Bell-type inequality of the previous section (see (\ref{bellineq})) in its quantum mechanical version, which is
\begin{align}\label{bellineqqm}
|\av{\mathfrak{B}}_\textrm{qm}|= |\av{\ara{A}\ara{B},\ara{C}\ara{D}}_\textrm{qm}&+
     \av{\ara{A}\ara{B},(\ara{C}\ara{D})'}_\textrm{qm}\nn\\&+
     \av{(\ara{A}\ara{B})',\ara{C}\ara{D}}_\textrm{qm}-\av{(\ara{A}\ara{B})',(\ara{C}\ara{D})'}_\textrm{qm}
   |\leq 2.
\end{align}
where $\mathfrak{B}$ is the corresponding Bell polynomial $\mathfrak{B}= AB\otimes CD +AB\otimes (CD)'+(AB)'\otimes CD -(AB)'\otimes (CD)'$.

Now that we have the quantum mechanical version of the
Bell-type inequality in terms of correlations between correlations, we will
provide an example of a violation of it.
Consider two sets of two dichotomic observables
represented by self-adjoint operators $X,~ X'$ and
$Y,~Y'$ for party $I$ and $II$ respectively. Each
observable acts on the subspace
$\H=\mathbb{C}^2\otimes\mathbb{C}^2$ of the two-qubit system held
by the respective party $I$ or $II$. These observables are chosen
to be dichotomous, i.e. to have possible outcomes in $\{-1,1\}$.
They are furthermore chosen to be
sums of projection
operators and thus give rise to unique joint probability
distributions on the set of quantum states. Measuring these
observables thus implies determining some quantum correlations.
For these observables  $X$, $X'$, $Y$ and $Y'$
the Bell operator $\mathfrak{B}$  on
$\H=\mathbb{C}^2\otimes\mathbb{C}^2\otimes\mathbb{C}^2\otimes\mathbb{C}^2$
becomes
$\mathfrak{B}=X\otimes Y+X\otimes Y'+X'\otimes Y-
X'\otimes Y'$.
The observables have the following form. Firstly, \begin{align}\label{observable1}
X=\hat{P}_{\psi^+}+\hat{P}_{\phi^+}-\hat{P}_{\psi^-}-\hat{P}_{\phi^-},
\end{align} which is a sum of four projections onto the Bell basis
$\ket{\psi^\pm}=1/\sqrt{2} (\ket{01} \pm
\ket{10})$ and $\ket{\phi^\pm}=1/\sqrt{2}
(\ket{00} \pm\ket{11})$. 
Secondly, \begin{align}
X'=\hat{P}_{\ket{00}}+\hat{P}_{\ket{01}}-\hat{P}_{\ket{10}}
-\hat{P}_{\ket{11}}, \end{align} where the projections
are onto the product states
$\ket{00},~\ket{01},~\ket{10},~\ket{11}$.
And finally, \begin{align} Y=
\hat{P}_{\ket{00}}+\hat{P}_{\ket{b+}}
-\hat{P}_{\ket{b-}}-\hat{P}_{\ket{11}}, \end{align}
\begin{align}\label{observable4}
Y'=\hat{P}_{\ket{11}}+\hat{P}_{\ket{b'+}}
-\hat{P}_{\ket{b'-}}-\hat{P}_{\ket{00}}, \end{align}
\noindent
where we have $\ket{b
\pm}=C^\pm(\ket{01} + (1\pm \sqrt{2})
\ket{10})$ and $\ket{b'
\pm}=C^\mp(\ket{01} +(-1\pm \sqrt{2})
\ket{10})$, with normalization coefficients
$C^\pm=(4\pm2\sqrt{2})^{-1/2}$
\footnote{
This particular choice of observables $X$, $X'$, $Y$, $Y'$ on
$\H=\mathbb{C}^2\otimes\mathbb{C}^2$ is motivated by a
well-known choice of single particle observables that gives a maximum violation of the original CHSH inequality when using the state $\ket{\phi^+}=1/\sqrt{2}
(\ket{00} +\ket{11})$. This choice is
$X=-\ara{\sigma_x},~X'=\ara{\sigma_z},~
Y=1/\sqrt{2}(\ara{-\sigma_z} +\ara{\sigma_x}),~
Y'=1/\sqrt{2}(\ara{\sigma_z} +\ara{\sigma_x})$
all on $\H=\mathbb{C}^2$.
The analogy can be seen by noting that in this latter choice the (unnormalized) eigenvectors of
$X$ are $\ket{0}+\ket{1},~\ket{0}-\ket{1}$,
of $X'$ they are $\ket{0},~ \ket{1}$,
of $Y$ they are $\ket{0} +(1+\sqrt{2})\ket{1},~
\ket{0} +(1-\sqrt{2})\ket{1}$
and finally of $Y'$ they are $\ket{0} +(-1+\sqrt{2})\ket{1},~
\ket{0} +(-1-\sqrt{2})\ket{1}$.
}.

Consider now the four particle entangled pure
state \begin{align}\label{state} \ket{\Psi}=\frac{1}{\sqrt{2}}
(\ket{0101}-\ket{1010}).
\end{align} The mean value of the Bell operator $\mathfrak{B}$ for
the above choice of $X$, $X'$, $Y$, $Y'$
in the state $\ket{\Psi}$ is equal to
\begin{align}
|\av{\mathfrak{B}}_\textrm{qm}|=|\,{\rm Tr} [\mathfrak{B}\ket{\Psi}\bra{\Psi}]\, |=2\sqrt{2}.
\end{align}
\noindent This gives us a violation of the Bell-type 
inequality (\ref{bellineqqm}) by a
factor of $\sqrt{2}$. This violation proves that
quantum correlations cannot be considered to be local elements of reality that pertain to a composite quantum system.

The violation is the maximum value because Tsirelson's inequality
\cite{cirelson} (i.e., $|\av{\mathfrak{B}}_\textrm{qm}|=|{\rm Tr} [\mathfrak{B}\rho] |\leq
2\sqrt{2}$ for all quantum states $\rho$) must hold for all
dichotomic observables $X,X',Y,Y'$ on
$\H=\mathbb{C}^2\otimes \mathbb{C}^2$ (possible outcomes in $\{-1,1\}$). One can easily see this
because for $X,X',Y,Y'$
we have that
$X^2= X'^2=Y^2=Y'^2=\1$,
 and it thus follows that the proof of \citet{landau} of  Tsirelson's inequality
 goes through.

\section{Entanglement is not ontologically robust}\label{ontol}

Entanglement is the fact that certain quantum states of a composite system exist
that are not convex sums of product states (cf. section \ref{sepentangintro}).\forget{\footnote{For completeness, a bi-partite state $W$
is \emph{entangled} iff $W \neq \sum_i p_i W_{i}^I\otimes W^{II}_i$,
where $W_{i}^I$ and $W^{II}_i$ are arbitrary states of the two subsystems
 and $\forall ~p_i>0$, $\sum_i p_i =1$.
 A state that is not entangled is called \emph{separable}.
 For the definition of
multi-partite entanglement see \cite{seevuff}.
}.
It gives rise to a special kind of quantum correlations, called
 non-classical correlations, which in the pure state case are able to violate a Bell-type inequality
(for mixed states it is not always the case that entanglement
implies violation of a Bell-type inequality \cite{werner}).  Furthermore, these correlations can also be used to
perform exceptional quantum information and computation tasks.}
 The SSC theorem of section \ref{qmcorr} tells us that
 quantum states, and thus also their entanglement, can be completely characterized by
the quantum correlations that it gives rise to.  Therefore the result of the previous section also applies to entanglement.
Then, if one considers the quantum state description to be complete,
entanglement cannot be viewed as ontologically robust
in the sense of being an objective local realistic property pertaining to some composite system.
  If one would do so nevertheless, one can construct a composite system
 that contains as a subsystem
 the entanglement (i.e. the entangled system) in question and
 which would allow for a violation of the
 Bell-type inequality (\ref{bellineqqm}). This implies
 (contra the assumption) that the entanglement cannot be regarded
 in a local realistic way, which we take to be
 a necessary condition for ontological robustness.

It is possible that one thinks that the requirement of local
 realism is too strong a requirement for ontological robustness.
 However, that one cannot think of entanglement as a property
which has some ontological robustness
 can already be seen using the following weaker requirement:
 anything which is ontologically robust can, without interaction, not
be mixed away, nor swapped to another object, nor
flowed irretrievably away into some environment.
 Precisely these features are possible in the case of entanglement
 and thus even the weaker requirement for ontological robustness does not hold.

 These features show up at the level of quantum states when considering a quantum system in
conjunction with other quantum systems: entanglement can (i) be created in previously
 non-interacting particles using swapping,
(ii) be mixed away and (iii) flow into some environment upon mixing, all without interaction
 between the subsystems in question.
It is this latter point, the fact that no interaction
 is necessary in these processes, that one cannot think of entanglement
 as ontologically robust.


To see that the above weaker requirement
   for ontological robustness of entanglement does
 not hold consider the following examples
 of the three above mentioned features.

  (i) Consider two maximally entangled pairs (e.g., two singlets) that are created
 at spacelike separation, where from each pair a particle is emitted
 such that these two meet and the other particle of each pair is emitted such
 that they fly away in opposite directions. Conditional on a suitable joint measurement performed on
 the pair of particles that will meet (a so called Bell-state measurement)
  the state of the remaining two particles, although they have never
  previously interacted nor are entangled initially, will be `thrown'
  into a maximally entangled state. The entanglement is swapped \cite{swap}.

 (ii) Equally mixing the two maximally entangled Bell states
$\ket{\psi^\pm}$
 gives the separable mixed state
 \begin{align} \label{Mix}\rho=
(P_{\ket{\psi^+}} +P_{\ket{\psi^-}}
)/2 =
(P_{\ket{01}} +P_{\ket{10}})/2.
\end{align}
   The entanglement is thus mixed away, without any necessary interaction between
   the subsystems.

  (iii) Equally mixing the following two states of three spin 1/2 particles,
where particles $2$ and $3$ are entangled in both states,
 \begin{align}\ket{\psi}=\ket{0}\otimes\ket{\psi^{-}},~~~~
  \ket{\phi}=\ket{1}\otimes\ket{\psi^{+}},
  \end{align}
gives the state
\begin{align} \rho= (\ket{\psi}\bra{\psi}+\ket{\phi}\bra{\phi})/2=(P_{0} \otimes P_{\ket{\psi^{-}}}+
 P_{1}\otimes P_{\ket{\psi^{+}}})/2. \end{align}
This three-particle state is two-particle entangled although it has no
two-particle subsystem whose (reduced) state is entangled (cf. section \ref{partsep}).
The bi-partite entanglement has thus irretrievably flowed into the three particle state,
again without any necessary interaction between the subsystems.

Another argument against the ontological robustness of entanglement -- not further studied here -- is that it is not relativistically invariant because  a state that is entangled in some inertial frame becomes less entangled (measured using the so-called logarithmic negativity) if the observers are relatively accelerated, and in the limit of infinite acceleration it can even vanish \cite{fuentes}.  

Does this lack of ontological robustness of entanglement question
the widespread idea of entanglement as a
resource for quantum information and computation tasks?
We think it does not. Quantum information theory is precisely a theory devised to deal with
the surprising characteristics of entanglement such as the
ontologically non-robustness here advocated
(and many other features, such as for example teleportation).
Entanglement is taken to a be specific type of correlation
that can be used as a resource for encoding and manipulating (quantum)
bits of information. For that purpose the ontological status of the information or of that which
bears the information does not matter. 
The only thing that matters is that one can manipulate systems that behave in a specific quantum-like way (of which it is said to be due to entanglement) according to certain information theoretic rules. 
 Whether the systems indeed contain entanglement in some ontologically robust sense is irrelevant.

To conclude this section we should mention that \citet{timpson} do argue for the ontological
robustness of entanglement in the mixing case (ii) above
by introducing ontological relevance
to the preparation procedure of a quantum state,
which supposedly can always be captured in the full quantum mechanical description.
 They introduce the distinction between
 `improper' and `proper separability', which is analogous
 to the well known distinction between proper and improper mixtures,
 to argue that one can retain
 an ontologically robust notion of entanglement.
They thus call the separable mixed state (\ref{Mix}) improperly separable
because the entanglement in the mixture becomes hidden on mixing
(i.e, it disappears), although there are
some extra facts of the matter
that tell that the separable state is in fact
composed out of an ensemble of entangled states.
Because of the existence of these extra facts of the matter ``there need be no mystery at the conceptual level
over the disappearance" \cite[sec.~2]{timpson}. We agree, and the introduced distinction between proper and improper
separability indeed shows this. However, we are not convinced
that their analysis of the improperly separable states indicates ontological robustness of entanglement.

The issue at stake hinges on what one takes to be necessary and/or sufficient conditions for ontological robustness. Consider a state that is improperly separable and which thus consists of a mixture of entangled states. If one would take the mere existence of the extra matters of fact (that tell that the state is improperly mixed) to be sufficient for ontological robustness of the entanglement, then the whole thing becomes circular, since that existence is guaranteed by definition in all states that are improperly separable. Other conditions are needed. 
Although Timpson and Brown do not explicitly give necessary or sufficient conditions for ontological robustness, they 
do argue that using the extra facts of the matter an observer is able to perform a place selection procedure that would allow the
ensemble to be separated out into the original statistically distinct sub-ensembles
[i.e., into the entangled states]. We take it that the existence of such a selection procedure is thus posited as a sufficient condition for ontological robustness: ``all that is required is access to these further facts" \cite[sec.~1]{timpson}.

We agree on this point, but we believe that it is very well
conceivable that according to quantum mechanics
we do not in principle always have access to these extra
facts. Perhaps the interactions between the object systems involved in
the preparation procedure and the environment are such that the
observer cannot become correlated to both
the extra facts and the objects states
in the right way for the facts to be accessible,
or, alternatively, the interactions could be such that no classical record of
the extra facts could possibly be left in the environment.
To put it differently, although we agree that in the case of improper separability
one can uphold an ignorance interpretation of the state in question
and that furthermore the ignorance is in principle about some extra
facts of the matter, we do not agree that it is certain that the ignorance about these extra facts
of the matter can be removed by the observer
in accordance with the dynamics of quantum mechanics in all
 conceivable preparation procedures. This issue thus awaits a (dis)proof of principle\footnote{Timpson (private communication) has informed us that  their argument is supposed to use the following clause (not mentioned in their original paper): ``A separating place selection procedure is in principle possible: \emph{given} access to the facts, the procedure could be performed.'' We do not question the latter. However, the issue at stake is if one can meet the conditional: it is by no means clear that quantum mechanics in principle allows one to have access to these facts and thus that such a place selection procedure is in principle possible.}.
 
 The existence of a selection procedure is indeed a sufficient condition for ensuring the ontological robustness of entanglement in improperly separable states. But since it is unclear whether one can indeed meet this condition it seems to be more fruitful to look for necessary conditions. We have proposed four different such conditions for ontological robustness and argued that they can not be met.

\section{Discussion}\label{conclu}
\noindent The Bell-type inequality violation of section \ref{qmbell}
tells us that despite the fact that a quantum state of a composite
system is determined by the correlations between
each of its possible subsystems, one cannot conclude that the quantum state can be given a local realistic account in terms of the correlations it gives rise to.\forget{they are determined (by the quantum state) in a local way.}
Just like
values of quantities correlations cannot be used to build up a world
consisting out of some local realistic structure.
We have that mathematically quantum correlations determine the quantum state,
but ontologically they cannot be considered to be
local realistic building blocks of the world.
Of course, if one wants to build up another world where the building blocks need not be constrained by local realism (e.g., a world consisting of unrestricted primitive intrinsic correlations  that do not supervene on intrinsic properties of the subsystems) then these results are not relevant for this.

A special type of quantum correlation is entanglement. Although entanglement
is taken to be a resource in quantum information theory,
we have argued that it cannot be considered ontologically robust
because it is not an objective local realistic property
and furthermore that without interaction it can be mixed away, swapped to another object,
and flowed irretrievably into some environment. 

The Bell-type inequality argument of section \ref{qmbell} was inspired by the work of
\citet{cabello} and \citet{jordan}
who obtain almost exactly the same conclusion,
although by different arguments.
The argument of Cabello differs the most from ours because
he uses a different conception of what a quantum correlation is. His argument
speaks of types of correlations which are associated with eigenvalues of
product observables. We believe this notion to be less general
than our notion of quantum correlation which only takes
joint probability distributions to be correlations.
  Jordan's argument, in contrast, does in effect use the same notion of quantum
 correlation as we do. He considers mean values of products
 of observables and since these are determined by mean values of
(sums of) products of projection operators he restricts himself to the latter.
Jordan thus uses the same notion as we do because the latter determine all
joint probability distributions.

However, Cabello and Jordan both need perfect
correlations for their argument to work because the state dependent
GHZ- or Hardy argument they use (Cabello uses both, Jordan only the latter
argument) need such strong correlations.
Our Bell-type inequality argument does not rely on this
specific type of correlation because non-perfect statistical correlations already suffice
to violate the Bell-type inequality here presented.
We therefore believe our argument
has an advantage over the one used by Cabello and Jordan,
because it is more amenable to experimental implementation.

     In fact, the Bell-type inequality argument here presented can be readily
implemented using current experimental technology.
Indeed, it is already possible to create fully four-particle-entangled states
\cite{sackett,zhao}
and measurement of the four observables $X$, $X'$, $Y$, $Y'$ of
(\ref{observable1})-(\ref{observable4}) seems not to be
problematic since they are sums of ordinary projections.
Furthermore, as said before, there is no need to produce perfect correlations;
non-perfect statistical ones will suffice. We therefore hope that in
the near future experiments testing our argument will be carried
out.

Lastly, returning to the questions stated in the introduction,
in so far as Mermin in his \cite{merminithaca1,merminithaca2,merminithaca3} is committed to
take correlations (as we have defined them here) to be interpreted local realistically
(which we think he is), his tentative interpretation is at odds
with predictions of quantum mechanics and would allow,
in view of the argument given here, for an experimental verdict.

\clearemptydoublepage
\thispagestyle{empty}

\noindent\chapter{Disentangling holism}
\label{chapter_holism}
\forget{\section{Abstract}
Motivated by the question what it is that makes quantum mechanics a 
holistic theory (if so), we try to define for general physical theories what 
we mean by `holism'. For this purpose we propose an epistemological criterion to 
decide whether or not a physical theory is holistic, namely: a physical theory 
is holistic if and only if it is impossible in principle to infer the global 
properties, as assigned in the theory, by local resources available to an agent.  
we propose that these resources include at least all local operations and 
classical communication. This approach is contrasted with the well-known 
approaches to holism in terms of supervenience. The criterion for holism 
proposed here involves a shift in emphasis from 
ontology to epistemology. We apply this epistemological criterion to 
classical physics and Bohmian mechanics as represented on a phase and 
configuration space respectively, and for quantum mechanics (in the orthodox 
interpretation) using the formalism of general quantum operations as completely 
positive trace non-increasing maps. Furthermore, we provide an interesting 
example from which one can conclude that quantum mechanics is holistic in the 
above mentioned sense, although, perhaps surprisingly, no entanglement is needed.
}
\noindent
This chapter is a slightly adapted version of \citet{seevhol}.

\section{Introduction}
\label{inleiding}

Holism is often taken to be the idea that the whole is more than the sum of its parts.
Because of being too vague, this idea has only served as a guideline
or intuition to various sharper formulations of holism.
Here we shall be concerned with the one relevant to physics, i.e., the doctrine of 
\ara{metaphysical holism}, which is the idea that properties or relations of a  
whole are not determined or cannot be determined by intrinsic properties or relations of the
parts\footnote{This \ara{metaphysical holism} (also called \ara{property holism})
 is to be contrasted  with \ara{explanatory holism} and \ara{meaning holism} \cite{healey91}.
The first is the idea that explanation of 
a certain behavior of an object cannot be given by analyzing the component 
parts of that object. Think of consciousness of which some claim that it cannot be
 fully explained in terms of physical and chemical laws obeyed by the molecules of 
 the brain. The second  is the idea that the meaning of a term cannot be given without regarding 
it within the full context of its possible functioning and usage in a language.}.
  This is taken to be opposed to a claim of supervenience \cite{healey91},
   to reductionism \cite{maudlin98}, to local physicalism \cite{teller86}, and 
   to particularism \cite{teller89}.
   In all these cases a common approach is used to define what metaphysical holism
is: via the notion of \ara{supervenience}\footnote{The notion of supervenience, as used here, is meant to describe 
a particular relationship between properties of a whole and properties of the parts of that whole.
The main intuition behind what particular kind of relationship is meant, is  
captured by the following impossibility claim. It is not possible that two things should
 be identical with respect to their subvenient or subjacent properties (i.e., the lower-level properties),
without also being identical with respect to their  supervening or upper-level properties.   
The first are the properties of the parts, the second are those of the whole. 
The idea is that there can be no relevant difference in the whole without 
a difference in the parts. (\citet{cleland84} uses a different definition in terms of
 modal logic.)}.
\label{supervenience}
According to this common approach metaphysical holism is the doctrine
 that some facts, properties, or relations of the whole do not supervene on 
intrinsic properties and relations of the parts, the latter together making up the\newpage\noindent so-called \ara{supervenience basis}. 
As applied to physical theories, quantum mechanics is then taken to be the paradigmatic
example of a holistic theory, since certain composite states
 (i.e., entangled states) do not supervene on subsystem states, a feature not 
 found in classical physical theories.

However, in this chapter we want to critically review the supervenience
approach to holism and propose a new criterion for deciding whether or not a physical theory is holistic. 
The criterion for whether 
or not a theory is holistic proposed here is an \ara{epistemological} one. 
It incorporates the idea that each physical theory 
(possibly supplemented with a property assignment rule via an interpretation) 
has the crucial feature that it tells us how to \ara{actually} infer 
properties of systems and subsystems.

The guiding idea of the approach here suggested, is that some property
of a whole would be holistic if, according to the theory in question, there is no 
way we can find out about it using \ara{only} local means,
i.e., by using only all possible non-holistic resources available to an agent.
In this case, access to the parts would not suffice for inferring the properties of the 
whole, not even via all possible subsystem property determinations that 
can be performed, and consequentially we would have some instantiation of holism, 
called \ara{epistemological holism}.
The set of non-holistic resources is called the \ara{resource basis}. We propose that this basis includes 
at least all local operations and classical communication of 
 the kind the theory in question allows for.
 
The approach suggested here thus focuses on the inference of  properties
instead of on the supervenience of properties. 
It can be viewed as a shift from ontology to 
epistemology\footnote{This difference is similar to the difference
between the two alternative
definitions of determinism. From an \ara{ontological} point of view,
determinism is the existence of a single possible
  future for every possible present.
  Alternatively, from an \ara{epistemological} point of view, it is the 
  possibility in principle of inferring the future from 
  the present.} and also as a shift that takes into account the full 
  potential of physical theories by including 
  what kind of property inferences or measurements are
   possible according to the theory in question. 
    The claim we make is that these two approaches are crucially different and that
 each have their own merits. We show the fruitfulness of the new approach 
 by illustrating 
it in classical physics, Bohmian mechanics and orthodox quantum mechanics. 
 
The structure of this chapter is as follows. First we will present in section \ref{supervenientie-sectie} a short review 
of the supervenience approach to holism. We especially look at the supervenience 
 basis used. To illustrate this approach 
 We consider what it has to say about classical physics and quantum mechanics. 
 Here we rigorously show that in this approach classical physics is non-holistic and furthermore
  that the orthodox interpretation of quantum mechanics is deemed holistic. 
  In the next section 
 (section \ref{criterion}) we will  give a different approach based on an epistemological 
 stance towards property determination within physical theories.
This approach is contrasted with the approach of the previous section and furthermore 
argued to be a very suitable one for addressing holism in physical theories.

In order to show its fruitfulness we will apply the epistemological approach to different physical theories. 
Indeed, in section \ref{QMsection} classical physics and Bohmian mechanics are 
proven not to be epistemologically holistic, whereas the orthodox interpretation 
of quantum mechanics is 
shown to be epistemologically holistic without making appeal to the feature of 
entanglement, a feature that was taken to be absolutely necessary in 
the supervenience approach for any holism to arise 
in the orthodox interpretation of quantum mechanics.
Finally in section \ref{conc} we will recapitulate, and argue this new approach to holism 
to be a fruit of the rise of the new field of quantum information theory.

\section{Supervenience approaches to holism}
\label{supervenientie-sectie}

The idea that holism in physical theories is opposed to supervenience of 
properties of the whole on intrinsic properties or relations
 of the parts, is worked out in detail by \citet{teller86} and by 
 \citet{healey91}, although others have used this idea as
  well, such as \citet{french89}\footnote{
\citet{french89} uses a slightly different approach to holism
 where supervenience is defined in terms of modal logic, following a proposal 
 by \citet{cleland84}. 
 However, for the present purposes, this approach leads essentially
 to the same results and we will not discuss it any further.}\label{voetnootfrech},
  \citet{maudlin98} and \citet{esfeld01}. We will review 
  the first two contributions in this section.

Before discussing the specific way in which part and whole are related,
\citet{healey91} clears the metaphysical ground of what 
it means for a system to be composed out of parts, so that the 
whole supervenience approach can get off the ground. We take this to be unproblematic here and say that a whole 
is composed if it has component parts. Using this notion of composition, 
holism is the claim that the whole has features that cannot be reduced to 
features of its component parts. Both \citet{healey91} and \citet{teller86} 
use the same kind of notion for the reduction relation, namely \ara{supervenience}. However, whereas Teller only
 speaks about relations of the whole and non-relational properties of the parts,
  Healey uses a broader view on what features of the whole should supervene on 
  what features of the parts. Because of its generality we take essentially 
  Healey's definition to be paradigmatic for the supervenience approach to 
holism\footnote{The exact definition by \citet[p.402]{healey91} is as follows.
``\ara{Pure physical holism}:
There is some set of physical objects from a domain $D$ subject
 only to processes of type $P$, not all of whose qualitative, intrinsic 
 physical properties and relations are supervenient upon the qualitative,
 intrinsic physical properties and relations of their basic physical parts 
 (relative to $D$ and $P$)''. The definition by \citet{teller86} is a
 restriction of this definition to solely relations of 
 the whole and intrinsic non-relational properties of the parts.
}\label{healey'sdef}. In this approach, holism in physical theories means 
that there are physical properties or relations of the whole  that are not
 supervenient on the intrinsic physical properties and relations of the 
 component parts. An essential feature of this approach is that the 
\ara{supervenience basis}, i.e., the properties or relations on which the whole 
 may or may not supervene,
are only the \ara{intrinsic} ones, which are those which the parts have at the time in question 
in and out of themselves, regardless of any other individuals.

We see that there are three different aspects involved in this approach.
The \ara{first} has to do with the metaphysical, or ontological effort of 
clarifying what it means that a whole is composed out of parts. We  took this to be 
unproblematic. The \ara{second} aspect gives us the type of dependence the whole 
should have to the parts in order to be able to speak of holism. 
This was taken to be supervenience.
\ara{Thirdly}, and very importantly for the rest of this chapter, the supervenience basis 
needs to be specified because  the supervenience criterion is 
 relativized to this basis. \citet[p.401]{healey91} takes this basis to be ``just the qualitative, intrinsic
properties and relations of the parts, i.e., the properties and relations that
these bear in and out of themselves, without regard to any other objects, and
irrespective of any further consequences of their bearing these properties
for the properties of any wholes they might compose.'' Similarly \citet[p.72]{teller86}
uses ``properties \ara{internal} to a thing, properties which a thing has independently
of the existence or state of other objects.''

Although the choice of supervenience basis is open to debate because it is hard to specify precisely, the idea is that 
we should not add global properties or relations 
to this basis. It is supposed to contain
only what we intuitively think to be \ara{non-holistic}.
However, as we aim to show in the next sections,  
an alternative basis exists to which a criterion for holism 
can be relativized. This alternative basis, the \ara{resource basis} as we call it,
 arises when one adopts a different view when considering physical theories.
 For such theories not only present us  an
 ontological picture of the world (although possibly only after an interpretation is provided),
  but also they present specific forms of property assignment and property
 determination. The idea then is that these latter processes, such as measurement or 
 classical communication, have 
 intuitively clear \ara{non-holistic features}, 
   which allow for an epistemological analysis 
   of whether or not a whole can be considered to be holistic or not.
   
However, before presenting this new approach, we discuss how the 
supervenience approach treats classical physics
and quantum mechanics (in the orthodox interpretation).  In treating these two theories we will first present 
some general aspects related to the structure of properties these theories allow 
for, since they are also needed in future sections.

\subsection{Classical physics in the supervenience approach}
\label{classical_sub}

Classical physics assigns two kinds of properties to a system.
State independent or fixed properties that remain unchanged (such as mass and charge)
 and dynamical properties associated with quantities called dynamical
  variables (such as position and momentum) \cite{healey91}. It is the
  latter we are concerned with in order to address holism in a theory
  since these are subject to the dynamical laws of the theory.
Thus in order to ask whether or not classical physics is holistic 
we need to specify how parts and wholes get assigned the dynamical properties in the
 theory\footnote{This presentation of the structure of properties in classical physics was inspired by \citet{isham95} although he gave 
no account of how the properties of a composite classical system are related to the properties of its subsystems.}. This \ara{ontological 
issue} is unproblematic in classical physics, for it views objects
as bearers of determinate properties (both fixed and dynamical ones). 
The \ara{epistemological issue} of how to gain knowledge of 
these properties is treated via the idea of \ara{measurement}. A measurement is any physical 
operation by which the value of a physical quantity can be inferred. 
Measurement reveals this value because it is assumed that the system has 
the property that the quantity in question has that value at the time of measurement. 
In classical physics there is no fundamental difference between 
measurement and any other
physical process. \citet[p.57]{isham95} puts it as follows: ``Properties are intrinsically 
attached to the object as it exists in the world, and measurement is 
nothing more than a particular type of physical interaction designed 
to display the value of a specific quantity.'' The bridge between ontology 
and epistemology, i.e., between property assignment (for any properties to 
exist at all (in the theory)) and property inference (to gain knowledge about them), 
is an easy and unproblematic one called measurement.

The specific way the dynamical properties of an object are encoded in the formalism of classical 
physics is in a state space $\Omega$ of physical states
 $x$ of a system. This is a phase space  where at each time a 
unique state $x$ can be assigned to the system. Systems or ensembles  can be 
 described by \ara{pure states} which are single points $x$ in $\Omega$ 
 or by  \ara{mixed states} which are unique convex combinations 
 of the pure states.
The set of dynamical properties determines the position of the system in the phase
 space $\Omega$ and conversely the dynamical properties of the system can
 be directly determined from the
 coordinates of the point in phase space. Thus, a \ara{one-to-one
 correspondence} exists between systems and their dynamical properties on the one hand, and 
 the mathematical representation in terms of points in phase space on the other.	 
 Furthermore, with observation of properties being unproblematic, the state corresponds 
 uniquely to the outcomes of the (ideal) measurements that 
 can be performed on the system.  The specific property 
 assignment rule for dynamical properties that captures the 
 above is the following.

A physical quantity $\mathfrak{A}$ is represented by a function
 $A:~\Omega \to  \mathbb{R}$
 such that $A(x)$ is the value $A$ possesses when the state is $x$. 
  To the property that the value of $A$ lies in the real-valued interval $\Delta$ there is associated
   a Borel-measurable subset
   \begin{align}\label{1system}
    \Omega_{A\in\Delta}= A^{-1} \{\Delta\}=    \{x\in \Omega|A(x) \in \Delta\},
\end{align}
of states in $\Omega$ for which the proposition that the system 
has this property is true. Thus dynamical properties are associated with
\ara{subsets} of the space of states $\Omega$, and we have the 
one-to-one correspondence mentioned above between properties and points in the state space  now as follows: 
$ A(x)\in\Delta \Leftrightarrow x\in\Omega_{A\in\Delta} $.
Furthermore, the logical structure of the propositions about the dynamical properties 
of the system is identified with the \ara{Boolean $\sigma$-algebra} $\mathcal{B}$ of subsets of the space of
states $\Omega$. This encodes the normal logical way  (i.e., Boolean logic) of dealing with 
propositions about properties\footnote{The relation of
\ara{conjunction} of propositions corresponds to the 
 set-theoretical \ara{intersection} (of subsets of the state space),
  that of  \ara{entailment} between propositions to the 
 set-theoretical \ara{inclusion}, that of \ara{negation} of a proposition
  to the set-theoretical \ara{complement} and finally that of \ara{disjunction} of propositions
  corresponds to the set-theoretical \ara{union}. 
  In classical physics the (countable) logic of propositions
     about properties is thus isomorphic
    to a Boolean $\sigma$-algebra of subsets of the state space.
  }.

In order to address holism we need to be able to speak about 
properties of composite systems in terms of properties of the subsystems.
The first we will call \ara{global} properties, the second \ara{local}
 properties\footnote{Note that `\ara{local}' has here nothing to do with the issue of locality or spatial separation.
 It is taken to be opposed to global, i.e., restricted to a subsystem.}\label{local}.
It is a crucial and almost defining feature of the state 
space of classical physics that the local dynamical properties \ara{suffice} 
for inferring \ara{all} global dynamical properties. 
This is formalized as follows\footnote{We have not been able to find elsewhere a formal treatment of how the properties of a composite classical system are related to the properties of its subsystems. Therefore we give such a formal treatment here.}.
Consider the simplest case of a composite system with two subsystems (labeled $1$ and $2$). 
Let the tuple $<\Omega_{12},\mathcal{B}_{12}>$ characterize 
the state space of the composite system and the Boolean $\sigma$-algebra 
of subsets of that state space. The latter is isomorphic 
to the logic of propositions about the global properties. This tuple is determined 
by the subsystems in the following way. Given the tuples $<\Omega_1,\mathcal{B}_1>$
and $<\Omega_2,\mathcal{B}_2>$ that characterize the subsystem state spaces and 
property structures, $\Omega_{12}$ is the Cartesian product space of $\Omega_{1}$ and  $\Omega_{2}$,
i.e.,
\begin{align} \label{cartesian}
\Omega_{12}=\Omega_1\times\Omega_2,
\end{align}
 and furthermore,
\begin{align}\label{algebras}
\mathcal{B}_{12}=\mathcal{A}(\mathcal{B}_1,\mathcal{B}_2),
\end{align}
where $\mathcal{A}(\mathcal{B}_1,\mathcal{B}_2)$ is the 
smallest $\sigma$-algebra generated by $\sigma$-algebras that contain
 Cartesian products as elements. This algebra is defined by the following three
properties \cite{halmos88}: (i) if $\mathcal{A}_1\in\mathcal{B}_1$, $\mathcal{A}_2\in\mathcal{B}_2$ then
$\mathcal{A}_1\times\mathcal{A}_2 \in \mathcal{A}(\mathcal{B}_1,\mathcal{B}_2)$,
 (ii) it is closed under countable conjunction, disjunction and taking differences, 
(iii) it is the smallest one generated in this way. The $\sigma$-algebra
 $\mathcal{B}_{12}$ thus contains by definition all sets that can be written as a countable conjunction
 of Cartesian product sets such as
  $\Lambda_{1}\times\Lambda_{2}\subset\Omega_{12}$ 
  (with $\Lambda_{1}\subset\Omega_{1}$, $\Lambda_{2}\subset\Omega_{2}$), also called \ara{rectangles}.
 
The above means that the Boolean $\sigma$-algebra of the properties of the composite system 
is in fact the product algebra of the subsystem algebras.
Thus propositions about global properties (e.g.,  global quantity $B$ 
having a certain value) 
 can be written as disjunctions of 
propositions which are conjunctions of propositions about local properties alone
(e.g., subsystem quantities $A_1$ and $A_2$ having certain values). In other words,
the truth value of all propositions about $B$ can be determined from the truth value 
of disjunctions of propositions about properties concerning $A_1$ and $A_2$ respectively. The first and the latter thus have the same extension.

On the phase space $\Omega_{12}$ all this gives rise to the following structure.
To the property that the value of $B$ of a composite 
system lies in $\Delta$ there is associated a Borel-measurable subset of $\Omega_{12}$, for which the proposition that the system has this property
  is true:  
  \begin{align}\label{samen}
    \{(x_{1}, x_{2})\in \Omega_{12}
|~B(x_{1},x_{2})\in \Delta\}\in\mathcal{B}_{12},
  \end{align} 
 where $(x_{1},x_{2})$ are the pure states (i.e., points) in the phase space of the composite system
  and $x_1$ and $x_2$ are the subsystem states that each lie in the state space $\Omega_1$ or $\Omega_2$ of the respective subsystem.
  The important thing to note is that this subset lies in the product algebra $\mathcal{B}_{12}$
  and therefore is determined by the subsystem algebras $\mathcal{B}_{1}$ 
  and $\mathcal{B}_{2}$ via the relation in (\ref{algebras}).
  
From the above we conclude, 
and so is concluded in the supervenience approaches mentioned 
in the introduction of section \ref {inleiding}, although on other non-formal grounds,
that classical physics is not holistic. For the global properties supervene 
on the local ones because the Boolean algebra structure of the global properties is 
determined by the Boolean algebra structures of the local ones.
Thus all quantities pertaining to the global properties 
defined on the composite phase space such as $B(x_{1},x_{2})$ 
 are supervening quantities. 

For concreteness consider two examples of such supervening quantities $B(x_{1},x_{2})$
of a composite system. The first is $q= \parallel\vec{q_1}-\vec{q_2}\parallel$ which gives us the 
global property of a system that specifies the distance 
between two subsystems.
The second is $\vec{F}= -\bm{\nabla} V(\parallel \vec{q_1}-\vec{q_2}\parallel)$\forget{ 
$\overrightarrow{F}= -\vec{\nabla} V(q)$ } 
which gives us the property of a system that indicates how strong the force is 
between its subsystems arising from the potential $V$. This could for example 
be the potential 
$\frac{m_1 m_2G}{\parallel\vec{q_1}-\vec{q_2}\parallel}$ \forget{$m_1 m_2G /q$} for  the Newtonian gravity force.
Although both examples are highly non-local and could involve action at a distance, 
no holism is involved since the global
properties supervene on the local ones. As \citet[p.76]{teller86} puts it: 
``Neither action at a distance nor distant spatial separation
threaten to enter the picture to spoil the idea of the world
working as a giant mechanism, understandable in terms of the individual parts.''

Some words about the issue of whether spatial relations are to be considered
  holistic, are in order here.
Although the spatial relation of relative distance of the whole
indicates the way in which the parts are related with respect to position, 
whereby it is not the case that each of the parts has a position 
independent of the other one, it is here nevertheless not regarded a holistic property
since it is supervening on spatial position. We have seen that the distance $q$ between
two systems is treated supervenient on the systems having positions $\vec{q_1}$ 
and $\vec{q_2}$ in the sense expressed by equation (\ref{samen}).
 However, the argumentation given here 
requires an \ara{absolutist} account of space  so that position can be regarded as
 an intrinsic property of a system.
But one can deny this and adopt a \ara{relational} account of space 
and then spatial relations become monadic and positions become derivative,
which has the consequence that one has to incorporate spatial relations 
in the supervenience basis\footnote{A more subtle example than the relative distance between two points 
would be the question whether or not the relative angle between two 
directions at different points in space is a supervening property, 
i.e., whether or not the relative angle is to be considered 
  holistic or not. This depends on
  whether or not one can consider local orientations
 as properties that are to be included in the supervenience basis.\label{angle}}.

On an absolutist account of space the spatial relation of relative distance 
  between the parts of a whole is shown to be supervenient upon local properties,
   and it is thus not to be included in the supervenience basis\footnote{\citet{teller87} for example takes spatial relations to be supervening on 
 intrinsic physical properties since for him the latter include spatiotemporal properties.}.
A relationist account, however, does include the spatial relations in the supervenience
 basis. The reason is that on this account they are to be regarded as 
\ara{intrinsically} relational, and therefore non-supervening on the subsystem properties. 
 \citet{cleland84} and \citet{french89} for example argue spatial relations
 to be non-supervening relations.  
 Furthermore, some hold that all other intrinsic relations can be regarded to be supervenient upon these.
The intuition is that wholes seem to be built out of their parts
 if arranged in the right spatial relations, and these spatial relations are taken 
 to be in some sense \ara{monadic} and therefore not holistic\footnote{\citet[p.409]{healey91} phrases this as follows: ``Spatial relations are of special significance because 
they seem to yield the only clear example of qualitative, intrinsic
relations required in the supervenience basis in addition to the
 qualitative intrinsic properties of the relata. Other intrinsic
  relations supervene on spatial relations.''}.

Thus we see that issues depend on what view one has about the 
nature of space (or space-time).
 Here we will not argue for any position,
but merely note that if one takes an absolutist stance towards space
 so that bodies are considered to have a particular position, then spatial relations can be considered to be supervening 
on the positions of the relata in the manner indicated by the decomposition
of (\ref{samen}).  This discussion about whether spatial relations 
are to be regarded as properties that should be included in the supervenience basis 
clearly indicates that the supervenience criterion must be relativized to the supervenience basis.
As we will see later on this is analogous to the fact that the 
epistemological criterion proposed here  must be relativized to the resource basis.

As a final note in this section, we mention that because of the 
one-to-one correspondence in classical physics 
between physical quantities on the one hand and states on the state space on the other hand,
 and because composite 
states are uniquely determined by subsystem states (as can be seen
 from (\ref{cartesian})), it suffices to consider
the state space of a system to answer the question whether or not 
some theory is holistic. 
The supervenience basis is thus determined by the state space (supplemented with the fixed properties).
However, this is a special case and it contrasts with the quantum mechanical case 
(as will be shown in the next subsection). The supervenience approach should 
take this into account.
Nevertheless, the supervenience approach mostly limits itself
to the quantum mechanical state space in determining whether or not 
quantum mechanics is holistic.
The epistemological
approach to be developed here uses also other relevant features 
of the formalism, such as property determination, and focuses
therefore primarily on the structure of the assigned properties and not on that of the state space.
This will be discussed in the following sections.
 
\subsection{Quantum physics in the supervenience approach}
\label{quantum_sub}

In this section we will first treat some general aspects of the 
quantum mechanical formalism before discussing how the supervenience approach
deals with this theory.

In quantum mechanics, just as in classical physics, systems are assigned two kinds of properties. 
On the one hand, the fixed properties that we find in classical physics 
supplemented with some new ones such as intrinsic spin.
On the other hand, dynamical properties such as components of spin \cite{healey91}.
These dynamical properties are, again just as in classical physics, determined in a 
certain way by values observables have when the system is in a particular state.
 However, the state space and observables are
 represented quite differently from what we have already seen in
 classical physics. In general, a quantum state \ara{does not} correspond
  uniquely to the outcomes of the measurements that can be
 performed on the system. Instead, the system is assigned a specific
 Hilbert space $\mathcal{H}$ as its state space and the physical state of the system is represented by a state vector 
$\left | \, \psi \right \rangle$ in the pure case and a density operator $\rho$ in the mixed case.
Any physical quantity  $\mathfrak{A}$ is represented by an observable or
self-adjoint operator\footnote{For clarity we denote the quantum mechanical operator that correspond to observable $\mathfrak{A}$  by $\hat{A}$ so as to distinguish it from the function $A$ which is used in classical physics to denote the  same observable.} $\hat{A}$.
  Furthermore, the spectrum of $\hat{A}$ is the set of possible values the quantity 
   $\mathfrak{A}$ can have upon measurement.  
    
The pure state $\left | \, \psi \right \rangle$ 
     can be considered to assign a probability distribution 
          $p_i=| \left \langle \,\psi\right |  i\rangle|^2  $ to an orthonormal 
     set of states $\verb|{|\left | \, i \right \rangle\verb|}|$.
     In the case where one of the states is the vector $\left | \, \psi \right \rangle$, it is
      completely concentrated onto this vector. The state $\left | \, \psi \right \rangle$ can thus be regarded as the analogon of a
	     $\delta$-distribution on the classical phase space $\Omega$, 
        as used in statistical physics. However
	     the radical difference is that the pure quantum states
         do not (in general) form an orthonormal set. This implies that the
	     pure state $\left | \, \psi \right \rangle$ will also assign a positive probability to a
        different state  $\left | \, \phi \right \rangle$ if they are non-orthogonal and thus have 
        overlap. 
        This is contrary to the
	     classical  case, where the pure state $\delta(q-q_0, p-p_0)$ concentrated
	      on $(p_0,q_0) \in \Omega$ will always give rise to a probability
        distribution that assigns probability zero to every other pure state, since pure states on $\Omega$ 
        cannot have overlap.
	      Furthermore, the probability that the value of an observable $\hat{B}$ lies 
        in the real interval $X$ when the system is in the quantum state $\rho$ is 
        $Tr\,(\rho P_{\hat{ B},X })$ where $P_{\hat{B},X}$ is the projector associated to 
        the pair $(\hat{B},X)$ by the spectral theorem for self-adjoint operators. 
        This probability is in general not concentrated in $\{0,1\}$ 
         even when $\rho$ is a pure state.  Only in the special case that 
          the state is an eigenstate of the  observable $\hat{B}$ 
           is it concentrated in $\{0,1\}$, and the system is assigned 
           the corresponding eigenvalue with certainty. 
	     From this we see that there is no one-to-one correspondence 
       between values an observable can obtain and states of the quantum system.
       
Because of this failure of a one-to-one correspondence there are \ara{interpretations}
 of quantum mechanics that postulate \ara{different} connections between 
 the state of the system and the dynamical properties it possesses. 
 Whereas in classical physics this was taken to be unproblematic and natural, 
 in quantum mechanics it turns out to be problematic and non-trivial. 
 But a connection must be given in order to ask about any holism, 
 since we have to be able to speak about possessed properties and 
  thus an interpretation that gives us a property assignment rule is necessary.
 Here we will consider the well-known \ara{orthodox interpretation} of 
 quantum mechanics that uses the so called \ara{eigenstate-eigenvalue link} for this connection:
 a physical system has the property that quantity $\mathfrak{A}$ has a particular 
 value if and only if its state is an eigenstate of the operator $\hat{A}$
 corresponding to $\mathfrak{A}$. This value is the eigenvalue associated with
 the particular eigenvector. Furthermore, in the orthodox interpretation 
 measurements are taken to be ideal \ara{von Neumann measurements}\footnote{These ideal von Neumann measurements use a projector valued measure (PVM) which is a set of projectors $\{ \hat{P}_i\}$ such that $\sum \hat{P}_i=\1$.}, whereby upon 
 measurement the system is projected into an eigenstate of the observable 
 being measured and the value found is the eigenvalue corresponding to 
 that particular eigenstate. The probability for this eigenvalue to
  occur is given by the well-known Born rule 
  $\left \langle \, i \right |\rho\left | \, i \right \rangle$, 
  with $\left | \, i \right \rangle$ the eigenstate that is projected upon and $\rho$ the state
   of the system before measurement.
Systems thus have properties \ara{only} if they are in an 
eigenstate of the corresponding observables, i.e., the system either already is or must first be projected into such an eigenstate
by the process of measurement. We thus see that the \ara{epistemological} 
scheme of how we gain knowledge of properties, i.e., the measurement process 
described above, serves also as an \ara{ontological} one defining what properties of a
 system can be regarded to exist at a given time at all.

Let us now go back to the supervenience approach to holism and ask what it says 
about quantum mechanics in the orthodox interpretation stated above.
According to all proponents of this approach mentioned 
in the Introduction quantum mechanics is holistic. The reason for this is supposed
 to be the feature of \ara{entanglement}, a feature absent in classical physics. In order to discuss
 the argument used, let us first recall some aspects
  of entanglement that were already treated in chapter \ref{definitionchapter}.  
 Entanglement is a property of composite quantum systems whereby the state 
 of the system cannot be derived from any combination of the subsystem states.
 It is due to the tensor product structure of a composite 
 Hilbert space and the linear superposition principle of quantum mechanics.
In the simplest case of two subsystems, the precise definition is 
that the composite state $\rho$ cannot be written as a convex sum
 of products of  single particle states, i.e., $\rho \neq \sum_i p_i \rho^1_i\otimes\rho^2_i$, with $p_i\in[0,1]$ and $ \sum_ip_i=1$. 
In the pure case, an entangled state is one that cannot 
be written as a product of single particle states.
Examples include  the so-called \ara{Bell states} $\left | \, \psi^{-} \right \rangle$ and $\left | \, \phi^{-} \right \rangle$
 of a spin-${\frac{1}{2}}$ particle. These states can be written as
 \begin{align}\label{entang}
 \left | \, \psi^{-} \right \rangle=\frac{1}{\sqrt{2}}(\left | \, 01 \right \rangle_z -\left | \, 10 \right \rangle_z),~~~~~~
 \left | \, \phi^{-} \right \rangle=\frac{1}{\sqrt{2}}(\left | \, 00 \right \rangle_z -\left | \, 11 \right \rangle_z),
 \end{align}
 with $\left | \, 0 \right \rangle_z$ and $\left | \, 1 \right \rangle_z$ eigenstates of the spin operator $\hat{S}_z=
\frac{\hbar}{2} \hat{\sigma}_z$, i.e., the spin up and down state in the $z$-direction respectively. These
 Bell states are eigenstates
  for total spin of the composite system given by 
the observable $\hat{S}^2=(\hat{S}_1+\hat{S}_2)^2$ with eigenvalue $0$ and $2\hbar^2$ 
  respectively. 
  
According to the orthodox
interpretation, if the composite system is in one of the states of 
(\ref{entang}), the system possesses one of two global properties for total spin  which are completely different, namely 
eigenvalue $0$ and eigenvalue $2\hbar^2$. The question now is whether 
or not this spin property is holistic, i.e., does it or does it not supervene on subsystem properties? According to 
the supervenience approach it does not and the argument goes 
as follows. 
Since the individual subsystems have the same reduced 
state, namely the completely mixed state $\mathds{1}/2$, and because these are not eigenstates of any spin observable, 
no spin property
at all can be assigned to them. So there is a difference in global properties to which no 
difference in the local properties of the subsystems corresponds. 
  Therefore there is no supervenience and we have an instantiation of
holism\footnote{This is the exact argument \citet{maudlin98} uses. 
 \citet{healey91} and \citet{esfeld01} also use an 
 entangled spin example whereas \citet{teller86,teller89}, \citet{french89} and \citet{howard89} use 
 different entangled states or some consequence of entanglement such as
  violation of the bi-partite local Bell inequalities that are to be obeyed by local correlations.}.
  It is the feature of entanglement in this example that is 
  held responsible for holism. \citet{maudlin98} even defines
 holism in quantum mechanics in terms of entanglement and \citet[p.205]{esfeld01} puts it as 
  follows: ``The entanglement of two or more states is the basis for the 
  discussion on holism in quantum physics.'' Also \citet[p.11]{french89},
 although using a different approach to supervenience (see footnote \ref{voetnootfrech}), shares this view: 
``Since the state function [...] is not a product of the separate state functions 
of the particles, one cannot [...] ascribe to each particle an individual state 
function. It is {\em{this}}, of course, which reveals the peculiar non-classical
holism of quantum mechanics.''

      We       would now like to make an observation of a crucial aspect of the reasoning
 the supervenience approach uses to conclude that quantum mechanics
  endorses holism. In the above and also in other cases the issue is treated via 
  the concept of entanglement of quantum states.
  This, however, is a notion primarily tied to the structure of the state space of 
  quantum mechanics, i.e., the Hilbert space, and not to the structure of the 
  properties assigned in the interpretation in question. There is no one-to-one 
  correspondence between states and assigned dynamical properties,
   contrary to what we have already seen in the classical case. Thus questions in terms of
   states, such as `is the state entangled?'  and in terms of properties such 
   as `is there non-supervenience?' are different \ara{in principle}. And although 
   there is some connection via the property assignment rule using the 
   eigenvalue-eigenstate link, we claim them to be relevantly different.
Holism is a thesis about the structure of properties assigned to a whole and to its parts,
not a thesis about the state space of a theory. 
The supervenience approach should carefully ensure that it takes this into account.
However, the epistemological approach of the next section naturally takes this into account 
since it focuses directly on property determination. It probes the structure
 of the assigned properties and  not just that of the state space.

The reason that in the supervenience approach one immediately and solely looks 
at the structure of the state space is because in its supervenience basis only 
the properties the subsystems have in and out of themselves at the time in 
question are regarded. 
This means that using the eigenstate-eigenvalue rule 
for the dynamical properties one focuses on properties the system has 
in so far as the state of the system implies them. Only eigenstates
give rise to properties, other states do not.
A different approach, still in the orthodox interpretation, would be to focus
 on properties the system can possess according 
 to the possible property determinations quantum mechanics allows for. It is the 
 structure of the properties that can be possibly assigned at all, which is then at the heart of 
 our investigations. In this view one could say that
 the physical state of a system is regarded more generally, as also 
 \citet{howard89} does, as a set of \ara{dispositions} for the system to manifest 
certain properties under certain (measurement) circumstances, whereby the 
eigenstates are a special case assigning properties with certainty.
This view is the one underlying the epistemological approach 
which will be proposed and worked out next.

\section[An epistemological criterion for holism in physical theories]{An epistemological criterion for holism in\\ physical theories}
\label{criterion}

Before presenting the new criterion for holism we would like to motivate
it by going back to the spin-${\frac{1}{2}}$ example of the last section.
Let us consider the example, which according to the supervenience 
approach gives an instantiation of holism, from a different point of view.  Instead of 
 solely considering state descriptions, let us look at what physical 
 processes can actually be performed according to the theory in question 
 in order to gain knowledge of the system. We call this an 
 \ara{epistemological stance}. We will show next that it then \ara{is} possible to 
 determine, using only non-holistic means (to be specified later on) whether 
 or not one is dealing with the Bell state  $\left | \, \psi^{-} \right \rangle$ or $\left | \, \phi^{-} \right \rangle$ of  (\ref{entang}). How?
First measure on each subsystem the spin in the $z$-direction. 
Next, compare these results using classical communication.
 If the results have the same parity, the composite system was in the
  state $\left | \, \phi^{-} \right \rangle$  with global spin property $2\hbar^2$. And if the results do not have the same parity,
 the system was in the state  $\left | \, \psi^{-} \right \rangle$ with global spin property $0$.

 Thus using local measurements and classical communication the different global
 properties can be inferred after all.
   There is \ara{no} indication of holism in this approach, 
which is different from what the supervenience approach
  told us in the previous section.
 Although it remains true that the mixed reduced states of the individual 
 subsystems do not determine the composite state and neither 
 a local observable (of which there is no eigenstate), enough information 
 can be nevertheless gathered by local operations 
 and classical communication to infer the global property.
 We see that from an epistemological point of view we should not get 
 stuck on the fact that the subsystems themselves
  have no spin property because they are not in an eigenstate of a 
  spin observable. We can assign them a state, and thus 
  can perform measurements and assign them some local properties, which in this case do
   determine the global property in question.
 
From this example we see that this approach to holism  
does not merely look at the state space of a theory, but focuses on the structure of properties assigned to a whole and 
to its parts, as argued before that it should do. 
Then how do we spot \ara{candidates} 
for holism in this approach? Two elements are crucial.
Firstly, the theory must contain global properties that cannot 
be inferred from the local properties assigned to the subsystems,
 while, secondly, we must take into account 
 \ara{non-holistic constraints} on the  determination of these properties. 
 These constraints are that we only use the resource basis 
 available to local agents (who each have 
 access to one of the subsystems).  The guiding intuition is that 
 using this resource basis will provide us with only
 non-holistic features of the whole.  From this we finally get 
the following criterion for holism in a physical theory:
\begin{quote}
A physical theory is holistic 
if and only if it is impossible in principle, 
for a set of local agents each having access to a single subsystem only,
to infer the global properties of a system as assigned in the theory 
(which can be inferred by global measurements), 
by using the resource basis available to the agents.
\end{quote}
Crucial is the specification of the resource basis.
The idea is that these are all non-holistic resources for property 
determination available to an agent. 
    However, just as in the case of the specification of the supervenience basis, 
  this basis probably cannot be uniquely specified, i.e., the exact content of the 
  basis is open to debate. Here we propose that these resources include at least 
all \ara{local operations and classical communication} (abbreviated as LOCC)\footnote
{Note again that `\ara{local}' has here nothing to do with the issue of locality or 
spatial separation, but that it is taken to be opposed to global,
 i.e., restricted to a subsystem.}.
 The motivation for this is the intuition that local operations, i.e., anything we do on the separate subsystems, 
and classically communicating whatever we find out about it,
 will only provide us with  non-holistic properties of a composite system. 
However it could be possible to include other, although more debatable,
 non-holistic resources.
  A good example of such a debatable resource we have already seen: Namely,
 whether or not an agent can consider the position 
 of a subsystem as a property of the subsystem, so that he can 
 calculate relative distances when he knows the fixed positions of other subsystems.
 Another example is provided by the discussion of footnote \ref{angle}
 which suggests the question whether or not an agent can use a shared 
 Cartesian reference frame,  or a channel that transmits objects with well-defined
orientations, as a resource for determining the relative 
angle between directions at different points in space.

We believe that the determination of these and other spatial relations 
should be nevertheless included in the resource basis, for we take these 
relations to be (spatially) non-local, yet not holistic. Furthermore, because we are dealing with epistemology
 in specifying the resource basis, we do not think that 
including them necessarily implies ontological commitment as to which view one
 must endorse  about space or space-time.
Therefore, when discussing different physical theories in the next section, 
we will use as the content of the resource basis, firstly, the determination 
of spatial relations, and secondly  LOCC (local operations and classical communication).
The latter can usually be unproblematically formalized within 
physical theories and do not depend on, for example, the ontological view one has about 
spacetime. We thus propose to study the physical realizability 
of measuring or determining global properties while taking as a constraint
that one uses LOCC supplemented with the determination of spatial relations.

Let us mention some aspects of this proposed approach 
before it is applied in the next section.
\ara{Firstly}, it tries to formalize the question of holism in the 
context of what modern physical theories are, taking them 
 to be (i) schemes to find out and predict what the results are of
 certain interventions, which can be possibly used for determination of assigned properties,
  and (ii), although not relevant here, possibly describing 
  physical reality. Theories are no longer taken to necessarily present
  us with an ontological picture of the world specified by the
  properties of all things possessed at a given time.    
  
\ara{Secondly}, the approach treats the concept of \ara{property}
physically and not ontologically (or metaphysically). We mean by this that the concept is treated
analogous to  the way Einstein treated space and time (as that what is given by measuring rods
  and revolutions of clocks), namely as that which can be attributed to a system
  when measuring it, or as that which determines the outcomes of interventions.

\ara{Thirdly}, by including classical communication, this 
approach considers the possibility of determining some intrinsic relations among 
the parts such as the parity of a pair of bits, as was seen in the previous 
spin-${\frac{1}{2}}$ example. The parts are considered as parts, i.e., as constituting a whole
  with other parts and therefore being related to each other. 
   But the idea is that they are nevertheless considered non-holistically
    by using only the resource basis each agent has for determining 
    properties and relations of the parts. 

\ara{Fourthly}, as mentioned before,  the epistemological criterion for holism
 is relativized to the resource basis. Note that this is analogous to the 
 supervenience criterion which is relativized to the supervenience basis.
 We believe this relativizing to be unavoidable and even desirable because 
 it, reflects the ambiguity and debatable aspect inherent in any discussion 
 about holism. Yet, in this way it is incorporated in a fair and clear way. 

 \ara{Lastly}, note that the epistemological criterion is logically
independent of the supervenience criterion. Thus whether or not a theory is holistic in the supervenience approach
 is independent of whether or not it is holistic in the 
 newly proposed epistemological approach. This is the case because not all 
 intrinsic properties and relations in the supervenience basis are necessarily 
 accessible using the resource basis, and conversely, some that are accessible using the resource 
 basis may not be included in the supervenience 
 one\footnote{Of the latter case an example was given using the 
 spin-${\frac{1}{2}}$ example, since the property that specifies whether the singlet state or the 
triplet state obtains is not supervening, but can be inferred using only 
LOCC. Of the first case an example will be given in the next section.}.

\section[Holism in classical physics and quantum mechanics; revisited]{Holism in classical physics and quantum\\ mechanics; revisited}
\label{QMsection}
In this section we will apply the epistemological criterion for holism 
to different physical theories, where we use as the content of the resource basis 
the determination of spatial relations supplemented with LOCC.

\subsection{Classical physics and Bohmian mechanics }
\label{classbohm}

In section \ref{classical_sub} classical physics on a phase space was deemed non-holistic 
in the supervenience approach because global properties in this 
theory were argued to be supervening on subsystem properties.
Using the epistemological criterion we again find that 
classical physics is deemed non-holistic\footnote{Note that in both cases only systems with finite many subsystems are considered.}. 
The reason is that because of the one-to-one relationship between
properties and the state space and the fact that a Cartesian product is
 used for combining subsystem state spaces,
 and because measurement in classical physics 
is unproblematic as a property determining process, the resource basis 
determines all subsystem properties. We thus
are able to infer the Boolean $\sigma$-algebra of the properties
of the subsystems. Finally, given this the global 
properties can be inferred from the local ones  (see section \ref{classical_sub}) because the Boolean algebra structure 
of the global properties is determined by the Boolean algebra structures of the local ones,
 as was given in (\ref{algebras}). Hence no epistemological holism can be found.

Another interesting theory that also uses a state space with 
a Cartesian product to combine state spaces of subsystems
 is \ara{Bohmian mechanics} (see e.g. \cite{durr95}). 
 It is not a phase space but a configuration space.
This theory has an ontology of 
particles with well defined positions on 
trajectories\footnote{\ara{Bohmian mechanics}, which has as 
ontologically existing only particles with well defined positions on 
trajectories, should be distinguished (although this is perhaps 
not common practice) from  the so-called the \ara{de Broglie-Bohm theory} 
 where besides particles also the wave function has ontological existence as a guiding field. 
 This contrasts with Bohmian mechanics since in this theory the wave function has 
 only nomological existence.  Whether or not de Broglie-Bohm theory is 
 holistic because of the different role assigned to the wave function 
   needs careful examination, which will here not be executed.
  }. Here we discuss the interpretation where this theory is supplemented with 
   a property assignment rule just as in 
classical physics (i.e., all functions on the state space correspond
 to possible properties that can all be measured). Indeed, pure physical 
 states of a system are given by single points $(\vec{q})$ 
 of the position variables $\vec{q}$  that together make up a configuration space.
  There is a one-to-one relationship between the set of properties a system has 
 and the state on the configuration space it is in, 
as was shown in section \ref{classical_sub}. 
    The dynamics is given by the 
possibly non-local quantum potential $U_{qm}(\vec{q})$ determined by
  the quantum mechanical state
  $\left | \, \psi \right \rangle$, supplemented with the ordinary classical potential 
 $V(\vec{q})$, such that the force on a particle is given by: 
 $\vec{F}:=\frac{d\vec{p}}{dt}=-\vec{\nabla}[V(\vec{q})+ U_{qm}(\vec{q})]$.
This theory can be considered to be a real mechanics, i.e., 
a Hamilton-Jacobi theory, although with a specific extra interaction term. 
This is the quantum potential in which 
 the wave function appears that has only nomological existence. (Although a
 Hamilton-Jacobi theory, it is not classical mechanics: the latter is a second order
  theory, whereas Bohmian mechanics is of first order, i.e.,
   velocity is not independent of position).
  
 In section \ref{classical_sub} all theories on a state space
 with a Cartesian product to combine subsystem state  spaces and
  using a property assignment rule just as in classical physics 
  were deemed  non-holistic by the supervenience approach and therefore we can conclude that 
 Bohmian mechanics is non-holistic in this approach.
 Perhaps not surprising, but the epistemological approach also 
deems this theory non-holistic. The reason why is the same as why classical physics 
as formulated on a phase space was argued above to be not holistic in this approach.

Because Bohmian mechanics and quantum mechanics in the orthodox interpretation
have the same empirical content, one might think that because the first is not holistic, neither is the latter.
However, this is not the case, as will be shown next. This illustrates the fact that an interpretation of a 
theory, in so far as a property assignment rule is to be given, 
is \ara{crucial} for the question of holism. A formalism on its own is not enough.

\subsection{Quantum operations and holism}
\label{qmholism}

In this section we will show that quantum mechanics in the orthodox 
interpretation is holistic using the epistemological criterion,
without using the feature of entanglement.
 In order to do this we need to specify  what the resource basis looks like
 in this theory. Thus we need to  formalize what a local operation is and what 
 is meant by classical communication  in the context of quantum mechanics. 
For the argument it is not necessary to deal
 with the determination of spatial relations and these will thus not be considered. 

  Let us first look  at a general quantum process $\mathcal{S}$ that takes 
  a state $\rho$ of a system  on a certain Hilbert space $\mathcal{H}_1$ to a different 
  state $\sigma$ on a possibly different Hilbert space $\mathcal{H}_2$, i.e., 
\begin{align}
\rho \rightarrow \sigma=\mathcal{S} (\rho) ,~~~~~~~~~~\rho \in \mathcal{H}_1,~~ \mathcal{S} (\rho)\in\mathcal{H}_2,
\end{align} 
where $\mathcal{S}: \mathcal{H}_1\rightarrow\mathcal{H}_2$ is a \ara{completely positive 
trace-nonincreasing map}. This is  an operator $\mathcal{S}$, positive and trace non-increasing,
 acting linearly on Hermitian matrices such that $\mathcal{S}\otimes\mathds{1}$ takes 
 states to states. These maps are  also called \ara{quantum operations}\footnote{See \citet{nielsen} for an introduction to the general formalism of
 quantum operations.}.  Any quantum process, such as for example unitary evolution or measurement, 
  can be represented by such a quantum operation.

We are now in the position to specify the class of LOCC operations that two parties $a$ and $b$ can perform. It is the 
class of local operations plus two-way classical communication. It
consists of compositions of elementary operations 
of the following two forms
\begin{align}
\mathcal{S}^a\otimes\mathds{1},~~~~~ \mathds{1}\otimes\mathcal{S}^{b},
\end{align}
with $\mathcal{S}^a$ and $\mathcal{S}^b$ arbitrary local quantum operations that can be performed by party $a$ and $b$ respectively.
 The class contains the identity and is closed under composition and taking tensor products.
As an example consider the case where  party $a$ performs a measurement and communicates 
her result $\alpha$ to party $b$, after which party $b$ performs his measurement. The state $\rho$ of the composite system held by party $a$ and $b$ will then be modified as follows:
\begin{align}
\mathcal{S}^{ab}(\rho)=(\mathds{1}\otimes\mathcal{S}^{a}_{\alpha})\circ(\mathcal{S}^b\otimes\mathds{1})(\rho).
\end{align} 
We see that party $b$ can \ara{condition} his measurement on the outcome that party $a$ obtained.
 This example can be extended to many such rounds in which party $a$ and party $b$ 
each perform certain local operations on their part of the system and condition their choices on what is communicated to them.

Suppose now that we have a physical quantity $\mathfrak{R}$ of a bi-partite system
with a corresponding operator $\hat{R}$ that has 
a set of nine eigenstates, $\left | \,\psi_1 \right \rangle$ to $\left | \,\psi_9 \right \rangle$, with eigenvalues $1$ to $9$.
The property assignment we consider is the following:
if the system is in an eigenstate $\left | \, \psi_{i} \right \rangle$
then it has the property that quantity $\mathfrak{R}$ has the fixed value $i$ 
(this is the eigenstate-eigenvalue link).
Suppose $\hat{R}$ works on $\mathcal{H}=\mathcal{H}_a\otimes\mathcal{H}_b$ (each three dimensions\footnote{We are thus considering two qutrits. This is the sole exception in this dissertation of not considering qubits.}) and
 has the following complete orthonormal set of \ara{non-entangled} eigenstates:
\begin{align}\label{productstates}
\left | \,\psi_1 \right \rangle &= \left | \, 1 \right \rangle\otimes\left | \, 1 \right \rangle,\nonumber\\
\left | \,\psi_{2,3} \right \rangle&= \left | \, 0 \right \rangle\otimes\left | \, 0\pm1 \right \rangle,\nonumber\\
\left | \,\psi_{4,5} \right \rangle&= \left | \, 2 \right \rangle\otimes\left | \, 1\pm2 \right \rangle,\nonumber\\
\left | \,\psi_{6,7} \right \rangle&= \left | \, 1\pm2 \right \rangle\otimes\left | \, 0 \right \rangle,\nonumber\\
\left | \,\psi_{8,9} \right \rangle&= \left | \, 0\pm1 \right \rangle\otimes\left | \, 2 \right \rangle,
\end{align}
with $\left | \, 0+1 \right \rangle=\frac{1}{\sqrt{2}}(\left | \, 0 \right \rangle+\left | \, 1 \right \rangle)$, etc.

We want to infer whether the composite system has the property that the value
of the observable $\mathfrak{R}$ is one of the numbers $1$ to $9$, using 
only LOCC operations performed by two parties  $a$ and $b$, who each have 
one of the individual subsystems.
Because the eigenstate-eigenvalue link
is the property assignment rule used here,
 this amounts to determining which eigenstate party $a$ and $b$
 have or project on during the LOCC measurement.
If party $a$ and $b$ project on eigenstate $\left | \, \psi_{i} \right \rangle$ 
then a quantum operation
\begin{align}\label{qmopera}
\mathcal{S}_i:\rho \rightarrow \frac{S_i(\rho)}{{\mathrm{Tr}}
 [S_i(\rho)]}
 \end{align}
 is associated to the measurement outcome $i$, with associated projection 
 operators $S_i=\left | \, i \right \rangle_a\left | \, i \right \rangle_b
 \left \langle \, \psi_i \right |$. This 
 is nothing but the well-known projection due to measurement 
(with additional renormalisation), but here written in the 
 language of local quantum operations\footnote{Instead of writing the projection operators as  
$S_i=\left | \, \psi_{i} \right \rangle \left \langle \, \psi_i \right |$, we write  $S_i=\left | \, i \right \rangle_a\left | \, i \right \rangle_b
\left \langle \, \psi_i \right |$ 
to show explicitly that \ara{only} local records are taken. Since the states $\left | \, i \right \rangle$ 
can be regarded eigenstates of some local observable, we
  can regard them to determine a local property using
   the property assignment rule in terms
    of the eigenvalue-eigenstate link of the orthodox interpretation.} .
   The state  $\left | \, i \right \rangle_a$ denotes 
 the classical record of the outcome of the measurement 
that party $a$ writes down, and similarly for $\left | \, i \right \rangle_b$.
These records can be considered to be \ara{local} properties of the subsystems 
held by party $a$ and $b$.

It follows from the theory of quantum operations \cite{nielsen} that 
 determining the \ara{global} property assignment given by $\hat{R}$ amounts to
implementing the quantum operation $\mathcal{S}(\rho)=\sum_i\mathcal{S}_i\rho\mathcal{S}_i^{\dagger}$, with $\mathcal{S}_i$ as in (\ref{qmopera}). 
Surprisingly, this cannot be done using LOCC, a result obtained by \citet{bennett99}.
For the complete proof see the original article by \citet{bennett99} or 
\citet{walgate02}\footnote{This result is a special case of the fact that 
some family of separable quantum operations (that all have a complete eigenbasis of separable states)
cannot be implemented by LOCC and von Neumann measurements. 
This is  proven by \citet{chen03}.}, but a sketch of it goes as follows. 
If $A$ or $B$ perform von Neumann measurements in any of 
 their operation and communication rounds then the distinguishability of the 
 states is spoiled. Spoiling occurs in \ara{any} local basis. The ensemble 
 of states as seen by $A$ or by $B$ alone is therefore non-orthogonal,
  although the composite states are in fact orthogonal.

  From this we see that a physical quantity, whose corresponding operator has only product 
eigenstates, gives a property assignment using the eigenvalue-eigenstate link that
is not measurable using LOCC. Furthermore, we see that the resource basis sketched before does not 
suffice in determining the global property assignment given by $\hat{R}$.
 Thus according to the epistemological criterion of the previous section  
quantum mechanics is holistic, although no entanglement is involved\footnote{\citet{groisman} have recently performed a more extended investigation of the non-classicality of separable (unentangled) quantum states.
They show many surprising non-classical aspects of sets of product states which includes amongst others  the product basis (\ref{productstates}).}.
 Examples of epistemological holism that do involve entanglement can of course be given. 
 For example, distinguishing the four (entangled) Bell states given by 
 $\left | \, \psi^{+} \right \rangle$, $\left | \, \psi^{-} \right \rangle$, $\left | \, \phi^{+} \right \rangle$ and $\left | \, \phi^{-} \right \rangle$
  (see (\ref{entang})) cannot be done by LOCC. 
  Thus entanglement is \ara{sufficient} to prove epistemological holism. However,
  this is hardly surprising. What is surprising is the fact that it 
  is \ara{not necessary}, i.e., that here a proof 
  of epistemological holism is given not involving entanglement.
  Furthermore because of the lack of entanglement in this example 
 it would not suffice for a proof of holism in the supervenience approach.
  Of course, it may well be that the resource basis used in this example is too limited,
 but we do not see other resources that may sensibly be included in this basis 
 so as to render this example epistemologically non-holistic.

\section{Discussion}
\label{conc}

We sketched an epistemological criterion for holism that determines,
once the resource basis has been specified, whether or not a physical
theory with a property assignment rule is holistic.
It was argued to be a suitable one for addressing holism in physical theories,
because it focuses on property determination
as specified by the physical theory in question 
(possibly equipped with a property assignment rule via an interpretation).
 We distinguished this criterion from the well-known supervenience criterion for
holism and showed them to be logically independent.
Furthermore, it was shown that both the epistemological and the supervenience
approaches require relativizing the criteria to respectively 
the resource basis and the supervenience basis. We argued that in general 
neither of these bases is determined by the state space of a physical theory. In other words, 
 holism is not a thesis about the state space a theory uses, 
 it is about the structure of properties and property 
 assignments to a whole and its parts that a theory or an interpretation 
 allows for.  And in investigating what it allows for 
 we need to try to formalize what we intuitively 
 think of as holistic and non-holistic.  
  Here, we hope to have given a satisfactory new epistemological
 formulation of this, that allows one to 
 go out into the world of physics and apply the new criterion to the theories or 
 interpretations one encounters.

In this chapter  we have only treated some specific physical theories.
It was shown that all theories on a state space using a Cartesian 
product to combine subsystem state spaces, 
 such as classical physics and Bohmian mechanics, are not holistic 
 in both the supervenience and epistemological approach.
  The reason for this is that the Boolean algebra structure of the global properties is 
determined by the Boolean algebra structures of the local ones.
 The orthodox interpretation of quantum mechanics, however, was found 
 to instantiate holism. This holds in both approaches,
 although on different grounds. For the supervenience approach it 
 is the feature of entanglement
 that leads to holism, whereas using only LOCC resources, one can have
epistemological holism in absence of any entanglement, i.e., 
when there is no holism according to the supervenience approach.
  
There are of course many open problems left.   What is it that we can 
single out to be the reason of the holism found?
The use of a Hilbert space with its feature of superposition? Perhaps, but not the kind of superposition
 that gives rise to entanglement, for we have argued that it is not entanglement
  that we should per se consider
 to be the paradigmatic example of holism.  Should we blame the property assignment rule which 
 the orthodox interpretation uses? We shall leave this an open problem.

The entangled Bell states $\left | \, \psi^{-} \right \rangle$ and $\left | \, \phi^{-} \right \rangle$  
of section \ref{quantum_sub} could,
 despite their entanglement, be distinguished after all using 
only LOCC, whereas this was not possible in the set of nine (non-entangled)
product states of (\ref{productstates}). These two quantum mechanical examples show us that 
we can do both more and less than quantum states at first seem to tell us. 
This is an insight gained from the new field of \ara{quantum information theory}.
 Its focus on what one \ara{can} or \ara{cannot do} do with quantum 
systems, although often from an engineering point of view, has produced a
 new and powerful way of dealing with questions in the foundations of quantum 
 mechanics that can lead to fundamental new insights or principles. 
 We hope the new criterion for holism in physical theories suggested 
 in this  chapter is an inspiring example of this.

%
%
%
%
%
\clearemptydoublepage
\thispagestyle{empty}
\part{Epilogue}

\clearemptydoublepage
\thispagestyle{empty}

\chapter{Summary 
and outlook}\label{sumout}

In this dissertation I have tried to understand different aspects of different kinds of correlations that can exist between the outcomes of measurement on subsystems of a larger system. Four different kinds of correlation have been investigated: local, partially-local, no-signaling and quantum mechanical.
Novel characteristics of these correlations  have been used to study how they are related and how they can be discerned. 
The main tool of this investigation has been the usage of Bell-type inequalities that give non-trivial bounds on the strength of the correlations. 
The study of quantum correlations has also prompted us to study the multi-partite qubit state space (i.e., for $N$ spin-$\frac{1}{2}$ particles) with respect to its entanglement and separability characteristics, and the differing strength of the correlations in separable and entangled qubit states.
\forget{Th
Furthermore what philosophical stance one can take to quantum correlations +  holism.
+monogamy + aspects of hidden-variable theories and how they relate to surface probs.}

Throughout this dissertation I have restricted myself to the case where only two dichotomous observables are measured on each subsystem.   
Comparing the different types of correlations for this case has provided us with many new results on the various strengths of the different types of correlation. Because of the generality of the investigation -- we have considered abstract general models, not some specific and particular ones -- these results have strong repercussions for different sorts of physical theories. I have commented on some of these repercussions, thereby obtaining foundational and philosophical results. These will be summarized below. 

Although each chapter ended with a summary and discussion of its own, I will nevertheless summarize each chapter individually in this final chapter. The reasons for doing so are twofold. Firstly, this allows a potential reader to get a good idea of what is obtained in this dissertation, and secondly, this allows me to provide the necessary background that is needed for a sound presentation of a number of open questions, conjectures and directions for future research that can be drawn from this dissertation. These will be presented throughout the summary in this final chapter, and, for clarity,  they will be denoted by the symbol $\rhd$.
\forget{OUDE FORMULERING
Because each chapter ended with a summary and discussion of its own, to avoid repetition we will not summarize each chapter at length nor will we discuss each chapter individually. Throughout the summary in  this final chapter we will present 
a number of open questions, conjectures and directions for future research that can be drawn from this dissertation.  For clarity, these will be denoted by the symbol $\rhd$.
}

 \subsubsection*{Part I}
After a historical and thematic introduction in {\bf chapter \ref{introduction_chapter}}, {\bf  chapter \ref{definitionchapter}} introduced the definitions of the different types of correlations, the notations used, as well as the mathematical methods that have been employed in the rest of the dissertation. 
 The different types of correlation have been identified with a set of positive and normalized joint probabilities for different outcomes of some experiment, conditional on the settings (observables) chosen, further restricted by (i) no-signaling, (ii) locality, (iii) partial locality and (iv) quantum mechanics respectively.  We have called such joint probabilities surface probabilities so as to distinguish them from subsurface probabilities that are further conditioned on some hidden variable. We have next employed a powerful geometrical interpretation of correlations, first introduced by \citet{pitowsky86},  where they are viewed as vectors in a certain real high dimensional space. This has enabled us to associate to each type of correlation
  a convex body in a high-dimensional probability space. Such a body is describable by bounding planes (halfplanes) called facets. These facets are identified by tight Bell-type inequalities.  All no-signaling, partially-local and local correlations are contained in some polytope that each has a finite number of vertices and bounding facets, but quantum correlations are constrained by an infinite set of bounding planes and is thus not a polytope. In later chapters we have studied some of the containment relationships that exist between these different correlation bodies, and this has provided us with many new results on how the different types of correlation are related.  Chapter 2 also paid special attention to quantum mechanics: we have introduced and discussed the distinction between entanglement and separability of quantum states.  After this introductory chapter we have presented our investigation and results in three parts, Part II, III and IV, which will be summarized below. 

 \subsubsection*{Part II}
Part II focused exclusively on bi-partite systems. In 
{\bf chapter \ref{chapter_CHSHclassical}}  it was investigated what assumptions suffice to derive the original CHSH inequality for the case of two parties and two dichotomous observables per party. We have reviewed the fact that the doctrine of Local Realism together with the assumption of free variables allows only local correlations and therefore obeys this inequality. 
However,  it  has been shown that one can relax all major physical assumptions and still derive the CHSH inequality.  Indeed, one can allow for explicit non-local setting and outcome dependence in the determination of the local outcomes of experiment as well as dependence of the hidden variables on the settings chosen (i.e.,  the observables are no longer free variables).   Therefore a larger class class of hidden-variable models than is commonly thought is ruled out by violations of the CHSH inequality.

This shows that the well-known conditions of Outcome Independence and Parameter Independence \cite{shimony}, that taken together imply the well-known conditions of Factorisability, can both be violated in deriving the inequality, i.e., they are not necessary for this inequality to obtain, but only sufficient. Therefore, we have no reason to expect either one of them to hold solely on the basis of the CHSH inequality.
Of course, when confronted with  experimental violations of the CHSH inequality one must still give up on at least one of the conditions of
Outcome Independence and Parameter Independence or that the settings are free variables. However, the crucial point is that giving up only one might not be sufficient. The CHSH inequality does not allow one to infer which of the conditions in fact holds.  Indeed, the results of this chapter have shown that even satisfaction of the inequality is not sufficient for claiming that either one holds. It could be that all conditions are violated in such a situation.

\forget{In the light of experimental violations of the CHSH inequality it has been argued that this shows that the question of whether it is Outcome Independence or Parameter Independence that should be abandoned is misplaced.  We have no reason to expect either one of them to hold solely on the basis of the CHSH inequality because they are not necessary conditions. Instead, we should look at the weakest set of assumptions that give this inequality and ask which one of these is to be abandoned.
}
\begin{quote}$\rhd$ 
 An open question remains what forms of non-local setting and outcome dependence at the subsurface (hidden-variable) level would be both necessary and sufficient for deriving the CHSH inequality. 
\end{quote}
\begin{quote}$\rhd$ How should we understand violations of the CHSH inequality?  This is a difficult question to answer, since no set of  
necessary and sufficient conditions is known such that a hidden-variable model obeys this inequality.  The list of options, available to us at present,  to answer this question is mentioned on page \pageref{listalternative}.  But this list is by no means definitive because no necessary and sufficient set of conditions has been found, and consequently we do not precisely know what a violation of the inequality amounts to. It is important to recognize this if we are to have a proper appreciation of the epistemological situation we are in when we attempt to glean metaphysical implications of the failure of the CHSH inequality.
\end{quote}

The Shimony conditions that give Factorisability have been shown to be non-unique by proving that the conjunction of Maudlin's conditions suffice too.  This has been first claimed by \citet{maudlin}, but since no proof has been offered in the literature, we have provided one ourselves. The Maudlin conditions need additional non-trivial assumptions, which are not needed by the Shimony ones, in order to be evaluated in quantum mechanics. The argument that one can equally well chose either set (Maudlin's or Shimony's) has therefore been argued to be false.

The non-local derivation of the CHSH inequality has been contrasted with the derivation of Leggett's inequality \cite{leggett} for Leggett-type models.  The discussion of Leggett's model has shown an interesting relationship between different conditions at different hidden-variable levels. It has been shown that in this model parameter dependence at the deeper hidden-variable level does not show up as parameter dependence at the higher hidden-variable level (where one integrates over a deeper level hidden variable), but only as outcome dependence, i.e., as a violation of Outcome Independence.  Conversely, for more general hidden-variable models it has been shown that violations of Outcome Independence can be a sign of a violation of deterministic Parameter Independence at a deeper hidden-variable level. This analysis  shows that which conditions are obeyed and which are not depends on the level of consideration. A conclusive picture therefore depends on which hidden-variable level is considered to be fundamental.
 \begin{quote}$\rhd$  
We leave as an interesting avenue for future research the investigation of the relationships between different assumptions at different hidden-variable levels in general hidden-variable models.
\end{quote}
We have  presented  analogies between different inferences that can be made on the level of surface and subsurface probabilities. The most interesting such an analogy is between, on the one hand, the subsurface inference that the condition of Parameter Independence and violation of Factorisability implies randomness at the hidden-variable level, and, on the other hand, the surface inference  that any deterministic no-signaling correlation must be local, as was recently proven by \citet{masanes06}. A corollary of this inference is that any deterministic hidden-variable theory that obeys no-signaling and gives non-local correlations must show randomness on the surface, i.e., the surface probabilities cannot be deterministic. Bohmian mechanics is a striking example of this. 
\begin{quote}$\rhd$  This result asks for a further investigation of the relationship between inferences and implications that exist at the levels of surface and subsurface probabilities.  
\end{quote}
In chapter \ref{definitionchapter} it was shown that the facets of the no-signaling polytope give non-trivial Bell-type inequalities for the marginal expectation values (e.g., $\av{A}^{B}\leq\av{A}^{B'}$, etc.), but not for the product expectation values such as $\av{AB}$. We have searched for non-trivial inequalities in terms of the latter too. These cannot be facets of the polytope, but they have been shown to be useful nevertheless.
We have first shown that an alleged no-signaling Bell-type inequality as proposed by \citet{roysingh} is in fact trivial. However, combining several such trivial inequalities has allowed us  to derive a new set of non-trivial no-signaling inequalities  in terms of bounds  on a linear sum of product and marginal expectation values.  In doing so we have had to go beyond the analysis used in deriving the CHSH inequality because this inequality is trivial for no-signaling correlations. 
\forget{
\begin{quote}$\rhd$   
Our inequality is not a facet of the no-signaling polytope.  It would be very interesting to find the full set of non-trivial facet inequalities of this polytope. This set would specify the necessary and sufficient conditions for a correlation to be reproducible using no-signaling correlations. Experimental investigations of this set would allow for tests whether nature is no-signaling.  
\end{quote}
}

 {\bf Chapter \ref{chapter_CHSHquantumorthogonal} and \ref{chapter_CHSHquantumtradeoff}} considered the CHSH inequality in quantum mechanics for the case of two qubits. 
 This inequality not only allows for discriminating quantum mechanics from local hidden-variable models, it also allows for discriminating separable from entangled states, i.e., the inequality is also a separability inequality. Chapter 4 has shown that significantly stronger bounds on the CHSH expression hold for separable states in the case of locally orthogonal observables. In the case of qubits such a choice amounts to choosing anti-commuting observables.   
  This was further strengthened using quadratic inequalities not of the CHSH form.    These new separability inequalities provide sharper tools for entanglement detection. 
We have shown that if they hold for all sets of locally orthogonal observables they are necessary and sufficient for
separability, so the violation of these separability inequalities is not only a sufficient but also
a necessary criterion for entanglement. They have been argued to improve upon other such criteria, and furthermore do not need 
a shared reference between the measurement apparata for each qubit because the orientation of the measurement basis has been shown to be irrelevant.
\begin{quote}$\rhd$  
An open problem is whether a \emph{finite} collection of orthogonal observables can be found for which the
satisfaction of these inequalities provides a necessary and sufficient condition for separability.
  For mixed states we have not been able to resolve this, but for pure states a set of six inequalities using only
three sets of orthogonal observables has been shown to be already necessary and sufficient for separability.
\end{quote}
These inequalities, however, have been shown not to be applicable to the original purpose of
testing local hidden-variable theories.  This has provided a more general example of
the fact first discovered by \citet{werner}, i.e.,  that some entangled
two-qubit states do allow a local realistic reconstruction for all correlations in a
standard Bell experiment. We have exhibited a `gap' between the
correlations that can be obtained by separable two-qubit quantum states and
those obtainable by local hidden-variable models. 
In fact, as is shown in chapter \ref{Npartsep_entanglement}, the gap between the correlations allowed for by local hidden-variable theories and those attainable  by separable qubit states increases exponentially with the number of particles. Therefore, local hidden-variable theories are able to give correlations for which quantum mechanics, in order to reproduce them in qubit states, needs recourse to entangled states; and even more and more so when the number of particles increases.
\begin{quote} 
$\rhd$  
This `non-classicality' of separable qubit states raises
interesting questions. 
 What is the relationship in quantum mechanics between the
independence notions derived from the principle of locality and from
a state being separable?  Are they fundamentally different and not equally strong requirements?
 Given the surprising results found here between separability of qubit states and local hidden-variable structures, the question what the classical correlations of quantum mechanics are, seems to still not be fully answered and open for new investigations.
\end{quote}
In chapter \ref{chapter_CHSHquantumtradeoff} we have relaxed the condition of anti-commutation (i.e., orthogonality) of the local observables and studied the bound on the CHSH inequality for the full spectrum of non-commuting observables, i.e., ranging from commuting to anti-commuting observables. Analytic expressions for the tight bounds for both entangled and separable qubit states have been provided. 

The results are complementary to the well-studied question what the maximum of the 
expectation value of the CHSH operator is when evaluated in a certain (entangled) state. Here the focus has not been on a certain given state, but instead on the observables chosen. Independent of the specific state of the system we have asked what the maximum of the expectation value of the CHSH operator 
is when using certain local observables. The answer found shows a diverging trade-off relation 
for the two classes of separable and non-separable two-qubit states. 
Apart from the purely theoretical interest of these bounds, they have been shown to have experimental relevance, namely that it is not necessary that one has exact knowledge about the observables one is implementing in the experimental procedure. Ordinary entanglement criteria do require such knowledge.

\begin{quote}$\rhd$ 
It would be interesting to look for similar bounds  that quantify what happens when the local observables range from commuting to anti-commuting in the case of non-linear separability inequalities, as well as  to extend this analysis to multi-qubit separability inequalities and entanglement criteria.
\end{quote}

 \subsubsection*{Part III}
In Part III we have extended our investigation from the bi-partite to the multi-partite case.  
{\bf Chapter \ref{Npartsep_entanglement}}  has been devoted to the investigation of multi-partite quantum correlations and of quantum entanglement and separability.
A classification of partial separability of multi-partite quantum states has been proposed that extends the one of \citet{duer2,duer22}.
The latter classification consists of  a hierarchy
of levels corresponding to the $k$-separable states for $k=1,
\ldots N$, and within each level different classes are
distinguished by specifying  under which partitions of the system
the state is $k$-separable or $k$-inseparable. We have argued that it is useful to extend this classification with one
more class at each level $k$, since the notion of $k$-separability with 
respect to a specific partition does not exhaust all partial separability properties.\forget{
This classification
consists of  a hierarchy of levels corresponding to the
$k$-separable states for $k=1, \ldots N$, and within each level
various classes are distinguished by specifying  under which
partitions of the system the state is separable or not.}\forget{  Here we have focused on partial separability of multi-partite quantum states 
by firstly distinguishing partial separability with respect to splits and partial separability \emph{simpliciter} and secondly by using these distinctions to classify the partial separability properties of multi-partite states in a hierarchical manner.  This has extended the classification proposed by \citet{duer2,duer22} by also including the classification with respect to partial separability \emph{simpliciter}.
} 
We have furthermore discussed the relationship of partial separability to multi-partite entanglement. This relation turned out
to be non-trivial and therefore the notions of a 
$k$-separable entangled state and a $m$-partite entangled
state has been distinguished. The interrelations of these kinds of
entanglement has turned out to be rather intricate.

 Next, we have discussed necessary conditions that distinguish all types of partial separability
  in the full hierarchic separability classification.
This has been done by generalizing the derivation of the two-qubit separability inequalities  of chapter \ref{chapter_CHSHquantumorthogonal} to the multi-qubit setting.  Violations of these inequalities provide, for all $N$-qubit states, strong criteria for
   the entire hierarchy of $k$-separable
entanglement, ranging from the levels $k$=1 (full or genuine
$N$-particle entanglement) to $k=N$ (full separability, no entanglement), as well as
for the specific classes within each level.

\begin{quote}$\rhd$
Although we believe we have resolved a large part of the  non-trivial entanglement and separability relations in multi-partite quantum systems,   
many open question remain unanswered.  Firstly, we have left completely untouched the relationship between the different separability levels and classes to the large variety of entanglement measures. 
 Secondly, it would be interesting to study how the different classes and levels are related under different kinds of quantum operations, in particular the so-called reversible LOCC operations. Thirdly, another avenue for future research is to find different stronger separability and entanglement criteria for partial separability and multi-partite entanglement than we have obtained here.  Lastly, it is worthwhile to investigate if the extension from qubits to qudits is possible. 
\end{quote}
The strength of these criteria has been demonstrated in two ways:
Firstly, by showing that they imply several other general
experimentally accessible entanglement criteria. 
We therefore believe these state-independent entanglement criteria to be the strongest experimentally accessible conditions for multi-qubit entanglement applicable to all multi-qubit states.  Secondly, the conditions have been compared to other state-specific
multi-qubit entanglement criteria both for their white noise robustness and for the number of
measurement settings required in their implementation. They performed well on both these issues.

{\bf Chapter \ref{chapter_monogamy}} has studied multi-partite correlations by investigating whether they can be shared to other parties. 
In case this is not possible the correlations exhibit monogamy constraints.   Here one focuses on subsets of the parties and whether their correlations can be extended to parties not in the original subsets. This can be done either directly in terms of joint probability distributions or in terms of relations between Bell-type inequalities that hold for different, but overlapping subsets of the parties involved.
Questions of monogamy and shareability were first studied for quantum entanglement and we have compared this to the question  of monogamy and shareability of correlations. It has been obtained that shareability of non-local (quantum) correlations implies shareability of entanglement (of mixed states), but not vice versa.

It has been proven that unrestricted general correlations can be shared to any number of parties (called $\infty$-shareable). In the case of no-signaling correlations it was already known that  such correlations can be  $\infty$-shareable iff the correlations are local. This has been shown to imply, firstly, that partially-local correlations are also $\infty$-shareable, since they are combinations of local and unrestricted correlations between subsets of the parties. Secondly, it  implies that both quantum and no-signaling correlations  that are non-local are not $\infty$-shareable and we have reviewed existing monogamy constraints for such correlations.  We have given an independent simpler proof of the monogamy relation of \citet{tonerverstraete}, and have  provided a different strengthening of this constraint than was already given by them.
  
For the case of two parties, the relationship between sharing non-local quantum correlations and sharing mixed entangled states has been investigated.  The Collins-Gisin Bell-type inequality, which has three dichotomous observables per party, has been found to indicate that non-local  quantum correlations can be shared and thus to indicate sharing of mixed state entanglement. The CHSH inequality does not indicate this. This shows that non-local correlations in a setup with two dichotomous observables per party cannot be shared, whereas this is possible in a setup with one observable per party more.

For no-signaling correlations the monogamy constraint of \citet{toner2} has been interpreted as a non-trivial bound on product expectation values attainable by three-partite no-signaling correlations.  We know of no other such discerning inequalities in terms of solely product expectation values 
for three or more parties. 

Lastly,  monogamy constraints of three-qubit bi-separable quantum correlations have been derived, which is a first example of investigating monogamy of quantum correlations using a three-partite Bell-type inequality.

\begin{quote}$\rhd$  
An avenue for new research is investigating the shareability and monogamy aspects of multi-partite quantum and no-signaling correlations. For $N>3$  this field is largely unexplored.  
\end{quote}
\begin{quote}$\rhd$ 
Another new direction for research is to further explore the relationships between shareability (monogamy) of entanglement  and  shareability (monogamy) of non-local quantum correlations.  
\end{quote}

{\bf Chapter \ref{chapter_svetlichny}} has been devoted to the task of discerning the different kinds of multi-partite correlations.\forget{by means of Bell-type inequalities.} 
New Bell-type inequalities have been constructed that discern partially-local from quantum mechanical and more general correlations. Also, the issue of discerning multi-partite no-signaling correlations has been discussed. 
  For the three-partite case \citet{svetlichny} derived a non-trivial Bell-type inequality for partially-local correlations. This inequality can thus  distinguish between full three-partite non-locality and two-partite non-locality in a three-partite system.
 We have shown that Svetlichny's inequalities generalize to the multi-partite case.
 
Quantum mechanics has been shown to violate these inequalities for some fully entangled multi-qubit states and these can thus be considered to be fully non-local.  In a recent four-particle experiment such a violation was observed, so full non-locality occurs in nature.
 However, any bi-separable state (i.e.,  which is not fully entangled) has correlations that are strictly weaker than those obtainable by partially-local hidden-variable models. Thus a `gap' has been shown to appear between the correlations
obtainable from bi-separable quantum states and those from partially-local hidden-variable models. It thus takes fully entangled states to retrieve all correlations obtainable by such a model. This is analogous to the results of chapter 4 and 6, where it was shown that one needs entangled states to give all the correlations that are producible by local hidden-variable models. 
 
\begin{quote}$\rhd$ 
  It is an open question whether all fully entangled states  imply full non-locality.  If they do, this cannot always be shown by violations of the multi-partite  Svetlichny inequalities, for we have shown  that fully entangled mixed states exist that do not violate any of them. 
\\\\$\rhd$ 
 It is  unknown if the Svetlichny inequalities give facets of the partially-local polytope. It would be interesting to obtain
 the full set of tight Svetlichny inequalities for $N$ parties, although this might be a computationally hard problem. 
 \\\\$\rhd$  
We have also observed that although the Svetlichny inequalities discern partially-local and quantum correlations from the most general correlations, they cannot do so for no-signaling correlations. Providing discriminating conditions for multi-partite no-signaling correlations ($N>3$) in terms of product expectation values  (possibly including some marginal expectation values)  is left as an open problem.
                       \forget{The question whether interesting inequalities exist, expressed solely in terms of expectation values, would discern the multi-partite no-signaling correlations, is left as an interesting open problem.} 

 \end{quote}

 \subsubsection*{Part IV}
 Part IV dealt with some philosophical aspects of quantum correlations. 
  {\bf Chapter \ref{chapter_quantumworldcorrelations}} has as a starting point the fact the global state of a composite quantum system can be
completely determined by specifying correlations between outcomes of measurements performed on subsystems only.
Although quantum correlations suffice to reconstruct the quantum state this does not justify the idea that they are objective local properties of the composite system in question. Using a Bell-type inequality argument it has been shown that they cannot be given a local realistic explanation.
 Such a latter view has been defended by \citet{merminithaca1,merminithaca3,merminithaca2}, although he has by now set this idea aside.

As a corollary to this result we have argued that entanglement cannot be considered to be ontologically robust. Four conditions have been presented  that were argued to be necessary conditions for ontological robustness of entanglement and it has been shown that they are all four violated by entangled states.
\begin{quote}$\rhd$ A problem left open is the ontological status of entanglement. Does entanglement require some separate metaphysical treatment? This is a particular instance of the bigger question what the ontological status of the quantum state itself is.  This problem reappears in the next chapter (see below). 
\end{quote}

In {\bf chapter \ref{chapter_holism}} we have considered two related questions that frequently come up when discussing entanglement in quantum mechanics, namely whether it forces this theory to be holistic, and whether the correlations entangled quantum states give rise to are holistic. In order to address these questions we have considered the idea of holism and have given two ways one might think of it in physics. The first is well-known and uses the supervenience approach to holism developed by \citet{teller86,teller89} and \citet{healey91}, the second has been proposed here and it uses an epistemological criterion to determine whether a theory is holistic, namely: a physical theory 
is holistic if and only if it is impossible in principle to infer some global 
properties (as assigned by the theory) by local resources available to an agent.  
We have proposed that these resources include at least all local operations and 
classical communication (LOCC). 

This approach has been contrasted with the supervenience approach, the latter having a greater emphasis on ontology.
Furthermore, it has been shown that both the epistemological and the supervenience
approaches require relativizing the criteria to respectively 
the resource basis and the supervenience basis. We have concluded that in general 
neither of these bases is determined by the state space of a physical theory. We have therefore argued that holism is not a thesis about the state space of a theory, but that it is a thesis about the structure of properties and property 
 assignments to a whole and its parts that a theory or an interpretation 
 allows for.  

In this chapter only some specific physical theories have been treated.
All theories on a state space using a Cartesian 
product to combine subsystem state spaces, 
 such as classical physics and Bohmian mechanics, have been shown not to be holistic 
 in both the supervenience and epistemological approach.
 The orthodox interpretation of quantum mechanics, however, has been found 
 to instantiate holism. This holds in both approaches,
 although on different grounds. For the supervenience approach it 
 is the feature of entanglement 
 that leads to holism, whereas using our epistemological criterion for holism we have shown  
 holism without using any entanglement. 

\begin{quote}$\rhd$ 
What is the non-classical part of quantum mechanics? This question has been asked before in chapter 4 and 6, but it reappears here. We have argued that entanglement is not required for the non-classical feature of holism to arise. Furthermore, what justification do we have for thinking  that a possible answer to this question can be read off from the quantum mechanical state space?  In assessing the metaphysical implications of quantum theory we propose that one should not focus solely on the state space but rather at the structure of properties and property assignments to a whole and its parts that this theory or an interpretation of it allows for. But to fully take this view home more work needs to be done.
\end{quote}



%
%
%
\newpage
\thispagestyle{empty}
\cleardoublepage

\pagestyle{thesisheadingsfullbackmatter}

\selectlanguage{american}
\cleardoublepage

\cleardoublepage
\markboth{\sc{Bibliography}}{\sc{Bibliography}}
\addcontentsline{toc}{chapter}{Bibliography}
\bibliographystyle{thesis}
{\small

  \addtocontents{toc}{\contentsline {chapter}{}{}}

%
%
%
%


\begin{thebibliography}{99}
%
%
\markboth{\textrm{}}{\textrm{}}


\bibitem[Ac\'in et al.(2001)Ac\'in, Bru\ss, Lewenstein and Sanpera]{acin}Ac\'in, A., Bru\ss, D., Lewenstein, M., \& Sanpera, A. (2001). Classification of mixed three-qubit states, Phys. Rev. Lett. {\bf 87}, 040401.
\bibitem[Ac\'in et al.(2002)Ac\'in, Scarani and Wolf]{acin2}Ac\'in, A., Scarani, V., \&  Wolf, M.M. (2002). Bell's Inequality and distillability in $N$-quantum-bit systems, Phys. Rev. A {\bf 66}, 042323.
\bibitem[Ac\'in et al.(2006a)Ac\'in, Gisin and Toner]{acingisintoner}Ac\'in, A., Gisin, N., \& Toner, B. (2006a). Grothendieck's constant and local models for noisy entangled quantum states, Phys. Rev. A {\bf 73}, 062105.
\bibitem[Ac\'in et al.(2006b)Ac\'in, Gisin and Masanes]{acin06}Ac\'in, A., Gisin, N., \& Masanes, Ll. (2006b). 
From Bell's theorem to secure quantum key distribution, Phys. Rev. Lett. {\bf 97}, 120405.
\bibitem[Ardehali(1992)]{ardehali92}Ardehali, M. (1992).  Bell inequalities with a magnitude of violation that grows exponentially with the number of particles,  Phys. Rev. A {\bf 46}, 5375. 
\bibitem[Arntzenius(1992)]{arntzenius}Arntzenius, F. (1992). Apparatus independence in proofs of non-locality, Found. Phys. Lett. {\bf 5}, 517.
\bibitem[Aspect et al.(1981)Aspect, Grangier and Roger]{aspect}Aspect, A., Grangier, P., \& Roger, G. (1981). Experimental test of realistic local theories via Bell's theorem. Phys. Rev. Lett. {\bf 47}, 460.
\bibitem[Audenaert and Plenio(2006)]{audenaert}Audenaert, K.M.R., \& Plenio, M.B. (2006). When are correlations quantum?--Verification and quantification of entanglement by simple measurements, New. J. Phys. {\bf 8}, 266.
%
%
\bibitem[Ballentine and Jarrett(1987)]{ballentinejarrett}Ballentine, L.E., \& Jarrett, J.P. (1987). Bell's theorem: Does quantum mechanics contradict relativity?, Am. J. Phys. {\bf 55}, 696.
\bibitem[Barrett(2007)]{barrett}Barrett, J. (2007). Information processing in generalized probabilistic theories, Phys. Rev. A {\bf 75}, 032304.
\bibitem[Barrett et al.(2005)Barrett, Linden, Massar, Pironio, Popescu and Roberts]{barrett05} Barrett, J., Linden, N., Massar, S., Pironio, S., Popescu, S., \& Roberts, D. (2005). Nonlocal correlations as an information theoretic resource, Phys. Rev. A {\bf 71}, 022101.
\bibitem[Bell(1964)]{bell64}Bell, J.S. (1964). On the Einstein-Podolski-Rosen Paradox, Physics {\bf 1}, 195.
Reprinted in \citep{bellspeakable}, chapter 1. 
\bibitem[Bell(1966)]{bell66}Bell, J.S. (1966). On the problem of hidden variables in quantum mechanics, Rev. Mod. Phys. {\bf 38}, 447. Reprinted in \citep{bellspeakable}, chapter 2.
\bibitem[Bell(1971)]{bell71}Bell, J.S. (1971). Introduction to the hidden-variable question, in \emph{Foundations
of Quantum Mechanics}  (pp. 171-181). Proceedings of the International School of Physics `Enrico Fermi', course IL.
New York: Academic. Reprinted in \citep{bellspeakable}, chapter 4.
\bibitem[Bell(1976)]{bell76}Bell, J.S. (1976). The theory of local beables, Epistemological Letters, vol. 9, March 1976. Reprinted in Dialectica {\bf  39}, 85 (1985) and in \citep{bellspeakable}, chapter 7.
\bibitem[Bell(1977)]{bell77}Bell, J.S. (1977). Free variables and local causality, Epistemological letters, February 1977. Reprinted in \citep{bellspeakable}, chapter 12.
\bibitem[Bell(1980)]{bell80}Bell, J.S. (1980). Atomic-cascade photons and quantum-mechanical nonlocality. Comments on Atomic and Molecular Physics {\bf 9}, 121.Reprinted in \citep{bellspeakable}, chapter 13.
\bibitem[Bell(1981)]{bell81}Bell, J.S. (1981). Bertlmann's socks and the nature of reality, Journal de Physique, Colloque C2 , suppl. au numero 3,  Tome 42, 41. Reprinted in \citep{bellspeakable}, chapter 16.
\bibitem[Bell(1982)]{bell82}Bell, J.S. (1982). On the impossible pilot wave. Found. Phys. {\bf 12}. 989. Reprinted in \citep{bellspeakable}, chapter 17.
\bibitem[Bell(1987)]{bellspeakable}Bell, J.S. (1987). \emph{Speakable and unspeakable
in quantum mechanics}. Cambridge: Cambridge University Press.
\bibitem[Bell(1990)]{bell90}Bell, J.S. (1990). La nouvelle cuisine. In A. Sarlemijn and P. Kroes (Eds.), \emph{Between Science and Technology} (pp.97-115). Elsevier (North-Holland).

\markboth{\textrm{Bibliography}}{\textrm{Bibliography}}

\bibitem[Belinskii and Klyshko(1993)]{belinskiiklyshko93}Belinskii, A.V., \& Klyshko, D.N. (1993). Interference of light and Bell's theorem, Usp. Fiz. Nauk {\bf 163-165}, 1 [Phys. Usp. {\bf 36}, 653].
\bibitem[Bennett and Brassard(1984)]{bb84}Bennett, C.H., Brassard, G. (1984). Quantum cryptography: public key distribution and coin tossing, Proc. of the IEEE Int. Conf. on Computers, Systems and Signal Processing, p. 175-179.
\bibitem[Bennett et al.(1999a)Bennett, DiVincenzo,  Fuchs, Mor, Rains, Shor, Smolin, and  Wootters]{bennett99}Bennett, C.H., DiVincenzo, D.P.,
   Fuchs, C.A., Mor, T., Rains, E., Shor, P.W., Smolin, J.A., \&  Wootters, W.K. (1999a). Quantum nonlocality without entanglement. Phys. Rev. A {\bf 59}, 1070.
\bibitem[Bennett et al.(1999b)Bennett, DiVincenzo, Mor, Shor, Smolin and Terhal]{bennett}Bennett, C.H., DiVincenzo, D.P., Mor, T., Shor, P.W., Smolin, J.A.,  Terhal, B.M. (1999b), Unextendible product bases and bound entanglement, Phys. Rev. Lett {\bf 82}, 5385.
\bibitem[Berkovitz(1998a)]{berkovitz1998a}Berkovitz, J. (1998a). Aspects of quantum non-locality I: superluminal signalling, action-at-a-distance,non-separability and holism. Studies in the History and Philosophy of Modern Physics {\bf 29}, 183.
\bibitem[Berkovitz(1998b)]{berkovitz1998b}Berkovitz, J. (1998b). Aspects of quantum non-locality II: superluminal causation and relativity. Studies in the History and Philosophy of Modern Physics {\bf 29}, 509.
\bibitem[Bohm(1952)]{bohm52}Bohm, D. (1952).  A suggested interpretation of the quantum theory in terms of ``hidden'' variables. I and II, Phys. Rev. {\bf 85}, 166 and 180.
\bibitem[Bohm and Hiley(1993)]{bohm}Bohm, D., Hiley, B.J. (1993). \emph{The Undivided Universe: An Ontological Interpretation of Quantum Theory}. London: Routledge.
\bibitem[Bourenanne et al.(2004)Bourennane, Eibl, Kurtsiefer, Gaertner, Weinfurter,  G\"uhne,  Hyllus,  Bru\ss,  Lewenstein and Sanpera]{setting}Bourennane, M., Eibl, M., Kurtsiefer, C., Gaertner, S., Weinfurter, H., G\"uhne, O., Hyllus, P., Bru\ss, D., Lewenstein, M., \& Sanpera, A. (2004). Experimental detection of multipartite entanglement using witness operators, Phys. Rev. Lett. {\bf 92}, 087902.
\bibitem[Bouwmeester et al.(1999)Bouwmeester, Pan, Daniell, Weinfurter and Zeilinger]{BOU1999}Bouwmeester, D., Pan, J.W., Daniell, M., Weinfurter, H., \& Zeilinger, A. (1999). Observation of three-photon Greenberger-Horne-Zeilinger entanglement, Phys. Rev. Lett. {\bf 82}, 1345.
\bibitem[Bouwmeester et al.(2000)Bouwmeester, Pan, Daniell, Weinfurter and Zeilinger]{BOU2000}Bouwmeester, D., Pan, J.W., Daniell, M., Weinfurter, H., \& Zeilinger, A. (2000).  Multi-particle entanglement. In D. Bouwmeester, A. Ekert and A. Zeilinger (Eds.), \emph{The Physics
of Quantum Information} (pp.197-209). Berlin: Springer.
 \bibitem[Braunstein  et al.(1992)Braunstein, Mann and Revzen]{braunstein}Braunstein, S.L., Mann, A., \& Revzen, M. (1992). Maximal violation of Bell inequalities for mixed states,  Phys. Rev. Lett. {\bf 68}, 3259.
\bibitem[Branciard et al.(2008)Branciard, Brunner, Gisin, Kurtsiefer, Lamas-Linares, Ling and Scarani]{branciard}Briancard, C., Brunner, N., Gisin, N., Kurtsiefer, C., Lamas-Linares, A., Ling, A., \& Scarani, V. (2008). A simple approach to test Leggett's model of nonlocal quantum correlations,  arXiv: 0801.2241.
 \bibitem[Brown(1991)]{brown} Brown, H.R. (1991). Nonlocality in Quantum Mechanics: part II, Arist. Soc. Sup.Vol. LXV, 141.
\bibitem[Brunner et al.(2006)Brunner, Scarani and Gisin]{brunner}Brunner, N., Scarani, V., \& Gisin, N. (2006). Bell-type inequalities for nonlocal resources, J. Math. Phys. {\bf 47}, 112101.
\bibitem[Brunner et al.(2008)Brunner, Gisin, Popescu and Scarani]{brunner0803}Brunner, N., Gisin, N., Popescu, S., \& Scarani, V. (2008). Simulation of partial entanglement with no-signaling resources, arXiv: 0803.2359. 
 \bibitem[Bru\ss~ et al.(2002)Bru\ss,  Cirac, Horodecki,  Hulpke,  Kraus,  Lewenstein and  Sanpera]{bruss02}Bru\ss, D., Cirac, J.I., Horodecki, P., Hulpke, F.,  Kraus, B.,  Lewenstein M., \& Sanpera, A. (2002). Reflections upon separability and distillability, J. Mod. Opt. {\bf 49}, 1399.
\bibitem[Bub(1997)]{bub}Bub, J. (1997). \emph{Interpreting the Quantum World}. Cambridge: Cambridge University press.
\bibitem[Butterfield(1989)]{butterfield1}Butterfield, J. (1989).  A space-time approach to the Bell Inequality. In J. Cushing and E. McMullin (Eds.), \emph{Philosophical
consequences of quantum theory: reflections on Bell's theorem} (pp. 114-153). Notre Dame:
University of Notre Dame Press.
\bibitem[Butterfield(1992)]{butterfield}Butterfield, J. (1992). Bell's theorem: What it takes, Brit. J. Phil. Sci. {\bf 43}, 41.
\bibitem[Butterfield(2007)]{butterfieldcommon}Butterfield, J. (2007). Stochastic Einstein Locality Revisited, Brit. J. Phil. Sci. {\bf 58}, 805.


 %
%
\bibitem[Cabello(1999)]{cabello}Cabello, A. (1999). Quantum correlations are not contained in the initial state, Phys. Rev. A {\bf 60}, 877.
\bibitem[Cabello et al.(2005)Cabello, Feito and  Lamas-Linares]{noise2}Cabello, A., Feito A., \& Lamas-Linares, A. (2005). Bell's inequalities with realistic noise for polarization-entangled photons, Phys. Rev. A {\bf 72}, 052112.
\bibitem[Caves et al.(2007)]{caves}Caves, C. M., Fuchs, C. A., and Schack, R. (2007). Subjective probability and quantum certainty, Stud. Hist. Phil. Mod. Phys., {\bf 38}, 255.
\bibitem[Cereceda(2002)]{CER} Cereceda, J.L. (2002). Three-particle entanglement versus three-particle nonlocality, Phys. Rev. A {\bf 66}, 024102.
\bibitem[Chen and Li(2003)]{chen03}Chen, P., \& Li, C. (2003). Distinguishing a 
set of full product bases needs only projective measurements and classical 
communication, arXiv: quant-ph/0311154.
\bibitem[Chen et al.(2006)Chen, Albeverio and Fei]{chenalbeveriofei}Chen, K., Albeverio, S., \& Fei, S-M. (2006). Two-setting Bell inequalities for many qubits, Phys. Rev. A {\bf 74}, 050101(R).
\bibitem[Chen(2006)]{chencritiq}Chen, Z. (2006). Variants of Bell inequalities, arXiv: quant-ph/0611126.
 \bibitem[Chen and Chen(2007)]{chen}Chen,~L.,~\&~Chen,~Y-X.~(2007). Multiqubit entanglement witness, Phys. Rev. A {\bf 76}, 022330.
 \bibitem[Clauser et al.(1969)Clauser,  Horne, Shimony and Holt]{chsh}Clauser, J.F., Horne, M.A., Shimony, A., \& Holt, R.A. (1969).
Proposed experiment to test local hidden-variable theories, Phys. Rev. Lett. {\bf 23}, 880.
\bibitem[Clauser and Horne(1974)]{ch}Clauser, J.F., \& Horne, M.A. (1974). Experimental consequences of objective local theories,
Phys. Rev. D. {\bf 10}, 526.
\bibitem[Cleland(1984)]{cleland84}Cleland, C.E. (1984). Space: an abstract system of non-supervenient relations.
Philosophical Studies, {\bf 46}, 19.
\bibitem[Clifton(1991)]{thesisclifton}Clifton, R.K. (1991). \emph{Nonlocality in quantum mechanics}, PhD thesis, Cambridge University.
\bibitem[Clifton et al.(1991)Clifton, Redhead and Butterfield]{clifton}Clifton, R.K., Redhead, M.L.G., \& Butterfield, J.N. (1991). Generalization of GHZ algebraic proof of nonlocality, Found. Phys. {\bf 21}, 149.
\bibitem[Clifton et al.(2003)Clifton, Bub and Halverson]{cbh}Clifton, R., Bub, J., Halverson, H. (2003). Characterizing quantum theory in terms of information-theoretic constraints, Found. Phys. {\bf 33}, 1561.
\bibitem[Coffman et al.(2000)Coffman, Kundu and Wootters]{ckw} Coffman, V., Kundu, J., \& Wootters, W.K. (2000). Distributed entanglement, Phys. Rev. A {\bf 61}, 052306.
\bibitem[Collins et al.(2002)Collins, Gisin, Popescu, Roberts and Scarani]{collins} Collins, D.,  Gisin, N., Popescu, S., Roberts, D., \& Scarani, V. (2002). Bell-type inequalities to detect true n-body nonseparability, Phys. Rev. Lett. {\bf 88}, 170405.
\bibitem[Collins and Gisin(2004)]{collinsgisin}Collins, D., \&  Gisin, N. (2004). A relevant two qubit Bell inequality inequivalent to the CHSH inequality, J. Phys. A {\bf 37}, 1775.
%
%
\bibitem[Degorre et al.(2005)]{degorre}Degorre, J., Laplante, S., Roland, J. (2005). Simulating quantum correlations as a distributed sampling problem. Phys. Rev. A {\bf 72}, 062314.
\bibitem[Dewdney et al.(1987)Dewdney,  Holland and  Kyprianidis]{dewdney}Dewdney, C., Holland, P.R., and  Kyprianidis, A. (1987).  A causal account of non-local Einstein-Podolsky-Rosen spin correlations, J. Phys. A {\bf 20}, 4717.
\bibitem[Dicke(1954)]{dicke}Dicke, R.H. (1954). Coherence in Spontaneous Radiation Processes, Phys. Rev. {\bf 93}, 99.
\bibitem[Dickson(1998)]{dickson}Dickson, M. (1998). \emph{Quantum chance and non-locality}. Cambridge: Cambridge University Press.
\bibitem[Dieks(2002)]{dieks}Dieks, D. (2002). Inequalities that test locality in quantum mechanics,  Phys. Rev. A. {\bf 66}, 062104.
\bibitem[Deutsch(1985)]{deutsch}Deutsch, D. (1985). Quantum theory, the Church-Turing principle, and the universal quantum Turing machine, Proc. Royal Society London {\bf A400}, 97.
\bibitem[D\"ur(2001a)]{durR}D\"ur, W. (2001a). Multipartite entanglement that is robust against disposal of particles, Phys. Rev.  A {\bf 63}, 020303(R).
\bibitem[D\"ur(2001b)]{dur}D\"ur, W. (2001b). Multipartite bound entangled states that violate Bell's inequality, Phys. Rev. Lett. {\bf 87}, 230402.
\bibitem[D\"ur et al.(1999)D\"ur, Cirac and Tarrach]{duer}D\"ur, W.,  Cirac, J.I., \& Tarrach, R. (1999). Separability and distillability of multiparticle uantum systems, Phys. Rev. Lett. {\bf 83}, 3562.
\bibitem[D\"ur  and Cirac(2000)]{duer2}D\"ur, W., \& Cirac, J.I. (2000). Classification of multiqubit mixed states: Separability and distillability properties, Phys. Rev. A {\bf 61}, 042314. 
\bibitem[D\"ur et al.(2000)D\"ur, Vidal and Cirac]{W}D\"ur, W., Vidal, G., \& Cirac, J.I. (2000). Three qubits can be entangled in two inequivalent ways, Phys. Rev. A {\bf 62}, 062314.
\bibitem[D\"ur and Cirac(2001)]{duer22}D\"ur, W., \& Cirac, J.I. (2001). Multiparticle entanglement and its experimental detection, J. Phys. A  {\bf 34}, 6837.
\bibitem[D\"urr et al.(1996)D\"urr, Goldstein and Zangh\`i]{durr95}
D\"urr, D., Goldstein, S., \& Zangh\`i, N. (1996). Bohmian mechanics as the foundation 
of quantum mechanics. In J.T. Cushing, A. Fine and S. Goldstein (Eds.), \emph{
Bohmian Mechanics and Quantum Theory: 
An Appraisal}. (pp. 21-44). Dordrecht: Kluwer Academic.
%
%
\bibitem[Ekert(1991)]{ekert}Ekert, A.K. (1991). Quantum cryptography based on Bell's theorem. Phys. Rev. Lett. {\bf 67}, 661.
\bibitem[Eggeling and Werner(2001)]{eggeling}Eggeling, T., \& Werner, R.F. (2001). Separability properties of tripartite states with $U\otimes U \otimes U$ symmetry, Phys. Rev. A. {\bf 63}, 042111.
\bibitem[van Enk et al.(2007)van Enk, L\"utkenhaus and Kimble]{enk}van Enk, S.J., L\"utkenhaus, N., \& Kimble, H.J. (2007). Experimental procedures for entanglement verification, Phys. Rev. A {\bf 75}, 052318.
\bibitem[Einstein(1949)]{schilp}Einstein, A. (1949). `Autobiographical Notes' and `Reply to Critics'. In P.A. Schilp (Ed.), \emph{Albert Einstein, Philosopher Scientist}. Library of Living Philosophers. Illinois: Evanston.
\bibitem[Einstein(1971)]{einsteinborn}Einstein, A. (1971). \emph{The Born-Einstein Letters; Correspondence between Albert Einstein and Max 
and Hedwig Born from 1916 to 1955}. New York: Walker. Translation of Einstein, A. (1969). \emph{Briefwechsel 1916 -- 1955 von Albert Einstein und Hedwig und Max Born}. M\"unchen: Nymphenburger Verlagshandlung.
\bibitem[Einstein et al.(1935)Einstein, Podolsky and Rosen]{epr}Einstein, A., Podolsky, B., \& Rosen, N. (1935). Can quantum-mechanical description of  physical reality be considered complete?, Phys. Rev. {\bf 47}, 777. 
\bibitem[Esfeld(2001)]{esfeld01}Esfeld, M. (2001). \emph{Holism in the philosophy of mind and philosophy of physics}. Dordrecht: Kluwer.
%
%
\bibitem[Fahmi(2005)]{fahmi1}Fahmi, A. (2005). Non-locality and Classical Communication of the
Hidden Variable Theories, arXiv: quant-ph/0511009.
\bibitem[Fahmi and Goldshani(2003)]{fahmi}Fahmi, A., \& Golshani, M. (2003). Locality, Bell's inequality,
and the GHZ theorem, Phys. Lett A {\bf 306}, 259.
\bibitem[Fahmi and Goldshani(2006)]{fahmi2}Fahmi, A., \& Golshani, M. (2006).  Locality and the Greenberger-Horne-Zeilinger theorem, arXiv: quant-ph/0608049.
\bibitem[Fannes et al.(1988)Fannes, Lewis and Verbeure]{fannes}Fannes, M., Lewis, J.T., \& Verbeure, A. (1988). Symmetric states of composite systems, Lett. Math. Phys. {\bf 15}, 255-260.
\bibitem[Fine(1982)]{fine}Fine, A. (1982). Hidden variables, joint probability, and the Bell inequalities, Phys. Rev. Lett. {\bf 48}, 291.
\bibitem[French(1989)]{french89}French, S. (1989). Individuality, supervenience and Bell's theorem. Philosophical Studies, {\bf{55}}, 1.
\bibitem[van Fraassen(1985)]{fraassen82}van Fraassen, B.C. (1985). EPR: When is a correlation not a mystery? In P. Lahti and P. Mittelstaedt (Eds.), \emph{Symposium on the foundations of modern physics.} (pp. 113-128). Singapore:  World Sci. Publ.
\bibitem[Fuentes-Schuller and Mann(2005)]{fuentes}Fuentes-Schuller, I., \& Mann, R.B. (2005). Alice falls into a black hole: entanglement in noninertial frames, Phys. Rev. Lett {\bf 95}, 120404.
%
%
\bibitem[Greenberger et al.(1989)Greenberger, Horne and Zeilinger]{ghz}Greenberger, D.M., Horne, M.A., \& Zeilinger, A. (1989). Going beyond Bell's theorem. In  M. Kafatos (Ed.), {\emph{Bell's theorem,
quantum theory and conceptions of the universe}} (pp. 69-76). Dordrecht: Kluwer Academic.
\bibitem[Greenberger et al.(1990)Greenberger, Horne and Zeilinger]{GHZ}Greenberger, D.M., Horne,  M.A., \& Zeilinger, A. (1990). Bell's theorem without inequalities, Am.\ J.\ Phys.\ {\bf 58}, 1131.
\bibitem[Gill et al.(2002)Gill,  Weihs,  Zeilinger and \.Zukowski]{gill}Gill, R.D., Weihs, G., Zeilinger A., \& \.Zukowski, M. (2002).
No time loophole in Bell's theorem; the Hess-Philipp model is non-local, 
Proc. Natl. Acad. Sci. USA {\bf 99}, 14632.
\bibitem[Gisin(1991)]{GISIN91}Gisin, N. (1991). Bell's inequality holds for all non-product states, Phys. Lett. A {\bf 154},  201.
\bibitem[Gisin(2007)]{gisinbell}Gisin, N. (2007). Bell inequalities many questions, a few answers, arXiv: quant-ph/0702021.
\bibitem[Gisin and Peres(1992)]{GisiN}Gisin, N. \& Peres, A. (1992). Maximal violation of Bell's inequality for arbitrarily large spin, Phys. Lett. A {\bf 162}, 15. 
\bibitem[Gisin and Bechmann-Pasquinucci(1998)]{gisin}Gisin, N., \& Bechmann-Pasquinucci, H. (1998). Bell inequality, Bell states and maximally entangled states for n qubits, Phys. Lett. A {\bf 246}, 1.
\bibitem[Gottesman(1996)]{gottesman}Gottesman, D., (1996). Class of quantum error-correcting codes saturating the quantum Hamming bound, Phys. Rev. A {\bf 54}, 1862.
\bibitem[Grinbaum(2007)]{grinbaum}Grinbaum, A. (2007). Reconstruction of Quantum Theory, Brit. J. Phil. Sci. {\bf 58}, 387.
\bibitem[Groisman et al.(2007)Groisman, Kenigsberg and Mor]{groisman}Groisman, B., Kenigsberg, D., \& Mor, T. (2007), ''Quantumness'' versus ''Classicality'' of Quantum States, arXiv: quant-ph/0703103.
\bibitem[Grunhaus et al.(1996)Grunhaus, Popescu and Rohrlich]{grunhaus}Grunhaus, J., Popescu, S., Rohrlich, D. (1996). Jamming nonlocal correlations, Phys. Rev.A {\bf 53}, 3781.
\bibitem[G\"uhne et al.(2002)G\"uhne,  Hyllus,  Bru\ss,  Ekert,  Lewenstein,   Macchiavello, and Sanpera]{guhnewitness1}G\"uhne, O., Hyllus, P., Bru\ss, D., Ekert, A., Lewenstein, M.,  Macchiavello, C., \& Sanpera, A. (2002). Detection of entanglement with few local measurements, Phys. Rev. A. {\bf 66}, 062305. 
\bibitem[G\"uhne et al.(2003)G\"uhne,  Hyllus,  Bru\ss,  Ekert,  Lewenstein,   Macchiavello, and Sanpera]{guhnewitness}G\"uhne, O., Hyllus, P., Bru\ss, D., Ekert, A., Lewenstein, M.,  Macchiavello, C., \& Sanpera, A. (2003).  Experimental detection of entanglement via witness operators and local measurements, J. Mod. Opt. {\bf 50}, 1079. 
\bibitem[G\"uhne and Hyllus(2003)]{guhnehyllus}G\"uhne, O.,  \& Hyllus, P. (2003). Investigating three qubit entanglement with local measurements, Int. J. Theor. Phys.  {\bf 42}, 1001.
\bibitem[G\"uhne et al.(2005)G\"uhne, T\'oth and  Briegel]{guhnetothbriegel}G\"uhne, O., T\'oth, G., \& Briegel, H.J. (2005). Multipartite entanglement in spin chains, New J. Phys. {\bf 7}, 229.
\bibitem[G\"uhne et al.(2006)G\"uhne, Mechler, T\'oth and  Adam]{nonlinear}G\"uhne, O., Mechler, M., T\'oth, G., \& Adam, P. (2006). Entanglement criteria based on local uncertainty relations are strictly stronger than the computable cross norm criterion, Phys. Rev. A {\bf 74}, 010301(R).
 \bibitem[G\"uhne et al.(2007)G\"uhne,  Lu,  Gao and  Pan]{guhne2007}G\"uhne, O., Lu, C-Y., Gao, W-B., \& Pan, J-W. (2007).  Toolbox for entanglement detection and fidelity estimation, Phys. Rev. A {\bf 76}, 030305(R).
%
%
\bibitem[H\"affner et al.(2005)]{haffner}H\"affner, H., H\"ansel, W.,  Roos, C.F.,  Benhelm, J., Chek-al-kar, D., Chwalla, M., K\"arber, T., Rapol, U.D.,  Riebe, M., Schmidt, P.O., Becher, C., G\"uhne, O., D\"ur, W., \& Blat, R. (2005). Scalable multiparticle entanglement of trapped ions, Nature (London) {\bf 438}, 643.
\bibitem[Halmos(1988)]{halmos88}Halmos, P. R. (1988). \emph{Measure theory} (Fourth printing). Berlin: Springer-Verlag.
\bibitem[Halverson(2001)]{halverson}Halverson, H.P. (2001). \emph{Locality, Localization and the Particle Concept: Topics in the Foundations of Quantum Field Theory}, PhD thesis. University of Pittsburgh.
\bibitem[Hardy(1993)]{hardy}Hardy, L. (1993). Nonlocality for two particles without inequalities for almost all states, Phys. Rev. Lett. {\bf 71}, 1665.
\bibitem[Healey(1991)]{healey91}Healey, R. A. (1991). Holism and nonseparability, Journal of Philosophy {\bf 88}, 393-421.
\bibitem[Hess and Philipp(2001)]{hess}Hess, K. \& Philipp, W. (2001). Bell's theorem and the problem of 
decidability between the views of Einstein and Bohr, Proc. Natl. Acad. Sci. USA {\bf 98}, 14228.
\bibitem[Horodecki et al.(1995)Horodecki, Horodecki and Horodecki]{H3}Horodecki, R., Horodecki, P. \& Horodecki, M. (1995). Violating Bell inequality by mixed spin-1/2 states: necessary and sufficient condition, Phys. Lett. A {\bf 200}, 340.
\bibitem[Horodecki and Horodecki(1996a)]{horodecki210}Horodecki, R., \& Horodecki, P. (1996a). Perfect correlations in the Einstein-Podolsky-Rosen experiment and Bell's inequalities, Phys. Lett. A {\bf 210}, 227.
\bibitem[Horodecki and Horodecki(1996b)]{H2}Horodecki, R., \& Horodecki, M. (1996b). Information-theoretic aspects of inseparability of mixed states, Phys. Rev. A {\bf 54}, 1838.
\bibitem[Horodecki et al.(1996)Horodecki, Horodecki and Horodecki]{horodeckiPPT}Horodecki, M., Horodecki, P., \& Horodecki, R. (1996). Separability of mixed states: necessary and sufficient conditions, Phys. Lett. A {\bf 223}, 1.
\bibitem[Horodecki et al.(1998)Horodecki, Horodecki and Horodecki]{bound}Horodecki, M., Horodecki, P. \& Horodecki, R., (1998). Mixed-state entanglement and distillation: Is there a ``bound'' entanglement in nature?, Phys. Rev. Lett. {\bf 80}, 5239.
\bibitem[Horodecki et al.(2007)Horodecki, Horodecki, Horodecki and Horodecki]{entanglement}Horodecki, R., Horodecki, P., Horodecki, M., \& Horodecki, K. (2007). Quantum entanglement, arXiv: quant-ph/0702225.
\bibitem[Howard(1989)]{howard89}Howard, D. (1989). Holism, separability, and the metaphysical
 implications of the Bell inequalities. In J.T. Cushing and E. McMullin (Eds.), \emph{Philosophical consequences of quantum theory: reflections
  on Bell's theorem} (pp. 224-253).  Notre Dame: University of Notre Dame Press. 
  %
%
\bibitem[Isham(1995)]{isham95} Isham, C.J. (1995). \emph{Lectures on quantum theory. Mathematical and structural foundations}. 
London: Imperial College Press.
%
%
\bibitem[Jang et al.(2006)Jang, Cheong, Kim and Lee]{noise}Jang, S., Cheong, Y., Kim, J., \& Lee, H-W. (2006).  Robustness of multiparty nonlocality to local decoherence, Phys. Rev. A {\bf 74}, 062112.
\bibitem[Jarrett(1984)]{jarrett}Jarrett, J.P.  (1984). On the physical significance of the Locality Conditions in the
Bell Arguments, No\^us {\bf 18}, 569.
\bibitem[Jordan(1999)]{jordan}Jordan, T.F. (1999). Quantum correlations violate Einstein-Podolsky-Rosen assumptions, Phys. Rev. A {\bf 60}, 2726.
 \bibitem[Jones and Clifton(1993)]{jonesclifton}Jones, M.R., \& Clifton, R.K. (1993). Against experimental metaphysics, Midwest Studies
In Philosophy, XVIII, 295.
\bibitem[Jones and Masanes(2005)]{jonesmasanes}Jones, N.S., \& Masanes, Ll. (2005). Interconversion of nonlocal correlations, Phys. Rev. A {\bf 72}, 052312.
\bibitem[Jones et al.(2005)Jones, Linden and Massar]{jones}Jones, N.S, Linden, N., \& Massar, S. (2005). Extent of multiparticle quantum nonlocality,
Phys. Rev. A {\bf 71}, 042329.



%
%
\bibitem[Khalfin and Tsirelson(1985)]{khalfin}Khalfin, L.A., \& Tsirelson, B.S. (1985). Quantum and quasi-classical analogs of Bell inequalities, 
in  P. Lahti  and P. Mittelstaedt (Eds.), \emph{Symposium on the Foundations of Modern Physics 1985} (pp. 441-460). Singapore:  World Sci. Publ.
\bibitem[Kronz(1990)]{kronz}Kronz, F.M. (1990). Hidden locality, conspiracy and superluminal signals. Phil. Sci. {\bf 57}, 420.
\bibitem[Kochen and Specker(1967)]{kochenspecker}Kochen, S., \& Specker, E.P. (1967). The problem of hidden variables
in quantum mechanics, J. Math. Mech. {\bf 17}, 59.
\bibitem[Ku\'s and \.Zyczkowski(2001)]{kus}Ku\'s, M., \& \.Zyczkowski, K. (2001). Geometry of entangled states, Phys. Rev. A {\bf 63}, 032307.
%
%
\bibitem[Landau(1987)]{landau}Landau, L.J. (1987). On the violation of Bell's inequality in quantum theory, Phys. Lett. A. {\bf 120}, 54.
\bibitem[Laskowski et al.(2004)Laskowski, Paterek,  \.Zukowski and Brukner]{laskowski}Laskowski, W., Paterek, T., \.Zukowski M., \& Brukner,  \v{C}. (2004). Tight multipartite Bell's inequalities involving many measurement settings, Phys. Rev. Lett. {\bf 93},  200401.
\bibitem[Laskowski and \.Zukowski(2005)]{laskowzukow}Laskowski, W., \& \.Zukowski, M. (2005). Detection of N-particle entanglement with generalized Bell inequalities, Phys. Rev. A. {\bf 72}, 062112.
\bibitem[Leggett(2003)]{leggett}Leggett, A.J. (2003). Nonlocal hidden-variable theories and quantum mechanics: an incompatibility theorem, Found. Phys. {\bf 33}, 1469.
\bibitem[Lewenstein et al.(2000)Lewenstein, Kraus,  Cirac and Horodecki]{lewenstein}Lewenstein, M., Kraus, B., Cirac, J.I., \& Horodecki, P. (2000). Optimization of entanglement witnesses, Phys. Rev. A {\bf 62},  052310.
\bibitem[Lewenstein et al.(2001)Lewenstein, Kraus,  Horedecki and  Cirac]{LEWENSTEIN}Lewenstein, M., Kraus,  B., Horedecki, P., \& Cirac, J.I. (2001). Characterization of separable states and entanglement witnesses, 
Phys. Rev. A {\bf 63}, 044304.
%
%
\bibitem[Marcovitch and Reznik(2007)]{marcovitch}Marcovitch, S., \& Reznik, B. (2007). Is Communication Complexity Physical?, arXiv: 0709.1602.
\bibitem[Masanes(2002)]{masanes02}Masanes, Ll. (2002). Tight Bell inequality for $d$-outcome measurement correlations, arXiv: quant-ph/0210073.
\bibitem[Masanes(2005)]{masanes05}Masanes, Ll. (2005). Extremal quantum correlations for N parties with two dichotomic observables per site, arXiv: quant-ph/0512100.
\bibitem[Masanes(2006)]{masanes}Masanes, Ll. (2006). Asymptotic violation of Bell inequalities and distillability, Phys. Rev. Lett. {\bf 97},  050503.
\bibitem[Masanes et al.(2006)Masanes, Ac\'in and Gisin]{masanes06}Masanes, Ll.,  Ac\'in, A., \& Gisin, N. (2006). General properties of nosignalling theories, Phys. Rev. A {\bf 73}, 012112.
\bibitem[Maudlin(1994)]{maudlin}Maudlin, T. (1994). \emph{Quantum non-locality and relativity},
Blackwell Publishers, Oxford.
\bibitem[Maudlin(1998)]{maudlin98}Maudlin, T. (1998). Part and whole in quantum mechanics.  
In  E. Castellani (Ed.), \emph{Interpreting bodies} (pp. 46-60).  Princeton: Princeton University Press. 
\bibitem[Mermin(1990)]{mermin}Mermin, N.D. (1990). Extreme quantum entanglement in a superposition of macroscopically distinct states, Phys. Rev. Lett. {\bf 65 }, 1838. 
\bibitem[Mermin(1998a)]{merminithaca1}Mermin, N.D. (1998a). What is quantum mechanics trying to tell us?, Am. J. Phys. {\bf 66}, 753.
\bibitem[Mermin(1998b)]{merminithaca3}Mermin, N.D. (1998b). The Ithaca interpretation of quantum mechanics, Pramana {\bf 51}, 549.
\bibitem[Mermin(1999)]{merminithaca2}Mermin, N.D. (1999). What do these correlations know about reality? Nonlocality and the absurd, Found. Phys. {\bf 29}, 571.
\bibitem[Mermin(2004)]{merminreply}Mermin, N.D. (2004). Reply to the Comment by K. hess and W. Philipp on ''Inclusion of time in the theorem of Bell'', Europhys. Lett {\bf 67}, 693.
\bibitem[Mintert(2006)]{mintert}Mintert, F. (2006).  Concurrence via entanglement witnesses, arXiv: quant-ph/0609024.

%
%
\bibitem[Nagata et al.(2002a)Nagata,  Koashi and  Imoto]{nagataPRL}Nagata, K.,  Koashi, M., \& Imoto, N. (2002a). Configuration of separability and tests for multipartite entanglement in Bell-type experiments, Phys. Rev. Lett. {\bf 89}, 260401.
\bibitem[Nagata et al.(2002b)Nagata,  Koashi and  Imoto]{nagata2002} Nagata, K.,  Koashi, M., \& Imoto, N. (2002b). Observables suitable for restricting the fidelity to multipartite maximally entangled states, Phys. Rev. A {\bf 65}, 042314.
\bibitem[Navascu\'es et al.(2007)Navascu\'es, Pironio and Ac\'in]{navascues}Navascu\'es, M., Pironio, S., \& Ac\'in, A. (2007). Bounding the set of quantum correlations, Phys. Rev. Lett. {\bf 98}, 010401.
\bibitem[von Neumann(1932)von Neumann's]{vonneumann}von Neumann, J. (1932). \emph{Mathematische grundlagen der Quanten-mechanik}, Berlin: Verlag Julius Springer. (English translation: Princeton: Princeton University Press (1955)).
\bibitem[Nielsen and Chuang(2000)]{nielsen}Nielsen M.A., \& Chuang, I.L. (2000). 
\emph{Quantum computation and quantum information}. Cambridge: Cambridge University Press. 
%
%
\bibitem[Osborne and Verstraete(2006)]{osborne}Osborne, T.J., \& Verstraete, F. (2006). General monogamy inequality for bipartite qubit entanglement, Phys. Rev. Lett. {\bf 96}, 220503.
\bibitem[Ota et al.(2007)Ota, Yoshida and Ohba]{ota}Ota, Y., Yoshida, M., \& Ohba, I. (2007). Decrease of entanglement by local operations in the D\"ur-Cirac method, arXiv: 0704.1375.
%
%
\bibitem[Pan et al.(2000)Pan,  Bouwmeester,  Daniell, Weinfurter and Zeilinger]{PAN}Pan, J-W., Bouwmeester, D., Daniell, M., Weinfurter, H., \& Zeilinger, A. (2000). Experimental test of quantum nonlocality in three-photon Greenberger-Horne-Zeilinger entanglement, Nature (London) {\bf 403}, 515
\bibitem[Pan et al.(2001)Pan, Daniell, Gasparoni, Weihs and  Zeilinger]{PAN2}Pan, J-W., Daniell, M., Gasparoni, S., Weihs, G.,  \& Zeilinger, A. (2001). Experimental demonstration of four-photon entanglement and high-fidelity teleportation, Phys. Rev. Lett. {\bf 86}, 4435.
\bibitem[Paterek(2007)]{paterek}Paterek, T. (2007). \emph{Quantum communication. Non-classical correlations and their applications}, PhD thesis. University of Gdansk.
\bibitem[Peres(1996)]{PPT}Peres, A. (1996).  Separability sriterion for density matrices,  Phys. Rev. Lett. {\bf 77}, 1413.
\bibitem[Pitowsky(1986)]{pitowsky86}Pitowsky, I. (1986). The range of quantum probability, J. Math. Phys. {\bf 27}, 1556.
\bibitem[Pitowsky(1989)]{pitowsky}Pitowsky, I. (1989). \emph{Quantum Probability--Quantum Logic}. Berlin:  Springer.
\bibitem[Pitowsky(1991)]{pitowsky91}Pitowsky, I. (1991). Correlation Polytopes. Their geometry and complexity, Math. Programming {\bf 50}, 395.
\bibitem[Pitowsky(2008)]{pitowsky08}Pitowsky, I. (2008). On the geometry of quantum correlations, arXiv: 0802.3632.
\bibitem[Popescu and Rohrlich(1992a)]{POPROHR}Popescu, S., \& Rohrlich, D. (1992a). Generic quantum nonlocality, Phys. Lett. A
{\bf 166}, 293.
\bibitem[Popescu and Rohrlich(1992b)]{popescu169}Popescu, S., \& Rohrlich, D. (1992b). Which states violate Bell's inequality maximally?,  Phys. Lett. A {\bf 169}, 411.
\bibitem[Popescu and Rohrlich(1994)]{prbox}Popescu, S., \& Rohrlich, D. (1994).  Quantum nonlocality as an axiom, Found. Phys. {\bf 24}, 379.
\bibitem[Popescu(1995)]{POPESCU}Popescu, S. (1995). Bell's inequalities and density matrices: revealing ``hidden'' nonlocality, Phys. Rev. Lett. {\bf 74}, 2619.
%
%

%
%
\bibitem[Raggio and Werner(1989)]{raggio}Raggio, G.A., \& Werner, R.F. (1989). Quantum statistical mechanics of general mean field theory, Helv. Phys. Acta {\bf 62}, 980-1003.
\bibitem[Rastall(1985)]{rastall}Rastall, P. (1985).  Locality, Bell's theorem and quantum mechanics, Found. Phys., {\bf 15}, 963.
\bibitem[Rauschenbeutel et al.(2000)]{RAUSCH}Rauschenbeutel, A., Nogues, G., Osnaghi, S., Bertet, P., Brune, M., Raimond, J., \&
Haroche, S. (2000). Step-by-step engineered multiparticle entanglement, Science {\bf 288}, 2024.
\bibitem[Redhead(1987)]{redhead}Redhead, M. (1987). \emph{Incompleteness, Nonlocality and Realism}. Oxford: Oxford University Press.
\bibitem[Roy(2005)]{roy}Roy, S.M. (2005). Multipartite separability inequalities exponentially stronger than local reality inequalities, Phys. Rev. Lett. {\bf 94}, 010402.
\bibitem[Roy and Singh(1989)]{roysingh}Roy, S.M., \& Singh, V. (1989). Hidden variable theories without non-local
signalling and their experimental tests, Phys. Lett. A {\bf 139}, 437.
\bibitem[Roy and Singh(1991)]{roysingh91}Roy, S.M., \&  Singh, V. (1991). Tests of signal locality and Einstein-Bell locality for multiparticle systems, Phys. Rev. Lett.  {\bf 67}, 2761. 

%
%
\bibitem[Sackett et al.(2000)]{sackett}Sackett, C.A., Kielpinski, D., King, B.E., Langer, C., Meyer, V., Myatt, C.J., 
Rowe, M., Turchette, Q.A., Itano, W.M., Wineland, D.J., \& Monroe, C. (2000). Experimental entanglement of four particles, Nature (London) {\bf 404}, 256.
\bibitem[Scarani and Gisin(2001)]{scarani}Scarani, V. \& Gisin, N. (2001).  Quantum communication between N partners and Bell's inequalities,  Phys. Rev. Lett. {\bf 87}, 117901.
\bibitem[Schr\"odinger(1935)]{schrodinger}Schr\"odinger, E. (1935). Die gegenw\"artige Situation in der Quantenmechanik. Naturwissenschaften {\bf 23}, 807-812, 823-828, 844-849.
\bibitem[Seevinck(2004)]{seevhol}Seevinck, M.P. (2004). Holism, Physical theories and quantum mechanics, Stud. Hist. Phil. Mod. Phys. {\bf 35B}, 693-712.
\bibitem[Seevinck(2006)]{seevcor}Seevinck, M.P. (2006). The quantum world is not built up from correlations, Found. Phys. {\bf 36}, 1573-1586.
\bibitem[Seevinck(2007a)]{seevmon}Seevinck, M.P. (2007a). Classification and monogamy of three-qubit biseparable Bell correlations, Phys. Rev. A {\bf 76}, 012106.
\bibitem[Seevinck(2007b)]{seevchen}Seevinck, M.P. (2007b). Separable quantum states do not have stronger correlations than local realism. A comment on quant-ph/0611126 by Z. Chen', arXiv: quant-ph/0701003.
\bibitem[Seevinck(2008a)]{seevnonlocal}Seevinck, M.P. (2008a). Deriving standard Bell inequalities from non-locality and its repercussions for the (im)possibility of doing experimental metaphysics. Draft.
\bibitem[Seevinck(2008b)]{seevdeep}Seevinck, M.P. (2008b). Completeness and deeper level hidden variables. Draft.
\bibitem[Seevinck and Uffink(2001)]{seevuff}Seevinck, M.P., \& Uffink, J. (2001).  Sufficient conditions for three-particle entanglement and their tests in recent experiments, Phys. Rev. A. {\bf 65}, 012107.
\bibitem[Seevinck and Svetlichny(2002)]{seevsvet}Seevinck, M.P., \& Svetlichny, G. (2002). Bell-type inequalities for partial separability in N-particle systems and quantum mechanical violations, Phys. Rev. Lett. {\bf 89}, 060401.
\bibitem[Seevinck and Uffink(2007)]{tradeoff}Seevinck, M.P., \& Uffink, J. (2007). Local commutativity versus Bell inequality violation for entangled states and versus non-violation for separable states, Phys. Rev. A {\bf 76}, 042105.
\bibitem[Seevinck and Uffink(2008)]{partsep}Seevinck, M.P., \& Uffink, J. (2008). Partial separability and entanglement criteria for multiqubit quantum states, Phys. Rev. A {\bf 78}, 032101. 
\bibitem[Shimony(1984)]{shimony84}Shimony, A., (1984). Controllable and uncontrollable nonlocality. In S. Kamef (Ed.), \emph{Proceedings of the International Symposium: Foundations of Quantum Mechanics in the light of New Technology} (pp.225-230). Tokyo: Physical Society of Japan.
\bibitem[Shimony(1986)]{shimony}Shimony, A. (1986). Events and processes in the quantum world. In  R. Penrose and C.J. Isham (Eds.), \emph{Quantum Concepts In Space
and Time} (pp. 182-203). Oxford: Oxford University Press. 
\bibitem[Shimony(1989)]{shimony1989}Shimony, A., (1989). Search for a worldview which can accomodate our knowledge of microphysics.  In J.T. Cushing and E. McMullin (Eds.), \emph{Philosophical consequences of quantum theory: reflections
  on Bell's theorem} (pp. 25-37). Notre Dame: University of Notre Dame Press. 
 \bibitem[Smolin(2001)]{smolin}Smolin, J.A. (2001). Four-party unlockable bound entangled state,  Phys. Rev. A {\bf 63}, 032306.
 \bibitem[Socolovsky(2003)]{socolovsky}Socolovsky, M. (2003). Bell inequality, nonlocality and analyticity, Phys. Lett A {\bf 316}, 10.
\bibitem[Spekkens(2004)]{spekkens}Spekkens, R. (2004). In defense of the epistemic view of quantum states: a toy theory, arXiv: quant-ph/0401052.
\bibitem[Stevenson et al.(2006)]{stevenson}Stevenson, R.M., Young, R.J., Atkinson, P., Cooper, K., Ritchie D.A., \& Shields, A.J. (2006). A semiconductor source of triggered entangled photon pairs, Nature (London) {\bf 439}, 179.
\bibitem[Sun and Fei(2006)]{sun}Sun, B-Z., \& Fei, S-M. (2006).  Bell inequalities classifying biseparable three-qubit states, Phys. Rev. A {\bf 74}, 032335.
\bibitem[Suppes and Zanotti(1976)]{suppes}Suppes, P., Zanotti, M. (1976). On the determinism of hidden variable theories with strict correlation and conditional statistical independence of observables. In P. Suppes (Ed.), \emph{Logic and Probability in Quantum Mechanics} (pp. 445-455). Dordrecht: D. Reidel Publishing Company.
\bibitem[Svetlichny(1987)]{svetlichny}Svetlichny, G. (1987). Distinguishing three-body from two-body nonseparability by a Bell-type inequality,  Phys. Rev. D {\bf 35}, 3066.
%
%
\bibitem[Teller(1986)]{teller86}Teller, P. (1986). Relational holism and quantum mechanics. 
Brit. J. Phil. Sci. {\bf 37}, 71.
\bibitem[Teller(1987)]{teller87}Teller, P. (1987). Space-time as a physical quantity.
In R. Kargon and P. Achinstein (Eds.), \emph{Kelvin's Baltimore Lectures and Modern 
Theoretical Physics} (pp. 425-448). Cambridge: MIT press. 
\bibitem[Teller(1989)]{teller89}Teller, P. (1989). Relativity, relational holism,
 and the Bell inequalities. In J.T. Cushing and E. McMullin (Eds.), \emph{Philosophical consequences of quantum theory: reflections
  on Bell's theorem} (pp. 208-223). Notre Dame: University of Notre Dame Press. 
  \bibitem[Terhal(1996)]{terhal96}Terhal, B.M. (1996). Bell inequalities and the separability criterion, Phys. Lett. A {\bf 271}, 319.
  \bibitem[Terhal(2002)]{terhalComputSci}Terhal, B.M. (2002).  Detecting quantum entanglement,  Theoret. Comput. Sci. {\bf 287}, 313.
  \bibitem[Terhal(2004)]{terhal}Terhal, B.M. (2004). Is entanglement monogamous?, IBM J. Res. \& Dev. {\bf 48}, 71.
   \bibitem[Timpson and Brown(2002)]{timpson2}Timpson, C.G., \& Brown, H.R. (2002). Entanglement and relativity. In R. Lupacchini and V. Fano (Eds.), \emph{Understanding Physical Knowledge} (pp. 147-166). Bologna:  University of Bologna, CLUEB. 
  \bibitem[Timpson and Brown(2005)]{timpson}Timpson, C.G., \& Brown, H.R. (2005). Proper and improper separability, Int. J. Quant. Inf. {\bf 3}, 679.
\bibitem[Toner(2006)]{toner2}Toner, B.F. (2006). Monogamy of nonlocal correlations, arXiv: quant-ph/0601172.
  \bibitem[Toner and Bacon(2003)]{tonerbacon} Toner, B.F., \& Bacon, D. (2003). Bell inequalities with auxiliary communication, Phys. Rev. Lett {\bf 90}, 157904.
  \bibitem[Toner and Bacon(2006)]{toner} Toner, B.F., \& Bacon, D. (2006). Communication cost of simulating Bell correlations, Phys. Rev. Lett {\bf 91}, 187904 (2003).
\bibitem[Toner and Verstraete(2006)]{tonerverstraete}Toner, B.F., \& Verstraete, F. (2006). Monogamy of Bell correlations and Tsirelson's bound,  arXiv: quant-ph/0611001.
\bibitem[T\'oth et al.(2005)T\'oth, G\"uhne, Seevinck and Uffink]{tothseev}T\'oth, G., G\"uhne, O., Seevinck, M.P., \& Uffink, J. (2005). Addendum to ``Sufficient conditions for three-particle entanglement and their tests in recent experiments'', Phys. Rev. A {\bf 72}, 014101.
\bibitem[T\'oth and G\"uhne(2005a)]{tothguhne2}T\'oth, G., \& G\"uhne, O. (2005a). Entanglement detection in the stabilizer formalism, Phys. Rev. A {\bf 72}, 022340.
\bibitem[T\'oth and G\"uhne(2005b)]{tothguhne3}T\'oth, G., \& G\"uhne, O. (2005b). Detecting genuine multipartite entanglement with two local  measurements, Phys. Rev. Lett. {\bf 94}, 060501.
\bibitem[T\'oth and G\"uhne(2006)]{tothguhne1}T\'oth, G., \& G\"uhne, O. (2006). Detection of multipartite entanglement with two-body correlations, Appl. Phys. B. {\bf 82}, 237.
\bibitem[Tsirelson(1980)]{cirelson}Tsirelson, B.S. (1980). Quantum generalizations of Bell's inequality,  Lett. Math. Phys. {\bf 4}, 93.
\bibitem[Tsirelson(1993)]{tsirelsonhodronic}Tsirelson, B.S. (1993). Some results and problems on quantum Bell-type inequalities, Hadr. J. Suppl. {\bf 8}, 329.

%
%

\bibitem[Uffink(2002)]{uffink}Uffink, J. (2002). Quadratic Bell inequalities as tests for multipartite entanglement, Phys. Rev. Lett. {\bf 88}, 230406.
\bibitem[Uffink and Seevinck(2008)]{uffseev}Uffink, J., \& Seevinck, M.P. (2008). Strengthened Bell inequalities for orthogonal spin directions, Phys. Lett. A. {\bf 372}, 1205.

%
%
\bibitem[Volz et al.(2006)]{volz}Volz, J., Weber, M., Schlenk, D., Rosenfeld, W., Vrana, J., Saucke, K., Kurtsiefer, C.,  \& Weinfurter, H. (2006). Observation of entanglement of a single photon with a trapped atom, Phys. Rev. Lett.  {\bf 96}, 030404. 

%
%
\bibitem[Walck and Lyons(2008)]{walck}Walck, S., \&  Lyons, D. (2008). Only n-qubit Greenberger-Horne-Zeilinger states are undetermined by their reduced density matrices,  Phys. Rev. Lett. {\bf 100}, 050501.
\bibitem[Walgate and Hardy(2002)]{walgate02}Walgate, J., \& Hardy, L. (2002). Nonlocality, asymmetry, and distinguishing bipartite states, 
Phys. Rev. Lett. {\bf 89}, 147901.
\bibitem[Werner(1989)]{werner}Werner, R.F. (1989). Quantum states with Einstein-Podolsky-Rosen
correlations admitting a hidden-variable model, Phys. Rev. A {\bf 40}, 4277.
\bibitem[Werner and Wolf(2000)]{wernerwolf}Werner, R.F.,  \& Wolf, M.M. (2000). Bell's inequalities for states with positive partial transpose, Phys. Rev. A {\bf 61}, 062102.
\bibitem[Werner and Wolf(2001)]{wernerwolf2}Werner, R.F.,  \& Wolf, M.M. (2001). All-multipartite Bell-correlation inequalities for two dichotomic observables per site, Phys. Rev. A {\bf 64}, 032112.
\bibitem[Werner and Wolf(2003)]{werwolf}Werner, R.F.,  \& Wolf, M.M. (2003). Bell inequalities and entanglement, Quant. Inf. \& Comp. {\bf} 1, no. 3, 1.
\bibitem[Wootters(1990)]{wootters}Wootters, W.K. (1990). Local accessibility of quantum states. In W.K. Zurek (Ed.), {\emph{Complexity, Entropy and the Physics of Information}} (pp. 39-46). Boston: Addison-Wesley.

%
%

%
%
\bibitem[Yu et al.(2003)Yu, Chen, Pan and Zhang]{yu03}Yu, S., Chen, Z-B., Pan, J-W., \& Zhang, Y-D. (2003). Classifying N-qubit entanglement via Bell's inequalities, Phys. Rev. Lett. {\bf 90}, 080401.
\bibitem[Yu et al.(2003)Yu, Pan, Chen and Zhang] {hefei}Yu, S., Pan, J-W., Chen, Z-B, \& Zhang, Y-D. (2003). Comprehensive test of entanglement for two-level systems via the indeterminacy relationship,  Phys. Rev. Lett.  {\bf 91},  217903.
\bibitem[Yu and Liu(2005)]{yu}Yu, S., \& Liu, N-L. (2005). Entanglement detection by local orthogonal observables, Phys. Rev. Lett. {\bf 95}, 150504.
%
%
\bibitem[Zeng et al.(2003)Zeng, Zhou, Zhang, Xu and You]{fidelity}Zeng, B., Zhou, D.L., Zhang, P., Xu, Z., \& You, L. (2003).  Criterion for testing multiparticle negative-partial-transpose entanglement,  Phys. Rev. A {\bf 68}, 042316.
\bibitem[Zhang et al.(2007)Zhang, Zhang, Zhang and Guo]{zhang}Zhang, C-J., Zhang, Y-S., Zhang, S., \& Guo, G-C. (2007). Optimal entanglement witnesses based on local orthogonal observables, Phys. Rev A. {\bf 76}, 012334.
\bibitem[Zhao et al.(2003)Zhao, Yang, Chen, Zhang, \.Zukowski and Pan]{zhao}Zhao,  Z., Yang, T., Chen, Y-A., Zhang, A-N., \.Zukowski, M., \& Pan, J-W. (2003). Experimental violation
  of local realism by four-photon Greenberger-Horne-Zeilinger entanglement,
   Phys. Rev. Lett. {\bf 91}, 180401.
   \bibitem[Ziegler(1995)]{ziegler}Ziegler, G.M. (1995). \emph{Lectures on Polytopes}. New York: Springer-Verlag.
\bibitem[\.Zukowski(2006)]{zukow1964}\.Zukowski, M. (2006). Separability of quantum states vs. original Bell (1964) inequalities,  Found. Phys.  {\bf 36}, 541.
\bibitem[\.Zukowski et al.(1993)\.Zukowski, Zeilinger, Horne and Ekert]{swap}\.Zukowski, M., Zeilinger, A., Horne, M.A., \& Ekert, A.K. (1993).
`Event-ready-detectors' Bell experiment via entanglement swapping, Phys. Rev. Lett. {\bf 71}, 4287.
\bibitem[\.Zukowski and Brukner(2002)]{zukowskibrukner}\.Zukowski, M., \& Brukner, \v C. (2002). Bell's theorem for general N-qubit states, Phys. Rev. Lett. {\bf 88}, 210401.
\bibitem[\.Zukowski et al.(2002)\.Zukowski, Brukner, Laskowski and Wiesniak]{zukow2002}\.Zukowski, M.,  Brukner, \v C., Laskowski, W., \& Wiesniak, M. (2002). Do all pure entangled states violate Bell's inequalities for correlation functions?, Phys. Rev. Lett {\bf 88}, 210402.
\end{thebibliography}
}

\selectlanguage{dutch}
\cleardoublepage
\markboth{\sc{Samenvatting}}{}
\addcontentsline{toc}{chapter}{Samenvatting}
\clearemptydoublepage
\markboth{\textrm{}}{\textrm{}}
\chapter*{Samenvatting} 

\noindent
Dit proefschrift onderzoekt verschillende aspecten waarin een geheel kan zijn samengesteld uit delen. 
Het onderzoek omvat drie onderwerpen. Allereerst de studie naar mogelijke correlaties tussen meetuitkomsten uitgevoerd aan deelsystemen van een samengesteld systeem zoals deze door een specifieke fysische theorie worden voorspeld. Ten tweede, de studie naar wat deze theorie voorspelt voor de verschillende relaties die de deelsystemen kunnen hebben met het samengestelde systeem waarvan ze een deel uitmaken. En ten derde, de vergelijking van verschillende fysische theorie\"en met betrekking tot deze twee aspecten. De bestudeerde fysische theorie\"en zijn niet-relativistische quantummechanica en gegeneraliseerde waarschijnlijkheidstheorie\"en in een quasi-klassiek raamwerk.

De motivatie achter dit onderzoek is dat de wijze waarop een fysische theorie de relatie tussen delen en gehelen beschrijft, 
bij uitstek aangeeft wat deze theorie over de wereld beweert.  Op deze wijze wordt, onafhankelijk van specifieke 
modellen, de essenti\"ele fysische vooronderstellingen en structurele aspecten van de bestudeerde theorie blootgelegd. Dit vergroot enerzijds het inzicht    in de verschillende bestudeerde fysische theorie\"en, en anderzijds in wat zij over de wereld beweren.

Vier verschillende soorten correlaties zijn bestudeerd in dit proefschrift: \mbox{lokale}, partieel-lokale, niet-seinen en quantummechanische correlaties. De onderlinge vergelijking van de verschillende correlaties heeft nieuwe resultaten opgeleverd over de relatieve sterkten van de verschillende correlaties,  alsmede hoe  deze van elkaar te onderscheiden. Met name de structuur van quantummechanische toestanden bleek verrassend complex te zijn.

De algemeenheid van het onderzoek -- er werd gekeken naar abstracte algemene modellen -- heeft het mogelijk gemaakt sterke conclusies af te leiden voor de verschillende fysische theorie\"en als geheel. Deze conclusies zijn van grondslagen en filosofisch gewicht, met name met betrekking tot de haalbaarheid van verborgen-variabelen-theorie\"en voor de quantummechanica, de klassiek-quantum dichotomie en de vraag naar holisme in fysische theorie\"en.
\\\\

{\bf Deel I} introduceert dit proefschrift. Na een historische en thematische inleiding in {\bf hoofdstuk 1}, wordt in {\bf hoofdstuk 2} de definities van de verschillende typen correlatie gegeven, alsmede de gebruikte notatie en wiskundige methoden.  
%
%

{\bf Deel II} behandelt uitsluitend samengestelde systemen bestaande uit  twee deelsystemen. 
In {\bf hoofdstuk 3} wordt onderzocht welke aannames volstaan om de zogenoemde Clauser-Horne-Shimony-Holt  (CHSH) ongelijkheid \cite{chsh} af te leiden voor het geval van twee deeltjes en twee dichotome observabelen per deeltje. 

Allereerst wordt in herinnering gebracht dat een lokale verborgen-variabelen-theorie die 
uitgaat van zogenoemde `vrije variabelen' noodzakelijkerwijs slechts lokale correlaties kan voortbrengen en derhalve in alle gevallen aan de CHSH  ongelijkheid moet voldoen. Vervolgens wordt, ondanks dat deze ongelijkheid \'e\'en van de bekendste en meest bestudeerde Bell-ongelijkheden is, aangetoond dat ons \mbox{begrip} ervan verre van volledig is en dat een nader onderzoek van deze ongelijkheid aanleiding geeft tot interessante, onverwachte beschouwingen.

Zo blijkt dat alle fysische aannames die men gewoonlijk maakt om de CHSH \mbox{ongelijkheid} af te leiden, kunnen worden afgezwakt. Onder deze zwakkere aannames, die onder andere verschillende vormen van niet-lokaliteit toestaan, geldt de CHSH ongelijkheid nog steeds. 
Bijgevolg geeft een schending van deze ongelijkheid aan dat een grotere klasse van verborgen-variabelen-theorie\"en uitgesloten is dan men gewoonlijk aanneemt.

Dit alles heeft sterke repercussies voor de interpretatie van schendingen van de CHSH ongelijkheid. Het is tot op heden onduidelijk wat een dergelijke schending precies inhoudt omdat er nog geen geheel van noodzakelijke en voldoende aannames is gevonden waaronder deze ongelijkheid moet gelden. Enkel voldoende aannames zijn bekend. Er wordt betoogd dat dit gegeven erkend moet worden willen we een juiste waardering verkrijgen van de kentheoretische situatie waarin we ons bevinden wanneer we proberen metafysische implicaties af te leiden uit de schending van de CHSH ongelijkheid.

Dit hoofdstuk onderzoekt tevens de onderlinge relaties van verschillende aan-names die leiden tot een bepaalde vorm van `factorisatie', ook wel `lokale causaliteit' of `Bell-lokaliteit' genoemd. Naast een vergelijk van de welbekende aannames opgesteld door \citet{jarrett} en \citet{shimony} wordt er stil gestaan bij de  aannames gemaakt door \citet{maudlin}. Van deze laatste wordt een bewijs gegeven dat zij daadwerkelijk de gewenste factorisatie geven -- een bewijs dat in de literatuur niet te vinden was. De toepassing in de quantummechanica van Maudlin's aannames vereist echter bijkomende niet-triviale aannames. Toepassing van de Shimony's aannames behoeft daarentegen geen supplementaire aannames. Dit haalt de stelling onderuit dat men net zo goed de \'e\'en als de ander kan kiezen.

 \markboth{\textrm{Samenvatting}}{\textrm{Samenvatting}}
 
Een analyse van de zogenoemde Leggett-ongelijkheid \cite{leggett} heeft ons een geheel nieuw gezichtspunt opgeleverd: dat van de vraag naar een mogelijk dieper liggend niveau van verborgen-variabelen. Het blijkt dat de geldigheid, dan wel ongeldigheid, van verschillende aannames omtrent verborgen-variabelen afhangt van welk niveau men beschouwt.  Een definitief oordeel hangt dus af van welk verborgen-variabelen-niveau als fundamenteel mag worden beschouwd. 

Dit hoofdstuk verdiept zich tot slot in de eis van `niet-seinen' (de eis dat \mbox{lokale} empirisch toegankelijke waarschijnlijkheden geen afhankelijkheid vertonen van \mbox{veraf} gelegen, niet-lokale vrijheidsgraden). Met deze eis in het achterhoofd worden verschillende analogi\"en  opgespoord tussen gevolgtrekkingen die gelden op verschillende verborgen-variabelen-niveaus. Een interessant uitvloeisel hiervan is de conclusie dat elke deterministische verborgen-variabelen-theorie,  die voldoet aan de eis van niet-seinen en die niet-lokale voorspellingen wil doen, aan de empirische oppervlakte indeterministisch moet zijn. Anders gezegd, een deterministische theorie  waarmee niet geseind kan worden, moet in haar empirisch toetsbare voorspellingen noodgedwongen probabilistisch van aard zijn.  De Bohmse mechanica is een treffend voorbeeld hiervan. Niet-seinen correlaties kunnen echter niet onderscheiden worden van meer algemene correlaties door gebruik te maken van de CHSH ongelijkheid.  \mbox{Nieuwe} ongelijkheden worden afgeleid die dit wel mogelijk maken, en deze hebben een verrassende gelijkenis met de aloude CHSH ongelijkheid.

{\bf Hoofdstuk 4 en 5} behandelen de CHSH ongelijkheid in de quantummechanica voor het geval van twee quantum bits, ook wel `qubits' geheten (dit zijn quantummechanische systemen met als toestandsruimte een twee-dimensionale Hilbert\mbox{ruimte} $\mathbb{C}^2$; bijvoorbeeld spin-1/2 
deeltjes).
Deze ongelijkheid is niet alleen geschikt om quantummechanische correlaties van lokale correlaties te onderscheiden,  maar ook separabele van niet-separabele (verstrengelde) quantummechanische toestanden. Hoofdstuk 4 laat zien dat voor separabele toestanden er een aanzienlijk sterkere grens op de CHSH uitdrukking moet gelden  wanneer er onderling loodrechte observabelen gemeten worden (gerepresenteerd door anti-commuterende operatoren). Dit resultaat wordt vervolgens versterkt met behulp van kwadratische ongelijkheden die niet van de CHSH-vorm zijn. Deze nieuwe separabiliteitsongelijkheden geven scherpere criteria voor de detectie van verstrengeling dan andere bestaande criteria. 

De separabiliteitsongelijkheden zijn echter niet geschikt om het oorspronkelijke doel van Bell-ongelijkheden te realiseren. Met andere woorden, zij kunnen geen lokaal verborgen-variabelen-theorie\"en toetsen. Dit verschijnsel is een meer algemeen voorbeeld van het feit, allereerst ontdekt door \citet{werner}, dat sommige verstrengelde twee-qubit toestanden een lokaal verborgen-variabelen-model toestaan. Een `gat' is blootgelegd tussen correlaties verkregen middels separabele twee-qubit toestanden en middels lokaal verborgen-variabelen-modellen. In hoofdstuk 6 wordt aangetoond dat het gat  
exponentieel toeneemt met het aantal qubits. Derhalve is het zo dat lokaal verborgen-variabelen-theori\"en in staat zijn correlaties te geven waarvoor de quantummechanica, wil het deze correlaties reproduceren middels \mbox{qubits}, een beroep moet doen op verstrengelde toestanden; en bij een stijgend aantal deeltjes zal dit beroep groter en groter moeten zijn.  De resultaten laten zien dat de vraag wat nu precies klassieke- en quantummechanische correlaties zijn, nog steeds niet definitief beantwoord is en dus nog nader onderzoek  vergt.
 
In hoofdstuk 5 wordt de eis van lokale orthogonaliteit (anti-commutativiteit) van de te meten observabelen (operatoren) afgezwakt. De grens op de CHSH uitdrukking  wordt bepaald voor het gehele spectrum van niet-commuterende operatoren; van commuterend tot anti-commuterend. Deze grens wordt voor zowel verstrengelde als separabele toestanden analytisch bepaald. Het gevonden resultaat laat een divergerende `trade-off' relatie  zien voor de twee klassen van separabele en ver\mbox{streng}elde toestanden. Afgezien van de puur theoretische relevantie,  heeft deze relatie een sterk experimenteel voordeel. Ze geeft een algemeen geldend en sterk criterium voor verstrengeling waarbij het niet noodzakelijk is dat precies bekend is welke observabelen er gemeten worden.

In {\bf deel III} is het onderzoek uitgebreid naar systemen met meer dan twee deelsystemen, zeg $N$ in totaal. Zulke systemen zullen we `meer-deeltjes systemen' noemen, of ook wel `$N$-deeltjes systemen'.  {\bf Hoofdstuk 6} is gewijd aan zowel het onderzoeken van quantummechanische correlaties in meer-deeltjes systemen,  als aan het bestuderen van meer-deeltjes verstrengeling en separabiliteit. We beperken ons weer tot qubits. Dit hoofdstuk laat zien dat de classificatie van parti\"ele separabiliteit van quantummechanische toestanden zoals gegeven door  \citet{duer2,duer22} op een belangrijk punt moet worden uitgebreid. Deze uitbreiding heeft gevolgen voor ons begrip van de relatie tussen  parti\"ele-separabiliteit enerzijds en meer-deeltjes verstrengeling anderzijds.  Deze relatie blijkt uiterst niet-triviaal te zijn en derhalve moeten we de noties van een $k$-separabel-verstrengelde toestand en een $m$-deeltjes-verstrengelde toestand introduceren en onderscheiden. 

Om meer grip op de verschillende klassen van quantummechanische toestanden te verkrijgen besteedt dit hoofdstuk veel aandacht aan het verkrijgen van noodzakelijke condities die deze klassen van elkaar kunnen onderscheiden. Daartoe wordt de analyse van hoofdstuk 4 gegeneraliseerd  van twee naar een willekeurig aantal deeltjes. Schendingen van deze condities geven experimenteel toegankelijke criteria voor de gehele hi\"erarchie van $k$-separabele verstrengeling, van $k=1$ (volledige verstrengeling) tot $k=N$ (volledige separabiliteit, geen verstrengeling). De sterkte van deze criteria wordt tweevoudig aangetoond. Ten eerste impliceren en versterken ze verscheidene andere criteria voor verstrengeling. Ten tweede hebben de criteria experimenteel gunstige eigenschappen. Ze bezitten een sterke robuustheid voor verschillende vormen van ruis en het noodzakelijk aantal te meten observabelen, vereist bij experimentele implementatie, is beperkt.

{\bf Hoofdstuk 7} onderzoekt de correlaties in meer-deeltjes systemen op geheel andere wijze. Er wordt onderzocht of deze correlaties kunnen worden gedeeld met andere deeltjes. Dit duiden we aan met de term `deelbaarheid'. (Een correlatie tussen twee systemen is deelbaar dan en slechts dan als een derde systeem dezelfde correlatie met \'e\'en van de twee oorspronkelijke systemen kan aannemen, zonder dat de oorspronkelijke correlatie tussen de eerste twee systemen verloren gaat.) 
 Indien deelbaarheid niet mogelijk is spreekt men van `monogamie' of `beperkte promiscu\"iteit' van correlaties. 
De onderzoeksmethode is in dit geval als volgt: men bestudeert deelverzamelingen van deeltjes en onderzoekt of hun correlaties kunnen worden gedeeld met deeltjes die niet in de oorspronkelijke deelverzameling zitten.  Monogamie van verstrengeling alsmede haar deelbaarheid zijn al eerder onderzocht en hier worden enkele resultaten van dat onderzoek vergeleken met behaalde resultaten betreffende monogamie en deelbaarheid van correlaties.   Er blijkt onder andere dat wanneer niet-lokale correlaties kunnen worden gedeeld, dit impliceert dat verstrengeling kan worden gedeeld. Het omgekeerde is echter niet het geval.

Tevens wordt  aangetoond dat algemene, niet nader ingeperkte correlaties te delen zijn met een willekeurig aantal andere deeltjes (dit noemen we `$\infty$-deelbaarheid'). Eerder was ontdekt dat niet-seinen correlaties  $\infty$-deelbaar zijn dan en slechts dan als de correlaties lokaal zijn. Hieruit volgt dat zowel quantummechanische als niet-seinen correlaties die niet-lokaal zijn, niet $\infty$-deelbaar zijn. Deze correlaties vertonen dus beperkte promiscu\"iteit. Naast bovenstaande bevat dit hoofdstuk vele idee\"en die nog slechts marginaal verkend zijn, maar ook nog enkele harde resultaten. Noemenswaardig is het eenvoudiger bewijs voor, en de versterking van, een zeer interessante monogamie-relatie van \citet{tonerverstraete}. Daarnaast wordt  een eerste voorbeeld gegeven van een onderzoek naar monogamie-eigenschappen van drie-deeltjes correlaties middels een drie-deeltjes 
Bell-ongelijkheid. 

{\bf Hoofdstuk 8} is gewijd aan de taak hoe verschillende meer-deeltjes correlaties van elkaar te onderscheiden.  Bell-ongelijkheden worden gegeven die partieel-lokale correlaties van quantummechanische en van meer algemene correlaties onderscheiden. De drie-deeltjes ongelijkheid zoals voor het eerst gegeven door \citet{svetlichny} wordt gegeneraliseerd naar een willekeurig aantal deeltjes. Deze ongelijkheid onderscheidt volledig niet-lokale van partieel-lokale correlaties. De quantummechanica schendt deze ongelijkheid voor sommige volledig verstrengelde toestanden. De correlaties in deze toestanden zijn dus volledig niet-lokaal. 

{\bf Deel IV}  gaat over filosofische aspecten van quantummechanische correlaties. {\bf Hoofdstuk 9} heeft als startpunt het feit dat de globale quantum toestand van een samengesteld systeem volledig kan worden gespecificeerd door te refereren naar correlaties tussen uitkomsten van metingen aan uitsluitend 
deelsystemen. Alhoewel quantummechanische correlaties dus volstaan om de toestand te reconstrueren, wordt er betoogd dat deze correlaties niet opgevat kunnen worden als objectieve lokale eigenschappen van de deelsystemen in kwestie. Met behulp van een argument dat gebruik maakt van een Bell-ongelijkheid wordt er aangetoond dat zij geen lokaal-realistische verklaring kunnen verkrijgen.    Dit resultaat wordt ten slotte gebruikt om de vraag naar de ontologische status van verstrengeling te stellen.  De beantwoording van deze vraag geeft  aan dat verstrengeling vier (weliswaar aanvechtbare) noodzakelijke condities voor ontologische robuustheid blijkt te schenden.  We moeten concluderen dat de ontologische status van verstrengeling verre van duidelijk is.

{\bf Hoofdstuk 10} behandelt een veelgestelde vraag omtrent de aard van quantummechanische verstrengeling: in hoeverre maakt verstrengeling de quantummechanica holistisch? Of anders gezegd, in hoeverre zijn de correlaties verkregen middels verstrengelde toestanden holistisch (`holistisch' als zou het geheel meer dan de som der delen zijn)?  Om deze vragen zinnig te behandelen worden twee verschillende opvattingen over  holisme in fysische theorie\"en behandeld. De eerste manier staat bekend als de superveni\"entie-benadering en is uitgewerkt door \citet{teller86,teller89} en \citet{healey91}, de tweede manier wordt hier voor het eerst uiteengezet en gebruikt een epistemologisch criterium om te bepalen of een theorie holistisch is. 
Beide manieren worden met elkaar gecontrasteerd en er wordt beargumenteerd -- \emph{contra communis opinio} -- dat holisme niet een these over de toestandsruimte van een theorie is, maar over de structuur van de eigenschapstoekenning aan een geheel en zijn delen zoals de theorie (of interpretatie) dat voorschrijft.
  
Toepassing van de analyse op verschillende fysische theorie\"en heeft de volgende conclusies opgeleverd. Elke theorie die een Cartesisch product gebruikt om toestandsruimten van deelsystemen te combineren ten einde de toestandsruimte van het totale systeem te verkrijgen, is nimmer holistisch. Denk bijvoorbeeld aan de klassieke en Bohmse mechanica. De quantummechanica is echter holistisch. Dit is zo in beide benaderingen, maar vanwege verschillende oorzaken.  Voor de superveni\"entie-benadering is het de aanwezigheid van verstrengeling die maakt dat de quantummechanica holistisch is, maar bij gebruik van het epistemologisch criterium is 
 de quantummechanica als holistisch aan te duiden zonder enig gebruik van verstrengeling.  Een onverwacht resultaat.

{\bf Deel V} beslaat slechts \'e\'en afsluitend hoofdstuk. {\bf Hoofdstuk 11} bevat  een conclusie en discussie van de behaalde resultaten, alsmede een vooruitblik waarin open vragen en voorstellen voor toekomstig onderzoek worden gegeven.

\newpage
\thispagestyle{empty}
   
   \cleardoublepage
\markboth{\sc{Publications}}{}
\addcontentsline{toc}{chapter}{Publications}
{\small
\markboth{\textrm{}}{\textrm{}}
\chapter*{Publications}\label{listpubli}

\noindent
\forget{Allemaal? Als deeluitmakend van CV?\\
Of alleen de 10 van dit proefschrift?}
{\bf Articles this dissertation is largely based on:}\\
\begin{itemize}
\item Uffink, J., \& Seevinck, M.P. (2008). Strengthened Bell inequalities for orthogonal spin directions, \emph{Physics Letters A} {\bf 372}, 1205.
\item Seevinck, M.P., \& Uffink, J. (2008). Partial separability and entanglement criteria for multiqubit quantum states, \emph{Physical Review A} {\bf 78}, 032101.  Selected for the September 2008 issue of Virtual Journal of Quantum Information. Selected for the September 15, 2008 issue of Virtual Journal of Nanoscale Science \& Technology.
\item Seevinck, M.P. (2008). Deriving standard Bell inequalities from non-locality and its repercussions for the (im)possibility of doing experimental metaphysics, \emph{Submitted}.
\item Seevinck, M.P., \& Uffink, J. (2007).  Local commutativity versus Bell inequality violation  for entangled states and versus non-violation for separable states, \emph{Physical Review A} {\bf 76}, 042105. Selected for the October 2007 issue of Virtual Journal of Quantum Information.
\item Seevinck, M.P. (2007). Classification and monogamy of three-qubit biseparable Bell correlations, \emph{Physical Review A} {\bf 76}, 012106.
Selected for the July 2007 issue of Virtual Journal of Quantum Information.
Selected for the July 23, 2007 issue of Virtual Journal of Nanoscale Science \& Technology.
\item Seevinck, M.P. (2007). Separable quantum states do not have stronger correlations than local realism. A comment on quant-ph/0611126 by Z. Chen, \emph{Available at arXiv}: quant-ph/0701003.
\item Seevinck, M.P. (2006). The quantum world is not built up from correlations, \emph{Foundations of Physics} {\bf 36}, 1573.
\item T\'oth, G., G\"uhne, O.,  Seevinck, M.P., \&  Uffink, J. (2005). Addendum to ``Sufficient conditions for three-particle entanglement and their tests in recent experiments'', \emph{Physical Review A} {\bf 72}, 014101. Selected for the July 2005 issue of Virtual Journal of Quantum Information. 
\item Seevinck, M.P. (2004). Holism, physical theories and quantum mechanics, \emph{Studies in the History and Philosophy of Modern Physics} {\bf 35B}, 693.
\item Seevinck, M.P., \& Svetlichny, G. (2002). Bell-type inequalities for partial separability in N-particle systems and quantum mechanical violations, \emph{Physical Review Letters} {\bf 89}, 060401.
\item Seevinck, M.P., \& Uffink, J. (2001). Sufficient conditions for three-particle entanglement and their tests in recent experiments, \emph{Physical Review A} {\bf 65}, 012107.
\end{itemize}\newpage
\noindent {\bf Articles not used in dissertation:}\\
\markboth{\textrm{Publications}}{\textrm{Publications}}
\begin{itemize}
\item Muller, F.A., \& Seevinck, M.P. (2008). Discerning Particles, \emph{Submitted}. 
\item Seevinck, M.P., \&  Larsson, J.-\AA. (2007). Comment on ``A local realist model for correlations of the singlet state'' (Eur. Phys. J. B 53:139-142, 2006), \emph{The European Physical Journal B} {\bf 58}, 51. 
\item Muller, F.A., \& Seevinck, M.P. (2007). Is standard quantum mechanics technologically inadequate?,  \emph{British Journal for the Philosophy of Science} {\bf 58}, 595. 
\item Seevinck, M.P. (2005). Belleterie van EPR, \emph{Nederlands Tijdschrift Voor Natuurkunde} {\bf 71}, 354. (in Dutch)
\end{itemize}
}

\addcontentsline{toc}{chapter}{Curriculum Vitae}
\cleardoublepage
\markboth{\sc{Curriculum Vitae}}{}
\markboth{\textrm{}}{\textrm{}}
\chapter*{Curriculum Vitae}
\noindent\begin{flushright}
\emph{Een groot dichter zijn en dan vallen.}\\ ---Nescio
\end{flushright}
\noindent\vskip1cm\noindent
The author\footnote{In 2006 he obtained a Ducati 900SS (1977).} was born in Pretoria, South Africa, on the 27$^{\textrm{th}}$ of February 1977.\\
High school: Rythovius College, Eersel, the Netherlands (1989-1995).\\
University education: After a preliminary start at the Technical University of Delft (1995), and another six months at  
Humboldt State University in California, USA (1996), the author decided in 1996 to study physics and philosophy at the University of Nijmegen.
He finally seemed to be heading in the right direction. Indeed, he soon 
enjoyed train rides taking him westwards to also study foundations of physics at Utrecht University; and in 2000 he continued his studies at the University of Oxford, UK, for two terms.  The author then stopped thinking for a while, but eventually regained himself. In 2002 he obtained the M.Sc degree in theoretical physics from the University of Nijmegen (\emph{cum laude}) and in foundations of physics from Utrecht University (\emph{cum laude}). 
A few months later he decided to accept a PhD position (AiO) at the Institute for History and Foundations of Science at Utrecht University.
 
%
%
%
%
\newpage
\thispagestyle{empty}
\cleardoublepage
            
\end{document}